\documentclass [12pt,notitlepage]{report}
\usepackage{amsmath,amssymb,cite}
\usepackage{appendix}
\usepackage{feynmp}
\usepackage{float}
\usepackage[vcentermath,enableskew]{youngtab}
\usepackage{tabularx}
\usepackage[english]{babel}
\usepackage{graphicx}
\usepackage{indentfirst}
\usepackage{epsfig}
\usepackage{epstopdf}
\usepackage[]{hyperref}
\usepackage[section]{placeins}
\usepackage[stable]{footmisc}
\usepackage{appendix}
\usepackage{tabularx}
\usepackage[english]{babel}
\usepackage{graphicx}
\usepackage{indentfirst}
\usepackage{epsfig}
\usepackage{slashed}
\usepackage{fancyhdr}
\numberwithin{equation}{section}

\fancypagestyle{plain}{\fancyhead{}
}

\begin{document}

\title{
%\hspace{15cm}{\small DIAS-05-06}\\
%Strings, AdS/CFT and Matrix Theory}%Everything is Matrix-Emergent And Related Topics}%Lectures on Matrix Field Theory II
Review of M(atrix)-Theory, Type IIB Matrix Model and Matrix String Theory\\
}
% Strings, Fields or  Matrices}

\author{Badis Ydri \footnote{ydri@stp.dias.ie.}, \\
Department of Physics,  Badji Mokhtar Annaba University,\\
Annaba, Algeria.\\
The Abdus Salam International Centre for Theoretical Physics,\\
Strada Costiera 11, I-34014, Trieste, Italy.
}

\maketitle

\begin{abstract}
A review of M-(atrix) theory (the BFFS matrix quantum mechanics), type IIB matrix model (the IKKT matrix model) and Matrix String Theory (the DVV matrix gauge theory) is presented.
%String theory provides one of the most deepest insights into quantum gravity. 
%Its single most central and profound result is the gauge/gravity duality, i.e. the emergence of gravity from gauge theory. The two examples of M(atrix)-theory and AdS/CFT correspondence are 
%of paramount importance to many fundamental problems including to the physics of black holes (in particular to the information loss paradox) and to the problem of reconciliation 
%of general relativity and quantum mechanics. 
\end{abstract}

\tableofcontents

\chapter{Introduction}
%\addcontentsline{toc}{chapter}{Introduction} 
%\section{Introducing the ten theories}
In string theory we are really dealing with a very complex conflation of several ideas and theories at once which we will attempt to list first as follows:
\begin{enumerate}
\item Gauge theory. The most famous prototype is Yang-Mills theory. The main symmetries playing a major role here besides local gauge invariance we mention supersymmetry and conformal invariance.
\item Theory of general relativity especially concerning black holes. 
\item Supergravity. Especially in $11$ and $10$ dimensions.
\item Bosonic string theory. It exists only in $26$ dimensions.

The fundamental objects in gauge theory are fields and particles. The fundamental objects in string theory are strings (open and close) while Dp-branes and the NS 5-branes are non-perturbative configurations in string theory. The Dp-branes are p-branes with Dirichlet boundary conditions. The p-branes are particles ($p=0$), strings ($p=1$), membranes ($p=2$), etc. They really play the role of electric and magnetic charges in ordinary physics.

The spectrum of string theory in Hilbert space is discrete with a mass gap and thus it can be mapped one-to-one with elementary particle-like states in the target space (spacetime).

\item Superstring theories. They exist in $10$ dimensions and they admit supergravity theories as low energy limits. They are: 
\begin{enumerate}
\item Type I is a theory of open strings with ${\cal N}=1$ supersymmetry. Only $SO(32)$ gauge charges are possible by the Green-Schwarz mechanism of anomaly cancellation. These charges are attached at the ends of the open strings by the Chan-Paton method. In this theory closed strings appear in the quantum theory as singlets under the gauge group.
\item Type II A is a theory of closed strings with ${\cal N}=2$ supersymmetry and where the two Majorana-Weyl spinors (or the corrsponding two conserved supercharges) are of opposite chirality. There is no allowed gauge group. 
\item Type II B  is a theory of closed strings with ${\cal N}=2$ supersymmetry and where the two Majorana-Weyl spinors (or the corrsponding two conserved supercharges) are of same chirality. There is no allowed gauge group.
\item Heterotic $SO(32)$. This is a theory of closed strings with ${\cal N}=1$ supersymmetry where the $SO(32)$ gauge charges are distributed on the closed strings.
\item Heterotic $E_8\times E_8$. The same as above except that the allowed gauge group by the Green-Schwarz mechanism is $E_8\times E_8$. 
\end{enumerate}
The local diffeomorphism symmetry or reparametrization invariance of the world sheet of the string plays a fundamental role. Only the above two local gauge groups $SO(32)$ and $E_8\times E_8$ are allowed in string theory by demanding the principle of anomaly cancellation. Roughly, we can think of open strings as gauge theories and closed strings as gravity theories. 

These superstrings are connected to each other via an intricate web of dualities which are generalizations of the electric-magnetic duality present in electromagnetism if magnetic monopoles exist.
\item M-theory. Mostly unknown. We only know for sure that it exists in $11$ dimensions and admits supergravity in $11$ dimension as a low energy limit. Objects of M-theory include the supergraviton, the M2-brane and the M5-brane. The $11-$dimensional supergravity contains membrane solutions.
\item Type IIB matrix model. This is the IKKT model which is the only known non-perturbative regularization of string theory.
\item M-(atrix) theory. In the context of string theory we mean by M-(atrix) theory the BFSS matrix quantum mechanics and its BMN pp-wave deformation.  The BFSS model can be obtained as 
\begin{enumerate}
\item the regularized and quantized $11-$dimensional supermembrane theory, or as 
\item the discrete light-cone quantization (DLCQ) of M-theory. 
\item More simply, it is the dimensional reduction of supersymmetric Yang-Mills theory in $10$ dimensions and as such it is a theory of D0-branes.
\item Or as the compactification of the IKKT matrix model on a circle. This in fact gives the finite temperature BFSS.
\end{enumerate}
Some authors have suggested that "M-(atrix) theory is perhaps even more powerful than string theory" \cite{Taylor:1999qk} in the sense that string theory gives only a first quantized theory in the target space whereas M-(atrix) theory gives possibly a second quantized theory.
\item Quantum gravity in two dimensions. The dynamical triangulation of quantum gravity in two dimensions and its matrix models are closely related to the IKKT ($D=0$) and BFSS ($D=1$) models. 
\item The gauge/gravity duality. The idea that every nonabelian gauge theory has a dual description as a quantum theory of gravity is not just an idea but it is in fact the most important idea which came out of string theory and this this idea is by now a theory in its own right. M(atrix) theory is one example but a more general scheme is given by the AdS/CFT correspondence.
\end{enumerate}

In the following, we will present three supersymmetric matrix theories in dimensions $0+0$ (type IIB Matrix Model), $1+0$ (M-(atrix) Theory), and $1+1$ (Matrix String Theory) which are of paramount importance to superstring theory and M-theory. 

These theories play a vital role in the emergence of geometry, gravity and cosmology, and in the description of gravitational instabilities such as the information loss problem and the black-hole/black-string transition. They are also featured prominently in the gauge/gravity duality which relates supersymmetric U(N) gauge theories in $p+1$ dimensions and type II string theories in $10$ dimensions around black p-brane solutions. These string models together with the techniques of random matrix theory, noncommutative geometry and lattice gauge theory provide a starting point for what we may call "computational string theory".

%\paragraph{Organization:}
This review is organized as follows. Chapter $2$ contains  a lightning introduction into string theory and some other related topics. Chapter $3$ deals mostly with the BFSS matrix quantum mechanics but also with the DVV matrix gauge theory and their physics. The two main applications which were discussed extensively are the black-hole/black-string (confinement-deconfinement) phase transition and the black hole information loss paradox and checks of the gauge/gravity duality. Chapter $4$ deals with the  IKKT matrix model. The two selected applications here are emergent matrix Yang-Mills cosmology from the Lorentzian IKKT matrix model and noncommutative/matrix emergent gravity from the Euclidean version. 

%\paragraph{About:} 
These notes are in some sense the second part of the LNP publication "Lectures on Matrix Field Theory" \cite{Ydri:2016dmy}.
%\paragraph{Acknowledgments:}

This research was supported by CNEPRU: "The National (Algerian) Commission for the Evaluation of University Research Projects"  under the contract number ${\rm DO} 11 20 13 00 09$. 

I woul like also to acknowledge the generous funding from the International Center for Theoretical Physics ICTP (Trieste) within the associate scheme $2015-2020$. 

I would like also to acknowledge generous funding and warm hospitality in previous years from the Dublin Institute for Advanced Studies DIAS (Dublin).

All illustrations found in this review are only sketches of original Monte Carlo and numerical results. They were created by Dr. Khaled Ramda and Dr. Ahlam Rouag.

%\paragraph{Disclaimer:}
Finally, I would like to apologize from the outset for any omission in references. It is certainly not my intention to cause any offence. The goal here is to simply cover a larger terrain of very interesting and difficult ideas in a pedagogical way. 

%\part{Part I: String Theory}
%%\input{old_bosonic_strings.tex}
%\input{bosonic.tex}
%\input{polyakov.tex}
%\input{conformal.tex}
%\input{the_five_superstrings.tex}
%\part{Part II: Matrix String Theory}

\chapter{A Lightning Introduction to String Theory and Some Related Topics}

\section{Quantum black holes}

String theory provides one of the most deepest insights into quantum gravity. 
Its single most central and profound result is the AdS/CFT correspondence or gauge/gravity duality 
\cite{Maldacena:1997re}. See \cite{Natsuume:2014sfa,Nastase:2007kj} for a pedagogical introduction. As it turns out, 
this duality allows us to study in novel ways: i) the physics of strongly coupled gauge theory (QCD in particular 
and the existence of Yang-Mills theories in $4$ dimensions), as well as ii) the physics of black holes 
(the information loss paradox and the problem of the reconciliation of general relativity and quantum mechanics). 
String theory reduces therefore for us to the study of the gauge/gravity duality and its most imporatnt example the AdS/CFT correspondence.

Indeed, the fundamental observation which drives the lectures in this chapter and \cite{ydri2017} is that: ``BFSS matrix model 
\cite{Banks:1996vh} and the AdS/CFT duality \cite{Maldacena:1997re,Gubser:1998bc,Witten:1998qj} relates string theory 
in certain backgrounds to quantum mechanical systems and quantum field theories'' which is a quotation taken 
from Polchinski \cite{Polchinski:2016hrw}. The basic problem which is of paramount interest to quantum gravity 
is Hawking radiation of a black hole and the consequent evaporation of the hole and corresponding information loss 
\cite{Hawking:1974sw,Hawking:1976ra}. The BFSS and the AdS/CFT imply that there is no information loss paradox in 
the Hawking radiation of a black hole. This is the central question we would like to understand in great detail.

Towards this end, we need to understand first quantum black holes, before we can even touch the gauge/gravity duality and the 
AdS/CFT correspondence, which require in any case a great deal of conformal field theory and string theory as crucial ingredients. Thus, 
in this section we will only worry about black hole radiation, black hole thermodynamics and the 
information problem following \cite{Polchinski:2016hrw,Susskind:2005js,Page:1993up,Harlow:2014yka,Jacobson:2003vx,Mukhanov:2007zz,
Carroll:2004st}. This section is a sort of a summary of the very detailed presentation \cite{ydri2017} which conatins an 
extensive list of references.

\subsection{Schwarzschild black hole}
We start by presenting the star of the show the so-called Schwarzschild eternel black hole given by the metric 
 \begin{eqnarray}
ds^2=-(1-\frac{2GM}{r})dt^2+(1-\frac{2GM}{r})^{-1}dr^2+r^2d\Omega^2.
\end{eqnarray}
Before we embark on the calculation of Hawking radiation it is very helpful to understand the physical origin behind this radiation in the 
most simple of terms.

\begin{figure}[H]
\begin{center}
\includegraphics[width=8.0cm,angle=0]{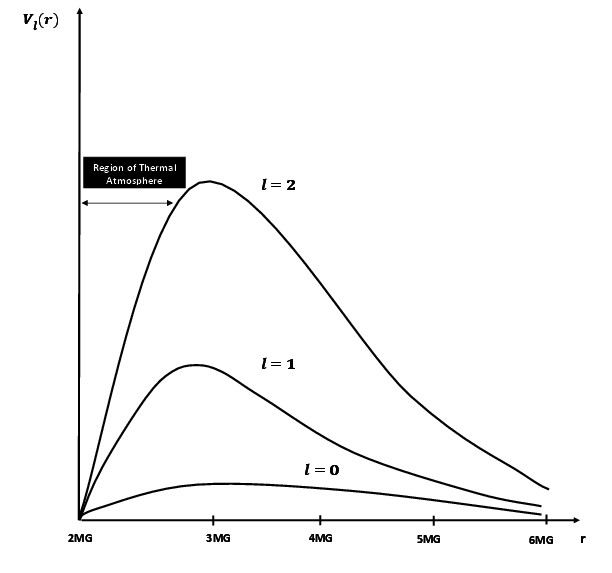}
\end{center}
\caption{Schwarzschild potential.}\label{QFTS}
\end{figure}

The motion of a scalar particle of energy $\nu$ and angular momentum $l$ in the background gravitational 
field of the Schwarzschild black hole is exactly equivalent to the motion of a quantum particle, i.e. a particle obeying the 
Schrodinger equation, with energy $E=\nu^2$ in a scattering potential given in the tortoise coordinate $r_*$ with the expression 
\begin{eqnarray}
V(r_*)=\frac{r-r_s}{r}\big(\frac{r_s}{r^3}+\frac{l(l+1)}{r^2}\big)~,~r_*=r+r_s\ln(\frac{r}{r_s}-1). 
\end{eqnarray}
This potential vanishes at infinity and at the event horizon $r_s$ and thus the particle is free at infinity and 
at the event horizon. See figure (\ref{QFTS}). This potential is characterized by a barrier at $r\sim 3r_s/2$ where the 
potential reaches its maximum and the height of this barrier is proportional to the square of the angular momentum, viz
\begin{eqnarray}
V_{\rm max}(r_*)\sim \frac{l^2+1}{G^2M^2}\sim (l^2+1)T_H,
\end{eqnarray}
where $T_H=1/(8\pi GM)$ is the Hawking temperature which we will compute shortly. 

On the other hand, these particles are in
thermal equilibrium at the Hawking temperature and thus the energy $\nu$ is proportional to $T_H$. Thus, we can immediately see from this simple argument that only particles with no angular momentum, i.e. $l=0$, can go through the potential barrier and 
escape from the black hole to infinity. These particles are precisely Hawking particles. The difference with the case of Rindler spacetime lies in the fact that in Rindler spacetime the potential barrier is infinite 
and thus no particles can go through and escape from the black hole to infinity. This is a very strong but simple 
physical description of Hawking radiation.

\subsection{Hawking temperature}
A systematic derivation of the Hawking radiation is given in three different ways. 

By employing the fact that the near-horizon geometry of Schwarzschild black hole is Rindler spacetime and then 
applying the Unruh effect in Rindler spacetime. Recall that Rindler spacetime is a uniformly accelerating observer with 
acceleration $a$ related to the event horizon $r_s$ by the relation 
\begin{eqnarray}
a=\frac{1}{2r_s}. 
\end{eqnarray} 
The Minkowski vacuum state is seen by the Rindler observer as a mixed thermal state at the temperature $T=a/2\pi$ which is 
  precisely the Enruh effect. This is in one sense what lies at the basis of Hawking radiation.

Secondly, by considering the eternal black hole geometry and studying the properties of the 
Kruskal vacuum state  $|0_K\rangle$ with respect to the Schwarzschild observer. The Kruskal state for Schwarzschild 
observer plays exactly the role of the Minkowski state for the Rindler observer  the Schwarzschild vacuum plays the 
role of the Rinlder vacuum. The Schwarzschild observer sees the Kruskal vacuum as a heath bath containing 
\begin{eqnarray}
    \langle 0_k |N_{\omega}|0_K\rangle=\frac{\delta(0)}{\exp(2\pi\omega/a)-1}
    \end{eqnarray}
particles. We can infer immediately from this result the correct value $T=a/2\pi$ of the Hawking temperature.

\begin{figure}[H]
\begin{center}
\includegraphics[width=7.0cm,angle=0]{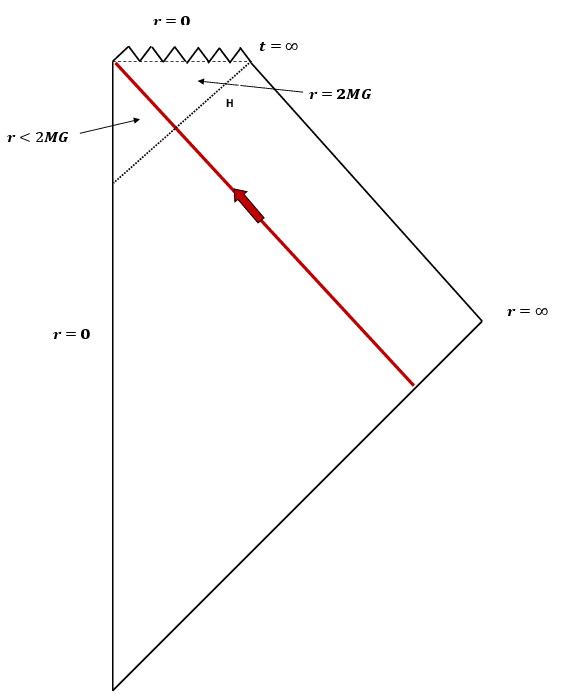}
\end{center}
\caption{Penrose diagram of the formation of a black hole from gravitational collapse}\label{PDG}
\end{figure}

Thirdly, we computed the Hawking temperature by considering the more realistic situation in which 
a Schwarzschild black hole is formed by gravitational collapse of a thin mass shell as in the Penrose diagram shown in figure (\ref{PDG}) (the
thin shell is the red line). Then by deriving the actual incoming state known as the Unruh vacuum state. The Unruh state is a
maximally entangled state describing a pair of particles with zero Killing energy. One of the pair $|n_R\rangle$ goes outside
the horizon and is seen as Hawking radiation whereas the other pair  $|n_L\rangle$  falls behind the horizon and goes into 
the singularity at the center and thus it corresponds to the information lost inside the black hole. This quantum state is 
given by the relation 

    \begin{eqnarray} 
    |U\rangle\sim \sum_n \exp(-\frac{n\pi\omega}{a})|n_R>|n_L>. 
    \end{eqnarray}
Although, the actual quantum state of the black hole is pure, the asymptotic Schwarzschild observer registers a 
thermal mixed state given by the density matrix
\begin{eqnarray} 
\rho_R={\rm Tr}_L|U\rangle\langle U|\sim \sum_n\exp(-\frac{2n\pi\omega}{a})|n_R\rangle\langle n_R|. 
\end{eqnarray}
Thus, the Schwarzschild observer registers a canonical ensemble with temperature 
\begin{eqnarray}
T=\frac{1}{8\pi GM}. 
\end{eqnarray}
%This is because the acceleration is $a=1/2r_s$ and the event horizon is $r_s=2GM$ where $M$ is the mass of the hole and 
%$G$ is Newton's gravitational constant.
In summary, a correlated entangled pure state near the horizon gives rise to a thermal mixed state outside the horizon.

\begin{figure}[H]
\begin{center}
\includegraphics[width=9.0cm,angle=0]{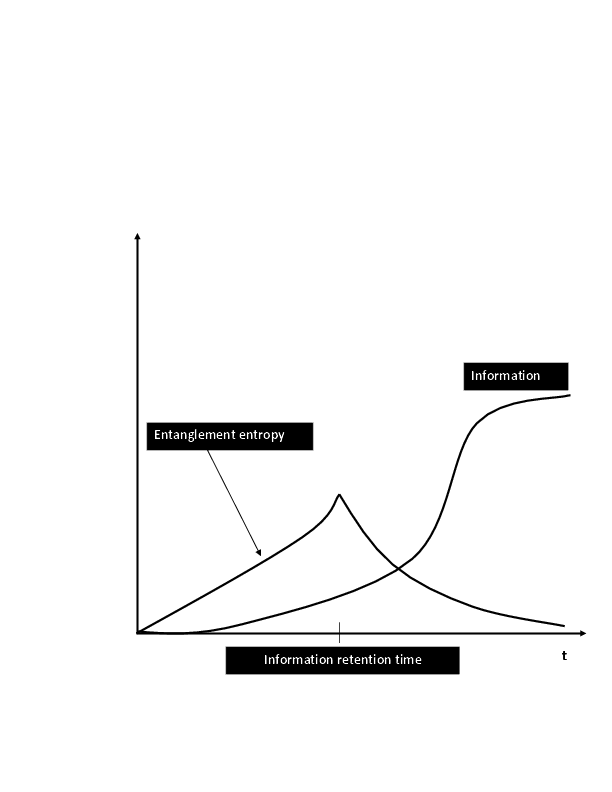}
\end{center}
\caption{Page time, the entanglement entropy and the information.}\label{information}
\end{figure}

\subsection{Page curve and unitarity}
The information loss problem can then be summarized as follows (see \cite{ydri2017} for an extensive discussion). 
The black hole starts in a pure state and after its complete evaporation the Hawking radiation is also in a pure state. 
This is the assumption of unitarity. The information is given by the difference between the thermal 
entropy of Boltzmann and the information entropy of Von Neumann whereas the entanglement entropy is the Von Neumann entropy 
in the case the system is described by a pure state. The entanglement entropy starts at zero value then it reaches a maximum value 
at the so-called Page time (also maybe called information retention time) then drops to zero again. The Page time is the
time at which the black hole evaporates around one half of its mass and the information starts to get out with the 
radiation. Before the Page time only energy gets out 
with the radiation with little or no information, while only at the Page time the information starts to get 
out, and it gets out completely at the moment of evaporation (see figure (\ref{information})). This is guaranteed to happen because of the second principle 
of thermodynamics and the assumption of unitarity. The behavior of the entanglement entropy with time is called the 
Page curve and a nice rough derivation of this curve can be outlined using the so-called Page theorem. The computation 
of the Page curve starting from first principles will provide, in some precise sense, the mathematical solution of the black hole information 
loss problem.

\subsection{Information loss problem}
Since this is a vital issue we state the information loss in different terms. 

We consider again a black hole formed by 
gravitational collapse as given by the above Penrose diagram. The Hilbert space ${\bf H}_{\rm in}$ of initial states $|\psi_{\rm in}\rangle$ is associated with null rays incoming from ${\cal J}^{-}$ at $r=\infty$, i.e. ${\bf H}_{\rm in}={\bf H}_{-}$. The Hilbert space ${\bf H}_{\rm out}$ of final states $|\psi_{\rm out}\rangle$ is clearly a tensor product of the Hilbert space ${\bf H}_{+}$ of the scattered outgoing radiation which escapes to the infinity ${\cal J}^+$ and the Hilbert space ${\bf H}_S$ of the transmitted radiation which falls behind the horizon into the singularity. This is the assumption of locality. Indeed, the outgoing Hawking particle and the lost quantum behind the horizon are maximally entangled, and thus they are space like separated, and as a consequence localized operators on ${\cal J}^+$ and $S$ must commute. We have then 
\begin{eqnarray}
H_{\rm in}=H_-~,~H_{\rm out}=H_+\otimes H_S.
\end{eqnarray}
From the perspective of observables at ${\cal J}^+$ (us), the outgoing Hawking particles can only be described by a reduced density matrix, even though the final state $|\psi_{\rm out}\rangle$ is obtained from the initial state $|\psi_{\rm in}\rangle$ by the action of a unitary $S$-matrix. This is the assumption of unitarity. This reduced density matrix is completely mixed despite the fact that the final state is a maximally entangled pure state. Eventually, the black hole will evaporate completely and it seems that we will end up only with the mixed state of the radiation. This the information paradox. There are six possibilities here:
\begin{enumerate}
\item Information is really lost which is Hawking original stand.
\item Evaporation stops at a Planck-mass remnant which contains all the information with extremely large entropy.
\item Information is recovered only at the end of the evaporation when the singularity at $r=0$ becomes a naked singularity. This contradicts the principle of information conservation with respect to the observe at ${\cal J}^+$ which states that by the time (Page or retention time) the black hole evaporates around one half of its mass the information must start coming out with the hawking radiation.

\item Information is not lost during the entire process of formation and evaporation. This is the assumption of unitarity. But how?
\item Horizon is like a brick wall which can not be penetrated. This contradicts the equivalence principle in an obvious way.
\item Horizon duplicates the information by sending one copy outside the horizon (as required by the principle of 
information conservation) while sending the other copy inside the horizon (as required by the equivalence principle). 
This is however forbidden by the linearity of quantum mechanics or the so-called quantum xerox principle. 
\end{enumerate}

\subsection{Thermodynamics}
The last point of primacy importance concerns black hole thermodynamics. The thermal entropy is the maximum amount of
information contained in the black hole. The entropy is mostly localized near the horizon, but quantum field theory 
(QFT) gives a divergent value, instead of the Bekenstein-Hawking value 
\begin{eqnarray}
S = A/4G, 
\end{eqnarray}
where $A$ is the surface area of the black hole. The number of accessible quantum microscopic states is determined by this
entropy via the formula 
\begin{eqnarray} 
n=\exp(S). 
\end{eqnarray}
Since QFT gives a divergent entropy instead of the Bekenstein-Hawking value it must be replaced by quantum 
gravity (QG) near the horizon and this separation of the QFT and QG degrees of freedom can be implemented by the 
stretched horizon which is a time like membrane, at a distance of one Planck length $l_P=\sqrt{G\hbar}$ from the actual
horizon, and where the proper temperature gets very large and most of the black hole entropy accumulates.

\section{Some string theory and conformal field theory}

%\subsection{Polyakov path integral and conformal anomaly}
%In this section we concentrate mostly on the bosonic string. 

The standard text remains the classic book by Green, Schwarz and Witten \cite{Green:1987sp,Green:1987mn}. Of course, also  a classic is the book by Polchinski \cite{Polchinski:1998rq,Polchinski:1998rr}. The modern text \cite{Becker:2007zj} is truly modern and as such it turns out to be extremely  useful. Also the books \cite{Johnson:2003gi}, \cite{Zwiebach:2004tj} and \cite{Szabo:2004uy} were used extensively. We also found the lectures \cite{Schellekens} on conformal field theory very illuminating and the seminal papers \cite{Lundholm,Goddard:1986ee,Friedan:1986kd} on the Virasoro algebra very helpful.

\subsection{The conformal anomaly}

The starting point is the path integral 
\begin{eqnarray}
&&Z=\int Dh(\sigma)DX(\sigma)~e^{iS [h,X]}~,~S[h,X]=-\frac{1}{2\pi}\int_{\Sigma} d^2\sigma \sqrt{h}~h^{ab}{\partial}_aX^{\mu}{\partial}_bX_{\mu}.
\end{eqnarray}
We use the two reparametrization invariances and the Weyl symmetry to impose the conformal gauge $h_{ab}=e^{\phi}{\eta}_{ab}$. The action $S[X,h]$ becomes 
\begin{eqnarray}
S [X]=-\frac{1}{2\pi}\int_{\Sigma} d^2\sigma {\partial}_aX^{\mu}{\partial}^aX_{\mu}.
\end{eqnarray}
The Fadeev-Popov gauge fixing procedure leads then to the ghost action  

\begin{eqnarray}
S[b,c]&=&\frac{i}{\pi}\int d^2\sigma
~c^-{\nabla}_+b_{--}+\frac{i}{\pi}\int d^2\sigma
~c^+{\nabla}_-b_{++}.
\end{eqnarray}
The ghost fields $c^+$ and $c^-$ are the components of a vector ghost field
$c^a$ whereas the antighosts $b_{++}$ and $b_{--}$ are the components of a
traceless symmetric antighost tensor field $b_{ab}$. The total gauge-fixed action $S[X,b,c]$ is given by
\begin{eqnarray}
S [X,b,c]=-\frac{1}{2\pi}\int_{\Sigma} d^2\sigma {\partial}_aX^{\mu}{\partial}^aX_{\mu}+\frac{i}{\pi}\int d^2\sigma ~c^-{\nabla}_+b_{--}+\frac{i}{\pi}\int d^2\sigma ~c^+{\nabla}_-b_{++}.
\end{eqnarray}
We recall the non-zero components of the energy-momentum tensor given by
\begin{eqnarray}
&&T_{++}={\partial}_+X^{\mu}{\partial}_+X_{\mu}\nonumber\\
&&T_{--}={\partial}_-X^{\mu}{\partial}_-X_{\mu}.
\end{eqnarray}
The contribution of the ghosts to the world-sheet energy-momentum tensor is given by 
\begin{eqnarray}
T_{++}^g&=&-i{\nabla}_+c^+ b_{++}-\frac{i}{2}c^+{\nabla}_+ b_{++}.
\end{eqnarray}
\begin{eqnarray}
T_{--}^g&=&-i{\nabla}_-c^- b_{--}-\frac{i}{2}c^-{\nabla}_- b_{--}.
\end{eqnarray}
The Virasoro generators $L_m^g$ are the Fourier modes of $T_{++}^g$ in the same that $L_m$ are the Fourier modes of $T_{++}$. They satisfy quantum mechanically the algebra 
\begin{eqnarray}
[L_m^g,L_n^g]=(m-n)L_{m+n}^g+(-\frac{13}{6}m^3+\frac{m}{6}){\delta}_{m+n,0}.
\end{eqnarray}
We recall the Virasoro algebra
\begin{eqnarray}
[L_m,L_{n}]=(m-n)L_{m+n}+\frac{D}{12}(m^3-m){\delta}_{m+n,0}.
\end{eqnarray}
The total Virasoro generators are defined by
\begin{eqnarray}
L_m^{\rm TOT}=L_m+L_m^g-a{\delta}_{m,0}.
\end{eqnarray}
The constraints read now
\begin{eqnarray}
L_m^{\rm TOT}=0.
\end{eqnarray}
The Virasoro algebra reads now
\begin{eqnarray}
[L_m^{\rm TOT},L_{n}^{\rm TOT}]&=&(m-n)L_{m+n}^{\rm
  TOT}+\bigg(2am+\frac{D}{12}(m^3-m)-\frac{13}{6}m^3+\frac{m}{6}\bigg){\delta}_{m+n,0}\nonumber\\
&=&(m-n)L_{m+n}^{\rm
  TOT}+\bigg(\frac{D-26}{12}m^3+\frac{2+24a-D}{12}m\bigg){\delta}_{m+n,0}.
\end{eqnarray}
The conformal anomaly vanishes iff
\begin{eqnarray}
D=26,a=1.
\end{eqnarray}
This means in particular that only for these values that the theory is truly conformally invariant.
\subsection{The operator product expansion}
We go now to Euclidean signature, i.e. $\sigma^2=i\sigma^0=i\tau$ and $\sigma^1=\sigma$, and we define the complex coordinates $z=\sigma^1+i\sigma^2$, $\bar{z}=\sigma^1-i\sigma^2$. The action $S[X]$ becomes 
\begin{eqnarray}
S=\frac{1}{\pi}\int d^2z \partial {X}^{\mu}\bar{\partial}{X}_{\mu}.
\end{eqnarray}
The right-moving solution $X_R^{\mu}$ becomes a holomorphic function, i.e. an analytic function of $z$, while the left-moving solution $X_L^{\mu}$ becomes an antiholomorphic function, i.e. an analytic function of $\bar{z}$. The residual symmetries of this Euclidean action are given by the conformal mappings 
\begin{eqnarray}
z\longrightarrow f(z)~,~\bar{z}\longrightarrow \bar{z}=\bar{f}(\bar{z}).
\end{eqnarray}
These are angle-preserving transformations when $f$ and its inverse are both holomorphic. For example, $z\longrightarrow z+a$ is a translation, $z\longrightarrow \zeta z$ where $|\zeta|=1$ is a rotation, and $z\longrightarrow \zeta z$ where $\zeta$ is real not equal to $1$ is a scale transformation called also dilatation.

We will work with the complex coordinates
\begin{eqnarray}
w=e^{-2iz}=e^{2(\sigma^2-i\sigma^1)}~,~\bar{w}=e^{-2i\bar{z}}=e^{2(\sigma^2+i\sigma^1)}.
\end{eqnarray}
The world sheet is now regarded as a Riemann surface. The Euclidean time $\sigma^2$ corresponds to the radial distance $r=\exp(2\sigma^2)$ on the complex plane, with the infinite past $\sigma^2=-\infty$ at $r=0$, and the infinite future $\sigma^2=+\infty$ is a circle at $r=\infty$. Thus, time ordered product of operators on the cylinder becomes radially ordered product of operators on the complex plane. We will rewrite in the following for simplicity 
\begin{eqnarray}
z=e^{2(\sigma^2-i\sigma^1)}~,~\bar{z}=e^{2(\sigma^2+i\sigma^1)}.
\end{eqnarray}
We compute then
\begin{eqnarray}
T_{\bar{z}\bar{z}}=-2\bar{\partial} X^{\mu}\bar{\partial}X^{\mu}=\sum_m \frac{\tilde{L}_m}{\bar{z}^{m+2}}~,~
T_{{z}{z}}=-2{\partial} X^{\mu}{\partial}X^{\mu}=\sum_m \frac{{L}_m}{{z}^{m+2}}.
\end{eqnarray}
The classical stress-energy tensor $T(z)=T_{{z}{z}}=-2\partial X^{\mu}(z)\partial X^{\mu}(z)$ is defined quantum mechanically by the normal ordered expression 
 \begin{eqnarray}
-2T(z)&=&4:\partial X^{\mu}(z)\partial X^{\mu}(z):\nonumber\\
&=&{\rm lim}_{z\longrightarrow w}\bigg(4R(\partial X^{\mu}(z)\partial X^{\mu}(w))+\frac{\delta^{\mu\mu}}{(z-w)^2}\bigg).
\end{eqnarray}
The radially ordered product is related to the normal ordered product by the relation 
\begin{eqnarray}
2^{n} R(X^{\mu}(z_1,\bar{z}_1)...X^{\nu}(z_n,\bar{z}_n))
&=&2^{n}:X^{\mu}(z_1,\bar{z}_1)...X^{\nu}(z_n,\bar{z}_n):+\sum{\rm contractions}.\nonumber\\
\end{eqnarray}
The sum runs over all ways of choosing one pair of fields (or two or more pairs in the case we have an arbitrary product of fields) from the product and replacing each term with the contraction 
\begin{eqnarray}
4R(X^{\mu}(z,\bar{z})X^{\nu}(w,\bar{w}))
&=&4:X^{\mu}(z,\bar{z})X^{\nu}(w,\bar{w}):-\delta^{\mu\nu}\bigg(\ln (z-w)+\ln(\bar{z}-\bar{w})\bigg).
\end{eqnarray}
A primary field is a conformal field of conformal dimension $(h,\bar{h})$. In other words, it is a tensor field of rank $n=h+\bar{h}$ under conformal transformations with components $\Phi_{z...z,\bar{z}...\bar{z}}(z,\bar{z})$ transforming under the conformal transformations $z\longrightarrow w(z)$  and $\bar{z}\longrightarrow \bar{w}(\bar{z})$ as  
\begin{eqnarray}
\Phi(z,\bar{z})\longrightarrow (\frac{\partial w}{\partial z})^{h}(\frac{\partial \bar{w}}{\partial \bar{z}})^{\bar{h}}\Phi(w,\bar{w}).
\end{eqnarray}
The rank $h+\bar{h}$ is called the dimension of $\Phi$ and it determines its behavior under scalings while $h-\bar{h}$ is the spin of $\Phi$ and it determines its behavior under rotations. 

Since $T_{zz}$ is holomorphic and $T_{\bar{z}\bar{z}}$ antiholomorphic the conserved currents in terms of the stress-energy tensor are given by (with $T(z)=T_{zz}=(\partial_z\Phi)^2$, $\bar{T}(\bar{z})=T_{\bar{z}\bar{z}}=(\partial_{\bar{z}}\Phi)^2$)
 \begin{eqnarray}
J_z=iT(z)\epsilon(z)~,~J_{\bar{z}}=i\bar{T}(\bar{z})\bar{\epsilon}(\bar{z}).
\end{eqnarray} 
We will only concentrate on the holomorphic part for simplicity. The conserved charge  $Q_{\epsilon}$ on the complex plane, associated with the infinitesimal conformal transformation $T(z)\epsilon(z)$, is then defined by
 \begin{eqnarray}
Q_{\epsilon}=\frac{1}{2\pi i}\oint dw T(w)\epsilon(w).
\end{eqnarray}
This generates the infinitesimal conformal transformation 
\begin{eqnarray}
\delta \Phi(w,\bar{w})=\epsilon(w)\partial_w\Phi+h\Phi \partial_{w}\epsilon(w).\label{infini0}
\end{eqnarray}
The quantum analogue of this equation is 
\begin{eqnarray}
\delta \Phi(w,\bar{w})&=&[Q_{\epsilon},\Phi(w,\bar{w})]\nonumber\\
&=&\frac{1}{2\pi i}\oint dz \epsilon(z)R(T(z)\Phi(w,\bar{w})).
\end{eqnarray}
The integration contour over $z$ is understood now to encircle the point $w$. The radially ordered product R must be analytic in the neighborhood of the point $w$ in order for the integral to make sense. Thus, one must have a Laurent expansion. We reproduce the infinitesimal transformations (\ref{infini0}) if and only if the radially ordered product is given by
\begin{eqnarray}
R(T(z)\Phi(w,\bar{w}))=\frac{h}{(z-w)^2}\Phi(w,\bar{w})+\frac{1}{z-w}\partial_w\Phi(w,\bar{w})+{\rm regular}~{\rm power}~{\rm series}~{\rm in}~(z-w).\nonumber\\
\end{eqnarray}
The conjugate analogue of this equation is
\begin{eqnarray}
R(\bar{T}(\bar{z})\Phi(w,\bar{w}))=\frac{\bar{h}}{(\bar{z}-\bar{w})^2}\Phi(w,\bar{w})+\frac{1}{\bar{z}-\bar{w}}\partial_{\bar{w}}\Phi(w,\bar{w})+{\rm regular}~{\rm power}~{\rm series}~{\rm in}~(\bar{z}-\bar{w}).\nonumber\\
\end{eqnarray}
As examples  we compute
\begin{eqnarray}
R(T(z) X^{\nu}(y))
&=&-2:\partial X^{\mu}(z)\partial X^{\mu}(w) X^{\nu}(y):+\frac{1}{z-y}\partial X^{\nu}(y)+...
\end{eqnarray}
\begin{eqnarray}
R(T(z)\partial X^{\nu}(y))
&=&-2:\partial X^{\mu}(z)\partial X^{\mu}(w)\partial X^{\nu}(y):+\frac{1}{(z-y)^2}\partial X^{\nu}(y)+\frac{1}{z-y}\partial^2X^{\nu}(y)+...\nonumber\\
\end{eqnarray}
The normal ordered objects behave as classical quantities and thus they are finite in the limit $z\longrightarrow w$. We rewrite these results as
\begin{eqnarray}
T(z) X^{\nu}(y)
&=&\frac{1}{z-y}\partial X^{\nu}(y)+...
\end{eqnarray}
\begin{eqnarray}
T(z)\partial X^{\nu}(y)
&=&\frac{1}{(z-y)^2}\partial X^{\nu}(y)+\frac{1}{z-y}\partial^2X^{\nu}(y)+...
\end{eqnarray}
This means in particular that $X^{\mu}$ and $\partial X^{\mu}$ are conformal fields of weights $h=0$ and $h=1$ respectively. These are examples of the operator product expansion.

The operator product expansion of the energy-momentum tensor with itself is found to be given by
\begin{eqnarray}
R(T(z)T(y))&=&:\partial X^{\mu}(z)\partial X^{\mu}(w)\partial X^{\nu}(y)\partial X^{\nu}(x):\nonumber\\
&+&\delta^{\mu\mu}\frac{1}{2(z-y)^4}+\frac{2}{(z-y)^2}T(y)+\frac{1}{z-y}\partial T(y).
\end{eqnarray}
We write this as
%\begin{eqnarray}
%T(z)T(y)&=&\delta^{\mu\mu}\frac{1}{2(z-y)^4}+\frac{2}{(z-y)^2}T(z)-\frac{1}{z-y}\partial T(z).
%\end{eqnarray}
%Or equivalently
\begin{eqnarray}
T(z)T(y)&=&\delta^{\mu\mu}\frac{1}{2(z-y)^4}+\frac{2}{(z-y)^2}T(y)+\frac{1}{z-y}\partial T(y).
\end{eqnarray}
We can immediately see that $T(z)$ is a conformal field of weight $h=2$, which is the classical value, if $\delta^{\mu\mu}=0$. 
%the operator product expansion of the energy-momentum tensor with itself is given by
%\begin{eqnarray}
%T(z)T(y)=\frac{c}{2(z-y)^4}+\frac{2}{(z-y)^2}T(y)+\frac{1}{z-y}\partial T(y).\label{OPET1}
%\end{eqnarray}
The constant $c=\delta^{\mu\mu}=D$ is called the central charge. The energy-momentum tensor is not conformal unless $c=0$. In this case it is a primary field of weight $(2,0)$. The central charge is therefore the conformal anomaly and it is due to quantum effects. Obviously, the central charge is equal one for a single scalar field.
\subsection{The bc CFT}

The gauge-fixed action is given in Euclidean world sheet by (with the scaling $b\longrightarrow b$ and $c^-\longrightarrow ic^-$ and $c^+\longrightarrow -ic^+$) 
\begin{eqnarray}
S_E=\frac{1}{2\pi}\int d^2z\bigg(2\partial X^{\mu}\bar{\partial}X_{\mu}+b_{--}\bar{\partial}c^{-}+ b_{++}\partial c^+\bigg).\label{oo1}
\end{eqnarray}
%\begin{eqnarray}
%S_E=\frac{1}{2\pi}\int d^2z\bigg(2\partial X^{\mu}\bar{\partial}X_{\mu}-ic^{-}\bar{\partial}b_{--}+ic^+\partial b_{++}\bigg).
%\end{eqnarray}
%\begin{eqnarray}
%S_E=\frac{1}{2\pi}\int d^2z\bigg(2\partial X^{\mu}\bar{\partial}X_{\mu}-ib_{--}\bar{\partial}c^{-}+i b_{++}\partial c^+\bigg).\label{oo}
%\end{eqnarray}
%We have used $i \int d^2\sigma\longrightarrow \int d^2z/2$, and $\tau+\sigma=\bar{z}=-i\sigma^2+\sigma^1$, $\tau-\sigma=-{z}=-i\sigma^2-\sigma^1$.  
 %We scale the ghosts fields as $b\longrightarrow b$ and $c^-\longrightarrow ic^-$ and $c^+\longrightarrow -ic^+$ to obtain 
%\begin{eqnarray}
%S_E=\frac{1}{2\pi}\int d^2z\bigg(2\partial X^{\mu}\bar{\partial}X_{\mu}+b_{--}\bar{\partial}c^{-}+ b_{++}\partial c^+\bigg).\label{oo1}
%\end{eqnarray}
The ghost energy-momentum tensor was found to be given by (including also Wick rotation)
%\begin{eqnarray}
%2T_{++}^g&=&2b_{++}{\partial}_+c^+ -c^+{\partial}_+ b_{++}.\label{tg1}
%\end{eqnarray}
%\begin{eqnarray}
%2T_{--}^g&=& 2b_{--}{\partial}_-c^--c^-{\partial}_- b_{--}.\label{tg2}
%\end{eqnarray}
\begin{eqnarray}
-2T_{++}^g=-2b_{++}\bar{\partial}c^+ +c^+\bar{\partial} b_{++}~,~-2T_{--}^g=-2b_{--}{\partial}c^-+c^-{\partial} b_{--}.\label{tg1}
\end{eqnarray}
%The basic anticommutation relations become 
%\begin{eqnarray}
%\{c^+(\sigma,\tau),b_{++}(\sigma^{'},\tau)\}=2\pi i\delta(\sigma-\sigma^{'})~,~\{c^-(\sigma,\tau),b_{--}(\sigma^{'},\tau)\}=-2\pi i\delta(\sigma-\sigma^{'}).
%\end{eqnarray}
%\begin{eqnarray}
%T_{++}^g&=&-{\partial}_+c^+ b_{++}-\frac{1}{2}c^+{\partial}_+ b_{++}.
%\end{eqnarray}
%\begin{eqnarray}
%T_{--}^g&=&-{\partial}_-c^- b_{--}-\frac{1}{2}c^-{\partial}_- b_{--}.
%\end{eqnarray}
The above theory involves a free fermionic conformal field theory termed bc CFT given in terms of anticommuting fields $b$ and $c$ by
\begin{eqnarray}
S_E=\frac{1}{2\pi}\int d^2zb\bar{\partial}c.
\end{eqnarray}
This is conformally invariant for all $b$ and $c$ transforming under conformal transformations as tensors of weights $(\lambda,0)$ and $(1-\lambda,0)$. In the quantization of the string action we have found that $\lambda=2$. Since the ghost fields $c^+$ and $c^-$ are the components of a vector ghost field $c^a$ whereas the antighosts $b_{++}$ and $b_{--}$ are the components of a traceless symmetric antighost tensor field $b_{ab}$. Thus, the ghost field $c$ has conformal dimension $-1$ whereas the antighost field $b$ has conformal dimension $2$. In other words, $b$ transforms as the energy-momentum tensor while $c$ transforms as the gauge transformations (diffeomorphism) parameter.

%We compute
%\begin{eqnarray}
%0=\int [db][dc]\frac{\delta}{\delta c(z,\bar{z})}\bigg[\exp(-S_E)c(0)\bigg]\Rightarrow <\bar{\partial}b(z).c(0)>=2\pi\delta^2(z,\bar{z}).\label{proF0}
%\end{eqnarray}
%Using the same method we obtain the quantum equations of motion
%\begin{eqnarray}
%<\bar{\partial}c(z)>=<\bar{\partial}b(z)>=0.
%\end{eqnarray}
We compute the radial ordering in terms of the normal ordering and a contraction (propagator) as follows
\begin{eqnarray}
R(b(z_1)c(z_2))=:b(z_1)c(z_2):+<b(z_1)c(z_2)>.
\end{eqnarray}
The propagator is given by
\begin{eqnarray}
<b(z_1)c(z_2)>=\frac{1}{z_1-z_2}.\label{proF}
\end{eqnarray}
%Let us redo this for the bosonic action 
%\begin{eqnarray}
%S=\frac{1}{2\pi\alpha^{'}}\int d^2z\partial\phi\bar{\partial}\phi.
%\end{eqnarray}
%We compute then 
%\begin{eqnarray}
%0=\int [d\phi]\frac{\delta}{\delta \phi(z,\bar{z})}\bigg[\exp(-S)\phi(0)\bigg]\Rightarrow \frac{1}{\pi\alpha^{'}}<\bar{\partial}\partial\phi(z).\phi(0)>=-\delta^2(z,\bar{z}).
%\end{eqnarray}
%We get the propagator 
%\begin{eqnarray}
%\frac{1}{\pi\alpha^{'}}\bar{\partial}\partial<\phi(z).\phi(0)>=-\delta^2(z,\bar{z}).
%\end{eqnarray}
%However, we obtain from (\ref{ope1} (after we insert the coefficient of $\alpha^{'}$)
%\begin{eqnarray}
%<\phi(z).\phi(0)>=<R(\phi(z)\phi(0)>=-\frac{\alpha^{'}}{2}\ln|z|^2.
%\end{eqnarray}
%Thus
%\begin{eqnarray}
%\bar{\partial}\partial\ln|z|^2=2\pi\delta^2(z,\bar{z}).
%\end{eqnarray}
%Equivalently 
%\begin{eqnarray}
%\bar{\partial}\frac{1}{z}=2\pi\delta^2(z,\bar{z}).\label{proF1}
%\end{eqnarray}
%This should be checked directly by integrating by part over a region containing the origin. By using (\ref{proF0}) and (\ref{proF1}) we can see that the fermion propagator is indeed given by (\ref{proF}). 
We get then the operator product expansions 
 \begin{eqnarray}
b(z_1)c(z_2)\sim \frac{1}{z_1-z_2}~,~c(z_1)b(z_2)\sim\frac{\epsilon}{z_1-z_2}.\label{fer}
\end{eqnarray}
\begin{eqnarray}
b(z_1)b(z_1)\sim O(z_1-z_2)~,~c(z_1)c(z_2)=O(z_1-z_2).
\end{eqnarray}
The $\epsilon$ in (\ref{fer}) is equal $+1$ in our case. But in the case where $b$ and $c$ satisfy Bose statistics we must set $\epsilon=-1$.

The generalization of the second equation of (\ref{tg1}) is given by (including also normal ordering to be precise)

\begin{eqnarray}
2T_{bc}&=&:-\lambda b(z){\partial}c(z) +\epsilon (\lambda-1)c(z){\partial} b(z):\nonumber\\
&=&{\rm lim}_{z\longrightarrow w}\bigg(-\lambda b(z){\partial}c(w) +\epsilon (\lambda-1)c(z){\partial} b(w)+\frac{1}{(z-w)^2}\bigg).
\end{eqnarray}
Indeed, we can check that
\begin{eqnarray}
R(T_{bc}(z)b(y))=:...:+\frac{\lambda}{(z-y)^2}b(y)+\frac{1}{z-y}\partial b(y).
\end{eqnarray}
\begin{eqnarray}
R(T_{bc}(z)c(y))=:...:+\frac{1-\lambda}{(z-y)^2}c(y)+\frac{1}{z-y}\partial c(y).
\end{eqnarray}
This shows explicitly that $b$ is of conformal dimension $\lambda$ and $c$ is of conformal weight equal $1-\lambda$, as it should be, and that $T(z)$ is then the correct form of the energy-momentum tensor.

The above formulas generalize, for a primary field  ${\cal O}$ with weights $(h,\bar{h})$, to the Ward identities
\begin{eqnarray}
R(T(z){\cal O}(y,\bar{y}))=:...:+\frac{h}{(z-w)^2}{\cal O}(y,\bar{y})+\frac{1}{z-y}\partial{\cal O}(y,\bar{y}).
\end{eqnarray}
\begin{eqnarray}
R(\bar{T}(\bar{z}){\cal O}(y,\bar{y}))=:...:+\frac{\bar{h}}{(\bar{z}-\bar{w})^2}{\cal O}(y,\bar{y})+\frac{1}{\bar{z}-\bar{y}}\partial{\cal O}(y,\bar{y}).
\end{eqnarray}
%for the general conformal transformations $z\longrightarrow z^{'}=f(z)$ under which the primary field  ${\cal O}$ with weights $(h,\tilde{h})$ transforms as 
%\begin{eqnarray}
%{\cal O}^{'}(z^{'},\bar{z}^{'})=(\partial_zz^{'})^{-h}(\partial_{\bar{z}}\bar{z}^{'})^{-\tilde{h}}{\cal O}(z,\bar{z}).
%\end{eqnarray}
%The sum $h+\tilde{h}$ is called the dimension of ${\cal O}$ and it determines its behavior under scalings while $h-\tilde{h}$ is the spin of ${\cal O}$ and it determines its behavior under rotations. 
For example, $X^{\mu}$ is of weights $(0,0)$, $\partial X^{\mu}$ is of weight $(1,0)$, $\bar{\partial}X^{\mu}$ is weight $(0,1)$, $\partial^2X^{\mu}$ is of weight $(2,0)$, and $:\exp(ikX):$ is of weights $(\alpha^{'}k^2/4,\alpha^{'}k^2/4)$. However, the operator product expansion of the energy-momentum tensor with itself is given in general by
\begin{eqnarray}
T(z)T(y)=\frac{c}{2(z-y)^4}+\frac{2}{(z-y)^2}T(y)+\frac{1}{z-y}\partial T(y).\label{OPET1}
\end{eqnarray}
The constant $c$ is called the central charge and the energy-momentum is therefore not a conformal field unless $c=0$. We have found that $c=\delta^{\mu\mu}=D$ for $D$ scalar fields.%In this case it is a primary field of weights $(2,0)$. The central charge is therefore the conformal anomaly. We have already computed that $c=\delta^{\mu\mu}=D$. Thus, the central charge is equal one for a single scalar field. 

What is the analogue result for the bc CFT above?

After some calculation we get the operator product expansion 
\begin{eqnarray}
T_{bc}(z)T_{bc}(y)&=&\frac{c}{2(z-y)^4}+2\frac{T_{bc}(y)}{(z-y)^2}+\frac{\partial T_{bc}(y)}{z-y}\nonumber\\
&& c=-2\epsilon (6\lambda^2-6\lambda+1).
\end{eqnarray}
%\begin{eqnarray}
%R(T_{bc}(z)T_{bc}(y))&=&:...:+\frac{c}{2(z-y)^4}+2\frac{T_{bc}(y)}{(z-y)^2}+\frac{\partial T_{bc}(y)}{z-y}\nonumber\\
%&& c=-2\epsilon (6\lambda^2-6\lambda+1).
%\end{eqnarray}
Th Faddeev-Popov ghosts of the gauge-fixed Polyakov action have weights corresponding to $\lambda=2$, viz $(2,0)$ for the antighost $b$ and $(-1,0)$ for the ghost $c$, with conformal anomaly $c=-26$. %Another important case is given by fermions  $b=\psi$ and $c=\bar{\psi}$ with equal weights $\lambda=1-\lambda=1/2$ and with conformal anomaly $c=1$ . 
Thus, the conformal anomaly $c(\lambda=2,\epsilon=1)=-26$ can be canceled by $26$ spacetime coordinates $X^{\mu}$ since the weight of every $X^{\mu}$ is $1$. 

\subsection{The super-conformal field theory}
We enhance conformal symmetry to the superconformal case. In this case we have $D$ scalars $X^{\mu}$ with conformal dimension $1$ and $D$ Majorana-Weyl spinors corresponding to  $b=\psi$ and $c=\bar{\psi}$ with equal conformal dimensions $\lambda=1-\lambda=1/2$. The gauge fixing in this case using the so-called superconformal gauge introduces, besides the two fermion ghosts $b$ and $c$ with $\lambda=2$, two boson ghosts with $\lambda=3/2$. Thus the total conformal anomaly coming from the ghosts in this case is $c(2,1)+c(3/2,-1)=-26+11=15$. This can be canceled by the contribution of the $D$ coordinates $X^{\mu}$ and their $D$ superpartners $\psi^{\mu}$ with $\lambda=1/2$. Indeed, the contribution of the dynamical fields to the conformal anomaly given by $c=D(1+1/2)=3D/2$ cancels exactly for $3D/2=15$, i.e. $D=10$. As it turns out, $c=3D/2$ is precisely the central charge of the energy-momentum tensor.

\subsection{Vertex operators}
In quantum field theory particles (or states) are created from the vacuum by quantum fields (operators). This provides a one-to-one map between states and operators. In closed string theory this map is given by vertex operators $V_{\phi}(z,\bar{z})$ which represent the absorption or emission of string states $|\phi\rangle$ from points $z$ on the world sheet. (Recall that the closed string cylinder is mapped to the complex plane under Wick rotation $\tau\longrightarrow i\tau$ and conformal mapping $z=\exp(\tau-i\sigma)$). The vertex operators $V_{\phi}(z,\bar{z})$ are clearly insertions of point like operators at the points $z$ on the complex plane. They are primary fields whereas the string states are highest weight states. By summing over all insertion points we obtain $g_s\int d^2z V_{\phi}(z,\bar{z})$. The vertex operators for closed strings have conformal dimension $(1,1)$.

The closed string ground state (tachyon) is the state with no oscillators excited  but with momentum $k$, viz $|\phi\rangle=|0,k\rangle$.  The corresponding vertex operator is $:\exp(ikX):$ which has a conformal dimension equal $({k^2}/{8},{k^2}/{8})=(1,1)$ (assuming $l_s=1$). We should then take an average over the absorption or emission point on the world sheet as $\int d^2z :\exp(ikX):$ since the state is independent of the insertion point.
 
The vertex operators for excited states will contain additional factors, of conformal dimension $(n,n)$ where $n$ is a positive integer, which are induced by the creation operators $\alpha_{-m}^{\mu}$, $m>0$. It is not difficult to convince ourselves that the desired rule to pass from the state to the operator is to replace  $\alpha_{-m}^{\mu}$ with $\partial^{m}X^{\mu}$.

The vertex operators for open strings are conformal fields of dimension $1$. The tachyon state is again associated with the operator $:\exp(ikX):$ which has a conformal dimension equal ${k^2}/{2}=1$ in open string theory. Also, by summing over all insertion points on the open string world sheet (upper complex half plane) we get an expression of the form $g_o\oint V_{\phi}(s)ds$ where $g_o$ is the open string coupling, viz $g_o^2=g_s$, and $s$ labels the boundary. 

The vertex operator for the vector gauge field (photon) state $|\phi\rangle=\xi_{\mu}(\alpha_{-1}^{\mu})^{\dagger}|0,k\rangle$ (the first excited state in open string theory) is given by 
\begin{eqnarray}
\int ds:\xi_{\mu}\partial_tX^{\mu} \exp(ikX):.
\end{eqnarray}
The insertions are located on the real axis, which is the boundary of the upper half complex plane, and $\partial_t$ is the derivative along the boundary. The boundary of the upper half complex plane corresponds (if we undo the conformal mapping) to the boundaries $\sigma=0$ and $\sigma=\pi$ of the open string world sheet which is a strip in spacetime. Thus the photon is associated with the end points of the open string.

\subsection{Background fields}
The spectrum of closed strings at the first excited level consists of a graviton, an antisymmetric second rank tensor field and a scalar field. The most important background fields which can couple to the string are precisely those fields which are associated with these massless bosonic degrees of freedom in the spectrum. Namely, the metric $g_{\mu\nu}(X)$, the antisymmetric two-form gauge field $B_{\mu\nu}(X)$, and the dilaton field $\Phi(x)$. The metric $g_{\mu\nu}$ couples in the obvious way
\begin{eqnarray}
S_1=-\frac{1}{4\pi\alpha^{'}}\int_M d^2\sigma\sqrt{-h}h^{\alpha\beta}g_{\mu\nu}(X)\partial_{\alpha}X^{\mu}\partial_{\beta}X^{\nu}.
\end{eqnarray}
The coupling of the two-form gauge field $B_{\mu\nu}$ is given by 
\begin{eqnarray}
S_2=-\frac{1}{4\pi\alpha^{'}}\int_M d^2\sigma\epsilon^{\alpha\beta}B_{\mu\nu}(X)\partial_{\alpha}X^{\mu}\partial_{\beta}X^{\nu}.
\end{eqnarray}
The epsilon symbol is defined such that $\epsilon^{01}=1$ and it is a tensor density, i.e. $\epsilon^{\alpha\beta}/\sqrt{-h}$ transforms as a tensor. The above action changes by a total divergence under the gauge transformations 
\begin{eqnarray}
\delta B_{\mu\nu}=\partial_{\mu}\Lambda_{\nu}-\partial_{\nu}\Lambda_{\mu}.\label{GT}
\end{eqnarray}
Thus, this term which couples the two-form gauge field $B_{\mu\nu}$ to the world sheet of the string is the analogue of the coupling of the one-form Maxwell field to the world line of a charged particle given by 
\begin{eqnarray}
S=q\int d\tau A_{\mu}\dot{x}^{\mu}.
\end{eqnarray}
Also, this term is only present for oriented strings, and it can be eliminated by a procedure called orientifold projection, i.e. a projection onto strings which are invariant under reversal of orientation. This term is also the source of much of the noncommutative geometry which appears from string theory.

The dilaton field is more interesting. It couples to the string via the scalar curvature $R^{(2)}(h)$ of the  metric $h_{\mu\nu}$ on the $2-$dimensional string world sheet. The action reads 
\begin{eqnarray}
S_3=\frac{1}{4\pi}\int_M d^2\sigma\sqrt{-h}\Phi(X)R^{(2)}(h).
\end{eqnarray}
Because of the absence of explicit factors of $X$ this action is one order higher than $S_1$ and $S_2$ in the  $\alpha^{'}$ expansion, i.e. $S_3$ should be thought of as an order $\alpha^{'}$ correction compared to the first two actions $S_1$ and $S_2$. For $\Phi=1$ we get The Hilbert-Einstein action in $2-$dimension. However,  in two dimensions this action is exactly equal the so-called Euler characteristic 
\begin{eqnarray}
\chi=\frac{1}{4\pi}\int_M d^2\sigma\sqrt{-h}R^{(2)}(h).\label{euler}
\end{eqnarray}
This is a topological invariant which gives no dynamics to the $2-$dimensional metric $h_{\mu\nu}$. The proof goes as follows. The variation of the Hilbert-Einstein action in any dimension is known to be given by
\begin{eqnarray}
\delta \chi= \frac{1}{4\pi}\int d^2\sigma\sqrt{-h}\delta h^{\alpha\beta}(R_{\alpha\beta}-\frac{1}{2}h_{\alpha\beta}R).
\end{eqnarray}
The Riemann tensor always satisfies $R_{\alpha\beta\gamma\delta}=-R_{\beta\alpha\gamma\delta}=-R_{\alpha\beta\delta\gamma}$. In two dimensions a second rank antisymmetric tensor can only be proportional to $\epsilon_{\alpha\beta}$. Thus, the Riemann tensor $R^{(2)}_{\alpha\beta\gamma\delta}$ must be proportional to the scalar curvature $R^{(2)}$. We find explicitly 
\begin{eqnarray}
\frac{1}{h}\epsilon_{\alpha\beta}\epsilon_{\gamma\delta}=h_{\alpha\gamma}h_{\beta\delta}-h_{\beta\gamma}h_{\alpha\delta}.
\end{eqnarray}
Hence
\begin{eqnarray}
R_{\alpha\beta\gamma\delta}=\frac{1}{2}(h_{\alpha\gamma}h_{\beta\gamma}-h_{\beta\gamma}h_{\alpha\delta})R^{(2)}.
\end{eqnarray}
We deduce immediately that 
\begin{eqnarray}
R_{\alpha\beta}^{(2)}-\frac{1}{2}h_{\alpha\beta}R^{(2)}=0\Rightarrow \delta\chi=0.
\end{eqnarray}
The Hilbert-Einstein action in two dimensions is therefore invariant under any continuous change in the metric. This does not mean that the Hilbert-Einstein action  vanishes in two dimensions but it means that it depends only on the global topology of the world sheet since it is in fact a boundary term, i.e. the integrand in (\ref{euler}) is a total derivative.

The action (\ref{euler}) is also invariant under Weyl rescalings for a world sheet without a boundary. In the presence of a boundary an additional boundary term is needed \cite{Polchinski:1998rq}. 

The action $S_1+\Phi\chi$, where $\Phi$ is here a constant, looks like the Hilbert-Einstein action for the $2-$dimensional metric $h_{\mu\nu}$ coupled to $D$ massless scalar field $X_{\mu}$ propagating on the world sheet. These scalar fields also define the embedding of the world sheet in a background target spacetime with metric $g_{\mu\nu}(X)$. 
\subsection{Beta function: finiteness and Weyl invariance}
We start by considering the action $S_1$. This is a nonlinear sigma model in the conformal gauge $h_{\alpha\beta}=\eta_{\alpha\beta}$. In this gauge one must also impose the Virasoro conditions $T_{\alpha\beta}=0$. In the critical dimension the two conditions $T_{++}=T_{--}=0$ are sufficient to define the Hilbert space of physical states without negative norm states. There remains the condition $T_{+-}=0$ which actually holds classically due to the invariance under rescalings. The goal now is to check whether or not there is an anomaly in $T_{+-}$.

The breakdown of scale invariance is due to the fact that we can not regularize the theory in a way which maintains scale or conformal invariance. Even dimensional regularization violates scale invariance. The breakdown of scale invariance can be described by the beta function of the theory which is related to the UV behavior of Feynman diagrams. The more fundamental question here is whether or not the nonlinear sigma model $S_1|_{h=\eta}$ is Weyl invariant. Indeed, Weyl invariance implies global scale invariance which in turns implies the vanishing of the beta function which is equivalent to UV finiteness. Also, we note that the beta function is in fact the trace of the energy-momentum tensor which shows explicitly why finiteness is equivalent to Weyl invariance.

Furthermore, we note that the quantum mechanical perturbation theory is an expansion around small $\alpha^{'}$, while the coupling constants of the theory are given by the metric components $g_{\mu\nu}(X)$ with corresponding beta functions $\beta_{\mu\nu}(X)$. The couplings $g_{\mu\nu}$ are actually functions and thus their associated beta functions are in fact functionals. In the explicit calculation, we will use dimensional regularization in $2+\epsilon$ dimensions.

We expand $X^{\mu}$ around inertial coordinates $X^{\mu}_0$, which do not depend on the world sheet coordinates $\sigma$ and $\tau$, with fluctuations given by Riemann normal coordinates $x^{\mu}$ as  
\begin{eqnarray}
X^{\mu}=X_{\mu}^0+x^{\mu}.
\end{eqnarray}
The metric $g_{\mu\nu}$ then starts as the flat metric $\eta_{\mu\nu}$ with corrections given in terms of the Riemann tensor of the target spacetime by 
\begin{eqnarray}
g_{\mu\nu}=\eta_{\mu\nu}-\frac{1}{3}R_{\mu\alpha\nu\beta}x^{\alpha}x^{\beta}+...\label{gexp}
\end{eqnarray}
The action $S_1$ organizes as an expansion in powers of $x$. The first term which is quadratic is the classical action. The second term which is quartic yields the one-loop correction by contracting two of the $x$'s. The beta function is related to the poles in dimensional regularization which originate from logarithmically divergent integrals. The counter term that must be subtracted from $S_1$ to cancel the logarithmic divergence of the theory is found to be given in terms of the Ricci tensor of the target spacetime by 
\begin{eqnarray}
S=-\frac{1}{12\pi\epsilon}\int d^2\sigma\partial_{\alpha}X^{\mu}\partial^{\alpha}X^{\nu}R_{\mu\nu}(X).
\end{eqnarray}
The $1/\epsilon$ pole is due to the logarithmic divergence of the propagator in two dimensions which behaves as 
\begin{eqnarray}
<x^{\mu}(\sigma)x^{\nu}(\sigma^{'})>=\frac{1}{2\epsilon}\eta^{\mu\nu}~,~\sigma^{'}\longrightarrow\sigma.
\end{eqnarray}
The beta function is extracted precisely from the $1/\epsilon$ pole. We get the beta functions
\begin{eqnarray}
\beta_{\mu\nu}(X)=-\frac{1}{4\pi}R_{\mu\nu}(X).
\end{eqnarray}
The condition for the vanishing of the beta functions is then precisely given by the Einstein equations 
\begin{eqnarray}
R_{\mu\nu}(X)=0.
\end{eqnarray}
We go back to the action $S_1$ and check directly that the condition of Weyl invariance leads to the same result. Again we work in $2+\epsilon$ dimensions. We substitute the conformal gauge $h_{\alpha\beta}=\exp(\phi)\eta_{\alpha\beta}$ in $S_1$. We also substitute the expansion (\ref{gexp}) in $S_1$. The $\phi$ dependence does not vanish in the limit $\epsilon\longrightarrow 0$. Indeed, we find (with $y^{\mu}=(1+\epsilon\phi/2)x^{\mu}$) \cite{Green:1987sp}
\begin{eqnarray}
S=-\frac{1}{2\pi}\int d^2\sigma \phi \partial^{\alpha}y^{\mu}\partial_{\alpha}y_{\mu}-\frac{1}{4\pi}\int d^2\sigma \phi R_{\mu\nu}\partial^{\alpha}y^{\mu}\partial_{\alpha}y^{\nu}.
\end{eqnarray}
This is $\phi$ independent, and hence the theory is Weyl invariant, if and only if $R_{\mu\nu}=0$.

By computing the two-loop correction to the beta function we can deduce the leading string-theoretic correction to general relativity. We find the beta function and the corrected Einstein equations
\begin{eqnarray}
\beta_{\mu\nu}(X)=-\frac{1}{4\pi}\bigg(R_{\mu\nu}+\frac{\alpha^{'}}{2}R_{\mu\lambda\alpha\beta}R_{\nu}^{\lambda\alpha\beta}\bigg)\Rightarrow R_{\mu\nu}+\frac{\alpha^{'}}{2}R_{\mu\lambda\alpha\beta}R_{\nu}^{\lambda\alpha\beta}=0.
\end{eqnarray}
We generalize this result to the nonlinear sigma model $S_1+S_2+S_3$. Again we work in $2+\epsilon$ dimensions in the conformal gauge $h_{\alpha\beta}=\exp(\phi)\eta_{\alpha\beta}$. Now the conditions for Weyl invariance in the limit $\epsilon\longrightarrow 0$ are precisely equivalent to the conditions of vanishing of the one-loop beta functions,  associated with the background fields $g_{\mu\nu}$, $B_{\mu\nu}$ and $\Phi$, and they are given explicitly by 
\begin{eqnarray}
0=R_{\mu\nu}+\frac{1}{4}H_{\mu}^{\lambda\rho}H_{\nu\lambda\rho}-2D_{\mu}D_{\nu}\Phi.
\end{eqnarray}
\begin{eqnarray}
0=D_{\lambda}H^{\lambda}_{\mu\nu}-2D_{\lambda}\Phi H^{\lambda}_{\mu\nu}.
\end{eqnarray}
\begin{eqnarray}
0=4(D_{\mu}\Phi)^2-4D_{\mu}D^{\mu}\Phi+R+\frac{1}{2}H_{\mu\nu\rho}H^{\mu\nu\rho}.
\end{eqnarray}
The $D_{\mu}$ is the covariant derivative with respect to the metric $g_{\mu\nu}$ of the target spacetime manifold, and $H_{\mu\nu\rho}$ is the third rank antisymmetric tensor field strength associated with the gauge field $B_{\mu\nu}$ defined by $H_{\mu\nu\rho}=\partial_{\mu}B_{\nu\rho}+\partial_{\rho}B_{\mu\nu}+\partial_{\nu}B_{\rho\mu}$, and therefore it must be invariant under the gauge transformations (\ref{GT}). The above conditions are in fact the equations of motion derived from the action 
\begin{eqnarray}
S=-\frac{1}{2\kappa^2}\int d^dx\sqrt{-g}\exp^{-2\Phi}\bigg(R-4D_{\mu}\Phi D^{\mu}\Phi+\frac{1}{12}H_{\mu\lambda\rho}H^{\mu\lambda\rho}\bigg).
\end{eqnarray}
%$\kappa$ is the string coupling constant which will be defined below.

\subsection{String perturbation expansions}
The Euler characteristic of a compact Riemann surface $M$ of genus $g$ is given,  in terms of the number $n_h$ of handles, the number $n_b$ of boundaries and the number of cross-caps $n_c$ of the surface, by 
\begin{eqnarray}
\chi&=&2(1-g)~,~2g=2n_h+n_b+n_c
\end{eqnarray}
If we choose $\Phi$ to be a constant, viz $\Phi=\lambda={\rm constant}$, then the Euclidean path integral with action $S_1+S_3$ on a surface of genus $g$ becomes 
\begin{eqnarray}
Z_g=\lambda^{-2(g-1)}\int [dXdh]\exp(-S_1).\label{pf}
\end{eqnarray}
The full partition function is then given by the sum over world sheets
\begin{eqnarray}
Z=\sum_gZ_g.
\end{eqnarray}
The coefficient $\lambda$ is the string coupling constant controlled by the Euler action or more precisely by the expectation value of the dilaton field, viz
\begin{eqnarray}
\kappa\equiv g_0^2\sim \exp(\lambda).
\end{eqnarray}
This is not a free parameter in the theory. Obviously the number of distinct topologies characterized by $g$ is very small compared to the number of Feynman diagrams.

We would like to show this result in a different way for its great importance. We consider a tree level scattering process involving $M$ gravitons. The $M$ external gravitons contribute $M-2$ interaction vertices while each loop contributes two vertices $\kappa$ (see figure $3.9$ of \cite{Green:1987sp}). The process is described by a Riemann surface $M$ of genus $g$. The genus $g$ of the surface is precisely the number of loops. This is because 

\begin{eqnarray}
(\kappa)^{M-2}(\kappa^2)^g=\kappa^M\kappa^{-2(1-g)}.
\end{eqnarray}
The factor $\kappa^M$ can be absorbed in the normalization of external vertex operators. Thus we obtain for the diagram or the Riemann surface the factor
\begin{eqnarray}
\kappa^{-\chi}.
\end{eqnarray}
This is exactly the factor appearing in the partition function (\ref{pf}).

There are four possible string theory perturbation expansions, corresponding to different sum over world sheets, depending on whether  the fundamental strings are oriented or unoriented and whether or not the theory contains open strings in addition to closed strings. Thus, we can not have a consistent string theory perturbation expansion with open strings alone. Strings can be unoriented because of the presence of the so-called orientifold plane whereas strings can be open because of the presence of $D-$branes. We list the four consistent string theories with their massless spectra and their allowed world sheet topologies:
\begin{itemize}
\item {\bf Closed oriented strings: $g_{\mu\nu}$, $B_{\mu\nu}$, $\Phi$. All oriented surfaces without boundaries.}

These are found in type II superstring theories and heterotic string theories. Here we can only have closed and oriented world sheets which have $n_b=n_c=$ and hence $g=n_h$. The genus is precisely the number of loops. Since $g=n_h$ there exists one single topology (one string Feynman graph) at each order of perturbation theory. This string Feynman graph contains all the field theory Feynman diagrams which are generated, with their enormous number,  in the singular limit where handles become too long compared to their circumferences and thus can be approximated by lines. The perturbation expansion of closed oriented strings are UV finite.

\item {\bf Closed unoriented strings: $g_{\mu\nu}$, $\Phi$. All surfaces without boundaries.}
\item {\bf Closed and open oriented strings: $g_{\mu\nu}$, $B_{\mu\nu}$, $\Phi$, $A_{\mu}$. All oriented surfaces with any number of boundaries.}
\item {\bf Closed and open unoriented strings: $g_{\mu\nu}$, $\Phi$. All surfaces with any number of boundaries.}

These are found in type I superstring theories.  In this case world sheets are Riemann surfaces with boundaries and cross-caps as well as handles. These theories are also UV finite but the cancellation between string Feynman diagrams of the same Euler class is more delicate.
\end{itemize}
%We discuss the first and third class in some more detail:
%\paragraph{Closed oriented strings:} 

%\paragraph{Closed and open unoriented strings:}

\subsection{Spectrum of type II string theory}

For the closed superstring we have left movers and right movers. 
The left movers can be in the Ramond R sector or the Neveu-Schwarz NS sector and similarly for the right movers. 
In other words, we have four sectors: R-R, NS-NS, R-NS and NS-R. In the NS sector the GSO condition means that 
we project on states with positive G-parity in order to eliminate the  tachyon whereas in the R sector we project 
on states with positive or negative G-parity depending on the chirality of the ground state. Thus, we can obtain two 
different theories depending on whether the left and right movers in the R sectors have the same or opposite G-parity. 

These two theories are:
\begin{enumerate}
\item{\bf Type IIA}: In this case the R sector of the left movers and the R sector of the right movers are distinct, i.e. they have opposite chiralities. Again, the ground state of the left handed movers in the R sector is a Majorana-Weyl spinor with positive chirality $|0,k,+\rangle_R$ while in the NS sector the ground state is the vector field $b_{-1/2}^i|0,k\rangle_{NS}$. Then, the ground states of the right handed movers are given by $|0,k,-\rangle_R$ and $b_{-1/2}^i|0,k\rangle_{NS}$. The ground states of the closed RNS superstring in its four sectors are then given by
\begin{eqnarray}
|0,k,+\rangle_R\otimes |0,k,-\rangle_R~,~{\rm R-}{\rm R}.
\end{eqnarray}  
\begin{eqnarray}
b_{-1/2}^i|0,k\rangle_{NS}\otimes b_{-1/2}^i|0,k\rangle_{NS}~,~{\rm NS-}{\rm NS}.
\end{eqnarray} 
\begin{eqnarray}
b_{-1/2}^i|0,k\rangle_{NS}\otimes |0,k,-\rangle_R~,~{\rm NS-}{\rm R}.
\end{eqnarray}  
\begin{eqnarray}
|0,k,+\rangle_R\otimes  b_{-1/2}^i|0,k\rangle_{NS}~,~{\rm R-}{\rm NS}.
\end{eqnarray} 
We write this tensor product as
\begin{eqnarray}
(8_v\oplus 8_{s_+})\otimes (8_v\oplus 8_{s_-})=(1\oplus 28\oplus 35_v\oplus 8_v\oplus 56_v)_B\oplus(8_{s+}\oplus 8_{s-}\oplus 56_{s+}\oplus 56_{s-})_F.\nonumber\\
\end{eqnarray}
Explicitly, we have
 \begin{eqnarray}
{\rm NS-}{\rm NS}&=&(1\oplus 28\oplus 35_v)_B\nonumber\\
&=&\phi({\rm scalar~dilaton})\oplus A_{\mu\nu}({\rm antisymmetric~2-form~gauge~field})\nonumber\\
&\oplus& g_{\mu\nu}({\rm symmetric~traceless~rank-two~tensor~graviton}).
\end{eqnarray}
 \begin{eqnarray}
{\rm R-}{\rm R}&=&(8_v\oplus 56_v)_B\nonumber\\
&=&A_{\mu}({\rm gauge~field})\oplus A_{\mu\nu\lambda}({\rm antisymmetric~3-form~gauge~field}).
\end{eqnarray}
\begin{eqnarray}
{\rm NS-}{\rm R}&=&(8_{s+}\oplus 56_{s+})_F\nonumber\\
&=&\psi_+({\rm spin~half~dilatino})\oplus \chi_{+}({\rm spin~three~half~gravitino}).
\end{eqnarray}
\begin{eqnarray}
{\rm R-}{\rm NS}&=&(8_{s-}\oplus 56_{s-})_F\nonumber\\
&=&\psi_-({\rm spin~half~dilatino})\oplus \chi_{-}({\rm spin~three~half~gravitino}).
\end{eqnarray}
This is the particle content of type II A supergravity in $10$ dimensions. And it is the particle content obtained from the dimensional reduction of $11-$dimensional supergravity. Since we have two dilatinos and two gravitinos we have ${\cal N}=2$ supersymmetry. 
\item{\bf Type IIB}: In this case the R sector of the left movers and the R sector of the right movers have the same chirality taken $+$ for concreteness. The ground state of the left handed movers in the R sector is a Majorana-Weyl spinor with positive chirality $|0,k,+\rangle_R$ while in the NS sector the ground state is the vector field $b_{-1/2}^i|0,k\rangle_{NS}$. Similarly, the ground states of the right handed movers are $|0,k,+\rangle_R$ and $b_{-1/2}^i|0,k\rangle_{NS}$. The ground states of the closed RNS superstring in its four sectors are then given by
\begin{eqnarray}
|0,k,+\rangle_R\otimes |0,k,+\rangle_R~,~{\rm R-}{\rm R}.
\end{eqnarray}  
\begin{eqnarray}
b_{-1/2}^i|0,k\rangle_{NS}\otimes b_{-1/2}^i|0,k\rangle_{NS}~,~{\rm NS-}{\rm NS}.
\end{eqnarray} 
\begin{eqnarray}
b_{-1/2}^i|0,k\rangle_{NS}\otimes |0,k,+\rangle_R~,~{\rm NS-}{\rm R}.
\end{eqnarray}  
\begin{eqnarray}
|0,k,+\rangle_R\otimes  b_{-1/2}^i|0,k\rangle_{NS}~,~{\rm R-}{\rm NS}.
\end{eqnarray} 
We write this tensor product as
\begin{eqnarray}
(8_v\oplus 8_{s_+})\otimes (8_v\oplus 8_{s_+})=(1\oplus 28\oplus 35_v\oplus 1\oplus 28\oplus 35_{+})_B\oplus(8_{s+}\oplus 8_{s+}\oplus 56_{s+}\oplus 56_{s+})_F.\nonumber\\
\end{eqnarray}
Explicitly, we have
 \begin{eqnarray}
{\rm NS-}{\rm NS}&=&(1\oplus 28\oplus 35_v)_B\nonumber\\
&=&\phi({\rm scalar~dilaton})\oplus A_{\mu\nu}({\rm antisymmetric~2-form~gauge~field})\nonumber\\
&\oplus& g_{\mu\nu}({\rm symmetric~traceless~rank-two~tensor~graviton}).
\end{eqnarray}
 \begin{eqnarray}
{\rm R-}{\rm R}&=&(1\oplus 28\oplus 35_{+})_B\nonumber\\
&=&\phi({\rm scalar~field})\oplus A_{\mu\nu}({\rm antisymmetric~2-form~gauge~field})\nonumber\\
&\oplus& A_{\mu\nu\alpha\lambda}({\rm antisymmetric~4-form~gauge~field~with~self-dual~field~strength}).\nonumber\\
\end{eqnarray}
\begin{eqnarray}
{\rm NS-}{\rm R}&=&(8_{s+}\oplus 56_{s+})_F\nonumber\\
&=&\psi_+({\rm spin~half~dilatino})\oplus \chi_{+}({\rm spin~three~half~gravitino}).
\end{eqnarray}
\begin{eqnarray}
{\rm R-}{\rm NS}&=&(8_{s+}\oplus 56_{s+})_F\nonumber\\
&=&\psi_-({\rm spin~half~dilatino})\oplus \chi_{-}({\rm spin~three~half~gravitino}).
\end{eqnarray}
This is the particle content of type II B supergravity in $10$ dimensions which can not be obtained by 
dimensional reduction of $11-$dimensional supergravity. 
\end{enumerate}

\section{On Dp-branes and T-duality}
\subsection{Introductory remarks} 
The p-branes are $p-$dimensional non-perturbative stable configurations which can carry generalized
conserved charges. These charges can obviously act as sources for antisymmetric tensor gauge
fields with $p + 1$ indices. For example the 0-brane which is a point
particle can have an electric and magnetic charges which generate the
 electromagnetic field.  The magnetic dual of a $p-$brane
is a $(D - p - 4)-$brane. For example in $D=10$ the magnetic dual of a 0-brane
is a 6-brane. The  Dirichlet p-branes or Dp-branes are p-branes which are
characterized by Dirichlet boundary conditions for open strings terminating on them. M-branes, and NS-branes are also p-branes.  The NS5-brane
is special since it is the magnetic dual of the fundamental string in the heterotic and type II superstring
theories. See the short review \cite{Schwarz:2008kd}.

We note that the D1-brane configuration is a string which we also call
 a D-string. D-strings carry an R-R charge and not an NS-NS charge as
opposed to the fundamental type IIB superstring which acts as a
source for the usual two-form B-field of the NS-NS sector and not as
 a source for the two-form C-field of the R-R sector. In general D-branes
will also carry an R-R charge, i.e. they act as sources for the corresponding R-R
 $(p+1)-$forms \cite{Polchinski:1995mt}.

The stable Dp-branes in type II superstring theory, which come with even values of $p$ in type IIA and odd values of $p$ in type IIB, preserve $16$ supersymmetries and as such they are called half-BPS Dp-branes. They carry conserved R-R charges and the corresponding open string spectrum is free of tachyons.

Let us now consider type II superstring theory which contains only closed strings. The presence of a Dp-brane modifies the allowed boundary conditions of the strings. Both Neumann and Dirichlet boundary conditions are now
allowed. Hence in addition to the closed strings we can have open strings whose endpoints are fixed on the Dp-brane.

These open strings describe therefore the  excitation of the Dp-brane and their
quantization is clearly identical to the quantization of the ordinary open
superstrings. An elementary exposition of this result can be found in \cite{Zwiebach:2004tj}.

On the other hand, the massless modes of type I open superstrings consist of a
Majorana-Weyl spinor $\psi $ coming from the Ramond sector and a gauge
field $A_a$ coming from the
Neveu-Schwarz sector. Their dynamics is given at low energy by a supersymmetric
U(1) gauge theory in $10$ dimensions. The field $A_a$ is a function of only the zero modes of the coordinates $x^0,x^1,x^2,...,x^p$ since, by the Dirichlet boundary conditions $x^{p+1}=...=x^{9}=0$ at $\sigma=0,\pi$, the zero modes of the other coordinates $x^{p+1}$,...,$x^{9}$ must vanish. The reduced massless vector field $A_a$, $a=0,1,...,p$, behaves therefore as a U(1) vector field on the p-brane
world volume while  ${X}_{a-p}\equiv A_a$ for $a=p+1,...,9$  behave as
scalar fields normal to the p-brane. Hence, these scalar fields describe
 fluctuations of the position of the p-brane. 

Therefore at low energy the theory on the $(p+1)-$dimensional world volume of the Dp-brane is the reduction to $p+1$
dimensions of $10-$dimensional supersymmetric U(1) gauge theory \cite{Witten:1995im}. Generalization of this fundamental result to the case of $N$ coincident Dp-branes is straightforward. At low energy the theory on the $(p+1)-$dimensional world-volume of  $N$ coincident Dp-branes is the reduction to $p+1$ dimensions of $10-$d supersymmetric U(N) gauge theory. When the velocities and/or the string coupling are not
small the minimal supersymmetric Yang-Mills is replaced by the
 supersymmetric Born-Infeld action \cite{Leigh:1989jq}.

\subsection{Coupling to abelian gauge fields}
We start by writing Maxwell's equations in the notation of differential forms. See for example the classic and pedagogical exposition \cite{Eguchi:1980jx}. The gauge field $A_{\mu}$ is a one-form $A_1=A_{\mu}dx^{\mu}$ whereas the field strength is a two-form $F_2=dA_1=F_{\mu\nu}dx^{\mu}\wedge dx^{\nu}/2$. Maxwell's equations in the presence of electric and magnetic charges given respectively by the electric and magnetic currents $J_{\mu}^e=(\rho_e,\vec{J}_e)$ and $J_{\mu}^m=(\rho_m,\vec{J}_m)$ are given by the equations 
\begin{eqnarray}
dF=*J_m~,~d*F=*J_e.\label{me}
\end{eqnarray}
The $J_e$ and $J_m$ are the electric and magnetic one-form currents defined by $J=J_{\mu}dx^{\mu}$. The $*$ is the duality transformation or Hodge star which converts $p-$forms into $(D-p)-$forms where $D$ is the dimension of spacetime. In the absence of magnetic charges, the first equation in (\ref{me}) is the homogeneous Bianchi identity which is a geometric equation whereas the second equation is Euler inhomogeneous equation which is a dynamical equation. 

The $U(1)$ gauge transformations in the notation of forms is given by $\delta A_n=d\Lambda_{n-1}$. In our case $n=1$ and thus $\Lambda_0$ is a zero-form (function) and $\delta$ is the adjoint of the exterior derivative $d$ which is given on $p-$forms by the relation $\delta=(-1)^{pD+d+1}*d*$. We have by construction $d^2=\delta^2=0$. The electric and magnetic charges in four dimensions are defined by the integrals
\begin{eqnarray}
e=\int_{S^2} *F~,~g=\int_{S^2} F.
\end{eqnarray}
The electric and magnetic charges are related by the celebrated Dirac quantization condition $eg/2\pi\in{\bf Z}$ (Dirac considered the motion of an electric charge in the field of a magnetic monopole and demanded that the wave function can be consistently defined). The electromagnetic duality is given by $F\leftrightarrow *F$ and $e\leftrightarrow g$.

Generalization to p-branes in $D$ dimensions is straightforward (the point particle is a 0-brane).  The world volume of a Dp-brane is a $(p+1)-$dimensional spacetime. The gauge field living on this world volume is an $n-$form $A_n$ where $n=p+1$ (the gauge field living on the world line of a particle is a one-form $A_1$). Thus the electric coupling of the $n-$form gauge field $A_n$ to the world volume of the Dp-brane is given by
\begin{eqnarray}
S=\mu_p\int A_{p+1}.
\end{eqnarray}
For a point particle this is the usual interaction $S=e\int A_{1}=e\int d\tau A_{\mu} dx^{\mu}/d\tau$. Thus $\mu_p$ is the electric charge of the Dp-brane. This electric charge is given in terms of the field strength $(n+1)-$form $F_{n+1}=dA_n$ by the obvious relation 
\begin{eqnarray}
\mu_p=\int * F_{p+2}.
\end{eqnarray}
The dual $*F_{p+2}$ of the $(p+2)-$form field strength $F$ is a $(D-p-2)-$form. Thus the integral in the above formula is over a sphere $S^{D-p-2}$ which is the surface that surrounds a p-brane in $D$ dimensions.

The magnetic dual of the electrically charged p-brane will carry a magnetic charge $\nu_p$ computed obviously by the integral 
\begin{eqnarray}
\nu_p=\int F_{p+2}.
\end{eqnarray}
The integral is over a sphere $S^{p+2}$. In the same way that the sphere $S^{D-p-2}$ surrounds a p-brane, the sphere $S^{D-q-2}=S^{p+2}$ must surround a q-brane where $q=D-p-4$. Hence the magnetic dual of a p-brane in $D$ dimensions is a $D-p-4$ brane. For example, the magnetic dual of a D0-brane is a $D-4$ brane (in $10$ dimensions this is the D6-brane).

The electric and magnetic charges $\mu_p$ and $\nu_p=\mu_{D-p-4}$ must also satisfy the Dirac quantization condition, viz $\mu_p\mu_{D-p-4}/2\pi\in {\bf Z}$. 

%\section{On T-duality and Dp-barnes}
In the remainder of this section we will simply follow the presentations \cite{Johnson:2003gi} and \cite{Becker:2007zj}.
\subsection{Symmetry under the exchange of momentum and winding}
Let us consider bosonic string theory in a spacetime compactified on a circle of radius $R$. A closed string wrapped around the circle can be contracted to a point and thus it is a topologically stable configuration characterized by a winding number $w\in Z$. The coordinate $X^{25}$ of the string along the circle must then be periodic such that
\begin{eqnarray}
X^{25}(\sigma+\pi,\tau)=X^{25}(\sigma,\tau)+2\pi R w.
\end{eqnarray}
For the other coordinates  $X^{\mu}$, $\mu=0,...,24$, the boundary condition is the usual periodic boundary condition with $w=0$. Thus the mode expansion along these directions remains unchanged whereas along the circular $25$th direction it becomes given by ($l_s^2=2\alpha^{\prime}$)

\begin{eqnarray}
X^{25}=x^{25}+2\alpha^{\prime} p^{25}\tau+2Rw\sigma +\frac{i}{2}l_s\sum_{n\neq 0}\frac{1}{n}e^{-2in\tau}\big({\alpha}_n^{25}e^{2in\sigma}+\tilde{\alpha}_n^{25}e^{-2in\sigma}\big).
\end{eqnarray}
Since the wave function will contain the factor $\exp(ip^{25}x^{25})$, and since $x^{25}$ is compact and periodic, we conclude that the momentum $p^{25}$ must be quantized as 
\begin{eqnarray}
p^{25}=\frac{k}{R}~,~k\in Z.
\end{eqnarray}
The quantum number $k$ is the so-called Kaluza-Klein excitation number. We split the above solution into left and right movers as usual, viz (with $\alpha_0^{25}=\tilde{\alpha}_0^{25}=l_sp^{25}/2$ and $\tilde{x}^{25}$ is an arbitrary constant)
\begin{eqnarray}
X_R^{25}(\tau-\sigma)=\frac{1}{2}(x^{25}-\tilde{x}^{25})+\sqrt{2\alpha^{\prime}}\alpha_0^{25}(\tau-\sigma)+\frac{i}{2}l_s\sum_{n\neq 0}\frac{1}{n}\alpha_n^{25}\exp(-2in(\tau-\sigma)).
\end{eqnarray}
\begin{eqnarray}
X_L^{25}(\tau+\sigma)=\frac{1}{2}(x^{25}+\tilde{x}^{25})+\sqrt{2{\alpha}^{\prime}}\tilde{\alpha}_0^{25}(\tau+\sigma)+\frac{i}{2}l_s\sum_{n\neq 0}\frac{1}{n}\alpha_n^{25}\exp(-2in(\tau-\sigma)).
\end{eqnarray}
The zero modes are given explicitly by 
\begin{eqnarray}
\sqrt{2\alpha^{\prime}}\alpha_0^{25}=\alpha^{\prime}\frac{k}{R}-wR~,~\sqrt{2\alpha^{\prime}}\tilde{\alpha}_0^{25}=\alpha^{\prime}\frac{k}{R}+wR.
\end{eqnarray}
The mass relation in the uncompactified $24+1$ dimensions (where now the winding number $k$ labels different particle species) is given by $M^2=-p^{\mu}p_{\mu}$, $\mu=0,...,24$. Thus
\begin{eqnarray}
M^2=\frac{2}{\alpha^{\prime}}(\alpha_0^{25})^2-p^{\mu}p_{\mu}&=&\frac{2}{\alpha^{\prime}}(\alpha_0^{25})^2+\frac{4}{\alpha^{\prime}}(N-1)\nonumber\\
&=&\frac{2}{{\alpha}^{\prime}}(\tilde{\alpha}_0^{25})^2+\frac{4}{\alpha^{\prime}}(\tilde{N}-1).
\end{eqnarray}
By taking the sum and the difference we get 
\begin{eqnarray}
\alpha^{\prime}M^2
&=&\frac{\alpha^{\prime}k^2}{R^2}+\frac{w^2R^2}{\alpha^{\prime}}+2(N+\tilde{N}-2).
\end{eqnarray}
\begin{eqnarray}
N-\tilde{N}=kw.
\end{eqnarray}
These formulas are invariant under the transformations 
\begin{eqnarray}
k\leftrightarrow w~,~R\leftrightarrow \tilde{R}=\alpha^{\prime}/R.
\end{eqnarray}
Thus the theory on a circle of radius $R$ with momentum modes $k$ and winding modes $w$ is equivalent to the theory on a circle of radius $\tilde{R}$ with momentum modes $w$ and winding modes $k$. This profound property is called T-duality and it goes beyond perturbative bosonic string theory to a non-perturbative symmetry of supersymmetric string theory. 

We remark that in the limit of large $R$ the momentum modes become lighter while winding modes become heavier. Thus, in the strict limit  $R\longrightarrow \infty$ only the states with zero winding $w=0$ and all values of momentum $k$ will survive forming a continuum. This is obviously the uncompactified theory.

Let us now consider the limit of small $R$ where momentum modes become heavier while winding modes become lighter. In this case only the states with zero momentum $k=0$ and all values of the winding number $w$ will survive, in the strict limit  $R\longrightarrow 0$, forming also a continuum. In other words, we end up with an effective uncompactified direction in the limit $R\longrightarrow 0$ as well. This is not the expected result of dimensional reduction found in field theory and also found in open string theory which should occur in the limit $R\longrightarrow 0$.

Under the above duality we have $\alpha_0^{25}\longrightarrow -\alpha_0^{25}$ and $\tilde{\alpha}_0^{25}\longrightarrow \tilde{\alpha}_0^{25}$. As it turns out, the theory is symmetric under the full exchange of the right moving and left moving parts of the compact direction $X^{25}$, viz
\begin{eqnarray}
X_L^{25}(\tau+\sigma)\longrightarrow X_L^{25}(\tau+\sigma)~,~X_R^{25}(\tau-\sigma)\longrightarrow -X_R^{25}(\tau-\sigma).
\end{eqnarray}
Thus the $25$th coordinate is mapped as 
\begin{eqnarray}
X^{25}=X_L^{25}+X_R^{25}=x^{25}+2\alpha^{\prime}\frac{k}{R}\tau+2Rw\sigma+...\longrightarrow \tilde{X}^{25}=X_L^{25}-X_R^{25}=\tilde{x}^{25}+2\alpha^{\prime}\frac{k}{R}\sigma+2Rw \tau+...\nonumber\\
\end{eqnarray}
The coordinate $\tilde{x}^{25}$ parametrizes the dual circle with period $2\pi\tilde{R}$ in the same way that the coordinate $x^{25}$ parametrizes the original circle with period $2\pi R$.
\subsection{Symmetry under the exchange of Neumann and Dirichlet}
We consider now bosonic open string theory in $26$ dimensions. As we know the requirement of Poincare invariance implies that the ends of the open strings must obey the Neumann boundary conditions 
\begin{eqnarray}
\frac{\partial}{\partial\sigma}X^{\mu}|_{\sigma=0,\pi}=0.
\end{eqnarray}
This holds in all directions, i.e. $\mu=0,...,25$. We compactify now the theory as before on a circle of radius $R$. In this case there is no winding modes attached to the open string. Indeed, the general solution for the $25$th coordinate satisfies  Neumann boundary conditions  and thus it must be given by the usual formula 
 \begin{eqnarray}
X^{25}=x^{25}+l_s^2p^{25}\tau+il_s\sum_{n\neq 0}\frac{1}{n}\alpha_n^{25}\exp(-in\tau)\cos n\sigma.
\end{eqnarray}
By splitting this solution into left moving and right moving parts $X_{R}^{25}$ and $X_{L}^{25}$, and then applying the T-duality transformation $X_R^{25}\longrightarrow \tilde{X}_R^{25}=-X_R^{25}$ and $X_L^{25}\longrightarrow \tilde{X}_L^{25}=X_L^{25}$, we obtain the solution 
 \begin{eqnarray}
\tilde{X}^{25}=\tilde{x}^{25}+l_s^2p^{25}\sigma+l_s\sum_{n\neq 0}\frac{1}{n}\alpha_n^{25}\exp(-in\tau)\sin n\sigma.
\end{eqnarray}
We note the following:
\begin{itemize}
\item The T-dual theory has no momentum in the $25$th direction since there is no $\tau$ dependence.
\item The T-dual string satisfies  Dirichlet boundary conditions in the $25$th direction, viz (where $p^{25}=k/R$ and $\tilde{R}=\alpha^{\prime}/R$)
\begin{eqnarray}
\tilde{X}^{25}|_{\sigma=0}=\tilde{x}^{25}~,~\tilde{X}^{25}|_{\sigma=\pi}=\tilde{x}^{25}+2\pi k\tilde{R}.
\end{eqnarray}
Thus, the ends of the open strings are fixed on the above wall on the dual circle of radius $\tilde{R}$. This corresponds to a hyperplane in spacetime which is precisely the so-called Dirichlet p-brane or Dp-brane for short. In the above case $p=24$ whereas in the original case where there was no compactified directions $p=25$. In the general  case where there are $n$ compactified directions we have $p=25-n$. 
\item The T-dual string wraps $k$ times around the dual circle of radius $\tilde{R}$ in the $25$th direction. The string remains open though since it lives in more dimensions. Again we observe that the momentum $k$ has become winding under T-duality. More importantly, this winding is topologically stable since the ends of the string are fixed. 
\end{itemize}
In summary, under T-duality a bosonic open string (with momentum and no winding) satisfying Neumann boundary conditions on a circle of radius $R$ is transformed into an open string (without momentum and with winding) satisfying Dirichlet boundary conditions on the dual circle of radius $\tilde{R}=\alpha^{\prime}/R$. The dual strings have their ends  fixed  on the D24-brane $\tilde{X}^{25}=\tilde{x}^{25}$ and they wrap around the dual circles an integer number of times.

Remark also that ordinary open string theory should be thought of as a theory of open strings ending on a D25-brane (a spacetime filling D-brane). T-duality acting (obviously along a parallel direction) on this D25-brane has produced a D24-brane. T-duality acting along the dual circle (which is the  perpendicular direction to the D24-brane) will take us back to the D25-brane. 

In general, T-duality acting along a parallel direction to a Dp-brane will produce a D(p$-1$)-brane while acting on a perpendicular direction will produce a D(p$+1$)-brane.  
\subsection{Chan-Paton factors}
Dp-branes carry background gauge fields on their world volumes. The end points of open strings terminating on Dp-branes are seen as charged particles by these gauge fields. We are thus led to the case of open strings with additional degrees of freedom at their end points which are called Chan-Paton charges.

The Chan-Paton charges are additional degrees of freedom carried by the open string at its end points which preserve spacetime Poincar\'e invariance and world sheet conformal invariance. They have zero Hamiltonian and hence they are background degrees of freedom, i.e. non dynamical. 

For example, if we want to describe oriented strings with $N$ additional degrees of freedom at their end points, we should then consider the gauge group $U(N)$. We can place at the end point $\sigma=0$ the fundamental representation ${\bf N}$ of the group $U(N)$ whereas at the end point $\sigma=\pi$ we place the antifundamental representation $\bar{\bf N}$. The open string states will then be labeled by the Fock space states $\phi$ and the momentum $k$, as usual, but also by two indices $i$ and $j$ running from $1$ to $N$ characterizing the Chan-Paton charges at the two ends $\sigma=0$ and $\sigma=\pi$ of the strings. We have then
\begin{eqnarray}
|\phi,k\rangle\longrightarrow |\phi,k,ij\rangle.
\end{eqnarray} 
By construction this state transforms under $U(1)_i$ as a quark of charge $+1$ whereas it transforms under $U(1)_j$ as an antiquark of charge $-1$. An arbitrary string state is then described by a linear combination of these states given by means of $N^2$ hermitian matrices $\lambda_{ij}^a$, $a=1,...,N^2$ (Chan-Paton matrices) as 
\begin{eqnarray}
|\phi,k,a\rangle=\sum_{i,j=1}^N|\phi,k,ij\rangle\lambda^a_{ij}.
\end{eqnarray} 
These states are called Chan-Paton factors. Since the Chan-Paton charges are non dynamical, in any open string scattering process, the right end of string number $a$ associated with the matrix $\lambda^a_{ij}$ is in the same state as the left end of string number $b$ associated with the matrix $\lambda^b_{kl}$ and so on, and hence summing over all possible values of the indices $i$, $j$, $k$, $l$,... will produce a trace of the product of  Chan-Paton factors, viz $Tr\lambda^a\lambda^b...$. These traces are clearly invariant under the global $U(N)$ world sheet symmetry
\begin{eqnarray}
\lambda^a\longrightarrow U\lambda^a U^{-1}.
\end{eqnarray} 
This shows explicitly that the index $i$ at the end point $\sigma=0$ of the open oriented string transforms like a quark under the fundamental representation ${\bf N}$ of $U(N)$ whereas the index $j$ at the end point $\sigma=\pi$ transforms like an antiquark under the antifundamental representation $\bar{\bf N}$. Hence string states become $N\times N$ matrices transforming in the adjoint representation of $U(N)$ which can be seen more clearly by going to vertex operators. We have therefore $N^2$ tachyons,  $N^2$ massless vector fields and so on labeled by the index $a$. In particular, each of the massless vector fields transforms under the adjoint representation ${\bf N}\otimes\bar{\bf N}$ of $U(N)$ and hence the global $U(N)$ world sheet symmetry is promoted to a local $U(N)$ spacetime symmetry.

\subsection{Electromagnetism on a circle and Wislon lines}
In this section we will follow mostly \cite{Zwiebach:2004tj,Szabo:2004uy}.

The Schrodinger equation of a particle with mass $m$ and charge $q$ is invariant under the gauge transformations $U=\exp(iq\chi)\in U(1)$ given explicitly by 
\begin{eqnarray}
\psi\longrightarrow \psi^{\prime}=U\psi~,~A_{\mu}\longrightarrow A_{\mu}^{\prime}=A_{\mu}-\frac{i}{q}(\partial_{\mu}U)U^{-1}.
\end{eqnarray} 
We assume now that there is a compact spatial direction $x$ which is assumed to be a circle and that the components of the vector potential $\vec{A}$ are all zero except the component $A_x$ along the circle. The Wilson line or holonomy of the gauge field is defined by 
\begin{eqnarray}
W=\exp(i w)=\exp(iq\oint dx A_x(x)).
\end{eqnarray} 
The gauge parameter $U$ must be periodic on the circle while the phase $\chi$ is quasi periodic, viz
\begin{eqnarray}
U(x+2\pi R)=U(x)\Rightarrow q\chi(x+2\pi R)=q\chi(x)+2\pi m~,~m\in {\bf Z}.\label{mv}
\end{eqnarray} 
Hence
\begin{eqnarray}
w^{\prime}=w+2\pi m.
\end{eqnarray} 
Thus $w$ is an angle $\theta\in [0,2\pi[$. The Wilson line $W=\exp(i\theta)$ is then gauge invariant. The solutions $\chi$ of (\ref{mv}) are not single valued. Among the infinitely many physically equivalent solutions we can take $\chi$ to be linear in the compact direction, i.e. 
\begin{eqnarray}
q\chi=(2\pi m)\frac{x}{2\pi R}.
\end{eqnarray} 
This solves by construction (\ref{mv}). The transformation of the gauge field becomes
\begin{eqnarray}
qRA_x\longrightarrow qRA_x^{\prime}=qRA_x+m.
\end{eqnarray} 
Thus in order to obtain a non trivial Wilson line we can simply choose constant backgrounds. In particular, we choose $qRA_x$ to be exactly equal to the holonomy angle $\theta$, viz
\begin{eqnarray}
qA_x=\frac{\theta}{2\pi R}.
\end{eqnarray} 
This is a trivial background since locally it is a pure gauge $A_x^{\prime}=0$ or equivalently ($U=\Lambda^{-1}$) 
 \begin{eqnarray}
qA_x=-i\Lambda^{-1}\partial_{\mu}\Lambda~,~\Lambda=\exp(iq \chi)~,~q\chi=\frac{\theta x}{2\pi R}.
\end{eqnarray} 
In other words, a constant gauge field on the circle can be gauged away by a suitable gauge transformation. However, this is only true locally. Since globally the constant background still have non trivial effect due to the compactness of the circle. This is exhibited by the fact that the gauge transformation $\Lambda$ is not single valued. Indeed, we have 
 \begin{eqnarray}
\Lambda(x+2\pi R)=W.\Lambda(x).
\end{eqnarray} 
Thus this constant background gauge field which corresponds to a zero magnetic field everywhere (flat potential $F=0$) and solves the source free Maxwell equations has concrete physical effects (Aharonov-Bohm effect). This effect lies precisely in the holonomy (the Wilson line $W$) which can not be set equal to $1$ by a local gauge transformation  $\Lambda$. Indeed, as the particle loops around the compact direction it picks up the phase factor $W$ due to the trivializing local gauge transformation  $\Lambda$.

Next we solve the Schrodinger equation on the circle. By demanding periodicity of the wave function $\psi(x+2\pi R)=\psi(x)$ we arrive at the solutions $\psi\sim \exp(i kx/R)$ where $k\in{\bf Z}$. The momentum along the compact direction is found to be fractional given by 
\begin{eqnarray}
p=\frac{k}{R}-\frac{\theta}{2\pi R}.
\end{eqnarray} 
This spectrum is invariant under $\theta\longrightarrow \theta+2\pi$ and $k\longrightarrow k+1$.

This result can also be seen by considering the coupling of a point particle of mass $m$ and charge $q$ to an electromagnetic field $A_x$ on a circle of radius $R$ given by the action 

\begin{eqnarray}
S=\int (-m\sqrt{-\dot{X}^{\mu}\dot{X}_{\mu}}+q\dot{X}^{\mu}{A}_{\mu})d\tau.
\end{eqnarray} 
We remark that 
\begin{eqnarray}
\exp(iS)\sim \exp(iq\oint dx A_x(x)).
\end{eqnarray} 
Thus the wave function $\exp(iS)$ is proportional to the Wilson line and hence when the particle moves along the circle the Wilson line calculates the corresponding phase factor. Since $w=q\oint dx A_x(x)$ is periodic we conclude that the wave function $\exp(iS)$ is gauge invariant as we loop around the compact direction a full circle.

In the gauge $\tau=X^0=t$ the above action reduces to ($v_i=\dot{x}_i$) 
\begin{eqnarray}
S=\int \big(-m\sqrt{1-\vec{v}^2}+q(A_0+\vec{v}\vec{A})\big)d\tau.
\end{eqnarray} 
The canonical momentum $P$ associated with the compact direction $x$ is given by 
\begin{eqnarray}
P&=&\frac{\delta S}{\delta \dot{x}}=\frac{m\dot{x}}{\sqrt{1-\vec{v}^2}}+qA_x=p+\frac{\theta}{2\pi R}.
\end{eqnarray} 
On the other hand, the physical wave function $\exp(iPx)$ is periodic and hence we must have $P=k/R$, i.e.
\begin{eqnarray}
\frac{k}{R}=p+\frac{\theta}{2\pi R}.
\end{eqnarray} 

\subsection{The D-branes on the dual circle}
We return now to the case of string theory and we consider a constant $U(N)$ background on the compact direction. More precisely we assume that only the component of the gauge potential along the circle takes a non zero constant value. As we have seen in the case of the point particle a flat potential on a compact direction can still have a non trivial effect analogous to the Aharanov-Bohm effect. Indeed, a constant background on the circle is locally trivial (since it can be removed by a gauge transformation) but globally it is non trivial (since the trivializing gauge transformation is not single valued). Such a topologically non trivial background can be characterized by its Wilson line.

Let us then consider the constant $U(N)$ background along the $25$th direction (bosonic open oriented stings) given by the pure gauge $A_{25}^{\prime}=0$ or equivalently
\begin{eqnarray}
A_{25}=\frac{1}{2\pi R}{\rm diag}(\theta_1,...,\theta_N)=-i\Lambda^{-1}\partial_{25}\Lambda.
\end{eqnarray} 
The wave function picks up a factor $\Lambda^{-1}$ under this gauge transformation. The trivializing local $U(N)$ gauge transformation is given by
\begin{eqnarray}
\Lambda={\rm diag}\bigg(\exp(\frac{iX^{25}\theta_1}{2\pi R}),...,\exp(\frac{iX^{25}\theta_N}{2\pi R})\bigg).
\end{eqnarray} 
This gauge transformation is not single valued since 
\begin{eqnarray}
\Lambda(X^{25}+2\pi R)=W\Lambda(X^{25}).
\end{eqnarray} 
The Wilson line is given explicitly by 
\begin{eqnarray}
W=\exp\big(i\int_0^{2\pi R}dX^{25} A_{25})={\rm diag}\bigg(\exp i\theta_1,...,\exp i\theta_N\bigg).
\end{eqnarray} 
Hence fields must pick a phase factor given precisely by the Wilson line $W$ as we loop around the compact direction.

The above $U(N)$ background gauge field configuration breaks the Chan-Paton $U(N)$ gauge symmetry at the end points of the open strings down to some abelian subgroup of $U(N)$ such as the maximal subgroup $U(1)^N$. The underlying reason is already what we have said which is the fact that we can not trivialize the Wilson line by setting it equal to $1$ by a local gauge transformation. The $U(N)$ group is broken to its maximal abelian subroup $U(1)^N$ when all the holonomy angles $\theta_a$ are distinct.

Let us now consider a string in the Chan-Paton state $|\phi,k,ij\rangle$. The end point $i$ (fundamental rep) of the string picks up a factor $\exp(-i\theta_i X^{25}/2\pi R)$ under the effect of the gauge transformation $\Lambda$ whereas the end point $j$ (antifundamental rep) picks up a factor $\exp(i\theta_j X^{25}/2\pi R)$. The string wave function acquires then a phase $\exp(-i(\theta_i-\theta_j) X^{25}/2\pi R)$ and hence as we loop around the compact direction $X^{25}\longrightarrow X^{25}+2\pi R$ it will pick up the Wilson line
\begin{eqnarray}
|\phi,k,ij\rangle\longrightarrow \exp(i(\theta_j-\theta_i))|\phi,k,ij\rangle.
\end{eqnarray} 
On the other hand, this wave function contains the plane wave $\exp(iPX^{25})$ and thus under the rotation $X^{25}\longrightarrow X^{25}+2\pi R$ it will acquire a phase equal to $\exp(iP 2\pi R)$. Thus
\begin{eqnarray}
P=\frac{k}{R}-\frac{\theta_i-\theta_j}{2\pi R}~,~k\in {\bf Z}.
\end{eqnarray} 
The momentum number is then fractional. When we apply now T-duality these fractional Kaluza-Klein excitation numbers on the circle will be mapped to fractional winding numbers on the dual circle. A fractional winding number means that the open string partially winds around the dual circle since it is connecting two separated D24-branes $i$ and $j$. The angles $\theta_i$ are thus interpreted as the angular positions of $N$ D24-branes on the dual circle. The D24-branes $i$ and $j$ are coincident only when $\theta_i=\theta_j$ in which case we get an integer winding. 

The mode expansions of the open string and its dual can now be found to be given by
\begin{eqnarray}
X^{25}=x^{25}+\theta_i\frac{\alpha^{\prime}}{R}+2\frac{\alpha^{\prime}}{R}\big(k-\frac{\theta_i-\theta_j}{2\pi}\big)\tau+...
\end{eqnarray} 
\begin{eqnarray}
\tilde{X}^{25}_{ij}=\tilde{x}^{25}+\theta_i\tilde{R}+2\tilde{R}\big(k-\frac{\theta_i-\theta_j}{2\pi}\big)\sigma+...
\end{eqnarray} 
The two end points of the dual string are at $\tilde{x}^{25}+\theta_i\tilde{R}$ and $\tilde{x}^{25}+\theta_j\tilde{R}$. These are precisely the locations of the $i$ and $j$ D24-branes respectively. Indeed, the Dirichlet boundary conditions become 
\begin{eqnarray}
\tilde{X}^{25}_{ij}|_{\sigma=\pi}-\tilde{X}^{25}_{ij}|{\sigma=0}=\tilde{R}(2\pi k+\theta_j-\theta_i).
\end{eqnarray} 
Thus the end points of the open strings in the gauge state $i$ are located on the hyperplanes (D24-branes) located at the positions 
\begin{eqnarray}
\tilde{X}^{25}_{ij}=2\pi\alpha^{\prime}(A_{25})_{ii}.
\end{eqnarray} 
The mass relation in the uncompactified $24+1$ dimensions for the open string is given by $M^2=k^2/R^2+(N-1)/\alpha^{\prime}$. In the presence of Wilson lines the spectrum of the $ij$ open string becomes  
\begin{eqnarray}
M_{ij}^2=\frac{1}{R^2}\big(k-\frac{\theta_i-\theta_j}{2\pi}\big)^2+\frac{N-1}{\alpha^{\prime}}.
\end{eqnarray} 
The main observation here is that only diagonal strings (strings starting and ending on the same D24-branes) contain in their spectrum a massless vector field. Hence if all the angles $\theta_i$ are different (no D24-branes coincide) the gauge group is $U(1)^N$ while if all the D24-branes coincide (the angles are all equal) the gauge group is $U(N)$. 

\section{Quantum gravity in two dimensions}
The standard reference for the matrix models of $D=0$ and $D=1$ string theories is the systematic reviews \cite{DiFrancesco:1993cyw,Ginsparg:1993is}. However, the short review \cite{Zarembo:1998uk} is an extremely useful concise description of the relevant points. 

%\subsection{'t Hooft planar limit}

\subsection{Dynamical triangulation}
The full string action includes the Hilbert-Einstein term as well as a cosmological constant for the world sheet metric. The action defines then a theory of $D$ scalar fields $X^{\mu}$ coupled minimally to the world sheet metric $h_{ab}$. Explicitly we have the Euclidean action
\begin{eqnarray}
S=\int d^2\sigma\sqrt{h}\big(\frac{1}{4\pi\alpha^{\prime}}h^{ab}\partial_aX^{\mu}\partial_bX_{\mu}+\frac{\lambda}{4\pi}R+\Lambda\big).
\end{eqnarray} 
The Hilbert-Einstein action in two dimensions is a topological term equals the Euler character, viz
\begin{eqnarray}
\chi=\int d^2\sigma\sqrt{h}\big(\frac{1}{4\pi}R\big)=2-2n,
\end{eqnarray} 
where $n$ is the genus of the world sheet surface ${\cal M}_n$ which is a sphere with $n$ handles. The parameter $\lambda$ is related to the expectation value of the dilaton field and thus it is determined by the string coupling constant, i.e.
\begin{eqnarray}
\exp(\lambda)=g_s=g_o^2.
\end{eqnarray} 
The partition function is then given by 
\begin{eqnarray}
Z=\sum_ng_s^{2n-2} \int [dX^{\mu}][dh_{ab}]\exp\bigg(-\int_{{\cal M}_n} d^2\sigma\sqrt{h}\big(\frac{1}{4\pi\alpha^{\prime}}h^{ab}\partial_aX^{\mu}\partial_bX_{\mu}+\Lambda\big)\bigg).
\end{eqnarray} 
A lattice-like regularization of this theory consists in discretizing the world sheet geometries by dynamical triangulations, i.e. we replace the integration over the world sheet metrics by a summation over world sheet triangulations \cite{Kazakov:1985ds,David:1984tx,Ambjorn:1985az}. Another approach, which we will not pursue here, is given by Liouville theory \cite{Polyakov:1981rd,DHoker:1988pdl}. See also \cite{DiFrancesco:1993cyw} and the extensive list of references therein. 

Let us for simplicity consider the $D=0$ theory which is a pure theory of surfaces with no conformal matter. We have 
\begin{eqnarray}
Z=\sum_ng_s^{2n-2} \int [dh_{ab}]\exp\bigg(-\int_{{\cal M}_n} d^2\sigma\sqrt{h}\Lambda\bigg)=\sum_n\int [dh]\exp(-\Lambda A-\lambda \chi).
\end{eqnarray} 
A random triangulation of a given surface is a discretization of the surface by equilateral triangles. On the plane at each vertex $i$ we find $N_i=0$ triangles meeting since there is no curvature. On a general curved surface at each vertex $i$ we find $N_i$ triangles. If $N_i=6$ there is no curvature at the vertex $i$, if $N_i<6$ there is a positive curvature, and if $N_i>6$ there is a negative curvature. Indeed, the Ricci scalar at the vertex $i$ is given by 
\begin{eqnarray}
R_i=2\pi \frac{6-N_i}{N_i}.
\end{eqnarray} 
This can be verified as follows. Each triangulation is characterized by a number of vertices $V$, a number of edges $E$ and a number of faces $F$. Thus $V=\sum_iN_i$ by construction. However, topologically since each edge is shared by two vertices we must have $2E=V$, and since each face has three edges and each edge is shared by two faces we must have $3F=2E$. The area of the triangulation is given by
\begin{eqnarray}
\int d^2\sigma \sqrt{h}=\sum_i\frac{N_i}{3}=\frac{V}{3}=\frac{2E}{3}=F.
\end{eqnarray} 
This is the total number of triangles. The Hilbert-Einstein term is given by 
\begin{eqnarray}
\int d^2\sigma \sqrt{h}R=\sum_i\frac{N_i}{3}2\pi \frac{6-N_i}{N_i}=4\pi(V-\frac{F}{2})=4\pi(V-E+F)=4\pi\chi.
\end{eqnarray} 
This is then the Euler character as it should be. The above partition function becomes in the discrete given by 
\begin{eqnarray}
Z_{DT}=\sum_{n=0}g_s^{2n-2} \sum_{T_n}\exp(-\Lambda F),\label{dtt}%=\sum_{T}\exp(-\Lambda F-\lambda(V-E+F)),
\end{eqnarray} 
where the summation is over the dynamical triangulations $T_n$ of the surfaces ${\cal M}_n$ which are explicitly constructed above. The continuum limit is defined by 
\begin{eqnarray}
\Lambda\longrightarrow\Lambda_c,
\end{eqnarray} 
where $\Lambda_c$ is independent of the genus $n$. The total number of graphs with a fixed genus $n$ and a fixed number of triangles $F$ increases with $F$ as $\exp(\Lambda_c F)/F^{b_n}$ \cite{Koplik:1977pf}. Thus, at $\Lambda=\Lambda_c$ the contribution from entropy (degeneracy increases greatly, i.e. the number of graphs with fixed $F$ becomes too large) dominates over the contribution from energy (the exponential convergent Boltzmann weight) and as a consequence the partition function diverges in a second order phase transition. This behavior can be characterized by the behavior of  the string susceptibility given by  
\begin{eqnarray}
f=\frac{\partial^2}{\partial\Lambda^2}Z_{DT}=\sum_{n=0}g_s^{2n-2} (\Lambda-\Lambda_c)^{-\gamma_n}~,~\gamma_n=-b_n+3.
\end{eqnarray} 
\subsection{Matrix models of $D=0$ string theory}
We consider now the cubic matrix model 
\begin{eqnarray}
Z_{\alpha}=\int [d\Phi]\exp(-N TrV(\Phi))~,~V=\frac{1}{2}\Phi^2-\frac{\alpha}{3}\phi^3.
\end{eqnarray} 
The propagator is given by 
\begin{eqnarray}
\langle\Phi_{ij}\Phi_{kl}\rangle=\frac{1}{Z_0}\frac{(2\pi)^{N^2/2}}{N}\delta_{il}\delta_{jk}.
\end{eqnarray} 
Thus, the propagator is represented by a double line. These two lines carry arrows in opposite directions because we are dealing with hermitian matrices which will correspond to orientable surfaces. Clearly, three such propagators come together in a matrix 3-point vertex. A typical Feynman diagram is then an oriented two-dimensional surface formed by polygons which are bounded by index loops. The so-called dual diagram is constructed by drawing lines through the centers of the polygons. It is seen that the dual diagram can also be obtained by placing the matrix 3-point vertices inside triangles  and thus by construction the dual diagram is composed of triangles.  In other words, the dual diagram is a dynamical triangulation of some Riemann surface.

Following 't Hooft \cite{tHooft:1973alw} we can organize the diagrammatic expansion in powers of $1/N$ where each order corresponds to a distinct topology. A given Feynman diagram of the cubic matrix model is characterized by $V$ vertex, $E$ propagators (edges) and $F$ loops (faces). The vertex is associated with a factor of $N$, the propagator is associated with a factor of $1/N$ and the loop is associated with a factor of $N$. Thus the Feynman diagram is of order $N^{V-E+F}=N^{\chi}$ where $\chi$ is the Euler character, viz $\chi=2-2n$. The free energy admits then the $1/N$ expansion  
\begin{eqnarray}
F_{\alpha}=\log Z_{\alpha}=\sum_{n=0}N^{2-2n}F_n.
\end{eqnarray} 
The free energy $F_n$  is given by the sum of the connected Feynman diagrams which can be drawn on a sphere with $n$ handles (obviously $Z_{\alpha}$ generates both connected and disconnected diagrams).
 
Since each matrix 3-point vertex is placed inside a triangle the number of vertices $v_G$ in a connected Feynman diagram $G$ of the cubic matrix model is equal to the number of triangles $F$ of the corresponding dynamical triangulation. Hence the area is $A=F=v_G$ (the area of each triangle is $1$). This Feynman diagram $G$ is obviously proportional to $\alpha^{v_G}$. We must also divide by the appropriate symmetry factor, i.e. $\alpha^{v_G}\longrightarrow\alpha^{v_G}/S_G$ where $S_G$ is the order of the discrete symmetry group of the diagram. Indeed, the symmetry group of the Feynman diagram or the dynamical triangulation is exactly the analogue of the isometry group of continuum manifolds \cite{Ginsparg:1993is}. The free energy $F_n$ is then given by 
\begin{eqnarray}
F_n=\sum_G \frac{\alpha^{v_G}}{S_G}.\label{fr0}
\end{eqnarray} 
Thus
\begin{eqnarray}
F_{\alpha}=\log Z_{\alpha}=\sum_{n=0}N^{2-2n}\sum_G \frac{\alpha^{v_G}}{S_G}.
\end{eqnarray}
By comparing with (\ref{dtt}) we get 
\begin{eqnarray}
N=\exp(-\lambda)=\frac{1}{g_s}~,~\exp(-\Lambda)=\alpha.
\end{eqnarray}
In other words, 
\begin{eqnarray}
F_{\alpha}=Z_{DT}.
\end{eqnarray}
The continuum limit is a double scaling limit defined by sending $N\longrightarrow \infty$ and $\alpha\longrightarrow \alpha_c$ keeping fixed the string coupling constant\cite{Brezin:1990rb,Douglas:1989ve,Gross:1989vs}
\begin{eqnarray}
\lambda=\frac{1}{N(\alpha_c-\alpha)^{5/4}}.
\end{eqnarray}
In this limit the partition function diverges signaling a second order phase transition. Indeed, the planar partition function $Z_{\alpha}=N^2Z_{\alpha}^{(0)}+...$ behaves in the limit $\alpha\longrightarrow \alpha_c$ as
\begin{eqnarray}
Z_{\alpha}^{(0)}\sim (\alpha_c-\alpha)^{2-\gamma}\sim -A^{\gamma-2},
\end{eqnarray}
where $A$ is the expectation value of the area, viz $A=\langle F\rangle=\langle v_G\rangle$ and $\gamma$ is the string
susceptibility exponent. For pure quantum gravity $\gamma=-1/2$ (see below).

Generalization of the above construction is straightforward. Instead of random triangulation by means of a cubic matrix model we can have  random polygonulations by means of a general potential of the form
\begin{eqnarray}
V=\sum_{j>1}{\alpha}_j\Phi^j.
\end{eqnarray}
\subsection{Matrix models of $D=1$ string theory}
We can generalize the above construction to strings in higher dimensions by considering multi-matrix models. For strings in $0<D\leq1$ dimension we consider the q-matrix model
\begin{eqnarray}
Z=\int \prod_{i=1}^q[d\Phi_i]\exp\big(-N\sum_i TrV(\Phi_i)+N\sum_{i}Tr\Phi_i\Phi_{i+1}\big).\label{lg}
\end{eqnarray}
The diagrammatic expansion of this model generates discretized surfaces with $q$ different states $\Phi_i$ existing at the vertices  \cite{Ginsparg:1993is}. More precisely, it describes bosonic strings in $0<D\leq 1$ or two-dimensional quantum gravity coupled to conformal matter in $D$ dimensions where $D$ is identified with the central charge of the Virasoro algebra. For example, in the unitary discrete series of conformal field theories which are labeled by an integer $m\geq 2$ we have \cite{DiFrancesco:1993cyw}
\begin{eqnarray}
D=c=1-\frac{6}{m(m+1)}.
\end{eqnarray}
The case $m=2$ gives $D=c=0$ (pure gravity) whereas the case $m=\infty$ gives $D=c=1$ (one boson). Fractional dimensions start with $m=3$ which gives $D=1/2$ (half boson!) corresponding to the Ising model.

As it turns out, the model (\ref{lg}) describes also a scalar field on a one-dimensional lattice when $q\longrightarrow\infty$. The coupling term is a nearest neighbor interaction and hence in the limit $q\longrightarrow\infty$ the partition function becomes

\begin{eqnarray}
Z=\int [d\Phi(t)]\exp\bigg(-N\int dt\big(\frac{1}{2}\dot{\Phi}^2+\frac{m^2}{2}\Phi^2+V_{\rm int}(\Phi)\big)\bigg).\label{d1}
\end{eqnarray}
This model has been solved in \cite{Brezin:1977sv}. For a cubic interaction the free energy $F_n$, given by the sum of the connected Feynman diagrams which can be drawn on a sphere with $n$ handles, is now given by (compare with (\ref{fr0}))
\begin{eqnarray}
F_n=\sum_G \frac{\alpha^{v_G}}{S_G}F_G.\label{fr1}
\end{eqnarray}
The Feynman integral $F_G$ is given explicitly by \cite{Zarembo:1998uk,Ginsparg:1993is}
\begin{eqnarray}
F_G=\int \prod_i\frac {dX_i}{2m}\exp(-m\sum_{\langle ij\rangle}|X_i-X_j|).
\end{eqnarray}
The variables $X_i$ are the values of the string coordinate $X$ at the vertex $i$ and the summation is over links $\langle ij\rangle$ between vertices. Thus the one-dimensional inverse propagator yields precisely the kinetic term for the bosonic field $X$.

In the continuum limit the partition function diverges as before with the leading singular behavior given by $Z(\alpha)\sim (\alpha_c-\alpha)^{2-\gamma}$ where the string susceptibility exponent is given by $\gamma=0$. 

In general, the string susceptibility exponent $\gamma$ for two-dimensional quantum gravity coupled with conformal matter is given in terms of the central charge $c=D$ by the formula 
\begin{eqnarray}
\gamma=\frac{1}{12}(D-1-\sqrt{(D-1)(D-25)})=-\frac{1}{m}.
\end{eqnarray}
Obviously $D=1$ is a barrier since the quantity under the square root becomes negative for $1<D<25$. The existence of this barrier is also related to the presence of a state with a mass squared proportional to $1-D$ in the string spectrum which becomes negative for $D>1$ (the tachyon). As a consequence the string phase is absent for $D>1$. 

\subsection{Preliminary synthesis}
In these notes we will replace the $D=0$ matrix model of string theory with Type IIB Matrix Model (IKKT Matrix Model) whereas the $D=1$ matrix model of string theory will be replaced with M-(atrix) Theory (BFSS Matrix Quantum Mechanics). Generalization to higher dimensions beyond the $D=1$ barrier exists and it starts with Matrix String Theory (DVV Matrix Quantum Gauge Theory). 

Another generalization beyond the $D=1$ barrier and even beyond two-dimensional quantum gravity is provided by our recent proposal on emergent quantum gravity from multitrace matrix models and noncommutative geometry \cite{Ydri:2017riq}.% (see also the last section of this review for a very brief description).

It seems also that a very natural generalization of the theory (\ref{d1}) is provided by the matrix and noncommutative scalar field theories considered in \cite{Ferretti:1995zn,Nishigaki:1996np,Ydri:2013zya}.

\chapter{M-(atrix) Theory and Matrix String Theory}

\section{The quantized membrane}
We start by writing down the action of a $p-$brane in $D-$dimensional spacetime with metric $g_{\mu\nu}=(-1,+1,...,+1)$. The $p-$brane is a $p-$dimensional object moving in spacetime with $p<D$. Thus the local coordinates will be denoted by $\sigma^{\alpha}$, $\alpha=0,...,p-1$, $\sigma^0=\tau$, with a local metric denoted by $h_{\alpha\beta}$.  The $p-$brane will sweep a $(p+1)-$dimensional hyper-volume called the world hyper-volume. The $0-$brane is a point, the $1-$brane is a string, the $2-$brane is a membrane,... which sweep a world line, a world sheet, a world volume,... respectively. The coordinates of the $p-$brane will be denoted by 

\begin{eqnarray}
X^{\mu}=X^{\mu}(\sigma^0,\sigma^1,...,\sigma^{p-1}).
\end{eqnarray} 
The induced metric is immediately given by 
\begin{eqnarray}
G_{\alpha\beta}=g_{\mu\nu}\partial_{\alpha}X^{\mu}\partial_{\beta}X^{\nu}.
\end{eqnarray} 
The Lorentz invariant infinitesimal hyper-volume element is given by 
 \begin{eqnarray}
d\mu_p=\sqrt{-{\rm det}G_{\alpha\beta}}d^{p+1}\sigma.
\end{eqnarray} 
The action of the $p-$brane is then given by (with $T_p$ the $p-$brane tension)
\begin{eqnarray}
S_p=-T_p\int d\mu_p=-T_p\int \sqrt{-{\rm det}G_{\alpha\beta}}d^{p+1}\sigma.
\end{eqnarray} 
The case of the membrane is given by 
\begin{eqnarray}
S_2=-T_2\int d\mu_2=-T_2\int \sqrt{-{\rm det}G_{\alpha\beta}}d^{3}\sigma.
\end{eqnarray} 
This is the Nambu-Goto action. The Polyakov action is given by (using the same symbol)
\begin{eqnarray}
S_2=-T_2^{\prime}\int d^{3}\sigma\sqrt{-h}(h^{\alpha\beta}G_{\alpha\beta}-\Lambda).
\end{eqnarray} 
The addition of the cosmological term is due to the fact that the membrane, as opposed to the string, is not scale invariant. The quantized supermembrane exists in $11$ dimensions in the same sense that the quantized superstring exists in $10$ dimensions \cite{Hoppe:1982,deWit:1988wri,Bergshoeff:1987cm}. The equation of motion with respect to $h_{\alpha\beta}$ is (with $\delta\sqrt{-h}=-\sqrt{-h}h_{\alpha\beta}\delta h^{\alpha\beta}/2$)
\begin{eqnarray}
G_{\alpha\beta}=\frac{1}{2}h_{\alpha\beta}(h^{\alpha\beta}G_{\alpha\beta}-\Lambda).
\end{eqnarray} 
By tracing we get 
\begin{eqnarray}
h^{\alpha\beta}G_{\alpha\beta}=3\Lambda\Leftrightarrow G_{\alpha\beta}=\Lambda h_{\alpha\beta}.
\end{eqnarray} 
Substituting this solution in the Polyakov we get the Nambu-Goto with the identification 
\begin{eqnarray}
2T_2^{\prime}=\sqrt{\Lambda}T_2.
\end{eqnarray} 
The metric $h_{\alpha\beta}$ contains $6$ independent components and the membrane action is invariant under three diffeomorphisms $\sigma^{\alpha}\longrightarrow \sigma^{\prime\alpha}=f^{\alpha}(\sigma)$. Thus three components of the metric can be fixed by a suitable gauge choice. If we suppose now that the topology of the membrane world volume is ${\bf R}\times \Sigma$ where the Riemann surface $\Sigma$ is of fixed topology, then we can fix the components $h_{00}$ and $h_{0i}$ as \cite{Taylor:1999qk}
 \begin{eqnarray}
h_{0i}=0~,~h_{00}=-\frac{4}{\rho^2\Lambda}{\rm det}G_{ij}.
\end{eqnarray} 
With this gauge choice the constraints become 
\begin{eqnarray}
g_{\mu\nu}\partial_0X^{\mu}\partial_iX^{\nu}=0~,~g_{\mu\nu}\partial_0X^{\mu}\partial_0X^{\nu}=-\frac{4}{\rho^2}{\rm det}G_{ij}.
\end{eqnarray} 
We get then 
\begin{eqnarray}
\sqrt{-{\rm det}G_{\alpha\beta}}=-\frac{\rho}{2}g_{\mu\nu}\partial_0X^{\mu}\partial_0X^{\nu}=\frac{2}{\rho}{\rm det}G_{ij}.
\end{eqnarray} 
Thus the membrane action becomes 
\begin{eqnarray}
S_2=\frac{T_2\rho}{4}\int d^2\sigma dt\bigg(g_{\mu\nu}\partial_0X^{\mu}\partial_0X^{\nu}-\frac{4}{\rho^2}{\rm det}G_{ij}\bigg).
\end{eqnarray} 
We introduce a canonical Poisson bracket on the membrane defined by (with $\epsilon^{12}=1$)  
 \begin{eqnarray}
\{f,g\}=\epsilon^{\alpha\beta}\partial_{\alpha}f\partial_{\beta}g.
\end{eqnarray} 
Then it is not difficult to show that 
\begin{eqnarray}
{\rm det}G_{ij}=\frac{1}{2}\{X^{\mu},X^{\nu}\}\{X_{\mu},X_{\nu}\}.
\end{eqnarray} 
The action becomes
\begin{eqnarray}
S_2=\frac{T_2\rho}{4}\int d^2\sigma dt\bigg(g_{\mu\nu}\partial_0X^{\mu}\partial_0X^{\nu}-\frac{2}{\rho^2}\{X^{\mu},X^{\nu}\}\{X_{\mu},X_{\nu}\}\bigg).
\end{eqnarray} 
The second constraint becomes 
\begin{eqnarray}
g_{\mu\nu}\partial_0X^{\mu}\partial_0X^{\nu}=-\frac{2}{\rho^2}\{X^{\mu},X^{\nu}\}\{X_{\mu},X_{\nu}\}.
\end{eqnarray} 
The first constraint $g_{\mu\nu}\partial_0X^{\mu}\partial_iX^{\nu}=0$ leads immediately to 
\begin{eqnarray}
g_{\mu\nu}\{\partial_0X^{\mu},X^{\nu}\}=0.
\end{eqnarray}
The equations of motion deriving from the above action reads 
  \begin{eqnarray}
\partial_0^2X^{\mu}=\frac{4}{\rho^2}\{\{X^{\mu},X^{\nu}\},X_{\nu}\}.
\end{eqnarray}
As in the case of the string, there is here a residual invariance which allows us to fix the gauge further. We choose the light-cone gauge
\begin{eqnarray}
X^{+}(\tau,\sigma_1,\sigma_2)=\tau.
\end{eqnarray}
The light-cone coordinates are defined by 
\begin{eqnarray}
X^{\pm}=\frac{X^0\pm X^{D-1}}{\sqrt{2}}.
\end{eqnarray}
In this gauge the number of degrees of freedom reduce from $D$ to $D-2$ since $X^{+}$ is fixed by the above condition while $X^{-}$ is obtained by solving the constraints which take the form 
\begin{eqnarray}
\partial_0X^{-}=\frac{1}{2}(\partial_0X_a)^2+\frac{1}{\rho^2}\{X_a,X_b\}^2~,~\partial_iX^{-}=\frac{1}{2}\partial_0X^a\partial_iX^a.
\end{eqnarray}
The indices $a$ and $b$ run from $1$ to $D-2$. The Hamiltonian of the remaining transverse degrees of freedom is computed as follows:
 \begin{eqnarray}
{\cal L}_2=\frac{T_2\rho}{4}\bigg(\partial_0X^{-}+\frac{1}{2}(\partial_0X_a)^2-\frac{1}{\rho^2}\{X_a,X_b\}^2\bigg).
\end{eqnarray}
\begin{eqnarray}
X^{-}\longrightarrow P^+=\frac{\delta {\cal L}_2}{\delta (\partial_0X^-)}=\frac{T_2\rho}{4}~,~X^{a}\longrightarrow P^a=\frac{\delta {\cal L}_2}{\delta (\partial_0X^a)}=\frac{T_2\rho}{4}\partial_0X_a.
\end{eqnarray}
\begin{eqnarray}
{\cal H}_2&=&P^+\partial_0X^-+P^a\partial_0X_a-{\cal L}_2\nonumber\\
&=&\frac{T_2\rho}{4}\bigg(\frac{1}{2}(\partial_0X_a)^2+\frac{1}{\rho^2}\{X_a,X_b\}^2\bigg).
\end{eqnarray}
\begin{eqnarray}
{H}
&=&\int d^2\sigma\bigg(\frac{2}{T_2\rho}P_a^2+\frac{T_2}{4\rho}\{X_a,X_b\}^2\bigg).
\end{eqnarray}
The remaining constraint is the statement 
\begin{eqnarray}
\{P^a,X^a\}=0.
\end{eqnarray}
This light-cone theory, as opposed to the case of string theory, is still very difficult to quantize. A solution due to Nicolai and Hoppe was found in the case of a spherical membrane $\xi_1^2+\xi_2^2+\xi_3^2=1$ in \cite{Hoppe:1982} where functions on the sphere are mapped to $N\times N$ matrices, the Poisson brackets are replaced by Dirac commutation rules, the integral is replaced by an appropriately normalized trace, derivations by adjoint commutators, and the coordinates $\xi_i$ are mapped to $SU(2)$ generators $L_i$ in the irreducible representation of spin $s=(N-1)/2$. The total number of degrees of freedom $N^2$ is equal to the number of linearly independent polarization tensors $T_{lm}$ with $l\leq N-1$ and they (these tensors) go in the large $N$ limit to the usual spherical harmonics $Y_{lm}$. Functions on the spherical membrane are expanded in terms of $Y_{lm}$ whereas functions on the regularized (fuzzy) spherical membrane are expanded in terms of $T_{lm}$. This is essentially the philosophy of fuzzy spaces and fuzzy physics. In summary, the dictionary for passing to the regularized theory is 
\begin{eqnarray}
\xi_i\longrightarrow \frac{2}{N}L_i~,~\{.,.\}\longrightarrow -\frac{iN}{2}[.,.]~,~{\cal L}_i=-i\epsilon_{ijk}x_j\partial_k\longrightarrow [L_i,.]~,~\int d^2\sigma\longrightarrow \frac{4\pi}{N}{\rm Tr}.
\end{eqnarray}
This can be generalized to membranes of arbitrary topology. But by supposing a spherical membrane for concreteness we obtain the Hamiltonian (with $\rho=N$ and $\pi T_2=1/2\pi l_p^3$) 
\begin{eqnarray}
{H}
&=&\frac{1}{2\pi l_p^3}{\rm Tr}\bigg(\frac{1}{2}(\partial_0X_a)^2-\frac{1}{4}[X_a,X_b]^2\bigg).
\end{eqnarray}
In terms of the momentum $P_a=\pi T_2\partial_0X_a$ we get 
 \begin{eqnarray}
{H}
&=&2\pi l_p^3{\rm Tr}(\frac{1}{2}P_a^2)-\frac{1}{2\pi l_p^3}{\rm Tr}(\frac{1}{4}[X_a,X_b]^2).
\end{eqnarray}
The constraint reads
\begin{eqnarray}
[P^a,X^a]=0.
\end{eqnarray}
This is Gauss constraint, i.e. observables must be $U(N)$ invariant. Thus, the remaining invariance of the un-regularized Hamiltonian,  which is time-independent area-preserving diffeomorphisms, is replaced in the regularized theory by $U(N)$ invariance. 

The quantization of this finite system is straightforward precisely because it is finite although it remains non-trivial in practice.

Three other points are worth mentioning:
\begin{itemize}
\item A quantum supermembrane with $16$ supercharges exists only in $D=11$ dimensions. The light-cone Hamiltonian of the regularized supermembrane can be found to be of the form  \cite{Taylor:1999qk,Duff:1996}
\begin{eqnarray}
{H}
&=&\frac{1}{2\pi l_p^3}{\rm Tr}\bigg(\frac{1}{2}(\partial_0X_a)^2-\frac{1}{4}[X_a,X_b]^2+\frac{1}{2}{\psi}^T\gamma^a[X^a,\psi]\bigg).
\end{eqnarray}
Recall that $a,b=1,...,D-2$. The $\gamma^a$ are $16\times 16$ Euclidean $SO(9)$ gamma matrices and $\psi$ is a $16-$component Majorana spinor of $SO(9)$.
\item The $\kappa-$ symmetry of the classical supermembrane action guarantees that the background geometry solves the equations of motion of $11$ dimensional supergravity \cite{Bergshoeff:1987cm,Duff:1996}.
\item The regularized supermembrane suffers from an instability due to flat directions which corresponds to a continuous spectrum \cite{WLN}. This problem is absent in string theory (where the spectrum is discrete) and also is absent in the bosonic regularized membrane where despite the presence of flat directions the spectrum is discrete. This means that we can interpret the states of the theory as a discrete particle spectrum. This issue and its proposed resolution in terms of viewing the quantum theory as a second quantized theory from the point of view of the target space is nicely discussed in \cite{Taylor:1999qk}.
\end{itemize}

\section{The IKKT model or type IIB matrix model}
The IKKT model is equivalent to Connes' approach to geometry!

A commutative/noncommutative space in Connes' approach to geometry is given in terms of a spectral triple $({\cal A},\Delta,{\cal H})$ rather than in terms of a set of points \cite{Connes:1996gi}. ${\cal A}$ is the algebra of functions or bounded operators on the space, $\Delta$  is the Laplace operator or, in the case of spinors, the Dirac operator, and ${\cal H}$ is the Hilbert space on which the algebra of bounded operators and the differential operator $\Delta$ are represented. 

In the IKKT model the geometry is in a precise sense emergent. And thus from this point of view it is obviously equivalent to Connes' noncommutative geometry. The algebra ${\cal A}$ is given, in the large $N$ limit, by Hermitian matrices with smooth eigenvalue distributions and bounded square traces \cite{Sochichiu:2000ud}. The Laplacian/Dirac operator is given in terms of the background solutions while the Hilbert space ${\cal H}$ is given by the adjoint representation of the gauge group $U(N)$.

I start immediately by presenting to you the fundamental model:
\begin{eqnarray}
S_{\rm IKKT}=\frac{1}{g^2}Tr\bigg(\frac{1}{4}[X_{\mu},X_{\nu}][X^{\mu},X^{\nu}]+\frac{1}{2}\bar{\Psi}^{\alpha}\Gamma^{\mu}_{\alpha\beta}[X_{\mu},\Psi^{\beta}]\bigg).
\end{eqnarray}
This is the IKKT or IIB matrix model discovered in 1996 by Ishibashi, Kawai, Kitazawa, and Tsuchiya \cite{Ishibashi:1996xs}. 

This has ${\cal N}=2$ supersymmetry between the Hermitian $N\times N$ bosonic matrices $X^{\mu}$, $\mu=1,...,D$, and the Hermitian $N\times N$ fermionic matrices $\Psi_{\alpha}$, $\alpha=1,...,2^{[D/2]}$. The first supersymmetry is inherited from the $D-$dimensional super Yang-Mills theory, while the second supersymmetry is a $U(1)-$shift of $U(1)-$components of fermionic matrices, which originates in the non-independence of the action on $Tr\Psi$ \cite{Zarembo:1998uk}. These two supersymmetries are given explicitly by
\begin{eqnarray}
\delta\Psi_{\alpha}^{IJ}=\frac{i}{2}[X^{\mu},X^{\nu}]^{IJ}(\Gamma_{\mu\nu}\epsilon)_{\alpha}+\delta^{IJ}\xi_{\alpha}~,~\delta X_{\mu}^{IJ}=i\bar{\epsilon}\Gamma_{\mu}\Psi^{IJ}.
\end{eqnarray}
Another term which is invariant under the above supersymmetry, and which is important in the limit $N\longrightarrow\infty$, is given by
\begin{eqnarray}
S_{2}=\gamma_{\mu\nu}Tr [X^{\mu},X^{\nu}].
\end{eqnarray}
In the above action, we have assumed implicitly Euclidean signature, viz $\eta^{\mu\nu}=\delta^{\mu\nu}$, and that $\Gamma$ are the Dirac matrices in $D-$dimensions in the Weyl representation. Indeed, the fermion $\Psi$ is a complex Weyl spinor which satisfies also the Majorana reality condition. The isometry group is $SO(D)$ whereas the gauge group is obviously $U(N)$.  In other words, the field/matrix $\Psi$ provides a Majorana-Weyl representation of ${\rm Spin}(D)$ whereas $X^{\mu}$ provides a vector representation of $SO(D)$. 

This model is the reduction to $0-$dimension of ${\cal N}=1$ super Yang-Mills theory in $D$ dimensions. ${\cal N}=1$ super Yang-Mills theories only exists in $D=10,6,4,3$. However, it can also be obtained from the Green-Schwarz action for IIB closed string theory, after gauge fixing in the Schild gauge, and with matrix regularization. In this latter case clearly $D$ must be equal to $10$. Alternatively, the IIB superstring action can be obtained from the IKKT model in the double scaling limit $N\longrightarrow\infty$, $g^2\longrightarrow 0$, keeping $Ng^2$ is kept fixed.

The theory is given by the partition function/path integral
\begin{eqnarray}
Z=\int dX d\Psi~\exp(-S_{\rm IKKT}).
\end{eqnarray}
The configuration space $(X,\Psi)$ defines a complex supermanifold, the action $S_{\rm IKKT}$ is a holomorphic function on this space, while physical observables are given by integrals with the weight $\exp(-S_{\rm IKKT})$.

The convergence properties of this path integral in various dimensions is studied in \cite{Krauth:1998xh}. The integral exists in $D=10,6,4$. 

But in Euclidean signature we have an important technical problem.

The fermion $\Psi$ is a complex Weyl spinor which satisfies the Majorana reality condition and thus it contains only $2^{[D/2]-2}$  independent degrees of freedom. However, Majoranan-Weyl fermions do not exist in Euclidean signature. The absence of Majorana-Weyl fermions means that there is no $SO(D,C)-$invariant real cycle or slice (reality condition) in the space of Weyl spinors, which is required to perform the integral over the complex supermanifold. Fortunately, this is irrelevant since the result of the integration does not depend on the choice of cycle for odd variables (spinors). This very technical point is discussed nicely in \cite{Connes:1997cr,Konechny:2000dp}.

Next, it is said that Euclidean IKKT model compactified on a circle gives the BFSS model at finite temperature \cite{Connes:1997cr}. This is the matrix quantum mechanics model discovered by Banks, Fischler, Shenker and Susskind \cite{Banks:1996vh}.

We define a restriction of the IKKT action functional to the subspace where some gauge equivalence relation holds. This subspace consists of all points $(X,\Psi)$ which remain in the same gauge class after a shift by a real number $2\pi R_0$ in the direction $X_0$ given explicitly in terms of unitary matrix $U$ by \cite{Connes:1997cr,Konechny:2000dp}
\begin{eqnarray}
&&UX_0U^{-1}=X_0+2\pi R_0 {\bf 1}\nonumber\\
&&UX_IU^{-1}=X_I, I\ne 0\nonumber\\
&&U\Psi^{\alpha}U^{-1}=\Psi^{\alpha}.\label{ger}
\end{eqnarray}
This can not be satisfied for finite matrices. 

We consider then the infinite-dimensional Hilbert space ${\cal H}=L_2(S^1)\otimes {\cal E}^{}$ where ${\cal E}^{}$ is some other Hilbert space which may or may not be finite-dimensional. The Hilbert space ${\cal H}$ is infinite dimensional because of the factor $L_2(S^1)$. Thus, ${\cal H}$ is the space of functions $f$ on the circle, i.e. $f=f(s)$, which take values in the Hilbert space ${\cal E}^{}$, i.e. they are states in this vector space. Thus ${\cal H}$ is a direct sum of an infinite number of copies of ${\cal E}$.

We assume now that $X$ and $\Psi$ are operators in this infinite-dimensional Hilbert space ${\cal H}$.

A solution of the above gauge equivalence relation (\ref{ger}) is
\begin{eqnarray}
&&X_0=2\pi iR_0\frac{\partial}{\partial\sigma}+A_0(\sigma)\nonumber\\
&&X_I=A_I(\sigma), I\ne 0\nonumber\nonumber\\
&&\Psi^{\alpha}=\chi^{\alpha}(\sigma)\nonumber\\
&&(Uf)(\sigma)=e^{i\sigma}f(\sigma).\label{ger1}
\end{eqnarray}
Indeed, we have
\begin{eqnarray}
&&UX_0U^{-1}=2\pi iR_0\frac{\partial}{\partial\sigma}+UA_0U^{-1}+2\pi i R_0 U\frac{\partial}{\partial\sigma}U^{-1}.
\end{eqnarray}
Obviously, $0\leq \sigma\leq 2\pi$ is a coordinate on $S^1=R/2\pi Z$, and $A_0(\sigma)$, $A_i(\sigma)$ are functions on the circle taking values in the space of operators which act on ${\cal E}^{}$. It can be shown that all other solutions to the gauge equivalence relation (\ref{ger}) are equivalent to the solution (\ref{ger1}).  See \cite{Connes:1997cr} for the elegant proof.

We want to return to finite dimensional matrices. Hence, we assume that ${\cal E}$ is of finite dimension $N$, and also regularize the circle by a lattice of spacing $a$, and thus $\partial/\partial\sigma$ must be understood as a finite difference operator. However, under these conditions, the solution (\ref{ger1}) becomes an approximate solution to (\ref{ger}). 

By substituting the above approximate solution in the IKKT action we get (with the scaling $A\longrightarrow 2\pi R_0 A$, $X_i\longrightarrow 2\pi R_0 A_i$, $\chi\longrightarrow i\sqrt{2}(2\pi R_0)^{3/2}\chi$, and defining $D_0=\partial/\partial\sigma -iA_0$) the action
\begin{eqnarray}
S_{\rm IKKT}=\frac{(2\pi R_0)^4}{g^2}Tr\bigg(\frac{1}{4}[{A}_I,{A}_J]^2+\frac{1}{2}(D_{0}A_I)^2+\bar{\chi}{\Gamma}^I[{A}_I,\chi]-i{\chi}^{\dagger}  D_0\chi\bigg).
\end{eqnarray}
By the Weyl and Majorana conditions we can rewrite the $32-$component spinor $\chi$ in terms of a $16-$component spinor $\psi$ satisfying  $\bar{\psi}={\psi}^+={\psi}^TC_9$ where $C_9$ is the charge conjugation in $9$ dimensions. Also by an appropriate choice of the $10-$dimensional Dirac matrices $\Gamma$ in terms of the $9-$dimensional Dirac matrices $\gamma$ (see next section) we can rewrite the above action as 
\begin{eqnarray}
S_{\rm IKKT}=\frac{(2\pi R_0)^4}{g^2}Tr\bigg(\frac{1}{4}[{A}_I,{A}_J]^2+\frac{1}{2}(D_{0}A_I)^2+\bar{\psi}{\gamma}^I[{A}_I,\psi]-i\bar{\psi}D_0\psi\bigg).
\end{eqnarray}
Of course, the trace over the circle is a sum over the lattice sites in the $\sigma$ direction. Finally we obtain, by taking the lattice spacing $a\longrightarrow 0$ while keeping the dimension $N$ of the Hilbert space ${\cal E}$ fixed, the action
\begin{eqnarray}
S_{\rm BFSS}=\frac{(2\pi R_0)^4}{g^2}\int d\sigma Tr\bigg(\frac{1}{4}[{A}_I,{A}_J]^2+\frac{1}{2}(D_{0}A_I)^2+\bar{\psi}{\gamma}^I[{A}_I,\psi]-i\bar{\psi}  D_0\psi\bigg).
\end{eqnarray} 
By Wick rotation we obtain the BFSS quantum mechanics also known as the M-(atrix) theory. Thus the circle is actually a compact time direction (finite temperature). %In the limit $a\longrightarrow 0$, keeping $N$ fixed, (\ref{ger1}) becomes an exact solution to (\ref{ger}) since the Hilbert space ${\cal H}$ becomes infinite dimensional. 

This exercise is an example of T-duality.

However, I should say that compactification which is an analogue of dimensional reduction does not really describe what has just happened. We know that dimensional reduction can be achieved by taking one dimension to be a circle and then keeping only the zero modes whereas in compactification we keep all modes in the lower dimensional theory. In a very clear sense the IKKT is the lower dimensional theory of the BFSS and thus the process of compactification should really be taking us from the BFSS to IKKT where the circle becomes a point in the infinite temperature limit. The description given above which escalated the IKKT to the BFSS is strictly speaking a decompactification where one extra dimension or coordinate has emerged from imposing an equivalence relation among gauge configurations along the $X^0$ direction.   
\section{The BFSS model from dimensional reduction}
But what exactly is the BFSS model? The above derivation in terms of compactification on the circle is not mathematically clean since it involves escalation of the problem to infinite dimensional matrices.

A closely related model to the IKKT matrix model is obtained by the reduction to one dimension of ${\cal N}=1$ super Yang-Mills theory in $D=10$ dimensions. This is a matrix quantum mechanics model which describes the low energy dynamics of a system of $N$ type IIA D0-branes \cite{Witten:1995im}. After this discovery, Banks, Fischler, Shenker, and Susskind  conjectured that the large $N$ limit of this model describes precisely M-theory in the infinite momentum frame ($P_z\longrightarrow\infty$) in the uncompactified limit ($R_s\longrightarrow\infty$). We remark that the infinite momentum limit is equivalent to the light cone quantization only in the limit $N\longrightarrow\infty$. This is the same model derived above from compactification.

Let us perform now the dimensional reduction explicitly \cite{Brink:1976bc}. 

We start from $D=10$ with metric ${\eta}^{MN}=(-1,+1,...,+1)$. The Clifford algebra is $32$ dimensional given by $\{G^M,G^N\}=2{\eta}^{MN}{\bf 1}_{32}$. The basic object of ${\cal N}=1$ SUSY in $10$ dimensions is a $32-$component complex spinor ${\Lambda}$ which satisfies the Majorana reality condition and the Weyl condition given by
\begin{eqnarray}
&&\bar{\Lambda}={\Lambda}^+{G}^0\equiv {\Lambda}^TC_{10}~,~C_{10}G^{M}C_{10}^{-1}=(G^{M})^T\nonumber\\
&&G_{11}{\Lambda}={\Lambda}~,~G_{11}=G^0...G^{9}.
\end{eqnarray}
The ${\cal N}=1$ supersymmetric
action in $d=10$ dimensions is given by 
\begin{eqnarray}
S=\frac{1}{g^2}\int d^{10}xTr \bigg[(-\frac{1}{4}F_{MN}F^{MN})|_{d=10}+\frac{i}{2} (\bar{\Lambda}G^{M}D_{M}\Lambda)|_{d=10}\bigg].
\end{eqnarray}
By using $D_{M}={\partial}_M-i[A_M,..]$, $A_i={X}_i$,
${\partial}_i=0$ we can immediately compute the reduction of the fermionic
action to $p+1$ dimensions  to be given by 
\begin{eqnarray}
 -\frac{1}{2} (\bar{\Lambda}G^{M}D_{M}\Lambda)|_{d=10}&=&-\frac{1}{2}
  (\bar{\Lambda}G^{\mu}D_{\mu}\Lambda)|_{p+1}+\frac{i}{2}\bar{\Lambda} G^i[{X}_i,\Lambda].
\end{eqnarray}
The reduction of the bosonic action to $p+1$ dimensions  is given by
\begin{eqnarray}
&&(-\frac{1}{4}F_{MN}F^{MN})|_{d=10}=(-\frac{1}{4}F_{\mu \nu}F^{\mu
    \nu})_{p+1}+\frac{1}{4}[{X}_i,{X}_j]^2-\frac{1}{2}(D_{\mu}X_i)(D^{\mu}{X}_i)\nonumber\\
&&F_{\mu \nu}={\partial}_{\mu}A_{\nu}-{\partial}_{\nu}A_{\mu}-i[A_{\mu},A_{\nu}].
\end{eqnarray}
The index $\mu$ runs over the values $0,1,...,p$ whereas the index $i$
runs over the values $p+1,p+2,...,9$. As an example we write down the  action on
the $1-$dimensional world-volume of $N$ coincident
D0-branes, i.e. $p=0$. This is given by
\begin{eqnarray}
S=\frac{1}{g^2}\int dt~ Tr\bigg(\frac{1}{4}[{X}_I,{X}_J]^2-\frac{1}{2}(D_{0}X_I)(D^{0}{X}_I)+\frac{1}{2}\bar{\Lambda} G^I[{X}_I,\Lambda]-\frac{i}{2}{\Lambda}^+D_0\Lambda\bigg).
\end{eqnarray}
We  choose the following representation 
\begin{eqnarray}
&&G^0=i{\sigma}_2{\otimes}{\bf
    1}_{16}~,~G^I={\sigma}_1{\otimes}{\Gamma}^I~,~I=1,...,9.
\end{eqnarray}
The charge conjugation decomposes as %$C_{10}={\sigma}_1 \otimes C_9$. 
\begin{eqnarray}
C_{10}={\sigma}_1 \otimes C_9.
\end{eqnarray}
The charge conjugation in $9$ dimensions satisfies
$C_{9}{\Gamma}^{I}C_{9}^{-1}=({\Gamma}^{I})^T$. By the Weyl and
Majorana conditions the $32-$component spinor $\Lambda$ can be
rewritten  in terms of a $16-$component spinor $\Psi$ as follows
\begin{eqnarray}
\Lambda=\sqrt{2}\left(
\begin{array}{c}
\Psi\\
0
\end{array}
\right)~,~{\Psi}^+={\Psi}^TC_9.
\end{eqnarray}
The action becomes
\begin{eqnarray}
&&S=\int dt~L\nonumber\\
&&L=\frac{1}{g^2}Tr\bigg(\frac{1}{4}[{X}_I,{X}_J]^2+\frac{1}{2}(D_{0}X_I)^2+{\Psi}^TC_9
{\Gamma}^I[{X}_I,\Psi]-i{\Psi}^T C_9 D_0\Psi\bigg).\label{S0}
\end{eqnarray}
The Clifford algebra in $D=9$ dimensions can be taken to be given by \cite{Kim:2003rza}
\begin{eqnarray}
&&{\Gamma}^a=\left(
\begin{array}{cc}
-{\sigma}_a\otimes{\bf 1}_4 &0\\
0&{\sigma}_a\otimes{\bf 1}_4
\end{array}
\right)~,~{\Gamma}^i=\left(
\begin{array}{cc}
0&{\bf 1}_2\otimes{\rho}_i\\
{\bf 1}\otimes ({\rho}_{i})^+&0
\end{array}
\right).
\end{eqnarray}
In above $a=1,...,3$, $i=4,...,9$ and 
\begin{eqnarray}
{\rho}_i({\rho}_j)^++{\rho}_j({\rho}_i)^+=({\rho}_i)^+{\rho}_j+({\rho}_j)^+{\rho}_i=2{\delta}_{ij}{\bf 1}_4.
\end{eqnarray}
The charge conjugation in $D=9$ is 
\begin{eqnarray}
C_9=\left(
\begin{array}{cc}
0 &-i{\sigma}_2\otimes {\bf 1}_4\\
i{\sigma}_2\otimes {\bf 1}_4&0
\end{array}
\right).
\end{eqnarray}
We can effectively work with a charge conjugation operator $C_9$ equal ${\bf 1}_8$ \cite{Kim:2002if}. Integration over $\Psi$ leads to the Pfaffian 
$ {\rm Pf}(C_9{\cal O})$ where ${\cal
  O}=-iD_0+{\Gamma}^I[X_I,...]$. Hence the operator $C_9$ will drop
from calculations. Thus we obtain the
action
\begin{eqnarray}
&&S=\int dt~L\nonumber\\
&&L=\frac{1}{g^2}Tr\bigg(\frac{1}{4}[{X}_I,{X}_J]^2+\frac{1}{2}(D_{0}X_I)^2+{\Psi}^T
{\Gamma}^I[{X}_I,\Psi]-i{\Psi}^T  D_0\Psi\bigg).\label{S}
\end{eqnarray}
Now we must have ${\Psi}^+={\Psi}^T$. In the gauge $A_0=0$ we get the
Lagrangian
\begin{eqnarray}
&&L=\frac{1}{g^2}Tr\bigg(\frac{1}{4}[{X}_I,{X}_J]^2+\frac{1}{2}({\partial}_{0}X_I)^2+{\Psi}^T
{\Gamma}^I[{X}_I,\Psi]-i{\Psi}^T  {\partial}_0\Psi\bigg).
\end{eqnarray}
The momentum conjugate corresponding to $X_I$ is $P_I={\partial}_0X_I$
whereas the momentum conjugate corresponding to ${\Psi}_{\alpha}$ is
${\Pi}_{\alpha}=-i{\Psi}_{\alpha}$. The Hamiltonian is given by
\begin{eqnarray}
H&=&\frac{1}{g^2}Tr\bigg(P_I{\partial}_0X_I+{\Pi}^T{\partial}_0\Psi -L\bigg)\nonumber\\
&=&\frac{1}{g^2}Tr\bigg(\frac{1}{2}P_I^2-\frac{1}{4}[{X}_I,{X}_J]^2-{\Psi}^T
{\Gamma}^I[{X}_I,\Psi]\bigg).\label{H}
\end{eqnarray}
As opposed to the IKKT model, the BFSS model does not enjoy the full Lorentz group $SO(1,9)$. It is invariant only under $SO(9)$ which rotates the $X_I$ among each other.
  
In order to obtain Euclidean signature we perform the Wick rotation $x^0=t\longrightarrow -ix^0=-it$. Thus $\partial_0\longrightarrow i\partial_0$, $A_0\longrightarrow iA_0$ and $D_0\longrightarrow iD_0$. Also we change $L\longrightarrow -L$. The action becomes
\begin{eqnarray}
&&S=\int dt~L\nonumber\\
&&L=\frac{1}{g^2}Tr\bigg(-\frac{1}{4}[{X}_I,{X}_J]^2+\frac{1}{2}(D_{0}X_I)^2-{\Psi}^T
{\Gamma}^I[{X}_I,\Psi]-{\Psi}^T  D_0\Psi\bigg).\label{S}
\end{eqnarray}

\section{Introducing gauge/gravity duality}
We will initially follow the excellent presentations of \cite{Horowitz:2006ct,Hanada:2016jok,OConnor:2016gbq}.

\subsection{Dimensional reduction to $p+1$ dimensions}
The starting point is the action of ${\cal N}=1$ supersymmetric Yang-Mills in $D=10$ dimensions given by the functional
\begin{eqnarray}
S=\frac{1}{g^2}\int d^D x\big[-\frac{1}{4}F_{\mu\nu}F^{\mu\nu}+\frac{i}{2}\bar{\Psi}\Gamma^{\mu}D_{\mu}\Psi\big].
\end{eqnarray}
The $g^2$ is the Yang-Mills coupling. Obviously, $\Gamma^{\mu}$ is the ten dimensional Clifford algebra which is $32$ dimensional. This theory consists of a gauge field $A_{\mu}$, $\mu=0,...,9$, with field strength $F_{\mu\nu}=\partial_{\mu}A_{\nu}-\partial_{\nu}A_{\mu}-i[A_{\mu},A_{\nu}]$  coupled in a supersymmetric fashion to a Majorana-Weyl spinor $\Psi$ in the adjoint representation of the gauge group. Since $\Psi$ is a Majorana-Weyl it can be rewritten as \footnote{In general a spinor in $D$ (even) dimension will have $2^{D/2}$ components. The Majorana condition is a reality condition which reduces the number of components by a factor of two whereas the Weyl condition is a chirality  condition (left-handed or right-handed) which reduces further the number of independent components down by another factor of two. In $4$ dimensions these two conditions are not mutually consistent because there is no a real irreducible representation of the spinor bundle whereas in $10$ dimensions we can impose both conditions at once. Hence in $10$ dimensions we have $2^{5}/4=16$ complex components or $16$ real components $\psi_{\alpha}$.} (see also previous section)
\begin{eqnarray}
\Psi=\left(
\begin{array}{c}
\psi\\
0
\end{array}
\right).
\end{eqnarray}
Hence the above action takes the form
\begin{eqnarray}
S=\frac{1}{g^2}\int d^D x\big[-\frac{1}{4}F_{\mu\nu}F^{\mu\nu}+\frac{i}{2}\bar{\psi}\gamma^{\mu}D_{\mu}\psi\big].
\end{eqnarray}
Now, $\gamma^{\mu}$ are the left-handed sector of the ten dimensional gamma matrices \footnote{In particular $\gamma^0=-1$, $\gamma^{a}$, $a=1,...,9$, are the Dirac matrices in $D=9$, and $\bar{\psi}=\psi^{T}C_9$. See the previous section.}. The gauge field $A$ and the Majorana-Weyl spinor $\psi$ are $N\times N$ Hermitian matrices which are represented by complex and Grassmann numbers respectively in the measure of the path integral.  Since $\Psi$ is in the adjoint representation the covariant derivative $D_{\mu}\psi$ is defined via commutator, i.e. $D_{\mu}\psi=\partial_{\mu}\psi-i[A_{\mu},\psi]$. The above action enjoys thus $U(N)$ gauge invariance given by 
\begin{eqnarray}
D_{\mu}\longrightarrow gD_{\mu}g^{\dagger}~,~\psi\longrightarrow g\psi g^{\dagger}.
\end{eqnarray}
But it does also enjoy ${\cal N}=1$ supersymmetry given by
\begin{eqnarray}
A_{\mu}\longrightarrow A_{\mu}+i\bar{\epsilon}\gamma_{\mu}\psi~,~\psi\longrightarrow\psi+\frac{1}{2}F_{\mu\nu}\gamma^{\mu\nu}\epsilon.
\end{eqnarray}
Hence, there are $16$ real supercharges corresponding to the $16$ components of $\epsilon$.

Dimensional reduction to $D=p+1$ dimensions consists in keeping only the coordinates $x_0$, $x_1$, ...$x_p$ and dropping all dependence on the coordinates $x_{p+1}$,...$x_9$. The gauge fields $A_a$ in the directions $a=p+1,...,9$ become scalar fields $X_I$ where $I=a-p=1,2,...,9-p$. The action becomes given immediately by 
\begin{eqnarray}
S=\frac{V_{9-p}}{g^2}\int d^{p+1} x\big[-\frac{1}{4}F_{\mu\nu}F^{\mu\nu}-\frac{1}{2}(D_{\mu}X_I)(D^{\mu}X_I)+\frac{1}{4}[X_I,X_J]^2+\frac{i}{2}\bar{\psi}\gamma^{\mu}D_{\mu}\psi+\frac{1}{2}\bar{\psi}\gamma^{I}[X_{I},\psi]\big].\label{run}\nonumber\\
\end{eqnarray}
%By Wick rotation to Euclidean signature we get the energy 
%\begin{eqnarray}
%S=\frac{V_{9-p}}{g^2}\int d^{p+1} x\big[\frac{1}{4}F_{\mu\nu}^2+\frac{1}{2}(D_{\mu}X_I)^2-\frac{1}{4}[X_I,X_J]^2+\frac{i}{2}\bar{\psi}\gamma^{\mu}D_{\mu}\psi+\frac{1}{2}\bar{\psi}\gamma^{I}[X_{I},\psi]\big].\label{run}
%\end{eqnarray}
In the above action the indices $\mu$ and $\nu$ run from $0$ to $p$ whereas the indices $I$ and $J$ run from $1$ to $D-1-p$ where $D=10$. This action is maximally supersymmetric enjoying $16$ supercharges. More importantly, it was established more than $20$ years ago that this action gives an effective description of Dp-branes in the low energy limit \cite{Witten:1995im,Dai:1989ua,Polchinski:1995mt}.

\subsection{Dp-branes revisited}

But what are Dp-branes?

In the words of Polchinski \cite{Polchinski:1995mt}: "..Dirichlet(p)-branes (are) extended objects defined by mixed Dirichlet-Neumann boundary conditions in string theory (which) break half of the supersymmetries of the type II superstring and carry a complete set of electric and magnetic Ramond-Ramond charges..".

Thus a Dp-brane is an extended object with $p$ spatial dimensions evolving in time. Let us consider type II superstring theory which contains only closed strings. The presence of a stable supersymmetric Dp-brane with $p$ even for type IIA and with $p$
odd for type IIB modifies the allowed boundary conditions of the strings. Both Neumann and Dirichlet boundary conditions are now
allowed. Hence in addition to the closed strings, which propagate in the bulk $(1+9)-$dimensional spacetime, we can have open
strings whose endpoints are fixed on the Dp-brane. 

Recall that for world sheets with boundaries there is a surface term in the variation of the Polyakov action which vanishes iff:
\begin{itemize}
\item{}
\begin{eqnarray}
\partial^{\sigma}x^{\mu}(\tau,0)=\partial^{\sigma}x^{\mu}(\tau,l)=0.
\end{eqnarray}
These are Neumann boundary conditions, i.e. the ends of the open string move freely in spacetime, and the component of the momentum normal to the boundary of the world sheet vanishes.
\item{}
\begin{eqnarray}
x^{\mu}(\tau,0)=x^{\mu}(\tau,l).
\end{eqnarray}
The fields are periodic which corresponds to  a closed string.
\end{itemize}
These are the only boundary conditions which are consistent with $D-$dimensional Poincar\'e invariance. 

However, in the presence of a Dp-brane defined for example by the condition
\begin{eqnarray}
x^{p+1}=...=x^{9}=0
\end{eqnarray}
the open strings must necessarily end on this surface and thus they must satisfy  Dirichlet boundary conditions in the directions $p+1$,..,$9$ while in the  directions $0$,...,$p$ they must satisfy the usual Neumann boundary conditions. In other words, we must have
\begin{eqnarray}
&&{\partial}_{\sigma}x^0=...={\partial}_{\sigma}x^p=0~,~{\rm at}~\sigma=0,\pi\nonumber\\
&&x^{p+1}=...=x^{9}=0~,~{\rm at}~\sigma=0,\pi.
\end{eqnarray}
Thus, the coordinates $x^0,x^1,...,x^p$ obey  Neumann boundary conditions while  the coordinates $x^{p+1},...,x^9$ obey  Dirichlet boundary conditions. As it turns out the reduced spacetime defined by the longitudinal coordinates $(x^0,x^1,...,x^p)$ corresponds to the world volume of the Dp-brane, i.e. the $U(N)$ gauge theory (\ref{run}) is living on the Dp-brane. However, this is really a theory of $N$ coincident branes and not a single brane. Indeed, the diagonal elements $(X_1^{ii},X_2^{ii},...,X_{9-p}^{ii})$ define the position of the $i$th Dp-brane along the transverse directions  $(x^{p+1},x^{p+2},...,x^9)$ whereas the off diagonal elements $(X_1^{ij},X_2^{ij},...,X_{9-p}^{ij})$ define open string excitations between the $i$th and $j$th Dp-branes. See for example \cite{Azeyanagi:2009zf}.

As we have already discussed in the previous chapter, this fundamental result can also be seen as follows. 

In addition to the closed strings of the type II superstring theory we can have one class of open strings starting and ending on the Dp-brane. The quantization of these open strings is equivalent  to the quantization of the ordinary open superstrings. See for example \cite{Zwiebach:2004tj}. 

On the other hand we know that the massless modes of type I open superstrings consist of a Majorana-Weyl spinor $\psi $ and a gauge field $A_a$ with dynamics given at low energy by a supersymmetric $U(1)$ gauge theory in $10$ dimensions. The gauge field  $A_a$ depends only on the zero modes of the coordinates $x^0,x^1,x^2,...,x^p$ since the zero modes of the other coordinates vanish due to the Dirichlet boundary condition $x^{p+1}=...=x^{9}=0$ at $\sigma=0,\pi$. The reduced gauge field $A_a$, $a=0,...,p$, behaves as a $U(1)$ vector on the p-brane world volume while the other components $A_a$ with $a=p+1,...,9$ behave as scalar fields normal to the p-brane. These scalar fields describe therefore fluctuations of the position of the p-brane.

%At low energy the theory on the $(p+1)-$dimensional world volume of the Dp-brane is the reduction to $p+1$ dimensions of $10-$dimensional supersymmetric $U(1)$ gauge theory \cite{Witten:1995im}. When the velocities and/or the string coupling are not small the minimal supersymmetric Yang-Mills is replaced by the  supersymmetric Born-Infeld action \cite{Leigh:1989jq}.

Therefore at low energy the theory on the $(p+1)-$dimensional world volume of the Dp-brane is the reduction to $p+1$
dimensions of $10-$dimensional supersymmetric $U(1)$ gauge theory \cite{Witten:1995im}. When the velocities and/or the string coupling are not
small the minimal supersymmetric Yang-Mills is replaced by the
 supersymmetric Born-Infeld action \cite{Leigh:1989jq}.

Let us consider now two parallel Dp-branes located at $X^{a}=0$ and
$X^{a}=Y^{a}$ respectively where  $a=p+1,...,9$. In addition to the
closed strings of the type II superstring theory we can have $4$ different
 classes of open strings. The first class consists of open strings
 which begin and end on the first Dp-brane  $X^{a}=0$, $a=p+1,...,9$. The second class consists of open strings
 which begin and end on the second Dp-brane
 $X^{a}=Y^a$, $a=p+1,...,9$. The third and fourth classes consist of
 open strings which begin on one Dp-brane and end on the other
 Dp-brane. These are stretched strings. These $4$ classes can be
 labeled by pairs $[ij]$ where $i,j=1,2$. These are the Chan-Paton
 indices. The open strings in the
 $[ij]$ sector extend from the Dp-brane $i$ to the Dp-brane
 $j$. As we have already seen the open string sectors $[11]$ and $[22]$ give
 rise to $U(1)\times U(1)$ gauge theory where the first $U(1)$ lives
 on the first brane and the second $U(1)$ lives on the second
 brane. The massless open string bosonic ground state includes for each
 factor of $U(1)$ a massless $(p+1)-$vector field and $9-p$ massless
 normal scalars propagating on the $(p+1)-$dimensional world volumes of the Dp-branes.

The  stretched strings in the sectors $[12]$ and $[21]$ will
 have a massive bosonic ground state.  The mass being equal to $T|Y|$ where $T$
 is the string tension. First, let us assume  that $Y^a=0$. Then the
 normal coordinates to the brane will correspond  to $9-p$ normal
 scalar fields while the tangent coordinates
 to the brane give rise to $(p+1)-2=p-1$  states. However, because
 $Y^a\neq 0$ we will have instead a massive $(p+1)-$vector field with $p$ (and not $p-1$) independent
 components and $8-p$ massive normal scalars. This is because the
 normal scalar field parallel to the unique normal direction $Y^{a}$
 which takes us from one brane to the other is in fact a part of the
 massive gauge boson. Let us also say that these $(p+1)-$dimensional
 fields live on a fixed $(p+1)-$dimensional space not necessarily
 identified with any of the Dp-branes. Alternatively, we can say that
 these fields live on the two Dp-branes at the same time. These
 fields have  clearly non-local interactions since the two Dp-branes
 are separated. The spacetime interpretaion of these fields seem to
 require noncommutative geometry \cite{Zwiebach:2004tj}.

Now we take the limit in which the separation between the two Dp-branes goes
to zero. The two Dp-branes remain distinguishable. In this
case the open string bosonic ground state which represents strings starting on
one brane and ending on the other becomes massless. Therefore the open
string bosonic ground state includes from each
 of the four sectors $[ij]$ a massless $(p+1)-$vector field and $9-p$
 massless normal scalar fields. In other words, when the two Dp-branes coincide we get $4$
 (and not $2$) massless $(p+1)-$vector fields giving a $U(2)$
 Yang-Mills gauge theory on the $(p+1)-$world volume of the coincident
 Dp-branes. In addition, we will have $9-p$ massless normal scalar fields in
 the adjoint representation of $U(2)$ which will play the role of
  position coordinates of the  coincident Dp-branes. The coordinates
  became noncommuting matrices. At low energy the theory on
the $(p+1)-$dimensional world volume of the two coincident Dp-branes is therefore the reduction to $p+1$
dimensions of $10-$d supersymmetric $U(2)$ gauge theory \cite{Witten:1995im}.  As before when the velocities and/or the string coupling are not
small this minimal supersymmetric Yang-Mills is replaced by a
non-abelian generalization of the supersymmetric Born-Infeld action
which is still not known.

Generalization to $N$ coincident Dp-branes is obvious: At low energy
(i.e. small velocities and small string coupling) the theory on
the $(p+1)-$dimensional world-volume of  $N$ coincident Dp-branes
is the reduction to $p+1$
dimensions of $10-$d supersymmetric $U(N)$ gauge theory. This will be
corrected by higher dimension operators when the velocities and/or the string coupling are not
small. In some sense the use of Yang-Mills
action instead of the Born-Infeld action is equivalent to a nonrelativistic approximation
to the dynamics of the coincident Dp-branes.

\subsection{The corresponding gravity dual}
The gauge/gravity duality in the words of \cite{Horowitz:2006ct} is the statement that "Hidden within every non-Abelian gauge theory, even within the weak and strong nuclear interactions, is a theory of quantum gravity". 

Let us then consider a system of $N$ coincident Dp-branes described by the $U(N)$ gauge theory (\ref{run}). This is a theory characterized by the gauge coupling constant $g_{\rm YM}^2\equiv g^2/V_{9-p}$ and the number of colors $N$. A gauge theory with an infinite number of degrees of freedom, which is the one relevant to supergravity and superstring and which must live in more dimensions by the holographic principle \cite{tHooft:1993dmi,Susskind:1994vu} in order to avoid the Weinberg-Witten no-go theorem \cite{Weinberg:1980kq}, is given by the t'Hooft planar limit in which $N$ is taken large and $g_{\rm YM}^2$ taken small keeping fixed the t'Hooft coupling $\lambda$ given by \cite{tHooft:1973alw}
\begin{eqnarray}
\lambda=g_{\rm YM}^2N.
\end{eqnarray}
There is another parameter in the gauge theory which is the temperature $T=1/\beta$. This is obtained by considering the finite temperature path integral over periodic Euclidean time paths with period $\beta$, i.e. by compactify the Euclidean time direction on a circle of length equal $\beta$. See for example \cite{Nastase:2007kj}.

We will study the Dp-branes in the field theory limit defined by 
\begin{eqnarray}
g_{\rm YM}^2=(2\pi)^{p-2}g_s(\alpha^{\prime})^{\frac{p-3}{2}}={\rm fixed}~,~\alpha^{\prime}\longrightarrow 0.\label{limit}
\end{eqnarray}
The $\alpha^{\prime}=l^2_s$ is the inverse of the string tension and $g_s$ is the string coupling constant related to the dilaton field by the expression 
 \begin{eqnarray}
g_s=\exp(\Phi).
\end{eqnarray}
The system of $N$ coincident Dp-branes is obviously a massive object which will curve the spacetime around it. As such it is called a black p-brane and is a higher dimensional analogue of the black hole. The singularity in a black p-brane solution is extended in $p-$spatial directions. The radial (transverse) distance from the center of mass of the black p-brane to a given point in spacetime will be denoted by $U$.  More precisely, $U=r/{\alpha}^{\prime}$. From the gauge theory point of view $U$ is the energy scale. The limit (\ref{limit}) will be taken in such a way that we keep $U$ fixed, i.e. we are interested in finite energy configurations in the gauge theory \cite{Itzhaki:1998dd}.

It is known that type II supergravity solution which describes $N$ coincident extremal ($T=0$) Dp-branes is given by the metric \cite{Gibbons:1987ps,Horowitz:1991cd,Itzhaki:1998dd} 
\begin{eqnarray}
ds^2=\frac{1}{\sqrt{f_p}}(-dt^2+dx_1^2+...+dx_p^2)+\sqrt{f_p}(dx_{p+1}^2+...+dx_{9}^2),
\end{eqnarray}
where
\begin{eqnarray}
f_p=1+\frac{d_pg_{\rm YM}^2N}{\alpha^{\prime 2}U^{7-p}}~,~d_p=2^{7-2p}\pi^{\frac{9-3p}{2}}\Gamma(\frac{7-p}{2}).
\end{eqnarray}
The near extremal p-brane solution in the limit (\ref{limit}) keeping the energy density on the brane finite, and which corresponds to the above finite temperature gauge theory, are given by the metric  \cite{Gibbons:1987ps,Horowitz:1991cd,Itzhaki:1998dd}

\begin{eqnarray}
ds^2=\alpha^{\prime}\bigg[\frac{U^{\frac{7-p}{2}}}{g_{\rm YM}\sqrt{d_pN}}\bigg(-(1-\frac{U_0^{7-p}}{U^{7-p}})dt^2+dy_{||}^2\bigg)+\frac{g_{\rm YM}\sqrt{d_pN}}{U^{\frac{7-p}{2}}}\frac{dU^2}{(1-\frac{U_0^{7-p}}{U^{7-p}})}+g_{\rm YM}\sqrt{d_pN}U^{\frac{p-3}{2}}d\Omega^2_{8-p}\bigg].\nonumber\\
\end{eqnarray}
This is the corresponding metric of the curved spacetime around the black p-brane. The $(t,y_{\parallel})$ are the coordinates along the p-brane world volume, $\Omega_{8-p}$ is the transverse solid angle associated with the transverse radius $U$. The $U_0$ corresponds to the radius of the horizon  and it is given in terms of the energy density of the brane $E$ (which is precisely the energy density of the gauge theory) by 
\begin{eqnarray}
U_0^{7-p}=a_pg_{\rm YM}^4E~,~a_p=\frac{\Gamma(\frac{9-p}{2})2^{11-2p}\pi^{\frac{13-3p}{2}}}{9-p}.
\end{eqnarray}
The string coupling constant in this limit is given by 
\begin{eqnarray}
\exp(\phi)=(2\pi)^{2-p}g_{\rm YM}^2(\frac{d_pg_{\rm YM}^2N}{U^{7-p}})^{\frac{3-p}{4}}.
\end{eqnarray}
In order to determine the Hawking temperature we introduce the Euclidean time $t_E=it$ and define the proper distance from the horizon by the relation 
\begin{eqnarray}
d\rho^2=\frac{g_{\rm YM}\sqrt{d_pN}}{U^{\frac{7-p}{2}}}\frac{dU^2}{(1-\frac{U_0^{7-p}}{U^{7-p}})}.
\end{eqnarray}
Near the horizon we write $U=U_0+\delta$ then
\begin{eqnarray}
\rho^2=\frac{4g_{\rm YM}\sqrt{d_pN}}{(7-p)U_0^{\frac{5-p}{2}}}\delta=4r_s\delta.
\end{eqnarray}
The relevant part of the metric becomes 
\begin{eqnarray}
ds^2=\alpha^{\prime}\bigg[\frac{\rho^2}{4r_s^2}dt^2+d\rho^2+...\bigg].\nonumber\\
\end{eqnarray}
The first two terms correspond to two-dimensional flat space, viz

\begin{eqnarray}
X=\rho\cos\frac{t_E}{2r_s}~,~Y=\rho\sin\frac{t_E}{2r_s}.
\end{eqnarray}
Hence in order for the Euclidean metric to be smooth the Euclidean time must be periodic with period $\beta=4\pi r_s$ otherwise the metric has a conical singularity at $\rho=0$. We get then the temperature 
\begin{eqnarray}
T=\frac{1}{4\pi r_s}=\frac{(7-p)U_0^{\frac{5-p}{2}}}{4\pi g_{\rm YM}\sqrt{d_pN}}.
\end{eqnarray}
The entropy density is given by the Bekenstein-Hawking formula
\begin{eqnarray}
{\cal S}=\frac{A}{4},
\end{eqnarray}
where $A$ is the area density of the horizon. This can be calculated from the first law of thermodynamics as follows  
\begin{eqnarray}
d{\cal S}&=&\frac{1}{T}dE\nonumber\\
&=&\frac{4\pi g_{\rm YM}\sqrt{d_pN}}{(7-p)U_0^{\frac{5-p}{2}}}dE\nonumber\\
&=&\frac{4\pi g_{\rm YM}\sqrt{d_pN}}{(7-p)(a_pg_{\rm YM}^4)^{\frac{5-p}{2(7-p)}}}\frac{dE}{E^{\frac{5-p}{2(7-p)}}}\nonumber\\
&=&\frac{4\pi g_{\rm YM}\sqrt{d_pN}}{(7-p)(a_pg_{\rm YM}^4)^{\frac{5-p}{2(7-p)}}}\frac{dE^{\frac{9-p}{2(7-p)}}}{{\frac{9-p}{2(7-p)}}}.
\end{eqnarray}
We get then the entropy 
\begin{eqnarray}
{\cal S}&=&\frac{8\pi \sqrt{d_pN}}{(9-p)a_p^{\frac{5-p}{2(7-p)}}}g_{\rm YM}^{\frac{p-3}{7-p}}E^{\frac{9-p}{2(7-p)}}.
\end{eqnarray}
In summary, we have found that maximally supersymmetric  $(p+1)-$dimensional $U(N)$ gauge theory is equivalent to type II superstring theory around black p-brane background spacetime. This is the original conjecture of Maldacena that weakly coupled super Yang-Mills theory and weakly coupled type II superstring theory both provide a description of $N$ coincident Dp-branes forming a black p-brane \cite{Maldacena:1997re}. This equivalence should be properly understood as a non-perturbative definition of string theory since the gauge theory is rigorously defined by a lattice \`a la Wilson \cite{Wilson:1974sk}. More precisely, we have \cite{Horowitz:2006ct,Hanada:2016jok,OConnor:2016gbq}
\begin{itemize}
\item The gauge theory in the limit $N\longrightarrow\infty$ (where extra dimensions will emerge) and $\lambda\longrightarrow\infty$ (where strongly quantum gauge fields give rise to effective classical gravitational fields) should be equivalent to classical type II supergravity around the p-brane spacetime.
\item The gauge theory with $1/N^2$ corrections should correspond to quantum loop corrections, i.e. corrections in $g_s$, in the gravity/string side.
\item The gauge theory with $1/\lambda$ corrections should correspond to stringy corrections, i.e. corrections in $l_s$, corresponding to the fact that degrees of freedom in the gravity/string side are really strings and not point particles.
\end{itemize}

\section{Black hole unitarity from M-theory}
\subsection{The black 0-brane}
The case of $p=0$ is of particular interest to us here. The gauge theory in this case is a maximally supersymmetric $U(N)$ quantum mechanics given by the Wick rotation to Euclidean signature of the action (\ref{run}) with $p=0$, viz
%\begin{eqnarray}
%S=\frac{1}{g_{\rm YM}^2}\int_0^{\beta} dt\big[\frac{1}{2}(D_{t}X_I)^2-\frac{1}{4}[X_I,X_J]^2+\frac{i}{2}\bar{\psi}\gamma^{0}D_{t}\psi+\frac{1}{2}\bar{\psi}\gamma^{I}[X_{I},\psi]\big].
%\end{eqnarray}
%\begin{eqnarray}
%S=\frac{1}{g_{\rm YM}^2}\int dt \big[-\frac{1}{2}(D_0X_I)(D^0X_I)+\frac{1}{4}[X_I,X_J]^2+\frac{i}{2}\bar{\psi}\gamma^{0}D_{0}\psi+\frac{1}{2}\bar{\psi}\gamma^{I}[X_{I},\psi]\big].
%\end{eqnarray}
\begin{eqnarray}
S=\frac{1}{g_{\rm YM}^2}\int_0^{\beta} dt \big[\frac{1}{2}(D_tX_I)^2-\frac{1}{4}[X_I,X_J]^2-\frac{1}{2}\bar{\psi}\gamma^{0}D_{0}\psi-\frac{1}{2}\bar{\psi}\gamma^{I}[X_{I},\psi]\big].
\end{eqnarray}
This is the BFSS quantum mechanics. This describes a system of $N$ coincident D0-branes forming a black 0-brane, i.e. a black hole. The corresponding metric is given by the type IIA supergravity solution 
\begin{eqnarray}
ds^2=\alpha^{\prime}\bigg[-\frac{U^{\frac{7}{2}}}{g_{\rm YM}\sqrt{d_0N}}(1-\frac{U_0^{7}}{U^{7}})dt^2+\frac{g_{\rm YM}\sqrt{d_0N}}{U^{\frac{7}{2}}}\frac{dU^2}{(1-\frac{U_0^{7}}{U^{7}})}+g_{\rm YM}\sqrt{d_0N}U^{\frac{-3}{2}}d\Omega^2_{8}\bigg].\nonumber\\
\end{eqnarray}
We write this as
\begin{eqnarray}
ds^2=\alpha^{\prime}\bigg[-\frac{F(U)}{\sqrt{H}}dt^2+\frac{\sqrt{H}}{F(U)}dU^2+\sqrt{H}U^{2}d\Omega^2_{8}\bigg],\label{0b}
\end{eqnarray}
where 
\begin{eqnarray}
F(U)=1-\frac{U_0^{7}}{U^{7}}~,~H=\frac{\lambda d_0}{U^7}~,~d_0=240\pi^5.
\end{eqnarray}
The temperature, the energy and the entropy are
\begin{eqnarray}
T=\frac{7U_0^{\frac{5}{2}}}{4\pi \sqrt{d_0\lambda}}.\label{tem}
\end{eqnarray}
\begin{eqnarray}
E=\frac{U_0^{7}}{a_0 g_{\rm YM}^4}\Rightarrow \frac{E}{N^2}=\frac{1}{a_0}\bigg(\frac{4\pi\sqrt{d_0}}{7}\bigg)^{14/5}\lambda^{-3/5}T^{14/5}.\label{delta}
\end{eqnarray}
\begin{eqnarray}
{\cal S}&=&\frac{8\pi \sqrt{d_0N}}{9a_0^{\frac{5}{14}}}g_{\rm YM}^{\frac{-3}{7}}E^{\frac{9}{14}}\Rightarrow  \frac{{\cal S}}{N^2}=\frac{14}{9a_0}\bigg(\frac{4\pi\sqrt{d_0}}{7}\bigg)^{14/5}\lambda^{-3/5}T^{9/5}.\label{tro}
\end{eqnarray}
This black 0-brane solution is one very important example of matrix black holes \cite{Horowitz:1997fr,Li:1998ci,Englert:1998vr,Banks:1997cm,Liu:1997gk,Das:1997tk,Li:1997iz,Banks:1997hz,Banks:1997tn,Klebanov:1997kv}. We will now show that this dual geometry for the BFSS model can be lifted to a solution to $11$-dimensional supergravity.
\subsection{Supergravity in $11$ dimensions and M-wave solution}
M-theory is not a field theory on an ordinary spacetime except approximately in the low energy perturbative region and it is not a string theory except via circle compactification. In
this theory, understood as M(atrix)-theory or AdS/CFT, there is the remarkable  possibility of dynamically
growing new dimensions of space, topology change, spontaneous emergence of geometry from $0$ dimension, and more dramatically we can have emergence of gravitational theories from gauge theories.

 The five superstring theories as well as $11-$d supergravity are
 believed to be limits of the same underlying $11-$d theory which came
 to be called M-theory.  At low energy M-theory is approximated by $11-$d
Supergravity, i.e. M-theory is by construction the UV-completion of
$11-$d supergravity.

It was shown in \cite{Nahm:1977tg} that the largest
number of dimensions in which supergravity can exist is $11$. This is the largest number of dimensions consistent with a single
graviton. Beyond $11$ dimensions spinors have at least
$64$ components which lead to massless fields with spins larger than $2$ which
have no consistent interactions. 

The field content of
$11-$d supergravity consists of a vielbein $e^A_M$ (or equivalently
the metric $g_{MN}$), a Majorana
gravitino ${\psi}_M$ of spin $3/2$, and a $3-$form gauge potential
$A_{MNP}\equiv A_3$ (with field strength $F_4=dA_3$). There is a
unique classical action \cite{Cremmer:1978km}. The bosonic part of this action reads
\begin{eqnarray}
S_{11}=\frac{1}{2k_{11}^2}\int
d^{11}x\sqrt{-g}(R-\frac{1}{2}|F_4|^4)-\frac{1}{12k_{11}^2}\int A_3\wedge
F_4\wedge F_4.\label{11Dsugra}
\end{eqnarray}
The form action is given explicitly by
\begin{eqnarray}
\int d^{11}x\sqrt{-g} |F_4|^2=\int  d^{11}x\sqrt{-g} F_4\wedge^{*}F_4=\int d^{11}x\sqrt{-g} \frac{1}{4!}g^{M_1N_1}...g^{M_4N_4}F_{M_1...M_4}F_{N_1...N_4}.\nonumber\\
\end{eqnarray}
Newton's constant in $11$ dimensions is given by
$2k_{11}^2/16\pi$, viz
\begin{eqnarray}
G_{11}=\frac{2k_{11}^2}{16\pi}.
\end{eqnarray}
In $D$ dimensions the gravitational force falls with the distance $r$ as $1/r^{D-2}$ and the Newton's and the Planck constants are related by the relation \cite{Zwiebach:2004tj}
\begin{eqnarray}
16\pi G_{D}=2k_{D}^2=\frac{1}{2\pi}(2\pi l_P)^{D-2}.
\end{eqnarray}
Let us count the number of degrees of freedom \cite{Cremmer:1978km,Polchinski:1998rr,Miemiec:2005ry}. The metric is a traceless symmetric tensor. It is contained in the symmetric part of the tensor product of two fundamental vector representations
of $SO(9)$. This is because each index of the metric is a vector index which can take only
$D-2=11-2=9$ values.  Since the metric tensor must also be a traceless
 tensor  it will contain in total $\frac{9^2-9}{2}+9-1=44$ degrees of  freedom.  

The gauge
potential $A_3$ must be contained in the antisymmetric part of the
tensor product of three fundamental vector representations
of $SO(9)$. The independent number of $A_{MNP}$ is obvioulsy given by
$\frac{9\times 8\times 7}{3!}=84$. The total number of bosonic
variables is thus given by $44+84=128$. 

In $11$
dimensions the Dirac matrices are $32\times 32$.  The gravitino is a
Majorana spin $\frac{3}{2}$ field which carries a vector and a spinor
indices. The
Rarita-Schwinger gravitino field must clearly be contained in the tensor product
of the  fundamental vector representation and the spinor representation of
$SO(9)$. There are only  $D-3=11-3=8$ propagating components for a spin
 $\frac{3}{2}$ field in the same way
that a spin $1$ field has only $D-2=11-2=9$ propagating components. The
independent number of fermionic degrees of freedom is therefore
$\frac{32\times 8}{2}=128$ where the division by $2$ is due to the
Majorana condition. The numbers of bosonic and fermionic degrees of freedom match.

Supergravity in $11$ dimensions contains
a $2-$brane solution which is an electrically charged configuration with
respect to  $A_{MNP}$ and a $5-$brane solution which is a magnetically
charged configuration with respect to  $A_{MNP}$. These supergravity
$2$-brane and $5$-brane solutions are the low-energy limits of the M$2-$ and M$5-$branes respectively of M-theory.

The equations of motion of 11-dimensional supergravity are given by the following Einstein-Maxwell system 
\begin{eqnarray}
R_{MN}-\frac{1}{2}g_{MN}R=\frac{1}{2}F_{MN}^2-\frac{1}{4}g_{MN}|F_4|^2.
\end{eqnarray}
\begin{eqnarray}
d*F_4+\frac{1}{2}F_4\wedge F_4=0~,~dF_4=0.
\end{eqnarray}
Type IIA supergravity theory can be constructed by dimensional reduction of supergravity in $11$ dimensions on a circle $S^1$. This as it turns out is a consequence of the deeper fact that M-theory compactified on a circle of radius $R$ is equivalent to type IIA superstring in ten dimensions with coupling constant $g_s=R/\sqrt{\alpha^{\prime}}$ (more on this below). Kaluza-Klein dimensional reduction is obtained by keeping only the zero modes in the Fourier expansions along the compact direction whereas compactification is obtained by keeping all Fourier modes in the lower dimensional theory. 

The dimensional reduction of the metric $g_{MN}$ is specified explicitly in terms of a ten-dimensional metric $g_{mn}$, a $U(1)$ gauge field $A_m$ and a scalar dilaton field $\Phi$, by \cite{Becker:2007zj}
\begin{eqnarray}
ds^2=g_{MN}dx^Mdx^N=e^{-\frac{2\Phi}{3}}g_{mn}dx^mdx^n+e^{\frac{4\Phi}{3}}(dx_{10}+A_mdx^m)^2.\label{bf}
\end{eqnarray}
Thus the distance $l_{11}$ between any two points  as viewed in the eleven-dimensional theory should be viewed as a distance $l_{10}$ in the ten-dimensional theory which are related by $l_{10}=g_s^{-1/3}l_{11}$ where $g_s=\langle \exp(\Phi)\rangle$ is the string coupling constant. Thus the Planck scale $l_P$ in eleven-dimension is related to the string length  $l_s=\sqrt{\alpha^{\prime}}$ \footnote{Note that the string length is sometimes defined by $l_s=\sqrt{\alpha^{\prime}}$ and sometimes by $l_s=\sqrt{2\alpha^{\prime}}$.} in ten-dimension by the relation 
\begin{eqnarray}
l_P=g_s^{1/3}l_s.
\end{eqnarray}
It is also not difficult to convince oneself that the $11-$dimensional coordinate transformation $x_{10}\longrightarrow x_{10} + f(x_{10})$ is 
precisely  equivalent to the $10-$dimensional $U(1)$ gauge transformation $A\longrightarrow A-df$.

The three-form $A_3$ in eleven dimensions gives a three-form $C_3$ and a two-form $B_2$ in ten dimensions given by
\begin{eqnarray}
C_3=A_{mnq}dx^m\wedge dx^n\wedge dx^q~,~B_2=A_{10 mn}dx^m\wedge dx^n.
\end{eqnarray}
From the action (\ref{11Dsugra}) we see that the ten-dimensional Newton's constant is given in terms of the eleven-dimensional Newton's constant by the relation 
\begin{eqnarray}
G_{10}=\frac{G_{11}}{2\pi R}=\frac{2k_{10}^2}{16\pi}.
\end{eqnarray}
The Newton's constant $G_{10}$, the gravitational coupling constant $k_{10}$, the string length $l_s$ and the coupling constant $g_s$ are related by the formula (see below)
\begin{eqnarray}
16\pi G_{10}=2k_{10}^2=\frac{1}{2\pi}g_s^2(2\pi l_s)^8.
\end{eqnarray}
We obtain immediately 
\begin{eqnarray}
\frac{2k_{11}^2}{2\pi R}=2k_0^2g_s^2~,~2k_0^2=\frac{1}{2\pi}(2\pi l_s)^8.
\end{eqnarray}
And
\begin{eqnarray}
R=g_sl_s.
\end{eqnarray}
Thus, the original metric $g_{MN}$ corresponds to the ten-dimensional fields $g_{mn}$, $\Phi$ and $A_1$, whereas the original three-form $A_3$ corresponds to the three-form $C_3$ and the two-form $B_2$. The leading $\alpha^{\prime}=l_s^2$ low energy effective action of the bosonic $D=11$ supergravity action, which is obtained by integrating over the compact direction, is given precisely by the action of type IIA supergravity in $D=10$ dimensions. This action consists of three terms 
  \begin{eqnarray}
S=S_{\rm NS}+S_{\rm R}+S_{\rm CS}.
\end{eqnarray}
The action $S_{\rm NS}$ describes the fields of the NS-NS sector, viz $(g_{mn},\Phi,B_2)$. The field strength of $B_2$ is denoted $H_3=dB_2$. It is given in the string frame explicitly by
  \begin{eqnarray}
S_{\rm NS}=\frac{1}{2k_0^2}\int d^{10}x\sqrt{-g}\exp(-2\Phi)\big(R+4\partial_{\mu}\Phi\partial^{\mu}\Phi-\frac{1}{2}|H_3|^2\big).
\end{eqnarray}
The action $S_{\rm R}$ describes the fields of the R-R sectors, viz $(A_1,C_3)$. The field strengths are given respectively by $F_2=dA_1$ and $\tilde{F}_4=dC_3+A_1\wedge H_3$. The action is given by 
 \begin{eqnarray}
S_{\rm R}=-\frac{1}{4k_0^2}\int d^{10}x\sqrt{-g}\big(|F_2|^2+|\tilde{F}_4|^2\big).
\end{eqnarray}
The Chern-Simons term is given obviously by (where $F_4=dA_3$)
\begin{eqnarray}
S_{\rm CS}=-\frac{1}{4k_{0}^2}\int B_2\wedge F_4\wedge F_4.
\end{eqnarray}
The solution of eleven-dimensional gravity which is relevant for the dual geometry to the BFSS quantum mechanics corresponds the non-extremal M-wave solution. This corresponds to $A_3=0$, i.e. $C_3=0$ and $B_2=0$. The $11-$dimensional metric corresponds to \cite{Hyakutake:2013vwa,Hyakutake:2014maa}
\begin{eqnarray}
ds^2_{11}&=&g_{MN}dx^Mdx^N\nonumber\\
&=&l_s^4\bigg(-H^{-1}Fdt^2+F^{-1}dU^2+U^2d\Omega_8^2+\big(l_s^{-4}H^{1/2}dx_{10}-(\frac{U_+}{U_-})^{7/2}H^{-1/2}dt\big)^2\bigg).\nonumber\\
\end{eqnarray}
\begin{eqnarray}
F=1-\frac{U_+^7-U_-^7}{U^7}=1-\frac{U_0^7}{U^7}~,~U_0^7=U_+^7-U_-^7~,~U=\frac{r}{\alpha^{\prime}}.
\end{eqnarray}
\begin{eqnarray}
H=l_s^4\frac{U_-^7}{U^7}.
\end{eqnarray}
By using the basic formula (\ref{bf}) we get immediately the $10-$dimensional metric 
\begin{eqnarray}
ds^2_{10}&=&g_{mn}dx^mdx^n\nonumber\\
&=&l_s^2\bigg(-H^{-1/2}Fdt^2+H^{1/2}F^{-1}dU^2+H^{1/2}U^2d\Omega_8^2\bigg).
\end{eqnarray}
This is the same solution as the non-extremal black 0-brane solution (\ref{0b}). The dilaton $\Phi$ and the one-form field $A=A_mdx^m$ are given by the identification
\begin{eqnarray}
\exp(\Phi)=l_s^{-3}H^{3/4}~,~A=-(\frac{U_+}{U_-})^{7/2}l_s^4H^{-1}dt.
\end{eqnarray}
The mass $M$ and the R-R charge $Q$ of this black 0-brane solution are given by (with $S^8$ being the volume of $S^8$)
\begin{eqnarray}
M=\frac{V_{S^8}\alpha^{\prime 7}}{2k_{10}^2}(8U_+^7-U_-^7)~,~Q=\frac{7V_{S^8}\alpha^{\prime 7}}{2k_{10}^2}(U_+U_-)^{7/2}.
\end{eqnarray}
The charge of $N$ coincident D0-branes in type IIA superstring is quantized as 
\begin{eqnarray}
Q=\frac{N}{l_sg_s}.
\end{eqnarray}
We write the radii $r_{\pm}$ as $r_{\pm}^7=(1+\delta)^{\pm 1} R$. Hence (with $V_{S^8}=2(2\pi)^4/(7.15)$)
\begin{eqnarray}
\frac{N}{l_sg_s}=\frac{7V_{S^8}}{2k_{10}^2}(r_+r_-)^{7/2}\Rightarrow R=\frac{N}{l_sg_s}\frac{2k_{10}^2}{7V_{S^8}}=15\pi N l_s^7g_s(2\pi)^2.
\end{eqnarray}
The above M-wave solution is non-extremal. The extremal solution corresponds to the limit $U_-\longrightarrow U_+$ or equivalently $\delta\longrightarrow 0$. The Schwarzschild limit corresponds to $U_-\longrightarrow 0$ and thus the charge $Q$ vanishes in this limit.

The third much more important limit is the near horizon limit $r\longrightarrow 0$ given by \cite{Itzhaki:1998dd}
\begin{eqnarray}
U=\frac{r}{\alpha^{\prime}}={\rm fixed}.
\end{eqnarray}
\begin{eqnarray}
U_0=(U_+^7-U_-^7)^{1/7}=U_-\alpha={\rm fixed}.
\end{eqnarray}
\begin{eqnarray}
\lambda=g_{\rm YM}^2N=\frac{g_s}{(2\pi)^2}\frac{1}{l_s^3}N={\rm fixed}.
\end{eqnarray}
$\alpha$ is a dimensionless parameter. Remark that the extremal limit corresponds to $\alpha\longrightarrow 0$. The near horizon limit  is also equivalent to the limit $\alpha\longrightarrow 0$ since $U_0/U=r_-\alpha/r={\rm fixed}$. Thus the near horizon limit is equivalent to the near extremal limit and it is given by  $\alpha\longrightarrow 0$ keeping fixed 
\begin{eqnarray}
\frac{U_0}{U}=\frac{r_-\alpha}{r}={\rm fixed}.
\end{eqnarray}
\begin{eqnarray}
\frac{1}{U_0}=\frac{\alpha^{\prime}}{r_-\alpha}={\rm fixed}.
\end{eqnarray}
\begin{eqnarray}
U_0^3\frac{g_s^2N^2}{l_s^6}=\frac{g_s^2N^2}{(r_-\alpha)^3}={\rm fixed}.
\end{eqnarray}
In summary, the M-wave solution which is a purely geometrical solution in $11$ dimensions corresponds to a charged black hole in $10$ dimensions smeared along the compact $x_{10}$ direction. In the extremal or near horizon limit the functions $F$ and $H$ characterizing this solution are given by 
\begin{eqnarray}
H=\frac{H_0}{U^7}~,~H_0=l_s^4U_-^7=\frac{1}{1+\delta}240\pi^5\lambda.
\end{eqnarray}
\begin{eqnarray}
F=1-\frac{U_0^7}{U^7}~,~U_0^7=\frac{\delta(2+\delta)}{1+\delta}\frac{15\pi\lambda(2\pi)^4}{l_s^4}.
\end{eqnarray}
By comparing with (\ref{delta}) we get that $\delta$ is proportional to the energy and that the extremal limit is also characterized by $l_s\longrightarrow 0$ keeping $\delta/l_s^4$ fixed given by 
\begin{eqnarray}
\frac{\delta}{l_s^4}=\frac{56\pi^2\lambda}{9}\frac{E}{N^2}.
\end{eqnarray}
The mass and the charge are given by 
\begin{eqnarray}
M=(1+\frac{9\delta}{7})Q~,~Q=\frac{N^2}{(2\pi)^2l_s^4\lambda}.
\end{eqnarray}
\subsection{Type IIA string theory at strong couplings is M-Theory }
 The theory of $11-$d supergravity arises also as a low energy
limit of type IIA superstring theory \cite{Witten:1995ex}. We will discuss
this point in some more detail following  \cite{Polchinski:1998rr}. We consider the behaviour of Dp-branes
 at strong coupling in type IIA string theory. The Dp-branes have
 tensions given by the Polchinski formula \cite{Polchinski:1995mt,deAlwis:1996ez}
\begin{eqnarray}
T^2_p=\frac{2\pi}{2k^2_{10}}(4{\pi}^2{\alpha}^{'})^{3-p}.\label{tp2}
\end{eqnarray}
The fundamental string tension is $1/(2\pi{\alpha}^{'})$ and Newton's
constant in $10$ dimensions is $2k^2_{10}/16\pi$. Type IIA string theory
contains only Dp-branes with $p$ even while type IIB string theory
contains  Dp-branes with $p$ odd. These two series are related by
T-duality. Indeed, it was shown using T-duality that \cite{Green:1996bh}
\begin{eqnarray}
T_{p-1} = 2\pi {\alpha}^{'\frac{1}{2}}T_p.
\end{eqnarray}
This is consistent with (\ref{tp2}). It was also shown  using
$SL(2,Z)$ duality that the D-string (i.e. $p=1$) in
type IIB string theory has a tension given by \cite{Schwarz:1995dk}
\begin{eqnarray}
T_1=g^{-1}(2\pi{\alpha}^{'})^{-1}.\label{t1}
\end{eqnarray}
The perturbative effective string coupling is given by $g=\langle e^{\Phi}\rangle+...$ where $\Phi$ is the dilaton field. From (\ref{tp2}) and
(\ref{t1}) we get that
\begin{eqnarray}
2k^2_{10}=(2\pi)^3g^2(2\pi {\alpha}^{'})^4.
\end{eqnarray}
Putting this last equation back in (\ref{tp2}) we obtain
\begin{eqnarray}
T_p=(2\pi)^{\frac{1-p}{2}}g^{-1}(2\pi {\alpha}^{'})^{-\frac{p+1}{2}}.
\end{eqnarray}
This tension translates into a mass scale \cite{Polchinski:1995mt}
\begin{eqnarray}
{T}_p^{\frac{1}{p+1}}=O(g^{-\frac{1}{p+1}}{\alpha}^{'-\frac{1}{2}}).
\end{eqnarray}
Remark that ${\alpha}^{'-1}$ has units of mass squared and $g$ is
dimensionless and thus ${T}_p^{\frac{1}{p+1}}$ has units of mass. Clearly, at strong string coupling $g$ the smallest $p$ (i.e.
$p=0$) 
gives the smallest scale, i.e the lowest modes. These are the D0-branes. Their mass is
\begin{eqnarray}
{T}_0=\frac{1}{g {\alpha}^{'\frac{1}{2}}}.
\end{eqnarray}
At weak coupling this is very heavy. Bound states of any number $n$ of D0-branes
   do exist. They have mass
\begin{eqnarray}
n{T}_0=\frac{n}{g {\alpha}^{'\frac{1}{2}}}.
\end{eqnarray}
As $g\longrightarrow \infty $ these states become massless and the
 above spectrum becomes a continuum identical to the continuum of momentum
Kaluza-Klein states which correspond to a periodic (i.e. compact) $11$ dimensions of radius 
\begin{eqnarray}
R_{10}=g {\alpha}^{'\frac{1}{2}}.\label{R10}
\end{eqnarray}
Thus, the limit $g\longrightarrow \infty $ is the decompactification
 limit $R_{10}\longrightarrow \infty$ where an extra $11$ dimensions
 appears. This extra dimension is invisible in string perturbation
 theory because if $g\longrightarrow 0$ it becomes very
 small (i.e.  $R_{10}\longrightarrow
 0$). 

In the  dimensional reduction of $11-$dimensional supergravity to
$10-$dimensional type IIA supergravity only the states with
momentum $p_{10}=0$ survives.  However, type IIA string theory (with type IIA supergravity as its low energy limit) contains also
states with $p_{10}\neq 0$ which are the D0-branes and their bound
states. In this dimensional reduction the Kaluza-Klein gauge field coupling to the
eleventh dimension is exactly the Ramond-Ramond gauge field coupling to
the D0-branes of type IIA supergravity.

From the action (\ref{11Dsugra}) we can immediately see that the
$11-$dimensional and $10-$dimensional gravitational couplings $k_{11}$
and $k_{10}$ are related by
\begin{eqnarray}
k_{11}^2=2\pi R_{10}k^2_{10}=\frac{1}{2}(2\pi)^8g^3{\alpha}^{'\frac{9}{2}}.
\end{eqnarray}
 The $11-$dimensional Planck mass is defined by
\begin{eqnarray}
2k_{11}^2\equiv (2\pi)^8M_{11}^{-9}
\end{eqnarray}
The power of $-9$ is dictated
by the fact that we have (by using the last two equations) 
\begin{eqnarray}
M_{11}=g^{-\frac{1}{3}}{\alpha}^{'-\frac{1}{2}}.\label{M11}
\end{eqnarray}
Since ${\alpha}^{'-\frac{1}{2}}$ has units of  mass the $M_{11}$ will have
units of mass. From
(\ref{R10}) and (\ref{M11}) we find that the parameters $g$ and
${\alpha}^{'}$ of type IIA
string theory are related to the parameters $R_{10}$ and $M_{11}$ of
the $11-$dimensional theory by
\begin{eqnarray}
g=(R_{10}M_{11})^{\frac{3}{2}}~,~{\alpha}^{'}=R_{10}^{-1}M_{11}^{-3}.
\end{eqnarray}
\section{M-theory prediction for quantum black holes}
Let us start by writing down the metric of the $10-$dimensional black 0-brane solution in type II A supergravity theory:
\begin{eqnarray}
ds^2=\alpha^{\prime}\bigg[-\frac{F(U)}{\sqrt{H}}dt^2+\frac{\sqrt{H}}{F(U)}dU^2+\sqrt{H}U^{2}d\Omega^2_{8}\bigg],\label{0b}
\end{eqnarray}
where 
\begin{eqnarray}
F(U)=1-\frac{U_0^{7}}{U^{7}}~,~H=\frac{240\pi^5\lambda }{U^7}.
\end{eqnarray}
The horizon is located at $U_0$ where
\begin{eqnarray}
U_0^{7}=a_0 \lambda^2\frac{E}{N^2}.
\end{eqnarray}
This solution is characterized by two parameters $U$ and $\lambda$ which correspond to a typical energy scale and 't Hooft coupling constant in the dual gauge theory. These two parameters define also the mass and the charge of the black hole.

By the gauge/gravity duality, classical type IIA  supergravity around this black 0-brane solution is equivalent to the limit $\lambda\longrightarrow \infty$, $\lambda\longrightarrow\infty$ of the $(0+1)-$dimensional gauge theory given by the BFSS model 
  \begin{eqnarray}
S=\frac{1}{g_{\rm YM}^2}\int_0^{\beta} dt \big[\frac{1}{2}(D_tX_I)^2-\frac{1}{4}[X_I,X_J]^2-\frac{1}{2}\bar{\psi}\gamma^{0}D_{0}\psi-\frac{1}{2}\bar{\psi}\gamma^{I}[X_{I},\psi]\big].
\end{eqnarray}
The quantum gravity corrections to the above black 0-brane solution were compute analytically in \cite{Hyakutake:2013vwa} whereas the corresponding non-perturbative gauge theory corrections using Monte Carlo simulations of the above gauge theory were computed in \cite{Hanada:2013rga}.This is an extremely important very concrete check of the gauge/gravity correspondence. In the following we will outline their main results.
\subsection{Quantum gravity corrections}
In this section we follow closely \cite{Hyakutake:2013vwa}.

The quantum gravity corrections are loop corrections in the string coupling $g_s$. From the dual gauge theory side this corresponds to corrections in $1/N^2$. The calculation of the effective action is done in eleven dimensions by demanding local supersymmetry. This effective action of M-theory is given by the action of $11-$dimensional supergravity with the addition of higher derivative terms. 

The effective action of M-theory becomes after dimensional reduction the one-loop effective action of type IIA superstring theory. Also, since we are only interested in the M-wave solution, which is a purely geometrical object in $11$ dimensions, we should only concentrate on the graviton field. The structure of the effective action of type IIA superstring theory must be such that it is consistent with the scattering amplitudes of strings. The leading corrections to type IIA supergravity involve four gravitons as asymptotic states and correspond to quartic terms of the Riemann tensor \cite{Gross:1986iv}. The corrections $[eR^4]$ when lifted to $11$ dimensions become $B_1=[eR^4]_7$ where the subscript $7$ indicates the number of potentially different contractions. The other corrections are determined by demanding local supersymmetry. In this way the combination of the quartic terms of the Riemann tensor in $B_1$ is determined uniquely  up to an overall factor. The effective action of the M-theory relevant to the graviton scattering is given explicitly by \cite{Hyakutake:2013vwa,Hyakutake:2006aq,Hyakutake:2007sm}

 \begin{eqnarray}
\Gamma_{11}=\frac{1}{2k_{11}^2}\int d^{11}x\sqrt{-g}\bigg(R+\gamma l_s^{12}(t_8t_8R^4-\frac{1}{4!}\epsilon_{11}\epsilon_{11}R^4)\bigg).
\end{eqnarray}
The tensor $t_8$ is the product of four Kronecker's symbols whereas the tensor $\epsilon_{11}$ is an antisymmetric tensor with eleven indices. The parameter $\gamma$ is given by
\begin{eqnarray}
\gamma=\frac{\pi^6}{2^73^2}\frac{\lambda^2}{N^2}
\end{eqnarray}
By dimensional reduction we get from $\Gamma_{11}$ the one-loop effective action of type IIA superstring theory. 

Next we derive the equations of motion corresponding to the action $\Gamma_{11}$ and solve them up to the linear order in $\gamma$. The near horizon geometry of the M-wave solution with quantum gravity corrections included takes the form \cite{Hyakutake:2013vwa}
\begin{eqnarray}
ds^2_{11}&=&l_s^4\bigg(-H_1^{-1}F_1dt^2+F_1^{-1}dU^2+U^2d\Omega_8^2+\big(l_s^{-4}H_2^{1/2}dx_{10}-H_3^{-1/2}dt\big)^2\bigg).\nonumber\\
\end{eqnarray}
The functions $H_i$ and $F_1$ are given by
\begin{eqnarray}
H_i=H+\frac{1}{N^2}\frac{5\pi^{16}\lambda^3}{24U_0^{13}}h_i~,~F_1=F+\frac{1}{N^2}\frac{\pi^{6}\lambda^2}{1152 U_0^{6}}f_1.
\end{eqnarray}
The functions $h_i$ and $f$ are functions of $x=U/U_0$ given by equation $(44)$ of \cite{Hyakutake:2013vwa}. The dimensional reduction of the above solution gives immediately the near horizon geometry of the black 0-brane solution with quantum gravity corrections. This is given explicitly by
\begin{eqnarray}
ds^2_{10}&=&l_s^2\bigg(-H_1^{-1}H_2^{1/2}F_1dt^2+H_2^{1/2}F_1^{-1}dU^2+H_2^{1/2}U^2d\Omega_8^2\bigg).
\end{eqnarray}
\begin{eqnarray}
\exp(\Phi)=l_s^{-3}H_2^{3/4}~,~A=-l_s^4H_2^{-1/2}H_3^{-1/2}dt.
\end{eqnarray}
The horizon becomes shifted given by 
\begin{eqnarray}
F_{1H}=0\Rightarrow \frac{U_H^7}{U_0^7}=1-\frac{1}{N^2}\frac{\pi^{6}\lambda^2}{1152 U_0^{6}}f_1.
\end{eqnarray}
The temperature of the black 0-brane is derived by performing Wick rotation and then removing the conical singularity by imposing periodicity in the Euclidean time direction. We define the proper distance from the horizon by 
\begin{eqnarray}
d\rho^2=F_1^{-1}dU^2\Rightarrow \rho=2F_{1H}^{\prime -\frac{1}{2}}\sqrt{U-U_H}.
\end{eqnarray}
Then 
\begin{eqnarray}
F_1=\frac{\rho^2}{4}F_1^{\prime 2}|_{H}.
\end{eqnarray}
The relevant part of the metric takes now the form 
\begin{eqnarray}
ds^2_{10}&=&l_s^2H_2^{1/2}\bigg(\frac{\rho^2}{4r_s^2}dt^2+d\rho^2+...\bigg)~,~\frac{1}{r_s}=H_{1H}^{-1/2}F_{1H}^{\prime}.
\end{eqnarray}
We get immediately the temperature 
\begin{eqnarray}
T=\frac{1}{4\pi r_s}=\frac{1}{4\pi}H_{1H}^{-1/2}F_{1H}^{\prime}.
\end{eqnarray}
In terms of $\tilde{U}_0=U_0/\lambda^{1/3}$, $\tilde{T}=T/\lambda^{1/3}$ and $\epsilon=\gamma/\lambda^2$ this formula reads 
\begin{eqnarray}
\tilde{T}=a_1\tilde{U}_0^{5/2}(1+\epsilon a_2\tilde{U}_0^{-6}).
\end{eqnarray}
The coefficient $a_1$ is precisely the coefficient computed before in equation (\ref{tem}). The coefficient $a_2$ is given by equation $(48)$ of  \cite{Hyakutake:2013vwa}. 

Next we compute the entropy of the black 0-brane solution by using Wald's formula \cite{Wald:1993nt,Iyer:1994ys} in order to maintain the first law of thermodynamics. This entropy is exactly equal to the entropy of the M-wave solution. The Wald's formula for the M-wave solution is given in terms of the effective action $\Gamma_{11}$ and an antisymmetric binormal tensor $N_{\mu\nu}$ by 
\begin{eqnarray}
{\cal S}=-2\pi\int_Hd\Omega_8dx_{10}\sqrt{h}\frac{\partial \Gamma_{11}}{\partial R_{\mu\nu\alpha\beta}}N_{\mu\nu}N_{\alpha\beta}.
\end{eqnarray}
The binormal tensor $N_{\mu\nu}$ satisfies $N_{\mu\nu}N^{\mu\nu}=-2$ with the only non-zero component being $N_{tU}=-l_s^4U_0H_1^{-1/2}$ and where $\sqrt{h}=(l_s^2U)^8H_2^{1/2}/l_s^2$ is the volume form at the horizon. The result of the calculation is \cite{Hyakutake:2013vwa}
\begin{eqnarray}
\frac{{\cal S}}{N^2}=a_3\tilde{T}^{9/5}(1+\epsilon a_4\tilde{T}^{-12/5}).
\end{eqnarray}
Again the coefficient $a_3$ is computed before in equation (\ref{tro}). The coefficient $a_4$ is given by equation $(53)$ of \cite{Hyakutake:2013vwa}. 

The energy $\tilde{E}=E/\lambda^{1/3}$ is given by the first law of thermodynamics, viz $d\tilde{E}=\tilde{T}d{\cal S}$. We get immediately 
\begin{eqnarray}
\frac{\tilde{E}}{N^2}=\frac{9a_3}{14}\tilde{T}^{14/5}-\frac{3\epsilon a_3a_4}{2}\tilde{T}^{2/5}.
\end{eqnarray}
The numerical coefficients are given by \cite{Hyakutake:2013vwa}
\begin{eqnarray}
\frac{9a_3}{14}=7.41~,~\frac{3\epsilon a_3a_4}{2}=\frac{5.77}{N^2}.
\end{eqnarray}
We also get the specific heat 
 \begin{eqnarray}
\frac{1}{N^2}\frac{d\tilde{E}}{dT}=\frac{9a_3}{5}\tilde{T}^{9/5}-\frac{3\epsilon a_3a_4}{5}\tilde{T}^{-3/5}.\label{grav0}
\end{eqnarray}
This can become negative at low temperature which means that the black 0-brane behaves as an evaporating Schwarzschild black hole. This instability is obviously removed in the limit $N\longrightarrow\infty$.
\subsection{Non-perturbative tests of the gauge/gravity duality}
The basic prediction coming from the dual gravity is the formula for the energy per one degree of freedom given by 
\begin{eqnarray}
{\cal E}_{\rm gravity}=\frac{\tilde{E}}{N^2}=7.41.\tilde{T}^{14/5}-\frac{5.77}{N^2}.\tilde{T}^{2/5}.\label{grav1}
\end{eqnarray}
This formula includes the quantum gravity (second term) corrections which correspond to loop corrections proportional to $g_s$ or equivalently $1/N^2$. We can also include stringy corrections which are proportional to $\alpha^{\prime}$ or equivalently $1/\lambda$ following \cite{Green:2006gt,Hanada:2008ez}. We get then the formula 
\begin{eqnarray}
{\cal E}_{\rm gravity}=\frac{\tilde{E}}{N^2}=(7.41.\tilde{T}^{2.8}+a.\tilde{T}^{4.6}+...)+(-5.77\tilde{T}^{0.4}+b\tilde{T}^{2.2}+...)\frac{1}{N^2}+O(\frac{1}{N^4}).\label{grav}
\end{eqnarray}
This result is reproduced non-perturbatively by Monte Carlo simulation of the dual gauge theory, here it is the BFSS quantum mechanics, at the level of classical supergravity (for $N=\infty$ with the result $a=-5.58(1)$) in \cite{Hanada:2008ez} and at the level of quantum gravity in \cite{Hanada:2013rga}. The gauge/gravity duality is also tested for D0-branes in \cite{Hanada:2016zxj,Hanada:2011fq}. Other non-perturbative original tests of the gauge/gravity duality can be found in \cite{Kabat:2000zv,Catterall:2008yz}. Here we will follow closely \cite{Hanada:2013rga}.

The BFSS quantum mechanics 
\begin{eqnarray}
S=\frac{1}{g_{\rm YM}^2}\int_0^{\beta} dt \big[\frac{1}{2}(D_tX_I)^2-\frac{1}{4}[X_I,X_J]^2-\frac{1}{2}\bar{\psi}\gamma^{0}D_{0}\psi-\frac{1}{2}\bar{\psi}\gamma^{I}[X_{I},\psi]\big]
\end{eqnarray}
is put on computer along the lines of  \cite{Hanada:2008ez,Anagnostopoulos:2007fw}. The goal is to compute the internal energy 
\begin{eqnarray}
{\cal E}_{\rm gauge}=\frac{E}{N^2}~,~E=-\frac{\partial}{\partial \beta}\ln Z~,~Z=\int {\cal D}X{\cal D}A{\cal D}\psi\exp(-S). 
\end{eqnarray}
The theory can be regularized in the time direction either by a momentum cutoff $\Lambda$ \cite{Hanada:2007ti,Hanada:2013rga}, which incidentally preserves gauge invariance in one dimension, or by a conventional lattice $a$ \cite{Catterall:2008yz} which really remains the preferred method because of the dire need of parallelization in  this sort of calculations. As usual the fermionic Pfaffian is complex and thus in practice only the absolute value can be taken into consideration in the Boltzmann probability which is a valid approximation in this context \cite{Hanada:2013rga,Hanada:2011fq}.

In order to be able to compare sensibly the quantum gauge theory corrections on the gauge theory side with the corresponding quantum gravity corrections on the gravity side given by the $1/N^2$ term in equation (\ref{grav}) the authors of \cite{Hanada:2013rga} simulated the gauge theory with small values of $N$. Here the problem of flat directions (commuting matrices) becomes quite acute which is a problem intimately related to the physical effect of Hawking radiation and the instability of the black hole solution encountered in equation (\ref{grav0}). In particular, it is observed that the eigenvalues of the bosonic matrices $X_I$, which describe the positions of the D-particles forming the black hole, start to diverge for small $N$ signifying that these particles are being radiated away out of the black hole due to quantum gravity effects. In other words, the black hole for small $N$ is a metastable bound state of the D0-branes and it becomes really stable only in the large $N$ limit when quantum gravity effects can be suppressed and classical gravity becomes an exact description. 

\begin{figure}[H]
\begin{center}
\includegraphics[width=10.0cm,angle=0]{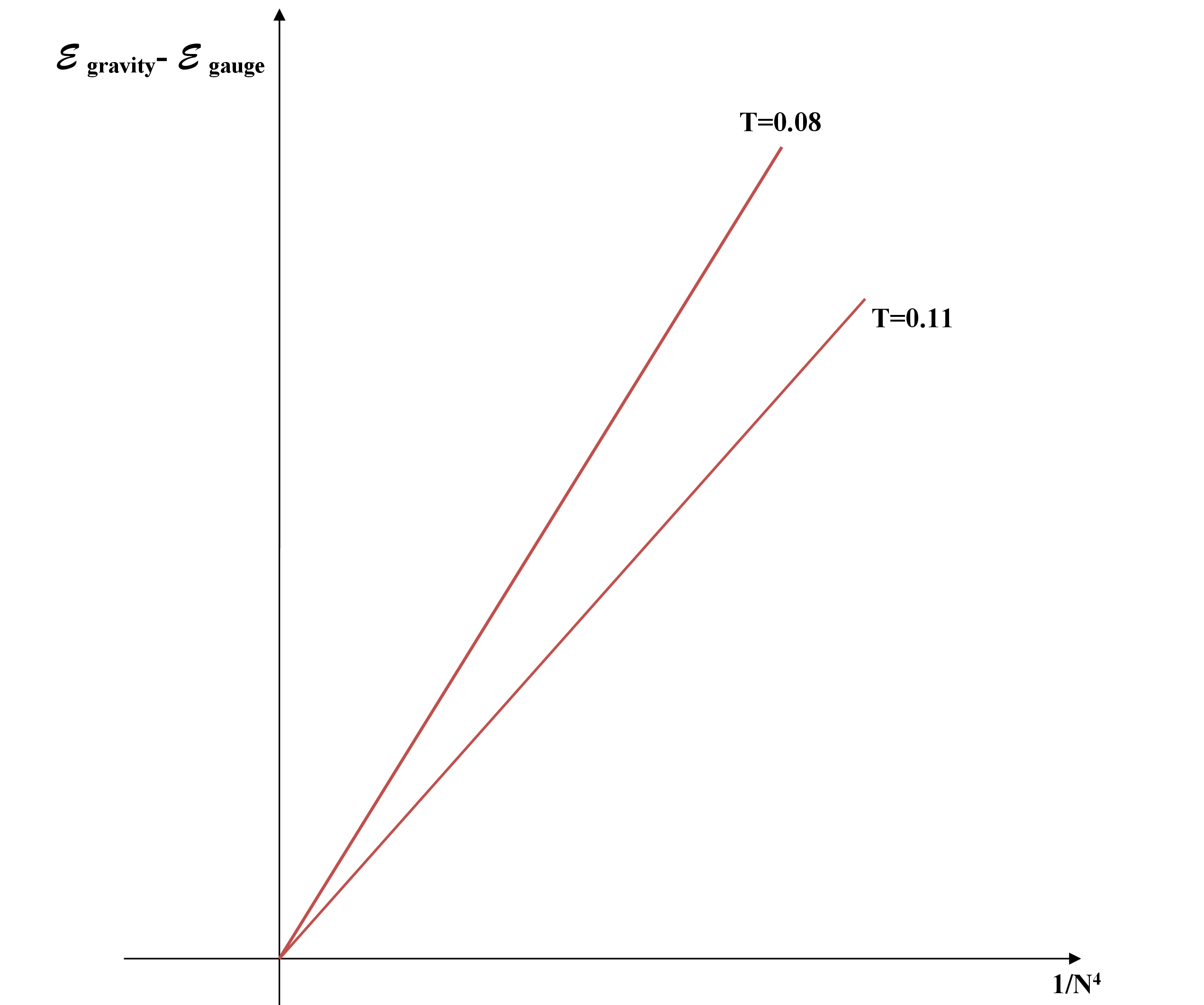}
\end{center}
\caption{Consistency of the gauge and gravity predictions.}\label{fd8}
\end{figure}

Hence, the black hole for small $N$ is only a metastable bound state of the D0-branes and quantum gravity is acting as a destabilizing effect. The energy of this metastable bound state is measured as follows:
\begin{itemize}
\item We introduce the extent of space
\begin{eqnarray}
R^2=\frac{1}{N\beta}\int_0^{\beta}dt\sum_{I=1}^9X_I(t)^2.
\end{eqnarray}
The histogram of $R^2$ presents a peak (bound state) and a tail (Hawking instability). See figure (\ref{fd6}).

\begin{figure}[H]
\begin{center}
\includegraphics[width=10.0cm,angle=0]{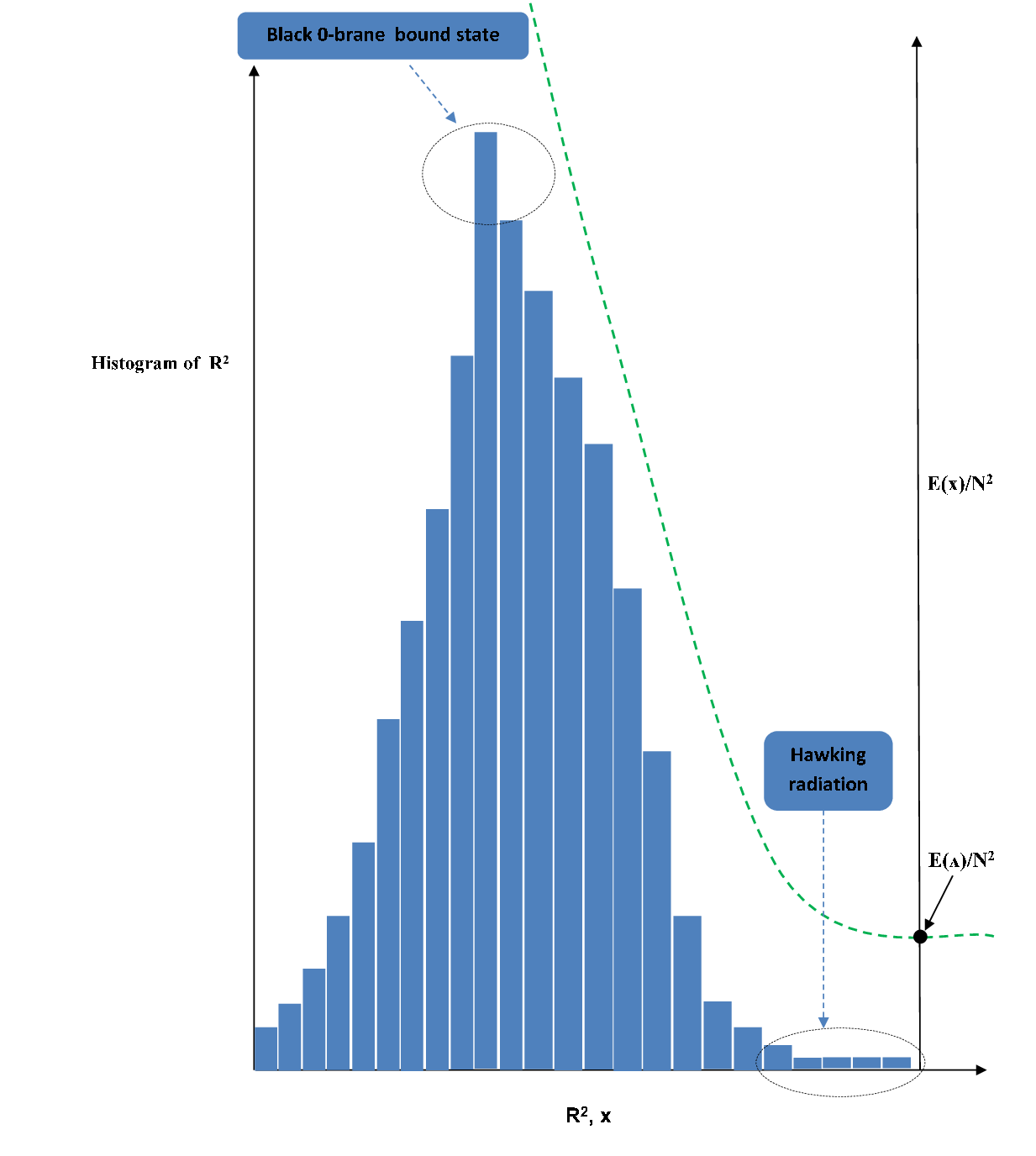}
\end{center}
\caption{Black 0-brane (black hole) and its Hawking radiation.}\label{fd6}
\end{figure}
\item The instability is tamed numerically by the addition of an appropriate potential.
\item We fix $T$, $N$ and $\Lambda$. In the measurement of the internal energy we only consider configurations which satisfy $R^2<x$. The corresponding estimation of the internal energy is denoted by ${\cal E}(x)=E(x)/N^2$. This estimation reaches a pleateau in the region of the tail of $R^2$, i.e. for all values of $x$ in the tail of the histogram of $R^2$ the energy takes on effectively the same value. This value is the measurement ${\cal E}(\Lambda)=E(\Lambda)/N^2$ of the internal energy for that particular set of values of $T$ and $N$ and $\Lambda$. 
\item We repeat the same analysis for other sets of values of $T$, $N$ and $\Lambda$. The internal energy in the continuum limit ${\cal E}_{\rm gauge}=E/N^2$ is obtained by fitting the results ${\cal E}(\Lambda)=E(\Lambda)/N^2$ for different $\Lambda$, but for the same $T$ and $N$, using the ansatz $E(\Lambda)=E+{\rm constant}/\Lambda$. We plot the energy ${\cal E}_{\rm gauge}$ as a function of $T$ for some fixed $N$. The authors of \cite{Hanada:2013rga} considered $N=3,4,5$. It is observed that the internal energy increases as $T$ decreases signaling that the specific heat is negative which is consistent with the result (\ref{grav0}) on the gravity side. See figure (\ref{fd7}).

\begin{figure}[H]
\begin{center}
\includegraphics[width=10.0cm,angle=0]{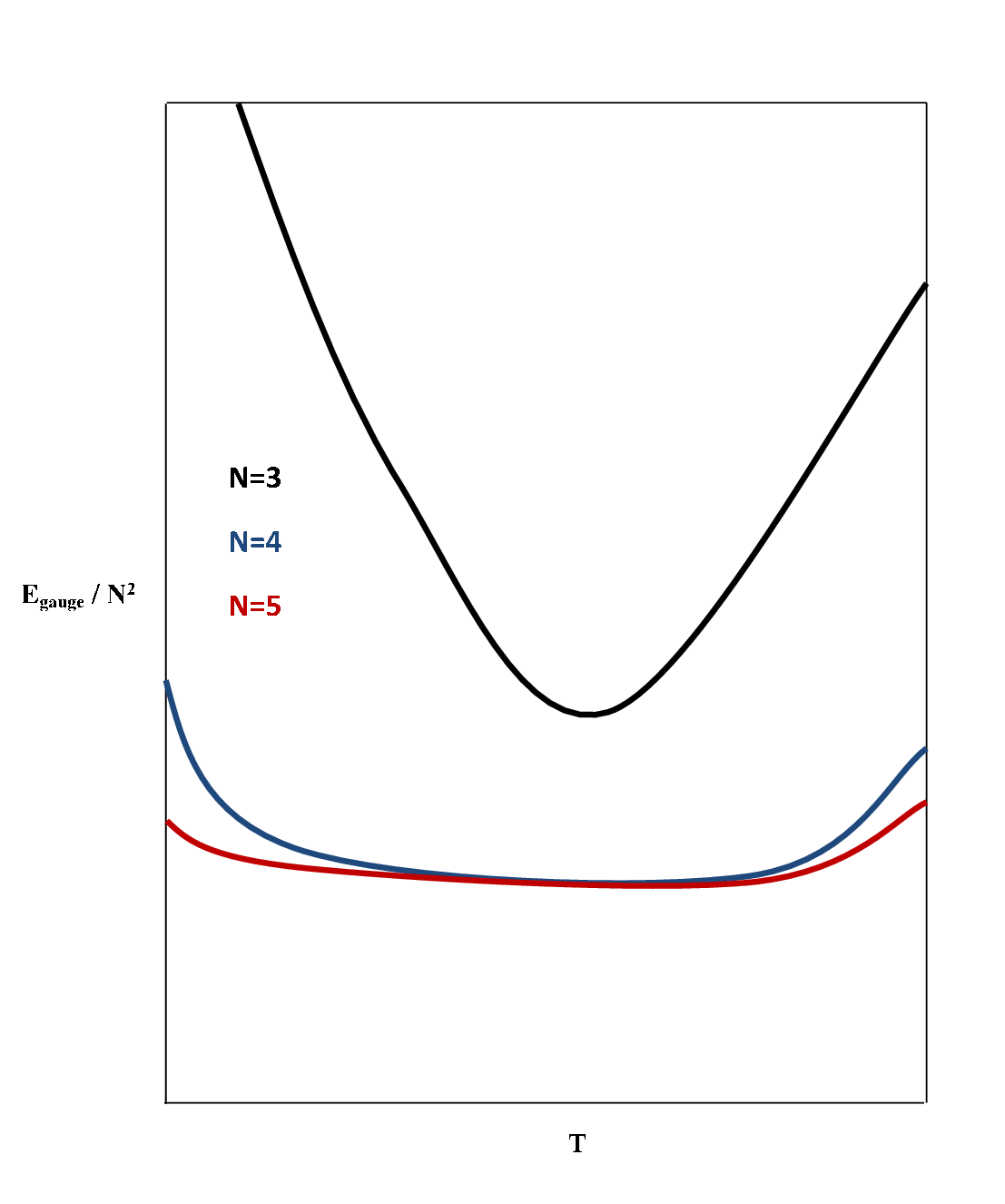}
\end{center}
\caption{Instability of the black hole: the specific heat increases as $T$ decreases.}\label{fd7}
\end{figure}
\item The stringy effects are found to be irrelevant in the temperature range $0.07\leq T\leq 0.12$ considered by the author of \cite{Hanada:2013rga}. The Monte Carlo results ${\cal E}_{\rm gauge}$ (figure $3$ of their paper) should then be compared with the gravity result (\ref{grav1}). They have provided convincing evidence that 
\begin{eqnarray}
{\cal E}_{\rm gravity}-{\cal E}_{\rm gauge}=\frac{{\rm constant}}{N^4}.
\end{eqnarray}
See figure (\ref{fd8}). This implies immediately that 
\begin{eqnarray}
{\cal E}_{\rm gauge}=\frac{{E}}{N^2}=7.41.{T}^{14/5}-\frac{5.77}{N^2}.{T}^{2/5}+\frac{c_2}{N^4}.\label{gauge}
\end{eqnarray}
The $T$ dependence of the coefficient $c_2$ is also found to be consistent with the prediction from the gravity side given by $c_2=cT^{-2.6}+...$.
\end{itemize}

%\section{T-duality and Dp-branes}

\section{Matrix string theory}
We consider now dimensional reduction to $D=2$ dimensions, i.e. $p=1$. Thus, we are dealing with a system of $N$ coincident D1-branes forming a black 1-brane solution with dynamics given by the maximally supersymmetric $U(N)$ gauge action (with $\mu,\nu=0,1$ and $I,J=1,...,8$)
\begin{eqnarray}
S=\frac{1}{g^2_{\rm YM}}\int d^{2} x\big[-\frac{1}{4}F_{\mu\nu}F^{\mu\nu}-\frac{1}{2}(D_{\mu}X_I)(D^{\mu}X_I)+\frac{1}{4}[X_I,X_J]^2+\frac{i}{2}\bar{\psi}\gamma^{\mu}D_{\mu}\psi+\frac{1}{2}\bar{\psi}\gamma^{I}[X_{I},\psi]\big].\nonumber\\\label{2dYM}
\end{eqnarray}
This is equivalent to type IIB superstring theory around the black 1-brane background spacetime. This black 1-brane solution can be mapped via S-duality to a black string solution \cite{Itzhaki:1998dd}. This theory is exactly equivalent to the so-called matrix string theory \cite{Dijkgraaf:1997vv} which can be thought of as a matrix gauge theory in the same way that M-(atrix) theory or the BFSS model is a matrix quantum mechanics. 

In the same way that the Euclidean IKKT matrix model decompactified on a circle ${S}^1$ gives the BFSS model at finite temperature, M-(atrix) theory decompactified on a circle ${S}^1$ should give the above matrix string theory. See for example \cite{Kawahara:2005an}. Indeed, compactification should always take us from a higher dimensional theory to a lower dimensional one since it is the analogue of dimensional reduction. Thus, we go from the $2-$dimensional matrix gauge theory (the DVV matrix string theory) to the $1-$dimensional matrix quantum mechanics (the BFSS M-(atrix) theory) and from the $1-$dimensional matrix quantum mechanics to the $0-$dimensional type IIB matrix model (the IKKT matrix model) via circle compactifications (assuming also Euclidean signature so there is no difference between time-like and space-like circles).

As we have discussed previously, the full dynamics of M-theory is captured by the large $N$ limit of the M-(atrix) theory or the BFSS quantum mechanics, whereas compactification of M-theory on a circle gives type IIA string theory. The fundamental degrees of freedom of M-(atrix) theory are D-particles. The D0-branes (D-particles) of type IIA string theory are described by collective coordinates given by $N\times N$ matrices and the limit $N\longrightarrow \infty$ is precisely the decompactification limit (recall that $\lambda=g^2_{YM}N=g_sN/(2\pi)^2l_s^3$). 

In this compactification, the strong limit of the string coupling constant $g_s$ determines the large radius of the circle by $R=g_s\sqrt{\alpha^{\prime}}$ while the type IIA D-particles are identified with non-zero Kaluza-Klein modes along the circle. 

Similarly, the circle decompactification of the BFSS quantum mechanics is achieved by reinterpreting the large $N\times N$ matrices $X_I$ as covariant derivatives of a $U(N)$ gauge field defined on a circle  ${S}^1$. Thus, a new compact coordinate along the circle emerges in M-(atrix) theory which then becomes matrix string theory. 

In summary, 
\begin{itemize}
\item 1) M-theory compactified on a circle gives type IIA string theory.
\item 2) M-theory is given by M-(atrix) theory.
\item 3) Matrix string theory compactified on a circle gives M-(atrix) theory. 
\end{itemize}
Hence, we conclude that dynamics of type IIA string theory must be captured by matrix string theory in the strong coupling region. See \cite{Dijkgraaf:1997vv} and also \cite{Banks:1996my,Motl:1997th,Sethi:1997sw}.
%But type IIA in this limit describes D-particles whereas matrix string theory describes D-strings. The connection is via T-duality.

However, matrix string theory describes D-strings, forming a black string, in type IIB string theory whereas M-(atrix) theory describes D-particles, forming a black hole, in type IIA string theory. These are related to each other via T-duality which maps type IIA string theory compactified on a circle of radius $R$ to type IIB string theory compactified on a circle of radius $\alpha^{\prime}/R$, and also simultaneously exchanging momentum modes with winding numbers.

\begin{figure}[H]
\begin{center}
\includegraphics[width=17.0cm,angle=0]{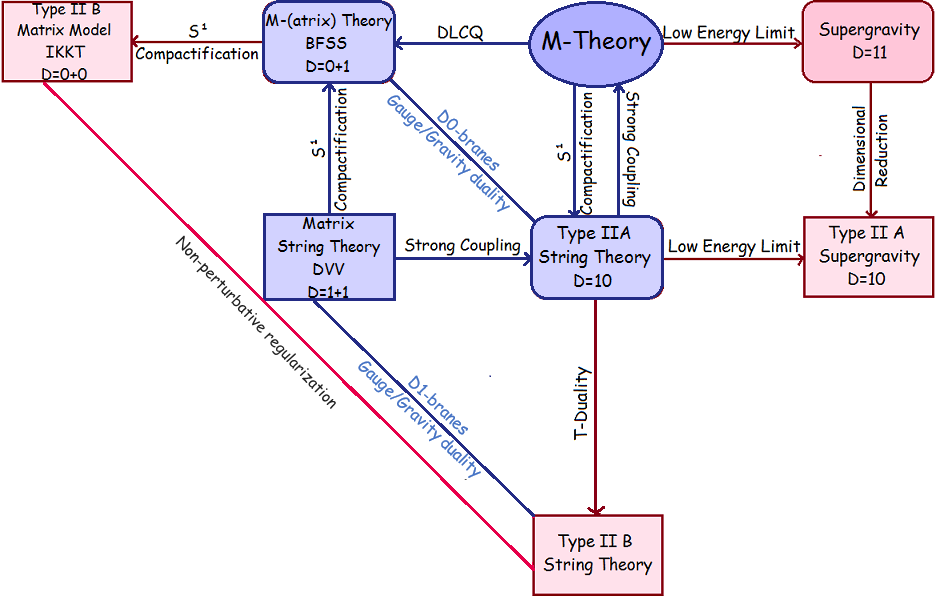}
\end{center}
\caption{M-theory et al.}\label{fd0}
\end{figure}

The mother of all theories M-theory and its relations to other theories is depicted in figure (\ref{fd0}).

\newpage
\section{Black-hole/black-string transition as the confinement/deconfinement transition}
\subsection{The black-hole/black-string phase transition}

In this section we follow the presentation of \cite{Hanada:2016jok,Kawahara:2007fn}.

We consider the action of $2-$dimensional maximally symmetric, i.e. ${\cal N}=8$, $U(N)$ Yang-Mills theory given by the action (\ref{2dYM}). We Wick rotate to Euclidean signature and denote the time by $\tilde{t}$. The theory is put at finite temperature via compactification of the Euclidean time on a circle of circumference given by the inverse temperature $\tilde{\beta}=1/\tilde{T}$. The spatial circle is also compactified on a circle of circumference $\tilde{L}$.

For very large compactification circumference $\tilde{L}$ this theory describes $N$ coincident D1-branes in type IIB string theory which are winding on the circle. When $\tilde{L}\longrightarrow 0$ the appropriate description becomes given by T-duality in terms of $N$ coincident D0-branes in type IIA string theory which are winding on a circle  of circumference $\alpha^{\prime}/\tilde{L}$. The positions of these D-particles on the T-dual circle are given by the eigenvalues of the Wilson loop winding on the circle, i.e. of the holonomy matrix
\begin{eqnarray}
W={\cal P}\exp(i\oint dx A_x).\label{Wl}
\end{eqnarray}
By an appropriate gauge transformation the Wilson line can be diagonalized as
\begin{eqnarray}
W={\rm diag}(\exp(i\theta_1),...,\exp(i\theta_N)).
\end{eqnarray}
The phase $\theta_i$ is precisely the position of the $i$th D0-brane on the T-dual circle. If all the angles $\theta_i$ accumulate at the same point then we obtain a black hole at that location whereas if they are distributed uniformly on the circle we obtain a uniform black string. We can also obtain a non-uniform black string phase or a phase with several black holes depending on the distribution of the eigenvalues $\theta_i$.  

In the high temperature limit a very nice reduction of this model occurs. In this limit $\tilde{\beta}\longrightarrow 0$ and as a consequence the temporal Kaluza-Klein mode, which have temporal momenta of the form $p=n/\tilde{\beta}$ for $n\in {\bf Z}$, become very heavy and thus decouple from the theory. Effectively the time direction is reduced to a point in the high temperature limit $\tilde{\beta}\longrightarrow 0$ and the theory reduces back to M-(atrix) theory. If we also assume that the fermions obey the anti-periodic boundary condition in the time direction, viz
\begin{eqnarray}
\psi(\tilde{t}+\tilde{\beta})=-\psi(\tilde{t}).
\end{eqnarray}
Then, in the high temperature limit $\tilde{\beta}\longrightarrow 0$ the fermions decouple and we end up with a bosonic theory, i.e. the bosonic part of the BFSS quantum mechanics given explicitly by 
 \begin{eqnarray}
S=\frac{1}{g_{\rm YM}^2}\int_0^{\beta} dt \big[\frac{1}{2}(D_tX_I)^2-\frac{1}{4}[X_I,X_J]^2\big].
\end{eqnarray}
However, note that the time direction of this $1-$dimensional model is the spatial direction of the $2-$dimensional model. More precisely, we have
\begin{eqnarray}
\frac{1}{g_{\rm YM}^2}=\frac{\tilde{\beta}}{\tilde{g}_{\rm YM}^2}\Rightarrow \lambda=\tilde{\lambda}\tilde{T}.
\end{eqnarray}
\begin{eqnarray}
\tilde{L}=\beta.
\end{eqnarray}
It is not difficult to see that all physical properties of the $1-$dimensional system depend only on the effective coupling constant 
\begin{eqnarray}
\lambda_{\rm eff}=\frac{\lambda}{T^3}=\tilde{\lambda}\tilde{T}\tilde{L}^3.
\end{eqnarray}
The Wilson loop (\ref{Wl}) winding around the spatial circle in the $2-$dimensional theory becomes in the $1-$dimensional theory the Polyakov line or holonomy
\begin{eqnarray}
P=\frac{1}{N}{\rm Tr}U~,~U={\cal P}\exp(i\int_0^{\beta}dt A(t)),\label{PL}
\end{eqnarray}
where $U$ is the holonomy matrix.

The high temperature $2-$dimensional Yang-Mills theory on a circle was studied in \cite{Aharony:2004ig,Harmark:2004ws} where a phase transition around $\lambda_{\rm eff}=1.4$ was observed. This result was made more precise by studying the $1-$dimensional matrix quantum mechanics in \cite{Kawahara:2007fn} where two phase transitions were identified of second and third order respectively at the values (see next section for detailed discussion) 

\begin{eqnarray}
\lambda_{\rm eff}=1.35(1)\Rightarrow \tilde{T}\tilde{L}=\frac{1.35(1)}{\tilde{\lambda}\tilde{L}^2}.
\end{eqnarray}
\begin{eqnarray}
\lambda_{\rm eff}=1.487(2)\Rightarrow \tilde{T}\tilde{L}=\frac{1.487(2)}{\tilde{\lambda}\tilde{L}^2}.
\end{eqnarray}
The second order transition separates between the gapped phase and the non-uniform phase whereas the third order separates between the non-uniform phase and the uniform phase. These phases, in the $2-$dimensional phase diagram with axes  given by the dimensionless parameters $\tilde{T}\tilde{L}$ and $\tilde{\lambda}\tilde{L}^2$, occur at high temperatures in the region where the $2-$dimensional Yang-Mills theory reduces to the bosonic part of the $1-$dimensional BFSS quantum mechanics. It is  conjectured in \cite{Kawahara:2007fn} that by continuing the above two lines to low temperatures we will reach a triple point where the two lines intersect and as a consequence the non-uniform phase ceases to exist below this  tri-critical point. 

\begin{figure}[H]
\begin{center}
\includegraphics[width=16.0cm,angle=0]{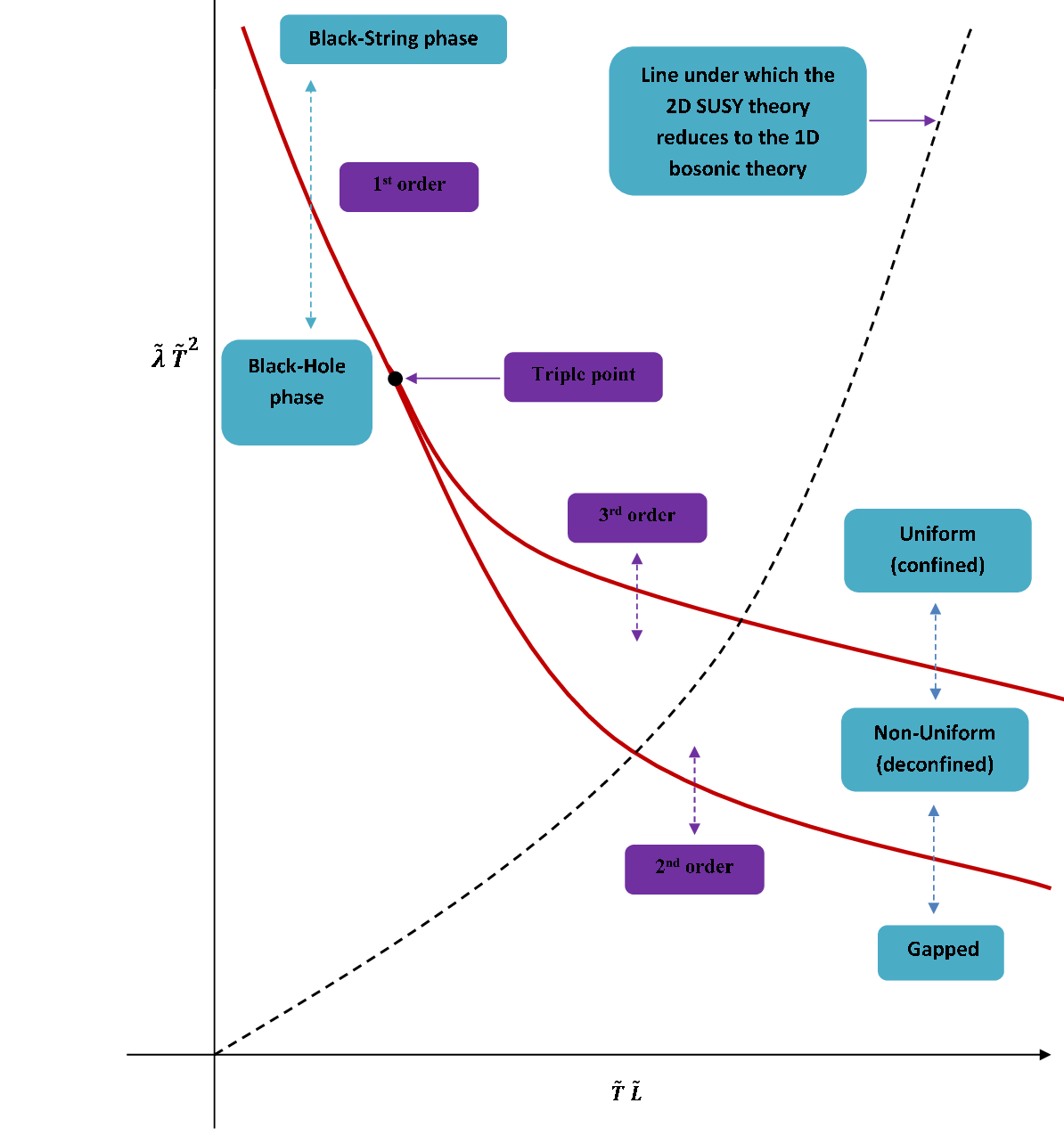}
\end{center}
\caption{The phase diagram of the $2-$d Yang-Mills theory on a circle and its relation to $1-$d Yang-Mills theory on a circle.}\label{fd}
\end{figure}

A phase diagram may look like the one on figure (\ref{fd}).

At small $\tilde{T}$ and large $\tilde{\lambda}$ it was shown in  \cite{Aharony:2004ig} that the  $2-$dimensional Yang-Mills theory exhibits a first order phase transition at the value 
\begin{eqnarray}
\tilde{T}\tilde{L}=\frac{2.29}{\sqrt{\tilde{\lambda}\tilde{L}^2}}.
\end{eqnarray}
Remark the extra square root. This corresponds in the dual gravity theory side to a transition between the black hole phase (gapped phase) and the black string phase (the uniform phase) \cite{Susskind:1997dr}. This black hole/black string first order phase transition is intimately related to Gregory-Laflamme instability \cite{Gregory:1993vy}. Thus, it seems that the first order black hole/black string transition seen at low temperatures splits at the triple point into the second order gapped/non-uniform and the third order non-uniform/uniform transitions seen at high temperatures, i.e. in the bosonic part of the $1-$dimensional BFSS quantum mechanics \cite{Kawahara:2007fn}.  See the above proposed phase diagram. The black hole/black string transition is a very important example of topology change transitions.
\subsection{The confinement/deconfinement phase transition}
In this section we follow the presentation of \cite{Kawahara:2007fn}.

The goal in this section is to describe in some detail the phase structure of the  $1-$dimensional BFSS quantum mechanics given by the model
\begin{eqnarray}
S=\frac{1}{g_{\rm YM}^2}\int_0^{\beta} dt \big[\frac{1}{2}(D_tX_I)^2-\frac{1}{4}[X_I,X_J]^2\big].
\end{eqnarray}
There is one single effective coupling constant $\lambda_{\rm eff}={\lambda}/{T^3}$ and thus without any loss of generality we can choose the t'Hooft coupling as $\lambda=g_{\rm YM}^2N=1$. A Monte Carlo study of the above model using the heat bath algorithm was performed in \cite{Kawahara:2007fn} and using the hybrid Monte Carlo algorithm was performed in \cite{OConnor:2016gbq}. The basic observables which we track in Monte Carlo simulation are the Polyakov line, the energy of the system and the extent of space given respectively by 
\begin{eqnarray}
\langle |P|\rangle=\langle \frac{1}{N}|{\rm Tr}U|\rangle~,~U={\cal P}\exp(i\int_0{\beta}dt A(t)).
\end{eqnarray}
\begin{eqnarray}
\frac{E}{N^2}=-\langle \frac{3T}{4N}\int_0^{\beta}dt {\rm Tr}[X_I,X_J]^2\rangle.
\end{eqnarray}
\begin{eqnarray}
R^2=\langle \frac{T}{N}\int_0^{\beta}dt {\rm Tr}X_I^2\rangle.
\end{eqnarray}
The order parameter is the Polyakov line. Some of the main results include:
\begin{itemize}
\item The Polyakov line is found to approach one in the deconfined (non-uniform) phase, then it starts changing quite fast at around $T\simeq 0.9$, then it goes to zero in the confined (uniform) phase. The data in the deconfined phase is well reproduced by the high temperature expansion \cite{Kawahara:2007ib} especially for $T\geq 2$. In the confined phase the Polyakov line goes to zero as $1/N$ as $T\longrightarrow 0$ which can be reproduced by generating the holonomy matrix $U$ with a probability given by the Haar measure $dU$.

\item  Thus $T=0.9$ marks the transition from the deconfined (non-uniform) to confined (uniform) phase transition. In the confining uniform phase the $U(1)$ symmetry 
\begin{eqnarray}
A(t)\longrightarrow A(t)+a.{\bf 1}
\end{eqnarray}
is not broken whereas in the deconfining non-uniform phase it is broken. Thus, the confining/deconfining phase transition at $T=0.9$ is associated with spontaneous symmetry breaking of the above $U(1)$ symmetry  \cite{Aharony:2004ig,Harmark:2004ws,Janik:2000tq,Bialas:2001fj,Kawahara:2007fn,Kawahara:2007nw,Azuma:2014cfa}. This transition is intimately related to the string theory Hagedorn transition \cite{Aharony:2003sx,Aharony:2005bq,Atick:1988si}.

\item The energy and the extent of space show a flat behavior in the confined (uniform) phase for $T< 0.9$. This can be interpreted following  \cite{Kawahara:2007fn} as due to the Eguchi-Kawai reduction \cite{Eguchi:1982nm} of $U(N)$ gauge theory on a lattice down to a $U(N)$ gauge theory on a point in the 't Hooft limit which is possible because in the planar approximation we find that Wilson loop amplitudes for disconnected diagrams enjoys factorization. Thus in  Eguchi-Kawai reduction only global invariance is left and expectation values of single trace operators are independent of the volume in the large $N$ limit if the central $U(1)$ symmetry is not broken \footnote{In $D$ dimensions the central symmetry is $U(1)^D$.}. See figure (\ref{fd2}).

\item The Polyakov line, the energy and the extent of space depend continuously on the temperature but their first derivatives is discontinuous at the critical temperature 
\begin{eqnarray}
T_{c1}=0.905(2).  
\end{eqnarray}
Thus, the transition from confining phase to deconfining phase is second order.

\begin{figure}[H]
\begin{center}
\includegraphics[width=9.0cm,angle=0]{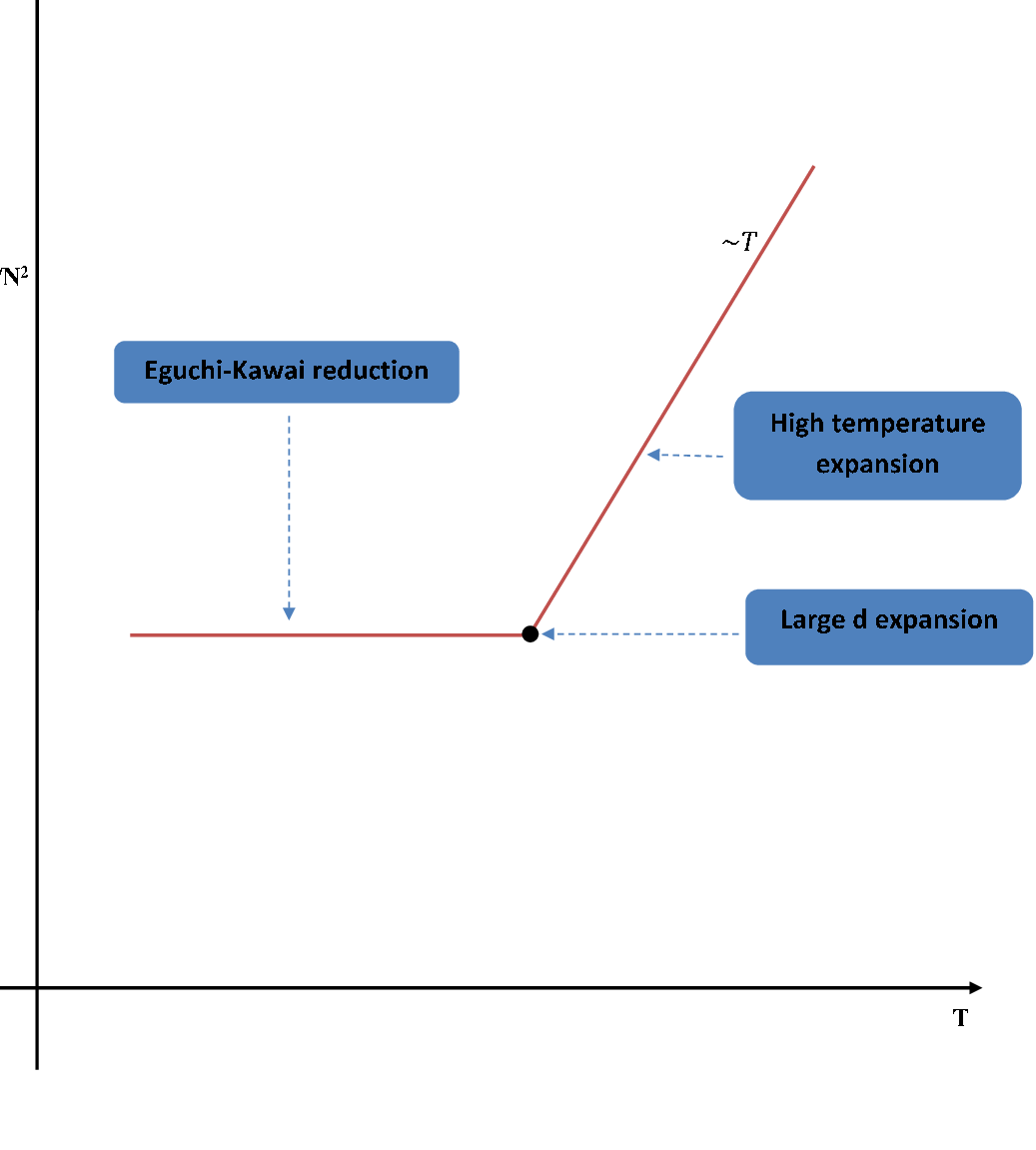}
\includegraphics[width=9.0cm,angle=0]{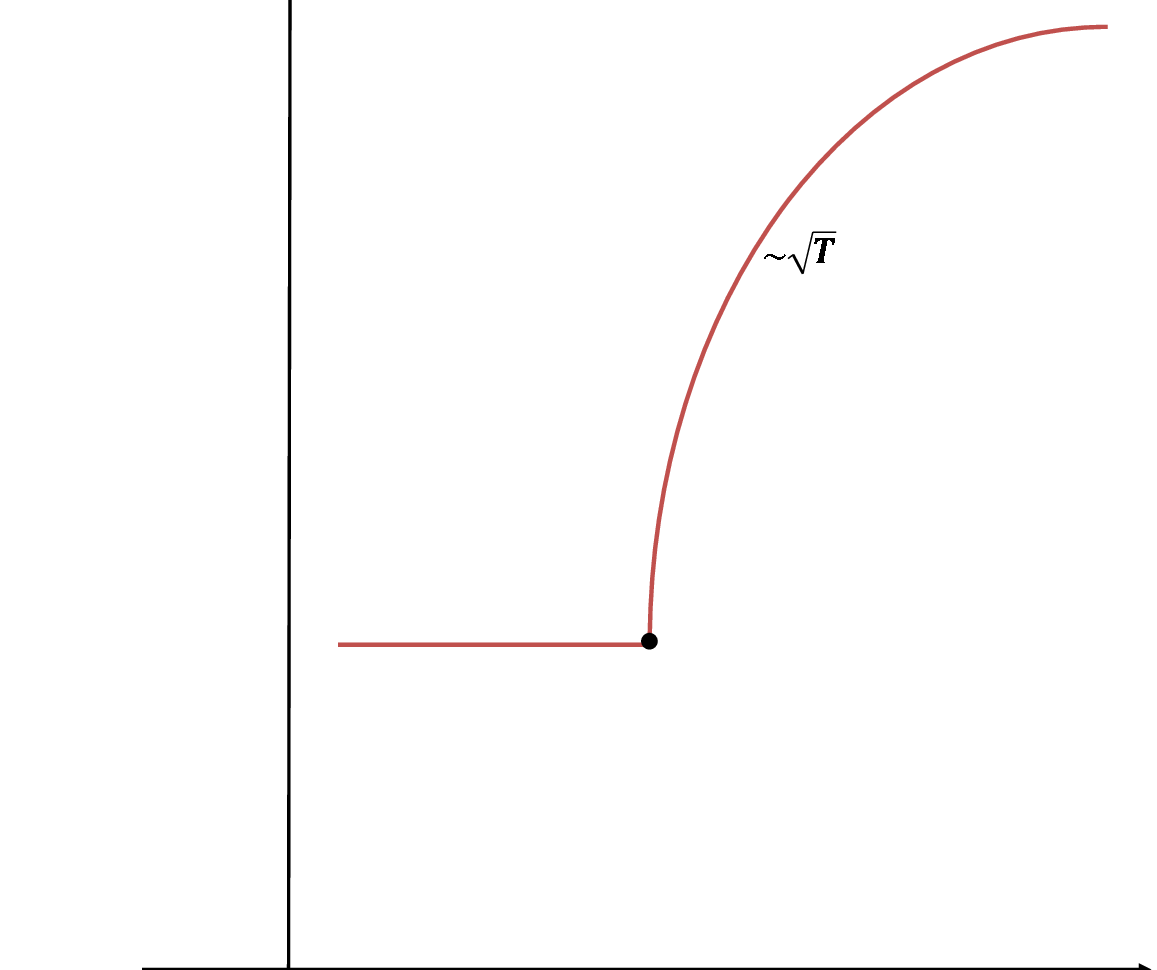}
\end{center}
\caption{The energy and the extent of space.}\label{fd2}
\end{figure}

\item A much more powerful order parameter is the eigenvalue distribution $\rho(\theta)$ of the eigenvalues $\exp(i\theta_i)$, $i=1,...,N$, of the holonomy matrix $U$. We must have   $\theta_i\in]-\pi,\pi]$ and $\sum_i\theta_i=0$.

The deconfinement phase where the central $U(1)$ symmetry is spontaneously broken is seen from the behavior of the eigenvalue distribution $\rho(\theta)$ to be divided into two distinct phases: the gapped phase and the non-uniform phase. As it turns out, these phases can be described by the Gross-Witten one-plaquette model \cite{Gross:1980he,Wadia:1980cp} 
\begin{eqnarray}
Z_{\rm GW}=\int dU\exp\bigg(\frac{N}{\kappa}({\rm Tr}U+{\rm Tr}U^{\dagger}\bigg).
\end{eqnarray}
The eigenvalue distributions in the large $N$ limit of this model are determined by the solutions  
\begin{eqnarray}
\rho_{\rm gapped}(\theta)=\frac{2}{\pi\kappa}\cos\frac{\theta}{2}\sqrt{\frac{\kappa}{2}-\sin^2\frac{\theta}{2}}~,~|\theta|\leq 2\sin^{-1}\sqrt{\kappa/2}~,~\kappa<2.
\end{eqnarray}
\begin{eqnarray}
\rho_{\rm non-uniform}(\theta)=\frac{1}{2\pi}(1+\frac{2}{\kappa}\cos{\theta})~,~|\theta|\leq \pi~,~\kappa\geq 2.
\end{eqnarray}
Thus, there exists in the Gross-Witten one-plaquette model a phase transition between the above two solutions occurring at $\kappa=2$ which is found to be of third order. This transition in the full model occurs at  $T_{1c}=0.905(2)$ between the gapped and the non-uniform phases and is of second order not third order yet the above distributions are still very good fits to the actual Monte Carlo data with some value of $\kappa$ for each $T$. The second order phase transition at $T_{1c}=0.905(2)$ is associated with the emergence of a gap in the spectrum.
\item In the confining phase the eigenvalue distribution is uniform. The $U(1)$ symmetry is unbroken and as a consequence the Eguchi-Kawai equivalence holds. Thus, the energy in this phase must be a constant proportional to $N^2$. The breaking of $U(1)$ symmetry and the resulting  Eguchi-Kawai equivalence, as we increase $T$, is of the order of $1/N^2$. The phase boundary at $T=T_{c2}$ between the confining uniform phase and the deconfining non-uniform phase can thus be determined by means of the Eguchi-Kawai equivalence instead of using the eigenvalue distribution. The behavior of the energy around $T_{c2}$ is found to be of the form \cite{Kawahara:2007fn}
\begin{eqnarray}
\frac{E}{N^2}=\epsilon_0~,~T\leq T_{c2}.
\end{eqnarray}
\begin{eqnarray}
\frac{E}{N^2}=\epsilon_0+c(T-T_{c2})^p~,~T> T_{c2}.
\end{eqnarray}
We find for $N=32$ \cite{Kawahara:2007fn}
\begin{eqnarray}
\epsilon_0=6.695(5)~,~c=413\pm 310~,~T_{c_2}=0.8758(9)~,~p=2.1(2).
\end{eqnarray}
Similarly, we get for the extent of space the values $T_{c_2}=0.8763(4)~,~p=1.9(2)$. The transition from the confining uniform to the deconfining non-uniform occurs then at the average value
\begin{eqnarray}
T_{c_2}=0.8761(3).
\end{eqnarray}
The power $p=2$ suggests that the second derivative of the energy is discontinuous and as a consequence the transition is third order.
\item Thus the transition from the confining uniform phase to the deconfining gapped phase goes through the deconfining non-uniform phase. See figure (\ref{fd1}). There is also the suggestion in \cite{Harmark:2004ws,Aharony:2004ig} that the transition is possibly a direct single first order phase transition which is the behavior observed in the plane-wave BMN matrix model \cite{Furuuchi:2003sy,Semenoff:2004bs,Kawahara:2006hs}.
\end{itemize}
\begin{figure}[H]
\begin{center}
\includegraphics[width=10.0cm,angle=0]{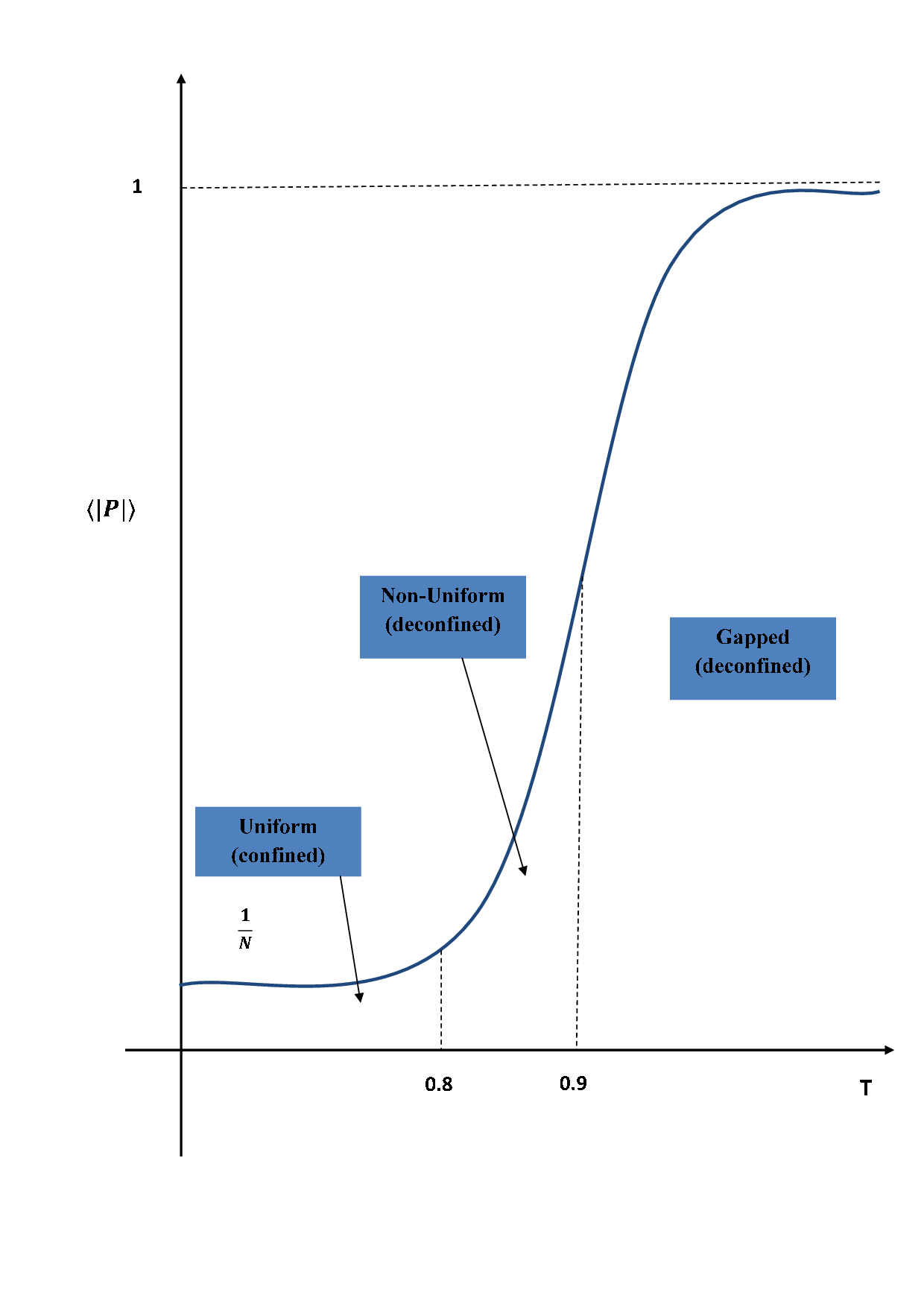}
\end{center}
\caption{The phase structure in terms of the Polyakov line.}\label{fd1}
\end{figure}
%\subsection{The mass gap, the $D=\infty$ limit and the Gaussian structure}
\subsection{The mass gap and the Gaussian structure}
More interesting results concerning the bosonic BFSS quantum mechanics can be found in \cite{Filev:2015hia,Filev:2015cmz,Asano:2016xsf,OConnor:2016gbq}. 

We only consider the theory at zero temperature $T=0$ or $\beta=\infty$. On a finite lattice $\beta=a\Lambda$ where $a$ is the lattice spacing and $\Lambda$ is the number of lattice points. The symmetric static gauge is given by the link variable $V=D^{1/\Lambda}$ where $D={\rm diag}(\exp(i\theta_1),...,\exp(i\theta_N))$. Thus, in the zero temperature limit $\Lambda\longrightarrow\infty$ and thus $V\longrightarrow {\bf 1}$. Hence in this limit the gauge field can be completely gauged away.

The mass gap can be captured at zero temperature by the large time behavior of the correlator 
\begin{eqnarray}
\langle {\rm Tr} X^1(0)X^1(t)\rangle\propto \exp(-m t)+...
\end{eqnarray}
The mass gap is given in terms of the energies of the first excited state $E_1$ and the ground state $E_0$ by
\begin{eqnarray}
m =E_1-E_0.
\end{eqnarray}
For finite temperature the above formula for the correlator is modified as 
\begin{eqnarray}
\langle {\rm Tr} X^1(0)X^1(t)\rangle\propto A(\exp(-m t)+\exp(-m (\beta-t)).\label{corre}
\end{eqnarray}
%This should be properly understood as the zero temperature correlator on the lattice where $\beta$ is identified as a the period of the time coordinate as opposed to inverse temperature \cite{Filev:2015hia}.
A measurement of this correlator for $N=30$, $a=0.25$ and $\beta=10$ yields precisely this behavior with the values   \cite{Filev:2015hia}  
\begin{eqnarray}
A=7.50\pm 0.2~,~m=(1.90\pm 0.01)\lambda^{1/3}.
\end{eqnarray}
The fundamental observation of  \cite{Filev:2015hia} is that this behavior of the correlator can be obtained from the Gaussian model of $d$ independent scalar fields with effective action
\begin{eqnarray}
S=N\int_{0}^{\beta} dt \big[\frac{1}{2}(D_tX_I)^2+\frac{m^2}{2}X_I^2\big].\label{ff}
\end{eqnarray}
At zero temperature where the gauge field can be set to zero this becomes 
\begin{eqnarray}
S=N\int_{0}^{\infty} dt \big[\frac{1}{2}(\partial_tX_I)^2+\frac{m^2}{2}X_I^2\big].
\end{eqnarray}
The eigenvalue distribution of any one of the $X_I$ is given by Wigner semicircle law of radius 
\begin{eqnarray}
R_{\lambda}=\sqrt{\frac{2}{m}}.\label{rad}
\end{eqnarray}
Also, the correlator $\langle {\rm Tr} X^1(0)X^1(t)\rangle$ in this theory is given exactly by the formula (\ref{corre}) with $A$ given by 
\begin{eqnarray}
A=\frac{N}{2m(1-\exp(-\beta m))}.
\end{eqnarray}
Monte Carlo measurement of the eigenvalue distribution of $X_1$ is indeed found to be given by a Wigner semicircle law with radius given, for the above parameters, by $R_{\lambda}=1.01$ \cite{Filev:2015hia} which is consistent with the above measured value of $m$ from the correlator.

The results of \cite{Filev:2015hia} suggest that for all values of the temperatures the eigenvalue distribution of any one of the $X_I$ is given by a semicircle law with a radius given in terms of the expectation value of the extent of space $ R^2$ by
\begin{eqnarray}
R_{\lambda}=\frac{4}{d}\langle R^2\rangle.
\end{eqnarray}
Thus the phase transition from the uniform confining phase at $T=T_{c2}=0.8..$ to the gapped deconfining phase at $T=T_{c1}=0.9..$, i.e. the emergence of a gap, is associated with a change in the radius of the eigenvalue distribution but not the shape which remains always given by a Wigner semicircle law. The effective action (\ref{ff}) works very well at low temperatures which can be motivated by a large $d$ expansion \cite{Hotta:1998en} of the bosonic BFSS model in which the quartic commutator term is replaced with a quadratic mass term with a specific value of the mass which depends on the dimension $d$, i.e. on the number of scalar fields.

\subsection{The large $d$ approximation}
To see this in more detail we go back to our original action 
\begin{eqnarray}
S=N\int_{0}^{\beta} dt \big[\frac{1}{2}(D_tX_I)^2-\frac{1}{4}[X_I,X_J]^2\big].
\end{eqnarray}
We introduce $SU(N)$ generators $t^a$ satisfying ${\rm Tr}t^at^b=\delta^{ab}$ and we expand the matrices $X_I$ as $X_I=t^aX_I^a$. The above action can be rewritten as 
 \begin{eqnarray}
S=N\int_{0}^{\beta} dt \frac{1}{2}(D_tX_I)^2-\frac{N}{4}\lambda^{abcd}\int_{0}^{\beta} dt X_I^aX_I^bX_J^cI_J^d,
\end{eqnarray}
where $\lambda^{abcd}$ is some $SU(N)$ tensor (see for example equation $(3.28)$ of \cite{Filev:2015hia}). We can add to the action without changing the dynamics any term $\Delta S$ depending on $X_I$ and other fields $k^{ab}$ such that $\int {\cal D}k^{ab}\exp(-\Delta S)={\rm constant}$. We add
\begin{eqnarray}
\Delta S=\frac{N}{4}\mu_{abcd}\int_{0}^{\beta} dt \big(k^{ab}+\lambda^{abef}X_I^eX_I^f\big)\big(k^{cd}+\lambda^{cdgh}X_J^gX_J^h\big),
\end{eqnarray}
where $\mu_{abcd}$ is the inverse kernel of $\lambda^{abcd}$. We consider then the action $S^{\prime}=S+\Delta S$ given by 
\begin{eqnarray}
S^{\prime}=N\int_{0}^{\beta} dt (\frac{1}{2}(D_tX_I)^2+\frac{k^{ab}}{2}X_I^aX_I^b)+\frac{N}{4}\mu_{abcd}\int_{0}^{\beta} dt k^{ab}k^{cd}.\label{333}
\end{eqnarray}
We will assume that the fields $k$'s are time independent, perform a Fourier transformation in the time direction with modes $n$, integrate out the fields $X$'s, and also choose for $k$ the ansatz 
\begin{eqnarray}
k_{ij}P_{ij,lm}=P_{ij,lm}k_{lm}=t_{ij}^ak^{ab}t_{lm}^b.
\end{eqnarray}
The $P$ is the projector on traceless matrices given by 
\begin{eqnarray}
P_{ij,lm}=t_{ij}^at_{lm}^a=\delta_{im}\delta_{jl}-\frac{1}{N}\delta_{ij}\delta_{lm}.
\end{eqnarray}
The above ansatz means in particular that $k_{ii}=k_{jj}$ for all $i$ and $j$ and we will also choose $k_{ij}$ to be symmetric. We obtain then the effective action for the fields $k$ given by
\begin{eqnarray}
S_{\rm eff}=\frac{d}{2}\sum_n \sum_{ij} P_{ij,ji}\log W(n)_{ij}+\frac{N\beta}{4}\mu_{abcd}k^{ab}k^{cd}.
\end{eqnarray}
The matrix $W(n)$ is given in terms of the holonomy angles $\theta_i$ by 
\begin{eqnarray}
W(n)_{ij}=\bigg(\frac{2\pi n+\theta_i-\theta_j}{\beta}\bigg)^2+k_{ij}.
\end{eqnarray}
Next we write down the saddle point equation $\partial S_{\rm eff}/\partial k_{ij}=0$ given explicitly by
\begin{eqnarray}
\frac{d}{2}\sum_n\frac{P_{ij,ji}}{\big(\frac{2\pi n+\theta_i-\theta_j}{\beta}\big)^2+k_{ij}}+\frac{\beta N}{2}\mu_{abcd}k^{ab}t_{ij}^ct_{ji}^d.
\end{eqnarray}
It is found that at low temperatures the effect of the holonomy is exponentially suppressed and thus the angles $\theta_i$ can be simply set equal to zero. The leading order is given by the ansatz $k_{ij}=m^2$ or equivalently $k^{ab}=m^2\delta^{ab}$ with
\begin{eqnarray}
m=d^{1/3}.
\end{eqnarray}
By substituting this solution in (\ref{333}) we get the Gaussian action (\ref{ff}). The radius (\ref{rad}) is then given by
\begin{eqnarray}
R_{\lambda}=\big(\frac{8}{d}\big)^{1/6}.
\end{eqnarray}
Next order corrections in $1/d$ were computed in \cite{Mandal:2011hb} and they are given by
\begin{eqnarray}
R_{\lambda}=\big(\frac{8}{d}\big)^{1/6}\bigg(1+\frac{1}{d}(\frac{7\sqrt{5}}{30}-\frac{9}{32})+...\bigg).
\end{eqnarray}
The Gaussian model (\ref{ff}) enjoys also a phase transition which occurs at the temperature \cite{Mandal:2009vz}
\begin{eqnarray}
T_c^{\rm gaussian}=\frac{m}{\ln d}=0.9467.
\end{eqnarray}
As we can see immediately, this value is in excellent agreement with the critical temperature $T_{c1}$ found in the full model of the phase transition to the gapped phase.

\subsection{High temperature limit}
At high temperatures the bosonic part of the BFSS quantum mechanics reduces to the bosonic part of the IKKT model \cite{Kawahara:2007ib}. The leading behavior of the various observables of interest at high temperatures can be obtained in terms of the corresponding expectation values in the IKKT model by the relations (with $D=d+1$ and $X$ are appropriately scaled)
\begin{eqnarray}
\langle R^2\rangle=\sqrt{T}\langle\frac{1}{N}{\rm Tr}X_I^2\rangle_{\rm IKKT}=\sqrt{T}\chi_1.
\end{eqnarray}
\begin{eqnarray}
\langle P\rangle=1-\frac{1}{2}{T}^{-3/2}\langle\frac{1}{N}{\rm Tr}X_D^2\rangle_{\rm IKKT}=1-\frac{1}{2d}{T}^{-3/2}\chi_1.
\end{eqnarray}
\begin{eqnarray}
\langle \frac{E}{N^2}\rangle=-\frac{3}{4}{T}\langle\frac{1}{N}{\rm Tr}[X_I,X_j]^2\rangle_{\rm IKKT}=\frac{3}{4}{T}\chi_2~,~\chi_2=(d-1)(1-\frac{1}{N^2}).
\end{eqnarray}
The coefficient $\chi_1$ for various $d$ and $N$ can be read off from table $1$ of \cite{Kawahara:2007ib} whereas the coefficient $\chi_2$ was determined exactly from the Schwinger-Dyson equation. The next-leading order corrections can also be computed from the reduced IKKT model along these lines  \cite{Kawahara:2007ib}.
\section{The discrete light-cone quantization (DLCQ) and infinite momentum frame (IMF)}

\subsection{Light-cone quantization and discrete light-cone quantization}

In this section we follow mostly \cite{Bigatti:1997jy} but also \cite{Banks:1996vh,Susskind:1997cw,Seiberg:1997ad,Taylor:2001vb,Sen:1997we}. We consider
$D=11-$dimensional spacetime with coordinates
$X^{\mu}=(t,z,X^1,...,X^{D-2})$ and metric $(1,-1,...,-1)$.  The coordinate $z$ is called
the longitudinal direction while  $X^1,...,X^{D-2}$ are the transverse
directions. The line-like coordiante $X^{\pm}$ are defined by
\begin{eqnarray}
X^{\pm}=\frac{X^0\pm X^{D-1}}{\sqrt{2}}=\frac{t\pm z}{\sqrt{2}}.
\end{eqnarray} 
We introduce the conjugate momenta
\begin{eqnarray}
P_{\pm}=\frac{P^0\mp P^{D-1}}{\sqrt{2}}=\frac{P_t\mp P_z}{\sqrt{2}}.
\end{eqnarray} 
The on-shell condition is
\begin{eqnarray}
P_{\mu}P^{\mu}=M^2\Leftrightarrow 2P_+P_--P_i^2=M^2.
\end{eqnarray}
In light-cone frame the coordinate $X^+$ plays the role of time and
thus $P_+$ is the Hamiltonian $H$. We have 
\begin{eqnarray}
H=\frac{P_i^2}{2P_-}+\frac{M^2}{2P_-}.
\end{eqnarray}
This is a non-relativistic expression where
$P_-$ plays the role of the mass. This can be made precise as
follows. 

The Poincare group is generated by $P_i$: the transverse
translations, $P_+=H$: the translation in the $X^+$ direction,
$P_-=\mu$: the translation in the $X^-$ direction, $L_{ij}=M_{ij}$: the
transverse rotations, $L_{iz}=M_{iz}$: rotations in the $(X^i,z)$ planes,
$K_{0z}=M_{0z}$: Lorentz boost along $z$ and $K_{0i}=M_{0i}$: Lorentz boosts along
$X^i$. The Poincaré algebra is 
\begin{eqnarray}
&&[P_\mu, P_\nu] = 0\nonumber\\
&&[M_{\mu\nu}, P_\rho] = \eta_{\mu\rho} P_\nu - \eta_{\nu\rho} P_\mu\nonumber\\
&&[M_{\mu\nu}, M_{\rho\sigma}] = \eta_{\mu\rho} M_{\nu\sigma} -
  \eta_{\mu\sigma} M_{\nu\rho} - \eta_{\nu\rho} M_{\mu\sigma} +
  \eta_{\nu\sigma} M_{\mu\rho}.
\end{eqnarray}
The Galilean group is essentially a subgroup of the Poincare group defined
as follows. We introduce
\begin{eqnarray}
\mu X_i^{C.M}=i\frac{K_{0i}-L_{zi}}{\sqrt{2}}.
\end{eqnarray}
We can immediately compute
\begin{eqnarray}
[P_i,\mu X_j^{C.M}]=i\mu {\eta}_{ij}.
\end{eqnarray}
The generators $\mu X_i^{C.M}$ are the Galilean boosts. Indeed we can
compute
\begin{eqnarray}
e^{-iV(\mu X^i)}P_ie^{iV(\mu X^i)}=P_i+\mu V.
\end{eqnarray}
The generator $P_-=\mu$ commutes with all the generators $P_i$, $P_+$,
$L_{ij}$ and $\mu X_i^{C.M}$. The mass is  therefore a central charge. The
Galilean group is isomorphic to the Poincare subroup generated by  $P_i$, $P_+$,
$L_{ij}$, $\mu X_i^{C.M}$ and $P_-$. Physics in the light-cone frame
is  Galilean invariant and hence the Hamiltonian must be of
the form
\begin{eqnarray}
H=\frac{P_i^2}{2P_-}+H_{\rm internal}.
\end{eqnarray}
The internal energy $H_{\rm internal}$ is Galilean invariant. The
generator $K_{0z}$ is not a part of the Galilean group. We have the
commutation relations
\begin{eqnarray}
[K_{0z},P_{\pm}]=\pm P_{\pm}.
\end{eqnarray}
Thus $[K_{0z},P_+P_-]=0$, i.e. $HP_-$ and as a consequence $H_{\rm internal}P_-={M^2}/{2}$
  are invaraint under longitudinal Lorentz boosts. We can also show
that $[\mu X_i^{C.M},H_{\rm internal}P_-]=0$. Hence $M^2$ is invariant
under Galilean boosts and Lorentz transformations. Remark also that  $H$ scales as $1/P_-$.  

Let us now consider the action
\begin{eqnarray}
S&=&\int dX^D
\bigg(\frac{1}{2}{\partial}_{\mu}\phi{\partial}^{\mu}\phi
-\frac{1}{2}m^2{\phi}^2-\lambda {\phi}^3\bigg)\nonumber\\
&=&\int dX^+ {\cal L}~,~{\cal L}=\int dX^- dX^i \bigg({\partial}_+\phi {\partial}_-\phi
-\frac{1}{2}({\partial}_i\phi)^2-\frac{m^2}{2}{\phi}^2-\lambda {\phi}^3\bigg).
\end{eqnarray}
The canonical momentum $\pi$ to the field $\phi$ is 
\begin{eqnarray}
\pi =\frac{{\partial}{\cal L}}{\partial ({\partial}_+\phi)}={\partial}_-\phi.
\end{eqnarray}
The equal time commutation relations
\begin{eqnarray}
[\phi(X^-,X^i),{\partial}_-\phi(Y^-,Y^i)]=i\delta (X^--Y^-) \delta (X^i-Y^i).
\end{eqnarray}
The field can be expanded as follows
\begin{eqnarray}
\phi(X^-,X^i)=\int_0^{\infty}dk_-\bigg[\frac{\phi(k_-,X^i)}{\sqrt{2\pi
      k_-}}e^{-ik_-X^-}+\frac{{\phi}^*(k_-,X^i)}{\sqrt{2\pi
      k_-}}e^{ik_-X^-}
\bigg].\label{exps}
\end{eqnarray}
Thus
\begin{eqnarray}
{\partial}_-\phi(X^-,X^i)=-i\int_0^{\infty}dk_-\bigg[\frac{\phi(k_-,X^i)}{\sqrt{2\pi
      }}\sqrt{k_-}e^{-ik_-X^-}-\frac{{\phi}^*(k_-,X^i)}{\sqrt{2\pi
      }}\sqrt{k_-}e^{ik_-X^-}
\bigg].
\end{eqnarray}
We can then compute the non relativistic commutation relations
\begin{eqnarray}
&&[{\phi}(k_-,X^i),\phi(l_-,Y^i)]=[{\phi}^*(k_-,X^i),{\phi}^*(l_-,Y^i)]=0\nonumber\\
&&[{\phi}(k_-,X^i),{\phi}^*(l_-,Y^i)]=\delta
(k_--l_-)\delta(X^i-Y^i).\label{coms}
\end{eqnarray}
The Hamiltonian in the light cone frame takes the non-relativistic
form
\begin{eqnarray}
H=H_0+H_I.
\end{eqnarray}
\begin{eqnarray}
H_0=\int_0^{\infty}
dk_- \int dX^i\bigg[\frac{{\partial}_i{\phi}^*(k_-,X^i){\partial}_i\phi(k_-,X^i)+m^2{\phi}^*(k_-,X^i)\phi(k_-,X^i)}{2k_-}+{\rm
    c.c}\bigg].
\end{eqnarray}
\begin{eqnarray}
H_I=\frac{3\lambda}{\sqrt{2\pi}}\int_0^{\infty} dk_-\int_0^{\infty}dl_-
\int dX^i\bigg[\frac{{\phi}^*(k_-,X^i)}{k_-}\frac{{\phi}^*(l_-,X^i)}{l_-}\frac{{\phi}(k_-+l_-,X^i)}{k_-+l_-}+{\rm c.c}\bigg]
\end{eqnarray}
We observe that $k_-$ can only take positive values and that it
is a conserved number at the interaction vertex. This is similar to
the positivity and the conservation of the mass in non-relativistic
quantum mechanics.

In discrete light-cone quantization we compactify the light-like
coordinate $X^-$ on a circle of radius $R$. The spectrum of the
 corresponding momentum $P_-$ becomes discrete given by
\begin{eqnarray}
P_-=\frac{N}{R}.
\end{eqnarray}
We have seen that $P_-$ plays the role of an invariant mass and that
 in the field theory it is a conserved quantum number. Hence in the  discrete light cone quantization we get an infinite number of
 superselection sectors defined by the positive integer $N$.

The expansion (\ref{exps}) becomes (with
$\phi(k_-,X^i)\equiv \sqrt{R}{\phi}_N(X^i)$)
\begin{eqnarray}
\phi(X^i,X^i)={\phi}_0(X^i)+\sum_{N=1}^{\infty}\frac{{\phi}_N(X^i)}{\sqrt{2\pi
    N}}e^{-i\frac{N}{R}X^-}+{\rm c.c}.
\end{eqnarray}
The mode ${\phi}_0(X^i)$ corresponds to $P_-=0$. The commutation relations (\ref{coms}) become 
\begin{eqnarray}
&&[{\phi}_N(X^i),{\phi}_M(Y^i)]=[{\phi}^*_N(X^i),{\phi}^*_M(Y^i)]=0\nonumber\\
&&[{\phi}_N(X^i),{\phi}^*_M(Y^i)]={\delta}_{NM}\delta(X^i-Y^i).
\end{eqnarray}
The Hamiltonians $H_0$ and $H_I$ become
\begin{eqnarray}
H_0=R\sum_{N=1}^{\infty}\int dX^i\bigg[\frac{{\partial}_i{\phi}^*_N(X^i){\partial}_i{\phi}_N(X^i)+m^2{\phi}^*_N(X^i){\phi}_N(X^i)}{2N}+{\rm
    c.c}\bigg].
\end{eqnarray}
\begin{eqnarray}
H_I=\frac{3\lambda \sqrt{R}}{\sqrt{2\pi}}R^2\sum_{N=1}^{\infty}\sum_{M=1}^{\infty}\int dX^i\bigg[\frac{{\phi}^*_N(X^i)}{N}\frac{{\phi}^*_M(X^i)}{M}\frac{{\phi}_{N+M}(X^i)}{N+M}+{\rm c.c}\bigg].
\end{eqnarray}
At the level of the action the zero mode
${\phi}_0$ can be shown to be non-dynamical 
because ${\partial}_-{\phi}_0=0$. Thus it can be integrated out yielding new
complicated terms in the Hamiltonian which will (by construction) conserve $P_-$ and the
Galilean symmetry.

The limit of physical interest is defined by $N\longrightarrow \infty$
and $R\longrightarrow \infty$ keeping the momentum $P_-=N/R$ fixed.

Note also that $H_0$ is the Hamiltonian of free particles
in non-relativistic quantum mechanics where ${\phi}_N(X^i)$ is the
second quantized Schrodinger field for the $N$th type of particle with
mass ${\mu}_N\equiv N/R=P_-$. 
 
\subsection{Infinite momentum frame and BFSS conjecture}
The original BFSS conjecture \cite{Banks:1996vh} relates  $M-$theory in the infinite
momentum frame (IMF) (and not  $M-$theory in the light cone
frame) to the theory of N D0-branes (\ref{S}) or (\ref{H}). This goes as follows.

$1)$ In the IMF formulation we boost the system along the
longitudinal direction $z$ until longitudinal momenta are much larger
than any other scale in the problem. The energy
$E=\sqrt{P_z^2+P^2+m^2}$ where $\vec{P}$ is the transverse spatial
momentum  becomes in the limit $P_z\longrightarrow \infty$ given by
\begin{eqnarray}
E=P_z+\frac{\vec{P}^2+m^2}{2P_z}.
\end{eqnarray}

$2)$ In the IMF formulation we compactify the space-like direction $z$ on a
 circle of radius $R_s$ and
 hence the momentum $P_z$ becomes quantized as $P_z=N/R_s$ in contrast 
 with the light-cone formulation where a light-like direction $X^-$  is
 compactified. However, as in the light-cone formulation all objects
  in the IMF formulation with vanishing and negative $P_z$  decouple. The limit $P_z=NR_s\longrightarrow \infty$ is equivalent to $N\longrightarrow \infty$ and/or $R_s\longrightarrow \infty$.

$3)$ M-theory with a compactified direction is by definition type IIA
string theory in the same way that $11-$dimensional supergravity
(which is the low energy limit of $M-$theory) with a compactified
direction is by definition type IIA supergravity (which is the low energy limit of
type IIA string theory). In the limit $R_s\longrightarrow 0$ the
only objects in $M-$theory (or equivalently type IIA string theory) which carry $P_z$ are the
D0-brane. In a sector with momentum $P_z=N/R_s$ the lowest
excitations  are $N$ D0-branes. The effective action is given  
by the supersymmetric Yang-Mills theory reduced to $1-$dimension
(\ref{S}). The corresponding Hamiltonian is given by the quantum
mechanical system (\ref{H}).

$4)$ Supergravitons carrying Kaluza-Klein momentum $P_z=1/R_s$ are the
elementary D0-branes. They carry the quantum numbers of the basic
$11-$dimensional supergravity multiplet which contains $44$ gravitons $G_{MN}$,
$128$ gravitinos ${\psi}_M$ and $84$ independent components of the $3-$form
field $A_3$. Supergravitons carrying Kaluza-Klein momentum $P_z=N/R_s$
are bound composites of N D0-branes. Perturbtaive string states
carry $P_z=0$ while elementary and bound composites of $N$ anti-D0-branes carry negative $P_z$ and as we
said they  decouple.

Thus the BFSS conjecture relates $M-$theory in the infinite momentum
frame ($P_z\longrightarrow \infty$)
in the uncompactified limit ($R_s\longrightarrow \infty$) to the large $N$ limit of the
supersymmetric quantum mechanical system (\ref{H}) describing $N$ D0-branes.

The infinite momentum frame and the light-cone quantization are
equivalent only in the limit $N\longrightarrow \infty$. For finite $N$
the light-cone quantization is superior since finite $N$ infinite
momentum  formulation does not have Galilean invariance and also in
this formulation negative
and vanishing $P_{z}$ do not decouple for finite $N$. 
 
A stronger BFSS conjecture \cite{Susskind:1997cw} relates discrete light-cone quantization
(DLCQ) of
$M-$theory to the
supersymmetric quantum mechanical system (\ref{H}). The DLCQ of
$M-$theory corresponds to the quantization of $M-$theory 
compactified on a light-like circle of radius $R$ in a sector with
momentum $P_-=N/R$. This theory is characterized by a finite
momentum $P_-$ and it is conjectured to be exactly given by the finite
$N$ supersymmetric matrix model (\ref{S}). Since for a
light-like circle the value of $R$ can not be changed via a boost the uncompactified theory is obtained by letting
$N,R\longrightarrow \infty$ keeping $P_-$ always fixed.

\subsection{More on light-like versus space-like compactifications} In this
section we follow \cite{Seiberg:1997ad}. The compactification of the light-like coordinate $X^-=\frac{t-z}{\sqrt{2}}$ on a circle of
radius $R$ corresponds to the identification $X^-\sim X^-+l$ where
$l=2\pi R$ or equivalently
\begin{eqnarray}
&&z\sim z-\frac{l}{\sqrt{2}}\nonumber\\
&&t\sim t+\frac{l}{\sqrt{2}}.\label{light}
\end{eqnarray}
Let us consider now the compactification on the space-like circle
\begin{eqnarray}
&&z\sim z-\sqrt{\frac{l^2}{2}+l_s^2}\nonumber\\
&&t\sim t+\frac{l}{\sqrt{2}}.\label{space}
\end{eqnarray}
For $l_s<<l$ this space-like compactification tends to the
previous light-like compactification, viz
\begin{eqnarray}
&&z\sim z-\frac{l}{\sqrt{2}}-\frac{l_s^2}{\sqrt{2}l}\nonumber\\
&&t\sim t+\frac{l}{\sqrt{2}}.
\end{eqnarray}
Consider now the simpler space-like compactification
\begin{eqnarray}
&&z\sim z-l_s\nonumber\\
&&t\sim t.\label{space1}
\end{eqnarray}
A Lorentz boost is given by
\begin{eqnarray}
&&z^{'} =\frac{1}{\sqrt{1-{\beta}^2}}(z-\beta t)\nonumber\\
&&t^{'}=\frac{1}{\sqrt{1-{\beta}^2}}(t-\beta z).
\end{eqnarray}
The point $(-l_s,0)$ is boosted to the point
$(-\sqrt{\frac{l^2}{2}+l_s^2},\frac{l}{\sqrt{2}})$ if we choose the
velocity $\beta$ to be
\begin{eqnarray}
\beta=\frac{l}{\sqrt{l^2+2l_s^2}}.\label{v}
\end{eqnarray}
 In other words, the space-like compactification (\ref{space}) is
 related by the above boost to the space-like compactification
 (\ref{space1}). For $l_s<<l$ the space-like compactification
 (\ref{space}) becomes the light-like compactification (\ref{light})
 and the velocity becomes large given by
 $\beta=1-\frac{l_s^2}{l^2}$. We can conclude that the light-like
 compactification (\ref{light}) is the $l_s\longrightarrow 0$ limit 
 of the almost light-like  compactification
 (\ref{space}). Equivalently,  the light-like
 compactification (\ref{light}) is  the $l_s\longrightarrow 0$ limit of the boosted space-like
 compactification (\ref{space1}) with a large velocity given by (\ref{v}).

The point $(-\frac{l}{\sqrt{2}},\frac{l}{\sqrt{2}})$ is boosted under
a Lorentz transformation to the point $\sqrt{\frac{1+\beta}{1-\beta}}(-\frac{l}{\sqrt{2}},\frac{l}{\sqrt{2}})$.
In other words, under a longitudinal boost of the light-like
compactification (\ref{light}) the radius $R$ of the compactification
 is rescaled as
\begin{eqnarray}
R^{'}=\sqrt{\frac{1+\beta}{1-\beta}}R.
\end{eqnarray}
The momenta $P_{\pm}$ transform as
\begin{eqnarray}
P_+^{'}=\sqrt{\frac{1+\beta}{1-\beta}}P_+=\frac{R^{'}}{R}P_+~,~P_-^{'}=\sqrt{\frac{1-\beta}{1+\beta}}P_-=\frac{R}{R^{'}}P_-.
\end{eqnarray}
It is then obvious that under a longitudinal boost of the light-like
compactification (\ref{light}) the light-cone energy $P_+$ is
 also rescaled. In fact we see that $P_+$ is proportional to $R$. For
 the space-like
compactification (\ref{space}) the light-cone energy $P_+$ is
 also proportional to $R$ in the limit $R_s\longrightarrow 0$ where
 this space-like
compactification becomes light-like. For $R_s\longrightarrow 0$ the
velocity (\ref{v}) becomes large given by $\beta=1-R_s^2/R^2$ and hence
$\sqrt{\frac{1+\beta}{1-\beta}}=\sqrt{2}\frac{R}{R_s}$. Since $P_+$ in
the almost light-like compactification (\ref{space}) is  proportional
to $R$ and since this compactification is obtained from the
 space-like compactification  (\ref{space1}) with the above large boost we can
immediately conclude that the value of $P_+$ in the space-like compactification
(\ref{space1}) must be  proportional
to $R_s$. In other words, the value of $P_+$ can be made independent of
$R$ and of order $R_s$ by an appropriate large boost.

From the above discussion we can see that the light-like compactification
of $M-$theory (i.e. the DLCQ and the stronger BFSS conjecture) is
related to the  space-like compactification of $M-$theory (i.e. the IMF
quantization and the original BFSS
conjecture). 

In the limit $R_s\longrightarrow 0$ the
space-like compactification (\ref{space1}) of $M-$theory yields weakly
coupled type IIA string theory where the parameters $g\equiv\tilde{g}_s$ (
$\tilde{g}_s$ is the string coupling) and
${\alpha}^{'}\equiv \tilde{M}_s^{-2}$ ($\tilde{M}_s$ is the string
scale) of type IIA
string theory are related to the parameters $R_{10}\equiv R_s$ and
$M_{11}\equiv \tilde{M}_p$ of
the $11-$dimensional theory by
\begin{eqnarray}
\tilde{g}_s=(R_{s}\tilde{M}_{p})^{\frac{3}{2}}~,~\tilde{M}_s^2=R_{s}\tilde{M}_{p}^{3}.
\end{eqnarray}
Clearly, when $R_s\longrightarrow 0$ we have
$\tilde{g}_s\longrightarrow 0$ and $\tilde{M}_s\longrightarrow
0$. This is a complicated theory. Next we apply the boost. As we have said the energy $P_+$ which is
proportional to $R_s$
becomes proportional to $R$. Since $P_+$ has
dimension of mass and $R_s$ has dimension of lenght we conclude that $P_+$ must be of the order of
$R_s\tilde{M}_p^2$ where $\tilde{M}_p^2$ is inserted on dimensional
grounds. In order to focus on the modes with these energies 
we replace the $M-$theory with  parameters $\tilde{M}_p$ and $R_s$
with a new  $M-$theory with parameters ${M}_p$ and $R$ such that the energy
$P_+$ is kept fixed in the double scaling limit  $R_s\longrightarrow
0$, $\tilde{M}_p\longrightarrow
\infty$, viz
\begin{eqnarray}
R_s\tilde{M}_p^2=R{M}_p^2.
\end{eqnarray}
In this limit the string coupling and the string scale become
\begin{eqnarray}
&&\tilde{g}_s=(R_s\tilde{M}_p)^{\frac{3}{2}}=R_s^{\frac{3}{4}}(RM_p^2)^{\frac{3}{4}}\longrightarrow
  0\nonumber\\
&&\tilde{M}_s^2=R_s\tilde{M}_p^3=R_s^{-\frac{1}{2}}(RM_p^2)^{\frac{3}{2}}\longrightarrow
  \infty.
\end{eqnarray}
This is weakly coupled string theory with large string scale which is
a very simple theory. Indeed the sector with $P_-=N/R_s$ is the theory of
$N$ D0-branes.

In summary, $M-$theory with Planck scale $M_p$ compactified on a light-like circle of radius $R$
and momentum $P_-=N/R$ can be mapped to the  $M-$theory with Planck
scale $\tilde{M}_p\longrightarrow \infty$ 
compactified on a space-like circle of radius
$R_s=(RM_p^2)/{\tilde{M}_p^2}\longrightarrow 0$ which is a theory
of $N$ D0-branes.

\section{M-(atrix) theory in pp-wave spacetimes}
 
\subsection{The pp-wave spacetimes and Penrose limit}
The BFSS model \cite{Banks:1996vh} is a matrix model associated with  the DLCQ
(discrete light-cone quantization) description of the maximally
supersymmetric $11$ dimensional flat background. The BMN model \cite{Berenstein:2002jq} is a matrix model which describes
the DLCQ compactification of
$M-$theory on the maximally supersymmetric pp-wave background of $11$d
supergravity. 

Supergravity in $11$ dimensions admits four types of maximally
supersymmetric solutions \cite{Blau:2001ne,KowalskiGlikman:1984wv}. These are 
\begin{itemize}
\item $1)$ the $11$d Minkowski space and its toroidal compactifications, 
\item $2)$ the $AdS^7 \times S^4$ (the M5-brane), 
\item $3)$ the $AdS^4 \times S^7$ (the M2-brane) and 
\item $4)$ the Kowalski-Glikman solution \cite{KowalskiGlikman:1984wv} which is a pp-wave metric.  
\end{itemize}
Similarly, type IIB supergravity admits three types of solutions 
\begin{itemize}
\item $1)$ $10$d Minkowski space,
\item $2)$ $AdS^5 \times S^5$ and 
\item $3)$ a pp-wave metric \cite{Blau:2001ne}. 
\end{itemize}
All maximally supersymmetric pp-wave geometries can arise as Penrose limits of $AdS_p\times S^q$
spaces \cite{Blau:2002dy}. The powerful Penrose theorem states that near null geodesics (which are
paths of light rays) any spacetime becomes a pp-wave spacetime, i.e. any metric near
 a null geodesic becomes a pp-wave metric \cite{penrose}.

First we discuss pp-wave geometries a little further. These  spaces  are solutions  of the Einstein equations which
correspond to perturbations moving at the speed of light with  plane
wave fronts. See \cite{Nastase:2007kj} and references therein. The bosonic content of $11$d supergravity
consists of the metric and a $4-$form $F_4$. The pp-wave solutions of $11$d supergravity are
given by \cite{Hull:1984vh,Nastase:2007kj,Blau:2001ne} 
\begin{eqnarray}
&&ds^2=2dx^+dx^{-}+(dx^+)^2H(x^+,x^{i})+\sum_{i=1}^9 (dx^i)^2\nonumber\\
&&F_4=dx^+\wedge \phi.\label{pp11}
\end{eqnarray}
In the above equation $x^{\pm}=(x\pm t)/\sqrt{2}$,$x\equiv x^{10}$. The $\phi$ is a $3-$form satisfying
\begin{eqnarray}
d\phi=d*\phi=0~,~{\partial}_i^2H=\frac{1}{12}|\phi|^2.
\end{eqnarray}
Recall that
\begin{eqnarray}
\phi={\phi}_{\mu\nu\rho}dx^{\mu}\wedge dx^{\nu}\wedge dx^{\rho}~,~|\phi|^2={\phi}_{\mu\nu\rho}{\phi}^{\mu\nu\rho}~,~(*\phi)_{{\mu}_1...{\mu}_8}={\epsilon}_{{\mu}_1...{\mu}_{11}}{\phi}^{{\mu}_9{\mu}_{10}{\mu}_{11}}.
\end{eqnarray}
The only non-zero component of the Ricci tensor of the above metric is
%\footnote{Exercise:verify this statement.}
\begin{eqnarray}
R_{++}=-\frac{1}{2}{\partial}_i^2H(x^+,x^{i})=-\frac{1}{24}|\phi|^2.
\end{eqnarray}
An interesting class of solutions is given by
\begin{eqnarray}
H=\sum_{i,j}A_{ij}x^{i}x^{j}~,~A_{ij}=A_{ji}~,~2tr A=\frac{1}{12}|\phi|^2.
\end{eqnarray}
For generic $(A,\phi)$ this solution will preserve $1/2$ of the
supersymmetry. Kowalski-Glikman showed in $1984$ that all supersymmetry
will be preserved for precisely one non-trivial choice of $A_{ij}$ and
$\phi$ given by
\begin{eqnarray}
H=\sum_{i,j}A_{ij}x^{i}x^{j}=-\sum_{i=1}^3\frac{{\mu}^2}{9}x_i^2-\sum_{a=4}^9\frac{{\mu}^2}{36}x_a^2~,~\phi=\mu
dx^1\wedge dx^2\wedge dx^3.
\end{eqnarray}
Similarly, the bosonic content of $10$d type IIB supergravity
consists of the metric and a $5-$form $F_5$. The pp-wave solutions of
$10$d type IIB supergravity are
given by  \cite{Nastase:2007kj}
\begin{eqnarray}
&&ds^2=2dx^+dx^{-}+(dx^+)^2H(x^+,x^{i})+\sum_{i=1}^8 (dx^i)^2\nonumber\\
&&F_5=dx^+\wedge (\omega +*{\omega}).\label{pp10}
\end{eqnarray}
The $\omega$ is a $4-$form satisfying
\begin{eqnarray}
d\omega=d*\omega=0~,~{\partial}_i^2H=-32|\omega|^2.
\end{eqnarray} 
Again the general metric preserves $1/2$ of the supersymmetry while all supersymmetry
will be preserved for precisely one non-trivial choice of $A_{ij}$ and
$\omega$ given by
\begin{eqnarray}
H=\sum_{i,j}A_{ij}x^{i}x^{j}={\mu}^2\sum_{i=1}^8x_i^2~,~\omega=\frac{\mu}{2}
dx^1\wedge dx^2\wedge dx^3\wedge dx^4.\label{pp10E}
\end{eqnarray}
The above maximally supersymmetric pp-waves are Penrose limits of maximally
supersymmetric $AdS_{p+2}\times S^n$ spaces. For $11$d supergravity  the pp-wave
metric (\ref{pp11}) arises as limit of $AdS_7\times S^4$ or
$AdS_4\times S^7$ where both spaces give the same metric. For $10$d
type IIB supergravity  the pp-wave
metric (\ref{pp10}) arises as limit of $AdS_5\times S^5$. Let us also
say that the near horizon geometry of M2-, M5- and D3-brane
solutions is $AdS_{p+2}\times S^{D-p-2}$ \cite{Blau:2002dy}. For the M2-
and the M5-brane solutions $D=11$ and $p=2$ and $5$
respectively. For the D3-brane solution $D=10$ and $p=3$. 

Define
$\rho=R_{AdS_{p+2}}/R_{S^{D-p-2}}$ then for the M2-brane $\rho=1/2$, for
the M5-brane $\rho=2$ while for the D3-brane $\rho=1$. The metrics
for $AdS_{p+2}$, $S^{D-p-2}$ and $AdS_{p+2}\times S^{D-p-2}$ are given
respectively by \cite{Blau:2001ne}
\begin{eqnarray}
ds_{AdS_{p+2}}^2=R^2_{AdS}\bigg[-(d\tau)^2+{\sin}^2 \tau\bigg(\frac{dr^2}{1+r^2}+r^2d{\Omega}_p^2\bigg)\bigg].
\end{eqnarray}
\begin{eqnarray}
ds_{S^n}^2=R^2_{S}\bigg[(d\psi)^2+{\sin}^2 \psi d{\Omega}_{n-1}^2\bigg].
\end{eqnarray}
\begin{eqnarray}
ds_{AdS_{p+2}\times S^n}^2=ds_{AdS_{p+2}}^2+ds_{S^n}^2.
\end{eqnarray}
In above $d{\Omega}_p^2$ is the $p-$sphere metric, $\psi$ is the
colatitude and  $d{\Omega}_{n-1}^2$ is the metric on the equatorial
$(n-1)-$sphere. Introduce the coordinates
\begin{eqnarray}
u=\psi +\rho \tau~,~v=\psi -\rho \tau.
\end{eqnarray}
The metric becomes
\begin{eqnarray}
ds_{AdS_{p+2}\times S^n}^2/R_S^2=dudv +{\rho}^2\sin^2\frac{u-v}{2\rho}\bigg(\frac{dr^2}{1+r^2}+r^2d{\Omega}_p^2\bigg)+\sin^2\frac{u+v}{2}d{\Omega}_{n-1}^2.
\end{eqnarray}
We consider Penrose limit along the null geodesic parametrised by
$u$. All coordinates with the exception of $u$ will be scaled to
$0$. The coordinate $v$ will be scaled to $0$ faster than all the
other coordinates and hence dependence on the coordinate $v$ will  be dropped. We
get
\begin{eqnarray}
ds_{AdS_{p+2}\times S^n}^2/R_S^2=dudv +{\rho}^2\sin^2\frac{u}{2\rho}ds^2_{E^{p+1}}+\sin^2\frac{u}{2}d{\Omega}_{n-1}^2.
\end{eqnarray}
Let $y^1,...,y^{p+1}$ be the coordinates of $E^{p+1}$ and
$z^{p+2},...,z^{D-2}$ be the coordinates of $S^{n-1}$. We
introduce
\begin{eqnarray}
&&x^{-}=\frac{u}{2}\nonumber\\
&&x^{+}=v-\frac{1}{4}\bigg[\rho \vec{y}^2 {\rm sin}\frac{u}{\rho}
    +\vec{z}^2{\rm sin}u\bigg]\nonumber\\
&&x^{a}=(\rho {\rm sin}\frac{u}{2\rho})y^a~,~a=1,...,p+1\nonumber\\
&&x^a=({\rm sin}\frac{u}{2}) z^a~,~a=p+2,...,D-2.
\end{eqnarray}
We compute
\begin{eqnarray}
\sum_{a=1}^{D-2}(dx^a)^2&=&{\rho}^2{\rm sin}^2\frac{u}{2\rho}(dy^a)^2+{\rm
    sin}^2\frac{u}{2}(dz^a)^2+\frac{1}{4}\bigg[\vec{y}^2{\rm
    cos}^2\frac{u}{2\rho}+\vec{z}^2{\rm
    cos}^2\frac{u}{2}\bigg](du)^2\nonumber\\
&+&\frac{1}{4}\bigg[\rho {\rm
    sin}\frac{u}{\rho} d (y^a)^2+{\rm sin} u d(z^a)^2\bigg]du.
\end{eqnarray}

\begin{eqnarray}
2dx^+dx^-=du dv-\frac{1}{4}\bigg[\rho {\rm
    sin}\frac{u}{\rho} d (y^a)^2+{\rm sin} u d(z^a)^2\bigg]du-\frac{1}{4}\bigg[\vec{y}^2{\rm
    cos}\frac{u}{\rho}+\vec{z}^2{\rm
      cos}u\bigg](du)^2.
\end{eqnarray}
The metric can be rewritten as
\begin{eqnarray}
ds_{AdS_{p+2}\times S^n}^2/R_S^2=2dx^+dx^--\bigg[\frac{1}{{\rho}^2}\sum_{a=1}^{p+1}(x^a)^2+\sum_{a=p+2}^{D-2}(x^a)^2\bigg](dx^-)^2+\sum_{a=1}^{D-2}(dx^a)^2.
\end{eqnarray}
This is a pp-wave metric as promised. The two cases $\rho=2$ and
$\rho=1/2$ are isometric. The corresponding difeomorphism is
\begin{eqnarray}
&&x^{-}\longrightarrow \frac{1}{2}x^{-}~,~x^+\longrightarrow
  2x^+\nonumber\\
&&(x^1,...,x^6,x^7,...,x^9)\longrightarrow (x^4,...,x^9,x^1,...,x^3).
\end{eqnarray}
Let us reconsider the case of $AdS_5\times S^5$ more
explicitly. In this case we have $R_{AdS}=R_S=R$ or equivalently $\rho=1$. The metric is
\begin{eqnarray}
ds_{AdS_{5}\times S^5}^2=R^2\bigg[-(d\tau)^2+{\sin}^2 \tau\bigg(\frac{dr^2}{1+r^2}+r^2d{\Omega}_3^2\bigg)+(d\psi)^2+{\sin}^2 \psi d{\Omega}_{4}^2\bigg].
\end{eqnarray}
This can also be put in the form \footnote{Exercise:verfiy this.  }
\begin{eqnarray}
ds^2=R^2\bigg[-dt^2{\cosh}^2\rho +d{\rho}^2+{\sinh}^2\rho
  d{\Omega}_3^2+d{\psi}^2{\cos}^2\theta +d{\theta}^2 +{\sin}^2\theta d{\Omega}_3^{'2}\bigg].
\end{eqnarray}
We consider a particle moving at the speed of light along an equator of $S^5$ ($\theta=0$)
 while staying in the center of $AdS_5$ ($\rho=0$). This is a null
geodesic (since it is the path of a light ray) parametrised by
$\psi$. The geometry near this trajectory is given by the metric
 \begin{eqnarray}
ds^2=R^2\bigg[-(1+\frac{{\rho}^2}{2})dt^2 +d{\rho}^2+{\rho}^2
  d{\Omega}_3^2+(1-\frac{{\theta}^2}{2})d{\psi}^2 +d{\theta}^2 +{\theta}^2 d{\Omega}_3^{'2}\bigg].
\end{eqnarray}
We define the null coordinates
\begin{eqnarray}
\tilde{x}^{\pm}=\frac{t\pm \psi}{2}.
\end{eqnarray}
Then we take the limit
\begin{eqnarray}
\tilde{x}^+=x^+~,~\tilde{x}^{-}=\frac{x^{-}}{R^2}~,~\rho=\frac{r}{R}~,~\theta=\frac{y}{R}.\label{PL}
\end{eqnarray}
The metric becomes
 \begin{eqnarray}
ds^2&=&-4dx^+dx^{-}-(r^2+y^2)(dx^+)^2+dr^2+dy^2+r^2d{\Omega}_3^2+y^2d{\Omega}_3^{'2}\nonumber\\
&=&-4dx^+dx^{-}-(\vec{r}^2+\vec{y}^2)(dx^+)^2+d\vec{r}^2+d\vec{y}^2.
\end{eqnarray}
The Penrose limit (\ref{PL}) can be understood as follows. We consider
the following boost along the equator of $S^5$ given by
\begin{eqnarray}
t={\cosh}\beta t^{'} +{\sinh}\beta {\psi}^{'}~,~ \psi={\sinh}\beta t^{'} +{\cosh}\beta {\psi}^{'}.
\end{eqnarray}
This is equivalent to
\begin{eqnarray}
\tilde{x}^{+}=e^{\beta}\tilde{x}^{+'}~,~\tilde{x}^{-}=e^{-\beta}\tilde{x}^{-'}.
\end{eqnarray}
If we make the identification $e^{\beta}=R$ and scale all coordinates
$t^{'},{\psi}^{'}$ and the rest by $1/R$ then we will obtain
(\ref{PL}).

The D3-brane carries fluxes with respect to a $D-p-2=5-$form field
strength. It is obvious that only the components of this $5-$form
field $F_5$ with an index $+$ will survive the above Penrose limit, viz
\begin{eqnarray}
F_{+1234}=F_{+5678}=\mu.
\end{eqnarray}
Thus the $5-$form field $F_5$ of the $AdS_5\times S^5$ solution
  matches in the Penrose limit the maximally supersymmetric pp-wave solution of IIB supergravity given in (\ref{pp10}) and
 (\ref{pp10E}).

The M2- and M5-brane solutions  carry fluxes with respect
to $D-p-2=7$ and $4-$form field strenghts respectively.

\subsection{The BMN matrix model}
As we have seen, the BFFS model is postulated to describe DLCQ quantization of M-theory in flat background spacetime. It is given by  
\begin{eqnarray}
S_{\rm BFFS}=\frac{1}{g^2}Tr\bigg(-\frac{1}{4}[{X}_I,{X}_J]^2+\frac{1}{2}(D_{0}X_I)^2-\frac{1}{2}{\psi}^TC_9
{\Gamma}^I[{X}_I,\psi]-\frac{1}{2}{\psi}^T C_9 D_0\psi\bigg).
\end{eqnarray}
Similarly, the BMN model is postulated to describe DLCQ quantization of M-theory in pp-wave background spacetime. It is in a precise sense a mass-deformation of the BFSS model given by the action \cite{Berenstein:2002jq} 
\begin{eqnarray}
S_{\rm BMN}=S_{\rm BFSS}+\Delta S_{\rm mas-def}.
\end{eqnarray}
\begin{eqnarray}
\Delta S_{\rm mas-def}&=&\frac{{\mu}^2}{2g^2}Tr\bigg(\sum_{i=1}^3X_i^2+\frac{1}{4}\sum_{a=4}^9X_a^2\bigg)-\frac{i\mu}{2g^2}{\epsilon}_{ijk}Tr[X_i,X_j]X_k-\frac{3i\mu}{8g^2}Tr{\psi}^TC_9{\gamma}^{123}\psi.\nonumber\\
\end{eqnarray}
This model as opposed to the BFSS model has as a solution a fuzzy sphere solution $X_i=\mu J_i$, $X_a=0$, $A_0=0$, $\psi=0$, where $[J_i,J_j]=i\epsilon_{ijk}J_k$, which preserves full supersymmetry.  Non-perturbative studies of the BMN and its relation to the gauge/gravity duality can be found in \cite{Hanada:2008gy,Catterall:2008yz,Anagnostopoulos:2007fw,Catterall:2009xn,Filev:2015hia,Hanada:2008ez,Catterall:2010fx,Kadoh:2015mka,Berkowitz:2016jlq}. For a concise summary of the results obtained for the BMN model and future prospects see \cite{Hanada:2016jok}.

\subsection{Construction of the BMN matrix model}
The original derivation of the BMN model consisted in showing that the $N=1$ mass-deformed BFSS model is given by the action of a superparticle in the above pp-wave background then they obtained the $N>1$ mass-deformed BFSS model by extending this result in a way consistent with supersymmetry.

The BMN model can be derived in a more direct way as follows \cite{Taylor:1999gq}. We start from
the  pp-wave solutions of $11$d supergravity 
given by 
\begin{eqnarray}
ds^2&=&2dx^+dx^{-}+H(dx^+)^2+(dx^I)^2\nonumber\\
&=&-(1-\frac{H}{2})dt^2+(1+\frac{H}{2})dx^2+Hdx dt +(dx^I)^2.
\end{eqnarray}
\begin{eqnarray}
H=-\sum_{i=1}^3\frac{{\mu}^2}{9}x_i^2-\sum_{a=4}^9\frac{{\mu}^2}{36}x_a^2.
\end{eqnarray}
\begin{eqnarray}
F_4=-\frac{\mu}{\sqrt{2}}dx^1\wedge dx^2\wedge
dx^3\wedge dx+\frac{\mu}{\sqrt{2}}dt\wedge dx^1\wedge dx^2\wedge dx^3.
\end{eqnarray}
If we  start with $\tilde{M}-$theory
compactified on a space-like
circle of radius $R_s$ then we apply a boost with velocity
$\beta=R/\sqrt{R^2+2R_s^2}$ and take the limit $R_s\longrightarrow
0$ we obtain $M-$theory compactified on an almost  light-like circle of radius
$R$. We are interested in the DLCQ of this $M-$theory on the above
pp-wave background with $N$ units of momentum, viz $P_-=N/R$. 

The
energy $P_+$ in the compactification on the almost light-like circle of radius $R$ is
proportional to $R$. The momentum $P_-^s$ and the energy $P_+^s$ in
the compactification on the space-like circle of radius $R_s$ are given
 through 
\begin{eqnarray}
P_-=\sqrt{\frac{1-\beta}{1+\beta}}P_-^s~,~P_+=\sqrt{\frac{1+\beta}{1-\beta}}P_+^s~,~ 
\end{eqnarray}
with 
\begin{eqnarray}
\sqrt{\frac{1-\beta}{1+\beta}}=\sqrt{2}\frac{R}{R_s}~,~R_s\longrightarrow 0. 
\end{eqnarray}
In other words, $P_-^s=\sqrt{2}\frac{N}{R_s}$
 and $P_+^s$ is proportional to $\frac{R_s}{\sqrt{2}}$ or equivalently
\begin{eqnarray}
P_t^s=\frac{N}{R_s}+\frac{R_s}{2}~,~P_x^s=\frac{N}{R_s}-\frac{R_s}{2}. 
\end{eqnarray}
In particular, we see that the
 momentum $P_x^s$ goes to $\infty$ when $R_s\longrightarrow
 0$. Furthermore, the string
 coupling and the string scale in the $\tilde{M}-$theory are $\tilde{g}_s=(R_s\tilde{M}_p)^{\frac{3}{2}}=R_s^{\frac{3}{4}}(RM_p^2)^{\frac{3}{4}}\longrightarrow
  0$ and $\tilde{M}_s^2=R_s\tilde{M}_p^3=R_s^{-\frac{1}{2}}(RM_p^2)^{\frac{3}{2}}\longrightarrow
  \infty$ when $R_s\longrightarrow 0$
  and $\tilde{M}_p\longrightarrow \infty$ keeping
  $R_s\tilde{M}_p^2$ fixed, viz $R_s\tilde{M}_p^2=RM_p^2$. 

The
  compactification on the space-like circle of
  radius $R_s\longrightarrow 0$ corresponds therefore to the quantization of $\tilde{M}-$theory in IMF with $N$ units
  of longitudinal momentum which we know is  weakly coupled type IIA
  string theory. The DLCQ of $M-$theory with $N$ units of momentum is exactly
  mapped to the theory of $N$ D0-branes. The $M-$theory on the above
  pp-wave background with mass $\mu$ corresponds to $\tilde{M}-$theory
  on the same pp-wave background with mass ${\mu}_s$ given through
\begin{eqnarray}
\mu=\sqrt{\frac{1-\beta}{1+\beta}}{\mu}_s=\sqrt{2}\frac{R}{R_s}{\mu}_s.
\end{eqnarray} 
In other words, ${\mu}_s=\frac{\mu}{\sqrt{2}}\frac{R_s}{R}$. Recall
that the energies $E_s=P_t^s-\frac{N}{R_s}$ of the D0-brane states
are proportional to $\frac{R_s}{{2}}$ so that they go to $0$ as
$R_s\longrightarrow 0$. The light-cone energies $P_+$
are proportional to $R$. In order that the energies of the D0-brane
states match the light-cone energies we must multiply $E_s$ by
$2\frac{R}{R_s}$. In other words, $E_s^{'}=R$. Multiplying ${\mu}_s$ by $2\frac{R}{R_s}$ we get
${\mu}_s^{'}=\sqrt{2}\mu$ or equivalently
$\mu=\frac{{\mu}_s^{'}}{\sqrt{2}}$. 

To take this rescaling into
account we make the replacing $\mu \longrightarrow
{\mu}/{\sqrt{2}}$ so that the $4-$form field becomes
\begin{eqnarray}
F_4=-\frac{\mu}{{2}}dx^1\wedge dx^2\wedge
dx^3\wedge dx+\frac{\mu}{{2}}dt\wedge dx^1\wedge dx^2\wedge dx^3.
\end{eqnarray}
We also take this rescaling into
account by replacing $\mu \longrightarrow
{\mu}/{\sqrt{2}}$ in $H$. In other words, we replace $H$ by $H/2$ in
the metric. The metric becomes (with $H\equiv -F^2$)
\begin{eqnarray}
ds^2
&=&-(1+\frac{F^2}{4})dt^2+(1-\frac{F^2}{4})dx^2-\frac{F^2}{2}dx dt +(dx^i)^2.
\end{eqnarray}
The most general metric which is invariant under translations in the
$10-$direction $x^{10}\equiv x$ is of the form \cite{Polchinski:1998rr}
\begin{eqnarray}
ds^2=G_{\mu\nu}^{10}(x^{\mu})dx^{\mu}dx^{\nu}+e^{2{\sigma}(x^{\mu})}(dx+A_{\nu}(x^{\mu})dx^{\nu})^2~,~\mu,\nu=0,...,9.
\end{eqnarray}
$A_{\mu}$ is the RR one-form. We can immediately find that
\begin{eqnarray}
&&e^{2\sigma}=1-\frac{F^2}{4}\nonumber\\
&&A_0=\frac{-\frac{F^2}{4}}{1-\frac{F^2}{4}}~,~A_i=0\nonumber\\
&&G_{00}^{10}=-e^{-2\sigma}~,~G_{0i}^{10}=0~,~G_{ij}^{10}={\delta}_{ij}.
\end{eqnarray}
The dilaton field is defined through
\begin{eqnarray}
\sigma=\frac{2\Phi}{3}.
\end{eqnarray}
The corresponding type IIA background is obtained by the redefinition
\begin{eqnarray}
ds^2({\rm new})=e^{\sigma}ds^2.
\end{eqnarray}
In other words
\begin{eqnarray}
&&G_{00}^{10}({\rm new})=-e^{-\sigma}~,~G_{0i}^{10}({\rm
    new})=0~,~G_{ij}^{10}({\rm new})=e^{\sigma} {\delta}_{ij}.
\end{eqnarray}
The $11-$dimensional pp-wave metric has zero scalar curvature. The
curvature of the $10-$dimensional metric $e^{\sigma
  (x^{\mu})}G_{\mu\nu}^{10}(x^{\mu})dx^{\mu}dx^{\nu}$ is not zero
given by %\footnote{Exercise:verify this statement.}
\begin{eqnarray}
{\cal R}\propto -\frac{{\mu}^2}{8}(1-\frac{F^2}{4})^{-\frac{3}{2}}.
\end{eqnarray}
This means that we must always have $F^2{\leq}4$.
 
The NS-NS and R-R three-form fields are given by
\begin{eqnarray}
H_{123}\equiv (F_4)_{12310}=-\frac{\mu}{2}~,~F_{0123}\equiv (F_4)_{0123}=\frac{\mu}{2}.
\end{eqnarray}

For small $F^2$ we have
\begin{eqnarray}
&&\Phi\sim -\frac{3F^2}{16}\nonumber\\
&&A_0\sim -\frac{F^2}{4}\nonumber\\
&&G_{00}^{10}({\rm new})=-1-\frac{F^2}{8}\nonumber\\
&&G_{ij}^{10}({\rm new})={\delta}_{ij}-\frac{F^2}{8}{\delta}_{ij}.
\end{eqnarray}
Aletrnatively we can write the metric as
\begin{eqnarray}
  ds^2({\rm new})={\eta}_{\mu\nu}dx^{\mu}dx^{\nu}+(dx^{10})^2+h_{\mu\nu}dx^{\mu}dx^{\nu}+h_{1010}(dx^{10})^2+2h_{010}dx^0dx^{10}.
\end{eqnarray}
\begin{eqnarray}
h_{\mu\nu}=\sigma {\delta}_{\mu\nu}=-\frac{F^2}{8}{\delta}_{\mu\nu}~,~h_{1010}=2\Phi=-\frac{3F^2}{8}~,~h_{010}=A_0=-\frac{F^2}{4}.
\end{eqnarray}
The matrix model corresponding to the flat metric $ds^2({\rm
  new})={\eta}_{\mu\nu}dx^{\mu}dx^{\nu}+(dx^{10})^2$ is given by the
  BFSS model in Minkowski signature given by the equation (\ref{S0}), namely
\begin{eqnarray}
&&S_0=\int dt~L_0\nonumber\\
&&L_0=\frac{1}{g^2}Tr\bigg(\frac{1}{4}[{X}_I,{X}_J]^2+\frac{1}{2}(D_{0}X_I)^2+{\psi}^TC_9
{\Gamma}^I[{X}_I,\psi]-i{\psi}^T C_9 D_0\psi\bigg).\label{S00}
\end{eqnarray}
In above $D_0={\partial}_0-i[A_0,..]$ and $I,J=1,...,9$. The coupling
constant $g^2$ is of dimension $L^2$, the matrices $X_I$ are of
dimension $L$, the operator $D_0$
is of dimension $L$ and the spinor $\psi$ is of dimension $L^{\frac{3}{2}}$. 

In the rest of this section  we will follow the derivation from D0-brane dynamics on compactified reduced pp-waves outlined in \cite{Dasgupta:2002hx} but very detailed in \cite{Taylor:1999gq}.
The correction to the matrix model $L_0$  corresponding to the metric
$h_{MN}$ are given by terms of the form \cite{Taylor:1999gq} (see also \cite{Taylor:1998tv})
\begin{eqnarray}
&&\Delta S_0[h]=\int dt~\Delta L_0[h].
\end{eqnarray}
\begin{eqnarray}
\Delta
  L_0[h]&=&\frac{1}{2}\sum_{n=0}^{\infty}\sum_{I_1,...,I_n}\frac{1}{n!}T^{MN(I_1...I_n)}{\partial}_{I_1}...{\partial}_{I_n}h_{MN}(0)\nonumber\\
&=&\frac{1}{4}\sum_{I_1,I_2}T^{MN(I_1I_2)}{\partial}_{I_1}{\partial}_{I_2}h_{MN}(0)\nonumber\\
&=&-\frac{1}{8}\bigg(\frac{1}{4}T^{\mu\mu(I_1I_2)}+T^{010(I_1I_2)}+\frac{3}{4}T^{1010(I_1I_2)}\bigg){\partial}_{I_1}{\partial}_{I_2}F^2.
\end{eqnarray}
In above $T^{MN(I_1,...,I_n)}$ are the matrix theory
forms of the multipole moments of the stress-energy tensor of 11D
supergravity which
couple to the derivatives of the background supergravity fields. Making use of equation $(17)$ of \cite{Taylor:1999gq} we have
%\footnote{Exercise:establish these identities.}
\begin{eqnarray}
&&T^{\mu\mu(I_1I_2)}=T^{00(I_1I_2)}+T^{II(I_1I_2)}=T^{++(I_1I_2)}+T^{+-(I_1I_2)}+T^{II(I_1I_2)}\nonumber\\
&&T^{010(I_1I_2)}=T^{++(I_1I_2)}\nonumber\\
&&T^{1010(I_1I_2)}=T^{++(I_1I_2)}-\frac{1}{3}T^{+-(I_1I_2)}-\frac{1}{3}T^{II(I_1I_2)}.
\end{eqnarray}
We obtain
\begin{eqnarray}
\Delta
  L_0[h]
&=&-\frac{1}{4}T^{++(I_1I_2)}{\partial}_{I_1}{\partial}_{I_2}F^2.
\end{eqnarray}
The zeroeth moment component $T^{++}$ of the stress-energy tensor is
given by
\begin{eqnarray}
T^{++} =\frac{1}{g^2}Tr ({\bf 1}).
\end{eqnarray}
This moment obviously corresponds to the momentum $P_-=N/R$. The
higher moments $T^{++(I_1I_2)}$ of the stress-energy tensor are
defined by \cite{Taylor:1999gq} %\cite{taylor1}
\begin{eqnarray}
T^{++(I_1I_2)} = {\rm Sym}(T^{++};X_{I_1},X_{I_2})=\frac{1}{g^2}{\rm
  Sym}(Tr{\bf 1};X_{I_1},X_{I_2}).\label{sym}
\end{eqnarray}
In general
\begin{eqnarray}
T^{IJ(I_1...I_n)} = {\rm Sym}(T^{IJ};X_{I_1},...,X_{I_n})+T^{IJ(I_1...I_n)}_{\rm fermion}.
\end{eqnarray}
The contributions ${\rm Sym}(STr (Y );X_{I_1}, . . . ,X_{I_n})$ (where $STr$
indicates a trace which is symmetrized over all orderings of terms
 under the trace) are the symmetrized average over
all possible orderings when the matrices $X_{I_k}$ are inserted into
the trace of the product $Y$. Thus
\begin{eqnarray}
T^{++(I_1I_2)}=\frac{1}{g^2}Tr X_{I_1}X_{I_2}.
\end{eqnarray}
The first correction due to the background metric $h$ (the $10$d metric, the dilaton field $\Phi$ and the R-R field $A$) takes then the form
\begin{eqnarray}
\Delta
  L_0[h]=-\frac{{\mu}^2}{18g^2}Tr\sum_{i=1}^3X_i^2-\frac{{\mu}^2}{72g^2}Tr\sum_{a=4}^9X_a^2.\label{S0c1}
\end{eqnarray}
The other degrees of freedom are the  NS-NS and R-R three-form fields
$H_{123}=-\frac{\mu}{2}$ and $F_{0123}=\frac{\mu}{2}$. The
corresponding potentials are
\begin{eqnarray}
B_{ij}=\frac{\mu}{6}{\epsilon}_{ijk}x_k~,~C_{0ij}=\frac{\mu}{6}{\epsilon}_{ijk}x_k.
\end{eqnarray}
In above we have used the fact that $H=-dB$, $F=dC$ or explicitly
$H_{kij}=-{\partial}_kB_{ij}-{\partial}_iB_{jk}-{\partial}_jB_{ki}$ and
$F_{0kij}={\partial}_kC_{0ij}+{\partial}_iC_{0jk}+{\partial}_jC_{0ki}$.
The correction to the matrix model $L_0$ given in equation (\ref{S0}) arising from the fields $B$
and $C$ are given by terms of the form \cite{Taylor:1999gq} 

\begin{eqnarray}
\Delta
  L_0[B,C]&=&\sum_{n=0}^{\infty}\sum_{{i}_1,...,{i}_n}\frac{1}{n!}T_B^{\mu\nu({i}_1...{i}_n)}{\partial}_{{i}_1}...{\partial}_{{i}_n}B_{\mu\nu}(0)+\sum_{n=0}^{\infty}\sum_{i_1,...,i_n}\frac{1}{n!}T_C^{\mu\nu\lambda({i}_1...{i}_n)}{\partial}_{i_1}...{\partial}_{i_n}C_{\mu\nu\lambda}(0)\nonumber\\
&=&\sum_{i_1}T_B^{\mu\nu(i_1)}{\partial}_{i_1}B_{\mu\nu}+\sum_{i_1}T_C^{\mu\nu\lambda(i_1)}{\partial}_{i_1}C_{\mu\nu\lambda}\nonumber\\
&=&\frac{\mu}{2}{\epsilon}_{ijk}(\frac{1}{3}T_B^{ij(k)}+T_C^{0ij(k)}).
\end{eqnarray}
The $T_{B}^{MN(I_1,...,I_n)}$ and $T_{B}^{MN(I_1,...,I_n)}$ (called $I_{s}$ and $I_2$ in \cite{Taylor:1999gq}) are the matrix theory
forms of the multipole moments of the membrane current of 11D
supergravity. By using equation $(19)$ of \cite{Taylor:1999gq}  we have the leading behavior %\footnote{Exercise:establish these identities.}
\begin{eqnarray}
&&T_B^{ij(k)}=3J^{+ij(k)}\nonumber\\%-3J^{-ij(k)}\nonumber\\
&&T_C^{0ij(k)}=J^{+ij(k)}.
\end{eqnarray}
\begin{eqnarray}
\Delta
  L_0[B,C]&=&\mu{\epsilon}_{ijk}J^{+ij(k)}.
\end{eqnarray}
We have (equation $(37)$ of \cite{Taylor:1999gq})
\begin{eqnarray}
J^{+ij}=\frac{i}{6g^2}STr[X_i,X_j].
\end{eqnarray}
Although these zeroeth moments are zero for finite $N$ the  higher moments
$J^{+ij(k)}$ are not zero given (by using equation
(\ref{sym})) precisely by the Chern-Simons action
\begin{eqnarray}
J^{+ij(k)}_B=\frac{i}{6g^2}Tr[X_i,X_j]X_k.
\end{eqnarray}
This is the Myers effect \cite{Myers:1999ps}. The fermionic contribution to
$J^{+ij(k)}$ is given by (see the appendix of \cite{Taylor:1999gq})
\begin{eqnarray}
J^{+ij(k)}_F=\frac{i}{24g^2}Tr{\psi}^TC_9{\gamma}^{[ijk]}\psi.
\end{eqnarray}
The correction ${\Delta}L_0[B,C]$ is then given by
\begin{eqnarray}
\Delta
  L_0[B,C]&=&\frac{i\mu}{6g^2}{\epsilon}_{ijk}Tr[X_i,X_j]X_k+\frac{i\mu}{4g^2}Tr{\psi}^TC_9{\gamma}^{123}\psi.\label{S0c2}
\end{eqnarray}
Putting (\ref{S00}), (\ref{S0c1}) and (\ref{S0c2}) together we get the total BMN
model, viz
\begin{eqnarray}
L&=&\frac{1}{g^2}Tr\bigg(\frac{1}{4}[{X}_I,{X}_J]^2+\frac{1}{2}(D_{0}X_I)^2+{\psi}^TC_9
{\Gamma}^I[{X}_I,\psi]-i{\psi}^T C_9 D_0\psi\bigg)\nonumber\\
&-&\frac{{\mu}^2}{18g^2}Tr\bigg(\sum_{i=1}^3X_i^2+\frac{1}{4}\sum_{a=4}^9X_a^2\bigg)+\frac{i\mu}{6g^2}{\epsilon}_{ijk}Tr[X_i,X_j]X_k+\frac{i\mu}{4g^2}Tr{\psi}^TC_9{\gamma}^{123}\psi.\nonumber\\
\end{eqnarray}
In Euclidean signature we get the Lagrangian 
\begin{eqnarray}
-L&=&\frac{1}{g^2}Tr\bigg(-\frac{1}{4}[{X}_I,{X}_J]^2+\frac{1}{2}(D_{0}X_I)^2-{\psi}^TC_9
{\Gamma}^I[{X}_I,\psi]-{\psi}^T C_9 D_0\psi\bigg)\nonumber\\
&+&\frac{{\mu}^2}{18g^2}Tr\bigg(\sum_{i=1}^3X_i^2+\frac{1}{4}\sum_{a=4}^9X_a^2\bigg)-\frac{i\mu}{6g^2}{\epsilon}_{ijk}Tr[X_i,X_j]X_k-\frac{i\mu}{4g^2}Tr{\psi}^TC_9{\gamma}^{123}\psi.\nonumber\\
\end{eqnarray}
We go back to the Minkowski signature and perform the scaling  $X_I\longrightarrow RX_I$, $\psi\longrightarrow
R^{\frac{3}{2}}\psi$ where $g^2=R^3$. We obtain
\begin{eqnarray}
L&=&Tr\bigg(\frac{R}{4}[{X}_I,{X}_J]^2+\frac{1}{2R}(D_{0}X_I)^2+R{\psi}^TC_9
{\Gamma}^I[{X}_I,\psi]-i{\psi}^T C_9 D_0\psi\bigg)\nonumber\\
&-&\frac{{\mu}^2}{18R}Tr\bigg(\sum_{i=1}^3X_i^2+\frac{1}{4}\sum_{a=4}^9X_a^2\bigg)+\frac{i\mu}{6}{\epsilon}_{ijk}Tr[X_i,X_j]X_k+\frac{i\mu}{4}Tr{\psi}^TC_9{\gamma}^{123}\psi.\label{bmn0}\nonumber\\
\end{eqnarray}
We set $R=1$, $\mu=m$, $A_0=X_0$, $X_a={\Phi}_a$, $a=4,...,9$, and we define
\begin{eqnarray}
\psi=\left(
\begin{array}{c}
{\theta}_{\alpha}^A\\
(i{\sigma}_2)_{\alpha \beta}{\theta}_{\beta}^{+A}.
\end{array}
\right)~,~\alpha=1,2,~A=1,...,4.
\end{eqnarray}
We also have
\begin{eqnarray}
&&{\gamma}^{123}=\left(
\begin{array}{cc}
-i{\bf 1}_2\otimes{\bf 1}_4 &0\\
0&i{\bf 1}_2\otimes{\bf 1}_4
\end{array}
\right).
\end{eqnarray}
The above Lagrangian becomes therefore (changing also the notation as $a,b=1,2,3$ and $i,j=4,...,9$)
\begin{eqnarray}
L&=&Tr\bigg[-\frac{1}{2}[X_0,X_a]^2-\frac{1}{2}[X_0,{\Phi}_i]^2+\frac{1}{4}[X_a,X_b]^2+\frac{1}{4}[{\Phi}_i,{\Phi}_j]^2+\frac{1}{2}[X_a,{\Phi}_i]^2\nonumber\\
&+&\frac{im}{3}{\epsilon}_{abc}X_aX_bX_c-\frac{m^2}{18}X_a^2-\frac{m^2}{72}{\Phi}_i^2\bigg]\nonumber\\
&+&Tr\bigg[-2{\theta}^+[X_0,\theta]-2{\theta}^+\bigg({\sigma}_a[X_a,\theta]-\frac{m}{4}\theta\bigg)+{\theta}^+i{\sigma}_2{\rho}_i[{\Phi}_i,({\theta}^+)^T]-{\theta}^Ti{\sigma}_2{\rho}_i^+[{\Phi}_i,\theta]\bigg]\nonumber\\
&+&Tr\bigg[\frac{1}{2}({\partial}_0X_a)^2+\frac{1}{2}({\partial}_0{\Phi}_i)^2-2i {\theta}^+{\partial}_0{\theta}-i{\partial}_0X_a[X_0,X_a]-i{\partial}_0{\Phi}_i[X_0,{\Phi}_i]\bigg].\label{bmn}
\end{eqnarray}

\subsection{Compactification on ${\bf R}\times {\bf S}^3$}% and ${\cal N}=4$ SYM theory in $4$ dimensions}

We give another derivation of the BMN model via dimensional reduction on ${\bf R}\times {\bf S}^3$ following \cite{Kim:2003rza,Kim:2002if}. 

We start from $D=10$ with metric ${\eta}^{\mu \nu}=(-1,+1,...,+1)$. The Clifford algebra is $32$ dimensional given by $\{G^M,G^N\}=2{\eta}^{MN}{\bf 1}_{32}$. The basic object of ${\cal N}=1$ SUSY in $10$ dimensions is a $32-$component complex spinor ${\Lambda}$ which satisfies the Majorana reality condition and the Weyl condition. We use the notation $I=1,...,9$, $\mu=0,...,3$, $a=1,..,3$, $i=4,..,9$. The Dirac matrices are given by
\begin{eqnarray}
&&G^0=i{\sigma}_2{\otimes}{\bf 1}_{16}~,~G^I={\sigma}_1{\otimes}{\Gamma}^I.
\end{eqnarray}
Explicitly we have 
\begin{eqnarray}
&&{\Gamma}^a=\left(
\begin{array}{cc}
-{\sigma}^a\otimes{\bf 1}_4 &0\\
0&{\sigma}^a\otimes{\bf 1}_4
\end{array}
\right)~,~{\Gamma}^i=\left(
\begin{array}{cc}
0&{\bf 1}_2\times{\rho}^i\\
{\bf 1}\otimes ({\rho}^{i})^+&0
\end{array}
\right).
\end{eqnarray}
In above the matrices ${\rho}_i$ satisfy
\begin{eqnarray}
{\rho}_i({\rho}_j)^++{\rho}_j({\rho}_i)^+=({\rho}_i)^+{\rho}_j+({\rho}_j)^+{\rho}_i=2{\delta}_{ij}{\bf 1}_4.
\end{eqnarray}
The matrices ${\Gamma}^I$ provide the Clifford algebra in $d=9$ dimensions. The ${\Gamma}^a$ provide  the $SO(3)$ Clifford algebra whereas ${\Gamma}^i$ provide  the $SO(6)$ Clifford algebra. The charge conjugation matrix $C_{10}$ in $10$ dimensions is related to the charge conjugation matrix $C_{9}$ in $9$ dimensions via the equation
\begin{eqnarray}
&&C_{10}={\sigma}_1\otimes C_9~,~C_9=\left(
\begin{array}{cc}
0 &-i{\sigma}_2\otimes {\bf 1}_4\\
i{\sigma}_2\otimes {\bf 1}_4&0
\end{array}
\right).
\end{eqnarray}
By the Weyl and Majorana conditions the $32-$component spinor $\Lambda$ can be put in the form
\begin{eqnarray}
\Lambda=\sqrt{2}\left(
\begin{array}{c}
\psi\\
0
\end{array}
\right)~,~{\psi}^+={\psi}^TC_9.
\end{eqnarray}
By the reality condition ${\psi}^+={\psi}^TC_9$ the $16-$component spinor $\psi$ can be written as
\begin{eqnarray}
\psi=\left(
\begin{array}{c}
s_{\alpha}^A\\
t_{\alpha}^A\equiv (i{\sigma}_2)_{\alpha \beta}s_{\beta}^{+A}.
\end{array}
\right)~,~\alpha=1,2,~A=1,...,4.
\end{eqnarray}
By using $D_{M}={\nabla}_M-i[A_M,..]$, $A_i={\phi}_i$, ${\partial}_i=0$ where ${\nabla}_M$ are spacetime covariant derivatives we can immediately compute the fermion action in $D=10$ to be given by
\begin{eqnarray}
 -\frac{1}{2} (\bar{\Lambda}G^{M}D_{M}\Lambda)|_{d=10}&=&\frac{1}{2}{\Lambda}^+D_0\Lambda-\frac{1}{2}{\Lambda}^+G^0G^aD_a\Lambda +\frac{i}{2}{\Lambda}^+G^0G^i[{\phi}_i,\Lambda]\nonumber\\
&=&2s^{+A}D_0s^A+2s^{+A}{\sigma}_aD_as^A+is^{+A}i{\sigma}_2({\rho}_i)_{AB}[{\phi}_i,(s^{+B})^T]\nonumber\\
&-&i(s^{A})^Ti{\sigma}_2({\rho}_i^+)_{AB}[{\phi}_i,s^{B}].
\end{eqnarray}
The Yang-Mills action takes the form
\begin{eqnarray}
(-\frac{1}{4}F_{MN}F^{MN})|_{d=10}=-\frac{1}{4}F_{\mu \nu}F^{\mu \nu}+\frac{1}{4}[{\phi}_i,{\phi}_j]^2-\frac{1}{2}(D_{\mu}\phi_i)(D^{\mu}{\phi}_i)~,~F_{\mu \nu}={\nabla}_{\mu}A_{\nu}-{\nabla}_{\nu}A_{\mu}-i[A_{\mu},A_{\nu}].\nonumber\\
\end{eqnarray}
We assume the ${\bf R}\times {\bf S}^3$ metric given by
\begin{eqnarray}
ds^2=h_{\mu \nu}dx^{\mu}dx^{\nu}&=&-dt^2+R^2(d{\theta}^2+{\rm sin}^2\theta d{\psi}^2 +{\rm sin}^2\theta~ {\rm sin}^2\psi ~d{\chi}^2)\nonumber\\
&=&+d{\tau}^2+R^2(d{\theta}^2+{\rm sin}^2\theta d{\psi}^2 +{\rm sin}^2\theta {\rm sin}^2\psi d{\chi}^2)~,~\tau=it.
\end{eqnarray}
${\bf R}\times {\bf S}^3$ is conformally flat because after the scaling $\tau=R\ln r$ the above metric becomes the flat metric on ${\bf R}^4$, viz
\begin{eqnarray}
\frac{e^{\frac{2\tau}{R}}}{R^2}ds^2
=d{r}^2+r^2(d{\theta}^2+{\rm sin}^2\theta d{\psi}^2 +{\rm sin}^2\theta {\rm sin}^2\psi d{\chi}^2)~,~\tau=it.
\end{eqnarray}
In $D$ dimension the conformally invariant Laplacian is ${\nabla}_M{\nabla}^M-\frac{d-2}{4(d-1)}{\cal R}$ where ${\cal R}$ is the Ricci scalar curvature. For ${\bf R}\times {\bf S}^3$ we have ${\cal R}=6/R^2$ and hence we replace the scalar quadratic term in the action as follows
\begin{eqnarray}
-\frac{1}{2}(D_{\mu}{\phi}_i)(D^{\mu}{\phi}_i)\longrightarrow -\frac{1}{2}(D_{\mu}{\phi}_i)(D^{\mu}{\phi}_i) -\frac{{\cal R}}{12}{\phi}_i^2.
\end{eqnarray} 
The ${\cal N}=1$ SYM action in $D=10$ is given by
\begin{eqnarray}
S=\frac{1}{g^2}\int d^4x\sqrt{h}Tr \bigg[(-\frac{1}{4}F_{MN}F^{MN})|_{d=10}+\frac{i}{2} (\bar{\Lambda}G^{M}D_{M}\Lambda)|_{d=10}\bigg]
\end{eqnarray}
The ${\cal N}=4$ SYM action in $D=4$ is given by (with ${\sigma}_0={\bf 1}_2$)
\begin{eqnarray}
S&=&\frac{1}{g^2}\int d^4x\sqrt{h}Tr\bigg[-\frac{1}{4}F_{\mu \nu}F^{\mu \nu}+\frac{1}{4}[{\phi}_i,{\phi}_j]^2-\frac{1}{2}(D_{\mu}\phi_i)(D^{\mu}{\phi}_i)-\frac{{\cal R}}{12}{\phi}_i^2\nonumber\\
&-&2is^{+A}{\sigma}_{\mu}D^{\mu}s^A+s^{+A}i{\sigma}_2({\rho}_i)_{AB}[{\phi}_i,(s^{+B})^T]-(s^{A})^Ti{\sigma}_2({\rho}_i^+)_{AB}[{\phi}_i,s^{B}]\bigg].\label{kaluza}
\end{eqnarray}
The supersymmetry transformations are given  by
\begin{eqnarray}
{\delta}A_0&=&-2i\bigg(s_{\alpha}^{+A}{\eta}_{\alpha}^A-{\eta}_{\alpha}^{+A}s_{\alpha}^A\bigg)\nonumber\\
{\delta}A_{a}&=&2i\bigg(s_{\alpha}^{+A}({\sigma}_a)_{\alpha \beta}{\eta}_{\beta}^A-{\eta}_{\alpha}^{+A}({\sigma}_a)_{\alpha \beta}s_{\beta}^A\bigg)\nonumber\\
%&&{\delta}{\phi}_i=-2i\bigg({s}_{\alpha}^{+A}{\rho}_i^{AB}(i{\sigma}_2{\eta}^{+B})_{\alpha}-{s}_{\alpha}^A({\rho}_i^+)^{AB}(i{\sigma}^2{\eta}^B)_{\alpha}\bigg).\label{sustr}\nonumber\\
{\delta}{\phi}_i&=&-2i\bigg({s}_{\alpha}^{+A}{\rho}_i^{AB}(i{\sigma}_2{\eta}^{+B})_{\alpha}+{s}_{\alpha}^A({\rho}_i^+)^{BA}(i{\sigma}^2{\eta}^B)_{\alpha}\bigg).\label{sustr}
\end{eqnarray}
Also
\begin{eqnarray}
{\delta}{s}^A_{\alpha}&=&\frac{i}{2}{\epsilon}_{abc}F^{ab}({\sigma}^c{\eta}^A)_{\alpha}-D_a{\phi}_i{\rho}_i^{AB}(i{\sigma}^a{\sigma}^2{\eta}^{+B})_{\alpha}-\frac{i}{2}[{\phi}_i,{\phi}_j]({\rho}_i{\rho}_j^+)^{AB}{\eta}_{\alpha}^B-\frac{1}{2}{\phi}_i{\rho}_i^{AB}(i{\sigma}^a{\sigma}^2{\nabla}_a{\eta}^{+B})_{\alpha}\nonumber\\
&+&D_aA_0({\sigma}^a{\eta}^A)_{\alpha}+[A_0,{\phi}_i]{\rho}_i^{AB}({\sigma}_2{\eta}^{+B})_{\alpha}\nonumber\\
&-&{\partial}_0A_a({\sigma}^a{\eta}^A)_{\alpha}+{\partial}_0{\phi}_i(i{\sigma}_2)_{\alpha\beta}{\rho}_i^{AB}{\eta}_{\beta}^{+B}+\frac{1}{2}{\phi}_i(i{\sigma}_2)_{\alpha \beta}{\rho}_i^{AB}{\partial}_0{\eta}_{\beta}^{+B}.
\label{sustrfermion}
\end{eqnarray}
The $4$th and $9$th terms will be absent in the case of flat space. In above the supersymmetry parameters ${\eta}_{\alpha}^A$ have the following dependence on time
\begin{eqnarray}
{\eta}_B\equiv {\eta}_B(t)=e^{-i\alpha t}{\eta}_B(0).
\end{eqnarray}
The supersymmetry parameters are $4$ Weyl spinors ${\eta}^A$ which satisfy one of the two conformal Killing spinors equations 
\begin{eqnarray}
{\nabla}_{\mu}{\eta}^{\pm}=\pm \frac{i}{2R}{\sigma}_{\mu}{\eta}^{\pm}. 
\end{eqnarray}
There are $4$ possible solutions and hence we have  ${\cal N}=4$ supersymmetry.

Next we expand the  fields $A_0$, $A_a$, ${s}_{\alpha}^A$, ${s}_{\alpha}^{+A}$ and ${\phi}_i$ and the supersymmetry  parameters ${\eta}_{\alpha}^{A}$, ${\eta}_{\alpha}^{+A}$ in terms of spherical harmonics on ${\bf S}^3$ and keep only the zero modes as follows
\begin{eqnarray}
&&{\phi}_i={\Phi}_i+...\nonumber\\
&&A_0=X_0+..\nonumber\\
&&A_a=\sum_{\hat{a}=1}^3X_{\hat{a}}V_{a}^{\hat{a}}+...\nonumber\\
&&{s}_{\alpha}^A=\sum_{\hat{\alpha}=1}^2{\theta}_{\hat{\alpha}}^AS_{\alpha}^{\hat{\alpha}+}+...\nonumber\\
&&{s}_{\alpha}^{+A}=\sum_{\hat{\alpha}=1}^2{\theta}_{\hat{\alpha}}^{+A}(S_{\alpha}^{\hat{\alpha}+})^++...\label{ansatz}
\end{eqnarray}
The supersymmetry  parameters are also expanded as
\begin{eqnarray}
&&{\eta}_{\alpha}^A=\sum_{\hat{\alpha}=1}^2{\epsilon}_{\hat{\alpha}}^AS_{\alpha}^{\hat{\alpha}+}+...\nonumber\\
&&{\eta}_{\alpha}^{+A}=\sum_{\hat{\alpha}=1}^2{\epsilon}_{\hat{\alpha}}^{+A}(S_{\alpha}^{\hat{\alpha}+})^++...
\end{eqnarray}
The  fields ${\phi}_i$, $X_0$ are scalar under the isometry group $SU(2)_L\times SU(2)_R$ of ${\bf R}\times {\bf S}^3$ so they transform as $(1,1)$ and hence the corresponding zero mode is the constant function. The  field ${s}_{\alpha}^A$ (for a given $A$) transforms as $(2,1)$ under  $SU(2)_L\times SU(2)_R$ and hence the two zero modes $S_{\alpha}^{\hat{\alpha}+}$ are the lowest spinor spherical harmonics of ${\bf S}^3$. They satisfy the Killing spinor equation
\begin{eqnarray}
{\nabla}_aS^{\hat{\alpha}+}= \frac{i}{2R}{\sigma}_{a}S^{\hat{\alpha}+}.
\end{eqnarray}
Similarly, the  field $A_a$ transforms as $(3,1)$ under  $SU(2)_L\times SU(2)_R$ and hence  the three zero modes $V_a^{\hat{a}+}$ are the lowest vector spherical harmonics of ${\bf S}^3$. They are given by
\begin{eqnarray}
(S^{\hat{\alpha}+})^{+}{\sigma}_aS^{\hat{\beta}+}=({\sigma}_{\hat{a}})^{\hat{\alpha}\hat{\beta}}V_a^{\hat{a}+}.
\end{eqnarray}
The zero modes $X_0$, $X_{\hat{a}}$, ${\Phi}_i$, ${\theta}_{\hat{\alpha}}^A$ and ${\theta}_{\hat{\alpha}}^{+A}$ are time-dependent matrices in the Lie algebra of the gauge group.

In terms of the zero modes $X_0$, $X_{\hat{a}}$, ${\Phi}_i$,  ${\theta}_{\hat{\alpha}}^A$, ${\theta}_{\hat{\alpha}}^{+A}$ and the parameters ${\epsilon}_{\hat{\alpha}}^A$, ${\epsilon}_{\hat{\alpha}}^{+A}$  the supersymmetry transformations (\ref{sustr}) take the form (by dropping also the hat on indices whenever is possible)
\begin{eqnarray}
&&{\delta}X_0=-2i\bigg({\theta}_{\alpha}^{+A}{\epsilon}^A_{\alpha}-{\epsilon}_{\alpha}^{+A}{\theta}^A_{\alpha}\bigg)\nonumber\\
&&{\delta}X_a=2i\bigg({\theta}_{\alpha}^{+A}({\sigma}_a{\epsilon}^A)_{\alpha}-{\epsilon}_{\alpha}^{+A}({\sigma}_a{\theta}^A)_{\alpha}\bigg)\nonumber\\
%&&{\delta}{\Phi}_i=-2i\bigg({\psi}_{\alpha}^{+A}{\rho}_i^{AB}(i{\sigma}_2{\epsilon}^{+B})_{\alpha}-{\psi}_{\alpha}^{A}({\rho}_i^+)^{AB}(i{\sigma}_2{\epsilon}^B)_{\alpha}\bigg).\nonumber\\
&&{\delta}{\Phi}_i=-2i\bigg({\theta}_{\alpha}^{+A}{\rho}_i^{AB}(i{\sigma}_2{\epsilon}^{+B})_{\alpha}+{\theta}_{\alpha}^{A}({\rho}_i^+)^{BA}(i{\sigma}_2{\epsilon}^B)_{\alpha}\bigg).\label{st1}
\end{eqnarray}
In above we have used the identities 
\begin{eqnarray}
&&(S_{\alpha}^{\hat{\alpha}})^+S_{\alpha}^{\hat{\beta}}={\delta}^{\hat{\alpha}\hat{\beta}}~,~(S_{\alpha}^{\hat{\alpha}})^+({\sigma}_a)_{\alpha \beta}S_{\beta}^{\hat{\beta}}=({\sigma}_{\hat{a}})^{\hat{\alpha}\hat{\beta}}V_a^{\hat{a}}\nonumber\\
&&S_{\alpha}^{\hat{\alpha}}(i{\sigma}^2)_{\alpha \beta}S_{\beta}^{\hat{\beta}}=(i{\sigma}^2)^{\hat{\alpha}\hat{\beta}}~,~S_{\alpha}^{\hat{\alpha}}(i{\sigma}^2{\sigma}^a)_{\alpha \beta}S_{\beta}^{\hat{\beta}}=(i{\sigma}^2{\sigma}^{\hat{a}})^{\hat{\alpha} \hat{\beta}}V^{\hat{a}a}.
\end{eqnarray}
By using other identities we can show that
\begin{eqnarray}
&&F_{ab}=-V_a^{\hat{a}}V_b^{\hat{b}}\tilde{F}_{\hat{a}\hat{b}}~,~\tilde{F}_{\hat{a}\hat{b}}=i[X_{\hat{a}},X_{\hat{b}}]-\frac{2}{R}{\epsilon}^{\hat{a}\hat{b}\hat{c}}X_{\hat{c}}\nonumber\\
&&D_a{\phi}_i=-i[X_{\hat{a}},{\Phi}_i]V_a^{\hat{a}}.
\end{eqnarray}
%For simplicity we will only compute the  supersymmetry transformations of the spinors ${\psi}_{\alpha}^A$ for time-independent fields. First we drop all dependence of fields on time and we set ${\partial}_0=0$. The supersymmetry transformations (\ref{sustrfermion}) become
%\begin{eqnarray}
%{\delta}{s}^A_{\alpha}&=&\frac{i}{2}{\epsilon}_{abc}F^{ab}({\sigma}^c{\eta}^A)_{\alpha}-D_a{\phi}_i{\rho}_i^{AB}(i{\sigma}^a{\sigma}^2{\eta}^{+B})_{\alpha}-\frac{i}{2}[{\phi}_i,{\phi}_j]({\rho}_i{\rho}_j^+)^{AB}{\eta}_{\alpha}^B-\frac{1}{2}{\phi}_i{\rho}_i^{AB}(i{\sigma}^a{\sigma}^2{\nabla}_a{\eta}^{+B})_{\alpha}\nonumber\\
%&+&D_aA_0({\sigma}^a{\eta}^A)_{\alpha}+[A_0,{\phi}_i]{\rho}_i^{AB}({\sigma}_2{\eta}^{+B})_{\alpha}+\frac{i\alpha}{2}{\phi}_i(i{\sigma}_2)_{\alpha \beta}{\rho}_i^{AB}{\eta}_{\beta}^{+B}.\label{sustr2}
%\end{eqnarray}
%We can immediately conclude that the supersymmetry transformations of the spinors ${\psi}_{\alpha}^A$ are given by
%\begin{eqnarray}
%{\delta}{\psi}_{\hat{\alpha}}^A&=&-\frac{i}{2}{\epsilon}^{\hat{a}\hat{b}\hat{c}}\tilde{F}_{\hat{a}\hat{b}}({\sigma}_{\hat{c}}{\epsilon}^A)_{\hat{\alpha}}-i[X_{\hat{a}},{\Phi}_i]{\rho}_i^{AB}\bigg(V_a^{\hat{a}}S_{\alpha}^{\hat{\alpha}}(i{\sigma}^2{\sigma}^a)_{\alpha \beta}S_{\beta}^{\hat{\beta}}\bigg)^+{\epsilon}_{\hat{\beta}}^{+B}+\frac{3}{4R}{\Phi}_i({\rho}_i)^{AB}({\sigma}^2)_{\hat{\alpha}\hat{\beta}}{\epsilon}_{\hat{\beta}}^{+B}\nonumber\\
%&-&\frac{i}{2}[{\Phi}_i,{\Phi}_j]({\rho}_i{\rho}_j^+)^{AB}{\epsilon}_{\hat{\alpha}}^B.
%\end{eqnarray}
Similarly,
\begin{eqnarray}
{\delta}{\theta}_{\hat{\alpha}}^A&=&-\frac{i}{2}{\epsilon}^{\hat{a}\hat{b}\hat{c}}\tilde{F}_{\hat{a}\hat{b}}({\sigma}_{\hat{c}}{\epsilon}^A)_{\hat{\alpha}}+i[X^{\hat{a}},{\Phi}_i]{\rho}_i^{AB}(i{\sigma}_{\hat{a}}{\sigma}_2)_{\hat{\alpha}\hat{\beta}}{\epsilon}_{\hat{\beta}}^{+B}+\frac{3}{4R}{\Phi}_i({\rho}_i)^{AB}({\sigma}_2)_{\hat{\alpha}\hat{\beta}}{\epsilon}_{\hat{\beta}}^{+B}\nonumber\\
&-&\frac{i}{2}[{\Phi}_i,{\Phi}_j]({\rho}_i{\rho}_j^+)^{AB}{\epsilon}_{\hat{\alpha}}^B-i[X^{\hat{a}},X_0]({\sigma}_{\hat{a}}{\epsilon}^A)^{\hat{\alpha}}+[X_0,{\Phi}_i]({\rho}_i)^{AB}({\sigma}_2)_{\hat{\alpha}\hat{\beta}}{\epsilon}^{+B}_{\hat{\beta}}\nonumber\\
&-&\frac{\alpha}{2}{\Phi}_i({\sigma}_2)_{\hat{\alpha} \hat{\beta}}{\rho}_i^{AB}{\epsilon}_{\hat{\beta}}^{+B}-{\partial}_0X^a({\sigma}_a{\epsilon}^A)_{\alpha}+i{\partial}_0{\Phi}_i({\sigma}_2)_{\alpha\beta}{\rho}_i^{AB}{\epsilon}_{\beta}^{+B}
.\label{st2}
%+\frac{1}{2}{\Phi}_i{\rho}_i^{AB}[X_0,({\sigma}_2)_{\alpha \beta}{\epsilon}^{+B}_{\beta}].
\end{eqnarray}
Also
\begin{eqnarray}
{\delta}{\theta}_{\hat{\alpha}}^{+A}&=&\frac{i}{2}{\epsilon}^{\hat{a}\hat{b}\hat{c}}\tilde{F}_{\hat{a}\hat{b}}({\epsilon}^{+A}{\sigma}_{\hat{c}})_{\hat{\alpha}}-i[X^{\hat{a}},{\Phi}_i]{\epsilon}_{\hat{\beta}}^{B}(i{\sigma}_2{\sigma}_{\hat{a}})_{\hat{\beta}\hat{\alpha}}({\rho}_i^+)^{BA}+\frac{3}{4R}{\Phi}_i{\epsilon}_{\hat{\beta}}^{B}({\sigma}_2)_{\hat{\beta}\hat{\alpha}}({\rho}_i^+)^{BA}\nonumber\\
&-&\frac{i}{2}[{\Phi}_i,{\Phi}_j]{\epsilon}_{\hat{\alpha}}^{+B}({\rho}_j{\rho}_i^+)^{BA}-i[X^{\hat{a}},X_0]({\epsilon}^{+A}{\sigma}_{\hat{a}})^{\hat{\alpha}}-[X_0,{\Phi}_i]{\epsilon}^{B}_{\hat{\beta}}({\sigma}_2)_{\hat{\beta}\hat{\alpha}}({\rho}_i^+)^{BA}\nonumber\\
&-&\frac{\alpha}{2}{\Phi}_i({\sigma}_2)_{\hat{\beta} \hat{\alpha}}({\rho}_i^+)^{BA}{\epsilon}_{\hat{\beta}}^{B}-{\partial}_0X^a({\epsilon}^{+A}{\sigma}_a)_{\alpha}-i{\partial}_0{\Phi}_i({\rho}_i^+)^{BA}{\epsilon}_{\beta}^{B}({\sigma}_2)_{\beta\alpha}.\label{st3}
%\nonumber\\
%&-&\frac{i\alpha}{2}{\Phi}_i(i{\sigma}_2)_{\alpha \beta}({\rho}_i^+)^{BA}{\epsilon}_{\beta}^{B}.
\end{eqnarray}
The equations of motion obtained by varying $A_{\mu}$ are
\begin{eqnarray}
D_{\mu}F^{\mu\nu}+i[{\phi}_i,D^{\nu}{\phi}_i]+2\{s_{\alpha}^{+A},({\sigma}^{\nu}s^A)_{\alpha}\}=0.
\end{eqnarray}
The other equations of motion are 
\begin{eqnarray}
&&D^{\mu}D_{\mu}\phi -\frac{1}{R^2}{\phi}_i+[{\phi}_j,[{\phi}_i,{\phi}_j]]-\{s_{\alpha}^{+A},{\rho}_i^{AB}(i{\sigma}_2(s^{+B})^T)_{\alpha}\}+\{s_{\alpha}^{A},({\rho}_i^+)^{AB}(i{\sigma}_2s^B)_{\alpha}\}=0\nonumber\\
&&i{\sigma}^{\mu}D_{\mu}s^A -i{\sigma}^2{\rho}_i^{AB}[{\phi}_i,(s^{+B})^T]=0.
\end{eqnarray}
Next we  insert (\ref{ansatz}) in the above equations of motion. The fields ${\phi}_i$, $A_0$, $A_a$, $s_{\alpha}^A$ and $s_{\alpha}^{+A}$ are found to solve these equations of motion provided that the zero modes ${\Phi}_i$, $X_0$, $X_a$, ${\theta}_{\alpha}^A$ and ${\theta}_{\alpha}^{+A}$ satisfy the equations of motion %\footnote{Exercise: derive these equations of motion.}
\begin{eqnarray}
&&[X_a,iD_0X_a]+[{\Phi}_i,iD_0{\Phi}_i]-2\{{\theta}^{+A}_{\alpha},{\theta}_{\alpha}^A\}=0\nonumber\\
&&D_0^2X_a+\frac{4}{R^2}X_a-\frac{6i}{R}{\epsilon}_{abc}X_bX_c-[{X}_b,[X_a,X_b]]-[{\Phi}_i,[X_a,{\Phi}_i]]-2\{{\theta}_{\alpha}^{+A},({\sigma}_a{\theta}^A)_{\alpha}\}=0\nonumber\\
&&D_0^2{\Phi}_i+\frac{1}{R^2}{\Phi}_i-[X_a,[{\Phi}_i,X_a]]-[{\Phi}_j,[{\Phi}_i,{\Phi}_j]]+\{{\theta}_{\alpha}^{+A},{\rho}_i^{AB}(i{\sigma}_2({\theta}^{+B})^T)_{\alpha}\}-\{{\theta}_{\alpha}^{A},({\rho}_i^+)^{AB}(i{\sigma}_2{\theta}^B)_{\alpha}\}=0\nonumber\\
&&iD_0{\theta}^A-\frac{3}{2R}{\theta}^A+[X_a,{\sigma}_a{\theta}^A]-i{\sigma}_2{\rho}_i^{AB}[{\Phi}_i,({\theta}^{+B})^T]=0.
\end{eqnarray}
As it turns out, these equations of motion can be derived from the following quantum mechanical model %\footnote{Exercise : verify this fact.}

\begin{eqnarray}
S&=&\int dt L~,~L=L_B+L_F+L_T.
\end{eqnarray}
\begin{eqnarray}
L_B&=&Tr\bigg[-\frac{1}{2}[X_0,X_a]^2-\frac{1}{2}[X_0,{\Phi}_i]^2+\frac{1}{4}[X_a,X_b]^2+\frac{1}{4}[{\Phi}_i,{\Phi}_j]^2+\frac{1}{2}[X_a,{\Phi}_i]^2\nonumber\\
&+&\frac{im}{3}{\epsilon}_{abc}X_aX_bX_c-\frac{m^2}{18}X_a^2-\frac{m^2}{72}{\Phi}_i^2\bigg]\nonumber\\
L_F&=&Tr\bigg[-2{\theta}^+[X_0,\theta]-2{\theta}^+\bigg({\sigma}_a[X_a,\theta]-\frac{m}{4}\theta\bigg)+{\theta}^+i{\sigma}_2{\rho}_i[{\Phi}_i,({\theta}^+)^T]-{\theta}^Ti{\sigma}_2{\rho}_i^+[{\Phi}_i,\theta]\bigg]\nonumber\\
L_T&=&Tr\bigg[\frac{1}{2}({\partial}_0X_a)^2+\frac{1}{2}({\partial}_0{\Phi}_i)^2-2i {\theta}^+{\partial}_0{\theta}-i{\partial}_0X_a[X_0,X_a]-i{\partial}_0{\Phi}_i[X_0,{\Phi}_i]\bigg].\label{bmn2}
\end{eqnarray}
This is exactly the BMN model (\ref{bmn}). In above 
\begin{eqnarray}
m=\frac{6}{R}. 
\end{eqnarray}
Before we conclude this section we define the Clebsch-Gordan coefficients ${\rho}_i^{AB}$ and $({\rho}_i^{+})^{AB}$. Let ${\Gamma}_{i}$, $i=4,5,6,7,8,9$ be the Clifford algebra in six dimensions, viz $\{{\Gamma}_i,{\Gamma}_j\}=2{\delta}_{ij}$. We also denote them by $\hat{\Gamma}_{a}={\Gamma}_{a+3}$, $\hat{\Gamma}_{a+3}={\Gamma}_{a+6}$, $a=1,2,3$. 
We work in the representation

\begin{eqnarray}
&&\hat{\Gamma}^{a}=(\hat{\Gamma}^{a})^+=\left(
\begin{array}{cc}
0 &\hat{\rho}^{a}\\
(\hat{\rho}^{a})^+&0
\end{array}
\right)~,~\hat{\Gamma}^{a+3}=(\hat{\Gamma}^{a+3})^+=\left(
\begin{array}{cc}
0 &\hat{\rho}^{a+3}\\
(\hat{\rho}^{a+3})^+&0
\end{array}
\right).
\end{eqnarray}
We will also introduce 
\begin{eqnarray}
&&{\Gamma}^{AB}=({\Gamma}^{AB})^+=-{\Gamma}^{BA}=\left(
\begin{array}{cc}
0 &{\gamma}^{AB}\\
({\gamma}^{AB})^+&0
\end{array}
\right)~,~A,B=1,2,3,4\nonumber\\
&&({\Gamma}^{AB})_{CD}={\delta}_{AC}{\delta}_{BD}-{\delta}_{AD}{\delta}_{BC}.
\end{eqnarray}
We define the gamma matrices $\hat{\Gamma}_{a}$ and $\hat{\Gamma}_{a+3}$  as follows
\begin{eqnarray}
\hat{\Gamma}_a=\frac{1}{2}{\epsilon}_{abc4}{\Gamma}_{bc}+{\Gamma}_{a4}~,~\hat{\Gamma}_{a+3}=\frac{i}{2}{\epsilon}_{abc4}{\Gamma}_{bc}-i{\Gamma}_{a4}~,~a=1,2,3.
\end{eqnarray}
We find immediately
\begin{eqnarray}
\hat{\rho}_a=-\hat{\rho}_a^+=\frac{1}{2}{\epsilon}_{abc4}{\gamma}_{bc}+{\gamma}_{a4}~,~\hat{\rho}_{a+3}=\hat{\rho}_{a+3}^+=\frac{i}{2}{\epsilon}_{abc4}{\gamma}_{bc}-i{\gamma}_{a4}~,~a=1,2,3.
\end{eqnarray}
Explicitly
\begin{eqnarray}
&&\hat{\rho}_1={\gamma}_{23}+{\gamma}_{14}=i{\sigma}_2\otimes {\sigma}_1\nonumber\\
&&\hat{\rho}_2=-{\gamma}_{13}+{\gamma}_{24}=-i{\sigma}_2\otimes {\sigma}_3\nonumber\\
&&\hat{\rho}_3={\gamma}_{12}+{\gamma}_{34}={\bf 1}\otimes i{\sigma}_2.
\end{eqnarray}

\begin{eqnarray}
&&\hat{\rho}_4=i({\gamma}_{23}-{\gamma}_{14})={\sigma}_1\otimes {\sigma}_2\nonumber\\
&&\hat{\rho}_5=i(-{\gamma}_{13}-{\gamma}_{24})={\sigma}_2\otimes {\bf 1}\nonumber\\
&&\hat{\rho}_6=i({\gamma}_{12}-{\gamma}_{34})=-{\sigma}_3\otimes {\sigma}_2.
\end{eqnarray}
We verify that these matrices satisfy ${\rho}_i^+{\rho}_j+{\rho}_j^+{\rho}_i={\rho}_i{\rho}_j^++{\rho}_j{\rho}_i^+=2{\delta}_{ij}{\bf 1}_4$. We also compute the identities
\begin{eqnarray}
&&\sum_{a=1}^6\hat{\rho}_a^{AB}\hat{\rho}_a^{CD}=\sum_{a=1}^6(\hat{\rho}_a^+)^{AB}(\hat{\rho}_a^+)^{CD}=2{\epsilon}_{ABCD}\nonumber\\
&&\sum_{a=1}^6\hat{\rho}_a^{AB}(\hat{\rho}_a^+)^{CD}=-{\gamma}_{EF}^{AB}{\gamma}_{EF}^{CD}=-2({\delta}_{AC}{\delta}_{BD}-{\delta}_{AD}{\delta}_{BC}).
\end{eqnarray}
\subsection{Dimensional reduction}
\subsubsection{Supersymmetry at fixed times}
In this section we go even further and turn compactification into dimensional reduction by dropping the time dependence of the matrices $X_0$,$X_a$,${\Phi}_i$ and ${\theta}_{\alpha}^A$, ${\theta}_{\alpha}^{+A}$. 

The action $S$ given by equation (\ref{bmn2}) remains supersymmetric under (\ref{st1}), (\ref{st2}), (\ref{st3}) provided we vary the fermion term  $-2i Tr{\theta}^+{\partial}_0{\theta}$ first  and then take into account that the supersymmetry parameters satisfy ${\partial}_0{\epsilon}_{\alpha}^A=-i\alpha {\epsilon}_{\alpha}^A$ and ${\partial}_0{\epsilon}_{\alpha}^{+A}=i\alpha {\epsilon}_{\alpha}^{+A}$ before we fix time. We use the notation $\theta=\psi$ and we  recall that the curvature is given by (with $v=2$)
\begin{eqnarray}
\tilde{F}_{ab}=i[X_a,X_b]-\frac{v}{R}{\epsilon}_{abc}X_c.
\end{eqnarray}
Right from the start we will drop the time dependence of the matrices $X_a$ and ${\Phi}_i$. Thus the terms containing time derivatives of $X_a$ and ${\Phi}_i$ in the supersymmetric transformations (\ref{st2}), (\ref{st3}) will be absent. We compute immediately the following variation of the fermion action

\begin{eqnarray}
{\delta}\bigg(-2Tr{\psi}_{\alpha}^{+A}({\sigma}_a)_{\alpha \beta}[X_a,{\psi}_{\beta}^A]\bigg)&=&2({\sigma}_a)_{\alpha \beta}Tr{\delta}X_a\{{\psi}_{\alpha}^{+A},{\psi}_{\beta}^A\}-iTr{\delta}X_b[X_a,\tilde{F}_{ab}]+Tr{\delta}{\Phi}_i[X_a,[X_a,{\Phi}_i]]\nonumber\\
&-&Tr\bigg(({\epsilon}^B)^T{\sigma}_2{\sigma}_c{\psi}^A({\rho}_i^+)^{BA}-{\psi}^{+A}{\sigma}_c{\sigma}_2({\epsilon}^{+B})^T{\rho}_i^{AB}\bigg)[{\epsilon}_{abc}\tilde{F}_{ab},{\Phi}_i]\nonumber\\
&-&Tr\bigg(({\epsilon}^B)^T{\sigma}_2{\sigma}_c{\psi}^A({\rho}_i^+)^{BA}-{\psi}^{+A}{\sigma}_c{\sigma}_2({\epsilon}^{+B})^T{\rho}_i^{AB}\bigg)[\frac{4v-3}{2R}X_c,{\Phi}_i]\nonumber\\
&-&iTr\bigg({\epsilon}^{+A}{\sigma}_a{\psi}^B+{\psi}^{+A}{\sigma}_a{\epsilon}^B\bigg)({\rho}_j{\rho}_i^+)^{AB}[X_a,[{\Phi}_i,{\Phi}_j]]\nonumber\\
%&-&2{\epsilon}_{abc}Tr\bigg({\epsilon}^{+A}{\sigma}_c{\psi}^A+{\psi}^{+A}{\sigma}_c{\epsilon}^A\bigg)[X_a,[X_b,X_0]]\nonumber\\
&-&Tr{\delta}X_0[X_a,[X_a,X_0]]\nonumber\\
&+&{\epsilon}_{abc}Tr\bigg({\epsilon}^{+A}{\sigma}_c{\psi}^A+{\psi}^{+A}{\sigma}_c{\epsilon}^A\bigg)[X_0,[X_a,X_b]]\nonumber\\
&-&2Tr\bigg(({\epsilon}^B)^T{\sigma}_2{\sigma}_a{\psi}^A({\rho}_i^+)^{BA}+{\psi}^{+A}{\sigma}_a{\sigma}_2({\epsilon}^{+B})^T{\rho}_i^{AB}\bigg)[X_a,[X_0,{\Phi}_i]]\nonumber\\
&-&\alpha Tr\bigg(({\epsilon}^B)^T{\sigma}_2{\sigma}_c{\psi}^A({\rho}_i^+)^{BA}-{\psi}^{+A}{\sigma}_c{\sigma}_2({\epsilon}^{+B})^T{\rho}_i^{AB}\bigg)[X_c,{\Phi}_i].
\label{vdf}\nonumber\\
\end{eqnarray}
Let $
{\delta}{\psi}_{{\alpha}}^A=-\frac{i}{2}{\epsilon}_{{a}{b}{c}}\tilde{F}_{{a}{b}}({\sigma}_{{c}}{\epsilon}^A)_{{\alpha}}$ then 
\begin{eqnarray}
-Tr{\delta}{\psi}_{\alpha}^A(i{\sigma}_2)_{\alpha \beta}({\rho}_i^+)^{AB}[{\Phi}_i,{\psi}_{\beta}^B]-Tr{\psi}_{\alpha}^A(i{\sigma}_2)_{\alpha \beta}({\rho}_i^+)^{AB}[{\Phi}_i,{\delta}{\psi}_{\beta}^B]+{\rm C.C}&=&\nonumber\\
\frac{1}{2}Tr\bigg[({\epsilon}^B)^T{\sigma}_2{\sigma}_c{\psi}^A\bigg(({\rho}_i^+)^{BA}-({\rho}_i^+)^{AB}\bigg)-{\psi}^{+A}{\sigma}_c{\sigma}_2({\epsilon}^{+B})^T\bigg({\rho}_i^{AB}-{\rho}_i^{BA}\bigg)\bigg][{\epsilon}_{abc}\tilde{F}_{ab},{\Phi}_i].\nonumber\\
\end{eqnarray}
If we assume that
\begin{eqnarray}
{\rho}_i^{AB}=-{\rho}_i^{BA}.\label{vdf1}
\end{eqnarray}
Then this variation will cancel the first term in the second line of (\ref{vdf}). 

Let ${\delta}{\psi}_{{\alpha}}^A=i[X_{{a}},{\Phi}_i]{\rho}_i^{AB}(i{\sigma}_{{a}}{\sigma}_2)_{{\alpha}{\beta}}{\epsilon}_{{\beta}}^{+B}$ then
\begin{eqnarray}
-Tr{\delta}{\psi}_{\alpha}^A(i{\sigma}_2)_{\alpha \beta}({\rho}_i^+)^{AB}[{\Phi}_i,{\psi}_{\beta}^B]-Tr{\psi}_{\alpha}^A(i{\sigma}_2)_{\alpha \beta}({\rho}_i^+)^{AB}[{\Phi}_i,{\delta}{\psi}_{\beta}^B]+{\rm C.C}&=&\nonumber\\
iTr\bigg[({\rho}_j^T{\rho}_i^*-{\rho}_j^T{\rho}_i^+âº)^{AB}{\epsilon}^{+A}{\sigma}_a{\psi}^B-({\rho}_i^T{\rho}_j^*-{\rho}_i{\rho}_j^*âº)^{AB}{\psi}^{+A}{\sigma}_a{\epsilon}^B\bigg][{\Phi}_i,[X_a,{\Phi}_j]]&=&\nonumber\\
iTr\bigg({\epsilon}^{+A}{\sigma}_a{\psi}^B+{\psi}^{+A}{\sigma}_a{\epsilon}^B\bigg)({\rho}_j{\rho}_i^+)^{AB}[X_a,[{\Phi}_i,{\Phi}_j]]-Tr{\delta}X_a[{\Phi}_i,[X_a,{\Phi}_i]].
\end{eqnarray}
The first term in this variation cancels the last line of (\ref{vdf}). In here we have used ${\rho}_i^T=-{\rho}_i$ and ${\rho}_i^+=-{\rho}_i^*$ which follow from (\ref{vdf1}), the Jacobi identity and the defining equation of the six $4\times 4$  matrices ${\rho}_i$ given by ${\rho}_i{\rho}_j^++{\rho}_j{\rho}_i^+={\rho}_i^+{\rho}_j+{\rho}_j^+{\rho}_i=2{\delta}_{ij}{\bf 1}_4$.

Let ${\delta}{\psi}_{{\alpha}}^A=\frac{3}{4R}{\Phi}_i({\rho}_i)^{AB}({\sigma}_2)_{{\alpha}{\beta}}{\epsilon}_{{\beta}}^{+B}$ then 
\begin{eqnarray}
-Tr{\delta}{\psi}_{\alpha}^A(i{\sigma}_2)_{\alpha \beta}({\rho}_i^+)^{AB}[{\Phi}_i,{\psi}_{\beta}^B]-Tr{\psi}_{\alpha}^A(i{\sigma}_2)_{\alpha \beta}({\rho}_i^+)^{AB}[{\Phi}_i,{\delta}{\psi}_{\beta}^B]+{\rm C.C}&=&\nonumber\\
\frac{3i}{4R}Tr\bigg[\big({\rho}_j^T{\rho}_i^*-{\rho}_j^T{\rho}_i^+\big)^{AB}{\epsilon}^{+A}{\psi}^B+
\big({\rho}_i^T{\rho}_j^*-{\rho}_i{\rho}_j^*\big)^{AB}{\psi}^{+A}{\epsilon}^B\bigg][{\Phi}_i,{\Phi}_j]&=&\nonumber\\
-\frac{3i}{2R}({\rho}_i{\rho}_j^+)^{AB}Tr\bigg({\epsilon}^{+A}{\psi}^B-{\psi}^{+A}{\epsilon}^B\bigg)[{\Phi}_i,{\Phi}_j].
\end{eqnarray}
Let  ${\delta}{\psi}_{{\alpha}}^A=-\frac{i}{2}[{\Phi}_i,{\Phi}_j]({\rho}_i{\rho}_j^+)^{AB}{\epsilon}_{{\alpha}}^B$ then 
\begin{eqnarray}
-Tr{\delta}{\psi}_{\alpha}^A(i{\sigma}_2)_{\alpha \beta}({\rho}_i^+)^{AB}[{\Phi}_i,{\psi}_{\beta}^B]-Tr{\psi}_{\alpha}^A(i{\sigma}_2)_{\alpha \beta}({\rho}_i^+)^{AB}[{\Phi}_i,{\delta}{\psi}_{\beta}^B]+{\rm C.C}&=&\nonumber\\
-\frac{i}{2}Tr\bigg[\bigg(({\rho}_k^*-{\rho}_k^+){\rho}_i{\rho}_j^+\bigg)^{AB}{\psi}^Ai{\sigma}_2{\epsilon}^B+\bigg(({\rho}_k-{\rho}_k^T){\rho}_i^*{\rho}_j^T\bigg)^{AB}{\psi}^{+A}i{\sigma}_2{\epsilon}^{+B}\bigg][{\Phi}_k,[{\Phi}_i,{\Phi}_j]]&=&\nonumber\\
Tr{\delta}{\Phi}_j[{\Phi}_i,[{\Phi}_i,{\Phi}_j]].\nonumber\\
\end{eqnarray}
The first term of (\ref{vdf}) can be put in the form
\begin{eqnarray}
2({\sigma}_a)_{\alpha \beta}Tr{\delta}X_a\{{\psi}_{\alpha}^{+A},{\psi}_{\beta}^A\}&=&8iTr\{{\psi}_{\alpha}^A,{\psi}_{\beta}^{+A}\}({\psi}_{\alpha}^{+B}{\epsilon}_{\beta}^B-{\epsilon}_{\alpha}^{+B}{\psi}_{\beta}^B)\nonumber\\
&-&4iTr\{{\psi}_{\alpha}^A,{\psi}_{\alpha}^{+A}\}({\psi}_{\beta}^{+B}{\epsilon}_{\beta}^B-{\epsilon}_{\beta}^{+B}{\psi}_{\beta}^B).
\end{eqnarray}
We also compute
\begin{eqnarray}
-Tr{\psi}_{\alpha}^A(i{\sigma}_2)_{\alpha \beta}({\rho}_i^+)^{AB}[{\delta}{\Phi}_i,{\psi}_{\beta}^B]+{\rm C.C}&=&Tr\{{\psi}_{\alpha}^A,{\psi}_{\beta}^{B}\}(i{\sigma}_2)_{\alpha \beta}({\rho}_i^+)^{AB}{\delta}{\Phi}_i+{\rm C.C}\nonumber\\
&=&8iTr\{{\psi}_1^A,{\psi}_1^{+A}\}({\psi}_2^{+B}{\epsilon}_2^B-{\epsilon}_2^{+B}{\psi}_2^B)\nonumber\\
&+&8iTr\{{\psi}_2^A,{\psi}_2^{+A}\}({\psi}_1^{+B}{\epsilon}_1^B-{\epsilon}_1^{+B}{\psi}_1^B)\nonumber\\
&-&8iTr\{{\psi}_1^A,{\psi}_2^{+A}\}({\psi}_1^{+B}{\epsilon}_2^B-{\epsilon}_1^{+B}{\psi}_2^B)\nonumber\\
&-&8iTr\{{\psi}_2^A,{\psi}_1^{+A}\}({\psi}_2^{+B}{\epsilon}_1^B-{\epsilon}_2^{+B}{\psi}_1^B).
\end{eqnarray}
In above we have used the identity
\begin{eqnarray}
{\epsilon}^{ABCD}(i{\sigma}_2)_{\alpha \beta}(i{\sigma}_2)_{\mu \nu}Tr{\psi}_{\alpha}^A{\psi}_{\beta}^B{\psi}_{\mu}^{C}{\epsilon}_{\nu}^D=0.
\end{eqnarray}
Thus we get
\begin{eqnarray}
2({\sigma}_a)_{\alpha \beta}Tr{\delta}X_a\{{\psi}_{\alpha}^{+A},{\psi}_{\beta}^A\}+\bigg[-Tr{\psi}_{\alpha}^A(i{\sigma}_2)_{\alpha \beta}({\rho}_i^+)^{AB}[{\delta}{\Phi}_i,{\psi}_{\beta}^B]+{\rm C.C}\bigg]&=&\nonumber\\
4iTr\{{\psi}_{\alpha}^A,{\psi}_{\alpha}^{+A}\}({\psi}_{\beta}^{+B}{\epsilon}_{\beta}^B-{\epsilon}_{\beta}^{+B}{\psi}_{\beta}^B)=-2Tr\{{\psi}_{\alpha}^A,{\psi}_{\alpha}^{+A}\}{\delta}X_0.
\end{eqnarray}
Let ${\delta}{\psi}_{\alpha}^A=-i[X_a,X_0]({\sigma}_a{\epsilon}^A)_{\alpha}$ then 
\begin{eqnarray}
-Tr{\delta}{\psi}_{\alpha}^A(i{\sigma}_2)_{\alpha \beta}({\rho}_i^+)^{AB}[{\Phi}_i,{\psi}_{\beta}^B]-Tr{\psi}_{\alpha}^A(i{\sigma}_2)_{\alpha \beta}({\rho}_i^+)^{AB}[{\Phi}_i,{\delta}{\psi}_{\beta}^B]+{\rm C.C}&=&\nonumber\\
-2Tr\bigg(({\epsilon}^B)^T{\sigma}_2{\sigma}_a{\psi}^A({\rho}_i^+)^{BA}+{\psi}^{+A}{\sigma}_a{\sigma}_2({\epsilon}^{+B})^T{\rho}_i^{AB}\bigg)[{\Phi}_i,[X_a,X_0]].
\end{eqnarray}
Let ${\delta}{\psi}_{\alpha}^A=[X_0,{\Phi}_i]{\rho}_i^{AB}({\sigma}_2)_{\alpha \beta}{\epsilon}^{+B}_{\beta}$ then 
\begin{eqnarray}
-Tr{\delta}{\psi}_{\alpha}^A(i{\sigma}_2)_{\alpha \beta}({\rho}_i^+)^{AB}[{\Phi}_i,{\psi}_{\beta}^B]-Tr{\psi}_{\alpha}^A(i{\sigma}_2)_{\alpha \beta}({\rho}_i^+)^{AB}[{\Phi}_i,{\delta}{\psi}_{\beta}^B]+{\rm C.C}&=&\nonumber\\
+i({\rho}_j{\rho}_i^+)^{AB}Tr\bigg({\psi}^{+A}{\epsilon}^B+{\epsilon}^{+A}{\psi}^B\bigg)[X_0,[{\Phi}_i,{\Phi}_j]]&&\nonumber\\
+Tr{\delta}X_0[{\Phi}_i,[X_0,{\Phi}_i]].
\end{eqnarray}
Let ${\delta}{\psi}_{\alpha}^A=\frac{i\alpha}{2}{\Phi}_i(i{\sigma}_2)_{\alpha \beta}{\rho}_i^{AB}{\epsilon}^{+B}_{\beta}$ then
\begin{eqnarray}
-Tr{\delta}{\psi}_{\alpha}^A(i{\sigma}_2)_{\alpha \beta}({\rho}_i^+)^{AB}[{\Phi}_i,{\psi}_{\beta}^B]-Tr{\psi}_{\alpha}^A(i{\sigma}_2)_{\alpha \beta}({\rho}_i^+)^{AB}[{\Phi}_i,{\delta}{\psi}_{\beta}^B]+{\rm C.C}&=&\nonumber\\
i\alpha ({\rho}_i{\rho}_j^+)^{AB}Tr\bigg({\epsilon}^{+A}{\psi}^B-{\psi}^{+A}{\epsilon}^B\bigg)[{\Phi}_i,{\Phi}_j].
\end{eqnarray}
Next we compute (with $3/R_1=m/2$)
\begin{eqnarray}
\frac{3}{R_1}{\delta}\bigg(Tr{\psi}^+{\psi}\bigg)&=&-\frac{3}{4R_1}{\epsilon}_{abc}Tr\tilde{F}_{ab}{\delta}X_c+\frac{3}{4R_1}(\frac{3}{2R}-\alpha)Tr{\Phi}_i{\delta}{\Phi}_i\nonumber\\
%+(\frac{9}{8R_1R}-\frac{3\alpha}{4R_1})Tr{\Phi}_i{\delta}{\Phi}_i\nonumber\\
&+&\frac{3}{R_1}Tr\bigg(({\epsilon}^B)^T{\sigma}_2{\sigma}_c{\psi}^A({\rho}_i^+)^{BA}-{\psi}^{+A}{\sigma}_c{\sigma}_2({\epsilon}^{+B})^T{\rho}_i^{AB}\bigg)[X_c,{\Phi}_i]\nonumber\\
&+&\frac{3i}{2R_1}({\rho}_i{\rho}_j^+)^{AB}Tr\bigg({\epsilon}^{+A}{\psi}^B-{\psi}^{+A}{\epsilon}^B\bigg)[{\Phi}_i,{\Phi}_j]\nonumber\\
&-&\frac{3i}{R_1}Tr\bigg({\epsilon}^{+A}{\sigma}_a{\psi}^A+{\psi}^{+A}{\sigma}_a{\epsilon}^A\bigg)[X_a,X_0]\nonumber\\
&+&\frac{3}{R_1}Tr\bigg({\psi}^{+A}{\sigma}_2({\epsilon}^{+B})^T{\rho}_i^{AB}+({\psi}^{A})^T{\sigma}_2{\epsilon}^B({\rho}_i^+)^{AB}\bigg)[X_0,{\Phi}_i].
\end{eqnarray}
Also
\begin{eqnarray}
{\delta}\bigg(-2Tr{\psi}_{\alpha}^{+A}[X_0,{\psi}_{\alpha}^A]\bigg)&=&i{\epsilon}_{abc}Tr\bigg({\epsilon}^{+A}{\sigma}_a{\psi}^A+{\psi}^{+A}{\sigma}_a{\epsilon}^A\bigg)[X_0,\tilde{F}_{ab}]+Tr{\delta}X_a[X_0,[X_a,X_0]]\nonumber\\
&-&i({\rho}_j{\rho}_i^+)^{AB}Tr\bigg({\psi}^{+A}{\epsilon}^B+{\epsilon}^{+A}{\psi}^B\bigg)[X_0,[{\Phi}_i,{\Phi}_j]]\nonumber\\
&-&\frac{3}{2R}Tr\bigg({\psi}^{+A}{\sigma}_2({\epsilon}^{+B})^T{\rho}_i^{AB}+({\psi}^{A})^T{\sigma}_2{\epsilon}^B({\rho}_i^+)^{AB}\bigg)[X_0,{\Phi}_i]\nonumber\\
&-&2Tr\bigg(({\epsilon}^B)^T{\sigma}_2{\sigma}_a{\psi}^A({\rho}_i^+)^{BA}+{\psi}^{+A}{\sigma}_a{\sigma}_2({\epsilon}^{+B})^T{\rho}_i^{AB}\bigg)[X_0,[{\Phi}_i,X_a]]\nonumber\\
&-&Tr{\delta}{\Phi}_i[X_0,[X_0,{\Phi}_i]]\nonumber\\
&+&\alpha Tr\bigg({\psi}^{+A}{\sigma}_2({\epsilon}^{+B})^T{\rho}_i^{AB}+({\psi}^{A})^T{\sigma}_2{\epsilon}^B({\rho}_i^+)^{AB}\bigg)[X_0,{\Phi}_i].
\end{eqnarray}
The full variation of the fermion action is then given by
\begin{eqnarray}
{\delta}S_F&=&-iTr{\delta}X_b[X_a,\tilde{F}_{ab}]+Tr{\delta}{\Phi}_i[X_a,[X_a,{\Phi}_i]]-Tr{\delta}X_a[{\Phi}_i,[X_a,{\Phi}_i]]-\frac{3}{4R_1}{\epsilon}_{abc}TrF_{ab}{\delta}X_c\nonumber\\
&+&\frac{3}{4R_1}(\frac{3}{2R}-\alpha)Tr{\Phi}_i{\delta}{\Phi}_i+Tr{\delta}{\Phi}_j[{\Phi}_i,[{\Phi}_i,{\Phi}_j]]+Tr{\delta}X_0[{\Phi}_i,[X_0,{\Phi}_i]]+Tr{\delta}X_a[X_0,[X_a,X_0]]\nonumber\\
&-&Tr{\delta}X_0[X_a,[X_a,X_0]]-Tr{\delta}{\Phi}_i[X_0,[X_0,{\Phi}_i]]\nonumber\\
&+&\bigg(-\frac{4v-3}{2R}-\alpha +\frac{3}{R_1}\bigg)Tr\bigg(({\epsilon}^B)^T{\sigma}_2{\sigma}_c{\psi}^A({\rho}_i^+)^{BA}-{\psi}^{+A}{\sigma}_c{\sigma}_2({\epsilon}^{+B})^T{\rho}_i^{AB}\bigg)[X_c,{\Phi}_i]\nonumber\\
&+&i\bigg(-\frac{3}{2R}+\frac{3}{2R_1}+\alpha\bigg) ({\rho}_i{\rho}_j^+)^{AB}Tr\bigg({\epsilon}^{+A}{\psi}^B-{\psi}^{+A}{\epsilon}^B\bigg)[{\Phi}_i,{\Phi}_j]\nonumber\\
&+&i\bigg(-\frac{3}{R_1}+\frac{2v}{R})Tr\bigg({\epsilon}^{+A}{\sigma}_a{\psi}^A+{\psi}^{+A}{\sigma}_a{\epsilon}^A\bigg)[X_a,X_0]\nonumber\\
&+&(\frac{3}{R_1}-\frac{3}{2R}+\alpha) Tr\bigg({\psi}^{+A}{\sigma}_2({\epsilon}^{+B})^T{\rho}_i^{AB}+({\psi}^{A})^T{\sigma}_2{\epsilon}^B({\rho}_i^+)^{AB}\bigg)[X_0,{\Phi}_i].
\end{eqnarray}
Let us now consider the Lagrangian
\begin{eqnarray}
&&L_T=Tr\bigg[\frac{1}{2}({\partial}_0X_a)^2+\frac{1}{2}({\partial}_0{\Phi}_i)^2-2i {\psi}^+{\partial}_0{\psi}-i{\partial}_0X_a[X_a,X_a]-i{\partial}_0{\Phi}_i[X_0,{\Phi}_i]\bigg].
\end{eqnarray}
Since the time dependence of the matrices $X_a$ and ${\Phi}_i$ is already dropped we have
\begin{eqnarray}
L_T=-2i Tr{\psi}^+{\partial}_0{\psi}.
\end{eqnarray}
Varying this action under full supersymmetry transformations then fixing time we obtain the variation
\begin{eqnarray}
-2i Tr\bigg[{\psi}_{\alpha}^{+A}({\partial}_0{\delta}{\psi}_{\alpha}^A)|_{t={\rm fixed}}-{\rm H.C}\bigg].
\end{eqnarray}
Since ${\partial}_0{\epsilon}_{\alpha}^A=-i{\alpha}{\epsilon}_{\alpha}^A$ and ${\partial}_0{\epsilon}_{\alpha}^{+A}=i{\alpha}{\epsilon}_{\alpha}^{+A}$ we have  
\begin{eqnarray}
({\partial}_0{\delta}{\psi}_{\hat{\alpha}}^A)|_{t={\rm fixed}}&=&i\alpha\bigg[\frac{i}{2}{\epsilon}^{\hat{a}\hat{b}\hat{c}}\tilde{F}_{\hat{a}\hat{b}}({\sigma}_{\hat{c}}{\epsilon}^A)_{\hat{\alpha}}+i[X_{\hat{a}},{\Phi}_i]{\rho}_i^{AB}(i{\sigma}^{\hat{a}}{\sigma}^2)_{\hat{\alpha}\hat{\beta}}{\epsilon}_{\hat{\beta}}^{+B}+\frac{3}{4R}{\Phi}_i({\rho}_i)^{AB}({\sigma}^2)_{\hat{\alpha}\hat{\beta}}{\epsilon}_{\hat{\beta}}^{+B}\nonumber\\
&+&\frac{i}{2}[{\Phi}_i,{\Phi}_j]({\rho}_i{\rho}_j^+)^{AB}{\epsilon}_{\hat{\alpha}}^B+i[X_{\hat{a}},X_0]({\sigma}_{\hat{a}}{\epsilon}^A)^{\hat{\alpha}}+[X_0,{\Phi}_i]({\rho}_i)^{AB}({\sigma}_2)_{\hat{\alpha}\hat{\beta}}{\epsilon}^{+B}_{\hat{\beta}}\nonumber\\
&-&\frac{\alpha}{2}{\Phi}_i({\sigma}_2)_{\alpha \beta}{\rho}_i^{AB}{\epsilon}_{\beta}^{+B}\bigg].
%+\frac{1}{2}{\Phi}_i{\rho}_i^{AB}[X_0,({\sigma}_2)_{\alpha \beta}{\epsilon}^{+B}_{\beta}].
\end{eqnarray}
We compute
\begin{eqnarray}
-2iTr\bigg[{\psi}_{\alpha}^{+A}({\partial}_0{\delta}{\psi}_{\alpha}^A)|_{t={\rm fixed}}-{\rm H.C}\bigg]&=&\frac{\alpha}{2}{\epsilon}_{abc}Tr\tilde{F}_{ab}{\delta}X_c+\alpha(\frac{3}{4R}-\frac{\alpha}{2})Tr{\Phi}_i{\delta}{\Phi}_i\nonumber\\
%+(\frac{9}{8R_1R}-\frac{3\alpha}{4R_1})Tr{\Phi}_i{\delta}{\Phi}_i\nonumber\\
&+&2\alpha Tr\bigg(({\epsilon}^B)^T{\sigma}_2{\sigma}_c{\psi}^A({\rho}_i^+)^{BA}-{\psi}^{+A}{\sigma}_c{\sigma}_2({\epsilon}^{+B})^T{\rho}_i^{AB}\bigg)[X_c,{\Phi}_i]\nonumber\\
&-&i\alpha ({\rho}_i{\rho}_j^+)^{AB}Tr\bigg({\epsilon}^{+A}{\psi}^B-{\psi}^{+A}{\epsilon}^B\bigg)[{\Phi}_i,{\Phi}_j]\nonumber\\
&+&2i\alpha Tr\bigg({\epsilon}^{+A}{\sigma}_a{\psi}^A+{\psi}^{+A}{\sigma}_a{\epsilon}^A\bigg)[X_a,X_0]\nonumber\\
&+&2\alpha Tr\bigg({\psi}^{+A}{\sigma}_2({\epsilon}^{+B})^T{\rho}_i^{AB}+({\psi}^{A})^T{\sigma}_2{\epsilon}^B({\rho}_i^+)^{AB}\bigg)[X_0,{\Phi}_i].\nonumber\\
\end{eqnarray}
The only pssible solution is given by
\begin{eqnarray}
v=2~,~R_1=R~,~\alpha =-\frac{1}{2R}.
\end{eqnarray}
The full variation of the fermion action is finally given by
\begin{eqnarray}
{\delta}S_F+{\delta}(-2i Tr{\psi}^+{\partial}_0{\psi})|_{t={\rm fixed}}&=&-iTr{\delta}X_b[X_a,\tilde{F}_{ab}]+Tr{\delta}{\Phi}_i[X_a,[X_a,{\Phi}_i]]-Tr{\delta}X_a[{\Phi}_i,[X_a,{\Phi}_i]]\nonumber\\
&+&\bigg[\frac{\alpha}{2}-\frac{3}{4R_1}\bigg]{\epsilon}_{abc}TrF_{ab}{\delta}X_c+\bigg[\alpha(\frac{3}{4R}-\frac{\alpha}{2})+\frac{3}{4R_1}(\frac{3}{2R}-\alpha)\bigg]Tr{\Phi}_i{\delta}{\Phi}_i\nonumber\\&+&Tr{\delta}{\Phi}_j[{\Phi}_i,[{\Phi}_i,{\Phi}_j]]+Tr{\delta}X_0[{\Phi}_i,[X_0,{\Phi}_i]]+Tr{\delta}X_a[X_0,[X_a,X_0]]\nonumber\\
&-&Tr{\delta}{\Phi}_i[X_0,[X_0,{\Phi}_i]]-Tr{\delta}X_0[X_a,[X_a,X_0]].
\end{eqnarray}
The variations of the bosonic terms are given by
\begin{eqnarray}
&&{\delta}\bigg(\frac{1}{4}Tr[X_a,X_b]^2\bigg)=Tr{\delta}X_a[X_b,[X_a,X_b]]=iTr{\delta}X_b[X_a,\tilde{F}_{ab}]-\frac{v}{R}{\epsilon}_{abc}Tr{\delta}X_c\tilde{F}_{ab}-\frac{v^2}{R^2}Tr{\delta}X_a^2\nonumber\\
&&{\delta}\bigg(\frac{1}{4}Tr[{\Phi}_i,{\Phi}_j]^2\bigg)=-Tr{\delta}{\Phi}_j[{\Phi}_i,[{\Phi}_i,{\Phi}_j]]\nonumber\\
&&{\delta}\bigg(\frac{1}{2}Tr[X_a,{\Phi}_i]^2\bigg)=-Tr{\delta}{\Phi}_i[X_a,[X_a,{\Phi}_i]]+Tr{\delta}X_a[{\Phi}_i,[X_a,{\Phi}_i]]\nonumber\\
&&{\delta}\bigg(-\frac{1}{2}Tr[X_0,{\Phi}_i]^2\bigg)=Tr{\delta}{\Phi}_i[X_0,[X_0,{\Phi}_i]]-Tr{\delta}X_0[{\Phi}_i,[X_0,{\Phi}_i]]\nonumber\\
&&{\delta}\bigg(-\frac{1}{2}Tr[X_0,X_a]^2\bigg)=Tr{\delta}X_a[X_0,[X_0,X_a]]-Tr{\delta}X_0[X_a,[X_0,X_a]]\nonumber\\
&&{\delta}\bigg(\frac{im}{3}{\epsilon}_{abc}TrX_aX_bX_c\bigg)=im{\epsilon}_{abc}Tr{\delta}X_aX_bX_c=\frac{m}{2}{\epsilon}_{abc}Tr{\delta}X_c\tilde{F}_{ab}+\frac{mv}{2R}Tr{\delta}X_a^2\nonumber\\
&&{\delta}\bigg(-\frac{1}{2}\big(\frac{m}{6}\big)^2Tr{\Phi}_i^2\bigg)=-\big(\frac{m}{6}\big)^2Tr{\Phi}_i{\delta}{\Phi}_i\nonumber\\
&&{\delta}\bigg(-\frac{1}{2}\big(\frac{m}{3}\big)^2TrX_a^2\bigg)=-\big(\frac{m}{3}\big)^2TrX_a{\delta}X_a.
\end{eqnarray} 
It is not difficult to convince ourselves that in order to have supersymmetric invariance we must have
\begin{eqnarray}
m=\frac{6}{R}.
\end{eqnarray}

\subsubsection{${\cal N}=4$ time-independent supersymmetry}
The aim is to construct an ${\cal N}=4$ supersymmetric action without time dependence. The total action is 
\begin{eqnarray}
S&=&S_B+S_F+S_T.
\end{eqnarray}
The bosonic action is given as before, viz
\begin{eqnarray}
S_B&=&Tr\bigg[-\frac{1}{2}[X_0,X_a]^2-\frac{1}{2}[X_0,{\Phi}_i]^2+\frac{1}{4}[X_a,X_b]^2+\frac{1}{4}[{\Phi}_i,{\Phi}_j]^2+\frac{1}{2}[X_a,{\Phi}_i]^2\nonumber\\
&+&\frac{im}{3}{\epsilon}_{abc}X_aX_bX_c-\frac{m^2}{18}X_a^2-\frac{m^2}{72}{\Phi}_i^2\bigg].
\end{eqnarray}
The fermionic action is given now by the action
\begin{eqnarray}
S_F&=&Tr\bigg[-2{\Psi}_{\dot{a}}^+[X_0,{\Psi}_{\dot{a}}]-2{\Psi}_{\dot{a}}^+\bigg({\sigma}_a[X_a,{\Psi}_{\dot{a}}]-\frac{m}{4}{\Psi}_{\dot{a}}\bigg)+{\Psi}_{\dot{a}}^+i{\sigma}_2{\rho}_i[{\Phi}_i,({\Psi}_{\dot{a}}^+)^T]\nonumber\\
&-&{\Psi}_{\dot{a}}^Ti{\sigma}_2{\rho}_i^+[{\Phi}_i,{\Psi}_{\dot{a}}]\bigg].
\end{eqnarray}
The extra index $\dot{a}$ takes two values $1$ and $2$ and therefore the spinor $\Psi$ is given by
\begin{eqnarray}
{\Psi}_{\dot{a};\alpha}^A=\left(
\begin{array}{c}
{\psi}_{\alpha}^A\\
{\chi}_{\alpha}^A
\end{array}
\right).
\end{eqnarray}
Similarly, the supersymmetric parameter will be given by the spinor
\begin{eqnarray}
{\Omega}_{\dot{a};\alpha}^A=\left(
\begin{array}{c}
{\epsilon}_{\alpha}^A\\
{\omega}_{\alpha}^A
\end{array}
\right).
\end{eqnarray}
We have the supersymmetry transformations
\begin{eqnarray}
&&{\delta}X_0=-2i\bigg({\Psi}_{\dot{a}}^+{\Omega}_{\dot{a}}-{\Omega}_{\dot{a}}^+{\Psi}_{\dot{a}}\bigg)\nonumber\\
&&{\delta}X_a=2i\bigg({\Psi}_{\dot{a}}^+{\sigma}_a{\Omega}_{\dot{a}}-{\Omega}_{\dot{a}}^+{\sigma}_a{\Psi}_{\dot{a}}\bigg)\nonumber\\
&&{\delta}{\Phi}_i=2\bigg({\Psi}_{\dot{a}}^+{\rho}_i{\sigma}_2({\Omega}_{\dot{a}}^+)^T+{\Psi}_{\dot{a}}^T{\rho}_i^*{\sigma}_2{\Omega}_{\dot{a}}\bigg).
\end{eqnarray}
\begin{eqnarray}
{\delta}{\Psi}_{\dot{a};{\alpha}}^A&=&-\frac{i}{2}{\epsilon}_{{a}{b}{c}}\tilde{F}_{{a}{b}}({\sigma}_{{c}})_{\alpha \beta}{\Omega}^A_{\dot{a};{\beta}}+i[X_{{a}},{\Phi}_i]{\rho}_i^{AB}(i{\sigma}_{{a}}{\sigma}_2)_{{\alpha}{\beta}}{\Omega}_{\dot{a};{\beta}}^{+B}+\frac{1}{R}{\Phi}_i({\rho}_i)^{AB}({\sigma}_2)_{{\alpha}{\beta}}{\Omega}_{\dot{a};{\beta}}^{+B}\nonumber\\
&-&\frac{i}{2}[{\Phi}_i,{\Phi}_j]({\rho}_i{\rho}_j^+)^{AB}{\Omega}_{\dot{a};{\alpha}}^B-i[X_{{a}},X_0]({\sigma}_{{a}})_{\alpha \beta}{\Omega}^A_{\dot{a};{\beta}}+[X_0,{\Phi}_i]({\rho}_i)^{AB}({\sigma}_2)_{{\alpha}{\beta}}{\Omega}^{+B}_{\dot{a};{\beta}}.\nonumber\\
%&+&\frac{1}{4R}{\Phi}_i({\sigma}_2)_{\alpha \beta}{\rho}_i^{AB}{\Omega}_{\dot{a};\beta}^{+B}.
\end{eqnarray}
In above $R=\frac{6}{m}$ and $\tilde{F}_{ab}=i[X_a,X_b]-\frac{2}{R}{\epsilon}_{abc}X_c$. 

The analogue of the action $-2i Tr{\psi}^+{\partial}_0{\psi}$ is given here by the term $S_T$ defined by the equation
\begin{eqnarray}
S_T=\frac{1}{R}Tr {\Psi}_{\dot{b}}^+({\tau}_2)_{\dot{b}\dot{a}}{\Psi}_{\dot{a}}~,~{\tau}_2=\left(
\begin{array}{cc}
0 &-i\\
i&0
\end{array}
\right).
\end{eqnarray}
Under supersymmetry this action changes by the amount
\begin{eqnarray}
{\delta}S_T=\frac{1}{R}Tr\bigg[{\Psi}_{\dot{b}}^+({\tau}_2)_{\dot{b}\dot{a}}{\delta}{\Psi}_{\dot{a}}+{\rm H.C}\bigg].
\end{eqnarray}
The analogue of the two conditions $i{\partial}_0{\epsilon}_{\alpha}^A=\alpha {\epsilon}_{\alpha}^A$ and $i{\partial}_0{\epsilon}_{\alpha}^{+A}=-\alpha {\epsilon}_{\alpha}^{+A}$ will be given by the conditions
\begin{eqnarray}
&&({\tau}_2)_{\dot{b}\dot{a}}{\Omega}_{\dot{a};\beta}^A={\Omega}_{\dot{b};\beta}^A\nonumber\\
&&({\tau}_2)_{\dot{b}\dot{a}}{\Omega}_{\dot{a};\beta}^{+B}=-{\Omega}_{\dot{b};\beta}^{+B}.
\end{eqnarray}
This means that the two supersymmetry parameters $\epsilon$ and $\omega$ are related by the equation
\begin{eqnarray}
\omega=i\epsilon.
\end{eqnarray}
Thus we compute
\begin{eqnarray}
({\tau}_2)_{\dot{b}\dot{a}}{\delta}{\Psi}_{\dot{a};{\alpha}}^A&=&-\frac{i}{2}{\epsilon}_{{a}{b}{c}}\tilde{F}_{{a}{b}}({\sigma}_{{c}})_{\alpha \beta}{\Omega}^A_{\dot{b};{\beta}}-i[X_{{a}},{\Phi}_i]{\rho}_i^{AB}(i{\sigma}_{{a}}{\sigma}_2)_{{\alpha}{\beta}}{\Omega}_{\dot{b};{\beta}}^{+B}-\frac{1}{R}{\Phi}_i({\rho}_i)^{AB}({\sigma}_2)_{{\alpha}{\beta}}{\Omega}_{\dot{b};{\beta}}^{+B}\nonumber\\
&-&\frac{i}{2}[{\Phi}_i,{\Phi}_j]({\rho}_i{\rho}_j^+)^{AB}{\Omega}_{\dot{b};{\alpha}}^B-i[X_{{a}},X_0]({\sigma}_{{a}})_{\alpha \beta}{\Omega}^A_{\dot{b};{\beta}}-[X_0,{\Phi}_i]({\rho}_i)^{AB}({\sigma}_2)_{{\alpha}{\beta}}{\Omega}^{+B}_{\dot{b};{\beta}}.\nonumber\\
%&-&\frac{1}{4R}{\Phi}_i({\sigma}_2)_{\alpha \beta}{\rho}_i^{AB}{\Omega}_{\dot{b};\beta}^{+B}.
\end{eqnarray}
We can immediately compute
\begin{eqnarray}
{\delta}S_T&=&-\frac{1}{4R}{\epsilon}_{abc}Tr\tilde{F}_{ab}{\delta}X_c-\frac{1}{2R^2}Tr{\Phi}_i{\delta}{\Phi}_i\nonumber\\
&-&\frac{1}{R} Tr\bigg({\Omega}_{\dot{b}}^T{\rho}_i^+{\sigma}_2{\sigma}_a{\Psi}_{\dot{b}}-{\Psi}^{+}_{\dot{b}}{\rho}_i{\sigma}_a{\sigma}_2({\Omega}_{\dot{b}}^+)^T\bigg)[X_a,{\Phi}_i]\nonumber\\
&+&\frac{i}{2R} Tr\bigg({\Omega}^{+}_{\dot{b}}{\rho}_i{\rho}_j^+{\Psi}_{\dot{b}}-{\Psi}^{+}_{\dot{b}}{\rho}_i{\rho}_j^+{\Omega}_{\dot{b}}\bigg)[{\Phi}_i,{\Phi}_j]\nonumber\\
&-&\frac{i}{R} Tr\bigg({\Omega}^{+}_{\dot{b}}{\sigma}_a{\Psi}_{\dot{b}}+{\Psi}^{+}_{\dot{b}}{\sigma}_a{\Omega}_{\dot{b}}\bigg)[X_a,X_0]\nonumber\\
&-&\frac{1}{R} Tr\bigg({\Psi}^{+}_{\dot{b}}{\rho}_i{\sigma}_2({\Omega}^{+}_{\dot{b}})^T+{\Psi}_{\dot{b}}^T{\rho}_i^+{\sigma}_2{\Omega}_{\dot{b}}\bigg)[X_0,{\Phi}_i].
\end{eqnarray}
This variation is exactely  equal to the variation of the term $-2i Tr{\psi}^+{\partial}_0{\psi}$ in the original time-dependent theory. Hence it must be equal to minus the variation of the action $S_B+S_F$. In other words
\begin{eqnarray}
{\delta}S_T=-{\delta}S_B-{\delta}S_F.
\end{eqnarray}
This establishes the ${\cal N}=4$ supersymmetry invariance.

\section{Other matrix models}
I just mention two more matrix models which are relevant to the gauge/gravity duality in lower dimensions:
\begin{itemize}
\item The BD  (Berkooz-Douglas) theory which is relevant to M2- and M5-branes \cite{Berkooz:1996is}. The M5-branes appear as additional fundamental hypermultiplets added to the BFSS model. A recent non-perturbative study of this model is given by \cite{Asano:2016xsf}. 
\item The ABJM theory in $(2+1)-$dimension which is relevant to the discussion of D2-branes (IIA superstring)  and M2-branes (M-theory) \cite{Aharony:2008ug}.  For a concise summary of the results obtained for the ABJM model and future prospects see \cite{Hanada:2016jok}.

\end{itemize}

\chapter{Type IIB Matrix Model}

\section{The IKKT model in the Gaussian expansion method}
We start by considering the IKKT matrix model in the Gaussian expansion method following mainly \cite{Nishimura:2011xy}. As we have already discussed the IKKT model also called the type IIB matrix model is the zero-volume limit, i.e. dimensional reduction to a point, of supersymmetric U(N) Yang-Mills gauge theory in $10$ dimensions where the components of the $10-$dimensional gauge field reduce to $10$ bosonic hermitian matrices $A_{\mu}$. This is an Euclidean $SO(10)$ model given by the action
 \begin{eqnarray}
S=-\frac{1}{4g^2}Tr [A_{\mu},A_{\nu}]^2-\frac{1}{2g^2}Tr{\psi}_{\alpha}({\cal C}\gamma^{\mu})_{\alpha\beta}[A_{\mu},\psi_{\beta}].
\end{eqnarray} 
The $A_{\mu}$ (vector) and $\psi_{\alpha}$ (Majorana-Weyl spinor) are $N\times N$ traceless hermitian matrices with complex and Grassmannian entries respectively. The $\gamma$ are the $16\times 16$ Dirac matrices in $10$ dimensions after Weyl projection and the charge conjugation operator is defined by $\gamma_{\mu}^T={\cal C}\gamma_{\mu}{\cal C}^{\dagger}$ and ${\cal C}^T={\cal C}$. We can without any loss of generality choose the coupling constant (scale parameter) $g$ such that $g^2N=1$. This model is supposed to give a non-perturbative regularization of type IIB superstring theory in the Schild gauge \cite{Ishibashi:1996xs}. The size of the matrices $N$ plays in this regularization the role of the cutoff. It is then supposed that in the limit $N\longrightarrow\infty$ (continuum limit) the $10$ bosonic hermitian matrices $A_{\mu}$ reproduces the $10-$dimensional target space of the string \cite{Aoki:1998vn}.

The above model corresponds to Yang-Mills matrix model in $D=10$ dimensions. The dimensional reduction to a point of supersymmetric U(N) Yang-Mills gauge theory in $D$ dimensions gives Yang-Mills matrix model in $D$ dimensions. We can also have supersymmetric models in $D=6$ and $D=4$. These three models $D=10,6,4$ enjoy a convergent partition function \cite{Krauth:1998xh,Austing:2001bd,Austing:2001pk}. The $D=3$ supersymmetric partition function is divergent while the bosonic one is convergent. The determinant in $D=4$ is real positive while the determinants in $D=6$ and $D=10$ are complex which means in particular that in $D=10$ and $D=6$ we can have spontaneous symmetry breaking of rotational invariance \cite{Anagnostopoulos:2001yb,Nishimura:2001sq,Nishimura:2004ts,Anagnostopoulos:2011cn} while in $D=4$ there is no spontaneous symmetry breaking \cite{Ambjorn:2000bf,Burda:2000mn,Ambjorn:2001xs}.

The main question we want to ask in the Gaussian expansion method is whether or not the rotational $SO(10)$ is spontaneously broken in the continuum large $N$ limit. Towards answering this question the authors of  \cite{Nishimura:2011xy} studied the $SO(d)$ symmetric vacua for all values of $d$ in the range between $2$ and $7$. They also impose on the shrunken $(10-d)$ directions an extra $\Sigma_d$ symmetry and thus the full symmetry imposed is actually $SO(d)\times\Sigma_d\in SO(10)$ which is much more stronger than simply $SO(d)$ which allowed them to reduce the number of free parameters in the Gaussian expansion method considerably (they exhausted all possible extra symmetries which leave not more that $5$ free parameters). This spontaneous symmetry breaking corresponds therefore to a dynamical compactification to $d$ dimensions.

The order parameter of the spontaneous symmetry breaking of $SO(10)$ is given by the set of the nine eigenvalues $\lambda_i$ of the moment of inertia tensor 
 \begin{eqnarray}
T_{\mu\nu}=\frac{1}{N} TrA_{\mu}A_{\nu}.
\end{eqnarray} 
The expectation values $\langle \lambda_i\rangle $ in the large $N$ limit are all equal if $SO(10)$ is not spontaneously broken. In the current Euclidean model it is found that the expectation values $\langle\lambda_1\rangle$,  $\langle\lambda_2\rangle$ and $\langle\lambda_3\rangle$ become much larger than the other ones in the large $N$ limit and hence $SO(10)$ is spontaneously broken down to $SO(3)$ due precisely to the phase of the Pfaffian \cite{Nishimura:2000ds,Nishimura:2000wf}. In the Lorentzian model as we will see the spontaneous symmetry breaking is due to the noncommutativity of the space coordinates \cite{Kim:2011cr}. 

The Gaussian expansion method is a nonperturbative scheme in which mostly perturbative calculations are performed \cite{Stevenson:1981vj}. We start by introducing a Gaussian action $S_0$ and split the action $S$ as
\begin{eqnarray}
S=(S-S_0)+S_0.
\end{eqnarray}
The Gaussian action $S_0$ is arbitrary since it contains many free parameters. It is expected that at any finite order the result of the Gaussian expansion method will depend on the parameters of the Gaussian action $S_0$. However, it is also known that there exists regions in the parameter space (plateau regions) for which the result at finite order will not depend on the choice of $S_0$ \cite{Nishimura:2011xy}. 

By requiring $SO(10)$ rotational invariance the most general U(N)-invariant Gaussian action $S_0$ takes the form 
\begin{eqnarray}
S_0=S_{0b}+S_{0f},
\end{eqnarray}
\begin{eqnarray}
S_{0b}=\frac{N}{2}\sum_{\mu=1}^{10}M_{\mu}Tr A_{\mu}^2~,~S_{0f}=\frac{N}{2}\sum_{\alpha,\beta=1}^{16}{\cal A}_{\alpha\beta}Tr\psi_{\alpha}\psi_{\beta}.
\end{eqnarray}
There are $10$ parameters $M_{\mu}$ and $120$ parameters ${\cal A}_{\alpha\beta}$. The $16\times 16$ complex matrix ${\cal A}$ can be given in terms of gamma matrices. The goal is to compute the free energy 
\begin{eqnarray}
F=-\log Z~,~Z=Z_0\langle\exp(-(S-S_0))\rangle_0~,~Z_0=\int dAd\psi\exp(-S_0).
\end{eqnarray}
This will be done by truncating the perturbative series at some finite order and as a consequence the free energy will depend on the parameters  $M_{\mu}$ and ${\cal A}_{\alpha\beta}$. We look for the regions in the parameter space for which the free energy $F$ is stationary in the sense that it solves the self-consistency conditions
 \begin{eqnarray}
\frac{\partial}{\partial M_{\mu}}F=0~,~\frac{\partial}{\partial {\cal A}_{\alpha\beta}}F=0.
\end{eqnarray}
The number of self-consistency conditions is equal to the number of parameters. The values of the free energy are obtained at the solutions of these conditions. Clearly, the number of solutions increases with the order. The sought after plateau in the parameter space is determined by the existence of multiple solutions with almost the same value of the free energy.

The above perturbative series corresponds to an ordinary loop expansion where the insertion of the $2-$point vertex $-S_0$ acts as if it is coming from a  one-loop counter term. Furthermore, due to the large $N$ limit, only planar Feynman diagram are of relevance. 

It is expected that $SO(10)$ symmetry will be spontaneously broken dynamically down to some $SO(d)$. Thus only $SO(d)$-symmetric vacua will be left and the Gaussian action in this case must only be required to be $SO(d)$-symmetric. As it turns out, the number of free parameters will be reduced considerably by imposing $SO(d)$ symmetry on the Gaussian action. Indeed, we get the total number of parameters to be $5, 9, 16, 27, 44, 73$ for $d = 7, 6, 5, 4, 3, 2$ respectively by imposing $SO(d)$ symmetry on the Gaussian action \cite{Nishimura:2011xy}. The Gaussian expansion method ceases to work properly for $d\geq 8$ but in these cases there exist also spontaneous symmetry breaking  and $SO(d)$-symmetric vacua. 

However, the self-consistency conditions can be solved only for not more than $5$ parameters. Therefore, an extra symmetry $\Sigma_d$ is imposed on the shrunken dimensions $x_{d+1},...,x_{10}$ in order to reduce the number of parameters to $5$ or less. The extra symmetry $\Sigma_d$ is a subgroup of $SO(10)$ formed out of cyclic permutations and reflections \cite{Nishimura:2011xy}. For each $SO(d)$ there are several choices $\Sigma_d$.

We fix then the ansatz $SO(d)\times\Sigma_d$ and we compute the free energy up to the third order as a function of the parameters of the Gaussian action. We differentiate the energy with respect to these parameters to obtain the self-consistency conditions and then solve these equations numerically. By substituting back in the free energy with the solution(s) we get the value(s) of the free energy at the solution(s). A given ansatz may correspond to several solutions. We locate the plateau region by the solutions for the various ansatz which have the same value of the free energy. By averaging over these physical solutions we get the value of the free energy for that particular $d$.

\begin{figure}[H]
\begin{center}
\includegraphics[width=9.0cm,angle=0]{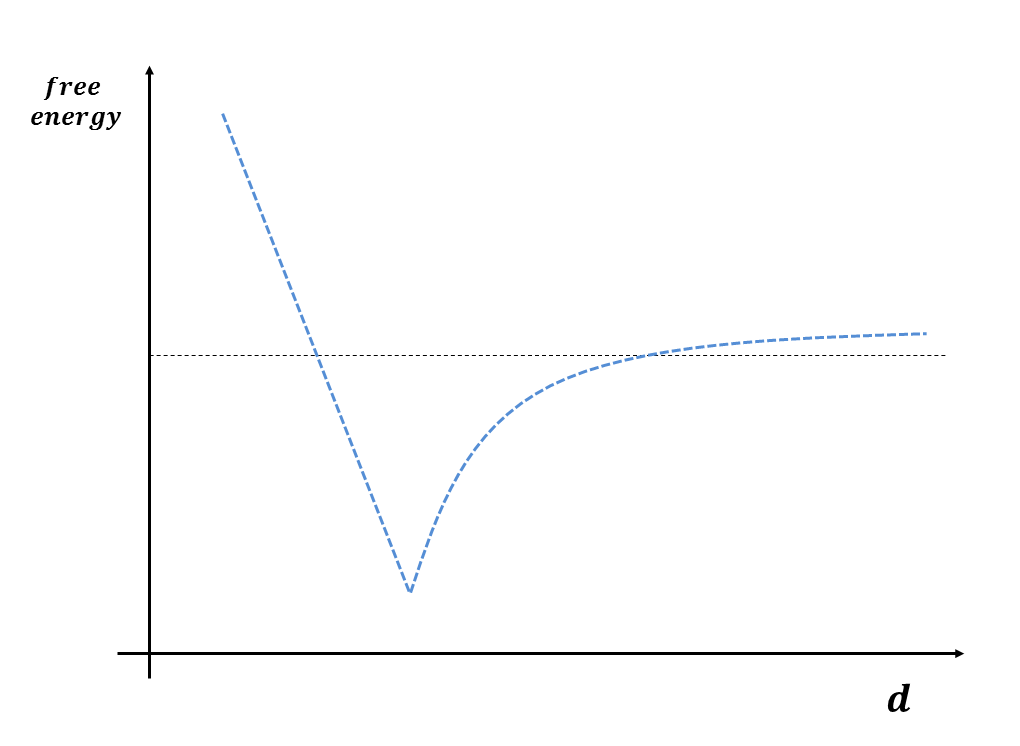}
\end{center}
\caption{The free energy of the IKKT model in the Gaussian expansion method. The minimum is at $d=3$.}\label{figill8}
\end{figure}

The main result of \cite{Nishimura:2011xy} is the following statement: {the free energy $F$ (averaged over the plateau region of physical solutions) takes its minimum value at $d=3$}. See figure (\ref{figill8}). This shows more or less explicitly that in the Euclidean model (and due to the phase of the Pfaffian) the stringy rotational symmetry $SO(10)$ must be spontaneously broken down to the physical rotational symmetry $SO(3)$.

\begin{figure}[H]
\begin{center}
\includegraphics[width=9.0cm,angle=0]{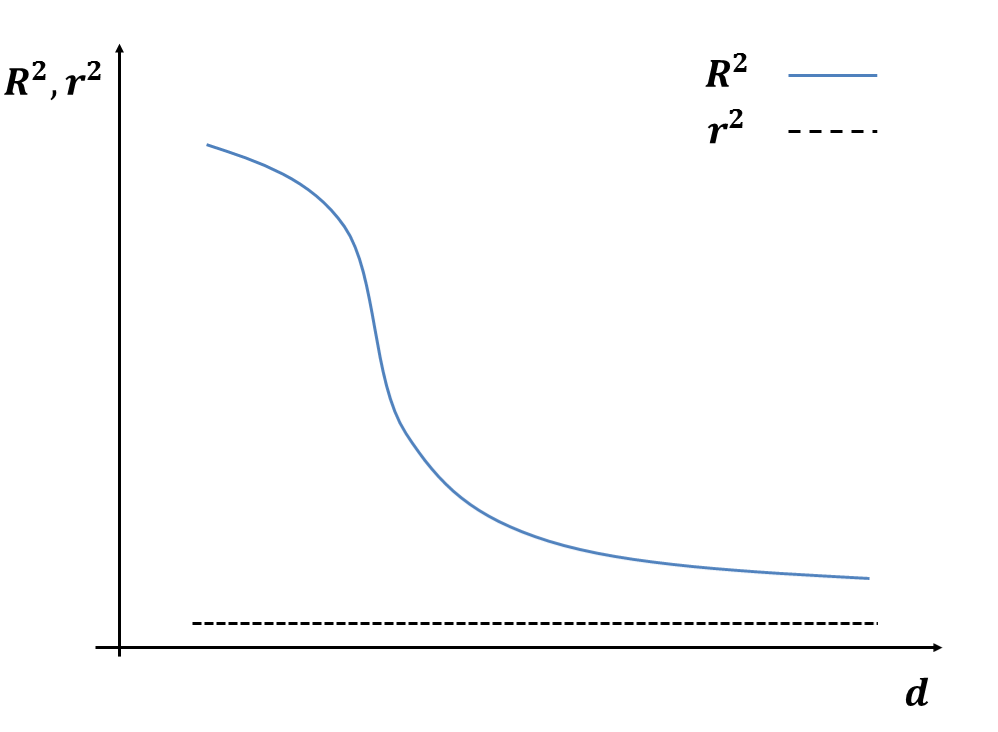}
\end{center}
\caption{The extent of space $R$ and the compactification scale $r$ for the IKKT model.}\label{figill9}
\end{figure}

The extent of space in the extended directions $R^2=\langle \lambda_1\rangle=...=\langle \lambda_d\rangle$ and the extent of space in the shrunken directions $r^2=\langle\lambda_{d+1}\rangle...=\langle\lambda_{10}\rangle$ can also be computed in the Gaussian expansion method with very illuminating results \cite{Nishimura:2011xy}. They found that $r$ remains almost constant for all $d$, and thus it is indeed a universal compactification scale, while $R$ becomes larger for smaller $d$. See figure (\ref{figill9}). This behavior is consistent with the so-called constant volume property given by \cite{Aoyama:2010ry}
\begin{eqnarray}
R^dr^{10-d}=l^{10}.
\end{eqnarray}

\section{Yang-Mills matrix cosmology}
\subsection{Lorentzian type IIB matrix model}

The Lorentzian type IIB matrix model allows us to study the real time dynamics of the emergence of $(1+3)-$dimensional Minkowski spacetime, the spontaneous symmetry breaking of SO(9) down to SO(3), as well as providing a mechanism for avoiding the Big Bang singularity, and allowing us to obtain the expansion (exponential at early times and power law, i.e. $\sqrt{t}$, at late times) of the Universe \cite{Kim:2011cr,Kim:2011ts,Nishimura:2012rs,Kim:2012mw,Nishimura:2013moa,Ito:2013ywa,Aoki:2014cya,Ito:2015mxa}. The model is given by (with $F_{\mu\nu}=i[A_{\mu},A_{\nu}]$ and $\bar{\psi}=\psi^T{\cal C}$, $\mu,...\nu=0,...,9$, $\alpha,...,\beta=1,...,16$ and the $16\times 16$ matrices $\gamma^{\mu}$ are the $9-$dimensional Dirac matrices)
\begin{eqnarray}
S=-\frac{1}{4g^2}Tr F_{\mu\nu}F^{\mu\nu}-\frac{1}{2g^2}Tr\bar{\psi}_{\alpha}(\gamma^{\mu})_{\alpha\beta}[A_{\mu},\psi_{\beta}].
\end{eqnarray}
The $A_{\mu}$ and $\psi_{\alpha}$ are $N\times N$ traceless hermitian matrices with complex and Grassmannian entries respectively. Because of the signature $\eta=(-1,+1,+1,...)$ the Yang-Mills term is not positive definite, i.e. the action is not bounded from below. As opposed to the Euclidean case the Pfaffian in the Lorentzian case is real (in fact it is positive definite at large $N$). The path integral is given by 
\begin{eqnarray}
Z=\int dA d\psi \exp(i S)=\int dA~ {\rm Pf}{\cal M}(A)~\exp(i S_B).
\end{eqnarray}
The Dirac operator is given by ${\cal M}(A)=({\cal C}\gamma^{\mu})_{\alpha\beta}[A_{\mu},...]$. 

This path integral is not finite as it stands (when $A_0$ diverges the action $S_B$ goes to $-\infty$). We introduce the SO(9,1) symmetric IR cutoff in the temporal direction as \cite{Kim:2011cr}
\begin{eqnarray}
\frac{1}{N}Tr A_0^2\leq \kappa\frac{1}{N}Tr A_a^2. \label{IR1}
\end{eqnarray}
This condition is reminiscent to what happens in causal dynamical triangulation.

\begin{figure}[H]
\begin{center}
\includegraphics[width=5.0cm,angle=0]{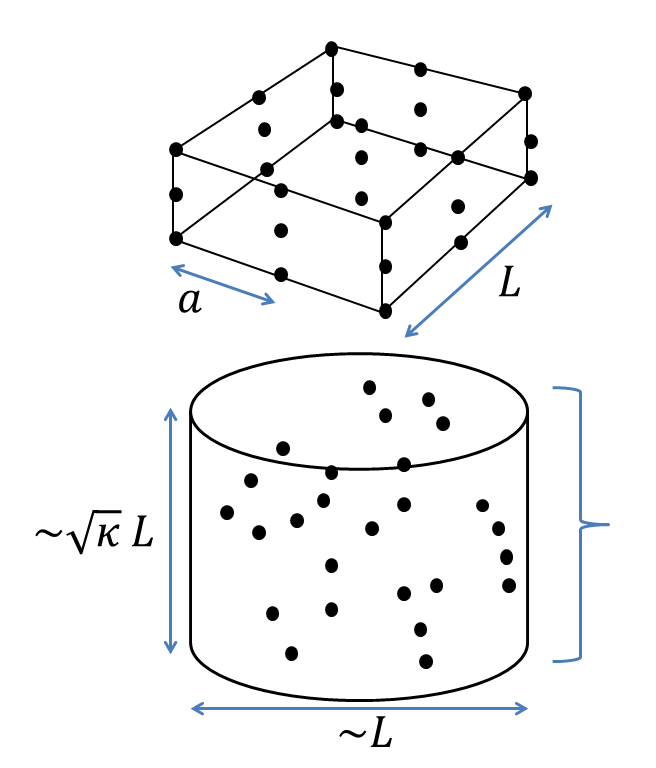}
\end{center}
\caption{The IR cutoffs $L$ and $\kappa$.}\label{figill1}
\end{figure}

The oscillating phase factor in the path integral is also regularized in the usual way by adding a damping factor $\exp(-\epsilon |S_B|)$ to the action where $\epsilon$ is some small positive number. We get then 
\begin{eqnarray}
Z_{}=\int dA~ {\rm Pf}{\cal M}(A)~\exp(i S_B-\epsilon|S_B|).
\end{eqnarray}
The IR cutoff (\ref{IR1}) is explicitly implemented by inserting in the path integral the expression 
 \begin{eqnarray}
\int_0^{\infty}dr~\delta(\frac{1}{N}Tr A_a^2-r)~\theta(\kappa r-\frac{1}{N}Tr A_0^2). 
\end{eqnarray}
The variable $r$ is effectively the scale factor. We scale the field as $A_{\mu}\longrightarrow \sqrt{r}A_{\mu}$ and perform the integral over $r$ (with $D=10$ and $d_F=8$). We find 
\begin{eqnarray}
\int_0^{\infty}dr~r^{\frac{D(N^2-1)+d_F(N^2-1)}{2}-1}~\exp(-r^2(\epsilon|x|-ix))\sim 1/|x|^{(D+d_F)(N^2-1)/4}.
\end{eqnarray}
This is a divergent integral which can be regularized by introducing a second IR cutoff
\begin{eqnarray}
\frac{1}{N}Tr A_a^2\leq L^2. \label{IR2}
\end{eqnarray}
We insert now in the path integral the condition 
\begin{eqnarray}
\int_0^{L^2}dr~\delta(\frac{1}{N}Tr A_a^2-r)~\theta(\kappa r-\frac{1}{N}Tr A_0^2). 
\end{eqnarray}
We get
\begin{eqnarray}
Z_{}&=&\int dA~ \int_0^{L^2}dr~(r^{1/2})^{D(N^2-1)+d_F(N^2-1)}~\frac{1}{r}\delta(\frac{1}{N}Tr A_a^2-1)~\theta(\kappa -\frac{1}{N}Tr A_0^2)~{\rm Pf}{\cal M}(A)\nonumber\\
&\times &\exp(-r^2(\epsilon|S_B|-iS_B)).
\end{eqnarray}
We use the result \cite{Ito:2013ywa}
\begin{eqnarray}
\int_0^{L^2}dr~r^{\frac{D(N^2-1)+d_F(N^2-1)}{2}-1}~\exp(-r^2(\epsilon|x|-ix))=\delta(x)~,~N,L\longrightarrow\infty.
\end{eqnarray}
We get then the path integral (by reinserting the scale parameter $L$ for later convenience)
\begin{eqnarray}
Z_{}&=&\int dA~ \delta(\frac{1}{N}Tr F_{\mu\nu}F^{\mu\nu})~\delta(\frac{1}{N}Tr A_a^2-L^2)~\theta(\kappa L^2 -\frac{1}{N}Tr A_0^2)~{\rm Pf}{\cal M}(A).
\end{eqnarray}
In this formulation $N$ plays the role of inverse lattice spacing and $\sqrt{\kappa} L^2$ plays the role of the volume ($L$ is the spacelike length and $\sqrt{\kappa}L$ is the timelike length). These two constraints can be removed in the continuum limit $N\longrightarrow\infty$ and the infinite volume limit $L\longrightarrow\infty$ and only one scale parameter $g$ remains (string coupling constant). See figure (\ref{figill1}).

The problem is then converted into a potential problem of the form
\begin{eqnarray}
Z_{}&=&\int dA~ \exp(-V_{\rm pot})~{\rm Pf}{\cal M}(A).
\end{eqnarray}
\begin{eqnarray}
V_{\rm pot}=\frac{1}{2}\gamma_C\big(\frac{1}{N}Tr F_{\mu\nu}F^{\mu\nu}\big)^2+\frac{1}{2}\gamma_L(\frac{1}{N}Tr A_a^2-L^2)^2+\frac{1}{2}\gamma_{\kappa}\big(\kappa L^2 -\frac{1}{N}Tr A_0^2\big)^2\theta(\frac{1}{N}Tr A_0^2-\kappa L^2).\nonumber\\
\end{eqnarray}
This theory enjoys SO(1,9) Lorentz symmetry, SO(9) rotational symmetry, U(N) gauge symmetry, ${\cal N}=1$ supersymmetry, translation symmetry given by the shift symmetry $A_{\mu}\longrightarrow A_{\mu}+\alpha_{\mu}{\bf 1}$. But it also enjoys an extended ${\cal N}=2$ supersymmetry and hence it includes implicitly gravity since ${\cal N}=1$ is the maximal supersymmetry without gravity.

The rotational symmetry SO(9) will be spontaneously broken which is the main goal in this model. Also, the shift symmetry $A_{\mu}\longrightarrow A_{\mu}+\alpha_{\mu}{\bf 1}$ will be spontaneously broken in the dynamics which causes problems in determining the origin of the time coordinate. This issue can be avoiding by adding a potential of the form 

\begin{eqnarray}
V_{\rm sym}=\frac{1}{2}\gamma_{\rm sym}\bigg(\frac{1}{N}Tr A^2|_{\rm left}-\frac{1}{N}Tr A^2|_{\rm right}\bigg)^2.
\end{eqnarray}
\begin{eqnarray}
Tr A^2|_{\rm left}=\sum_{i=1}^d\sum_{a+b<N+1}|(A_i)_{ab}|^2~,~Tr A^2|_{\rm right}=\sum_{i=1}^d\sum_{a+b>N+1}|(A_i)_{ab}|^2.
\end{eqnarray}
The parameters $\gamma_{C,L,\kappa,{\rm sym}}$ are chosen as \cite{Ito:2013ywa}
\begin{eqnarray}
\gamma_C=N^2~,~\gamma_L=\gamma_{\kappa}=100N^2~,~\gamma_{\rm sym}=100.
\end{eqnarray}

\subsection{Spontaneous symmetry breaking and continuum and infinite volume limits}
\begin{figure}[H]
\begin{center}
\includegraphics[width=6.0cm,angle=0]{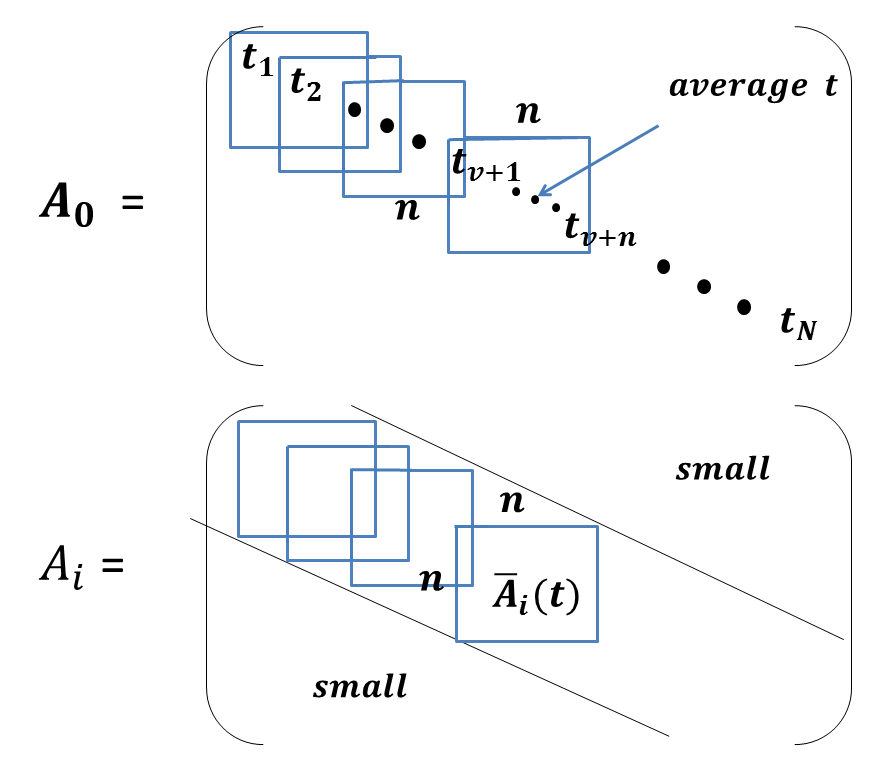}
\end{center}
\caption{The band structure.}\label{figill2}
\end{figure}

We will now employ SU(N) invariance to diagonalize the timelike matrix $A_0$, producing also a Vandermonde determinant, as 
\begin{eqnarray}
A_0={\rm diag}(t_1,...,t_N)~,~t_1<...<t_N.
\end{eqnarray}
Thus the measure becomes 
\begin{eqnarray}
\int dA=\int dA_a\int \prod_i dt_i \Delta(t)^2~,~\Delta(t)=\prod_{i>j}(t_i-t_j).
\end{eqnarray}
The effect of the Vandermonde determinant $\Delta(t)$ cancels exactly at one-loop order due to supersymmetry (more on this below). Indeed, at one-loop the repulsive effective action of the eigenvalues $t_i$ is given by 
\begin{eqnarray}
S_{\rm eff}=(D-2-d_F)\sum_{i\ne j}\ln (t_i-t_j)^2=0.
\end{eqnarray}
Thus the spectrum of $A_0$, i.e. time, extends to infinity even for finite $N$. Locality in time is also guaranteed as follows. Instants of time will be defined by 
\begin{eqnarray}
t=\frac{1}{n}\sum_{i=1}^nt_{\nu+i}.\label{time}
\end{eqnarray}
In the eigenbasis of $A_0$ the spacelike matrices $A_a$ have a band diagonal structure. This highly non-trivial property is determined dynamically. This band diagonal structure means in particular that the off diagonal elements $(A_a)_{ab}$ for $|a-b|\geq n$ are very small for some integer $n$. But $n$ should also be sufficiently large that it includes non negligible off diagonal elements. Thus we can consider $n\times n$ block matrices 
 \begin{eqnarray}
(\bar{A}_a)_{IJ}(t)=(A_a)_{I+\nu,J+\nu}.
\end{eqnarray}
The indices $I,J$ run from $1$ to $n$ and thus the index $\nu$ runs from $0$ to $N-n$. The time $t$ appearing in this equation is the one defined in equation (\ref{time}). The matrices $\bar{A}_a$ represent the state of the Universe at time $t$. The progression of $t$ is encoded in the index $\nu$. See figure (\ref{figill2}).

The block size $n$ is determined as follows \cite{Kim:2011ts}. We take $N$ even and consider the $N\times N$ matrix $(Q)_{IJ}=(A_1^2)_{IJ}$. We plot  for a fixed $L=(I+J)/2$ the quantity $\sqrt{|Q_{IJ}/Q_{N/2N/2}|}$ as a function of $I-J=2(I-L)$. It is sufficient to consider only the values $L=2,4,6,...,N/2$. It is found that the quantity $\sqrt{|Q_{IJ}/Q_{N/2N/2}|}$ decreases exponentially with $|I-J|$ and that the half width is maximum for $L=N/2$. The block size $n$ is given precisely by the largest half width. %See figure (\ref{figill7}).

%\begin{figure}[H]
%\begin{center}
%\includegraphics[width=9.0cm,angle=0]{illus7.png}
%\end{center}
%\caption{Dynamical determination of the block size.}\label{figill7}
%\end{figure} 

\begin{figure}[H]
\begin{center}
\includegraphics[width=9.0cm,angle=0]{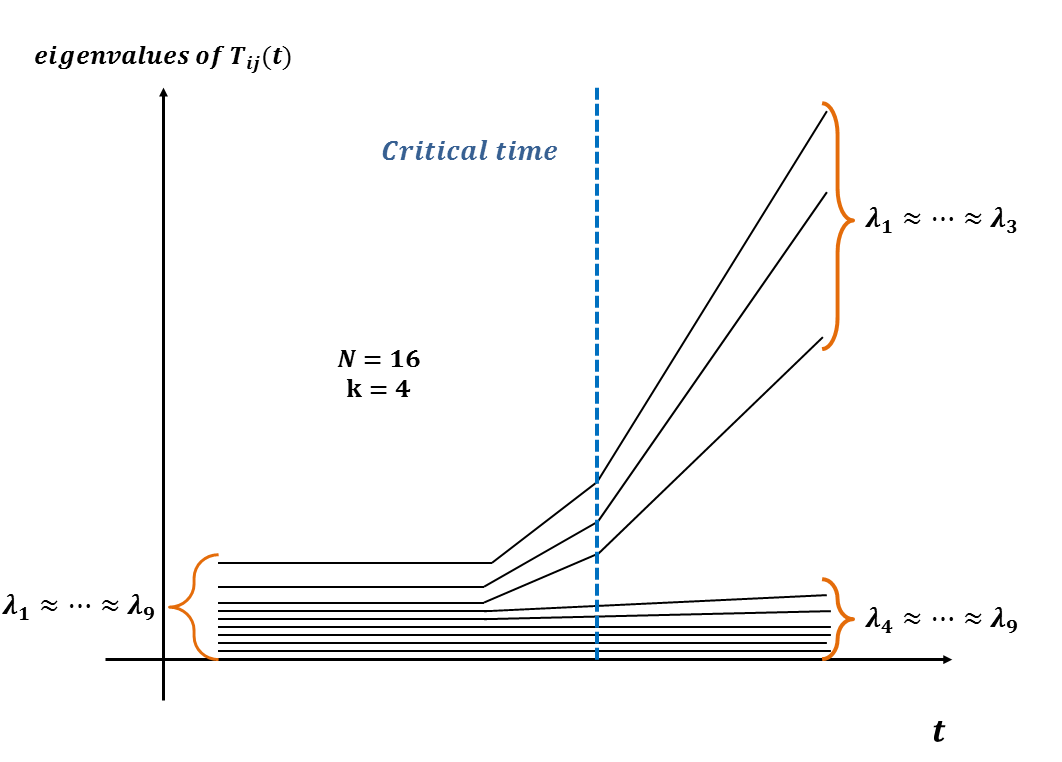}
\end{center}
\caption{SSB of SO(9).}\label{figill3}
\end{figure}

At time $t$ the square of the extent of space should be defined in terms of  $\bar{A}_a$ as follows (where tr is the trace taken over the $n\times n$ block)
\begin{eqnarray}
R^2(t)=\langle \frac{1}{n}tr \bar{A}_a^2(t)\rangle.
\end{eqnarray}
The order parameter for the spontaneous symmetry breaking (SSB) of rotational symmetry SO(9) is given by the moment of inertia tensor (symmetric under $t\longrightarrow -t$ and thus we may only take $t<0$)
\begin{eqnarray}
T_{ab}(t)=\frac{1}{n}tr \bar{A}_a(t)\bar{A}_b(t).
\end{eqnarray}
This is a real symmetric $9\times 9$ matrix with eigenvalues denoted by $\lambda_i(t)$ where $\lambda_1(t)>...>\lambda_9(t)$. We have then the behavior at early times in the large $N$ and $n$ limits
\begin{eqnarray}
\langle \lambda_1(t)\rangle=...=\langle\lambda_9(t)\rangle~,~{\rm Exact}~SO(9).
\end{eqnarray}
For late times we have instead 
\begin{eqnarray}
\langle \lambda_1(t)\rangle=\langle\lambda_2(t)\rangle=\langle\lambda_3(t)\rangle>>\langle\lambda_i(t)\rangle~,~i\ne 1,2,3~,~{\rm SSB~of}~SO(9).
\end{eqnarray}
The spontaneous symmetry breaking of SO(9) down to SO(3) occurs at some critical time $t_c$. See figure (\ref{figill3}). The mechanism behind this breaking is noncommutative geometry and not the complex Pfaffian as is the case in Euclidean signature.

The scaling limit is achieved as follows. The square of the extent of space $R^2(t)$ is given by the sum of the eigenvalues of the moment of inertia tensor $\langle T_{ab}(t)\rangle$. The extend of space $R(t)$ is found to scale in the large $N$ limit by sending the infrared cutoffs $\kappa$ and $L$ to infinity in a prescribed way \cite{Kim:2011cr}. This is done in two steps: 
\begin{itemize}
\item Continuum limit: We send $\kappa\longrightarrow\infty$ as $\kappa=\beta N^{p}$, $p=1/4$. The extent of space $R(t)$ for different values of $N$ is then seen to collapse to a single curve depending on $\beta$.
\item Infinite volume limit: We fix $N$ then we send $\beta\longrightarrow\infty$ (equivalently $\kappa$) with $L\longrightarrow\infty$. The extent of space $R(t)$ for various values of $\beta$ is then seen to collapse to a single curve.
\end{itemize}
Thus the two IR cutoffs $\kappa$ and $L$ are then removed in the large $N$ limit and the theory depends only on one single parameter $\kappa$ (which should be thought of as the string coupling constant). The only scale parameter is the size of the Universe $R(t_c)$ at the critical time.

We would like to discuss the infinite volume limit further following  \cite{Kim:2011cr}. We fix $N$ and $\kappa$ and calculate the extent of space $R(t)$ for $L=1$. They observe that the constraint (\ref{IR2}) is saturated for the dominant configurations. Thus they only need to scale $A_{\mu}\longrightarrow L A_{\mu}$ to reinstate the IR cutoff $L$. Then they choose $L$ such that the extent of space at the critical time $t_c$ is one, viz $R(t_c)=1$. They repeat for different values of $\beta$ and each time they determine implicitly the corresponding value of $L$ in this way. Thus increasing values of $\beta$ is equivalent to increasing values of $L$ and the extent of space  $R(t)$ is seen to scale with $\beta$.

To summarize, it is seen that the extent of space  $R(t)$ for different values of $N$ and $\kappa$ converge to a single scaled curve. This achieves the non-trivial continuum limit and infinite volume limit of the theory.

\subsection{Expansion}
The first fundamental observation is that the birth of the Universe at the critical time $t_c$ emerges without any singularity, i.e. the problem of the initial singularity is completely avoided, and the underlying mechanism behind it can be determined to be the noncommutativity of the space.

\begin{figure}[H]
\begin{center}
\includegraphics[width=8.0cm,angle=0]{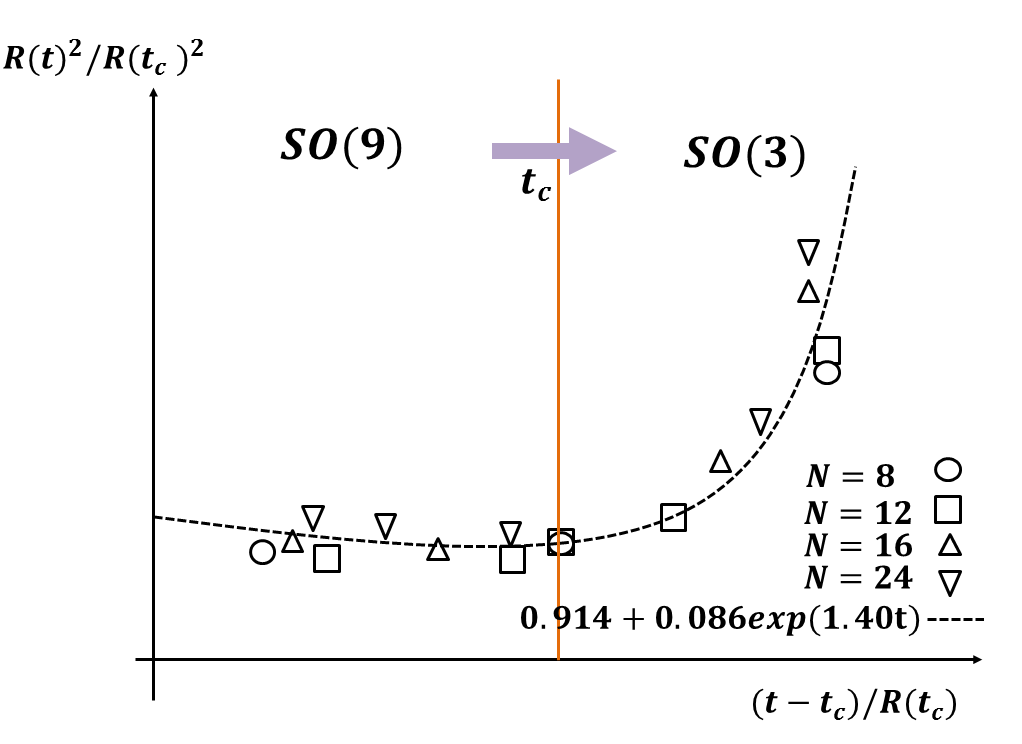}
\end{center}
%\caption{Exponential expansion in the type IIB model (left) and in the renormalized VDM model (right).}\label{figill4}
\caption{Exponential expansion in the type IIB model.}\label{figill4}
\end{figure}

After the birth of the Universe three coordinates start to expand and the other six shrink. It can be verified from the measurement of the extent of space $R(t)$ that for the very early times after $t_c$ the expansion of the three spatial coordinates is indeed exponential (inflation). See figure (\ref{figill4}). However, at later times the expansion is expected to become a power-law $\sqrt{t}$ behavior (radiation-dominated FLRW Universe)  which is a fact that has been explicitly checked in the bosonic model in \cite{Ito:2015mxa}.

The early and late times can be approximated by the Vandermonde and bosonic models respectively as follows. We write the fermion action as 
\begin{eqnarray}
S_F=Tr\bar{\Psi}{\Gamma}^{\mu}[A_{\mu},\Psi]=Tr\bar{\Psi}{\Gamma}^{0}[A_{0},\Psi]+Tr\bar{\Psi}{\Gamma}^{a}[A_{a},\Psi].
\end{eqnarray}
Thus for early times $A_0>>A_a$ and the first term dominates while at late time $A_0<<A_i$ and it is the second time that dominates. The bosonic and fermionic actions can be expanded as follows
\begin{eqnarray}
S_B=-\frac{1}{4g^2}(t_i-t_j)^2|(A_a)_{ij}|^2+...
\end{eqnarray}
\begin{eqnarray}
S_F=-\frac{1}{2g^2}(t_i-t_j)(\Psi_{\alpha})_{ji}({\cal C}\Gamma^{0})_{\alpha\beta}(\Psi_{\beta})_{ij}+...
\end{eqnarray}
The subleading terms are small for large $|t_i-t_j|$. The one-loop integration over bosonic degrees of freedom gives $\Delta(t)^{-2(D-1)}$ whereas the integration over fermionic degrees of freedom gives $\Delta(t)^{2d_F}$. Thus the effective potential at one-loop vanishes since $D-2-d_F=0$ and the spectrum of $A_0$ extends due to supersymmetry to infinity even for finite $N$ in the limit $\kappa\longrightarrow\infty$. In the bosonic model the eigenvalues $t_i$ are attracted to each other and the spectrum has a finite extent without any cutoff.

At early times we quench the model by including the repulsive force between the eigenvalues $t_i$ given by the fermion determinant $\Delta(t)^{2d_F}$. The Pfaffian is then approximated by 
\begin{eqnarray}
{\rm Pf}{\cal M}(A)=\Delta(t)^{2d_F}=\prod_{i>j}(t_i-t_j)^{2d_F}.
\end{eqnarray}
The corresponding model is called the Vandermonde (VDM) model in \cite{Ito:2013ywa}. It shares with the original supersymmetric model some crucial features such as spontaneous symmetry breaking and exponential expansion at very early times. It can also be accessed via Monte Carlo simulation as easily as the bosonic model which is valid at late times when it is possible to fully quench the fermions and the Pfaffian is approximated by 
\begin{eqnarray}
{\rm Pf}{\cal M}(A)=1.
\end{eqnarray}
In this bosonic model the IR cutoff (\ref{IR1}) is not required. We can observe the emergence of an exponentially expanding Universe only after some critical value of $N$ given by $N_c=110$ \cite{Ito:2015mxa}.  Also by studying this bosonic model we find the extent of space behavior %(see figure (\ref{figill6}))
\begin{eqnarray}
R^2(t)\sim t.
\end{eqnarray}
In the VDM model with $D=6$ the extent of space $R(t)$ was found for very early times to be given by the exponential fit \cite{Ito:2013ywa}
\begin{eqnarray}
\frac{R^2(t)}{R^2(t_c)}=C+(1-C)\exp(-bx)~,~x=\frac{t-t_c}{R(t_c)}.
\end{eqnarray}
This is the inflationary behavior seen also in the Lorentzian type IIB matrix model \cite{Ito:2015mem}. 

The calculation time caused by the Pfaffian in the supersymmetric model is of order $N^5$ whereas in the quenched model is of order $N^3$.

The late time behavior in the VDM model can still be studied (in fact very carefully) by using the renormalization group method. First, we note that the late time behavior is described by the inner part of the matrices $A_{\mu}$. By integrating out the outer part of the matrices $A_{\mu}$ corresponding to early times we get thus a renormalized theory with a smaller number of degrees of freedom which can be studied more efficiently by means of Monte Carlo. This ingenious idea with very interesting results for the late time behavior of the Vandermonde model is reported in  \cite{Ito:2013ywa}.

It is also found in the Lorentzian type IIB matrix model that the space-time noncommutativity (given by the double commutator $[A_0,A_a]^2$) is of order $O(1)$ only at $t=0$ (end of expansion) then it decreases at $|t|^{-1.7}$ at large $t$. The space-space noncommutativity plays a crucial role in the SSB of SO(9) at early times and dynamically disappears at later times (possibly marking the end of inflation) \cite{Kim:2011ts}.
 
The extent of time is defined by 
\begin{eqnarray}
\Delta=\frac{t_p-t_c}{R(t_c)},
\end{eqnarray}
where $t_p$ is the instant at which the extent of space becomes maximum which is by the symmetry $t\longrightarrow -t$ ($A_0\longrightarrow -A_0$) must be zero. It is a dynamical question to show whether or not $\Delta\longrightarrow \infty$ (no big crunch) in the limit $N\longrightarrow\infty$. If $\Delta$ does not diverge in the continuum limit then the extent of space has a genuine maximum and a consequence there will be a recollpase of the Universe.

%\begin{figure}[H]
%\begin{center}
%\includegraphics[width=9.0cm,angle=0]{illus6.png}
%\end{center}
%\caption{Power-law expansion in the bosonic model.}\label{figill6}
%\end{figure} 
\subsection{Role of noncommutativity}
The fundamental role played by noncommutativity in the spontaneous symmetry breaking of rotational symmetry, the emergence of an expanding Universe, and the end of inflation can be found discussed in great detail in \cite{Kim:2011cr,Kim:2011ts,Kim:2012mw}. This is done by writing down explicit Lie algebra solutions of the classical equations of motion and studying their properties. Here we will mainly follow this discussion.

We start by the simpler situation of the large $\kappa$ limit. Due to the IR cutoff $TrA_0^2/N\geq \kappa L^2$,  the eigenvalues of the configuration $A_0$ in the limit $\kappa\longrightarrow\infty$ tend to become larger and thus the first term in $TrF_{\mu\nu}F^{\mu\nu}=-2TrF_{0i}^2+TrF_{ij}^2$ becomes very large negative quantity. As a consequence the first term $TrF_{ij}^2$ must become large positive quantity in order to maintain the condition $TrF_{\mu\nu}F^{\mu\nu}=0$ \footnote{If we compare here between Lorentzian and Euclidean we find that $TrF_{\mu\nu}F^{\mu\nu}=0\Rightarrow 2TrF_{0i}^2=TrF_{ij}^2\neq 0$ (Lorentzian, noncommutative) and $TrF_{\mu\nu}F^{\mu\nu}=0\Rightarrow 2TrF_{0i}^2=TrF_{ij}^2=0$ (Euclidean, commutative).}. But we also have to remember the other IR cutoff $TrA_i^2/N=L^2$. Thus we should maximize $TrF_{ij}^2$ with the constraint  $TrA_i^2/N=L^2$. This should be done more efficiently at $t=0$ where the first term in  $TrF_{\mu\nu}F^{\mu\nu}$ has its least value (this is why the peak of $R(t)$ at $t=0$ grows with $\kappa$).

The problem now is to minimize the Lagrangian  (we set $L=1$ and $\lambda$ is a Lagrange multiplier)
\begin{eqnarray}
L=-\frac{1}{4N}TrF_{ij}^2+\frac{\lambda}{2}(\frac{1}{N}TrA_i^2-1).
\end{eqnarray}
The equations of motion with respect to $A_i$ are 
\begin{eqnarray}
[A_j,[A_j,A_i]]=\lambda A_i.
\end{eqnarray}
We consider Lie algebra solutions given by the ansatz 
\begin{eqnarray}
A_i=\chi L_i~,~i\leq d~,~A_j=0~,~j\geq d+1.
\end{eqnarray}
The $L_i$ are the generators of a compact semi-simple $d-$dimensional Lie algebra in a unitary representation. The Jacobi identity guarantees the solution of the equations of motion (it determines the value of the Lagrange multiplier $\lambda$).  The coefficient $\chi$ is determined as $\chi=\sqrt{N/Tr L_i^2}$ and as a consequence 
\begin{eqnarray}
Tr F_{ij}^2=\frac{N^2}{(Tr L_i^2)^2}Tr (f_{ijk}L_k)^2.
\end{eqnarray}
The maximum of this quantity is achieved for SU(2) Lie algebra ($L_i$ are angular momentum operators in the representation given by the direct sum of a spin $1/2$ representations and $N-2$ copies of the trivial representation and $f_{ijk}=\epsilon_{ijk}$) \cite{Kim:2011cr}
\begin{eqnarray}
\frac{1}{N}Tr F_{ij}^2=\frac{N}{(Tr L_i^2)^2}Tr (f_{ijk}L_k)^2\leq \frac{N}{(Tr L_i^2)^2}Tr (\epsilon_{ijk}L_k)^2=\frac{4N}{3}.
\end{eqnarray}
However, it can be checked that the spectrum of the $n\times n$ matrix $Q(t)=\sum_i\bar{A}_i^2(t)$ is continuous and therefore the space is actually not a sphere.  This classical picture is confirmed in the quantum theory where $\kappa$ goes to $\infty$ as $N^{1/4}$.

In general we should consider both $A_i$ and $A_0$ and extremize the action $S_B$ with the constraints $TrA_i^2/N={\rm fixed}$ and $TrA_0^2/N={\rm fixed}$. We consider then two Lagrange multipliers $\lambda$ and $\tilde{\lambda}$ and the Lagrangian 
\begin{eqnarray}
L=-\frac{1}{4N}Tr[A_{\mu},A_{\nu}][A^{\mu},A^{\nu}]-\frac{\lambda}{2}(\frac{1}{N}TrA_i^2-L^2)+\frac{\tilde{\lambda}}{2}(\frac{1}{N}TrA_0^2-\kappa L^2).
\end{eqnarray}
The equations of motion are now given by 

\begin{eqnarray}
-[A_0,[A_0,A_i]]+[A_j,[A_j,A_i]]-\lambda A_i=0~,~[A_j,[A_j,A_0]]-\tilde{\lambda}A_0=0.
\end{eqnarray}
Again we look for a solution given by the generators of a compact semi-simple Lie algebra in  unitary representations. We look for a $d-$dimensional solution which is of the form 
\begin{eqnarray}
A_i\neq 0~,~i\leq d~,~A_i=0~,~i\geq d+1.
\end{eqnarray}
We assume spatially commutative solutions, viz
\begin{eqnarray}
[A_i,A_j]=0.\label{li1}
\end{eqnarray}
We will denote the time-space commutator by
\begin{eqnarray}
[A_0,A_i]=iE_i.\label{li2}
\end{eqnarray}
We need to compute  $[A_0,E_i]=iF_i$ and $[A_i,E_j]=iG_{ij}$. By substituting in the first equation of motion we get immediately $F_i=\lambda A_i$. By substituting in the second equation of motion we get $G_{jj}=\tilde{\lambda}A_0$. Also by using the Jacobi identity between $A_i$, $A_j$ and $A_0$ we can show that $G_{ij}$ is a symmetric tensor. We split it then into a symmetric traceless tensor $M_{ij}$ and a diagonal part $\delta_{ij} H/d$ with $H=\tilde{\lambda}A_0$. Furthermore, we will assume that $M_{ij}$ is diagonal, viz $M_{ij}=M_i\delta_{ij}$ where $\sum_iM_i=0$. We have then the commutation relations 
\begin{eqnarray}
[A_0,E_i]=i\lambda A_i~,~[A_i,E_j]=i\delta_{ij}(M_i+\frac{\tilde{\lambda}}{d}A_0).\label{li3}
\end{eqnarray}
By computing the commutator $[A_0,G_{ij}]$ we get $\delta_{ij}[A_0,M_i]=[E_i,E_j]$. Thus we must have
\begin{eqnarray}
[A_0,M_i]=0~,~[E_i,E_j]=0.\label{li4}
\end{eqnarray}
Next from the two identities $[A_i,G_{jk}]=[A_j,G_{ki}]$ and $[E_i,G_{jk}]=[E_k,G_{ji}]$ and by employing Jacobi identities we derive for $k=j\neq i$ the two commutators 
\begin{eqnarray}
[A_i,M_j]=i\frac{\tilde{\lambda}}{d}E_i~,~[E_i,M_j]=i\frac{\lambda\tilde{\lambda}}{d}A_i.
\end{eqnarray}
These commutators for any $i$ and $j$ are given by 
\begin{eqnarray}
[A_i,M_j]=i\frac{\tilde{\lambda}}{d}(1-d\delta_{ij})E_i~,~[E_i,M_j]=i\frac{\lambda\tilde{\lambda}}{d}(1-d\delta_{ij})A_i.\label{li5}
\end{eqnarray}
We can then check that for $i\neq j$ and by using once again the Jacobi identity we have the commutator $[A_k,[E_i,M_j]]=-i\delta_{ik}[M_j,M_k]=0$. Thus we have the commutator  
\begin{eqnarray}
[M_i,M_j]=0.\label{li6}
\end{eqnarray}
Our spatially commutative solution is then given by the generators $A_i$, $A_0$, $E_i$ and $M_i$ satisfying the Lie algebra (\ref{li1}, (\ref{li2}), (\ref{li3}), (\ref{li4}), (\ref{li5}) and (\ref{li6}).

The solution for $d=2$ corresponds either to $SO(2,2)$ or $SO(4)$ \cite{Kim:2012mw}. There is a single solution for $d=3$ corresponding to the unique  $4-$dimensional real Lie algebra with $SO(3)$ symmetry \cite{Patera:1976ud}. In this solution $\tilde{\lambda}=M_i=0$ and $E_i=\pm \sqrt{\lambda}A_i$. However $\lambda\neq 0$ is crucial for this solution to describe the expanding behavior \cite{Kim:2011ts}. The solutions with $\lambda=\tilde{\lambda}=0$ corresponding to a Minkowski spacetime noncommuting with extra dimensions given by fuzzy spheres are given in \cite{Steinacker:2011wb}. The solutions of the corresponding Euclidean equations of motion are given in \cite{Chatzistavrakidis:2011su}.

In here we consider as a very illustrative explicit solution of the above Lie algebra the simple case of $d=1$. In this case $M_1=0$ and we get the Lie algebra 
\begin{eqnarray}
[A_0,A_1]=iE~,~[A_0,E]=i\lambda A_1~,~[A_1,E]=i\tilde{\lambda}A_0.
\end{eqnarray}
This is either and $SU(1,1)$ Lie algebra or an $SU(2)$ Lie algebra depending on the signs of $\lambda$ and $\tilde{\lambda}$. This very simple $d=1$ solution can also be used to construct higher dimensional solutions. We rotate by an $SO(9)$ transformation the above solution into a solution with only the $i$th spatial matrix non-zero given by $r_iA_1$. Obviously we must have $r_i^2=1$. Let us then consider the general solution \cite{Kim:2012mw}
\begin{eqnarray}
A_0^{\prime}=A_0\otimes {\bf 1}_K~,~A_i^{\prime}=A_1\otimes {\rm diag}(r_i^{(1)},...,r_i^{(K)})~,~\vec{r}^{(j)}.\vec{r}^{(j)}=1.
\end{eqnarray}
We can make this solution $SO(D)$ symmetric by requiring the vectors $\vec{r}{(j)}$ to lie on a sphere $S^{D-1}$, $d=1,...,9$. In other words, $A_i^{\prime}\neq 0$ only for $i\leq 4$. The geometry of spacetime is then $R\times S^{D-1}$. We are interested obviosuly in $SO(4)$ symmetric solutions.

In the $SU(1,1)$ ($SL(2,R)$) Lie algebra with $\lambda<0$ and $\tilde{\lambda}<0$ ($d=1$ solution (b) of \cite{Kim:2012mw}) we have $A_0=aT_0$, $A_1=bT_1$, $E=cT_2$ with $a^2=-\lambda$, $b^2=-\tilde{\lambda}$ and $ab=c$. Thus
\begin{eqnarray}
[T_0,T_1]=iT_2~,~[T_2,T_0]=iT_1~,~[T_1,T_2]=-iT_0.
\end{eqnarray}
We consider the primary unitary series representation (PUSR) of this algebra (with $\epsilon=0$ and $\rho$ positive) given by \cite{VilKli}
\begin{eqnarray}
&&(T_0)_{mn}=n\delta_{mn}\nonumber\\
&&(T_1)_{mn}=-\frac{i}{2}(n-i\rho+\frac{1}{2})\delta_{m,n+1}-\frac{i}{2}(n+i\rho-\frac{1}{2})\delta_{m,n-1}\nonumber\\
&&(T_2)_{mn}=-\frac{1}{2}(n-i\rho+\frac{1}{2})\delta_{m,n+1}-\frac{1}{2}(n+i\rho-\frac{1}{2})\delta_{m,n-1}
\end{eqnarray}
These are infinite dimensional unitary representations (since $SL(2,R)$ is noncompact). 

Since $A_0=aT_0$ is diagonal and $A_1=bT_1$ is tri-diagonal we can extract the time evolution of the space by considering the $3K\times 3K$ diagonal blocks defined by the $3\times 3$ submatrices $\bar{A}_0(n)$ and $\bar{A}_1(n)$ of $A_0$ and $A_1$ given explicitly by
\begin{eqnarray}
\bar{A}_0^{\prime}(n)=\bar{A}_0(n)\otimes{\bf 1}_K=a\left( \begin{array}{ccc}
n-1 & 0 & 0 \\
0 & n & 0 \\
0 & 0 & n+1 \end{array} \right)\otimes{\bf 1}_K.
\end{eqnarray}
\begin{eqnarray}
\bar{A}_i^{\prime}(n)&=&\bar{A}_i(n)\otimes{\rm diag}(r_i^{(1)},...,r_i^{(K)})\nonumber\\
&=&\frac{ib}{2}\left( \begin{array}{ccc}
0 & n+i\rho-\frac{1}{2} & 0 \\
-n+i\rho+\frac{1}{2} & 0 & n+i\rho+\frac{1}{2} \\
0 & -n+i\rho-\frac{1}{2} & 0 \end{array} \right)\otimes{\rm diag}(r_i^{(1)},...,r_i^{(K)}).
\end{eqnarray}
This is an $SO(4)$ symmetric solution. We think of $n$ as a discrete time and thus the matrices $\bar{A}_i^{\prime}(n)$ provide the state of the Universe at time $n$. The space-time noncommutativity disappears in the continuum limit by construction. The extent of space is then defined by 
\begin{eqnarray}
R^2(n)=\frac{1}{3K}Tr\bar{A}_i^{\prime 2}(n).
\end{eqnarray}
We compute immediately 
\begin{eqnarray}
R(n)=\sqrt{\frac{b^2}{3}(n^2+\rho^2+\frac{1}{4})}.
\end{eqnarray}
The continuum limit is defined by $a\longrightarrow 0$ and $\rho\longrightarrow \infty$ such that $t_0=a\rho$ (present time) and $b/a=\alpha$ are kept fixed. The time $t$ is then defined by $t=na$. We get
\begin{eqnarray}
R(t)=\sqrt{\frac{\alpha^2}{3}(t^2+t_0^2)}.
\end{eqnarray}
In the continuum limit the above Lie algebra solution becomes commutative $R\times S^3$. The cutoffs $L$ and $\kappa$ are determined to be $L\sim N/a$ and $\kappa=1/\alpha$. Thus $L\longrightarrow \infty$ if we send $N\longrightarrow \infty$ faster than $1/a$ whereas $\kappa$ remains finite in the continuum limit. We identify $R(t)$ with scale factor and compute the Hubble constant 
\begin{eqnarray}
H(t)=\frac{\dot{R}(t)}{R(t)}&=&\frac{t}{t^2+t_0^2}\nonumber\\
&=&\frac{\alpha}{\sqrt{3}R^2}\sqrt{R^2-\frac{\alpha^2}{3}t_0^2}.
\end{eqnarray}
We can also compute the parameter $w$ as
\begin{eqnarray}
w&=&-\frac{2R}{3}\frac{d\ln H}{dR}-1\nonumber\\
&=&-\frac{2 t_0^2}{3t^2}-\frac{1}{3}.
\end{eqnarray}
In particular we compute that $w$ at the present time is given by $w(t_0)=-1$. This value corresponds to a vacuum density, i.e. a cosmological constant, and hence it explains the current observed acceleration of the expansion of the Universe. Since $H=1/2t_0$ at $t=t_0$ the cosmological constant is given today by $\Lambda\sim H^2\sim 1/t_0^2$ which explains its smallness. In the future $t\longrightarrow \infty$ we have $w\longrightarrow -1/3$ and $H\longrightarrow 0$. Hence the cosmological constant will vanish in the future according to this model.
\subsection{Other related work}
The main idea behind the Lorentzian matrix model is to use matrix regularization to avoid the big bang singularity and to have the Universe with three expanding directions emerge in a phase transition associated with spontaneous symmetry breaking of rotational invariance.

There are so many good ideas out there which try also to use matrices to reproduce cosmology with or without singularity. It is practically impossible to review all these excellent ideas here. But we can mention for example the cosmology from matrix string theory in \cite{Das:2005vd,Craps:2005wd}, the cosmology from the BFSS and BFSS-type models in \cite{Freedman:2004xg,Matsuo:2008yd,Ishino:2006nx,Martinec:2006ak,Chen:2005mga,Li:2005sz},  the cosmology from IKKT-type models of noncommutative gauge theories and emergent gravity in\cite{Klammer:2009ku,Lee:2010zf,She:2005mt}, and the cosmology from the IKKT-type models of fuzzy spaces and emergent cosmology in \cite{Stern:2014aqa,Chaney:2015mfa,Chaney:2016npa,Chaney:2015ktw}. All these intimately related approaches can be called Yang-Mills matrix cosmology or perhaps emergent cosmology.

\section{Emergent gravity: introductory remarks}
\subsection{Noncommutative electromagnetism is a gravity theory}
The idea of emergent gravity from noncommutative gauge theory was first put forward by Rivelles \cite{Rivelles:2002ez}. It was then pursued vigorously by Yang \cite{Yang:2008fb,Yang:2006dk} and by Steinacker \cite{Steinacker:2010rh,Steinacker:2016vgf,Steinacker:2007dq} and . See also for example \cite{Yang:2009pm,Kawai:2016wfh}.

Emergent gravity is one of the most important development in recent years, if not the most important one,  in noncommutative field theory and noncommutative geometry and their underlying matrix models.

In the rest of this chapter we will give a deconstruction, then hopefully a reconstruction, of the emergent gravity approach of Steinacker in which $4$-dimensional Einstein (and other) gravity(ies) emerge from the "mother of all noncommutative geometry": The IKKT Yang-Mills matrix model. Emergent gravity, which was initiated as we just mentioned by Rivelles and Yang, is an approach to quantum gravity, very similar to the AdS/CFT and to the BFSS quantum mechanics, in which we maintain that gravity is equivalent to a gauge theory. More precisely, it states that noncommutative $U(1)$ gauge theory behaves in many respect as a gravitational theory. This is a profound idea which merits systematic pursuing. The work of Steinacker on obtaining Einstein equation from the IKKT model, which is deconstructed in the remainder of this chapter, is very promising although the calculation is still very semi-classical and not as rigorous as we would like it to be.

In a nutshell, noncommutative Abelian gauge theory, i.e. a noncommutative gauge theory based on a $U(1)$ group, can be reinterpreted as a gravitational theory.  In other words, gravity is equivalent to a gauge theory albeit noncommutative. This is exactly in the same spirit as that of the AdS/CFT correspondence which states that gravity is equivalent to a gauge theory albeit conformal and in one lower dimension. In contrast, the equivalence between the emergent gravity and the noncommutative gauge theory occurs in the same number of dimensions (well almost!)

On the other hand, recall that on commutative spaces the cherished Maxwell theory of electromagnetism is a $U(1)$ gauge theory. Thus, in physical terms, the essence of emergent gravity is the statement that gravity is equivalent to noncommutative electromganetism, and as such it has a dual meaning of being a theory for the (Riemannian, exclusively external) geometry of spacetime as usual, or of being a theory of the (symplectic, internal as well as external) geometry of noncommutative spacetime.

Emergent gravity is not necessarily general relativity but it is a theory of spin $2$ field with many interesting effects. And the equivalence between noncommutative electromganetism and emergent gravity is only expected to hold in the semiclassical limit defined here by the limit of noncommutativity of spacetime going to zero. Interestingly enough, if we stick to this limit the emergent metric will also satisfy Einstein equations in vacuum.

There are always issues regarding Lorentz invariance and renormalizability, which can be avoided if we go to two dimensions which might not be very interesting to many people, but the overall picture presented in this scenario is still very compelling.

Thus, electromagnetism on a noncommutative space is actually gravity. We would like to elaborate on this point a little further using the celebrated Seiberg-Witten map.  However,  before getting into that we will need some background on noncommutative field theory.

First, let us recall that a noncommutative space is a space which implements Heisenberg uncertainty principle, and thus its coordinates $x^{\mu}$ satisfy the Dirac canonical quantization  relations, viz

\begin{eqnarray}
[\hat{x}^{\mu},\hat{x}^{\nu}]=i\theta^{\mu\nu},
\end{eqnarray}
where $\theta^{\mu\nu}$ is the noncommutativity parameter. This Heisenberg algebra defines a Hilbert space ${\cal H}$ in the usual way.

The plane waves $\exp(ikx)$, the preferred basis in QFT on flat backgrounds, where $k$ is the momentum, becomes plane wave opeartors $\exp(ik\hat{x})$ which satisfy by the Baker-Cambell-Hausdorff formula the relation

\begin{eqnarray}
\exp(ik\hat{x})\exp(ip\hat{x})=\exp(i(k+p)\hat{x}-\frac{i}{2}k\theta p).
\end{eqnarray}
This is the torus algebra. This algebra can be mapped back to the commutative plane waves $\exp(ikx)$ by means of the so-called Weyl map, which maps the coordinates operators $\hat{x}$ back to the commutative coordinates $x^{\mu}$, and by utilizing the so-called Moyal-Weyl star product $*$, in place of the pointwise multiplication of operators, defined for any two functions $f$ and $g$, by the formula \cite{Ydri:2016dmy}

\begin{eqnarray}
f*g(x)=\exp(\frac{i}{2}\theta^{\mu\nu}\partial_{\mu}^y\partial_{\nu}^z)f(y)g(z)|_{y=z=x}.
\end{eqnarray}
Indeed, we can easily check that the commutative plane waves $\exp(ikx)$ satisfy with the Moyal-Weyl star product the same torus algebra satisfied by the plane wave operators, viz

\begin{eqnarray}
\exp(ik{x})*\exp(ip{x})=\exp(i(k+p)x-\frac{i}{2}k\theta p).
\end{eqnarray}
The Weyl map $\hat{x}^{\mu}\longrightarrow x^{\mu}$ extends to all operators on the Hilbert space $\hat{O}$ which are mapped to functions $O$ by the relation

\begin{eqnarray}
O(x)=\langle x|\hat{O}|x\rangle,
\end{eqnarray}
where $|x\rangle$ is an appropriate coherent state which provides a basis for the Hilbert space ${\cal H}$.

Furthermore, the derivative and the integral in the noncommutative setting are also defined by the almost obvious relations

\begin{eqnarray}
\partial_{\mu}\longrightarrow \hat{\partial}_{\mu}=-i(\theta^{-1})_{\mu\nu}\hat{x}^{\nu}.
\end{eqnarray}

\begin{eqnarray}
\int d^Dx\longrightarrow \sqrt{{\rm det} (2\pi\theta)}{\rm Tr}_{\cal H}.
\end{eqnarray}
We need now to write down dynamics, i.e. action functionals which describe completely the classical behavior of the physical system and also they are crucial ingredients in the description of the corresponding quantum behavior through path integrals.

Towards this end,  we can invoke a kind of "minimal coupling principle", which can be verified explicitly using the above formalism,  allowing us to generate correctly how gauge and matter fields couple to the spacetime noncommutativity by the simple rule:

\begin{itemize}
\item To obtain noncommutative action functionals and their noncommutative symmetries, we simply replace everywhere in the commutative action functionals and their commutative symmetries, the pointwise multiplication of functions by the Moyal-Weyl star product.
\end{itemize}

Let us give the example of massless scalar electrodynamics. In other words, electromagnetism coupled to a charged spin $0$ particle. The commutative action functional is given by

\begin{eqnarray}
S=-\frac{1}{4g^2}\int d^Dx F_{\mu\nu}F^{\mu\nu}+\frac{1}{2}\int d^Dx (D_{\mu}\phi)^{\dagger}(D^{\mu}\phi).
\end{eqnarray}
The first term is precisely Maxwell action and $\phi$ is the complex scalar field describing the charged spin $0$ particle. The field strength $F$ and the covariant derivative $D$ are given in terms of the electromagnetic (photon) field $A^{\mu}$ by the equations

\begin{eqnarray}
D_{\mu}=\partial_{\mu}-iA_{\mu}~,~F_{\mu\nu}=i[D_{\mu},D_{\nu}]=\partial_{\mu}A_{\nu}-\partial_{\nu}A_{\mu}.
\end{eqnarray}
Thus

\begin{eqnarray}
D_{\mu}\phi=\partial_{\mu}\phi-iA_{\mu}\phi.
\end{eqnarray}
This is a gauge theory meaning it is invariant under the following gauge transformations (labeled by the gauge parameter $\lambda$)

\begin{eqnarray}
&&A_{\mu}\longrightarrow A_{\mu}^{\prime}=A_{\mu}+[D_{\mu},{\lambda}]=A_{\mu}+\partial_{\mu}\lambda\nonumber\\
&&\phi\longrightarrow \phi^{'}=\phi+i\lambda\phi\nonumber\\
&&\phi^{\dagger}\longrightarrow\phi^{'\dagger}=\phi^{\dagger}-i\phi^{\dagger}\lambda.
\end{eqnarray}
The electromagnetic photon (gauge) field $A^{\mu}$ transforms in the adjoint representation of the gauge group $U(1)$ (similarly for the gauge parameter $\lambda$), the complex scalar field $\phi$ transforms in the fundamental representation of the gauge group $U(1)$ while $\phi^{\dagger}$ transforms in the anti-fundamental representation.

By invoking the above minimal coupling principle we obtain immediately the noncommutative action functional as (we may also place a hat over the various functions to distinguish them from their commutative counterparts)

\begin{eqnarray}
S=-\frac{1}{4g^2}\int d^Dx \hat{F}_{\mu\nu}*\hat{F}^{\mu\nu}+\frac{1}{2}\int d^Dx (\hat{D}_{\mu}\hat{\phi})^{\dagger}*(\hat{D}^{\mu}\hat{\phi}),
\end{eqnarray}
where
\begin{eqnarray}
\hat{D}_{\mu}={\partial}_{\mu}-i\hat{A}_{\mu}* \Rightarrow \hat{D}_{\mu}\hat{\phi}=\partial_{\mu}\hat{\phi}-i\hat{A}_{\mu}*\hat{\phi}.
\end{eqnarray}

\begin{eqnarray}
\hat{F}_{\mu\nu}=i[\hat{D}_{\mu},\hat{D}_{\nu}]_*=\partial_{\mu}\hat{A}_{\nu}-\partial_{\nu}\hat{A}_{\mu}-i[\hat{A}_{\mu},\hat{A}_{\nu}]_*.
\end{eqnarray}
The last term is very similar to the commutator term in the Yang-Mills gauge theory. As it turns out noncommutative $U(1)$ gauge theory is actually a large $U(N)$ gauge theory in a precise sense \cite{Ydri:2004vq,Yang:2007rg,Bietenholz:2004as}.

The noncommutative gauge transformations are also obtained in the same way, i.e. by replacing everywhere pointwise multiplication of functions by the Moyal-Weyl star product. We get

\begin{eqnarray}
&&\hat{A}_{\mu}\longrightarrow \hat{A}_{\mu}^{\prime}=\hat{A}_{\mu}+[\hat{D}_{\mu},\hat{\lambda}]_*=\hat{A}_{\mu}+\partial_{\mu}\hat{\lambda}-i[\hat{A}_{\mu},\hat{\lambda}]_*\nonumber\\
&&\hat{\phi}\longrightarrow \hat{\phi}^{\prime}=\hat{\phi}+i\hat{\lambda}*\hat{\phi}\nonumber\\
&&\hat{\phi}^{\dagger}\longrightarrow\hat{\phi}^{\prime\dagger}=\hat{\phi}^{\dagger}-i\hat{\phi}^{\dagger}*\hat{\lambda}.
\end{eqnarray}
This minimal coupling principle is the first indication that noncommutative geometry behaves somehow similarly to gravity in the sense that all fields regardless of their charges will couple to the noncommutativity parameter in this way, i.e. via the same prescription of replacing pointwise multiplication of functions by the Moyal-Weyl star product. Consider for example a real scalar field ${{\Phi}}$ in the adjoint representation of the gauge group $U(1)$. In the commutative setting, this field because it is neutral it can not couple to the gauge field $A^{\mu}$. In the noncommutative setting there is a coupling given by the action

\begin{eqnarray}
S=\frac{1}{2}\int d^Dx (\hat{D}_{\mu}\hat{{\Phi}})*(\hat{D}^{\mu}\hat{{\Phi}}).
\end{eqnarray}
Since $\hat{{\Phi}}$ is in the adjoint representation its covariant derivative and its gauge transformation are given respectively by

\begin{eqnarray}
\hat{D}_{\mu}\hat{{\Phi}}=\partial_{\mu}\hat{{\Phi}}-i[\hat{A}_{\mu},\hat{{\Phi}}]_*.
\end{eqnarray}

\begin{eqnarray}
&&\hat{{\Phi}}\longrightarrow \hat{{\Phi}}^{\prime}=\hat{{\Phi}}+i[\hat{\lambda},\hat{{\Phi}}]_*.
\end{eqnarray}
The star product reflecting spacetime noncommutativity allows therefore all fields to couple to the spacetime symplectic geometry in the same way. This is very reminiscent of the equivalence principle of general relativity.

\subsection{Seiberg-Witten map}
In order to exhibit the hidden gravity in noncommutative $U(1)$ gauge theory we apply the Seiberg-Witten map \cite{Seiberg:1999vs,Douglas:2001ba}.
We start with noncommutative Moyal-Weyl $U(1)$ gauge theory coupled to a complex scalar field in the fundamental representation given by the action 
\begin{eqnarray}
S=-\frac{1}{4g^2}\int d^Dx \hat{F}_{\mu\nu}*\hat{F}^{\mu\nu}+\frac{1}{2}\int d^Dx(\hat{D}_{\mu}\hat{\phi})^{\dagger}*(\hat{D}^{\mu}\hat{\phi}). 
\end{eqnarray}
The star $U(1)$ gauge transformations are given explicitly by
\begin{eqnarray}
\hat{A}_{\mu}\longrightarrow \hat{A}_{\mu}^{\prime}=\hat{A}_{\mu}+\partial_{\mu}\hat{\lambda}-i[\hat{A}_{\mu},\hat{\lambda}]_*~,~\hat{\phi}\longrightarrow \hat{\phi}^{'}=\hat{\phi}+i\hat{\lambda}*\hat{\phi}.
\end{eqnarray}
Following Seiberg and Witten we will now construct an explicit map between the noncommutative vector potential ${\hat{A}}_{\mu}$ and a commutative vector potential $A_{\mu}$ which will implement explicitly the perturbative equivalence of the above noncommutative gauge theory to a conventional gauge theory. Clearly, this map must depend both on the gauge parameter as well as on the vector potential in order to be able to achieve equivalence between the physical orbits in the two theories. Also, since the gauge field is coupled to a scalar field, the noncommutative scalar field will also be mapped to a conventional scalar field. We write then 
\begin{eqnarray}
&&\hat{A}_{\mu}(A+\delta_{\lambda}A)=\hat{A}_{\mu}(A)+\delta_{\hat{\lambda}}\hat{A}_{\mu}(A)\nonumber\\
&&\hat{\phi}(A+\delta_{\lambda}A,\phi+\delta_{\lambda}\phi)=\hat{\phi}(A,\phi)+\delta_{\hat{\lambda}}\hat{\phi}(A,\phi).
\end{eqnarray}
\begin{eqnarray}
\delta_{\hat{\lambda}}\hat{A}_{\mu}=\partial_{\mu}\hat{\lambda}-i[\hat{A}_{\mu},\hat{\lambda}]_*~,~\delta_{\hat{\lambda}}\hat{\phi}=i\hat{\lambda}*\hat{\phi}.
\end{eqnarray}
\begin{eqnarray}
\delta_{{\lambda}}{A}_{\mu}=\partial_{\mu}{\lambda}~,~\delta_{{\lambda}}{\phi}=i{\lambda}{\phi}.
\end{eqnarray}
To solve the above equation we write 
\begin{eqnarray}
\hat{A}^{\mu}=A^{\mu}+A^{\mu\prime}(A)~,~\hat{\lambda}=\lambda+\lambda^{\prime}(\lambda,A).
\end{eqnarray}
It reduces then to
\begin{eqnarray}
A_{\mu}^{\prime}(A+\delta A)-A_{\mu}^{\prime}(A)-\partial_{\mu}\lambda^{\prime}&=&-i[A_{\mu},\lambda]_*\nonumber\\
&=&\theta^{\alpha\beta}\partial_{\alpha}A_{\mu}\partial_{\beta}\lambda.
\end{eqnarray}
We have
\begin{eqnarray}
A^{\mu\prime}(A+\delta A)=A^{\mu\prime}(A)+\frac{\delta}{\delta A_{\beta}}A^{\mu\prime}.\partial_{\beta}\lambda+\frac{\delta}{\delta (\partial_{\nu}A_{\beta})}A^{\mu\prime}.\partial_{\nu}\partial_{\beta}\lambda.
\end{eqnarray}
$A^{\mu\prime}$ and $\lambda^{\prime}$ must be both of order $\theta$. Furthermore, by thinking along the lines of a derivative expansion, we know that $A^{\mu\prime}$ is quadratic in $A^{\mu}$ of the form $A\partial A$ whereas $\lambda^{\prime}$ is linear in $A^{\mu}$. The last equation above also suggests that $\lambda^{\prime}$ is proportional to $\partial_{\alpha}\lambda$. For constant $A$ the above condition reduces then to
\begin{eqnarray}
A_{\mu}(A+\delta A)-A_{\mu}(A)-\partial_{\mu}\lambda^{\prime}&=&0\Rightarrow
\frac{\delta}{\delta (\partial_{\nu}A_{\beta})}A_{\mu}^{\prime}.\partial_{\nu}\partial_{\beta}\lambda+\frac{1}{2}\theta^{\alpha\beta}\partial_{\mu}\partial_{\beta}\lambda.A_{\alpha}=0,\nonumber\\
\end{eqnarray}
where we had set
\begin{eqnarray}
\lambda^{\prime}=\frac{1}{2}\theta^{\alpha\beta}\partial_{\alpha}\lambda A_{\beta}.
\end{eqnarray}
The coefficient $1/2$ is fixed by the requirement that the second derivative in $\lambda$ cancels. The condition becomes
\begin{eqnarray}
\frac{1}{2}\frac{\delta}{\delta
(\partial_{\nu}A_{\beta})}A_{\mu}^{\prime}+\frac{1}{2}\frac{\delta}{\delta 
(\partial_{\beta}A_{\nu})}A_{\mu}^{\prime}+\frac{1}{4}\theta^{\alpha\beta}\eta^{\nu}_{\mu}A_{\alpha}+\frac{1}{4}\theta^{\alpha\nu}\eta^{\beta}_{\mu}A_{\alpha}=0.
\end{eqnarray}
A solution is given by

\begin{eqnarray}
A_{\mu}^{\prime}=-\frac{1}{2}\theta^{\alpha\beta}A_{\alpha}(2\partial_{\beta}A_{\mu}-\partial_{\mu}A_{\beta}).
\end{eqnarray}
It is now very easy to verify that this solves the Seiberg-Witten condition also for non-constant $A$.
We do now the same for the scalar field. We write 
\begin{eqnarray}
\hat{\phi}=\phi+\phi^{\prime}(A,\phi).
\end{eqnarray}
We get immediately the condition 
\begin{eqnarray}
\phi^{\prime}(A+\delta A,\phi+\delta \phi)-\phi^{\prime}(A,\phi)&=&i\hat{\lambda}*\hat{\phi}-\delta_{\lambda}\phi\nonumber\\
&=&i{\lambda}{\phi}^{\prime}+\frac{i}{2}\theta^{\mu\nu}\partial_{\mu}\lambda A_{\nu}\phi-\frac{1}{2}\theta^{\mu\nu}\partial_{\mu}\lambda\partial_{\nu}\phi. 
\end{eqnarray}
Again, we note that $\phi^{\prime}$ is of order $\theta$ and it must be proportional to $A\partial\phi$. It is not difficult to convince ourselves that the solution is given by

\begin{eqnarray}
\phi^{\prime}=-\frac{1}{2}\theta^{\mu\nu}A_{\mu}\partial_{\nu}\phi.
\end{eqnarray}
We compute the expression of the action in the new variables. We compute first
\begin{eqnarray}
\hat{F}_{\mu\nu}&=&\partial_{\mu}\hat{A}_{\nu}-\partial_{\nu}\hat{A}_{\mu}-i[\hat{A}_{\mu},\hat{A}_{\nu}]_*\nonumber\\
&=&F_{\mu\nu}+\partial_{\mu}A_{\nu}^{\prime}-\partial_{\nu}A_{\mu}^{\prime}+\theta^{\alpha\beta}\partial_{\alpha}A_{\mu}\partial_{\beta}A_{\nu}\nonumber\\ 
&=&F_{\mu\nu}+\theta^{\alpha\beta}F_{\mu\alpha}F_{\nu\beta}-\theta^{\alpha\beta}A_{\alpha}\partial_{\beta}F_{\mu\nu}.
\end{eqnarray}
We get immediately the action
\begin{eqnarray}
S_F&=&-\frac{1}{4g^2}\int d^Dx \hat{F}_{\mu\nu}\hat{F}^{\mu\nu}\nonumber\\
&=&-\frac{1}{4g^2}\int d^Dx \bigg[{F}_{\mu\nu}{F}^{\mu\nu}+2\theta^{\alpha\beta}F_{\beta}~^{\nu}\bigg(F_{\alpha}~^{\sigma}F_{\sigma\nu}+\frac{1}{4}\eta_{\nu\alpha}F^{\rho\sigma}F_{\rho\sigma}\bigg)\bigg].
\end{eqnarray}
Also we compute the covariant derivative 
\begin{eqnarray}
\hat{D}_{\mu}\hat{\phi}&=&\partial_{\mu}\hat{\phi}-i\hat{A}_{\mu}*\hat{\phi}\nonumber\\
&=&D_{\mu}\phi-\frac{1}{2}\theta^{\alpha\beta}A_{\alpha}\partial_{\beta}D_{\mu}\phi-\frac{\theta^{\alpha\beta}}{2}D_{\alpha}\phi F_{\beta\mu}.
\end{eqnarray}
\begin{eqnarray}
(\hat{D}_{\mu}\hat{\phi})^{\dagger}&=&\partial_{\mu}\hat{\phi}^{\dagger}+i\hat{\phi}^{\dagger}*\hat{A}_{\mu}\nonumber\\
&=&(D_{\mu}\phi)^{\dagger}-\frac{1}{2}\theta^{\alpha\beta}A_{\alpha}\partial_{\beta}(D_{\mu}\phi)^{\dagger}-\frac{\theta^{\alpha\beta}}{2}(D_{\alpha}\phi)^{\dagger} F_{\beta\mu}.
\end{eqnarray}
The charged scalar action becomes 
\begin{eqnarray}
S_{\phi}&=&\frac{1}{2}\int d^Dx(\hat{D}_{\mu}\hat{\phi})^{\dagger}*(\hat{D}^{\mu}\hat{\phi})\nonumber\\
&=&\frac{1}{2}\int d^Dx\bigg[({D}_{\mu}{\phi})^{\dagger}({D}^{\mu}{\phi})-\frac{1}{2}\bigg(\theta^{\mu\alpha}F_{\alpha}~^{\nu}+\theta^{\nu\alpha}F_{\alpha}~^{\mu}+\frac{1}{2}\eta^{\mu\nu}\theta^{\alpha\beta}F_{\alpha\beta}\bigg)({D}_{\mu}{\phi})^{\dagger}({D}_{\nu}{\phi})\bigg].\nonumber\\
\end{eqnarray}
The main observation of Rivelles \cite{Rivelles:2002ez} is that we can rewrite the above $\theta$-expanded actions of the noncommutative $U(1)$ gauge field  $\hat{A}_{\mu}$ and the noncommutative charged scalar field $\hat{\phi}$ as a coupling of a commutative $U(1)$ gauge field $A_{\mu}$ and a commutative charge scalar field $\phi$ to a metric $g_{\mu\nu}=\eta_{\mu\nu}+h_{\mu\nu}+\eta_{\mu\nu} h$ with $h_{\mu\nu}\eta^{\mu\nu}=0$. This metric itself is determined by the commutative $U(1)$ gauge field $A_{\mu}$ and the noncommutativity structure $\theta_{\mu\nu}$. Indeed, the dynamics of a commutative Maxwell field $A_{\mu}$ and the charged scalar field $\phi$ in a linearized gravitational field is given by the actions 

\begin{eqnarray}
S_F&=&-\frac{1}{4g^2}\int d^Dx \sqrt{-g}{F}_{\mu\nu}{F}^{\mu\nu}\nonumber\\
&=&-\frac{1}{4g^2}\int d^Dx \bigg[{F}_{\mu\nu}{F}^{\mu\nu}+2h^{\mu\nu}F_{\mu}~^{\rho}F_{\rho\nu}\bigg].
\end{eqnarray}
\begin{eqnarray}
S_{\phi}&=&\frac{1}{2}\int \sqrt{-g}d^Dx({D}_{\mu}{\phi})^{\dagger}*({D}^{\mu}{\phi})\nonumber\\
&=&\frac{1}{2}\int d^Dx\bigg[({D}_{\mu}{\phi})^{\dagger}({D}^{\mu}{\phi})-h^{\mu\nu}({D}_{\mu}{\phi})^{\dagger}({D}_{\nu}{\phi})+2h({D}_{\mu}{\phi})^{\dagger}({D}^{\mu}{\phi})\bigg].
\end{eqnarray}
We get immediately the traceless metric 
\begin{eqnarray}
h^{\mu\nu}=\frac{1}{2}(\theta^{\mu\beta}F_{\beta}~^{\nu}+\theta^{\nu\beta}F_{\beta}~^{\mu})+\frac{1}{4}\theta^{\alpha\beta}F_{\alpha\beta}\eta^{\mu\nu}.
\end{eqnarray}
The gauge field has in this setting a dual role. It couples minimally to the charged scalar field as usual but also it sources the gravitational field. In other words, the gravitational field is not just a background field since it is determined by the dynamical gauge field. We write the interval
\begin{eqnarray}
ds^2=(1+\frac{1}{4}\theta^{\alpha\beta}F_{\alpha\beta})dx_{\mu}dx^{\mu}+\theta^{\nu\beta}F_{\beta\mu}dx_{\mu}dx^{\nu}.
\end{eqnarray}
We can compute the Riemann tensor, the Ricci tensor and the Ricci scalar and find them of order one, two and two in $\theta$  respectively. Thus, to the linear order in $\theta$, the Riemann tensor is non zero while the Ricci tensor and the Ricci scalar are zero. In other words, the above metric can not describe a flat spacetime. In fact it describes a gravitational plane wave since at zero order in $\theta$ we must have ordinary electromagnetism and thus $F_{\mu\nu}$ (and hence the metric $h_{\mu\nu}$) depends on the plane wave $\exp(ikx)$ with $k^2=0$. At first order a plane wave solution $A_{\mu}=F_{\mu}\exp(ikx)$ with $k^2=0$ and $k^{\mu}F_{\mu}=0$ can also be constructed explicitly \cite{Rivelles:2002ez}. 

On the other hand, we can see that this metric describes a spacetime with a covariantly constant symplectic form $\theta$ which is also null, since to the linear order in $\theta$ we must have the equations $D_{\mu}\theta^{\alpha\beta}=0$ and $\theta_{\mu\nu}\theta^{\mu\nu}=0$, and as a consequence this spacetime is indeed a pp-wave spacetime \cite{EK}, which is precisely a gravitational plane wave as we have checked explicitly above to the first order in $\theta$. 

I
\section{Fuzzy spheres and fuzzy ${\bf CP}^n$}
The topics of this section are slightly off the main line of development of this chapter and may be skipped.
\subsection{Co-adjoint orbits}
Fuzzy spaces and their field theories and fuzzy physics are discussed for example in \cite{Balachandran:2005ew,Ydri:2001pv,Kurkcuoglu:2004gf,Balachandran:2002ig,O'Connor:2003aj,Steinacker:2004mq,Abe:2010an,Karabali:2004xq}. Fuzzy spaces are finite dimensional approximations to the algebra of functions on continuous manifolds which preserve the isometries and (super)symmetries  of the underlying manifolds. Thus by construction the corresponding field theories contain a finite number of degrees of freedom. The basic and original motivation behind fuzzy spaces is non-perturbative regularization of quantum field theory similar to the familiar technique of lattice regularization. See for example \cite{Grosse:1996mz,Grosse:1995ar}. 

Another very important motivation lies in the fact that string theory suggests that spacetime may be fuzzy and noncommutative at its fundamental level \cite{Ahluwalia:1993dd,Alekseev:1999bs,Hikida:2001py}. For older and other more recent motivations see \cite{Snyder:1946qz,Yang:1947ud,Doplicher:1994tu}.

It is well established that the specification of fuzzy spaces requires the language of Connes' noncommutative geometry. In particular, following Connes \cite{Connes:1994yd} and Fr\"ohlich and Gaw\c{e}dzki \cite{Frohlich:1993es}, the geometry of a riemannian manifold ${\cal M}$ can be reconstructed from the so-called spectral triple $({\cal A},{\cal H},{\cal D})$ where ${\cal A}=C^{\infty}(\infty)$ is the algebra of smooth bounded functions on the manifold, ${\cal H}$ is the Hilbert space of square-integrable spinor functions on ${\cal M}$, and ${\cal D}$ is the Dirac operator on ${\cal M}$ which encodes all the information about the metric aspects of the manifold ${\cal M}$. In the case of the absence of spinors the Dirac operator can be replaced by the Laplace-Beltrami operator ${\Delta}$ on ${\cal M}$. Similarly, a fuzzy space will be given by a sequences of triples
\begin{eqnarray}
({\rm Mat}_{d_L},{\cal H}_L,\Delta_L),\label{triple}
\end{eqnarray}
where ${\rm Mat}_{d_L}$ is the algebra of $d_L\times d_L$ hermitian matrices with inner product $(A,B)=TrA^+B/d_L$, ${\cal H}_L={\bf C}^{d_L^2}$ is the Hilbert space of the $d_L-$dimensional matrix algebra ${\rm Mat}_{d_L}$, and $\Delta_L$ is an appropriate Laplacian acting on these matrices which encodes in a precise sense the metric aspects of the fuzzy space. For example, the dimension of the space is given by the growth of the number of eigenvalues of the Laplacian. In the limit $L\longrightarrow \infty$ we obtain the spectral triple $({\cal A},{\cal H},\Delta)$ associated with the corresponding commutative manifold. The fuzzy Laplacian $\Delta_L$ is typically a truncated version of the commutative Laplacian $\Delta$ with the same isometries and symmetries. The fuzzy geometry can be mapped to an algebra of functions with an appropriate star product by constructing the corresponding Weyl map.

% By the Cayley-Hamilton theorem, there is a natural cutoff on the number of degrees of freedom contained in functions on fuzzy spaces and these functions turn out to be in one-to-one correspondence with matrices in   ${\rm Mat}_N$. 
Obviously, the corresponding fuzzy field theories are non-perturbatively regularized since they contain a finite number of degrees of freedom given precisely by $d_L^2$ with the correct limiting commutative behavior. The fuzzy approach is thus the most natural one in Monte Carlo simulations of noncommutative field theories or in Monte Carlo simulations of commutative field theories where (super)symmetry plays a crucial role.
  
Construction of fuzzy spaces by quantizing compact symplectic manifolds, via geometric quantization \cite{Sniatycki:1995sn,Woodhouse:1992de,Nair:2005iw,Hu:2001ue},  is equivalent to the construction of a quantum Hilbert space from a classical phase space. More explicitly, quantization is the construction of a correspondence between the algebra of Poisson brackets, represented by real functions generating canonical transformations on the phase space or symplectic manifold, and the algebra of commutators, represented by hermitian operators generating an irreducible unitary transformations on the Hilbert space.  The irreducibility of the representation is equivalent, in the context of geometric quantization, to the holomorphic or polarization condition. For example, in the usual one-dimensional quantum mechanics given by $[\hat{x},\hat{p}]=i\hbar$ the wave functions depend only on $x$ and not $x$ and $p$. In general, within geometric quantization where $1)$ we consider a prequantum line bundle on the phase space with curvature given by the symplectic $2-$form and $2)$ impose a complex structure on the phase space in which the symplectic $2-$form is identified as a K\"ahler form, the resulting Hilbert space is given by sections of a polarized line bundle satisfying a holomorphic condition. This is intimately related to the Borel-Weil-Bott theorem \cite{Sniatycki:1995sn,Woodhouse:1992de,Nair:2005iw,Hu:2001ue} which states that all unitary irreducible representations of a compact Lie group $G$ are realized by holomorphic sections of a complex line bundle on a coset space $G/T$, where $T$ is the maximal torus of $G$,  $G/T$ is a K\"ahler manifold, and where the group $G$ acts on holomorphic sections by right translations.

Examples of compact symplectic manifolds are the so-called co-adjoint or adjoint orbits of compact semi-simple Lie groups $G$ which can be geometrically quantized by quantizing their underlying symplectic $2-$forms when they satisfy the Dirac quantization condition. Co-adjoint orbits are coset spaces $G/H_t=\{gtg^{-1}:g\in G,t\in \underline{G}\}$ where $\underline{G}$ is the Lie algebra of $G$ and $H_t\subset G$ is the stabilizer of $t$. The fuzzy co-adjoint orbits corresponding to  $G/H_t$ are such that their Hilbert spaces consist of holomorphic sections of a complex line bundle over $G/H_t$ associated with  unitary irreducible representations of $G$. The fuzzy coset spaces constructed so far satisfying (\ref{triple}) are fuzzy complex projective spaces, which are mostly degenerate co-adjoint orbits with dimensions ${\rm dim}G/H_t={\rm dim}G-{\rm dim}H_t$, given by
\begin{eqnarray}
{\bf CP}^k=G/H_t=SU(k+1)/U(k).
\end{eqnarray}
Another class is given by flag manifolds in which $H$ coincides with the maximal torus $H$ of $G$ and the dimension of the co-adjoint orbits becomes maximal given by ${\rm dim}G/H_t={\rm dim}G-{\rm rank}G$. See for example \cite{Murray:2006pi}.

%In previous chapters, we have already discussed in great detail the fuzzy sphere ${\bf CP}^1={\bf S}^2$ and with some considerable detail fuzzy ${\bf CP}^k$ as well as fuzzy tori. 
In the remainder of this section we will discuss further fuzzy ${\bf CP}^k$ and write down their Yang-Mills matrix models. The four-dimensional case of fuzzy ${\bf S}^2\times {\bf S}^2$ is discussed in \cite{CastroVillarreal:2005uu,Azuma:2005pm,Behr:2005wp,Vaidya:2003ew,Imai:2003ja,Kaneko:2005pw} while the case of fuzzy ${\bf CP}^2$ is also discussed in \cite{Alexanian:2001qj,Azuma:2004qe,Grosse:2004wm,Dou:2007in,Grosse:1999ci} and \cite{Kitazawa:2002xj,Imai:2003jb}. The physics on a single fuzzy sphere is studied for example in \cite{Anagnostopoulos:2005cy,Valtancoli:2002rx,Azuma:2004ie,Azuma:2005bj,Azuma:2004zq,Ishii:2008tm,Ishiki:2009vr,Ishiki:2008vf,DelgadilloBlando:2008vi,DelgadilloBlando:2007vx,OConnor:2006iny}. For related topics see also \cite{Dolan:2006tx,Balachandran:2001dd} and \cite{Karabali:2002im,Karabali:2003bt,Karabali:2004km,Karabali:2006eg,Ho:2000br}. Fuzzy projective spaces  are examples of co-adjoint orbits and homogeneous spaces  which, among many other properties, admit an underlying symplectic structure. The existence of fuzzy co-adjoint orbits and similar fuzzy spaces satisfying (\ref{triple}) relies on the single fact that the corresponding symplectic structure is quantizable.  For recent work on fuzzy physics in $2$ and $4$ dimensions see \cite{Ydri:2016daf,Ydri:2016osu,Ydri:2016kua,Ydri:2015vba,Ydri:2015zsa}. As a concert example here we will consider in the following the case of fuzzy  ${\bf CP}^2$ (and fuzzy ${\bf CP}^k$) and more importantly the case of fuzzy ${\bf S}^4$ which is relevant to the emergence of Einstein gravity from the IKKT matrix model. %The next two sections are mostly taken from  \cite{Dou:2007in}.

%\subsection{The fuzzy spheres ${\bf S}^2$ and ${\bf S}^2\times{\bf S}^2$}

\subsection{Fuzzy projective space ${\bf CP}^2$}
Let $T_a$, $a=1,...,8$, be the generators of $SU(3)$ in the symmetric irreducible
 representation $(n,0)$ of dimension $N=\frac{1}{2}(n+1)(n+2)$. They satisfy
\begin{eqnarray}
[T_a,T_b]=if_{abc}T_c~\label{comm}
\end{eqnarray}
and
\begin{eqnarray}
T_a^2=\frac{1}{3}n(n+3)\equiv
|n|^2~,~d_{abc}T_aT_b=\frac{2n+3}{6}T_c.\label{idd}
\end{eqnarray}
Let $t_a={{\lambda}_a}/{2}$, where ${\lambda}_a$ are the usual Gell-Mann matrices,   be the generators of $SU(3)$ in the fundamental
 representation $(1,0)$ of dimension $N=3$. They also satisfy
\begin{eqnarray}
&&2t_at_b=\frac{1}{3}{\delta}_{ab}+(d_{abc}+if_{abc})t_c\nonumber\\
&&tr_3t_at_b=\frac{1}{2}{\delta}_{ab}~,~tr_3t_at_bt_c=\frac{1}{4}(d_{abc}+if_{abc}).
\end{eqnarray}
The $N-$dimensonal generator $T_a$ can be obtained by taking the symmetric product of $n$ copies of the fundamental $3-$dimensional generator $t_a$, viz
\begin{eqnarray}
T_a=(t_a{\otimes}{\bf 1}{\otimes}...{\otimes}{\bf 1}+{\bf 1}{\otimes}t_a{\otimes}...{\otimes}{\bf 1}+...+{\bf 1}{\otimes}{\bf 1}{\otimes}...{\otimes}t_a)_{\rm symmetric}.
\end{eqnarray}
The commutative ${\bf CP}^2$ is the space of all unit vectors $|\psi>$ in ${\bf C}^3$ modulo the phase. Thus $e^{i\theta}|\psi>$,
for all $\theta {\in}[0,2\pi[$, define the same point on ${\bf CP}^2$. It is obvious that all these vectors $e^{i\theta}|\psi>$ correspond to
 the same projector $P=|\psi><\psi|$. Hence ${\bf CP}^2$ is the space of all projection operators of rank one on ${\bf C}^3$. Let ${\bf H}_N$
  and ${\bf H}_3$ be the Hilbert spaces of the $SU(3)$ representations  $(n,0)$ and $(1,0)$ respectively.
We will define fuzzy ${\bf CP}^2$ through  the canonical $SU(3)$ coherent states as follows. Let $\vec{n}$ be a vector in ${\bf R}^8$ and 
 we define the projector
\begin{eqnarray}
P_3=\frac{1}{3}{\bf 1}+n_at_a.
\end{eqnarray}
The requirement $P_3^2=P_3$ leads to the condition that $\vec{n}$ is a point on ${\bf CP}^2$ satisfying the equations

\begin{eqnarray}
[n_a,n_b]=0~,~n_a^2=\frac{4}{3}~,~d_{abc}n_an_b=\frac{2}{3}n_c.
\end{eqnarray}
We can write
\begin{eqnarray}
P_3=|\vec{n},3><3,\vec{n}|.
\end{eqnarray}
We think of $|\vec{n},3>$ as the coherent state in ${\bf H}_3$, level $3\times 3$ matrices,  which is localized at
the point $\vec{n}$ of  ${\bf CP}^2$. Therefore the coherent state $|\vec{n},N>$ in ${\bf H}_N$, level $N\times N$ matrices,
which is localized around the point $\vec{n}$ of  ${\bf CP}^2$ is defined by the projector
\begin{eqnarray}
P_N=|\vec{n},N><N,\vec{n}|=(P_3{\otimes}P_3{\otimes}...{\otimes}P_3)_{\rm symmetric}.
\end{eqnarray}
We compute that
\begin{eqnarray}
tr_3t_aP_3=<\vec{n},3|t_a|\vec{n},3>=\frac{1}{2}n_a~,~
tr_NT_aP_N=<\vec{n},N|T_a|\vec{n},N>=\frac{n}{2}n_a.
\end{eqnarray}
Hence it is natural to identify fuzzy ${\bf CP}^2$ at level $N=\frac{1}{2}(n+1)(n+2)$ by the coordinate operators

\begin{eqnarray}
x_a=\frac{2}{n}T_a.
\end{eqnarray}
They satisfy

\begin{eqnarray}
[x_a,x_b]=\frac{2i}{n}f_{abc}x_c~,~x_a^2=\frac{4}{3}(1+\frac{3}{n})~,~d_{abc}x_ax_b=\frac{2}{3}(1+\frac{3}{2n})x_c.
\end{eqnarray}
Therefore in the large $N$ limit we can see that the algebra of $x_a$ reduces to the continuum algebra of $n_a$. Hence $x_a{\longrightarrow}n_a$
 in the commutative limit $N{\longrightarrow}{\infty}$.

The algebra of ~functions on fuzzy  ${\bf CP}^2$ is identified with the algebra of $N{\times}N$ matrices ${\rm Mat}_N$ generated by
 all polynomials in the coordinate operators $x_a$. Recall that $N=\frac{1}{2}(n+1)(n+2)$. The left action of $SU(3)$
  on this algebra is generated by $(n,0)$ whereas the right action is generated by $(0,n)$. Thus the algebra ${\rm Mat}_N$ decomposes
  under the action of $SU(3)$ as
\begin{eqnarray}
(n,0){\otimes}(0,n)={\otimes}_{p=0}^n(p,p).
\end{eqnarray}
A general function on fuzzy  ${\bf CP}^2$ is therefore written as

\begin{eqnarray}
F=\sum_{p=0}^nF_{I^2,I_3,Y}^{(p)}T_{I^2,I_3,Y}^{(p,p)}
\end{eqnarray}
$T_{I^2,I_3,Y}^{(p,p)}$ are $SU(3)$ polarization tensors in the irreducible representation $(p,p)$. $I^2,I_3$ and $Y$ are the square of the isospin,
the third component of the isospin and the hypercharge quantum numbers which characterize $SU(3)$ representations.

The derivations on fuzzy  ${\bf CP}^2$ are defined by the commutators $[T_a,..]$. The Laplacian is then obviously given by ${\Delta}_N=[T_a,[T_a,...]]$.
Fuzzy ${\bf CP}^2$ is completely determined by the sequence of spectral triples $(Mat_N,{\Delta}_N,{\bf H}_N)$. 

\subsection{Tangent projective module on fuzzy ${\bf CP}^2$}
We will introduce fuzzy gauge fields $A_a$, $a=1,...,8$, through the covariant derivatives $D_a$, $a=1,...,8$, as follows
\begin{eqnarray}
D_a=T_a+A_a.
\end{eqnarray}
The $D_a$ are $N{\times}N$ hermitian matrices which transform covariantly under the action
of $U(N)$. % as follows $D_a{\longrightarrow}UD_aU^{+}$ where
%$U{\in}U(N)$. Hence $A_a$ are   $N{\times}N$ matrices which
%transform as $A_a{\longrightarrow}UA_aU^{+}+U[T_a,U^{+}]$. 
In order for the field $\vec{A}$ to be a $U(1)$ gauge field on fuzzy
${\bf CP}^2$ it must satisfy some additional constraints so
that only four of its components are non-zero. These are the
tangent
 components to ${\bf CP}^2_n$. The other four components of $\vec{A}$ are normal to ${\bf CP}^2_n$ and in general they will be projected out from the model.

Let us go back to the commutative ${\bf CP}^2$ and let us consider a gauge field $A_a$, $a=1,...,8$,
 which is strictly tangent to ${\bf CP}^2$ . By construction this gauge field must satisfy
\begin{eqnarray}
A_a=P_{ab}^TA_b~,~P^T=(n_aAdt_a)^2.\label{cond}
\end{eqnarray}
The $P^T$ is the projector which defines the tangent bundle over ${\bf CP}^2$. The normal bundle over ${\bf CP}^2$ will be defined by the projector
$P^N=1-P^T$. Explicitly these are  given by
\begin{eqnarray}
P^T_{ab}=n_cn_d(Adt_c)_{ae}(Adt_d)_{eb}=n_cn_df_{cae}f_{dbe}~,~P^N_{ab}={\delta}_{ab}-n_cn_df_{cae}f_{dbe}.
\end{eqnarray}
In above we have used the fact that the generators in the adjoint representation $(1,1)$ satisfy $(Adt_a)_{bc}=-if_{abc}$.
 Remark that we have the identities $n_aP^T_{ab}=n_bP^T_{ab}=0$. Hence the condition (\ref{cond}) takes the natural form
\begin{eqnarray}
n_aA_a=0.\label{cond1}
\end{eqnarray}
This is one condition which allows us to reduce the number of independent components of $A_a$ by one. We know that there must be three
 more independent constraints which the tangent field $A_a$ must satisfy since it has only $4$ independent components. To find them we
  start from the identity 

\begin{eqnarray}
d_{abk}d_{cdk}=\frac{1}{3}\bigg[{\delta}_{ac}{\delta}_{bd}+{\delta}_{bc}{\delta}_{ad}-{\delta}_{ab}{\delta}_{cd}+f_{cak}f_{dbk}+f_{dak}f_{cbk}\bigg].\label{24}
\end{eqnarray}
Thus
\begin{eqnarray}
n_cn_dd_{abk}d_{cdk}=\frac{2}{3}\bigg[{n}_{a}{n}_{b}-\frac{2}{3}{\delta}_{ab}+n_cn_df_{cak}f_{dbk}\bigg].
\end{eqnarray}
By using the fact that $d_{cdk}n_cn_d=\frac{2}{3}n_k$ we obtain

\begin{eqnarray}
d_{abk}n_k={n}_{a}{n}_{b}-\frac{2}{3}{\delta}_{ab}+n_cn_df_{cak}f_{dbk}.\label{plk55}
\end{eqnarray}
Hence it is a straightforward calculation to find that the gauge
field $A_a$ must also satisfy the conditions
\begin{eqnarray}
d_{abk}n_kA_b=\frac{1}{3}A_a.\label{cond2}
\end{eqnarray}
In the case of ${\bf S}^2$ the projector $P^T$ takes the simpler form $P^T_{ab}={\delta}_{ab}-n_an_b$ and hence $P^N_{ab}=n_an_b$.
 From equation (\ref{plk55}) we have on ${\bf CP}^2$
\begin{eqnarray}
P_{ab}^T=d_{abc}n_c-n_an_b+\frac{2}{3}{\delta}_{ab}~,~P_{ab}^N=-d_{abc}n_c+n_an_b+\frac{1}{3}{\delta}_{ab}.
\end{eqnarray}
%If we choose to sit on the ``north pole'' of ${\bf CP}^2$, i.e $\vec{n}=(0,0,0,0,0,0,0,-\frac{2}{\sqrt{3}})$
%then we can find that $P^T=diag(0,0,0,1,1,1,1,0)$ and as a consequence $P^N=(1,1,1,0,0,0,0,1)$ . So $Adt_a$, $a=1,2,3,8$ correspond to
% the normal directions while $Adt_a$, $a=4,5,6,7$ correspond to the tangent directions.
%Indeed by substituting $\vec{n}=(0,0,0,0,0,0,0,-\frac{2}{\sqrt{3}})$ in equation (\ref{cond2})
% and using $d_{8ij}=\frac{1}{\sqrt{3}}{\delta}_{ij}$ where $i,j=1,2,3$ and $d_{8\alpha \alpha}=-\frac{1}{2\sqrt{3}}$
% where $\alpha =4,5,6,7$ and $d_{888}=-\frac{1}{\sqrt{3}}$ we get $A_1=A_2=A_3=A_8=0$ which is what we want. 
As it turns out, the constraint (\ref{cond2})
 already contains (\ref{cond1}). In other words, it contains exactly the correct number of equations needed to project out the gauge field
 $A_a$ onto the tangent bundle of ${\bf CP}^2$.
Let us also say that given any commutative gauge field $A_a$ which does not satisfy the constraints (\ref{cond1}) and (\ref{cond2})
 we can always make it tangent by applying the projector $P^T$. Thus we will have the tangent gauge field
\begin{eqnarray}
A_a^T=P_{ab}^TA_b=d_{abc}n_cA_b-n_a(n_bA_b)+\frac{2}{3}A_a.
\end{eqnarray}
Similarly, the fuzzy gauge field must satisfy some conditions  which should reduce to (\ref{cond1}) and (\ref{cond2}).
As it turns out, constructing a tangent fuzzy gauge field using an expression like (\ref{cond}) is a highly non-trivial task due to   $1)$
 gauge covariance problems and  $2)$ operator ordering problems. However, implementing (\ref{cond1}) and (\ref{cond2}) in the fuzzy setting
 is quite easy  since we will only need to return to the covariant derivatives $D_a$ and require them to satisfy the $SU(3)$ identities  (\ref{idd}), viz
\begin{eqnarray}
&&D_a^2=\frac{1}{3}n(n+3)\nonumber\\
&&d_{abc}D_aD_b=\frac{2n+3}{6}D_c.\label{cond3}
\end{eqnarray}
So $D_a$ are almost the $SU(3)$ generators except that they fail to satisfy the  fundamental commutation relations of $SU(3)$ given by equation (\ref{comm}).
This failure is precisely measured by the curvature of the gauge field $A_a$, namely
\begin{eqnarray}
F_{ab}&=&i[D_a,D_b]+f_{abc}D_c\nonumber\\
&=&i[T_a,A_b]-i[T_b,A_a]+f_{abc}A_c+i[A_a,A_b].
\end{eqnarray}
This has the correct commutative limit which is clearly given by the usual
curvature on ${\bf CP}^2$, viz by $F_{ab}=i{\cal L}_aA_b-i{\cal
L}_bA_a+f_{abc}A_c+i[A_a,A_b]$. 
%To check that this fuzzy gauge
%field $A_a$ has the correct degrees of freedom we need to check
%that the identities (\ref{cond3}) reduce to (\ref{cond1}) and
%(\ref{cond2}) in the continuum limit $n{\longrightarrow}\infty $.
%This fact is quite straightforward to verify and we leave it as an
%exercise.
\subsection{Yang-Mills matrix models for fuzzy ${\bf CP}^k$}
Next, we need to write down actions on fuzzy ${\bf CP}^2$. The first piece is the usual Yang-Mills action
\begin{eqnarray}
S_{\rm YM}=\frac{1}{4g^2N}TrF_{ab}^2.
\end{eqnarray}
By construction it has the correct commutative limit.

The second  piece in the action is a potential term which has to
implement the constraints (\ref{cond3}) in some limit. Indeed, we
will not impose these constraints rigidly on the path integral but
we will include their effects by adding to the action a very
special potential term. In other words, we will not assume that
$D_a$ satisfy (\ref{cond3}). To the end of writing this potential
term  we will  introduce the four normal scalar fields on fuzzy
${\bf CP}^2_n$ by the equations (see equations (\ref{cond3}))

\begin{eqnarray}
&&{\Phi}=\frac{1}{n}(D_a^2-\frac{1}{3}n(n+3))=\frac{1}{2}x_aA_a+\frac{1}{2}A_ax_a+\frac{1}{n}A_a^2{\longrightarrow}n_aA_a,
\end{eqnarray}
and
\begin{eqnarray}
{\Phi}_c=\frac{1}{n}(d_{abc}D_aD_b-\frac{2n+3}{6}D_c)&=&\frac{1}{2}d_{abc}x_aA_b+\frac{1}{2}d_{abc}A_ax_b-\frac{2n+3}{6n}A_c+\frac{1}{n}d_{abc}A_aA_b\nonumber\\
&{\longrightarrow}&d_{abc}n_aA_b-\frac{1}{3}A_c.\label{cond4}
\end{eqnarray}
We add to the Yang-Mills action the potential term
\begin{eqnarray}
V=\frac{M_0^2}{N} Tr{\Phi}^2 +\frac{M^2}{N} Tr_N{\Phi}_a^2.
\end{eqnarray}
In the limit where the parameters $M_0^2$ and $M^2$ are taken to be very large positive numbers we can see that only configurations $A_a$, or equivalently $D_a$, such that $\Phi=0$ and ${\Phi}_c=0$ dominate the path integral which is precisely what we want. 

The total action is then given by
\begin{eqnarray}
S&=&\frac{1}{2g^2}Tr_NF_{ab}^2+\beta Tr_N{\Phi}^2+M^2 Tr_N{\Phi}_a^2\nonumber\\
&=&\frac{1}{g^2N}Tr\bigg[-\frac{1}{4}[D_a,D_b]^2+if_{abc}D_aD_bD_c\bigg]+\frac{3n}{4g^2N}Tr\Phi +\frac{M_0^2}{N}  Tr_N{\Phi}^2+ \frac{M^2}{N} Tr_N{\Phi}_a^2.\nonumber\\
\end{eqnarray}
This is the desired Yang-Mills matrix model in which fuzzy ${\bf CP}^2$ is described as a noncommutative brane solution of the equations of motion. To obtain the corresponding Yang-Mills matrix models for fuzzy ${\bf CP}^k$  we simply replace the $SU(3)$ constants $f$ and $d$ by their $SU(k+1)$ values and extend the indices $a,b,c,...$ from $1$ to $k(k+2)$. The case of the sphere is much simpler since $d=0$ for $SU(2)$. We obtain in this case (with $f=\epsilon$)
\begin{eqnarray}
S
&=&\frac{1}{g^2N}Tr\bigg[-\frac{1}{4}[D_a,D_b]^2+i\epsilon_{abc}D_aD_bD_c\bigg]+\frac{3n}{4g^2N}Tr\Phi +\frac{M_0^2}{N}  Tr{\Phi}^2+ \frac{M^2}{N}\frac{(2n+3)^2}{36n} Tr{\Phi}.\nonumber\\
\end{eqnarray}
This action will be studied in great detail in subsequent sections. Extension of this action to fuzzy ${\bf S}^2\times{\bf S^2}$ and higher cartesian products of the fuzzy sphere is straightforward.

%\section{Star Product and Coherent States on ${\bf CP}^k$}
%We will closely follow \cite{Balachandran:2001dd}. Here Latin indices refer to $SU(k+1)$, viz $a,b=1,...,k(k+2)$. This section can be skipped by experts. 

\subsection{Coherent states}
We will closely follow \cite{Balachandran:2001dd}. Here Latin indices refer to $SU(k+1)$, viz $a,b=1,...,k(k+2)$. This section can be skipped by experts. 

Classical ${\bf CP}^{k}$ can be given by the projector

\begin{equation}
P=\frac{1}{k+1}{\bf 1}+\alpha_{k}n^at_a. \label{projector}
\end{equation}
The requirement that $P^2=P$ will lead to the three equations
\begin{eqnarray}
\vec{n}^2&=&1\nonumber\\
d_{abc}n^an^b&=&\frac{2}{\alpha_k}\frac{k-1}{k+1}n^c,\nonumber\\
{\alpha}_k&=&{\pm}\sqrt{\frac{2k}{k+1}}.
\end{eqnarray}
This defines ${\bf CP}^{k}$ as embedded in ${\bf R}^{k+2}$. First, let us specialize the projector (\ref{projector}) to the "north" pole of ${\bf CP}^{k}$:
\begin{equation}
\vec{n}_0=(0,0,...,1).
\end{equation}
We have then the projector
\begin{equation}
P_0=\frac{1}{k+1}{\bf 1}+\alpha_k t_{k(k+2)}.
\end{equation}
Now, by using the result
\begin{equation}
t_{k(k+2)}=\frac{1}{\sqrt{2k(k+1)}}{\rm diag}(1,1,...,1,-k),
\end{equation}
we get
\begin{equation}
P_0={\rm diag}(0,0,...,1),
\end{equation}
if we choose the minus sign for ${\alpha}_N$, namely
\begin{equation}
{\alpha}_k=-\sqrt{\frac{2k}{k+1}}.
\end{equation}
So at the "north" pole , our projector projects down to the state
\begin{equation}
|{\psi}_0>=(0,0,...,1)\label{fiducialnorth}
\end{equation}
of the Hilbert space ${\bf C}^{k+1}$ on which the defining representation of $SU(k+1)$ is acting.

A general point $\vec{n}{\in}{\bf CP}^{k}$ can be obtained from $\vec{n}_0$ by the action of a certain
element $g{\in}SU(k+1)$
\begin{equation}
\vec{n}=g\vec{n}_0.
\end{equation}
$P$ will then project down to the state
\begin{equation}
|\psi>=g|{\psi}_0>\label{fiducialgeneral}
\end{equation}
of ${\bf C}^{k+1}$. One can show that
\begin{equation}
P=|\psi><\psi|=g|{\psi}_0><{\psi}_0|g^{+}=gP_0g^{+},
\end{equation}
provided
\begin{equation}
gt_{k(k+2)}g^{+}=n^{a}t_a.
\end{equation}
This last equation is the usual definition of ${\bf C}{\bf P}^{k}$. Under $g{\longrightarrow}gh$ where
$h{\in}U(k)$ we have $ht_{k(k+2)}h^{+}=t_{k(k+2)}$, i.e. $U(k)$ is the stability group of $t_{k(k+2)}$ and hence
\begin{equation}
{\bf C}{\bf P}^{k}=SU(k+1)/U(k).
\end{equation}
Points $\vec{n}$ of ${\bf CP}^{k}$ are then equivalent classes $[g]=[gh],h{\in}U(k)$ .

In the case of $SU(2)$, fuzzy ${\bf S}^2_N$ is the algebra of operators generated by the orbital
angular momenta $L_i,i=1,2,3$, where $[L_i,L_j]=i{\epsilon}_{ijk}L_k$, and $\sum_{i=1}^3L_i^2=l(l+1)$. Since these
operators define the IRR $l$ of $SU(2)$, fuzzy ${\bf S}^2$ will act on the Hilbert space $H_l^{(2)}$, which is the
$d_l^{(2)}=(2l+1)-$dimensional irreducible representation of $SU(2)$, i.e. $N=2l+1$. This representation can be obtained from the
symmetric product of $2l$ fundamental representations ${\bf 2}$ of $SU(2)$. Given an element $g{\in}SU(2)$, its
$l-$representation matrix $U^{(l)}(g)$ is given as follows
\begin{equation}
U^{(l)}(g)=U^{(\bf 2)}(g){\otimes}_s...{\otimes}_sU^{(\bf 2)}(g),2l-{\rm times}.
\end{equation}
$U^{(\bf 2)}(g)$ is the spin $\frac{1}{2}$ representation of $g{\in}SU(2)$ .

Similarly, fuzzy ${\bf C}{\bf P}^{k}$ is the algebra of all operators which act on the Hilbert space $H_l^{(k+1)}$, where $H_l^{(k+1)}$ is the $d_l^{(k)}-$ dimensional irreducible representation of $SU(k+1)$ obtained from the symmetric product of $2l$ fundamental representations ${\bf N}$ of $SU(N)$, where
\begin{eqnarray}
d_l^{(k)}=\frac{(k+2l)!}{k!(2l)!}.
\end{eqnarray}
Remark that for $l=\frac{1}{2}$ we have $d_{\frac{1}{2}}^{(k)}=k+1$ and therefore $H_{\frac{1}{2}}^{(k+1)}={\bf C}^{k+1}$ is the
fundamental representation of $SU(k+1)$. 

Clearly, the states $|{\psi}_0>$ and $|\psi>$ of $H_{\frac{1}{2}}^{(k+1)}$, given by equations  (\ref{fiducialnorth})
and (\ref{fiducialgeneral}), will correspond in $H_l^{(k+1)}$ to the states $|\vec{n}_0,l>$ and $|\vec{n},l>$ respectively, so that $|{\psi}_0>=|\vec{n}_0,\frac{1}{2}>$ and $|\psi>=|\vec{n},\frac{1}{2}>$. Equation (\ref{fiducialgeneral}) becomes
\begin{equation}
|\vec{n},l>=U^{(l)}(g)|\vec{n}_0,l>.\label{fundamental}
\end{equation}
$U^{(l)}(g)$ , where $g{\in}SU(k+1)$ , is the representation given by
\begin{equation}
U^{(l)}(g)=U^{(\bf k+1)}(g){\otimes}_s...{\otimes}_sU^{(\bf k+1)}(g),2l-{\rm times}.
\end{equation}
To any operator $\hat{F}$ on $H_l^{(k+1)}$, which can be thought of as a fuzzy function on fuzzy ${\bf CP}^{k}$, we associate a "classical" function $F_l(\vec{n})$ on a classical ${\bf CP}^{k}$  by
\begin{equation}
F_l(\vec{n})=<\vec{n},l|\hat{F}|\vec{n},l>,\label{map}
\end{equation}
such that the product of two such operators $\hat{F}$ and $\hat{G}$ is mapped to the star product of the
corresponding two functions by the relation
\begin{equation}
F_l*G_l(\vec{n})=<\vec{n},l|\hat{F}\hat{G}|\vec{n},l>.\label{starproduct1}
\end{equation}

\subsection{Star product}

This is a very long calculation which I would like to do once and for all \cite{Balachandran:2001dd}. First we use the result that any
operator $\hat{F}$ on the Hilbert space $H_l^{(N)}$ admits the expansion
\begin{equation}
\hat{F}=\int_{SU(k+1)}d{\mu}(h)\tilde{F}(h)U^{(l)}(h),\label{expansion}
\end{equation}
where $U^{(l)}(h)$ are taken to satisfy the normalization
\begin{equation}
TrU^{(l)}(h)U^{(l)}(h^{'})=d_l^{(N)}{\delta}(h^{-1}-h^{'}).
\end{equation}
Using the above two equations, one can derive the value of the coefficient $\tilde{F}(h)$ to be
\begin{equation}
\tilde{F}(h)=\frac{1}{d_l^{(k+1)}}Tr\hat{F}U^{(l)}(h^{-1}).
\end{equation}
Using the expansion (\ref{expansion}) in (\ref{map}) we get
\begin{eqnarray}
F_l(\vec{n})&=&\int_{SU(k+1)}d{\mu}(h)\tilde{F}(h){\omega}^{(l)}(\vec{n},h)~,~{\omega}^{(l)}(\vec{n},h)=<\vec{n},l|U^{(l)}(h)|\vec{n},l>.
\end{eqnarray}
On the other hand, using the expansion (\ref{expansion}) in (\ref{starproduct1}) will give
\begin{equation}
F_l*G_l(\vec{n})=\int \int_{SU(k+1)}d{\mu}(h)d{\mu}(h^{'})\tilde{F}(h)\tilde{G}(h^{'}){\omega}^{(l)}(\vec{n},hh^{'}).
\end{equation}
The computation of this star product boils down to the computation of ${\omega}^{(l)}(\vec{n},hh^{'})$. We have
\begin{eqnarray}
{\omega}^{(l)}(\vec{n},h)&=&<\vec{n},l|U^{(l)}(h)|\vec{n},l>\nonumber\\
&=&\bigg[<\vec{n},\frac{1}{2}|{\otimes}_s...{\otimes}_s<\vec{n},\frac{1}{2}|\bigg]\bigg[U^{(\bf
k+1)}(h){\otimes}_s...{\otimes}_sU^{(\bf k+1)}(h)\bigg]\bigg[|\vec{n},\frac{1}{2}>{\otimes}_s...{\otimes}_s|\vec{n},\frac{1}{2}>\bigg]\nonumber\\
&=&[{\omega}^{(\frac{1}{2})}(\vec{n},h)]^{2l},
\end{eqnarray}
where
\begin{eqnarray}
{\omega}^{(\frac{1}{2})}(\vec{n},h)&=&<\vec{n},\frac{1}{2}|U^{(\bf k+1)}(h)|\vec{n},\frac{1}{2}>=<\psi|U^{(\bf k+1)}(h)|\psi>.
\end{eqnarray}
In the fundamental representation ${\bf k+1}$ of $SU(k+1)$ we have $U^{(\bf k+1)}(h)=exp(im^at_a)=c(m){\bf 1}+is^a(m)t_a$ and therefore
\begin{eqnarray}
{\omega}^{(\frac{1}{2})}(\vec{n},h)&=&<\psi|c(m){\bf 1}+is^a(m)t_a|\psi>=c(m)+is^a(m)<\psi|t_a|\psi>,
\end{eqnarray}
where
\begin{eqnarray}
{\omega}^{(\frac{1}{2})}(\vec{n},hh^{'})&=&<\psi|U^{(\bf N)}(hh^{'})|\psi>\nonumber\\
&=&<\psi|(c(m){\bf 1}+is^a(m)t_a)(c(m^{'}){\bf 1}+is^a(m^{'})t_a)|\psi>\nonumber\\
&=&c(m)c(m^{'})+i[c(m)s^a(m^{'})+c(m^{'})s^a(m)]<\psi|t_a|\psi>-s^a(m)s^b(m^{'})<\psi|t_at_b|\psi>.\nonumber\\
\end{eqnarray}
Now, it is not difficult to check that
\begin{eqnarray}
<\psi|t_a|\psi>&=&Trt_aP=\frac{\alpha_k}{2}n^a\nonumber\\
<\psi|t_at_b|\psi>&=&Trt_at_bP=\frac{1}{2(k+1)}{\delta}_{ab}+\frac{\alpha_k}{4}(d_{abc}+if_{abc})n^c.\label{nice}
\end{eqnarray}
Hence, we obtain
\begin{eqnarray}
{\omega}^{(\frac{1}{2})}(\vec{n},h)&=&c(m)+i\frac{\alpha_k}{2}\vec{s}(m).\vec{n}\nonumber\\
{\omega}^{(\frac{1}{2})}(\vec{n},hh^{'})&=&c(m)c(m^{'})-\frac{1}{2(k+1)}\vec{s}(m).\vec{s}(m^{'})+i\frac{\alpha_k}{2}\bigg[c(m)s^a(m^{'})+c(m^{'})s^a(m)\bigg]n^a\nonumber\\
&-&\frac{\alpha_k}{4}(d_{abc}+if_{abc})n^cs^a(m)s^b(m^{'}).
\end{eqnarray}
These two last equations can be combined to get the pre-final result
\begin{eqnarray}
{\omega}^{(\frac{1}{2})}(\vec{n},hh^{'})-{\omega}^{(\frac{1}{2})}(\vec{n},h){\omega}^{(\frac{1}{2})}(\vec{n},h^{'})&=&
-\frac{1}{2(k+1)}\vec{s}(m).\vec{s}(m^{'})-\frac{\alpha_k}{4}(d_{abc}+if_{abc})n^cs^a(m)s^b(m^{'})\nonumber\\
&+&\frac{{\alpha}_k^2}{4}n^an^bs_a(m)s_b(m^{'}).
\end{eqnarray}
We can remark that in this last equation, we have got rid of all reference to $c$'s. We would like also to get
ride of all reference to $s$'s. This can be achieved by using the formula
\begin{equation}
s_a(m)=\frac{2}{i\alpha_N}\frac{\partial}{{\partial}n^a}{\omega}^{(\frac{1}{2})}(\vec{n},h).
\end{equation}
By using this formula , we get the final result
\begin{eqnarray}
{\omega}^{(\frac{1}{2})}(\vec{n},hh^{'})-{\omega}^{(\frac{1}{2})}(\vec{n},h){\omega}^{(\frac{1}{2})}(\vec{n},h^{'})&=&
K_{ab}\frac{\partial}{{\partial}n^a}{\omega}^{(\frac{1}{2})}(\vec{n},h)\frac{\partial}{{\partial}n^b}{\omega}^{(\frac{1}{2})}(\vec{n},h^{'}),\nonumber\\
&&
\end{eqnarray}
where
\begin{eqnarray}
K_{ab}&=&\frac{2}{(k+1){\alpha}_k^2}{\delta}_{ab}-n_an_b+\frac{1}{\alpha_k}(d_{abc}+if_{abc})n^c.
\end{eqnarray}
Therefore
\begin{eqnarray}
F_l*G_l(\vec{n})&=&\int
\int_{SU(k+1)}d{\mu}(h)d{\mu}(h^{'})\tilde{F}(h)\tilde{G}(h^{'}){\omega}^{(l)}(\vec{n},hh^{'})\nonumber\\
&=&\int
\int_{SU(k+1)}d{\mu}(h)d{\mu}(h^{'})\tilde{F}(h)\tilde{G}(h^{'})[{\omega}^{(\frac{1}{2})}(\vec{n},hh^{'})]^{2l}.
\end{eqnarray}
More explicitly 
\begin{eqnarray}
F_l*G_l(\vec{n})
&=&
\int_{SU(k+1)}d{\mu}(h)d{\mu}(h^{'})\tilde{F}(h)\tilde{G}(h^{'})\sum_{q=0}^{2l}\frac{(2l)!}{q!(2l-q)!}K_{a_1b_1}...K_{a_qb_q}[{\omega}^{(\frac{1}{2})}(\vec{n},h)]^{2l-q}\nonumber\\
&{\times}&[{\omega}^{(\frac{1}{2})}(\vec{n},h^{'})]^{2l-q}
\frac{\partial}{{\partial}n^{a_1}}{\omega}^{(\frac{1}{2})}(\vec{n},h)...\frac{\partial}{{\partial}n^{a_q}}{\omega}^{(\frac{1}{2})}(\vec{n},h)\frac{\partial}{{\partial}n^{b_1}}{\omega}^{(\frac{1}{2})}(\vec{n},h^{'})....\frac{\partial}{{\partial}n^{b_q}}{\omega}^{(\frac{1}{2})}(\vec{n},h^{'})\nonumber\\
&=&\sum_{q=0}^{2l}\frac{(2l)!}{q!(2l-q)!}K_{a_1b_1}....K_{a_qb_q}\int_{SU(k+1)}d{\mu}(h)\tilde{F}(h)[{\omega}^{(\frac{1}{2})}(\vec{n},h)]^{2l-q}\frac{\partial}{{\partial}n^{a_1}}{\omega}^{(\frac{1}{2})}(\vec{n},h)...\frac{\partial}{{\partial}n^{a_q}}{\omega}^{(\frac{1}{2})}(\vec{n},h)\nonumber\\
&{\times}&\int_{SU(k+1)}d{\mu}(h^{'})\tilde{G}(h^{'})[{\omega}^{(\frac{1}{2})}(\vec{n},h^{'})]^{2l-q}\frac{\partial}{{\partial}n^{b_1}}{\omega}^{(\frac{1}{2})}(\vec{n},h^{'})...\frac{\partial}{{\partial}n^{b_q}}{\omega}^{(\frac{1}{2})}(\vec{n},h^{'}).
\end{eqnarray}
Next, we use the formula
\begin{eqnarray}
\frac{(2l-q)!}{(2l)!}\frac{\partial}{{\partial}n^{a_1}}...\frac{\partial}{{\partial}n^{a_q}}F_l(\vec{n})=
\int_{SU(k+1)}d{\mu}(h)\tilde{F}(h)[{\omega}^{(\frac{1}{2})}(\vec{n},h)]^{2l-q}\frac{\partial}{{\partial}n^{a_1}}{\omega}^{(\frac{1}{2})}(\vec{n},h)...\frac{\partial}{{\partial}n^{a_q}}{\omega}^{(\frac{1}{2})}(\vec{n},h)\nonumber\\
\end{eqnarray}
to get the final result
\begin{equation}
F_l*G_l(\vec{n})=\sum_{q=0}^{2l}\frac{(2l-q)!}{q!(2l)!}K_{a_1b_1}....K_{a_qb_q}\frac{\partial}{{\partial}n^{a_1}}...\frac{\partial}{{\partial}n^{a_q}}F_j(\vec{n})\frac{\partial}{{\partial}n^{b_1}}...\frac{\partial}{{\partial}n^{b_q}}G_j(\vec{n}).\label{starproduct}
\end{equation}
\subsection{Fuzzy derivatives}
Derivations on ${\bf C}{\bf P}^k$ are generated by the vector fields
\begin{eqnarray}
{\cal L}_a&=&-if_{abc}n_b\frac{\partial}{{\partial}n_c}~,~
[{\cal L}_a,{\cal L}_b]=if_{abc}{\cal L}_c.
\end{eqnarray}
The corresponding adjoint action on the Hilbert space $H_l^{(N)}$ is generated by $L_a$, $[L_a,L_b]=if_{abc}L_c$, and is
given by
\begin{equation}
<\vec{n},l|U^{(l)}(h^{-1})\hat{F}U^{(l)}(h)|\vec{n},l>=<\vec{n}_0,l|U^{(l)}(g^{-1}h^{-1})\hat{F}U^{(l)}(hg)|\vec{n}_0,l>,
\end{equation}
where we have used equation (\ref{fundamental}), and such that $U^{(l)}(h)$ is given by
$U^{(l)}(h)=exp(i{\eta}_aL_a)$. Now if we take ${\eta}$ to be small, then one computes
\begin{equation}
<\vec{n},l|U^{(l)}(h)|\vec{n},l>=1+i{\eta}_a<\vec{n},l|L_a|\vec{n},l>.
\end{equation}
On the other hand, we know that the representation $U^{(l)}(h)$ is obtained by taking the symmetric product of
$2l$ fundamental representations ${\bf k+1}$ of $SU(k+1)$, and hence
\begin{eqnarray}
<\vec{n},l|U^{(l)}(h)|\vec{n},l>&=&(<\vec{n},\frac{1}{2}|1+i{\eta}_at_a|\vec{n},\frac{1}{2}>)^{2l}\nonumber\\
&=&1+i(2l){\eta}_a<\vec{n},\frac{1}{2}|t_a|\vec{n},\frac{1}{2}>\nonumber\\
&=&1+i(2l){\eta}_a\frac{\alpha_k}{2}n_a,
\end{eqnarray}
where we have used the facts, $L_a=t_a{\otimes}_s....{\otimes}_st_a$,
$|\vec{n},l>=|\vec{n},\frac{1}{2}>{\otimes}_s...{\otimes}_s|\vec{n},\frac{1}{2}>$, and the first equation of
(\ref{nice}). Hence we get the important result
\begin{equation}
<\vec{n},l|L_a|\vec{n},l>=l{\alpha}_kn_a.
\end{equation}
We should define the fuzzy derivative $[{L}_a,\hat{F}]$ by
\begin{eqnarray}
({\cal L}_aF)_l(\vec{n})&=&<\vec{n},l|[L_a,\hat{F}]|\vec{n},l>.
\end{eqnarray}
Indeed, we have
\begin{eqnarray}
<\vec{n},l|[L_a,\hat{F}]|\vec{n},l>&=&<\vec{n},l|L_a{\hat{F}}|\vec{n},l>-<\vec{n},l|\hat{F}L_a|\vec{n},l>\nonumber\\
&=&l{\alpha}_k\big(n_a*F_l(\vec{n})-F_l*n_a(\vec{n})\big).\nonumber\\
\end{eqnarray}
But we can compute
\begin{eqnarray}
n_a*F_l(\vec{n})&=&n_aF_{l}(\vec{n})+\frac{1}{2l}K_{ab}\frac{\partial}{{\partial}n^b}F_l(\vec{n})~,~F_l*n_a(\vec{n})=n_aF_{l}(\vec{n})+\frac{1}{2l}K_{ba}\frac{\partial}{{\partial}n^b}F_l(\vec{n}).
\end{eqnarray}
In other words,
\begin{eqnarray}
n_a*F_l(\vec{n})-F_l*n_a(\vec{n})&=&\frac{1}{2l}(K_{ab}-K_{ba})\frac{\partial}{{\partial}n^b}F_l(\vec{n})=\frac{1}{2l}(\frac{2i}{\alpha_k}f_{abc}n^c)\frac{\partial}{{\partial}n^b}F_l(\vec{n}).
\end{eqnarray}
Therefore
\begin{eqnarray}
<\vec{n},l|[L_a,\hat{F}]|\vec{n},l>&=&if_{abc}n^c\frac{\partial}{{\partial}n^b}F_l(\vec{n})=({\cal L}_aF_l)(\vec{n}).
\end{eqnarray}

\section{Fuzzy ${\bf S}^4_N$: symplectic and Poisson structures} 
%\section{Fuzzy ${\bf S}^4_N$: First Look}
\subsection{The spectral triple and fuzzy ${\bf CP}^k_N$: another look}
In order to avoid the string theory landscape it is has been argued in \cite{Steinacker:2007dq,Steinacker:2010rh,Yang:2006dk,Yang:2008fb,Yang:2009pm} that gravity should emerge in the IKKT or IIB matrix model from the noncommutative physics of $4-$dimensional brane solutions and not from the $10-$dimensional physics of the bulk. From the other hand, although noncommutative gauge theories behave similarly to gravity theories they are generically different from Einstein theory \cite{Rivelles:2002ez,Szabo:2006wx,Szabo:2009tn}.

We consider covariant noncommutative spaces such as fuzzy ${\bf S}^4_N$ which is a compact Euclidean version of Snyder space \cite{Snyder:1946qz,Yang:1947ud}. Indeed, there are extra generators here denoted by $\theta^{\mu\nu}$ and ${\cal P}^{\mu}$, and the non-commutativity $\theta^{\mu\nu}$ is not central here as opposed to the DFR theory \cite{Doplicher:1994tu} which is the source of gravity \cite{Steinacker:2016vgf}. However, as in the case of DFR quantum spacetime the non-commutativity $\theta$ is averaged over the extra dimensions, which is here a fuzzy ${\bf S}^2_N$, in order to recover $SO(5)$ invariance.

We will follow \cite{Dolan:2003kq, Medina:2012cs,Medina:2002pc} and \cite{Steinacker:2015dra,Steinacker:2016vgf}. Very closely related constructions are found in \cite{Grosse:1996mz,Castelino:1997rv,Kimura:2002nq,Ramgoolam:2001zx,Abe:2004sa,Valtancoli:2002sm,CarowWatamura:2004ct,Balachandran:2001dd}. See also \cite{Kimura:2003ab,Aoyama:2002jd,Azuma:2004yg,Ramgoolam:2002wb}.

%\subsection{The Spectral Triple and Fuzzy ${\bf CP}^k_N$: Another Look}
The fuzzy four-sphere ${\bf S}^4_N$, similarly to all fuzzy spaces, is specified by a sequence of triples \cite{Frohlich:1993es,Connes:1994yd}

\begin{eqnarray}
{\cal M}_n=({\cal H},{\cal A},\Delta).%(H_N,{\rm Mat}_{d_N},\Delta_N).
\end{eqnarray}
All fuzzy spaces, the fuzzy four-sphere ${\bf S}^4_N$ included, are given in terms of matrix algebras. Thus, the algebra ${\cal A}$ is the algebra of $d_n\times d_n$ matrices with the obvious inner product
\begin{eqnarray}
<M,N>=\frac{1}{d_n}Tr M^+N.
\end{eqnarray}
In other words,
\begin{eqnarray}
{\cal A}={\rm Mat}_{d_n}.
\end{eqnarray}
And
\begin{eqnarray}
{\cal H}=H_{n}={\bf C}^{d_n}
\end{eqnarray}
is the Hilbert space on which the algebra of matrices acts in a natural way, whereas $\Delta=\Delta_n$ is an appropriate Laplacian acting the matrices.

The data contained in the above triple, which defines the fuzzy space, can be specified completely by giving a scalar action on the fuzzy space.

Fuzzy ${\bf S}^4_N$ is really fuzzy ${\bf CP}^3_N$. More precisely, ${\bf CP}^3$ is an ${\bf S}^2$ bundle over ${\bf S}^4$. Since ${\bf CP}^3$ is a coadjoint orbit given by ${\bf CP}^3=SU(4)/U(3)$ it can be subjected to fuzzification by quantization in the usual way to obtain a matrix approximation which is fuzzy ${\bf CP}^3_N$. %We start from the generators of $SU(4)={\rm spin}(6)$:

The space ${\bf CP}^{k}=SU(k+1)/U(k)$ can be thought of as a brane surface embedded in ${\bf R}^{k(k+2)}$. The fundamental representation of $SU(k+1)$ is $(k+1)-$dimensional and is denoted ${\bf k+1}$.  Let $\Lambda_{\mu}$, $\mu=1,...,k(k+2)$, be the generators of 
$SU(k+1)$ in the fundamental representation. These can be given by the Gell-Mann matrices $t_{\mu}=\lambda_{\mu}/2$ satisfying 
\begin{eqnarray}
[t_{\mu},t_{\nu}]=if_{\mu\nu\lambda}t_{\lambda}~,~2t_{\mu}t_{\nu}=\frac{1}{k+1}\delta_{\mu\nu}+(d_{\mu\nu\lambda}+if_{\mu\nu\lambda})t_{\lambda}~,~Trt_{\mu}t_{\nu}=\frac{1}{2}\delta_{\mu\nu}.
\end{eqnarray}
We take the $n-$fold symmetric tensor product of the fundamental representation of $SU(k+1)$ to obtain the $d_n^k-$dimensional irreducible representation of $SU(k+1)$, viz
\begin{eqnarray}
{\bf d}_n^k=\big({\bf k+1}\otimes...\otimes{\bf k+1}\big)_{\rm sym}~,~n-{\rm times}.
\end{eqnarray}
It is not difficult to show that
\begin{eqnarray}
d_n^k=\frac{(k+n)!}{k!n!}.
\end{eqnarray}
The dimension of the space is given by
\begin{eqnarray}
{\rm dimension}=2.{\rm lim}_{n\longrightarrow\infty}\frac{\ln d_n^k}{\ln n}=2k.
\end{eqnarray}
It is  immediately seen that for fuzzy ${\bf CP}^1_N={\bf S}^2_N$ we obtain $d_n^1=n+1$ and thus $n/2$ is the spin quantum number characterizing the irreducible representations of $SU(2)$. Let $T_{\mu}$  be the generators of $SU(k+1)$ in the $d_n^k-$dimensional irreducible representation and $-T_{\mu}^R$ be the generators in the complex conjugate representation, viz 
\begin{eqnarray}
T_{\mu}=\bigg(t_{\mu}\otimes {\bf 1}\otimes...\otimes {\bf 1}+{\bf 1}\otimes t_{\mu}\otimes...\otimes {\bf 1}+{\bf 1}\otimes{\bf 1}\otimes ....\otimes t_{\mu}\bigg)_{\rm sym}~,~n-{\rm times}.\label{T}
\end{eqnarray}
We know that functions on fuzzy ${\bf CP}^k_N$ are matrices in ${\rm Mat}_{d_n^k}$ which transforms under the action of $SU(k+1)$ as the tensor product ${\bf d}_n^k\otimes \bar{\bf d}_n^k$ and thus can be expanded in terms of $SU(k+1)$ polarization tensors. These representations can be found as follows. First, we note that
\begin{eqnarray}
{\bf d}_n^k=(n,0,...,0)~,~\bar{\bf d}_n^k=(0,0,...,n).
\end{eqnarray}
Then, we use the result for $SU(k+1)$, from \cite{Fulton}, that 
%that , for $SU(M)$
%, the tensor product of the two representations
%$(n_1,n_2,...,n_{M-1})$ and $(k,0,...,0)$ is given by
\begin{eqnarray}
&&(n_1,n_2,...,n_{k}){\otimes}(n,0,...,0)={\bigoplus}(b_1,b_2,...,b_{k})\nonumber\\
&&b_i=n_i+c_i-c_{i+1}~,~c_{i+1}{\leq}n_i~,~1{\leq}i{\leq}k,
\end{eqnarray}
where $c_i$'s are non-negative integers satisfying
$c_1+c_2+....+c_{k+1}=n$. We compute immediately that
\begin{eqnarray}
&&(0,0,...,n){\otimes}(n,0,...,0)={\bigoplus}(c_1,0,...,0,c_1).
\end{eqnarray}
Thus for ${\bf CP}^3$ or $SU(4)$ we have
\begin{eqnarray}
&&(0,0,n){\otimes}(n,0,0)={\bigoplus}(c_1,0,c_1).
\end{eqnarray}
These representations exist only for $c_1=0,...,n$ with corresponding dimensions  \cite{Fulton}
\begin{eqnarray}
{\rm dim}(c_1,0,c_1)=\frac{1}{12}(c_1+1)^2(c_1+2)^2(2c_1+3).
\end{eqnarray}
Hence, functions on fuzzy ${\bf CP}^k_N$ are given by $N\times N$ matrices $M$ in ${\rm Mat}_{N}$ where 
\begin{eqnarray}
N\equiv d_n^k=\frac{(k+n)!}{k!n!},
\end{eqnarray}
with permitted $SU(k+1)$ representations $(l,0,...,0,l)$~, with $k$~factors, such that $l\leq n$. The commutative limit is given by $n\longrightarrow \infty$ where

The Laplacian on fuzzy ${\bf CP}^k_N$ is given by 
\begin{eqnarray}
\Delta_{n,k}=({\rm Ad} T_{\mu})^2=[T_{\mu},[T_{\mu},...]].
\end{eqnarray}
We need also to find the coordinate operators on fuzzy ${\bf CP}^k_N$. We consider the tensor product
\begin{eqnarray}
(n,0,...,0)\otimes (1,0,...,0)=(n+1,0,....,0)\oplus (n-1,1,0,...,0).
\end{eqnarray}
Then we consider the intertwiner on the above vector space given by the operator \cite{CarowWatamura:2004ct}
\begin{eqnarray}
2X=2T_{\mu}t_{\mu}=(T_{\mu}+t_{\mu})^2-T_{\mu}^2-t_{\mu}^2.
\end{eqnarray}
For simplicity we consider $SU(4)$ first. The Casimir operators and the dimensions of $SU(4)$ irreducible representations with highest weight $\Lambda=(\lambda_1,\lambda_2,\lambda_3)$ are given by
\begin{eqnarray}
C_2(\Lambda)=\frac{1}{8}\lambda_1(3\lambda_1+2\lambda_2+\lambda_3+12)+\frac{1}{4}\lambda_2(\lambda_1+2\lambda_2+\lambda_3+8)+\frac{1}{8}\lambda_3(\lambda_1+2\lambda_2+3\lambda_3+12).\nonumber\\
\end{eqnarray}
\begin{eqnarray}
{\rm dim}(\Lambda)=\frac{1}{12}(\lambda_1+1)(\lambda_2+1)(\lambda_3+1)(\lambda_1+\lambda_2+2)(\lambda_2+\lambda_3+2)(\lambda_1+\lambda_2+\lambda_3+3).
\end{eqnarray}
The relevant Casimirs are 
\begin{eqnarray}
T_{\mu}^2=C_2(n,0,0)=\frac{3n(n+4)}{8}.
\end{eqnarray}
\begin{eqnarray}
t_{\mu}^2=C_2(1,0,0)=\frac{15}{8}.
\end{eqnarray}
\begin{eqnarray}
(T_{\mu}+t_{\mu})^2=C_2(n+1,0,0)=\frac{3(n+1)(n+5)}{8}~,~(T_{\mu}+t_{\mu})^2=C_2(n-1,1,0)=\frac{1}{8}(3n^2+10n+7).\nonumber\\
\end{eqnarray}
The eigenvalues of $X$ are therefore given by
\begin{eqnarray}
2X|_{(n+1,0,0)}=\frac{3n}{4}~,~2X|_{(n-1,1,0)}=-\frac{n}{4}-1.
\end{eqnarray}
The characteristic equation for $X$ is given by
 \begin{eqnarray}
(2X-\frac{3n}{4})(2X+\frac{n}{4}+1)=0.
\end{eqnarray}
The generalization to $SU(k+1)$ is immediately given by
 \begin{eqnarray}
(2X-\frac{kn}{k+1})(2X+\frac{n}{k+1}+1)=0\Rightarrow X^2=\frac{1}{4}\frac{kn}{k+1}\big(\frac{n}{k+1}+1\big)+\frac{1}{2}\bigg(\frac{kn}{k+1}-\frac{n}{k+1}-1\bigg)T_{\lambda}t_{\lambda}.\nonumber\\
\end{eqnarray}
From the other hand, we compute
 \begin{eqnarray}
X^2=\frac{1}{2(k+1)}T_{\mu}^2+\frac{1}{2}(d_{\mu\nu\lambda}+if_{\mu\nu\lambda})T_{\mu}T_{\nu}t_{\lambda}.
\end{eqnarray}
By identification, we get
\begin{eqnarray}
T_{\mu}^2=\frac{kn}{2(k+1)}(n+k+1).
\end{eqnarray}
\begin{eqnarray}
d_{\mu\nu\lambda}T_{\mu}T_{\nu}+if_{\mu\nu\lambda}T_{\mu}T_{\nu}=\frac{1}{k+1}\bigg((k-1)n-k-1\bigg)T_{\lambda}.
\end{eqnarray}
The adjoint representation of $SU(4)$ is $(1,0,1)$  and has dimension and Casimir given by $15$ and $C_2^{\rm ad}=4$ respectively. For $SU(N)$ the adjoint representation is $(1,0,...,0,1)$ with dimension and Casimir equal to $k(k+2)$ and $C_2^{\rm ad}=k+1$. The generators in the adjoint representation are given by $({\rm ad}t_{\mu})_{\alpha\beta}=-if_{\mu\alpha\beta}$. Thus we have
 \begin{eqnarray}
if_{\mu\nu\alpha}T_{\mu}T_{\nu}&=&-\frac{1}{2}f_{\mu\nu\alpha}f_{\mu\nu\beta}T_{\beta}\nonumber\\
&=&-\frac{1}{2}({\rm ad}t_{\mu}{\rm ad}t_{\mu})_{\alpha\beta}T_{\beta}\nonumber\\
&=&-\frac{1}{2}C_2^{\rm ad}T_{\alpha}\Rightarrow if_{\mu\nu\alpha}T_{\mu}T_{\nu}=-\frac{1}{2}(k+1)T_{\alpha}.
\end{eqnarray}
In other words, the other defining equations of fuzzy ${\bf CP}^k_N$ are given by 
\begin{eqnarray}
d_{\mu\nu\lambda}T_{\mu}T_{\nu}=(k-1)\bigg(\frac{n}{k+1}+\frac{1}{2}\bigg)T_{\lambda}.
\end{eqnarray}

\subsection{Fuzzy ${\bf S}^4_N$}

Now we construct fuzzy ${\bf S}^4_N$. Although ${\bf S}^4$ does not admit a symplectic structure.

Let $\Gamma_i$, $i=1,...,4$, be the Dirac matrices in $4$ dimensions and let $\Gamma_5=\Gamma_1\Gamma_2\Gamma_3\Gamma_4$. Collectively we write them $\Gamma_a$, $a=1,...,5$, and they satisfy $\{\Gamma_a,\Gamma_a\}=2\delta_{ab}$.These are the gamma matrices associated to $SO(5)$.  We define 
\begin{eqnarray}
X_a=\frac{R}{\sqrt{5}}\Gamma_a.
\end{eqnarray}
These satisfy 
\begin{eqnarray}
X_a^2=R^2.
\end{eqnarray}
This is a fuzzy ${\bf S}^4_N$ with $N=4$. This corresponds to the four dimensional representation $(1/2,1/2)$ of $SO(5)$ or ${\rm spin}(5)$ (or equivalently $Sp(2)$).  ${\rm spin}(5)$ is a $2-$to$-1$ cover of ${\rm SO}(5)$ and therefore they have the same Lie algebra and the same representation theory. The generators of ${\rm spin}(5)$ in the fundamental representation are not $\Gamma_a$ but they are given by

\begin{eqnarray}
\frac{\sigma_{ab}}{2}=\frac{1}{4i}[\Gamma_a,\Gamma_b].\label{analogue}
\end{eqnarray}
Any $4\times 4$ matrix, i.e. any function on  ${\bf S}^4_4$, can be expanded in terms of the $16$ matrices $\Gamma_a$ and $\sigma_{ab}$ and the identity  as follows 

\begin{eqnarray}
M=M_0{\bf 1}+M_a\Gamma_a+M_{ab}\sigma_{ab}.
\end{eqnarray}
Fuzzy  ${\bf S}^4_4$ corresponds only to the first two terms in the above expansion. 

Higher approximations of the four-sphere ${\bf S}^4$ are obtained as follows. We consider the irreducible representation obtained by taking the $n-$fold symmetric tensor product of the fundamental representation $(1/2,1/2)$ of ${\rm spin}(5)$. Thus
\begin{eqnarray}
J_a=\frac{1}{2}\bigg(\Gamma_a\otimes {\bf 1}\otimes...\otimes {\bf 1}+{\bf 1}\otimes \Gamma_a\otimes...\otimes {\bf 1}+{\bf 1}\otimes{\bf 1}\otimes ....\otimes \Gamma_a\bigg)_{\rm sym}~,~n-{\rm times}.
\end{eqnarray}
This corresponds to the spin $(n/2,n/2)$ irreducible representation of ${\rm spin}(5)$ or $SO(5)$. The corresponding Dynkin labels are $\lambda_1=n_1-n_2=0$ and $\lambda_2=2n_2=n$. The gamma matrices are taken in the representation 
\begin{eqnarray}
 \Gamma_i=\left( \begin{array}{cc}
0 & \sigma_i  \\
\sigma_i & 0  \end{array} \right)~,~\Gamma_4=\left( \begin{array}{cc}
0 & i  \\
-i & 0  \end{array} \right)~,~\Gamma_5=\left( \begin{array}{cc}
1 & 0  \\
0 & -1  \end{array} \right).
\end{eqnarray}
The irreducible representations of $SO(5)$ are characterized by the 
highest weight vectors $\Lambda=({n}_1,{n}_2)$ with
${n}_1{\geq}{n}_2{\geq}0$ with dimensions and Casimirs 
\begin{eqnarray}
{\rm dim}(\Lambda)=\frac{1}{6}(2n_1+3)(2n_2+1)(n_1-n_2+1)(n_1+n_2+2).
\end{eqnarray}
\begin{eqnarray}
C_2(\Lambda)=\frac{1}{2}n_1(n_1+3)+\frac{1}{2}n_2(n_2+1).
\end{eqnarray}
Thus,
\begin{eqnarray}
{\rm dim}(\frac{n}{2},\frac{n}{2})=\frac{1}{6}(n+1)(n+2)(n+3)~,~C_2(\frac{n}{2},\frac{n}{2})=\frac{1}{4}n(n+4).
\end{eqnarray}
This can be given an explicit construction in terms of creation and annihilation operators \cite{Steinacker:2015dra}. We write
\begin{eqnarray}
J_a=\frac{1}{2}a_{\alpha}^+(\Gamma_a)_{\alpha\beta}a_{\beta}~,~[a_{\alpha},a_{\beta}^+]=\delta_{\alpha\beta}.
\end{eqnarray}
We compute
\begin{eqnarray}
a_{\alpha}^+(\frac{\sigma_{ab}}{2})_{\alpha\beta}a_{\beta}=\frac{1}{i}[J_a,J_b].
\end{eqnarray}
This is the analogue of (\ref{analogue}). Thus the generators of ${\rm spin}(5)$ in the irreducible representation $(n/2,n/2)$ are given by
\begin{eqnarray}
{\cal M}_{ab}=a_{\alpha}^+(\frac{\sigma_{ab}}{2})_{\alpha\beta}a_{\beta}.
\end{eqnarray}
Also we compute
\begin{eqnarray}
[{\cal M}_{ab},J_c]&=&\frac{1}{2}a_{\alpha}^+[\frac{\sigma_{ab}}{2},\Gamma_c]_{\alpha\beta}a_{\beta}\nonumber\\
&=&\frac{i}{2}a_{\alpha}^+\big(\delta_{ac}\Gamma_b-\delta_{bc}\Gamma_a\big)_{\alpha\beta}a_{\beta}\nonumber\\
&=&i\big(\delta_{ac}J_b-\delta_{bc}J_a\big).
\end{eqnarray}
And
\begin{eqnarray}
[{\cal M}_{ab},{\cal M}_{cd}]&=&a_{\alpha}^+[\frac{\sigma_{ab}}{2},\frac{\sigma_{cd}}{2}]_{\alpha\beta}a_{\beta}\nonumber\\
&=&i a_{\alpha}^+\big(\delta_{ac}\frac{\sigma_{bd}}{2}-\delta_{ad}\frac{\sigma_{bc}}{2}-\delta_{bc}\frac{\sigma_{ad}}{2}+\delta_{bd}\frac{\sigma_{ac}}{2}\big)_{\alpha\beta}a_{\beta}\nonumber\\
&=&i \big(\delta_{ac}{\cal M}_{bd}-\delta_{ad}{\cal M}_{bc}-\delta_{bc}{\cal M}_{ad}+\delta_{bd}{\cal M}_{ac}\big).
\end{eqnarray}
$J_a$ transforms as a vector in the irreducible representation $(1,0)$ under ${\rm spin}(5)$. Thus $\sum_a J_a^2$ is invariant under $SO(5)$, and since $(n/2,n/2)$ is an irreducible representation, the quantity $\sum_a J_a^2$ must be proportional to the identity \cite{Castelino:1997rv}. We show this explicitly as follows.

We go on now to the groups ${\rm spin}(6)$, $SO(5)$ and $SU(4)$. ${\rm spin}(6)$ is a $2-$to$-1$ cover of ${\rm SO}(6)$ and it is locally isomorphic to $SU(4)$.  They have the same Lie algebra. Irreducible representations of $SO(6)$ are characterized by the 
highest weight vectors $\Lambda=({n}_1,n_2,{n}_3)$ with
${n}_1{\geq}{n}_{2}{\geq}|{n}_3|{\geq}0$ with dimensions and Casimirs 
\begin{eqnarray}
{\rm dim}(\Lambda)=\frac{1}{12}((n_1+2)^2-n_3^2)((n_1+2)^2-(n_2+1)^2)((n_2+1)^2-n_3^2).
\end{eqnarray}
\begin{eqnarray}
C_2(\Lambda)=\frac{1}{2}n_1(n_1+4)+\frac{1}{2}n_2(n_2+2)+\frac{1}{2}n_3^2.
\end{eqnarray}
The fundamental is $(1/2,1/2,1/2)$ whereas the anti fundamental is $\overline{(1/2,1/2,1/2)}=(1/2,1/2,-1/2)$. We have the identification with $SU(4)$ representations
\begin{eqnarray}
&&(n,0,0)\leftrightarrow (\frac{n}{2},\frac{n}{2},\frac{n}{2})~,~(0,0,n)\leftrightarrow \overline{(\frac{n}{2},\frac{n}{2},\frac{n}{2})}=(\frac{n}{2},\frac{n}{2},-\frac{n}{2}).\nonumber\\
\end{eqnarray}
The generators of ${\rm spin}(6)$ in the fundamental representation $(1/2,1/2,1/2)$ are given by $\sigma_{ab}/2$ and $\sigma_{a6}/2=-\sigma_{6a}/2=\Gamma_a/2$. They are written collectively as $\sigma_{AB}/2$, $A,B=1,...,6$, and they satisfy 
\begin{eqnarray}
[\frac{{\sigma}_{AB}}{2},\frac{{\sigma}_{CD}}{2}]
&=&i \big(\delta_{AC}\frac{{\sigma}_{BD}}{2}-\delta_{AD}\frac{{\sigma}_{BC}}{2}-\delta_{BC}\frac{{\sigma}_{AD}}{2}+\delta_{BD}\frac{{\sigma}_{AC}}{2}\big).
\end{eqnarray}
The irreducible representations $(n/2,n/2,n/2)$ of ${\rm spin}(6)$ are obtained by taking the $n-$fold symmetric tensor product of the fundamental representation $(1/2,1/2,1/2)$. The corresponding dimension and Casimir are 
\begin{eqnarray}
{\rm dim}(\frac{n}{2},\frac{n}{2},\frac{n}{2})=\frac{1}{6}(n+1)(n+2)(n+3)~,~C_2(\frac{n}{2},\frac{n}{2},\frac{n}{2})=\frac{3}{8}n(n+4).
\end{eqnarray}
The corresponding Dynkin labels are $\lambda_1=n_1-n_2=0$, $\lambda_2=n_2-n_3=0$, $\lambda_3=n_2+n_3=n$. The generators are exactly given by $J_a={\cal M}_{a6}=-{\cal M}_{6a}$ and ${\cal M}_{ab}$ which are denoted collectively ${\cal M}_{AB}$, $A,B=1,...,6$, and they satisfy 
\begin{eqnarray}
[{\cal M}_{AB},{\cal M}_{CD}]
&=&i \big(\delta_{AC}{\cal M}_{BD}-\delta_{AD}{\cal M}_{BC}-\delta_{BC}{\cal M}_{AD}+\delta_{BD}{\cal M}_{AC}\big).
\end{eqnarray}
We know now the Casimirs 
\begin{eqnarray}
\frac{1}{4}{\cal M}_{ab}{\cal M}_{ab}=\frac{1}{4}n(n+4).
\end{eqnarray}
\begin{eqnarray}
\frac{1}{4}{\cal M}_{AB}{\cal M}_{AB}=\frac{1}{4}{\cal M}_{ab}{\cal M}_{ab}+\frac{1}{2}J_a^2=\frac{3}{8}n(n+4).
\end{eqnarray}
Thus we get 
\begin{eqnarray}
J_a^2=\frac{1}{4}n(n+4).
\end{eqnarray}
In summary, the defining equations of fuzzy ${\bf S}^4_N$ are 
\begin{eqnarray}
X_a^2=R^2.
\end{eqnarray}
\begin{eqnarray}
X_a=rJ_a~,~r^2=\frac{4R^2}{n(n+4)}.
\end{eqnarray}
The coordinate operators do not commute, viz
\begin{eqnarray}
[X_a,X_b]=i\Theta_{ab}~,~\Theta_{ab}=r^2{\cal M}_{ab}.
\end{eqnarray}
The coordinate operators $X_a$ are covariant under $SO(5)$, viz
\begin{eqnarray}
[{\cal M}_{ab},X_c]
&=&i\big(\delta_{ac}X_b-\delta_{bc}X_a\big).\label{comm2}
\end{eqnarray}
\begin{eqnarray}
[{\cal M}_{ab},{\cal M}_{cd}]
&=&i \big(\delta_{ac}{\cal M}_{bd}-\delta_{ad}{\cal M}_{bc}-\delta_{bc}{\cal M}_{ad}+\delta_{bd}{\cal M}_{ac}\big).\label{comm1}
\end{eqnarray}
The fuzzy ${\bf S}^4_N$ is therefore a covariant quantum space very similar to Snyder quantum spacetime  \cite{Snyder:1946qz,Yang:1947ud}. It is also similar to the DFR spacetime \cite{Doplicher:1994tu} but the $\Theta$ are not central.

A general function on fuzzy ${\bf S}^4_N$ is an $N\times N$ matrix where 
\begin{eqnarray}
N=d_N^3=\frac{1}{6}(n+1)(n+2)(n+3). 
\end{eqnarray}
Thus we are dealing with a sequence of matrix algebras ${\rm Mat}_{N}$. The basis is given by the polarization tensors corresponding to the tensor products
\begin{eqnarray}
(\frac{n}{2},\frac{n}{2})\otimes(\frac{n}{2},\frac{n}{2})=\sum_{l=0}^n\sum_{k=0}^l(l,k)~,~{\rm spin}(5).\label{funcs4}
\end{eqnarray}
\begin{eqnarray}
(\frac{n}{2},\frac{n}{2},\frac{n}{2})\otimes(\frac{n}{2},\frac{n}{2},-\frac{n}{2})=\sum_{l=0}^n(l,l,0)~,~{\rm spin}(6).\label{funcs4e}
\end{eqnarray}
\begin{eqnarray}
(0,0,n)\otimes(n,0,0)=\sum_{l=0}^n(l,0,l)~,~SU(4).\label{funcs4ee}
\end{eqnarray}
Thus functions on fuzzy ${\bf S}^4_N$ obviously involve $X_a$ and ${\cal M}_{ab}$, and because of the constraints, the dimension of this space is actually six and not four (we are really dealing with fuzzy ${\bf CP}^3_N$). We can see from (\ref{funcs4}), (\ref{funcs4e}) and (\ref{funcs4ee}) that the $(l,l,0)$ representation of $SO(6)$ and the $(l,0,l)$ representation of $SU(4)$ decomposes in terms of $SO(5)$ representations as
\begin{eqnarray}
(l,l,0)_{SO(6)}=\sum_{k=0}^l(l,k)_{SO(5)}.
\end{eqnarray}
\begin{eqnarray}
(l,0,l)_{SU(4)}=\sum_{k=0}^l(l,k)_{SO(5)}.
\end{eqnarray}
In terms of representation theory functions on fuzzy ${\bf S}^4_N$ correspond only to the representation $(l,0)$ in (\ref{funcs4}). This can be seen as follows. A general $N\times N$ matrix $M$ can be expanded in terms of $SO(5)$ polarization tensors $T^{(l,k)}_{a_1,...,a_{l+k}}$ as follows
\begin{eqnarray}
M=\sum_{l=0}^n\sum_{k=0}^lM^{(l,k)}_{a_1,...,a_{l+k}}T^{(l,k)}_{a_1,...,a_{l+k}}.
\end{eqnarray}
The  polarization tensors $T^{(l,k)}_{a_1,...,a_{l+k}}$ are symmetrized $n-$th order polynomials of $X_a$ and ${\cal M}_{ab}$ where $l$ is the order of $X_a$ and $k$ is the order of ${\cal M}_{ab}$. Thus fuzzy ${\bf S}^4_N$ corresponds to $k=0$, i.e. to the $(l,0)$ representations.

These symmetric representations $(l,0)$ which occur in the expansion of functions on ${\bf S}^4$ is also due to the fact that ${\bf S}^4$ is the coadjoint orbit ${\bf S}^4=SO(5)/SO(4)$. Thus the harmonic expansion of functions on ${\bf S}^4$ requires irreducible representations of $SO(5)$ which contain singlets of $SO(4)$ under the decomposition $SO(5)\longrightarrow SO(4)$ \cite{Dolan:2003kq}. This is analogous to the statement that since ${\bf CP}^3$ is the coadjoint orbit ${\bf CP}^3_N=SU(4)/U(3)$ harmonic expansion of functions on ${\bf CP}^3$ requires irreducible representations of $SU(4)$ which contain singlets of $U(3)$ under the decomposition $SU(4)\longrightarrow SU(3)\times U(1)$. But  ${\bf CP}^3$ is also the coadjoint orbit ${\bf CP}^3=SO(5)/(SU(2)\times U(1))$ and thus harmonic expansion of functions on ${\bf CP}^3$ requires irreducible representations of $SO(5)$ which contain singlets of $SU(2)\times U(1)$ under the decomposition $SO(5)\longrightarrow SU(2)\times U(1)$. In the first case the Laplacian on ${\bf CP}^3$ is $SO(6)-$invariant whereas in the second case the Laplacian is $SO(5)-$invariant although not unique. This is in fact why we can extract fuzzy ${\bf S}^4_N$ from fuzzy ${\bf CP}^3_N$ with a fiber given by fuzzy ${\bf S}^2_N$. Indeed, schematically we have
\begin{eqnarray}
SO(5)/(SU(2)\times U(1))&=&\big(SO(5)/SO(4)\big)\times \big(SO(4)/(SU(2)\times U(1))\big)\nonumber\\
&=&\big(SO(5)/SO(4)\big)\times \big(SU(2)/U(1)\big).
\end{eqnarray}
This will be given a precise meaning below.

Thus, fuzzy ${\bf S}^4_N$ have in some sort an internal structure given by fuzzy ${\bf S}^2_N$. If we simply project out the unwanted degrees of freedom we obtain a non-associative algebra \cite{Ramgoolam:2001zx}. Another more elegant approach is due to O'Connor et al and goes as follows. The eigenvalues of the ${\rm spin}(5)$ and ${\rm spin}(6)$ Casimirs on the polarization tensors  $T^{(l,k)}$ are
\begin{eqnarray}
C_2^{SO(5)}T^{(l,k)}=\frac{1}{2}\big(l(l+3)+k(k+1)\big)T^{(l,k)}.
\end{eqnarray}
\begin{eqnarray}
C_2^{SO(6)}T^{(l,k)}=l(l+3)T^{(l,k)}.
\end{eqnarray}
In the last line we have used the fact that all $T^{(l,k)}$,  $k=0,...,l$, correspond to the $SO(6)$ representation $(l,l,0)$. We remark then immediately that the operator 
\begin{eqnarray}
C_I=2C_2^{SO(5)}-C_2^{SO(6)},
\end{eqnarray}
has eigenvalues 
\begin{eqnarray}
C_IT^{(l,k)}=k(k+1)T^{(l,k)}.
\end{eqnarray}
Since it only depends on $k$ it can be used to penalize the representations with $k\neq 0$ in order  these unwanted zero modes on fuzzy ${\bf S}^4_N$. The Laplacian on the fuzzy ${\bf S}^4_N$ is then given by 
\begin{eqnarray}
\Delta_n=\frac{1}{R^2}\big(C_2^{SO(6)}+hC_I\big).
\end{eqnarray}
The parameter $h$ will be taken to $\infty$ but the theory is stable for all $h\in[-1,\infty]$. We also write explicitly these Laplacians as

\begin{eqnarray}
C_2^{SO(5)}=\frac{1}{4}[{\cal M}_{ab},[{\cal M}_{ab},...]].
\end{eqnarray}
\begin{eqnarray}
C_2^{SO(6)}=\frac{1}{4}[{\cal M}_{AB},[{\cal M}_{AB},...]]=\frac{1}{4}[{\cal M}_{ab},[{\cal M}_{ab},...]]+\frac{1}{2}[J_a,[J_a,...]].
\end{eqnarray}
A non-commutative scalar field theory on fuzzy ${\bf S}^4_N$ is given by 
\begin{eqnarray}
S=\frac{R^4}{N}Tr \bigg(\Phi\Delta_n\Phi+V(\Phi)\bigg).
\end{eqnarray}

%\section{Symplectic and Poisson Structures} 

%In this section we will mostly follow \cite{Steinacker:2015dra,Steinacker:2016vgf}. There is no symplectic $2-$form on ${\bf S}^4$ since ${\bf H}^2({\bf S}^4)=0$. Thus, we insist that fuzzy ${\bf S}^4_N$ should be viewed as a squashed fuzzy ${\bf CP}^3_N$ with degenerate fiber fuzzy ${\bf S}^2_{N}\equiv {\bf S}^2_{n+1}$ (see below). We explain this in some detail.

\subsection{Hopf map}
In this section and the rest of this chapter we will mostly follow \cite{Steinacker:2015dra,Steinacker:2016vgf}. There is no symplectic $2-$form on ${\bf S}^4$ since ${\bf H}^2({\bf S}^4)=0$. Thus, we insist that fuzzy ${\bf S}^4_N$ should be viewed as a squashed fuzzy ${\bf CP}^3_N$ with degenerate fiber fuzzy ${\bf S}^2_{N}\equiv {\bf S}^2_{n+1}$ (see below). We explain this in some detail.

We consider the fundamental representation $(1,0,0)$ of $SU(4)$. This is a $4-$dimensional representation which we will view as ${\bf C}^4$. Let $z_0=(1,0,0,0)$ be some reference point in ${\bf C}^4$. Obviously, $SU(4)$ will act on this point giving $z=Uz_0$, where $U\in SU(4)$, in such a way that $z^+z=1$, i.e. $z\in {\bf S}^7\in {\bf R}^8= {\bf C}^4$. Further, starting from the $4\times 4$ gamma matrices $\Gamma_a$ of $SO(5)$, we define
\begin{eqnarray}
x_a=z_{\alpha}^*(\Gamma_a)_{\alpha\beta}z_{\beta}=<z|\Gamma_a|z>.
\end{eqnarray}
By using the result \cite{Castelino:1997rv,badis}
\begin{eqnarray}
\sum_a(\Gamma_a\otimes\Gamma_a)_{\rm sym}=({\bf 1}\otimes{\bf 1})_{\rm sym},
\end{eqnarray}
we can show that $x_a^2=1$, i.e. $x_a\in{\bf S}^4$. We can then define the Hopf map
\begin{eqnarray}
&&{\bf S}^7\longrightarrow{\bf S}^4\nonumber\\
&&z_{\alpha}\longrightarrow x_a.
\end{eqnarray}
But since the phase of $z$ drops in a trivial way, this Hopf map is actually a map from ${\bf CP}^3={\bf S}^7/U(1)$ into ${\bf S}^4$, viz
\begin{eqnarray}
&&x_a~:~{\bf CP}^3\longrightarrow{\bf S}^4\nonumber\\
&&|z><z|\longrightarrow <z|\Gamma_a|z>=x_a.
\end{eqnarray}
Here, we have identified ${\bf CP}^3$ with the space of rank one projectors $|z><z|$.

We consider the gamma matrices in the Weyl basis where $\Gamma_5$ is diagonal with eigenvalues $+1$ and $-1$ and degeneracy equal $2$ for each. The reference point in ${\bf C}^4$ is $z_0=(\hat{z}_0,\tilde{z}_0)$ where $\hat{z}_0=(1,0)$ and $\tilde{z}_0=0$. The coordinates of the point $p_0\in {\bf S}^4\in{\bf R}^5$ at the reference point are
\begin{eqnarray}
x_i=0~,~x_5=1=\hat{z}_0^+\hat{z}_0=z_0^+\Gamma_5 z_0.
\end{eqnarray}
This is essentially the north pole. But recall that ${\bf CP}^3=SO(5)/(SU(2)\times U(1))$.  Therefore, the stabilizer at $p_0$ is obviously given by $SO(4)\in SO(5)$, viz
\begin{eqnarray}
H=\{h\in SO(5); [h,\Gamma_5]=0\}=SO(4)=SU(2)_L\times SU(2)_R.
\end{eqnarray}
This can also be seen from the fact that ${\bf S}^4$ is the $SO(4)$ orbit of $SO(5)$ through $\Gamma_5$ given by 
\begin{eqnarray}
x_a\Gamma_a=g\Gamma_5g^{-1}~,~g\in SO(5).
\end{eqnarray}
The $SU(2)_L$ acts on the eigenspace of $\Gamma_5$ with eigenvalue $+1$. The fiber over $p_0\in {\bf S}^4$ is clearly given by the condition (with $z=(\hat{z},0)$)
\begin{eqnarray}
{z}^+\Gamma_5 {z}=1\Rightarrow |z_1|^2+|z_2|^2=1.
\end{eqnarray}
This is ${\bf S}^3$ and because the phase of $\hat{z}$ drops we get ${\bf S}^3/U(1)={\bf S}^2$, i.e. the fiber is ${\bf S}^2$. In other words, ${\bf CP}^3$ in an ${\bf S}^2-$bundle over ${\bf S}^4$. On the other hand, the action of $SU(2)_R$ in ${\bf CP}^3$ is trivial.

What is the matrix analogue of the above Hopf map?

The quantization of the classical Hopf map $x_a:{\bf CP}^3\longrightarrow {\bf S}^4$ is clearly given by $X_a=rJ_a$ where $r^2=4R^2/n(n+4)$ and $J_a=a_{\alpha}^+(\Gamma_a)_{\alpha\beta}a_{\beta}/2$. The generators of $SO(5)$ are ${\cal M}_{ab}=a_{\alpha}^+(\sigma_{ab})_{\alpha\beta}a_{\beta}/2$ whereas the generators of $SO(6)=SU(4)$ are ${\cal M}_{AB}=\{{\cal M}_{ab},{\cal M}_{a6}=J_a\}$ and they act on the Hilbert space $(n,0,0)_{SU(4)}=(n/2,n/2,n/2)_{SO(6)}=(n/2,n/2)_{SO(5)}$. The generators of $SO(6)=SU(4)$ can also be given by 
\begin{eqnarray}
J_{\mu}=\frac{1}{2}a_{\alpha}^+(\Gamma_{\mu})_{\alpha\beta}a_{\beta}~,~\mu=1,...,15.
\end{eqnarray}
Essentially $J_{\mu}$ are the $T_{\mu}$ in (\ref{T}) and the correspondence between $\Gamma_{\mu}$ and $\{\Gamma_a, \sigma_{ab}\}$ is obvious. Strictly speaking, $X_{\mu}=rJ_{\mu}$ is the quantization of the classical Hopf map  $x_a:{\bf CP}^3\longrightarrow {\bf S}^4$ since fuzzy ${\bf S}^4_N$ is a squashed ${\bf S}^2-$bundle over ${\bf S}^4$ given by ${\bf CP}^3$ where the fiber is degenerate. 

\subsection{Poisson structure}
This then should be viewed as the quantization of the Kirillov-Kostant symplectic form corresponding to the Poisson structure 
\begin{eqnarray}
\{J_{\mu},J_{\nu}\}=f_{\mu\nu\lambda}J_{\lambda}
\end{eqnarray}
giving the commutators 
\begin{eqnarray}
[J_{\mu},J_{\nu}]=if_{\mu\nu\lambda}J_{\lambda}.
\end{eqnarray}
Thus $X_a=rJ_a\sim x_a$, $a=1,...,5$, is a subset of the quantized embedding functions $X_{\mu}=rJ_{\mu}\sim x_{\mu}:{\bf CP}^3\hookrightarrow su(4)={\bf R}^{15}$. %In other words, fuzzy ${\bf S}^4_N$ is a squashed fuzzy ${\bf CP}^3_N$ with a degenerate fiber given by a fuzzy sphere ${\bf S}^2_N$. 

The noncommutativity is given by ${\cal M}_{ab}$ since
\begin{eqnarray}
[X_a,X_b]=i\Theta_{ab}=ir^2{\cal M}_{ab},
\end{eqnarray}
and it arises from the Poisson structure on ${\bf CP}^3$. In other words, $\Theta^{ab}$ is the quantization of the embedding function  defined on ${\bf CP}^3$ by $\theta_{ab}:{\bf CP}^3\hookrightarrow so(5)$ where the antisymmetric tensor $\theta_{ab}$ is given by the Poisson bracket $\theta_{ab}=\{x_a,x_b\}$. This embedding is by construction not constant along the fiber, and thus it does not define a Poisson bracket on ${\bf S}^4$ ,and each point on the fiber ${\bf S}^2$ corresponds to a different choice of the noncommutativity $\theta_{ab}$ on ${\bf S}^4$.  This kind of averaging over ${\bf S}^2$ of the noncommutativity is also what guarantees $SO(5)$ invariance on this noncommutative space. The embedding function $\theta_{ab}$ therefore resolves completely the ${\bf S}^2$ fiber over ${\bf S}^4$.

Thus, the local noncommutativity is ${\cal M}_{ij}$ which also generate the local $SO(4)$ rotations. Indeed, by going to the north pole, we find that the stabilizer group is $SO(4)=SU(2)_L\times SU(2)_R$. We can decompose there the $SO(5)$ generators ${\cal M}_{ab}$ into the $SO(4)$ generators ${\cal M}_{ij}$ and the translations 
\begin{eqnarray}
P_i=\frac{1}{R}{\cal M}_{i5}.
\end{eqnarray}
They satisfy 
\begin{eqnarray}
[P_i,X_j]=\frac{i}{R}\delta_{ij}X_5~,~[P_i,P_j]=\frac{i}{R^2}{\cal M}_{ij}~,~[P_i,X_5]=-\frac{i}{R}X_i.
\end{eqnarray}
$X_i$ and $P_i$ are vectors whereas $X_5$ is a scalar under the rotations ${\cal M}_{ij}$. The $P_i$ reduce to ordinary translations if we set $X_5=R$, since we are at the north pole, and take the limit $R\longrightarrow\infty$. This is an Inonu-Wigner contraction of $SO(5)$ yielding the full Poincare group in $4$ dimensions. Thus the generators $P_i$ will allow us to move around ${\bf S}^4$. The  local noncommutativity is given by
\begin{eqnarray}
[X_i,X_j]=ir^2{\cal M}_{ij}~,~[X_i,X_5]=ir^2RP_i.
\end{eqnarray}
Obviously, $r^2{\cal M}_{ij}$ is the noncommutativity parameter $\Theta_{ij}$ and $r^2\longrightarrow 0$ in the limit since ${\cal M}\sim n$. In other words, the noncommutativity scale is given by 
\begin{eqnarray}
L_{\rm NC}^2=r^2n=\frac{4R^2}{n}.
\end{eqnarray}
\subsection{Coherent state}
This can also be seen as follows. The coherent state on ${\bf CP}^3=SO(5)/(SU(2)\times U(1))$ is given by the orbit $P=gP_0g^+=|x,\xi><x,\xi|$, where $g\in SO(5)$, $P_0=|\Lambda><\Lambda|$ where $|\Lambda>$ is the highest weight state of the irreducible representation $(n/2,n/2)$ of $SO(5)$,  and $|x,\xi>$ is the coherent state with $x\in{\bf S}^4$ and $\xi\in{\bf S}^2$. The north pole corresponds to the highest weight state. We have then
\begin{eqnarray}
\mu x^a=<x,\xi|X^a|x,\xi>=<X^a>.
\end{eqnarray}
This is consistent with
\begin{eqnarray}
\frac{x^a}{R}=<x,\xi|\Gamma^a|x,\xi>
\end{eqnarray}
with $\mu=1/\sqrt{1+4/n}$. The spread is given by
\begin{eqnarray}
\Delta_x^2=<(X^a-<X^a>)^2>=<(X^a)^2>-<X^a>^2=R^2-\mu^2 R^2=L_{\rm NC}^2.
\end{eqnarray}
The coherent state $|x,\xi>$ is therefore optimally localized since it minimizes the uncertainty relation. Functions on ${\bf CP}^3$ are associated to operators on $(n/2,n/2)$ by means of the coherent state by the usual formula 
\begin{eqnarray}
\phi(x,\xi)=<x,\xi|\hat{\Phi}|x,\xi>.
\end{eqnarray}
The coherent states are in one-to-one correspondence with point on ${\bf CP}^3$ up to a $U(1)$ factor. Thus, the coherent state is a $U(1)$ bundle over ${\bf CP}^3$. The curvature of the corresponding connection is the symplectic form $\omega$ on ${\bf CP}^3$  associated with the Poisson structure
\begin{eqnarray}
\theta_{ab}(x,\xi)=<x,\xi|[X_a,X_b]|x,\xi>=i<x,\xi|\Theta_{ab}|x,\xi>.
\end{eqnarray}
\subsection{Local Flatness}
The matrices $X_a$ describe a matrix or quantized membrane $4-$sphere which appears locally to be an ${\rm L}5-$brane in matrix theory with the correct charge \cite{Castelino:1997rv}. The rotational invariance under $SO(5)$ is given by the condition 
\begin{eqnarray}
R_{ab}X_b=UX_aU^{-1}.
\end{eqnarray}
Thus, the noncommutativity tensor will transform correspondingly as
\begin{eqnarray}
R_{aa^{'}}R_{bb^{'}}\Theta_{ab}=U\Theta_{a^{'}b^{'}}U^{-1}.
\end{eqnarray}
Rotations are then implemented by gauge transformations, i.e. as local transformations, in the spirit of gravity.

The matrices $J_a$ satisfy, among other things,  the so-called local flatness condition given by 
\begin{eqnarray}
\epsilon_{abcde}J_aJ_bJ_cJ_d=\alpha J_e.
\end{eqnarray}
By rotational invariance this needs only to be checked for $e=5$. We compute
\begin{eqnarray}
\epsilon_{abcd5}J_aJ_bJ_cJ_d=\{[J_1,J_2],[J_3,J_4]\}-\{[J_1,J_3],[J_2,J_4]\}+\{[J_1,J_4],[J_2,J_3]\}.
\end{eqnarray}
For the smallest possible representation $J_a=\Gamma_a/2$ we get $3J_5$ and thus $\alpha=3$. For any $J_a$ in the irreducible representation $(n,0,0)$ of $SU(4)$ we obtain instead \cite{badis}
\begin{eqnarray}
\epsilon_{abcde}J_aJ_bJ_cJ_d=(n+2)J_e.
\end{eqnarray}
By using $J_a^2=n(n+4)/4$ this can be rewritten as
\begin{eqnarray}
\epsilon_{abcde}J_aJ_bJ_cJ_dJ_e=\frac{1}{4}n(n+2)(n+4).\label{lf}
\end{eqnarray}
We can also rewrite this as (remember that $J_e={\cal M}_{e6}$)
\begin{eqnarray}
-\frac{1}{4}\epsilon_{abcde6}{\cal M}_{ab}{\cal M}_{cd}{\cal M}_{e6}=\frac{1}{4}n(n+2)(n+4).
\end{eqnarray}
Or
\begin{eqnarray}
-\frac{1}{24}\epsilon_{ABCDEF}{\cal M}_{AB}{\cal M}_{CD}{\cal M}_{EF}=\frac{1}{4}n(n+2)(n+4).
\end{eqnarray}
We recognize this to be the cubic Casimir of $SU(4)$, viz
\begin{eqnarray}
d_{\mu\nu\lambda}T_{\mu}T_{\nu}T_{\lambda}=-\frac{1}{32}\epsilon_{ABCDEF}{\cal M}_{AB}{\cal M}_{CD}{\cal M}_{EF}=\frac{3}{16}n(n+2)(n+4).
\end{eqnarray}
This with the quadratic Casimir
\begin{eqnarray}
T_{\mu}^2=\frac{1}{4}{\cal M}_{AB}{\cal M}_{AB}=\frac{3}{8}n(n+4)
\end{eqnarray}
provide the defining equations of ${\bf CP}^3_N$ as we know.
\subsection{Noncommutativity scale}
The noncommutativity $\Theta_{ij}=r^2{\cal M}_{ij}$, in the semi-classical limit, is a self-dual antisymmetric tensor $\theta_{ij}(x,\xi)$, transforming as $(1,0)$ under the local group $SO(4)=SU(2)_L\times SU(2)_R$. It corresponds to a bundle of self-dual frames over ${\bf S}^4$ which averages out over ${\bf S}^2$ \cite{Steinacker:2016vgf}.  The generators of $SU(2)_L$ denoted by $J_{\hat{i}}^L$ (the index $\hat{i}$ runs from $1$ to $3$) generate the local fiber given by the fuzzy sphere ${\bf S}^2_{N}\equiv {\bf S}^2_{n+1}$. In other words, the self-dual antisymmetric tensor ${\cal M}_{ij}$ is the flux of a $U(n+1)-$valued noncommutative gauge field given by the non-trivial instanton configuration 
\begin{eqnarray}
{\cal M}_{ij}=\epsilon_{ij}^{\hat{k}}J_{\hat{k}}^L.
\end{eqnarray}
Thus, locally (north pole) the four-sphere ${\bf S}^4$ is characterized by the $4$ coordinates $X_i$ whereas the fiber ${\bf S}^2$ is characterized by two of the three components of the self-dual tensor ${\cal M}_{ij}$.

This can also be interpreted as arising from a twisted stack  of $n+1$ noncommutative spherical branes carrying $U(n+1)-$valued gauge field \cite{Ho:2001as,Karczmarek:2015gda}. We identify these branes by choosing $n+1$ coherent states $|i>$ on ${\bf S}^2_{n+1}$. Thus every point on ${\bf S}^4$ is covered by $n+1$ sheets and these sheets are connected by the modes $|i><j|\in U(n+1)$. Also we can assign a Poisson structure to these sheets or leaves by the usual formula $<i|\Theta_{ij}|i>$. We will apply the semi-classical formula 
\begin{eqnarray}
tr\sim \frac{1}{\pi^2}\int \frac{1}{2}\omega\wedge\omega
\end{eqnarray}
for symplectic $4-$dimensional spaces to the $n+1$ leaves. We consider the local flatness condition (\ref{lf}) at the north pole which takes in the semi-classical limit the form 
 \begin{eqnarray}
\frac{1}{4}n(n+2)(n+4)&\sim &\frac{1}{4r^5}\epsilon_{abcde}\theta_{ab}\theta_{cd}x_e\nonumber\\
&\sim &\frac{R}{2r^5}{\rm Pf}\theta.
\end{eqnarray}
The Pfaffian is defined by 
  \begin{eqnarray}
{\rm Pf}\theta=\frac{1}{2}\epsilon_{ijkl}\theta_{ij}\theta_{kl}.
\end{eqnarray}
Thus,
 \begin{eqnarray}
\epsilon_{abcde}tr J_aJ_bJ_cJ_dJ_e\sim \frac{1}{24}n^6.
\end{eqnarray}
On the other hand, by applying the trace to the $n+1$ leaves we obtain 
 \begin{eqnarray}
\epsilon_{abcde}tr J_aJ_bJ_cJ_dJ_e&\sim& \frac{n+1}{\pi^2}\int \frac{1}{2}\omega^2\bigg(\frac{R}{2r^5}{\rm Pf}\theta\bigg)\nonumber\\
&\sim& \frac{n+1}{\pi^2}\frac{R}{2r^5}V_{{\bf S}^4}\nonumber\\
&\sim &\frac{n+1}{\pi^2}\frac{R}{2r^5}\frac{8\pi^2}{3}R^4.
\end{eqnarray}
By comparing we get $r=2R/n$ which is correct. Then, fuzzy ${\bf S}^4_N$ consists of $n^3/6$ cells of volume $L_{\rm NC}^4$. The volume of these cells is equal to $n+1$ (due to the fiber structure given by fuzzy ${\bf S}^2_{n+1}$) times the volume of the  ${\bf S}^4$, viz
\begin{eqnarray}
\frac{n^3}{6}L_{\rm NC}^4=(n+1).\frac{8\pi^2}{3}R^4\Rightarrow  L_{\rm NC}^2=\frac{4\pi R^2}{n}.
\end{eqnarray}
This is consistent with our previous estimate.
\subsection{Matrix model}

We have already discussed the harmonic expansion of functions on fuzzy ${\bf CP}^3_N$ given by the $SO(5)$ tensor product 
\begin{eqnarray}
(\frac{n}{2},\frac{n}{2})\otimes(\frac{n}{2},\frac{n}{2})=\sum_{l=0}^n\sum_{k=0}^l(l,k).
\end{eqnarray}
In terms of the Dynkin labels the representations $(l,k)$ read $(\lambda_1,\lambda_2)=(l-k,2k)$. Fuzzy ${\bf S}^4_N$ is given by the symmetric representations $(l,0)$ which correspond to totally symmetric polynomials of degree $l$ in $J_a$.  The space of functions on fuzzy ${\bf S}_N^4$ is written as 
\begin{eqnarray}
C_N({\bf S}^4)=\sum_{l=0}^n(l,0).
\end{eqnarray}
The representations $(l-k,2k)$ with $k\ne 0$ correspond to bosonic higher spin modes of $SO(5)$ on ${\bf S}^4_N$ which are spanned by the polynomials 
\begin{eqnarray}
(l-k,2k)={\rm span}\{P_{l-k}(J_a)P_k({\cal M}_{ab})\}.
\end{eqnarray}
 Indeed, the bosonic modes $(l-2k,2k)$ are functions on fuzzy ${\bf CP}^3_N$, which have a non-trivial dependence along the fiber ${\bf S}^2$, i.e. they transform non-trivially under the local group $SU(2)_L\times SU(2)_R$, and thus they are higher spin modes on ${\bf S}^4_N$ and not Kaluza-Klein modes. The low spin modes correspond to small $k$ and the spin $2$ sector is precisely the graviton \cite{Steinacker:2016vgf}. It is also known that projecting out these modes leads to a non-associative algebra \cite{Ramgoolam:2001zx}. Furthermore, the fermionic higher spin modes will involve representations $(l,k)$ with $k$ odd \cite{Steinacker:2015dra}.

The $SO(5)$ generators ${\cal M}_{ab}$ are tangential to the $4-$sphere since
 \begin{eqnarray}
i{\cal M}_{ab}J_b+iJ_b{\cal M}_{ab}=[J_a,J_b^2]=0.
\end{eqnarray}
Furthermore, we have
\begin{eqnarray}
\Box J_b=[J_a,[J_a,J_b]]=i[J_a,{\cal M}_{ab}]=4J_b.\label{fun1}
\end{eqnarray}
And
\begin{eqnarray}
X_aX_dX_a=R^2 X_d-\frac{1}{2}[X_a,[X_a,X_d]]=(R^2-2r^2)X_d.\label{fun1e}
\end{eqnarray}
A crucial property is given by 
\begin{eqnarray}
-g_{ac}\{{\Theta}^{ab},{\Theta}^{cd}\}&=&r^2\big(\{X^b,X^d\}-2R^2g^{bd}\big).\label{fun2}
\end{eqnarray}
The proof goes as follows. The ambient metric is given by
\begin{eqnarray}
g_{ab}=\delta_{ab}.
\end{eqnarray}
By using (\ref{fun1e}) we get immediately 
\begin{eqnarray}
-r^4g_{ac}\{{\cal M}^{ab},{\cal M}^{cd}\}&=&(R^2-4r^2)\{X_b,X_d\}-X_a\{X_b,X_d\}X_a\nonumber\\
&=&-4r^2\{X_b,X_d\}-[X_a,\{X_b,X_d\}]X_a.
\end{eqnarray}
Then (using Jacobi identity and (\ref{comm2}) and (\ref{fun1}))
\begin{eqnarray}
[X_a,\{X_b,X_d\}]X_a&=&[[X_a,X_d],X_bX_a]+[[X_a,X_b],X_dX_a]\nonumber\\
&=&[X_d,[X_b,X_a]]X_a+[X_b,[X_d,X_a]]X_a-X_b[X_a,[X_a,X_d]]-X_d[X_a,[X_a,X_b]]\nonumber\\
&=&r^2\big(2\delta_{bd}R^2-\{X_b,X_d\}\big)-4r^2\{X_b,X_d\}.
\end{eqnarray}
This leads to the desired result (\ref{fun2}). The semi-classical limit of this equation (\ref{fun2}) is given immediately by
\begin{eqnarray}
g_{ac}{\theta}^{ab}{\theta}^{cd}&=&-r^2R^2(P^T)^{bd}\sim -\frac{1}{4}\Delta_x^4 (P^T)^{bd},
\end{eqnarray}
where $\Delta_x=L_{\rm NC}$ is the non-commutativity scale and $(P^T)^{ab}$ is the tangent projector 
\begin{eqnarray}
(P^T)^{ab}=g^{ab}-\frac{x^ax^b}{R^2}.
\end{eqnarray}
We remark that (\ref{fun2}), as opposed to the non-commutativity itself, is a tensor living on fuzzy ${\bf S}^4_N$. As it turns out, the effective background metric around any point $p\in{\bf S}^4$ (north pole) is given exactly by \cite{Steinacker:2016vgf}
\begin{eqnarray}
\gamma^{jl}=g_{ik}{\theta}^{ij}{\theta}^{kl}=-\frac{1}{4}\Delta_x^4 g^{jl}.\label{metric}
\end{eqnarray}
This is a very important result and can also be seen alternatively as follows. [The minus sign is an error due to our correspondence $\theta_{ij}\longrightarrow i\Theta_{ij}$ which should be corrected].

The result (\ref{fun1}) allows us to immediately write down a five matrix model with a ground state given by fuzzy ${\bf S}^4_N$. This is given explicitly by \cite{Steinacker:2015dra}
\begin{eqnarray}
S=\frac{1}{g^2}Tr\bigg(-[D_a,D_b][D^a,D^b]+\mu^2 D_aD^a\bigg).
\end{eqnarray}
The equations of motion read
\begin{eqnarray}
[D_b,[D^b,D^a]]+\frac{\mu^2}{2}D^a=0.
\end{eqnarray}
Clearly, $D_a=J_a$ is a solution if $\mu^2=-8$.

The above matrix model involves the Laplacian
\begin{eqnarray}
\bar{\Box}=C_2^{SO(6)}-C_2^{SO(5)}=\frac{1}{2}[J_a,[J^a,...]]
\end{eqnarray}
with eigenvalues given by 
\begin{eqnarray}
\frac{1}{2}\bigg(l(l+3)-k(k+1)\bigg)~,~k\leq l.
\end{eqnarray}
We recall the $SO(5)$ Casimir and its eigenvalues (with ${\cal M}_{i5}=RP_i$)
\begin{eqnarray}
C_2^{SO(5)}=\frac{1}{4}{\cal M}_{ab}^2=\frac{R^2}{2}\big(P_i^2+\frac{1}{2R^2}{\cal M}_{ij}^2\big)\longrightarrow \frac{1}{2}\bigg(l(l+3)+k(k+1)\bigg).
\end{eqnarray}
At the north pole $p\in{\bf S}^4$ we can neglect the angular momentum contribution compared the translational contribution to get 
\begin{eqnarray}
C_2^{SO(5)}\simeq \frac{R^2}{2}P_i^2=\frac{1}{8r^2}\Delta_x^4 P_i^2.
\end{eqnarray}
In the local frame at $p$ we can also replace $P_i=i\partial_i$. Thus
\begin{eqnarray}
C_2^{SO(5)}\simeq -\frac{1}{2r^2}\gamma^{ij}\partial_i\partial_j.
\end{eqnarray}
Also we note that for the low spin modes $m=0,1,2$ we can make the approximation 
\begin{eqnarray}
\bar{\Box}\simeq C_2^{SO(5)}\longrightarrow \frac{1}{2}l(l+3).
\end{eqnarray}
We have then
\begin{eqnarray}
{\Box}=2r^2\bar{\Box}=[X_a,[X_a,...]]\simeq 2r^2C_2^{SO(5)}=-\gamma^{ij}\partial_i\partial_j.
\end{eqnarray}
Thus, $\gamma^{ij}$ is indeed the effective background metric.

There are other important results that can be derived from (\ref{metric}). First, we introduce a frame $\theta_{\mu\nu}^a$ on the bundle of self-dual tensors $\theta_{\mu\nu}$ over ${\bf S}^4$ normalized such that 
  \begin{eqnarray}
\theta^{\mu\nu}_a\theta^b_{\mu\nu}=4\delta_a^b.
\end{eqnarray}
This must be a self-dual frame and therefore
 \begin{eqnarray}
\theta_{\alpha\beta}^a=\frac{1}{2}\epsilon_{\mu\nu\alpha\beta}\theta^{\mu\nu a}.
\end{eqnarray}
This can be shown by using $\epsilon_{\mu\nu\alpha\beta}\epsilon^{\mu\nu\rho\sigma}=2(\delta_{\alpha}^{\rho}\delta_{\beta}^{\sigma}-\delta_{\beta}^{\rho}\delta_{\alpha}^{\sigma})$. We write $\theta_{\mu\nu}$ in terms of the frame $\theta_{\mu\nu}^a$ and in terms of the generators $J^a$ of the internal fuzzy sphere ${\bf S}^2_{n+1}$ as 
 \begin{eqnarray}
\theta^{\mu\nu}=r^2\theta^{\mu\nu}_aJ^a.
\end{eqnarray}
The normalization is fixed by (\ref{metric}) or equivalently $\theta_{\mu\nu}\theta^{\mu\nu}=\Delta_x^4\sim n^2r^4$. This background flux $\theta$ is a function on ${\bf CP}^3$, viz $\theta=\theta(x,\xi)$. We want to compute the average over the sphere, i.e. over $\xi\in {\bf S}^2$, of various objects constructed form $\theta$. For example, the average of the product $\theta^{\mu\nu}\theta^{\rho\sigma}$ over ${\bf S}^2$ obviously can only depend on the two constant tensors on ${\bf S}^4$: the Levi-Civita tensor $\epsilon^{\mu\nu\rho\sigma}$ and the generalized Kronecker delta $\delta^{\mu\rho}\delta^{\nu\sigma}-\delta^{\nu\rho}\delta^{\mu\rho}$. Thus
\begin{eqnarray}
[\theta^{\mu\nu}\theta^{\rho\sigma}]_{{\bf S}^2}=a\big(\delta^{\mu\rho}\delta^{\nu\sigma}-\delta^{\nu\rho}\delta^{\mu\rho}\big)+b\epsilon^{\mu\nu\rho\sigma}.
\end{eqnarray}
By contracting with $\epsilon_{\mu\nu\rho\sigma}$ and using  $\epsilon_{\mu\nu\rho\sigma}\epsilon^{\mu\nu\rho\sigma}=4!$ we find $b=\Delta_x^4/12$, while by contracting with $g_{\mu\rho}$ we obtain $a=\Delta_x^4/12$. Thus we get 
\begin{eqnarray}
[\theta^{\mu\nu}\theta^{\rho\sigma}]_{{\bf S}^2}=\frac{\Delta_x^4}{12}\bigg(\delta^{\mu\rho}\delta^{\nu\sigma}-\delta^{\nu\rho}\delta^{\mu\sigma}+\epsilon^{\mu\nu\rho\sigma}\bigg).\label{metric1}
\end{eqnarray}
From the fundamental result (\ref{fun2}) we obtain by substituting $b=d=5$ and using the semi-classical result $X^{\mu}\longrightarrow \mu x^{\mu}$ where $\mu^2=1/(1+4/n)$ the result 
\begin{eqnarray}
P^{\mu}P_{\mu}=\frac{4}{\Delta^2_x}.
\end{eqnarray}
This gives us the normalization of the $P^a$, which should be thought of as function on ${\bf CP}^3$, although they vanish in the semi-classical limit at the north pole $p$. We get then immediately the average on the fiber ${\bf S}^2$ given by 
\begin{eqnarray}
[P^{\mu}P^{\nu}]_{{\bf S^2}}=\frac{1}{\Delta^2_x}g^{\mu\nu}.\label{momentum1}
\end{eqnarray}
Since the semi-classical limit of $g_{ac}\{{\Theta}^{a\mu},\Theta^{c5}\}$ is zero we get also the average
  \begin{eqnarray}
[P^{\mu}{\cal M}^{\alpha\beta}]_{{\bf S^2}}=0.\label{a_momentum}
\end{eqnarray}
%\begin{figure}%[h]
%\begin{center}
%\includegraphics[width=7cm,angle=-90]{fig2_phase_diagram.ps}
%\caption{} 
%\label{phasediag}
%\end{center}
%\end{figure}

%\begin{figure}[htbp]
%\begin{center}
%\includegraphics[width=7.0cm,angle=0]{various_deformations.png}
%\end{center}
%\caption{Figure taken from \cite{Yang:2008fb}.}\label{ev7}
%\end{figure}

%\begin{figure}[htbp]
%\begin{center}
%\includegraphics[width=7.0cm,angle=0]{symplectic_vs_riemannian.png}
%\end{center}
%\caption{Figure taken from \cite{Yang:2008fb}.}\label{ev7}
%\end{figure}

%\begin{figure}[htbp]
%\begin{center}
%\includegraphics[width=7.0cm,angle=0]{IKKT.png}
%\end{center}
%\caption{Figure taken from \cite{Steinacker}.}\label{ev7}
%\end{figure}

%\paragraph{Acknowledgments}
\section{Emergent matrix gravity}
In this section we will follow mainly \cite{Steinacker:2016vgf}. We will also use \cite{Steinacker:2007dq,Steinacker:2010rh,Steinacker:2015dra}.
\subsection{Fluctuations on fuzzy ${\bf S}^4_N$}
We return to the matrix model
\begin{eqnarray}
S[D]=\frac{1}{g^2}Tr\bigg(-[D_a,D_b][D^a,D^b]+\mu^2 D_aD^a\bigg).\label{fund3}
\end{eqnarray}
We will allow now $a=1,...,10$. The equations of motion read
\begin{eqnarray}
[D_b,[D^b,D^a]]+\frac{\mu^2}{2}D^a=0.
\end{eqnarray}
We will expand around the background $D_a=J_a$, $a=1,...,5$ and $D_a=J_a=0$, $i=6,...,10$, with fluctuations ${\cal A}_a$ as
\begin{eqnarray}
D_a=J_a+{\cal A}_a.
\end{eqnarray}
The corresponding fluctuations of the flux are given by 
 \begin{eqnarray}
r^2[D^a,D^b]=i\Theta^{ab}_{(D)}=iF^{ab}+i\Theta^{ab}~,~iF^{ab}=r[J^a,{\cal A}^b]-r[J^b,{\cal A}^a]+r^2[{\cal A}^a,{\cal A}^b].
\end{eqnarray}
The definition of $\Theta^{ab}_{(J)}=\Theta^{ab}$ is obvious. Obviously, this will lead to a noncommutative gauge theory on the noncommutative brane ${\cal M}$ defined by the $J_a$ or equivalently to a geometric deformation of $x_a:{\cal M}\hookrightarrow{\bf R}^{10}$. 

We expand the fluctuation ${\cal A}_a$, keeping only tensors of rank up to $3$, into tangential and radial components as 
\begin{eqnarray}
{\cal A}^a=\hat{\kappa} \frac{X^a}{R}+\hat{\xi}^a+i{\Theta}^{ab}\hat{A}_b.\label{exfund}
\end{eqnarray}
The fields $\hat{\kappa}$ (radial fluctuation) and $\hat{\xi}^a$, $\hat{A}_b$ (tangential deformations) are  given by
\begin{eqnarray}
\hat{\kappa}=\kappa+\kappa_{bc}{\cal M}^{bc}+....
\end{eqnarray}
\begin{eqnarray}
\hat{\xi}^a=\xi^a+\xi^{abc}{\cal M}_{bc}+....
\end{eqnarray}
\begin{eqnarray}
\hat{A}_b=A_b+A_{bcd}{\cal M}^{cd}+....
\end{eqnarray}
The tensor fields ${\kappa}$, $\kappa_{bc}$, ${\xi}^a$, $\xi^{abc}$, ${A}_b$ and $A_{bcd}$ are functions on fuzzy ${\bf S}^4_N$, i.e. $\in C_N({\bf S}^4)$. The modes ${\xi}^a$ are redundant with the trace sector of the $A_{bcd}$ modes. The tangential deformation $\hat{A}_a$ corresponds to noncommutative gauge theory.  Indeed,  $\hat{A}_a$ is a $u(1)\times so(5)-$valued gauge field corresponding to symplectomorphisms on the bundle ${\bf CP}^3$. More generally, the full expansion into higher spin modes is captured by allowing $\hat{A}_a$, $\hat{\xi}^a$ and $\hat{\kappa}$ to take value in the universal enveloping algebra  $U(so(5))$.

Let us consider these fields near the north pole $p\in{\bf S}^4$. We will change notation here so that Greek indices refer now to $4$ dimensions $\mu,\nu=1,2,3,4$. Recall that $P_{\mu}={\cal M}_{\mu 5}/R$ and $\Theta_{\mu\nu}=r^2{\cal M}_{\mu\nu}$. The above expansion reads
\begin{eqnarray}
{\cal A}^{\mu}=\hat{\kappa} \frac{X^{\mu}}{R}+\hat{\xi}^{\mu}+i{\Theta}^{\mu\nu}\hat{A}_{\nu}+ir^2RP^{\mu}\hat{A}_5.
\end{eqnarray}
\begin{eqnarray}
{\cal A}^{5}=\hat{\kappa} \frac{X^{5}}{R}+\hat{\xi}^{5}-ir^2RP^{\mu}\hat{A}_{\mu}.
\end{eqnarray}
We consider the semi-classical limit of this formula. First, at the north pole $X^5=R$. Second, $\hat{\xi}^a$ is a tangential deformation and hence $\hat{\xi}^5=0$. Then, $r^2R\sim L_{\rm NC}^4/R\longrightarrow 0$. The semi-classical limit of the above expansion becomes then
\begin{eqnarray}
{\cal A}^{\mu}=\hat{\kappa} \frac{x^{\mu}}{R}+\hat{\xi}^{\mu}+{\theta}^{\mu\nu}\hat{A}_{\nu}.
\end{eqnarray}
\begin{eqnarray}
{\cal A}^{5}=\hat{\kappa}.
\end{eqnarray}
The expansions of the various fields become (with  ${\kappa}_{\mu}=2R{\kappa}_{\mu 5}$, $\xi^{\mu\nu}=3R\xi^{\mu\nu 5}$, $A_{\mu\nu}=2RA_{\mu\nu 5}$)
\begin{eqnarray}
\hat{\kappa}=\kappa+{\kappa}_{\mu}P^{\mu}+{\kappa}_{\mu\nu}{\cal M}^{\mu\nu}+...
\end{eqnarray}
\begin{eqnarray}
\hat{\xi}^{\mu}=\xi^{\mu}+\xi^{\mu\nu}P_{\nu}+\xi^{\mu\nu\rho}{\cal M}_{\nu\rho}+...
\end{eqnarray}
\begin{eqnarray}
\hat{A}_{\mu}=A_{\mu}+A_{\mu\nu}P^{\nu}+A_{\mu\nu\rho}{\cal M}^{\nu\rho}+....
\end{eqnarray}
The radial deformation $\kappa$ is the only mode which modifies the embedding of ${\bf S}^4$ in target space. This mode contributes also to the conformal metric. The metric $h_{\mu\nu}$ is the symmetric part of  $A_{\mu\nu}$, viz
\begin{eqnarray}
A_{\mu\nu}=\frac{1}{2}(h_{\mu\nu}+a_{\mu\nu}).
\end{eqnarray}
The mode $A_{\mu\nu\rho}$ is antisymmetric in the last two indices. The trace part of this mode is then given by $A_{\mu\nu\rho}=g_{\mu\nu}B_{\rho}-g_{\mu\rho}B_{\nu}$. The contribution of this trace part to ${\cal A}^{\mu}$ is given by
\begin{eqnarray}
{\cal A}^{\mu}=\theta^{\mu\nu}A_{\nu\alpha\beta}{\cal M}^{\alpha\beta}=\frac{2i}{r^2}g_{\nu\alpha}\theta^{\nu\mu}\theta^{\alpha\beta}B_{\beta}=2iR^2(P^TB)^{\mu}.
\end{eqnarray}
The trace part is then redundant with $\xi^{\mu}$. The modes $A_{\mu\nu}$ and $A_{\mu\nu\rho}$ are tangential in the first index similarly to $A_{\mu}$. The field $A_5$ leads to a tangential contribution but it drops out. The mode $A_{\mu}$ is a $U(1)$ gauge field, the symmetric part $h_{\mu\nu}$ of the mode $A_{\mu\nu}$ is the metric, and $A_{\mu\nu\rho}$ is an $SO(4)-$spin connection. These modes are local degrees of freedom on ${\bf S}^4$, i.e. their averages on the fiber ${\bf S}^2$ vanish.
\subsection{Gauge transformations}
Finite gauge transformations are given by 
\begin{eqnarray}
D^a\longrightarrow UD^aU^{-1}~,~U\in {\cal U}({\cal H}).
\end{eqnarray}
Recall that the Hilbert space ${\cal H}$ corresponds to the irreducible representation $(n/2,n/2)$ of $SO(5)$. We can write an infinitesimal gauge transformation as
\begin{eqnarray}
U=1+i\Lambda+...
\end{eqnarray}
Obviously the gauge parameter is function on fuzzy ${\bf CP}^3_N$, i.e.
\begin{eqnarray}
\Lambda\in{\rm End}({\cal H})=\oplus_{0\leq k\leq l\leq n}(l-k,2k)\oplus_k\Gamma^k{\bf S}^4.\label{bundle}
\end{eqnarray}
$\Gamma^k{\bf S}^4$ can be viewed as a higher-spin tensor bundle over ${\bf S}^4$. The local rotation group $SO(4)$ acts on these tensors via $-i[{\cal M}^{\mu\nu},...]$ while the gauge group ${\cal U}({\cal H})$ acts non-locally and mixes these tensors together. On the other hand, $X^{\mu}$ act on this bundle as derivative operators. For example, they act on functions as
\begin{eqnarray}
[X^{\mu},\phi]=\theta^{\mu\nu}\partial_{\nu}\phi.
\end{eqnarray}
We explain these things and related issues in detail now.

The infinitesimal gauge transformations are given by 
\begin{eqnarray}
D^a\longrightarrow D^a+i[\Lambda,D^a]\Rightarrow {\cal A}^a\longrightarrow {\cal A}^a+i[\Lambda,J^a]+i[\Lambda,{\cal A}^a].
\end{eqnarray}
Again, we expand the gauge parameter keeping only tensors of rank up to $3$ as
\begin{eqnarray}
\Lambda&=&\Lambda_0+\frac{1}{2}\Lambda_{ab}{\cal M}^{ab}+...\nonumber\\
&=&\Lambda_0+v_{\mu}P^{\mu}+\frac{1}{2}\Lambda_{\mu\nu}{\cal M}^{\mu\nu}+...
\end{eqnarray}
The gauge parameters $\Lambda_0$, $\Lambda_{ab}$, $v_{\mu}=R\Lambda_{\mu5}$, etc are functions on fuzzy ${\bf S}^4_N$, i.e. they are $\in C_N({\bf S}^4)$. The complete gauge field ${\cal A}_a$ should be decomposed similarly as (\ref{exfund}) or equivalently as
\begin{eqnarray}
{\cal A}^a={\cal A}^a_0+ir^2{\cal M}^{ab}\hat{A}_b+...,
\end{eqnarray}
where the definition of ${\cal A}^a_0$ is obvious and the gauge field is given by
\begin{eqnarray}
\hat{A}_b=A_b+A_{bcd}{\cal M}^{cd}+...
\end{eqnarray}
The gauge parameter $\Lambda_{ab}$ generates an $SO(5)$ transformation of the gauge field $\hat{A}_a$ which depends on where we are on ${\bf S}^4$. This is therefore a local $SO(5)$ symmetry, and the field $\hat{A}_a$ is a noncommutative $u(1)\times so(5)-$valued gauge field.

We consider the semi-classical limit at the north pole $x^{\mu}=0$, $x^5=R$ where commutators are replaced by Poisson brackets. Around the north pole we can assume that the noncommutativity $\theta_{\mu\nu}(x,\xi)=i<x,\xi|\Theta_{\mu\nu}|x,\xi>$ is independent of $x$, viz $\partial_{\mu}\theta^{\mu\nu}=0$. We can approximate the fuzzy ${\bf S}^4_N$ here with the Moyal-Weyl noncommutativity $[X^{\mu},X^{\nu}]=i\Theta^{\mu\nu}$ with $\Theta$ constant. In particular, derivations $P^{\mu}\sim i\partial_{\mu}$ are approximated by
\begin{eqnarray}
P^{\mu}=(\Theta^{-1})^{\mu\nu}X_{\nu}\Rightarrow<x,\xi|[X_{\mu},f]|x,\xi>=\{x_{\mu},f\}=\theta_{\mu\nu}\partial^{\nu}f.
\end{eqnarray}
We have also used the usual star product to compute
\begin{eqnarray}
<x,\xi|[f,g]|x,\xi>=\{f,g\}=\theta_{\mu\nu}\partial^{\mu}f\partial^{\nu}g.
\end{eqnarray}
We compute then immediately 
\begin{eqnarray}
\delta_{\Lambda_0}X^{\mu}\equiv i[\Lambda_0,X^{\mu}]\sim -i\theta^{\mu\nu}\partial_{\nu}\Lambda_0.
\end{eqnarray}
\begin{eqnarray}
\delta_{\Lambda}X^{\mu}\equiv \frac{i}{2}[\Lambda_{\rho\sigma}{\cal M}^{\rho\sigma},X^{\mu}]=\frac{i}{2}[\Lambda_{\rho\sigma},X^{\mu}]{\cal M}^{\rho\sigma}+\frac{i}{2}\Lambda_{\rho\sigma}[{\cal M}^{\rho\sigma},X^{\mu}]\sim -\frac{i}{2}\theta^{\mu\nu}\partial_{\nu}\Lambda_{\rho\sigma}{\cal M}^{\rho\sigma}.
\end{eqnarray}
\begin{eqnarray}
\delta_{v}X^{\mu}\equiv i[v_{\nu}P^{\nu},X^{\mu}]= -iv^{\mu}\frac{X_5}{R}+i[v_{\nu},X^{\mu}]P^{\nu}\sim -iv^{\mu}-i\theta^{\mu\nu}\partial_{\nu}v_{\lambda}P^{\lambda}.
\end{eqnarray}
\begin{eqnarray}
\delta_{v}\phi\equiv i[v_{\nu}P^{\nu},\phi]= i[v_{\nu},\phi]P^{\nu}+iv_{\nu}[P^{\nu},\phi]\sim i\theta^{\mu\nu}\partial_{\mu}v_{\rho}\partial_{\nu}\phi P^{\rho}-v_{\nu}\partial^{\nu}\phi.
\end{eqnarray}
\begin{eqnarray}
\delta_{\Lambda}\phi\equiv \frac{i}{2}[\Lambda_{\rho\sigma}{\cal M}^{\rho\sigma},\phi]=\frac{i}{2}[\Lambda_{\rho\sigma},\phi]{\cal M}^{\rho\sigma}+\frac{i}{2}\Lambda_{\rho\sigma}[{\cal M}^{\rho\sigma},\phi]\sim \frac{i}{2}\theta^{\mu\nu}\partial_{\mu}\Lambda_{\rho\sigma}\partial_{\nu}\phi{\cal M}^{\rho\sigma}.
\end{eqnarray}
Thus, $\delta_{\Lambda}$ corresponds to the action of local $SO(4)$ rotations on tensors whereas $\delta_v$ corresponds to the action of diffeomorphisms. We obtain then the gauge transformations
\begin{eqnarray}
\delta{\cal A}^{\mu}=\delta_{\Lambda_0}{\cal A}^{\mu}+\delta_v{\cal A}^{\mu}+\delta_{\Lambda}{\cal A}^{\mu}-\frac{i}{r}\theta^{\mu\nu}\partial_{\nu}\big(\Lambda_0+v_{\lambda}P^{\lambda}+\frac{1}{2}\Lambda_{\rho\sigma}{\cal M}^{\rho\sigma}\big)-\frac{i}{r}v^{\mu}.
\end{eqnarray}
Similarly we have
\begin{eqnarray}
\delta{\cal A}^{5}=\delta_{\Lambda_0}\hat{\kappa}+\delta_v\hat{\kappa}+\delta_{\Lambda}\hat{\kappa}.
\end{eqnarray}
Also we compute
\begin{eqnarray}
\delta_{\Lambda_0}(\Theta^{\mu\nu}\hat{A}_{\nu})\equiv i[\Lambda_{0},\Theta^{\mu\nu}\hat{A}_{\nu}]\sim -i\theta^{\mu\nu}\delta_{\Lambda_0}\hat{A}_{\nu}\sim 0.
\end{eqnarray}
\begin{eqnarray}
\delta_v(\Theta^{\mu\nu}\hat{A}_{\nu})\equiv i[v_{\alpha}P^{\alpha},\Theta^{\mu\nu}\hat{A}_{\nu}]\sim -r^2v_{\rho}(-g^{\rho\mu}P^{\nu}+g^{\rho\nu}P^{\mu})\hat{A}_{\nu}-i\theta^{\mu\nu}\delta_v\hat{A}_{\nu}.
\end{eqnarray}
The first term will be dropped in the semi-classical limit $r^2\sim L_{\rm NC}^2/n\longrightarrow 0$ whereas the second term yields $i\theta^{\mu\nu}v^{\alpha}\partial_{\alpha}\hat{A}_{\nu}$. Also we compute
\begin{eqnarray}
\delta_{\Lambda}(\Theta^{\mu\nu}\hat{A}_{\nu})\equiv \frac{i}{2}[\Lambda_{\rho\sigma}{\cal M}^{\rho\sigma},\Theta^{\mu\nu}\hat{A}_{\nu}]&=&\frac{i}{2}[\Lambda_{\rho\sigma}{\cal M}^{\rho\sigma},\Theta^{\mu\nu}]\hat{A}_{\nu}+\Theta^{\mu\nu}\delta_{\Lambda}\hat{A}_{\nu}\nonumber\\
&\sim & i\Lambda_{\rho\sigma}\big(g^{\rho\mu}\theta^{\sigma\nu}-g^{\rho\nu}\theta^{\sigma\mu}\big)\hat{A}_{\nu}-i\theta^{\mu\nu}\delta_{\Lambda}\hat{A}_{\nu}\nonumber\\
&\sim & -i(\Lambda.\theta\hat{A})^{\mu}+i\theta^{\mu\nu}(\Lambda.\hat{A})_{\nu}.
\end{eqnarray}
The action of the local rotation $\Lambda\in so(4)$ on the $4-$dimensional gauge field $\hat{A}^{\mu}$ is defined by $(\Lambda.A)_{\mu}=-\Lambda_{\mu\nu}A^{\nu}$. We can then write down the full infinitesimal gauge transformation as
\begin{eqnarray}
\delta{\cal A}^{\mu}=\delta{\cal A}_0^{\mu}+\theta^{\mu\nu}\delta\hat{A}_{\nu}.
\end{eqnarray}
\begin{eqnarray}
\delta\hat{A}_{\nu}=-\frac{i}{r}\partial_{\nu}\big(\Lambda_0+v_{\lambda}P^{\lambda}+\frac{1}{2}\Lambda_{\rho\sigma}{\cal M}^{\rho\sigma}\big)-v^{\alpha}\partial_{\alpha}\hat{A}_{\nu}-(\Lambda.\hat{A})_{\nu}-\frac{4}{L_{\rm NC}^4}\theta_{\nu\alpha}(\Lambda.\theta\hat{A})^{\alpha}.
\end{eqnarray}
\begin{eqnarray}
\delta {\cal A}_0^{\mu}=\delta \hat{\xi}^{\mu}=-\frac{i}{r}v^{\mu}-v^{\alpha}\partial_{\alpha}\xi^{\mu}.
\end{eqnarray}
These equations are slightly different from those found originally in \cite{Steinacker:2016vgf}. In the second equation we have used $g_{\mu\nu}\theta^{\mu\alpha}\theta^{\nu\beta}=\Delta_x^4g^{\alpha\beta}/4$. We will drop the second term in this equation for the only reason that it is quadratic in the non-commutativity. 

Explicitly, the infinitesimal gauge transformation read then
\begin{eqnarray}
\delta\hat{A}_{\nu}=\delta A_{\nu}+\frac{1}{2}(\delta h_{\nu\mu}+\delta a_{\nu\mu})P^{\mu}+\delta A_{\nu\rho\sigma}{\cal M}^{\rho\sigma},
\end{eqnarray}
where
\begin{eqnarray}
\delta A_{\nu}=-\frac{i}{r}\partial_{\nu}\Lambda_0-v^{\alpha}\partial_{\alpha}{A}_{\nu}-(\Lambda.{A})_{\nu}.
\end{eqnarray}
\begin{eqnarray}
\delta h_{\nu\mu}=-\frac{i}{r}(\partial_{\nu}v_{\mu}+\partial_{\mu}v_{\nu})-v^{\alpha}\partial_{\alpha}h_{\nu\mu}-(\Lambda.h)_{\nu\mu}.
\end{eqnarray}
\begin{eqnarray}
\delta a_{\nu\mu}=-\frac{i}{r}(\partial_{\nu}v_{\mu}-\partial_{\mu}v_{\nu})-v^{\alpha}\partial_{\alpha}a_{\nu\mu}-(\Lambda.a)_{\nu\mu}.
\end{eqnarray}
\begin{eqnarray}
\delta A_{\nu\rho\sigma}=-\frac{i}{2r}\partial_{\nu}\Lambda_{\rho\sigma}-v^{\alpha}\partial_{\alpha}A_{\nu\rho\sigma}-(\Lambda.A)_{\nu\rho\sigma}.
\end{eqnarray}
This is a combined effect of local $SO(4)$ gauge transformations generated by $\Lambda_{\mu\nu}$, local $U(1)$ gauge transformations generated by $\Lambda_0$, and diffeomorphism transformations generated by $-v^{\rho}\partial_{\rho}$. The $U(1)$ gauge field is $A_{\mu}$, the $SO(4)$ gauge field is the spin connection $A_{\mu\nu\rho}$, while the gauge fields associated with the diffeomorphisms are $h_{\mu\nu}$ and $a_{\mu\nu}$. The metric fluctuation is identified with the symmetric rank two tensor $h_{\mu\nu}$. 
\subsection{Emergent geometry}
We rewrite the metric (\ref{metric}) on ${\bf S}^4_N$ in the semi-classical limit as
\begin{eqnarray}
\bar{\gamma}^{\mu\nu}=g_{\alpha\beta}e^{\alpha\mu}e^{\beta\nu}=\frac{1}{4}\Delta_x^4g^{\mu\nu}.
\end{eqnarray}
The vielbein $e^{\alpha\mu}$ is defined by
\begin{eqnarray}
e^{\alpha\mu}=\theta^{\alpha\mu}~,~e^{\alpha}=e^{\alpha\mu}\partial_{\mu}\leftarrow X^{\alpha}=\Theta^{\alpha\mu}P_{\mu}.
\end{eqnarray}
This is not a fixed frame on ${\bf S}^4$ since it corresponds to the bundle of self-dual tensors $\theta^{\mu\nu}$, which transforms under the local $SO(4)$ in the $(1,0)$ representation along the fiber ${\bf S}^2_N$, and thus it averages out over the fiber, viz
\begin{eqnarray}
[e^{\alpha\mu}]_{{\bf S}^2}=[\theta^{\alpha\mu}]_{{\bf S}^2}=0.
\end{eqnarray}
The metric $\bar{\gamma}$ is however fixed on ${\bf S}^4$ and well defined since
\begin{eqnarray}
[\bar{\gamma}^{\mu\nu}]_{{\bf S}^2}=\bar{\gamma}^{\mu\nu}.
\end{eqnarray}
As we have discussed the derivative operators acting on the higher-spin tensor bundle ${\rm End}({\cal H})$ given in equation (\ref{bundle}) are given by $X^a$. Indeed, the derivative of a general tensor $\phi\in{\rm End}({\cal H})$ is given by
\begin{eqnarray}
{\cal D}^{\mu}\phi=-i[X^{\mu},\phi]=-ie^{\mu\nu}\partial_{\nu}\phi+{\rm non-derivative~ terms}.
\end{eqnarray}
The non-derivative terms arise from commutators between $X$, from one hand, and the $P$ and ${\cal M}$, which appear in the expansion of $\phi$, from the other hand. However, the metric is always obtained from the leading derivative term. Indeed
\begin{eqnarray}
{\cal D}^{a}\phi {\cal D}_{a}\phi=-\bar{\gamma}^{\alpha\beta}\partial_{\alpha}\phi\partial_{\beta}\phi=-[X^{a},\phi][X_{a},\phi].
\end{eqnarray}
This generalizes when fluctuations are included to
\begin{eqnarray}
{\cal D}^{a}\phi {\cal D}_{a}\phi=-{\gamma}^{\alpha\beta}\partial_{\alpha}\phi\partial_{\beta}\phi=-[Y^{a},\phi][Y_{a},\phi].
\end{eqnarray}
The covariant derivative $Y$ is defined by
\begin{eqnarray}
Y^{a}=rD^{a}=X^{a}+r{\cal A}^{a}.
\end{eqnarray}
The metric can then be given by
\begin{eqnarray}
-{\cal D}^{a}x^{\alpha} {\cal D}_{a}x^{\beta}={\gamma}^{\alpha\beta}=[Y^{a},X^{\alpha}][Y_{a},X^{\beta}].
\end{eqnarray}
In summary, the curved over-complete basis is now defined by
\begin{eqnarray}
{\cal D}^{a}=-i[Y^a,...]=-ie^a[{\cal A}]=-ie^{a\nu}[{\cal A}]\partial_{\nu}.
\end{eqnarray}
After another messy calculation we find the covariant derivative (using $\hat{\phi}=\phi+\phi_{\mu}P^{\mu}+\phi_{\rho\sigma}{\cal M}^{\rho\sigma}$, ${\cal A}^{\mu(2)}=\theta^{\mu\nu}A_{\nu\rho}P^{\rho}$, ${\cal A}^{\mu(3)}=\theta^{\mu\nu}A_{\nu\rho\sigma}{\cal M}^{\rho\sigma}$, and setting $x^{\mu}=0$ at the north pole, and also neglecting there quadratic terms in $\theta$)
  \begin{eqnarray}
{\cal D}^{\mu}\hat{\phi}=\partial_{\nu}\phi\bigg[-i(1+r\frac{\hat{\kappa}}{R})\theta^{\mu\nu}-ir\theta^{\alpha\nu}\partial_{\alpha}\hat{\xi}^{\mu}+r\theta^{\mu\lambda}A^{\rho\nu}g_{\lambda\rho}\bigg]-(1+r\frac{\hat{\kappa}}{R})\phi^{\mu}-r\phi_{\nu}\partial^{\nu}\hat{\xi}^{\mu}-ir[{\cal A}_{\mu}^{(1)}+{\cal A}_{\mu}^{(3)},\hat{\phi}].\nonumber\\
\end{eqnarray}
Since we must have ${\cal D}^{\mu}\hat{\phi}=-ie^{\mu\nu}[{\cal A}]\partial_{\nu}\phi+...$ we obtain the tangential contribution to the  vielbein $e^{\mu\nu}$ (dropping also higher modes in $\hat{\xi}$)
 \begin{eqnarray}
e^{\mu\nu}[{\cal A}] = (1+r\frac{\hat{\kappa}}{R})\theta^{\mu\nu}+\delta e^{\mu\nu}~,~\delta e^{\mu\nu}=ir\theta^{\mu\lambda}A_{\lambda\rho}g^{\rho\nu}+r\partial_{\alpha}\xi^{\mu}\theta^{\alpha\nu}.
\end{eqnarray}
We can drop  $\hat{\kappa}$ and re-incorporate it by the replacement $h^{\mu\nu}\longrightarrow \tilde{h}^{\mu\nu}=h^{\mu\nu}+2r\hat{\kappa}g_{\mu\nu}/R$. In any case the contribution of $\hat{\kappa}$ is subleading.
 
A very neat calculation gives now the metric fluctuation on fuzzy ${\bf S}^4_N$: 
 \begin{eqnarray}
\gamma^{\mu\nu}=g_{\alpha\beta}e^{\alpha\mu}e^{\beta\nu}=\bar{\gamma}^{\mu\nu}+\delta\gamma^{\mu\nu}\sim [Y^a,X^{\mu}][Y_a,X^{\mu}],
\end{eqnarray}
where
\begin{eqnarray}
\delta\gamma^{\mu\nu}&=&g_{\alpha\beta}\big(\theta^{\alpha\mu}\delta e^{\beta\nu}+\delta e^{\alpha\mu}\theta^{\beta\nu}\big)\nonumber\\
&=&\frac{ir}{4}\Delta_x^4h^{\mu\nu}+r\theta^{\alpha\mu}\theta^{\beta\nu}\big(\partial_{\alpha}\xi_{\beta}+\partial_{\beta}\xi_{\alpha}\big).                          
\end{eqnarray}
The crucial observation here is that the metric $h^{\mu\nu}$ arises from the commutators of the $P$ modes in the gauge field $\hat{A}_{\mu}$, i.e. from the term $A_{\mu\nu}P^{\nu}$. This crucial property is absent on most other noncommutative spaces as seen as noncommutative branes in the matrix model \cite{Steinacker:2016vgf}. However, as we have seen the $\xi^{\mu}$ is redundant and in fact it can be gauged away, and thus the metric fluctuation does really consist only of the term $h$. By averaging over the internal sphere ${\bf S}^2_N$ (using (\ref{metric1})) we obtain the metric 
\begin{eqnarray}
[\delta\gamma^{\mu\nu}]_{{\bf S}^2}
&=&\frac{ir}{4}\Delta_x^4h^{\mu\nu}+\frac{r\Delta_x^4}{12}\bigg(\delta^{\alpha\beta}\delta^{\mu\nu}-\delta^{\mu\beta}\delta^{\alpha\nu}\bigg)\big(\partial_{\alpha}\xi_{\beta}+\partial_{\beta}\xi_{\alpha}\big).                          
\end{eqnarray}
We can now write down the action of a scalar field on fuzzy ${\bf S}^4_N$ as
\begin{eqnarray}
S=-\frac{2}{g^2}Tr[Y^a,\phi][Y_a,\phi]\sim -\frac{2}{g^2}\frac{{\rm dim}{\cal H}}{{\rm Vol}({\cal M}^4)}\int_{{\cal M}}d^4x\gamma^{\mu\nu}\partial_{\mu}\phi\partial_{\nu}\phi.
\end{eqnarray}
The covariant effective metric $G$ should be defined by 
\begin{eqnarray}
\gamma^{\mu\nu}=\frac{\Delta_x^4}{4}\sqrt{{\rm det}G_{\mu\nu}}G^{\mu\nu}.
\end{eqnarray}
But we have 
\begin{eqnarray}
\gamma^{\mu\nu}&=&\frac{1}{4}{\Delta_x^4}g^{\mu\nu}+\delta\gamma^{\mu\nu}\nonumber\\
&=&\frac{\Delta_x^4}{4}(g^{\mu\nu}+irh^{\mu\nu}+...)\Rightarrow \gamma_{\mu\nu}=\frac{4}{\Delta_x^4}(g_{\mu\nu}-irh_{\mu\nu}).
\end{eqnarray}
We define 
\begin{eqnarray}
G^{\mu\nu}=\alpha\frac{4}{\Delta_x^4}\gamma^{\mu\nu}\Rightarrow \frac{\Delta_x^4}{4}\sqrt{{\rm det}G_{\mu\nu}}G^{\mu\nu}=(\frac{\Delta_x^4}{4})^2\frac{1}{\alpha}\sqrt{{\rm det}\gamma_{\mu\nu}}\gamma^{\mu\nu}.
\end{eqnarray}
Thus
\begin{eqnarray}
\alpha=\bigg((\frac{\Delta_x^4}{4})^4{{\rm det}\gamma_{\mu\nu}}\bigg)^{1/2}.
\end{eqnarray}
We compute
\begin{eqnarray}
{\rm det}\gamma_{\mu\nu}=(\frac{4}{\Delta_x^4})^4(1-irh)\Rightarrow \alpha=1-\frac{ir}{2}h.
\end{eqnarray}
We write 
\begin{eqnarray}
G^{\mu\nu}=g^{\mu\nu}+ir H^{\mu\nu}.
\end{eqnarray}
We find
\begin{eqnarray}
H^{\mu\nu}=h^{\mu\nu}-\frac{1}{2}hg^{\mu\nu}.
\end{eqnarray}
We will impose the so-called De Donder gauge
\begin{eqnarray}
\partial_{\mu}h^{\mu\nu}=0\Rightarrow \partial_{\mu}H^{\mu\nu}=\frac{1}{2}\partial^{\nu}H.\label{donder}
\end{eqnarray}
The action becomes (with ${{\Phi}}=\Delta_x^2\phi/2$) given by 
\begin{eqnarray}
S=-\frac{2}{g^2}Tr[Y^a,\phi][Y_a,\phi]\sim -\frac{1}{g^2}\frac{\Delta_x^8}{8}\frac{{\rm dim}{\cal H}}{{\rm Vol}({\cal M}^4)}\int_{{\cal M}}d^4x\sqrt{{\rm det}G}G^{\mu\nu} \partial_{\mu}{{\Phi}}\partial_{\nu}{{\Phi}}.
\end{eqnarray}
The metric $G^{\mu\nu}=g^{\mu\nu}+irH^{\mu\nu}$ at the linearized level is thus obtained by the replacemnt $h^{\mu\nu}\longrightarrow H^{\mu\nu}=h^{\mu\nu}-hg^{\mu\nu}/2$. We have
  \begin{eqnarray}
H_{\mu\nu}=A_{\mu\nu}+A_{\nu\mu}-\frac{1}{2}hg^{\mu\nu}=A_{\mu\nu}^{'}+A_{\nu\mu}^{'}~,~A_{\mu\nu}^{'}=A_{\mu\nu}-\frac{1}{4}hg_{\mu\nu}.
\end{eqnarray}
We also introduce the effective vielbeins 
 \begin{eqnarray}
G^{\mu\nu}=g_{\alpha\beta}\tilde{e}^{\alpha\mu}[{\cal A}]\tilde{e}^{\beta\nu}[{\cal A}],\label{stringmetric}
\end{eqnarray}
where (using ${\rm det}G=1+irh/3$)
\begin{eqnarray}
\tilde{e}^{\alpha\nu}[{\cal A}]&=&\frac{2}{\Delta_x^2}\frac{1}{({\rm det}G)^{1/4}}e^{\alpha\nu}[{\cal A}]=\frac{2}{\Delta_x^2}\theta^{\alpha\lambda}\bigg(\delta^{\nu}_{\lambda}+irA^{'}_{\lambda\rho}g^{\rho\nu}\bigg).\label{veilbein}
\end{eqnarray}
This is essentially the open string metric on the noncommutative $D-$branes in a strong magnetic field  in the limit $\tilde{\alpha}^{'}\longrightarrow 0$ considered in \cite{Seiberg:1999vs}.

The inverse vielbeins defined by $\tilde{e}^{\alpha\nu}\tilde{e}_{\nu\beta}=\delta^{\alpha}_{\beta}$ is given by
\begin{eqnarray}
\tilde{e}_{\alpha\nu}[{\cal A}]&=&\frac{\Delta_x^2}{2}(\theta^{-1})_{\alpha\lambda}\bigg(\delta^{\nu}_{\lambda}-irA^{'}_{\lambda\rho}g^{\rho\nu}\bigg).
\end{eqnarray}
The torsion-free spin connection corresponding to the vielbein $\tilde{e}^{\alpha\mu}$ is given by \cite{Steinacker:2016vgf}
\begin{eqnarray}
A_{\mu;\alpha\beta}=\frac{1}{4}(-\partial_{\alpha}H_{\mu\beta}+\partial_{\beta}H_{\alpha\mu}).\label{spin-connection}
\end{eqnarray}
We compute further 

\begin{eqnarray}
\tilde{e}^{\alpha\nu}[{\cal A}]-\tilde{e}^{\nu\alpha}[{\cal A}]&=&\frac{2}{\Delta_x^2}(\theta^{\alpha\nu}+\tilde{\theta}^{\alpha\nu}).\label{veilbein1}
\end{eqnarray}
\begin{eqnarray}
\tilde{\theta}^{\alpha\nu}=\theta^{\alpha\nu}-ir\theta^{\nu\lambda}{A}_{\lambda\rho}^{'}g^{\rho\alpha}+ir\theta^{\alpha\lambda}{A}_{\lambda\rho}^{'}g^{\rho\nu}.
\end{eqnarray}
The bit $rh\theta^{\alpha\nu}$ in $r\theta^{\nu\lambda}{A}_{\lambda\rho}^{'}g^{\rho\alpha}$ can be neglected.
\subsection{Emergent gauge theory}
Recall that the gauge field is given by ${\cal A}_{\mu}=\xi_{\mu}+\theta_{\mu\nu}\hat{A}^{\nu}$, $\hat{A}^{\lambda}=A^{\lambda}+A^{\lambda\rho}P_{\rho}+A^{\lambda\rho\sigma}{\cal M}_{\rho\sigma}$. The corresponding fluctuations of the flux are given by 
 \begin{eqnarray}
i\Theta^{\mu\nu}_{(Y)}&=&[Y^{\mu},Y^{\nu}]\nonumber\\
&=&\theta^{\mu\nu}+r{\cal F}^{\mu\nu},
\end{eqnarray}
where 
\begin{eqnarray}
 {\cal F}_{\mu\nu}&=&[X_{\mu},\hat{\cal A}_{\nu}]-[X_{\nu},\hat{\cal A}_{\mu}]+r[\hat{\cal A}_{\mu},\hat{\cal A}_{\nu}]\nonumber\\
&=&\theta_{\mu\lambda}\partial^{\lambda}\xi_{\nu}-\theta_{\nu\lambda}\partial^{\lambda}\xi_{\mu}+\theta_{\nu\lambda}\theta_{\mu\rho}(\partial^{\rho}\hat{A}^{\lambda}-\partial^{\lambda}\hat{A}^{\rho})-i\theta_{\nu\lambda}A^{\lambda\mu}+i\theta_{\mu\lambda}A^{\lambda\nu}+r[\hat{\cal A}_{\mu},\hat{\cal A}_{\nu}]\nonumber\\
&=&\theta_{\nu\lambda}\theta_{\mu\rho}(\partial^{\rho}\hat{A}^{\lambda}-\partial^{\lambda}\hat{A}^{\rho})-i\theta_{\nu\lambda}\tilde{A}^{\lambda\mu}+i\theta_{\mu\lambda}\tilde{A}^{\lambda\nu}+r[\hat{\cal A}_{\mu},\hat{\cal A}_{\nu}].
\end{eqnarray}
The shifted gauge field is invariant under diffeomorphisms and is given by
\begin{eqnarray}
 \tilde{A}^{\lambda\nu}=A^{\lambda\nu}-i\partial^{\lambda}\xi^{\nu}.
\end{eqnarray}
The most important term in $[\hat{\cal A}_{\mu},\hat{\cal A}_{\nu}]$ is $\theta_{\mu\rho}\theta_{\nu\lambda}[\hat{A}_{\mu},\hat{A}_{\nu}]$. We get then the curvature
\begin{eqnarray}
 {\cal F}_{\mu\nu}
&=&\theta_{\mu\rho}\theta_{\nu\lambda}(\partial^{\rho}\hat{A}^{\lambda}-\partial^{\lambda}\hat{A}^{\rho}+r [\hat{A}_{\mu},\hat{A}_{\nu}])-i\theta_{\nu\lambda}\tilde{A}^{\lambda\mu}+i\theta_{\mu\lambda}\tilde{A}^{\lambda\nu}\nonumber\\
&=&\theta_{\mu\rho}\theta_{\nu\lambda}\hat{F}^{\rho\lambda}-i\theta_{\nu\lambda}\tilde{A}^{\lambda\mu}+i\theta_{\mu\lambda}\tilde{A}^{\lambda\nu}.
\end{eqnarray}
Obviously, $\hat{F}$ decomposes in terms of the $so(4)\times u(1)$ components as follows 
\begin{eqnarray}
\hat{F}^{\rho\lambda}=F^{\rho\lambda}+R^{\rho\lambda}+T^{\rho\lambda},
\end{eqnarray}
where the $U(1)$ field strength $F$, the Riemann curvature $R$ of the $SO(4)$ connection $\omega^{\mu}=A^{\lambda\rho\sigma}{\cal M}_{\rho\sigma}$, the linearized spin connection  $T$ are given by (with $\alpha_{\mu}=A_{\mu\lambda}P^{\lambda}$)
\begin{eqnarray}
{F}^{\rho\lambda}=\partial^{\rho}{A}^{\lambda}-\partial^{\lambda}{A}^{\rho}.
\end{eqnarray}
\begin{eqnarray}
{R}^{\rho\lambda}=\partial^{\rho}{\omega}^{\lambda}-\partial^{\lambda}{\omega}^{\rho}+r [{\omega}^{\rho},{\omega}^{\lambda}].
\end{eqnarray}
\begin{eqnarray}
{T}^{\rho\lambda}=\partial^{\rho}{\alpha}^{\lambda}-\partial^{\lambda}{\alpha}^{\rho}+r [{\omega}^{\rho},{\alpha}^{\lambda}]-r [{\omega}^{\lambda},{\alpha}^{\rho}].
\end{eqnarray}
The linearized form of $T$ is precisely given by the spin-connection (\ref{spin-connection}), viz (by dropping $a_{\mu\nu}$ and $h=g_{\mu\nu}h^{\mu\nu}$)
 \begin{eqnarray}
{T}_{\mu\nu}=-P^{\rho}(\partial_{\mu}A_{\rho\nu}-\partial_{\nu}A_{\rho\mu})=2P^{\rho}A_{\rho;\mu\nu}+...
\end{eqnarray}
The geometric deformation of the background $\theta$ is then given by $\theta\longrightarrow\tilde{\theta}$ where
 \begin{eqnarray}
i\Theta^{\mu\nu}_{(Y)}
&=&\theta_{\mu\nu}+r{\cal F}_{\mu\nu}\nonumber\\
&=&\tilde{\theta}_{\mu\nu}+r\theta_{\mu\rho}\theta_{\nu\lambda}\hat{F}^{\rho\lambda},
\end{eqnarray}
where
\begin{eqnarray}
\tilde{\theta}_{\mu\nu}=\theta_{\mu\nu}-ir\theta_{\nu\lambda}\tilde{A}^{\lambda\mu}+ir\theta_{\mu\lambda}\tilde{A}^{\lambda\nu}.
\end{eqnarray}
The mode $\xi^{\alpha}$ in $\tilde{A}^{\rho\alpha}$ can be dropped since it is unphysical. Thus, $\tilde{\theta}$ can be viewed as deformation of $\theta$ in the background $Y$ and this deformation is provided by $\tilde{A}^{\mu\nu}$ or $A^{\mu\nu}$ which its symmetric part encodes also the metric. The vielbein $\tilde{e}^{\mu\nu}$ given by (\ref{veilbein}) is then viewed as the deformation of the vielbein $e^{\alpha\mu}=\theta^{\alpha\mu}$ (see also (\ref{veilbein1}))

We also rewrite the above flux fluctuation as
\begin{eqnarray}
i\Theta^{\mu\nu}_{(Y)}=[Y^{\mu},Y^{\nu}]=\theta^{\mu\rho}\theta^{\nu\lambda}\bar{F}_{\rho\lambda}~,~\bar{F}_{\rho\lambda}=(\theta^{-1})_{\rho\rho_1}(\theta^{-1})_{\nu_1\lambda}\tilde{\theta}^{\rho_1\nu_1}+r\hat{F}_{\rho\lambda}.
\end{eqnarray}
Thus the Yang-Mills action is given by
\begin{eqnarray}
-\frac{1}{g^2}Tr[Y^{\mu},Y^{\nu}][Y_{\mu},Y_{\nu}]&=&-\frac{1}{g^2}(\frac{\Delta_x^4}{4})^2Tr g^{\rho\rho_1}g^{\lambda\lambda_1}\bar{F}_{\rho_1\lambda_1}\bar{F}_{\rho\lambda}\nonumber\\
&=&-\frac{1}{g^2}\frac{{\rm dim}{\cal H}}{{\rm Vol}({\cal M}^4)}\int_{\cal M}d^4x\bar {\gamma}^{\rho\rho_1}\bar{\gamma}^{\lambda\lambda_1}\bar{F}_{\rho_1\lambda_1}\bar{F}_{\rho\lambda}.
\end{eqnarray}
We concentrate on the gauge field ${A}^{\lambda}$. Then
\begin{eqnarray}
-\frac{1}{g^2}Tr[Y^{\mu},Y^{\nu}][Y_{\mu},Y_{\nu}]
&=&-\frac{r^2}{g^2}\frac{{\rm dim}{\cal H}}{{\rm Vol}({\cal M}^4)}(\frac{\Delta_x^4}{4})^2\int_{\cal M}d^4x g^{\rho\rho_1} g^{\lambda\lambda_1}{F}_{\rho_1\lambda_1}{F}_{\rho\lambda}.
\end{eqnarray}
The covariant form of this action is (which can be shown using the Seiberg-Witten map \cite{Seiberg:1999vs,Steinacker:2007dq})
\begin{eqnarray}
-\frac{1}{g^2}Tr[Y^{\mu},Y^{\nu}][Y_{\mu},Y_{\nu}]
&=&-\frac{r^2}{g^2}\frac{{\rm dim}{\cal H}}{{\rm Vol}({\cal M}^4)}(\frac{\Delta_x^4}{4})^2\int_{\cal M}d^4x \sqrt{{\rm det}G}G^{\rho\rho_1} G^{\lambda\lambda_1}{F}_{\rho_1\lambda_1}{F}_{\rho\lambda}.
\end{eqnarray}
However, we should insist that the gauge field considered so far is a $U(1)$ gauge field which contributes really to the gravity sector in the matrix theory. We should therefore consider the addition of $SU(n)$ gauge field with the correct scaling $\bar{A}^{\mu}=\tilde{e}^{\mu\nu}A_{\nu}$, which mimics $[Y^{\mu},...]=e^{\mu\nu}[{\cal A}]\partial_{\nu}$ and thus leads to $i\Theta_{\mu\nu}=r\tilde{e}_{\mu\rho}\tilde{e}_{\nu\lambda}\hat{F}^{\rho\lambda}+...$, to be able to reproduce the above action. The metric $G$ in the above equation is precisely the string metric  (\ref{stringmetric}).

We compute also the gauge condition 
\begin{eqnarray}
f({\cal A})=-i[X^a,{\cal A}_a]&=&-i[X^5,{\cal A}_5]-i[X^{\mu},{\cal A}_{\mu}]\nonumber\\
&=&-i\theta^{5\mu}\partial_{\mu}\hat{\kappa}-i[X^{\mu},{\cal A}_{\mu}]\nonumber\\
&=&r^2RP^{\mu}\partial_{\mu}\hat{\kappa}-i[X^{\mu},{\cal A}_{\mu}]\nonumber\\
%&=&r^2RP^{\mu}\partial_{\mu}\hat{\kappa}-i[X^{\mu},{\cal A}_{\mu}]\nonumber\\
&=&\frac{r\Delta_x^2}{2}P^{\mu}\partial_{\mu}\hat{\kappa}-i\theta^{\mu\nu}\partial_{\nu}\xi_{\mu}+\theta_{\mu\nu}A^{\mu\nu}-\frac{i}{4}\Delta_x^4\partial_{\nu}\hat{A}^{\nu}.
\end{eqnarray}
By inspection we get then the detailed gauge conditions 
\begin{eqnarray}
-\frac{i}{4}\Delta_x^4\partial_{\nu}A^{\nu}=0.
\end{eqnarray}
\begin{eqnarray}
-\frac{i}{4}\Delta_x^2\partial_{\nu}(h^{\nu\rho}+a^{\nu\rho})+r\partial^{\rho}\hat{\kappa}=0.
\end{eqnarray}
\begin{eqnarray}
\theta_{\rho\sigma}\bigg(-R^2\partial_{\nu}A^{\nu\rho\sigma}-i\partial^{\rho}\xi^{\sigma}+a^{\rho\sigma}\bigg)=0.
\end{eqnarray}
The first equation gives the Lorentz condition. Since $r\hat{\kappa}/\Delta_x^2\sim 0$ we have the solution $\partial^{\mu}h_{\mu\nu}=0$, $a_{\mu\nu}=0$, $\xi=0$, $\partial_{\nu}A^{\nu\rho\sigma}=0$.

\subsection{Emergent gravity: Einstein equations}
By expanding the action (\ref{fund3}) up to second order in ${\cal A}^a$, and using Jacobi identity appropriately, we get \cite{CastroVillarreal:2004vh}
\begin{eqnarray}
S[D]=S[J]+\frac{2}{g^2}Tr\bigg(2{\cal A}^a(\Box+\frac{\mu^2}{2})J_a+{\cal A}_a(\Box+\frac{\mu^2}{2}){\cal A}^a-2[J_a,J_b][{\cal A}^a,{\cal A}^b]-f^2\bigg).
\end{eqnarray}
We have redefined the Laplacian on the space of matrix configurations by the formula
\begin{eqnarray}
\Box=[J_a,[J^a,...]].
\end{eqnarray}
Also $f$ is given by
\begin{eqnarray}
f=i[{\cal A}^a,J_a].
\end{eqnarray}
By assuming that $J_a$ solves the classical equations of motion and also adding a suitable Faddeev-Popov gauge fixing term in the Feynman gauge we get \cite{CastroVillarreal:2004vh,Blaschke:2011qu}
%\begin{eqnarray}
%S[D]=S[J]+\frac{2}{g^2}Tr\bigg(A_a(\Box+\frac{\mu^2}{2})A^a-2[J_a,J_b][A^a,A^b]\bigg).
%\end{eqnarray}
\begin{eqnarray}
S[D]=S[J]+\frac{2}{g^2}Tr{\cal A}^a\bigg((\Box+\frac{\mu^2}{2})g_{ab}+2i[{\cal M}_{ab},...]\bigg){\cal A}^b.
\end{eqnarray}
Recall that $i{\cal M}^{ab}=[J^a,J^b]$. We define the vector-matrix Laplacian acting on gauge configurations by
 \begin{eqnarray}
({\cal D}^2{\cal A})_a=\bigg(\Box+\frac{\mu^2}{2}-M_{rs}^{(A)}[{\cal M}^{rs},...]\bigg)_{ab}{\cal A}^b.
\end{eqnarray}
We have introduced in this last equation the vector representation of $SO(5)$ generators given by 
\begin{eqnarray}
(M_{rs}^{(A)})_{ab}=i(\delta_{rb}\delta_{sa}-\delta_{ra}\delta_{sb}).
\end{eqnarray}
The action of the quadratic fluctuations takes then the form
\begin{eqnarray}
S[D]=S[J]+\frac{2}{g^2}Tr{\cal A}^a({\cal D}^2{\cal A})_a.
\end{eqnarray}
We consider now the gravitational ansatz (by setting the redundant mode $\xi^{\mu}$ to zero and dropping the $U(1)$ gauge field for simplicity)
\begin{eqnarray}
{\cal A}_{\rm gr}^{\mu}%&=&\theta^{\mu\nu}(A_{\nu\sigma}P^{\sigma}+A_{\nu\rho\sigma}{\cal M}^{\rho\sigma})\nonumber\\
&=&\theta^{\mu\nu}(P^{\sigma}A_{\nu\sigma}+{\cal M}^{\rho\sigma}A_{\nu\rho\sigma})~,~{\cal A}_{\rm gr}^{5}={\kappa}.%-i\theta^{\mu\nu}\partial^{\sigma}A_{\nu\sigma}.
\end{eqnarray}
We have used that $[P_{\mu},g]=i\partial_{\mu}g$, $[\Theta^{\mu\nu},g]\sim x\partial_x g\sim 0$. We will also use $[f,g]=\theta^{\mu\nu}\partial_{\mu}f\partial_{\nu}g$. 
We compute
%\begin{eqnarray}
%\Box(-i\theta^{\mu\nu}\partial^{\sigma}A_{\nu\sigma})&=&-2i\theta^{\mu\nu}\partial^{\sigma}A_{\nu\sigma}+...
%\end{eqnarray}
\begin{eqnarray}
\Box(\theta_{\mu\nu}P_{\sigma}A^{\nu\sigma})&=&\Box(\theta_{\mu\nu}P_{\sigma})A^{\nu\sigma}+(\theta_{\mu\nu}P_{\sigma})\Box A^{\nu\sigma}+2[J^a,\theta_{\mu\nu}P_{\sigma}][J_a,A^{\nu\sigma}]\nonumber\\
&\sim & 4\theta_{\mu\nu}P_{\sigma}.A^{\nu\sigma}+(\theta_{\mu\nu}P_{\sigma})\Box A^{\nu\sigma}-2i\theta_{\mu\nu}{\cal M}_{\sigma\rho}\partial^{\sigma}A^{\nu\rho}.
\end{eqnarray}
\begin{eqnarray}
\Box(\theta_{\mu\nu}{\cal M}_{\rho\sigma}A^{\nu\rho\sigma})&=&\Box(\theta_{\mu\nu}{\cal M}_{\rho\sigma})A^{\nu\rho\sigma}+(\theta_{\mu\nu}{\cal M}_{\rho\sigma})\Box A^{\nu\rho\sigma}+2[J^a,\theta_{\mu\nu}{\cal M}_{\rho\sigma}][J_a,A^{\nu\rho\sigma}]\nonumber\\
&\sim & 4\theta_{\mu\nu}{\cal M}_{\rho\sigma}.A^{\nu\rho\sigma}+(\theta_{\mu\nu}{\cal M}_{\rho\sigma})\Box A^{\nu\rho\sigma}.
\end{eqnarray}
We have used $\Box \theta_{\mu\nu}=2\theta_{\mu\nu}$ and $\Box P_{\sigma}=2P_{\sigma}$. Thus
 \begin{eqnarray}
\big(\Box+\frac{\mu^2}{2}\big)_{\mu\nu}{\cal A}^{\nu}_{\rm gr}=\theta_{\mu\nu}P_{\sigma}(\Box+\frac{1}{2}\mu^2+4)A^{\nu\sigma}+\theta_{\mu\nu}{\cal M}_{\rho\sigma}(\Box+\frac{1}{2}\mu^2+4)A^{\nu\rho\sigma}.
\end{eqnarray}
Also
\begin{eqnarray}
\bigg(M_{rs}^{(A)}[{\cal M}^{rs},...]\bigg)_{\mu\nu}{\cal A}^{\nu}_{\rm grad}=2i\bigg([{\cal M}_{\nu\mu},\theta^{\nu\lambda}P^{\sigma}]A_{\lambda\sigma}+[{\cal M}_{\nu\mu},\theta^{\nu\lambda}{\cal M}^{\sigma\rho}]A_{\lambda\sigma\rho}\bigg).
\end{eqnarray}
We compute (using $\theta^{\mu\nu}P_{\nu}=0$ at the point $p$)
\begin{eqnarray}
[{\cal M}_{\nu \mu},\theta^{\nu\lambda}P^{\sigma}]=2i\theta_{\mu}^{\lambda}P^{\sigma}+i\theta^{\sigma\lambda}P_{\mu}.
\end{eqnarray}
\begin{eqnarray}
[{\cal M}_{\nu\mu},\theta^{\nu\lambda}{\cal M}^{\sigma\rho}]=\frac{1}{r^2}\bigg(2\theta_{\mu}^{\lambda}\theta^{\sigma\rho}+\theta^{\sigma\lambda}\theta_{\mu}^{\rho}-\theta^{\rho\lambda}\theta_{\mu}^{\sigma}+\bar{\gamma}^{\lambda\sigma}g_{\mu}^{\rho}-\bar{\gamma}^{\lambda\rho}g_{\mu}^{\sigma}\bigg).
\end{eqnarray}
Thus
\begin{eqnarray}
\bigg(M_{rs}^{(A)}[{\cal M}^{rs},...]\bigg)_{\mu\nu}{\cal A}^{\nu}_{\rm grad}&=&-4\theta_{\mu\nu}P_{\sigma}A^{\nu\sigma}-2\theta^{\sigma\lambda}P_{\mu}A_{\lambda\sigma}-4\theta_{\mu\lambda}{\cal M}_{\rho\sigma}A^{\lambda\rho\sigma}-2{\cal M}^{\sigma\rho}\theta^{\mu\nu}(A_{\rho\sigma\nu}-A_{\rho\nu\sigma})\nonumber\\
&+&\frac{4i}{r^2}\bar{\gamma}^{\mu\rho}g^{\nu\sigma}A_{\nu\sigma\rho}.
\end{eqnarray}
Hence
\begin{eqnarray}
{\cal D}^2_{\mu\nu}{\cal A}^{\nu}_{\rm gr}&=&\theta_{\mu\nu}P_{\sigma}(\Box+\frac{1}{2}\mu^2+8)A^{\nu\sigma}+\theta_{\mu\nu}{\cal M}_{\rho\sigma}(\Box+\frac{1}{2}\mu^2+8)A^{\nu\rho\sigma}+2\theta^{\sigma\lambda}P_{\mu}A_{\lambda\sigma}+2{\cal M}_{\sigma\rho}\theta_{\mu\nu}(A^{\rho\sigma\nu}-A^{\rho\nu\sigma})\nonumber\\
&-&\frac{4i}{r^2}\bar{\gamma}_{\mu\rho}g_{\nu\sigma}A^{\nu\sigma\rho}-2i\theta_{\mu\nu}{\cal M}_{\sigma\rho}\partial^{\sigma}A^{\nu\rho}.
\end{eqnarray}
We use also $\bar{\gamma}_{\mu\rho}g_{\nu\sigma}A^{\nu\sigma\rho}=-ir^2\theta_{\mu\nu}g^{\nu\sigma}A^{\alpha\beta\rho}g_{\alpha\beta}{\cal M}_{\sigma\rho}$. The term $2\theta^{\sigma\lambda}P_{\mu}A_{\lambda\sigma}$ is a higher mode of ${\cal A}^{\mu}_{\rm gr}$ and thus can be dropped. We get
\begin{eqnarray}
{\cal D}^2_{\mu\nu}{\cal A}^{\nu}_{\rm gr}&=&\theta_{\mu\nu}P_{\sigma}(\Box+\frac{1}{2}\mu^2+8)A^{\nu\sigma}+\theta_{\mu\nu}{\cal M}_{\rho\sigma}(\Box+\frac{1}{2}\mu^2+8)A^{\nu\rho\sigma}+2{\cal M}_{\sigma\rho}\theta_{\mu\nu}(A^{\rho\sigma\nu}-A^{\rho\nu\sigma})\nonumber\\
&-&4 \theta_{\mu\nu}g^{\nu\sigma}A^{\alpha\beta\rho}g_{\alpha\beta}{\cal M}_{\sigma\rho}-2i\theta_{\mu\nu}{\cal M}_{\sigma\rho}\partial^{\sigma}A^{\nu\rho}.
\end{eqnarray}
Similarly, we compute
\begin{eqnarray}
({\cal D}^2)_{55}{\cal A}_{\rm gr}^5&=&(\Box+\frac{1}{2}\mu^2+4)\kappa.
\end{eqnarray}
The equations of motion $({\cal D}^2{\cal A}_{\rm gr})_{a}=0$ give then the detailed equations of motion 
\begin{eqnarray}
(\Box+\frac{1}{2}\mu^2+4)\kappa=0.
\end{eqnarray}
The linear term will drop under averaging over the fiber. Also
\begin{eqnarray}
(\Box+\frac{1}{2}\mu^2+8)A^{\nu\sigma}=0.\label{first}
\end{eqnarray}
\begin{eqnarray}
(\Box+\frac{1}{2}\mu^2+8)A^{\nu\rho\sigma}+P_{\rm mix}A^{\nu\rho\sigma}+2 g^{\nu\sigma}A^{\alpha\beta\rho}g_{\alpha\beta}-2 g^{\nu\rho}A^{\alpha\beta\sigma}g_{\alpha\beta}=-i\partial^{\sigma}A^{\nu\rho}+i\partial^{\rho}A^{\nu\sigma},\label{second}
\end{eqnarray}
where
\begin{eqnarray}
P_{\rm mix}A^{\nu\rho\sigma}=-(A^{\rho\sigma\nu}-A^{\rho\nu\sigma})+(A^{\sigma\rho\nu}-A^{\sigma\nu\rho}).
\end{eqnarray}
Because of (\ref{first}) a solution of (\ref{second}) is given by a solution of the inhomogeneous equation  
\begin{eqnarray}
P_{\rm mix}A^{\nu\rho\sigma}+2 g^{\nu\sigma}A^{\alpha\beta\rho}g_{\alpha\beta}-2 g^{\nu\rho}A^{\alpha\beta\sigma}g_{\alpha\beta}=-i\partial^{\sigma}A^{\nu\rho}+i\partial^{\rho}A^{\nu\sigma},
\end{eqnarray}
We find explicitly \cite{Steinacker:2007dq}
\begin{eqnarray}
A_{\nu\sigma\rho}=\frac{1}{4}(\partial_{\rho}H_{\nu\sigma}-\partial_{\sigma}H_{\nu\rho})=A_{\nu;\sigma\rho}.
\end{eqnarray}
Thus $A_{\nu\sigma\rho}$ is the torsion-free spin connection of $H_{\mu\nu}$. The general solution will be given by this torsion-free spin connection plus a torsion wave solution of the free wave  equation, i.e. the equation (\ref{second}) without the $\partial_{\mu}A_{\alpha\beta}$ terms.

We need to compute the quadratic action for the fluctuation
\begin{eqnarray}
S[{\cal A}]=\frac{2}{g^2}Tr{\cal A}^a\bigg((\Box+\frac{\mu^2}{2})g_{ab}+2i[{\cal M}_{ab},...]\bigg){\cal A}^b.
\end{eqnarray}
By using the averages (\ref{metric1}), (\ref{momentum1}) and (\ref{a_momentum}) we compute the average over the fiber ${\bf S}^2$ of the following term:
\begin{eqnarray}
[{\cal A}^{\mu}_{\rm gr}.\theta_{\mu\nu}P_{\sigma}(\Box+\frac{\mu^2}{2}+8)A^{\nu\sigma}]_{{\bf S}^2}&=&[\theta^{\mu\lambda}P^{\rho}A_{\lambda\rho}.\theta_{\mu\nu}P_{\sigma}(\Box+\frac{\mu^2}{2}+8)A^{\nu\sigma}]_{{\bf S}^2}\nonumber\\
&=&\frac{\Delta_x^2}{4}A^{\lambda\sigma}(\Box+\frac{\mu^2}{2}+8)A_{\lambda\sigma}\nonumber\\
&=&\frac{\Delta_x^2}{16}h^{\lambda\sigma}(\Box+\frac{\mu^2}{2}+8)h_{\lambda\sigma}+\frac{\Delta_x^2}{16}a^{\lambda\sigma}(\Box+\frac{\mu^2}{2}+8)a_{\lambda\sigma}.\nonumber\\
\end{eqnarray}
\begin{eqnarray}
[{\cal A}^{5}_{\rm gr}({\cal D}^2{\cal A}_{\rm gr})_{5}]_{{\bf S}^2}&=&\kappa (\Box+\frac{\mu^2}{2}+4)\kappa.
\end{eqnarray}
\begin{eqnarray}
[{\cal A}^{\mu}_{\rm gr}.\theta_{\mu\nu}{\cal M}_{\rho\sigma}\bigg((\Box+\frac{\mu^2}{2}+8)A^{\nu\rho\sigma}+...\bigg)]_{{\bf S}^2}&=&[\theta^{\mu\lambda}{\cal M}^{\alpha\beta}A_{\lambda\alpha\beta}.\theta_{\mu\nu}{\cal M}_{\rho\sigma}\bigg((\Box+\frac{\mu^2}{2}+8)A^{\nu\rho\sigma}+...\bigg)]_{{\bf S}^2}\nonumber\\
&=&-\frac{\Delta_x^4}{4r^4}[A_{\nu\alpha\beta}.\theta^{\alpha\beta}{\theta}_{\rho\sigma}\bigg((\Box+\frac{\mu^2}{2}+8)A^{\nu\rho\sigma}+...\bigg)]_{{\bf S}^2}\nonumber\\
&=&-\frac{\Delta_x^8}{48r^4}\epsilon^{\alpha\beta\rho\sigma}g^{\nu\mu}A_{\nu\alpha\beta}.\bigg((\Box+\frac{\mu^2}{2}+8)A_{\mu\rho\sigma}+...\bigg)]_{{\bf S}^2}.\nonumber\\
\end{eqnarray}
We will neglect this last term. The quadratic action for the fluctuation fields $h_{\mu\nu}$, $a_{\mu\nu}$ and $\kappa$ is given by
\begin{eqnarray}
S[{\cal A}]=\frac{2}{g^2}\frac{{\rm dim}{\cal H}}{{\rm Vol}({\cal M}^4)}\int_{{\cal M}} d^4x\bigg[\frac{\Delta_x^2}{16}h^{\lambda\sigma}(\Box+\frac{\mu^2}{2}+8)h_{\lambda\sigma}+\frac{\Delta_x^2}{16}a^{\lambda\sigma}(\Box+\frac{\mu^2}{2}+8)a_{\lambda\sigma}+\kappa (\Box+\frac{\mu^2}{2}+4)\kappa \bigg].\nonumber\\
\end{eqnarray}
Coupling to matter will be of the canonical form (re-incorporating also the contribution of the radial component $\kappa$ by the replacement $h_{\mu\nu}\longrightarrow\tilde{h}_{\mu\nu}=h_{\mu\nu}+2r\kappa g_{\mu\nu}/R$) 
\begin{eqnarray}
S[\Phi]=\frac{{\rm dim}{\cal H}}{{\rm Vol}({\cal M}^4)}\frac{\Delta_x^8}{8g^2}\int_{{\cal M}} d^4x\tilde{h}^{\mu\nu}T_{\mu\nu}[\Phi].
\end{eqnarray}
The equations of motion are then given by 
\begin{eqnarray}
(\Box+\frac{\mu^2}{2}+8)h_{\lambda\sigma}=-\frac{\Delta_x^6}{2}T_{\mu\nu}.\label{EOMh}
\end{eqnarray}
\begin{eqnarray}
(\Box+\frac{\mu^2}{2}+8)a_{\lambda\sigma}=0.
\end{eqnarray}
\begin{eqnarray}
(\Box+\frac{\mu^2}{2}+4)\kappa=-\frac{r\Delta_x^8}{16R}T.
\end{eqnarray}
Thus we can see that we can solve with $a_{\mu\nu}=0$. Also the gauge condition $\partial_{\mu}h^{\mu\nu}=0$ is consistent with the equation of motion of $h^{\mu\nu}$ with a conserved energy-momentum tensor. 

We have the metric
\begin{eqnarray}
G^{\mu\nu}=g^{\mu\nu}+ir H^{\mu\nu}.
\end{eqnarray}
The corresponding Ricci tensor is given by (see \cite{Wald:1984rg} equation $(4.4.4)$) 
\begin{eqnarray}
R^{\mu\nu}[G]=R^{\mu\nu}[g]+ir\bigg(\frac{1}{2}\partial^{\mu}\partial^{\nu}H+\frac{1}{2}\partial_{\alpha}\partial^{\alpha}H^{\mu\nu}-\frac{1}{2}\partial^{\mu}\partial_{\rho}H^{\nu\rho}-\frac{1}{2}\partial^{\nu}\partial_{\rho}H^{\mu\rho}\bigg).
\end{eqnarray}
The Ricci tensor on the sphere ${\bf S}^4$ is given by
\begin{eqnarray}
R^{\mu\nu}[g]=\frac{3}{R^2}g_{\mu\nu}.
\end{eqnarray}
In the De Donder gauge (\ref{donder}) we find the Ricci tensor and the Ricci scalar
\begin{eqnarray}
R^{\mu\nu}[G]=\frac{3}{R^2}g^{\mu\nu}+\frac{ir}{2}\partial_{\alpha}\partial^{\alpha}H^{\mu\nu}.
\end{eqnarray}
\begin{eqnarray}
R[G]=G_{\mu\nu}R^{\mu\nu}[G]=\frac{12}{R^2}+\frac{ir}{2}\partial_{\alpha}\partial^{\alpha}H-\frac{3ir}{R^2}H.
\end{eqnarray}
The Einstein tensor is then
\begin{eqnarray}
{\cal G}^{\mu\nu}[G]&=&R^{\mu\nu}[G]-\frac{1}{2}G_{\mu\nu}R[G]\nonumber\\
&=&-\frac{3}{R^2}g^{\mu\nu}+\frac{1}{2}\partial_{\alpha}\partial^{\alpha}h^{\mu\nu}+\frac{3ir}{2R^2}(4H_{\mu\nu}+g_{\mu\nu}H).
\end{eqnarray}
By using $[X^{\mu},\phi]=i\theta^{\mu\nu}\partial_{\nu}\phi$ we can rewrite the equation of motion (\ref{EOMh}) as
 \begin{eqnarray}
(\frac{1}{4}\partial_{\alpha}\partial^{\alpha}-m_h^2)h_{\mu\nu}=r^2\Delta_x^2T_{\mu\nu}.
\end{eqnarray}
The mass $m_h^2$ is given by
 \begin{eqnarray}
m_h^2=\frac{r^2}{\Delta_x^4}(\frac{{\mu^2}}{2}+8)=\frac{2}{R^2}~,~\mu^2\longrightarrow 0.
\end{eqnarray}
By neglecting this mass we obtain the equation of motion  
 \begin{eqnarray}
\frac{1}{4}\partial_{\alpha}\partial^{\alpha}h_{\mu\nu}=r^2\Delta_x^2T_{\mu\nu}.
\end{eqnarray}
By neglecting also the other terms which are proportional to the background curvature $1/R^2$ we obtain the Einstein tensor 
\begin{eqnarray}
{\cal G}^{\mu\nu}=2r^2\Delta_x^2T_{\mu\nu}=8\pi G r^2T_{\mu\nu}.
\end{eqnarray}
The Newton constant is given by
\begin{eqnarray}
G=\frac{\Delta_x^2}{4\pi}=\frac{R^2}{\pi n}.
\end{eqnarray}
The Planck scale is thus suppressed compared to the cosmological scale by the quantization integer $n$.

\section{Emergent quantum gravity from multitrace matrix models}
In summary, there exists hidden inside any noncommutative $U(1)$ gauge theory a gravitational theory. This is the idea of emergent noncommutative/matrix gravity which is essentially the same idea (but certainly much simpler) as the one found in the AdS/CFT correspondence and gauge/gravity duality in general. 

The idea that matrices can capture curvature is also discussed in \cite{Hanada:2005vr}. This certainly works in the case of quantum gravity in $2$ dimensions which is known to emerge from random Riemannian surfaces \cite{DiFrancesco:1993cyw}.  By analogy, a proposal for quantum gravity in higher dimensions emerging from random spaces is recently put forward in \cite{Ydri:2017riq}. This works for spaces which are given by finite spectral triples (for example fuzzy spaces) and the underlying matrix models are necessarily multitrace.


\begin{thebibliography}{10}

%\cite{Green:1987sp}
\bibitem{Green:1987sp} 
  M.~B.~Green, J.~H.~Schwarz and E.~Witten,
  ``Superstring Theory. Vol. 1: Introduction.''c
  %Submitted to: Cambridge Monogr.Math.Phys..
  %153 citations counted in INSPIRE as of 22 Mar 2016

%\cite{Green:1987mn}
\bibitem{Green:1987mn} 
  M.~B.~Green, J.~H.~Schwarz and E.~Witten,
  ``Superstring Theory. Vol. 2: Loop Amplitudes, Anomalies And Phenomenology.''
  %Cambridge, Uk: Univ. Pr. ( 1987) 596 P. ( Cambridge Monographs On Mathematical Physics)
  %117 citations counted in INSPIRE as of 22 Mar 2016

%\cite{Becker:2007zj}
\bibitem{Becker:2007zj} 
  K.~Becker, M.~Becker and J.~H.~Schwarz,
  ``String theory and M-theory: A modern introduction.''c
  %%CITATION = INSPIRE-744404;%%
  %59 citations counted in INSPIRE as of 07 Apr 2016



%\cite{Polchinski:1998rq}
\bibitem{Polchinski:1998rq} 
  J.~Polchinski,
  ``String theory. Vol. 1: An introduction to the bosonic string.''
  %%CITATION = INSPIRE-487240;%%
  %172 citations counted in INSPIRE as of 18 May 2016

%\cite{Polchinski:1998rr}
\bibitem{Polchinski:1998rr} 
  J.~Polchinski,
  ``String theory. Vol. 2: Superstring theory and beyond.''c
  %%CITATION = INSPIRE-487241;%%
  %175 citations counted in INSPIRE as of 18 May 2016

%\cite{Johnson:2003gi}
\bibitem{Johnson:2003gi} 
  C.~V.~Johnson,
  ``D-branes,''
  Cambridge, USA: Univ. Pr. (2003) 548 p.
  %33 citations counted in INSPIRE as of 01 Mar 2017


%\cite{Schellekens}
\bibitem{Schellekens} 
  A.N.~Schellekens,
  ``Conformal field theory.''

%\cite{Zwiebach:2004tj}
\bibitem{Zwiebach:2004tj} 
  B.~Zwiebach,
  ``A first course in string theory,''
  Cambridge, UK: Univ. Pr. (2009) 673 p.c
  %61 citations counted in INSPIRE as of 03 May 2017

%\cite{Szabo:2004uy}
\bibitem{Szabo:2004uy} 
  R.~J.~Szabo,
  ``An Introduction to String Theory and D-Brane Dynamics,''
  Imperial College Press, 2004. ISBN 1-86094-427-2. 140p.c
  %3 citations counted in INSPIRE as of 27 Jul 2017






%\cite{Lundholm}
\bibitem{Lundholm} 
  D.~Lundholm,
  ``The Virasoro algebra and its representations in physics.''


%\cite{Goddard:1986ee}
\bibitem{Goddard:1986ee} 
  P.~Goddard, A.~Kent and D.~I.~Olive,
  ``Unitary Representations of the Virasoro and Supervirasoro Algebras,''
  Commun.\ Math.\ Phys.\  {\bf 103}, 105 (1986).
%  doi:10.1007/BF01464283
  %%CITATION = doi:10.1007/BF01464283;%%
  %719 citations counted in INSPIRE as of 08 Sep 2016

%\cite{Friedan:1986kd}
\bibitem{Friedan:1986kd} 
  D.~Friedan, S.~H.~Shenker and Z.~a.~Qiu,
  ``Details of the Nonunitarity Proof for Highest Weight Representations of the Virasoro Algebra,''
  Commun.\ Math.\ Phys.\  {\bf 107}, 535 (1986).
%  doi:10.1007/BF01205483
  %%CITATION = doi:10.1007/BF01205483;%%
  %93 citations counted in INSPIRE as of 08 Sep 2016

%\cite{Eguchi:1980jx}
\bibitem{Eguchi:1980jx} 
  T.~Eguchi, P.~B.~Gilkey and A.~J.~Hanson,
  ``Gravitation, Gauge Theories and Differential Geometry,''
  Phys.\ Rept.\  {\bf 66}, 213 (1980).
  doi:10.1016/0370-1573(80)90130-1
  %%CITATION = doi:10.1016/0370-1573(80)90130-1;%%
  %1190 citations counted in INSPIRE as of 31 Jul 2017



%\cite{Hoppe:1982}
\bibitem{Hoppe:1982} 
  J.~Hoppe,
  ``Quantum theory of a massless relativistic surface and a two-dimensional bound state problem,''
  PhD Thesis, Massachusetts Institute of Technology (MIT), Cambridge, MA, USA. http://dspace.mit.edu/handle/1721.1/15717.c
  
%\cite{deWit:1988wri}
\bibitem{deWit:1988wri} 
  B.~de Wit, J.~Hoppe and H.~Nicolai,
  ``On the Quantum Mechanics of Supermembranes,''
  Nucl.\ Phys.\ B {\bf 305}, 545 (1988).c
%  doi:10.1016/0550-3213(88)90116-2
  %%CITATION = doi:10.1016/0550-3213(88)90116-2;%%
  %719 citations counted in INSPIRE as of 01 Mar 2017

%\cite{Bergshoeff:1987cm}
\bibitem{Bergshoeff:1987cm} 
  E.~Bergshoeff, E.~Sezgin and P.~K.~Townsend,
  ``Supermembranes and Eleven-Dimensional Supergravity,''
  Phys.\ Lett.\ B {\bf 189}, 75 (1987).c
%  doi:10.1016/0370-2693(87)91272-X
  %%CITATION = doi:10.1016/0370-2693(87)91272-X;%%
  %750 citations counted in INSPIRE as of 01 Mar 2017



%\cite{Taylor:1999qk}
\bibitem{Taylor:1999qk} 
  W.~Taylor,
  ``The M(atrix) model of M theory,''
  NATO Sci.\ Ser.\ C {\bf 556}, 91 (2000).
%  doi:10.1007/978-94-011-4303-5_3
  [hep-th/0002016].c
  %%CITATION = doi:10.1007/978-94-011-4303-5_3;%%
  %82 citations counted in INSPIRE as of 01 Mar 2017


%\cite{OConnor:2016gbq}
\bibitem{OConnor:2016gbq} 
  D.~O'Connor and V.~G.~Filev,
  ``Membrane Matrix models and non-perturbative checks of gauge/gravity duality,''
  PoS CORFU {\bf 2015}, 111 (2016)
  [arXiv:1605.01611 [hep-th]].
  %%CITATION = ARXIV:1605.01611;%%
  %1 citations counted in INSPIRE as of 01 Mar 2017

%\cite{Hanada:2016jok}
\bibitem{Hanada:2016jok} 
  M.~Hanada,
  ``What lattice theorists can do for superstring/M-theory,''
  Int.\ J.\ Mod.\ Phys.\ A {\bf 31}, no. 22, 1643006 (2016),
%  doi:10.1142/S0217751X16430065
  [arXiv:1604.05421 [hep-lat]].
  %%CITATION = doi:10.1142/S0217751X16430065;%%
  %10 citations counted in INSPIRE as of 01 Mar 2017



%\cite{Volkholz:1900zz}
\bibitem{Volkholz:1900zz} 
  J.~Volkholz and W.~Bietenholz,
  ``Simulations of a supersymmetry inspired model on a fuzzy sphere,''
  PoS LAT {\bf 2007}, 283 (2007)
  [arXiv:0808.2387 [hep-th]].
  %%CITATION = ARXIV:0808.2387;%%
  %3 citations counted in INSPIRE as of 02 Mar 2017

%\cite{Kawahara:2007fn}
\bibitem{Kawahara:2007fn} 
  N.~Kawahara, J.~Nishimura and S.~Takeuchi,
  ``Phase structure of matrix quantum mechanics at finite temperature,''
  JHEP {\bf 0710}, 097 (2007),
%  doi:10.1088/1126-6708/2007/10/097
  [arXiv:0706.3517 [hep-th]].
  %%CITATION = doi:10.1088/1126-6708/2007/10/097;%%
  %37 citations counted in INSPIRE as of 02 Mar 2017



%\cite{Ishibashi:1996xs}
\bibitem{Ishibashi:1996xs} 
  N.~Ishibashi, H.~Kawai, Y.~Kitazawa and A.~Tsuchiya,
  ``A Large N reduced model as superstring,''
  Nucl.\ Phys.\ B {\bf 498}, 467 (1997),
%  doi:10.1016/S0550-3213(97)00290-3
  [hep-th/9612115].c
  %%CITATION = doi:10.1016/S0550-3213(97)00290-3;%%
  %858 citations counted in INSPIRE as of 11 Jun 2016

%\cite{Banks:1996vh}
\bibitem{Banks:1996vh} 
  T.~Banks, W.~Fischler, S.~H.~Shenker and L.~Susskind,
  ``M theory as a matrix model: A Conjecture,''
  Phys.\ Rev.\ D {\bf 55}, 5112 (1997),
%  doi:10.1103/PhysRevD.55.5112
  [hep-th/9610043].c
  %%CITATION = doi:10.1103/PhysRevD.55.5112;%%
  %2514 citations counted in INSPIRE as of 15 Jun 2016


%\cite{Connes:1997cr}
\bibitem{Connes:1997cr} 
  A.~Connes, M.~R.~Douglas and A.~S.~Schwarz,
  ``Noncommutative geometry and matrix theory: Compactification on tori,''
  JHEP {\bf 9802}, 003 (1998),
%  doi:10.1088/1126-6708/1998/02/003
  [hep-th/9711162].c
  %%CITATION = doi:10.1088/1126-6708/1998/02/003;%%
  %1593 citations counted in INSPIRE as of 14 Jun 2016

%\cite{Witten:1995im}
\bibitem{Witten:1995im} 
  E.~Witten,
  ``Bound states of strings and p-branes,''
  Nucl.\ Phys.\ B {\bf 460}, 335 (1996),
%  doi:10.1016/0550-3213(95)00610-9
  [hep-th/9510135].c
  %%CITATION = doi:10.1016/0550-3213(95)00610-9;%%
  %1371 citations counted in INSPIRE as of 15 Jun 2016



%\cite{Konechny:2000dp}
\bibitem{Konechny:2000dp} 
  A.~Konechny and A.~S.~Schwarz,
  ``Introduction to M(atrix) theory and noncommutative geometry,''
  Phys.\ Rept.\  {\bf 360}, 353 (2002),
%  doi:10.1016/S0370-1573(01)00096-5
  [hep-th/0012145].c
  %%CITATION = doi:10.1016/S0370-1573(01)00096-5;%%
  %165 citations counted in INSPIRE as of 14 Jun 2016

%\cite{Connes:1996gi}
\bibitem{Connes:1996gi} 
  A.~Connes,
  ``Gravity coupled with matter and foundation of noncommutative geometry,''
  Commun.\ Math.\ Phys.\  {\bf 182}, 155 (1996),
%  doi:10.1007/BF02506388
  [hep-th/9603053].c
  %%CITATION = doi:10.1007/BF02506388;%%
  %377 citations counted in INSPIRE as of 15 Jun 2016




%\cite{Sochichiu:2000ud}
\bibitem{Sochichiu:2000ud} 
  C.~Sochichiu,
  ``M[any] vacua of IIB,''
  JHEP {\bf 0005}, 026 (2000),
%  doi:10.1088/1126-6708/2000/05/026
  [hep-th/0004062].c
  %%CITATION = doi:10.1088/1126-6708/2000/05/026;%%
  %21 citations counted in INSPIRE as of 15 Jun 2016

%\cite{Zarembo:1998uk}
\bibitem{Zarembo:1998uk} 
  K.~L.~Zarembo and Y.~M.~Makeenko,
  ``An introduction to matrix superstring models,''
  Phys.\ Usp.\  {\bf 41}, 1 (1998)
  [Usp.\ Fiz.\ Nauk {\bf 168}, 3 (1998)].c
%  doi:10.1070/PU1998v041n01ABEH000327
  %%CITATION = doi:10.1070/PU1998v041n01ABEH000327;%%
  %3 citations counted in INSPIRE as of 15 Jun 2016

%\cite{Brink:1976bc}
\bibitem{Brink:1976bc} 
  L.~Brink, J.~H.~Schwarz and J.~Scherk,
  ``Supersymmetric Yang-Mills Theories,''
  Nucl.\ Phys.\ B {\bf 121}, 77 (1977).
%  doi:10.1016/0550-3213(77)90328-5
  %%CITATION = doi:10.1016/0550-3213(77)90328-5;%%
  %726 citations counted in INSPIRE as of 15 Jun 2016

%\cite{Kim:2003rza}
\bibitem{Kim:2003rza} 
  N.~Kim, T.~Klose and J.~Plefka,
  ``Plane wave matrix theory from N=4 superYang-Mills on R x S**3,''
  Nucl.\ Phys.\ B {\bf 671}, 359 (2003),
%  doi:10.1016/j.nuclphysb.2003.08.019
  [hep-th/0306054].
  %%CITATION = doi:10.1016/j.nuclphysb.2003.08.019;%%
  %83 citations counted in INSPIRE as of 15 Jun 2016

%\cite{Kim:2002if}
\bibitem{Kim:2002if} 
  N.~Kim and J.~Plefka,
  ``On the spectrum of PP wave matrix theory,''
  Nucl.\ Phys.\ B {\bf 643}, 31 (2002),
%  doi:10.1016/S0550-3213(02)00738-1
  [hep-th/0207034].
  %%CITATION = doi:10.1016/S0550-3213(02)00738-1;%%
  %70 citations counted in INSPIRE as of 15 Jun 2016





%\cite{Aharony:2004ig}
\bibitem{Aharony:2004ig} 
  O.~Aharony, J.~Marsano, S.~Minwalla and T.~Wiseman,
  ``Black hole-black string phase transitions in thermal 1+1 dimensional supersymmetric Yang-Mills theory on a circle,''
  Class.\ Quant.\ Grav.\  {\bf 21}, 5169 (2004),
%  doi:10.1088/0264-9381/21/22/010
  [hep-th/0406210].
  %%CITATION = doi:10.1088/0264-9381/21/22/010;%%
  %123 citations counted in INSPIRE as of 06 Mar 2017

%\cite{Kol:2004ww}
\bibitem{Kol:2004ww} 
  B.~Kol,
  ``The Phase transition between caged black holes and black strings: A Review,''
  Phys.\ Rept.\  {\bf 422}, 119 (2006),
%  doi:10.1016/j.physrep.2005.10.001
  [hep-th/0411240].
  %%CITATION = doi:10.1016/j.physrep.2005.10.001;%%
  %153 citations counted in INSPIRE as of 06 Mar 2017

%\cite{Itzhaki:1998dd}
\bibitem{Itzhaki:1998dd} 
  N.~Itzhaki, J.~M.~Maldacena, J.~Sonnenschein and S.~Yankielowicz,
  ``Supergravity and the large N limit of theories with sixteen supercharges,''
  Phys.\ Rev.\ D {\bf 58}, 046004 (1998),
%  doi:10.1103/PhysRevD.58.046004
  [hep-th/9802042].
  %%CITATION = doi:10.1103/PhysRevD.58.046004;%%
  %850 citations counted in INSPIRE as of 06 Mar 2017


%\cite{Hyakutake:2013vwa}
\bibitem{Hyakutake:2013vwa} 
  Y.~Hyakutake,
  ``Quantum near-horizon geometry of a black 0-brane,''
  PTEP {\bf 2014}, 033B04 (2014),
%  doi:10.1093/ptep/ptu028
  [arXiv:1311.7526 [hep-th]].
  %%CITATION = doi:10.1093/ptep/ptu028;%%
  %15 citations counted in INSPIRE as of 07 Mar 2017


%\cite{Hanada:2016zxj}
\bibitem{Hanada:2016zxj} 
  M.~Hanada, Y.~Hyakutake, G.~Ishiki and J.~Nishimura,
  ``Numerical tests of the gauge/gravity duality conjecture for D0-branes at finite temperature and finite N,''
  Phys.\ Rev.\ D {\bf 94}, no. 8, 086010 (2016),
%  doi:10.1103/PhysRevD.94.086010
  [arXiv:1603.00538 [hep-th]].
  %%CITATION = doi:10.1103/PhysRevD.94.086010;%%
  %7 citations counted in INSPIRE as of 07 Mar 2017


%\cite{Duff:1996}
\bibitem{Duff:1996} 
  M.J.~Duff,
  ``Supermembranes.''
  [hep-th/9611203].c

%\cite{WLN}
\bibitem{WLN} 
  B.~de Wit, M.~Luscher, H.~Nicolai,
  Nucl.Phys. {\bf B}320(1989)135.c
  

%\cite{Witten:1995im}
\bibitem{Witten:1995im} 
  E.~Witten,
  ``Bound states of strings and p-branes,''
  Nucl.\ Phys.\ B {\bf 460}, 335 (1996),
%  doi:10.1016/0550-3213(95)00610-9
  [hep-th/9510135].c
  %%CITATION = doi:10.1016/0550-3213(95)00610-9;%%
  %1403 citations counted in INSPIRE as of 02 May 2017

%\cite{Dai:1989ua}
\bibitem{Dai:1989ua} 
  J.~Dai, R.~G.~Leigh and J.~Polchinski,
  ``New Connections Between String Theories,''
  Mod.\ Phys.\ Lett.\ A {\bf 4}, 2073 (1989).c
%  doi:10.1142/S0217732389002331.c
  %%CITATION = doi:10.1142/S0217732389002331;%%
  %802 citations counted in INSPIRE as of 02 May 2017

%\cite{Polchinski:1995mt}
\bibitem{Polchinski:1995mt} 
  J.~Polchinski,
  ``Dirichlet Branes and Ramond-Ramond charges,''
  Phys.\ Rev.\ Lett.\  {\bf 75}, 4724 (1995),
%  doi:10.1103/PhysRevLett.75.4724
  [hep-th/9510017].c
  %%CITATION = doi:10.1103/PhysRevLett.75.4724;%%
  %2418 citations counted in INSPIRE as of 02 May 2017


%\cite{Schwarz:2008kd}
\bibitem{Schwarz:2008kd} 
  J.~H.~Schwarz,
  ``Status of Superstring and M-Theory,''
  Int.\ J.\ Mod.\ Phys.\ A {\bf 25}, 4703 (2010), 
  [Subnucl.\ Ser.\  {\bf 46}, 335 (2011)], 
  [arXiv:0812.1372 [hep-th]].c
  %%CITATION = doi:10.1142/S0217751X1005072X, 10.1142/9789814340212_0012;%%
  %8 citations counted in INSPIRE as of 02 May 2017


%\cite{Azeyanagi:2009zf}
\bibitem{Azeyanagi:2009zf} 
  T.~Azeyanagi, M.~Hanada, T.~Hirata and H.~Shimada,
  ``On the shape of a D-brane bound state and its topology change,''
  JHEP {\bf 0903}, 121 (2009),
%  doi:10.1088/1126-6708/2009/03/121
  [arXiv:0901.4073 [hep-th]].c
  %%CITATION = doi:10.1088/1126-6708/2009/03/121;%%
  %18 citations counted in INSPIRE as of 02 May 2017

%\cite{Leigh:1989jq}
\bibitem{Leigh:1989jq} 
  R.~G.~Leigh,
  ``Dirac-Born-Infeld Action from Dirichlet Sigma Model,''
  Mod.\ Phys.\ Lett.\ A {\bf 4}, 2767 (1989).c
%  doi:10.1142/S0217732389003099
  %%CITATION = doi:10.1142/S0217732389003099;%%
  %781 citations counted in INSPIRE as of 03 May 2017


%\cite{Gibbons:1987ps}
\bibitem{Gibbons:1987ps} 
  G.~W.~Gibbons and K.~i.~Maeda,
  ``Black Holes and Membranes in Higher Dimensional Theories with Dilaton Fields,''
  Nucl.\ Phys.\ B {\bf 298}, 741 (1988).c
%  doi:10.1016/0550-3213(88)90006-5
  %%CITATION = doi:10.1016/0550-3213(88)90006-5;%%
  %1039 citations counted in INSPIRE as of 07 May 2017

%\cite{Horowitz:1991cd}
\bibitem{Horowitz:1991cd} 
  G.~T.~Horowitz and A.~Strominger,
  ``Black strings and P-branes,''
  Nucl.\ Phys.\ B {\bf 360}, 197 (1991).c
%  doi:10.1016/0550-3213(91)90440-9
  %%CITATION = doi:10.1016/0550-3213(91)90440-9;%%
  %1002 citations counted in INSPIRE as of 07 May 2017

%\cite{Itzhaki:1998dd}
\bibitem{Itzhaki:1998dd} 
  N.~Itzhaki, J.~M.~Maldacena, J.~Sonnenschein and S.~Yankielowicz,
  ``Supergravity and the large N limit of theories with sixteen supercharges,''
  Phys.\ Rev.\ D {\bf 58}, 046004 (1998).
%  doi:10.1103/PhysRevD.58.046004
  [hep-th/9802042].c
  %%CITATION = doi:10.1103/PhysRevD.58.046004;%%
  %856 citations counted in INSPIRE as of 07 May 2017

%\cite{Horowitz:2006ct}
\bibitem{Horowitz:2006ct} 
  G.~T.~Horowitz and J.~Polchinski,
  ``Gauge/gravity duality,''
  In *Oriti, D. (ed.): Approaches to quantum gravity* 169-186
  [gr-qc/0602037].c
  %%CITATION = GR-QC/0602037;%%
  %106 citations counted in INSPIRE as of 07 May 2017


%\cite{tHooft:1993dmi}
\bibitem{tHooft:1993dmi} 
  G.~'t Hooft,
  ``Dimensional reduction in quantum gravity,''
  Salamfest 1993:0284-296
  [gr-qc/9310026]. c
  %%CITATION = GR-QC/9310026;%%
  %1877 citations counted in INSPIRE as of 07 May 2017

%\cite{Susskind:1994vu}
\bibitem{Susskind:1994vu} 
  L.~Susskind,
  ``The World as a hologram,''
  J.\ Math.\ Phys.\  {\bf 36}, 6377 (1995),
%  doi:10.1063/1.531249
  [hep-th/9409089].c
  %%CITATION = doi:10.1063/1.531249;%%
  %2284 citations counted in INSPIRE as of 07 May 2017

%\cite{Weinberg:1980kq}
\bibitem{Weinberg:1980kq} 
  S.~Weinberg and E.~Witten,
  ``Limits on Massless Particles,''
  Phys.\ Lett.\  {\bf 96B}, 59 (1980).c
%  doi:10.1016/0370-2693(80)90212-9
  %%CITATION = doi:10.1016/0370-2693(80)90212-9;%%
  %423 citations counted in INSPIRE as of 07 May 2017

%\cite{tHooft:1973alw}
\bibitem{tHooft:1973alw} 
  G.~'t Hooft,
  ``A Planar Diagram Theory for Strong Interactions,''
  Nucl.\ Phys.\ B {\bf 72}, 461 (1974).c
%  doi:10.1016/0550-3213(74)90154-0
  %%CITATION = doi:10.1016/0550-3213(74)90154-0;%%
  %4279 citations counted in INSPIRE as of 07 May 2017

%\cite{Nastase:2007kj}
\bibitem{Nastase:2007kj} 
  H.~Nastase,
  ``Introduction to AdS-CFT,''
  arXiv:0712.0689 [hep-th].c
  %%CITATION = ARXIV:0712.0689;%%
  %82 citations counted in INSPIRE as of 08 May 2017

%\cite{Maldacena:1997re}
\bibitem{Maldacena:1997re} 
  J.~M.~Maldacena,
  ``The Large N limit of superconformal field theories and supergravity,''
  Int.\ J.\ Theor.\ Phys.\  {\bf 38}, 1113 (1999),
  [Adv.\ Theor.\ Math.\ Phys.\  {\bf 2}, 231 (1998)],
%  doi:10.1023/A:1026654312961
  [hep-th/9711200].c
  %%CITATION = doi:10.1023/A:1026654312961;%%
  %12727 citations counted in INSPIRE as of 08 May 2017

%\cite{Wilson:1974sk}
\bibitem{Wilson:1974sk} 
  K.~G.~Wilson,
  ``Confinement of Quarks,''
  Phys.\ Rev.\ D {\bf 10}, 2445 (1974).c
%  doi:10.1103/PhysRevD.10.2445.
  %%CITATION = doi:10.1103/PhysRevD.10.2445;%%
  %4458 citations counted in INSPIRE as of 08 May 2017

%\cite{Nahm:1977tg}
\bibitem{Nahm:1977tg} 
  W.~Nahm,
  ``Supersymmetries and their Representations,''
  Nucl.\ Phys.\ B {\bf 135}, 149 (1978).c
%  doi:10.1016/0550-3213(78)90218-3
  %%CITATION = doi:10.1016/0550-3213(78)90218-3;%%
  %604 citations counted in INSPIRE as of 11 May 2017

%\cite{Cremmer:1978km}
\bibitem{Cremmer:1978km} 
  E.~Cremmer, B.~Julia and J.~Scherk,
  ``Supergravity Theory in Eleven-Dimensions,''
  Phys.\ Lett.\  {\bf 76B}, 409 (1978).c
%  doi:10.1016/0370-2693(78)90894-8
  %%CITATION = doi:10.1016/0370-2693(78)90894-8;%%
  %1507 citations counted in INSPIRE as of 11 May 2017

%\cite{Miemiec:2005ry}
\bibitem{Miemiec:2005ry} 
  A.~Miemiec and I.~Schnakenburg,
  ``Basics of M-theory,''
  Fortsch.\ Phys.\  {\bf 54}, 5 (2006),
%  doi:10.1002/prop.200510256
  [hep-th/0509137].c
  %%CITATION = doi:10.1002/prop.200510256;%%
  %16 citations counted in INSPIRE as of 11 May 2017


%\cite{Witten:1995ex}
\bibitem{Witten:1995ex} 
  E.~Witten,
  ``String theory dynamics in various dimensions,''
  Nucl.\ Phys.\ B {\bf 443}, 85 (1995).
%  doi:10.1016/0550-3213(95)00158-O
  [hep-th/9503124].c
  %%CITATION = doi:10.1016/0550-3213(95)00158-O;%%
  %2114 citations counted in INSPIRE as of 11 May 2017

%\cite{deAlwis:1996ez}
\bibitem{deAlwis:1996ez} 
  S.~P.~de Alwis,
  ``A Note on brane tension and M theory,''
  Phys.\ Lett.\ B {\bf 388}, 291 (1996),
%  doi:10.1016/S0370-2693(96)01172-0
  [hep-th/9607011].c
  %%CITATION = doi:10.1016/S0370-2693(96)01172-0;%%
  %73 citations counted in INSPIRE as of 11 May 2017

%\cite{Green:1996bh}
\bibitem{Green:1996bh} 
  M.~B.~Green, C.~M.~Hull and P.~K.~Townsend,
  ``D-brane Wess-Zumino actions, t duality and the cosmological constant,''
  Phys.\ Lett.\ B {\bf 382}, 65 (1996),
%  doi:10.1016/0370-2693(96)00643-0
  [hep-th/9604119].c
  %%CITATION = doi:10.1016/0370-2693(96)00643-0;%%
  %179 citations counted in INSPIRE as of 11 May 2017


%\cite{Schwarz:1995dk}
\bibitem{Schwarz:1995dk} 
  J.~H.~Schwarz,
  ``An SL(2,Z) multiplet of type IIB superstrings,''
  Phys.\ Lett.\ B {\bf 360}, 13 (1995)
  Erratum: [Phys.\ Lett.\ B {\bf 364}, 252 (1995)],
%  doi:10.1016/0370-2693(95)01405-5, 10.1016/0370-2693(95)01138-G
  [hep-th/9508143].c
  %%CITATION = doi:10.1016/0370-2693(95)01405-5, 10.1016/0370-2693(95)01138-G;%%
  %582 citations counted in INSPIRE as of 11 May 2017

%\cite{Hyakutake:2014maa}
\bibitem{Hyakutake:2014maa} 
  Y.~Hyakutake,
  ``Quantum M-wave and Black 0-brane,''
  JHEP {\bf 1409}, 075 (2014),
%  doi:10.1007/JHEP09(2014)075
  [arXiv:1407.6023 [hep-th]].c
  %%CITATION = doi:10.1007/JHEP09(2014)075;%%
  %5 citations counted in INSPIRE as of 14 May 2017



%\cite{Horowitz:1997fr}
\bibitem{Horowitz:1997fr} 
  G.~T.~Horowitz and E.~J.~Martinec,
  ``Comments on black holes in matrix theory,''
  Phys.\ Rev.\ D {\bf 57}, 4935 (1998),
%  doi:10.1103/PhysRevD.57.4935
  [hep-th/9710217].c
  %%CITATION = doi:10.1103/PhysRevD.57.4935;%%
  %91 citations counted in INSPIRE as of 15 May 2017



%\cite{Li:1998ci}
\bibitem{Li:1998ci} 
  M.~Li and E.~J.~Martinec,
  ``Probing matrix black holes,''
  hep-th/9801070.c
  %%CITATION = HEP-TH/9801070;%%
  %35 citations counted in INSPIRE as of 15 May 2017


%\cite{Englert:1998vr}
\bibitem{Englert:1998vr} 
  F.~Englert and E.~Rabinovici,
  ``Statistical entropy of Schwarzschild black holes,''
  Phys.\ Lett.\ B {\bf 426}, 269 (1998),
%  doi:10.1016/S0370-2693(98)00293-7
  [hep-th/9801048].c
  %%CITATION = doi:10.1016/S0370-2693(98)00293-7;%%
  %21 citations counted in INSPIRE as of 15 May 2017




%\cite{Banks:1997cm}
\bibitem{Banks:1997cm} 
  T.~Banks, W.~Fischler and I.~R.~Klebanov,
  ``Evaporation of Schwarzschild black holes in matrix theory,''
  Phys.\ Lett.\ B {\bf 423}, 54 (1998),
%  doi:10.1016/S0370-2693(98)00118-X
  [hep-th/9712236].c
  %%CITATION = doi:10.1016/S0370-2693(98)00118-X;%%
  %43 citations counted in INSPIRE as of 15 May 2017

%\cite{Liu:1997gk}
\bibitem{Liu:1997gk} 
  H.~Liu and A.~A.~Tseytlin,
  ``Statistical mechanics of D0-branes and black hole thermodynamics,''
  JHEP {\bf 9801}, 010 (1998),
%  doi:10.1088/1126-6708/1998/01/010
  [hep-th/9712063].c
  %%CITATION = doi:10.1088/1126-6708/1998/01/010;%%
  %57 citations counted in INSPIRE as of 15 May 2017

%\cite{Das:1997tk}
\bibitem{Das:1997tk} 
  S.~R.~Das, S.~D.~Mathur, S.~Kalyana Rama and P.~Ramadevi,
  ``Boosts, Schwarzschild black holes and absorption cross-sections in M theory,''
  Nucl.\ Phys.\ B {\bf 527}, 187 (1998),
%  doi:10.1016/S0550-3213(98)00229-6
  [hep-th/9711003].c
  %%CITATION = doi:10.1016/S0550-3213(98)00229-6;%%
  %41 citations counted in INSPIRE as of 15 May 2017


%\cite{Li:1997iz}
\bibitem{Li:1997iz} 
  M.~Li,
  ``Matrix Schwarzschild black holes in large N limit,''
  JHEP {\bf 9801}, 009 (1998),
%  doi:10.1088/1126-6708/1998/01/009
  [hep-th/9710226].c
  %%CITATION = doi:10.1088/1126-6708/1998/01/009;%%
  %62 citations counted in INSPIRE as of 15 May 2017


%\cite{Banks:1997hz}
\bibitem{Banks:1997hz} 
  T.~Banks, W.~Fischler, I.~R.~Klebanov and L.~Susskind,
  ``Schwarzschild black holes from matrix theory,''
  Phys.\ Rev.\ Lett.\  {\bf 80}, 226 (1998),
%  doi:10.1103/PhysRevLett.80.226
  [hep-th/9709091].c
  %%CITATION = doi:10.1103/PhysRevLett.80.226;%%
  %145 citations counted in INSPIRE as of 15 May 2017



%\cite{Banks:1997tn}
\bibitem{Banks:1997tn} 
  T.~Banks, W.~Fischler, I.~R.~Klebanov and L.~Susskind,
  ``Schwarzschild black holes in matrix theory. 2.,''
  JHEP {\bf 9801}, 008 (1998),
%  doi:10.1088/1126-6708/1998/01/008
  [hep-th/9711005].c
  %%CITATION = doi:10.1088/1126-6708/1998/01/008;%%
  %126 citations counted in INSPIRE as of 15 May 2017

%\cite{Klebanov:1997kv}
\bibitem{Klebanov:1997kv} 
  I.~R.~Klebanov and L.~Susskind,
  ``Schwarzschild black holes in various dimensions from matrix theory,''
  Phys.\ Lett.\ B {\bf 416}, 62 (1998),
%  doi:10.1016/S0370-2693(97)01318-X
  [hep-th/9709108].c
  %%CITATION = doi:10.1016/S0370-2693(97)01318-X;%%
  %92 citations counted in INSPIRE as of 15 May 2017

%\cite{Gross:1986iv}
\bibitem{Gross:1986iv} 
  D.~J.~Gross and E.~Witten,
  ``Superstring Modifications of Einstein's Equations,''
  Nucl.\ Phys.\ B {\bf 277}, 1 (1986).c
%  doi:10.1016/0550-3213(86)90429-3
  %%CITATION = doi:10.1016/0550-3213(86)90429-3;%%
  %692 citations counted in INSPIRE as of 16 May 2017


%\cite{Hyakutake:2006aq}
\bibitem{Hyakutake:2006aq} 
  Y.~Hyakutake and S.~Ogushi,
  ``Higher derivative corrections to eleven dimensional supergravity via local supersymmetry,''
  JHEP {\bf 0602}, 068 (2006),
%  doi:10.1088/1126-6708/2006/02/068
  [hep-th/0601092].c
  %%CITATION = doi:10.1088/1126-6708/2006/02/068;%%
  %26 citations counted in INSPIRE as of 16 May 2017

%\cite{Hyakutake:2007sm}
\bibitem{Hyakutake:2007sm} 
  Y.~Hyakutake,
  ``Toward the Determination of R**3 F**2 Terms in M-theory,''
  Prog.\ Theor.\ Phys.\  {\bf 118}, 109 (2007),
%  doi:10.1143/PTP.118.109
  [hep-th/0703154 [HEP-TH]].c
  %%CITATION = doi:10.1143/PTP.118.109;%%
  %17 citations counted in INSPIRE as of 16 May 2017




%\cite{Wald:1993nt}
\bibitem{Wald:1993nt} 
  R.~M.~Wald,
  ``Black hole entropy is the Noether charge,''
  Phys.\ Rev.\ D {\bf 48}, no. 8, R3427 (1993),
%  doi:10.1103/PhysRevD.48.R3427
  [gr-qc/9307038].c
  %%CITATION = doi:10.1103/PhysRevD.48.R3427;%%
  %1262 citations counted in INSPIRE as of 19 May 2017

%\cite{Iyer:1994ys}
\bibitem{Iyer:1994ys} 
  V.~Iyer and R.~M.~Wald,
  ``Some properties of Noether charge and a proposal for dynamical black hole entropy,''
  Phys.\ Rev.\ D {\bf 50}, 846 (1994),
%  doi:10.1103/PhysRevD.50.846
  [gr-qc/9403028].c
  %%CITATION = doi:10.1103/PhysRevD.50.846;%%
  %1044 citations counted in INSPIRE as of 19 May 2017

%\cite{Hanada:2013rga}
\bibitem{Hanada:2013rga} 
  M.~Hanada, Y.~Hyakutake, G.~Ishiki and J.~Nishimura,
  ``Holographic description of quantum black hole on a computer,''
  Science {\bf 344}, 882 (2014),
%  doi:10.1126/science.1250122
  [arXiv:1311.5607 [hep-th]].c
  %%CITATION = doi:10.1126/science.1250122;%%
  %52 citations counted in INSPIRE as of 19 May 2017



%\cite{Hanada:2008ez}
\bibitem{Hanada:2008ez} 
  M.~Hanada, Y.~Hyakutake, J.~Nishimura and S.~Takeuchi,
  ``Higher derivative corrections to black hole thermodynamics from supersymmetric matrix quantum mechanics,''
  Phys.\ Rev.\ Lett.\  {\bf 102}, 191602 (2009),
%  doi:10.1103/PhysRevLett.102.191602
  [arXiv:0811.3102 [hep-th]].c
  %%CITATION = doi:10.1103/PhysRevLett.102.191602;%%
  %98 citations counted in INSPIRE as of 19 May 2017

%\cite{Anagnostopoulos:2007fw}
\bibitem{Anagnostopoulos:2007fw} 
  K.~N.~Anagnostopoulos, M.~Hanada, J.~Nishimura and S.~Takeuchi,
  ``Monte Carlo studies of supersymmetric matrix quantum mechanics with sixteen supercharges at finite temperature,''
  Phys.\ Rev.\ Lett.\  {\bf 100}, 021601 (2008),
%  doi:10.1103/PhysRevLett.100.021601
  [arXiv:0707.4454 [hep-th]].c
  %%CITATION = doi:10.1103/PhysRevLett.100.021601;%%
  %143 citations counted in INSPIRE as of 19 May 2017


%\cite{Hanada:2007ti}
\bibitem{Hanada:2007ti} 
  M.~Hanada, J.~Nishimura and S.~Takeuchi,
  ``Non-lattice simulation for supersymmetric gauge theories in one dimension,''
  Phys.\ Rev.\ Lett.\  {\bf 99}, 161602 (2007),
%  doi:10.1103/PhysRevLett.99.161602
  [arXiv:0706.1647 [hep-lat]].c
  %%CITATION = doi:10.1103/PhysRevLett.99.161602;%%
  %93 citations counted in INSPIRE as of 19 May 2017


%\cite{Catterall:2008yz}
\bibitem{Catterall:2008yz} 
  S.~Catterall and T.~Wiseman,
  ``Black hole thermodynamics from simulations of lattice Yang-Mills theory,''
  Phys.\ Rev.\ D {\bf 78}, 041502 (2008),
%  doi:10.1103/PhysRevD.78.041502
  [arXiv:0803.4273 [hep-th]].
  %%CITATION = doi:10.1103/PhysRevD.78.041502;%%
  %113 citations counted in INSPIRE as of 19 May 2017


%\cite{Hanada:2011fq}
\bibitem{Hanada:2011fq} 
  M.~Hanada, J.~Nishimura, Y.~Sekino and T.~Yoneya,
  ``Direct test of the gauge-gravity correspondence for Matrix theory correlation functions,''
  JHEP {\bf 1112}, 020 (2011),
%  doi:10.1007/JHEP12(2011)020
  [arXiv:1108.5153 [hep-th]].c
  %%CITATION = doi:10.1007/JHEP12(2011)020;%%
  %34 citations counted in INSPIRE as of 19 May 2017


%\cite{Green:2006gt}
\bibitem{Green:2006gt} 
  M.~B.~Green, J.~G.~Russo and P.~Vanhove,
  ``Non-renormalisation conditions in type II string theory and maximal supergravity,''
  JHEP {\bf 0702}, 099 (2007),
%  doi:10.1088/1126-6708/2007/02/099
  [hep-th/0610299].c
  %%CITATION = doi:10.1088/1126-6708/2007/02/099;%%
  %128 citations counted in INSPIRE as of 19 May 2017


%\cite{Kabat:2000zv}
\bibitem{Kabat:2000zv} 
  D.~N.~Kabat, G.~Lifschytz and D.~A.~Lowe,
  ``Black hole thermodynamics from calculations in strongly coupled gauge theory,''
  Int.\ J.\ Mod.\ Phys.\ A {\bf 16}, 856 (2001),
  [Phys.\ Rev.\ Lett.\  {\bf 86}, 1426 (2001)],
%  doi:10.1103/PhysRevLett.86.1426
  [hep-th/0007051].c
  %%CITATION = doi:10.1103/PhysRevLett.86.1426;%%
  %63 citations counted in INSPIRE as of 19 May 2017


%\cite{Dijkgraaf:1997vv}
\bibitem{Dijkgraaf:1997vv} 
  R.~Dijkgraaf, E.~P.~Verlinde and H.~L.~Verlinde,
  ``Matrix string theory,''
  Nucl.\ Phys.\ B {\bf 500}, 43 (1997)
  doi:10.1016/S0550-3213(97)00326-X
  [hep-th/9703030].
  %%CITATION = doi:10.1016/S0550-3213(97)00326-X;%%
  %623 citations counted in INSPIRE as of 20 May 2017

%\cite{Krauth:1998xh}
\bibitem{Krauth:1998xh} 
  W.~Krauth, H.~Nicolai and M.~Staudacher,
  ``Monte Carlo approach to M theory,''
  Phys.\ Lett.\ B {\bf 431}, 31 (1998)
  doi:10.1016/S0370-2693(98)00557-7
  [hep-th/9803117].c
  %%CITATION = doi:10.1016/S0370-2693(98)00557-7;%%
  %140 citations counted in INSPIRE as of 21 May 2017

%\cite{Banks:1996my}
\bibitem{Banks:1996my} 
  T.~Banks and N.~Seiberg,
  ``Strings from matrices,''
  Nucl.\ Phys.\ B {\bf 497}, 41 (1997),
%  doi:10.1016/S0550-3213(97)00278-2
  [hep-th/9702187].c
  %%CITATION = doi:10.1016/S0550-3213(97)00278-2;%%
  %313 citations counted in INSPIRE as of 23 May 2017

%\cite{Motl:1997th}
\bibitem{Motl:1997th} 
  L.~Motl,
  ``Proposals on nonperturbative superstring interactions,''
  hep-th/9701025.c
  %%CITATION = HEP-TH/9701025;%%
  %294 citations counted in INSPIRE as of 23 May 2017

%\cite{Sethi:1997sw}
\bibitem{Sethi:1997sw} 
  S.~Sethi and L.~Susskind,
  ``Rotational invariance in the M(atrix) formulation of type IIB theory,''
  Phys.\ Lett.\ B {\bf 400}, 265 (1997),
%  doi:10.1016/S0370-2693(97)00359-6
  [hep-th/9702101].c
  %%CITATION = doi:10.1016/S0370-2693(97)00359-6;%%
  %119 citations counted in INSPIRE as of 23 May 2017




%\cite{Harmark:2004ws}
\bibitem{Harmark:2004ws} 
  T.~Harmark and N.~A.~Obers,
  ``New phases of near-extremal branes on a circle,''
  JHEP {\bf 0409}, 022 (2004),
%  doi:10.1088/1126-6708/2004/09/022
  [hep-th/0407094].c
  %%CITATION = doi:10.1088/1126-6708/2004/09/022;%%
  %52 citations counted in INSPIRE as of 23 May 2017

%\cite{Susskind:1997dr}
\bibitem{Susskind:1997dr} 
  L.~Susskind,
  ``Matrix theory black holes and the Gross-Witten transition,''
  hep-th/9805115.c
  %%CITATION = HEP-TH/9805115;%%
  %42 citations counted in INSPIRE as of 23 May 2017

%\cite{Gregory:1993vy}
\bibitem{Gregory:1993vy} 
  R.~Gregory and R.~Laflamme,
  ``Black strings and p-branes are unstable,''
  Phys.\ Rev.\ Lett.\  {\bf 70}, 2837 (1993),
%  doi:10.1103/PhysRevLett.70.2837
  [hep-th/9301052].c
  %%CITATION = doi:10.1103/PhysRevLett.70.2837;%%
  %795 citations counted in INSPIRE as of 23 May 2017


%\cite{Kawahara:2007ib}
\bibitem{Kawahara:2007ib} 
  N.~Kawahara, J.~Nishimura and S.~Takeuchi,
  ``High temperature expansion in supersymmetric matrix quantum mechanics,''
  JHEP {\bf 0712}, 103 (2007),
%  doi:10.1088/1126-6708/2007/12/103
  [arXiv:0710.2188 [hep-th]].c
  %%CITATION = doi:10.1088/1126-6708/2007/12/103;%%
  %34 citations counted in INSPIRE as of 24 May 2017


%\cite{Eguchi:1982nm}
\bibitem{Eguchi:1982nm} 
  T.~Eguchi and H.~Kawai,
  ``Reduction of Dynamical Degrees of Freedom in the Large N Gauge Theory,''
  Phys.\ Rev.\ Lett.\  {\bf 48}, 1063 (1982).
%  doi:10.1103/PhysRevLett.48.1063
  %%CITATION = doi:10.1103/PhysRevLett.48.1063;%%
  %653 citations counted in INSPIRE as of 24 May 2017


%\cite{Janik:2000tq}
\bibitem{Janik:2000tq} 
  R.~A.~Janik and J.~Wosiek,
  ``Towards the matrix model of M theory on a lattice,''
  Acta Phys.\ Polon.\ B {\bf 32}, 2143 (2001),
  [hep-th/0003121].c
  %%CITATION = HEP-TH/0003121;%%
  %30 citations counted in INSPIRE as of 24 May 2017

%\cite{Bialas:2001fj}
\bibitem{Bialas:2001fj} 
  P.~Bialas and J.~Wosiek,
  ``Towards the lattice study of M theory. 2.,''
  Nucl.\ Phys.\ Proc.\ Suppl.\  {\bf 106}, 968 (2002),
%  doi:10.1016/S0920-5632(01)01900-4
  [hep-lat/0111034].c
  %%CITATION = doi:10.1016/S0920-5632(01)01900-4;%%
  %11 citations counted in INSPIRE as of 24 May 2017

%\cite{Kawahara:2007nw}
\bibitem{Kawahara:2007nw} 
  N.~Kawahara, J.~Nishimura and S.~Takeuchi,
  ``Exact fuzzy sphere thermodynamics in matrix quantum mechanics,''
  JHEP {\bf 0705}, 091 (2007),
%  doi:10.1088/1126-6708/2007/05/091
  [arXiv:0704.3183 [hep-th]].
  %%CITATION = doi:10.1088/1126-6708/2007/05/091;%%
  %16 citations counted in INSPIRE as of 24 May 2017


%\cite{Aharony:2003sx}
\bibitem{Aharony:2003sx} 
  O.~Aharony, J.~Marsano, S.~Minwalla, K.~Papadodimas and M.~Van Raamsdonk,
  ``The Hagedorn - deconfinement phase transition in weakly coupled large N gauge theories,''
  Adv.\ Theor.\ Math.\ Phys.\  {\bf 8}, 603 (2004),
%  doi:10.4310/ATMP.2004.v8.n4.a1
  [hep-th/0310285].c
  %%CITATION = doi:10.4310/ATMP.2004.v8.n4.a1;%%
  %363 citations counted in INSPIRE as of 24 May 2017

%\cite{Aharony:2005bq}
\bibitem{Aharony:2005bq} 
  O.~Aharony, J.~Marsano, S.~Minwalla, K.~Papadodimas and M.~Van Raamsdonk,
  ``A First order deconfinement transition in large N Yang-Mills theory on a small S**3,''
  Phys.\ Rev.\ D {\bf 71}, 125018 (2005),
%  doi:10.1103/PhysRevD.71.125018
  [hep-th/0502149].c
  %%CITATION = doi:10.1103/PhysRevD.71.125018;%%
  %134 citations counted in INSPIRE as of 24 May 2017


%\cite{Atick:1988si}
\bibitem{Atick:1988si} 
  J.~J.~Atick and E.~Witten,
  ``The Hagedorn Transition and the Number of Degrees of Freedom of String Theory,''
  Nucl.\ Phys.\ B {\bf 310}, 291 (1988).c
%  doi:10.1016/0550-3213(88)90151-4
  %%CITATION = doi:10.1016/0550-3213(88)90151-4;%%
  %648 citations counted in INSPIRE as of 24 May 2017



%\cite{Gross:1980he}
\bibitem{Gross:1980he} 
  D.~J.~Gross and E.~Witten,
  ``Possible Third Order Phase Transition in the Large N Lattice Gauge Theory,''
  Phys.\ Rev.\ D {\bf 21}, 446 (1980).c
%  doi:10.1103/PhysRevD.21.446
  %%CITATION = doi:10.1103/PhysRevD.21.446;%%
  %656 citations counted in INSPIRE as of 24 May 2017

%\cite{Wadia:1980cp}
\bibitem{Wadia:1980cp} 
  S.~R.~Wadia,
  ``$N$ = Infinity Phase Transition in a Class of Exactly Soluble Model Lattice Gauge Theories,''
  Phys.\ Lett.\  {\bf 93B}, 403 (1980).c
%  doi:10.1016/0370-2693(80)90353-6
  %%CITATION = doi:10.1016/0370-2693(80)90353-6;%%
  %149 citations counted in INSPIRE as of 24 May 2017

%\cite{Furuuchi:2003sy}
\bibitem{Furuuchi:2003sy} 
  K.~Furuuchi, E.~Schreiber and G.~W.~Semenoff,
  ``Five-brane thermodynamics from the matrix model,''
  hep-th/0310286.c
  %%CITATION = HEP-TH/0310286;%%
  %29 citations counted in INSPIRE as of 24 May 2017

%\cite{Semenoff:2004bs}
\bibitem{Semenoff:2004bs} 
  G.~W.~Semenoff,
  ``Matrix model thermodynamics,''
  hep-th/0405107.c
  %%CITATION = HEP-TH/0405107;%%
  %16 citations counted in INSPIRE as of 24 May 2017


%\cite{Kawahara:2006hs}
\bibitem{Kawahara:2006hs} 
  N.~Kawahara, J.~Nishimura and K.~Yoshida,
  ``Dynamical aspects of the plane-wave matrix model at finite temperature,''
  JHEP {\bf 0606}, 052 (2006),
%  doi:10.1088/1126-6708/2006/06/052
  [hep-th/0601170].c
  %%CITATION = doi:10.1088/1126-6708/2006/06/052;%%
  %18 citations counted in INSPIRE as of 24 May 2017


%\cite{Filev:2015hia}
\bibitem{Filev:2015hia} 
  V.~G.~Filev and D.~O'Connor,
  ``The BFSS model on the lattice,''
  JHEP {\bf 1605}, 167 (2016),
%  doi:10.1007/JHEP05(2016)167
  [arXiv:1506.01366 [hep-th]].c
  %%CITATION = doi:10.1007/JHEP05(2016)167;%%
  %16 citations counted in INSPIRE as of 26 May 2017

%\cite{Filev:2015cmz}
\bibitem{Filev:2015cmz} 
  V.~G.~Filev and D.~O'Connor,
  ``A Computer Test of Holographic Flavour Dynamics,''
  JHEP {\bf 1605}, 122 (2016),
%  doi:10.1007/JHEP05(2016)122
  [arXiv:1512.02536 [hep-th]].c
  %%CITATION = doi:10.1007/JHEP05(2016)122;%%
  %5 citations counted in INSPIRE as of 26 May 2017

%\cite{Asano:2016xsf}
\bibitem{Asano:2016xsf} 
  Y.~Asano, V.~G.~Filev, S.~Kováčik and D.~O'Connor,
  ``The Flavoured BFSS model at high temperature,''
  JHEP {\bf 1701}, 113 (2017),
%  doi:10.1007/JHEP01(2017)113
  [arXiv:1605.05597 [hep-th]].c
  %%CITATION = doi:10.1007/JHEP01(2017)113;%%
  %1 citations counted in INSPIRE as of 26 May 2017


%\cite{Brezin:1977sv}
\bibitem{Brezin:1977sv} 
  E.~Brezin, C.~Itzykson, G.~Parisi and J.~B.~Zuber,
  ``Planar Diagrams,''
  Commun.\ Math.\ Phys.\  {\bf 59}, 35 (1978).c
%  doi:10.1007/BF01614153
  %%CITATION = doi:10.1007/BF01614153;%%
  %1189 citations counted in INSPIRE as of 27 May 2017


%\cite{Hotta:1998en}
\bibitem{Hotta:1998en} 
  T.~Hotta, J.~Nishimura and A.~Tsuchiya,
  ``Dynamical aspects of large N reduced models,''
  Nucl.\ Phys.\ B {\bf 545}, 543 (1999),
%  doi:10.1016/S0550-3213(99)00056-5
  [hep-th/9811220].c
  %%CITATION = doi:10.1016/S0550-3213(99)00056-5;%%
  %114 citations counted in INSPIRE as of 27 May 2017

%\cite{Mandal:2011hb}
\bibitem{Mandal:2011hb} 
  G.~Mandal and T.~Morita,
  ``Phases of a two dimensional large N gauge theory on a torus,''
  Phys.\ Rev.\ D {\bf 84}, 085007 (2011),
%  doi:10.1103/PhysRevD.84.085007
  [arXiv:1103.1558 [hep-th]].c
  %%CITATION = doi:10.1103/PhysRevD.84.085007;%%
  %9 citations counted in INSPIRE as of 27 May 2017


%\cite{Mandal:2009vz}
\bibitem{Mandal:2009vz} 
  G.~Mandal, M.~Mahato and T.~Morita,
  ``Phases of one dimensional large N gauge theory in a 1/D expansion,''
  JHEP {\bf 1002}, 034 (2010),
%  doi:10.1007/JHEP02(2010)034
  [arXiv:0910.4526 [hep-th]].c
  %%CITATION = doi:10.1007/JHEP02(2010)034;%%
  %26 citations counted in INSPIRE as of 27 May 2017

%\cite{Natsuume:2014sfa}
\bibitem{Natsuume:2014sfa} 
  M.~Natsuume,
  ``AdS/CFT Duality User Guide,''
  Lect.\ Notes Phys.\  {\bf 903}, pp.1 (2015),
%  doi:10.1007/978-4-431-55441-7
  [arXiv:1409.3575 [hep-th]].c
  %%CITATION = doi:10.1007/978-4-431-55441-7;%%
  %22 citations counted in INSPIRE as of 18 Sep 2016


%\cite{ydri2017}
\bibitem{ydri2017} 
  B.~Ydri,
  ``Quantum Black Holes,''
  [arXiv:17.. [hep-th]].c

%\cite{Gubser:1998bc}
\bibitem{Gubser:1998bc} 
  S.~S.~Gubser, I.~R.~Klebanov and A.~M.~Polyakov,
  ``Gauge theory correlators from noncritical string theory,''
  Phys.\ Lett.\ B {\bf 428}, 105 (1998),
%  doi:10.1016/S0370-2693(98)00377-3
  [hep-th/9802109].c
  %%CITATION = doi:10.1016/S0370-2693(98)00377-3;%%
  %6885 citations counted in INSPIRE as of 15 Sep 2016

%\cite{Witten:1998qj}
\bibitem{Witten:1998qj} 
  E.~Witten,
  ``Anti-de Sitter space and holography,''
  Adv.\ Theor.\ Math.\ Phys.\  {\bf 2}, 253 (1998)
  [hep-th/9802150].c
  %%CITATION = HEP-TH/9802150;%%
  %7959 citations counted in INSPIRE as of 15 Sep 2016

%\cite{Polchinski:2016hrw}
\bibitem{Polchinski:2016hrw} 
  J.~Polchinski,
  ``The Black Hole Information Problem,''
  arXiv:1609.04036 [hep-th].c
  %%CITATION = ARXIV:1609.04036;%%

%\cite{Hawking:1974sw}
\bibitem{Hawking:1974sw} 
  S.~W.~Hawking,
  ``Particle Creation by Black Holes,''
  Commun.\ Math.\ Phys.\  {\bf 43}, 199 (1975)
  Erratum: [Commun.\ Math.\ Phys.\  {\bf 46}, 206 (1976)].c
%  doi:10.1007/BF02345020
  %%CITATION = doi:10.1007/BF02345020;%%
  %6179 citations counted in INSPIRE as of 15 Sep 2016


%\cite{Hawking:1976ra}
\bibitem{Hawking:1976ra} 
  S.~W.~Hawking,
  ``Breakdown of Predictability in Gravitational Collapse,''
  Phys.\ Rev.\ D {\bf 14}, 2460 (1976).c
%  doi:10.1103/PhysRevD.14.2460
  %%CITATION = doi:10.1103/PhysRevD.14.2460;%%
  %1237 citations counted in INSPIRE as of 15 Sep 2016


%\cite{Susskind:2005js}
\bibitem{Susskind:2005js} 
  L.~Susskind and J.~Lindesay,
  ``An introduction to black holes, information and the string theory revolution: The holographic universe,''
  Hackensack, USA: World Scientific (2005) 183 p.c
  %23 citations counted in INSPIRE as of 18 Sep 2016



%\cite{Page:1993up}
\bibitem{Page:1993up} 
  D.~N.~Page,
  ``Black hole information,''
  hep-th/9305040.c
  %%CITATION = HEP-TH/9305040;%%
  %126 citations counted in INSPIRE as of 15 Sep 2016



%\cite{Harlow:2014yka}
\bibitem{Harlow:2014yka} 
  D.~Harlow,
  ``Jerusalem Lectures on Black Holes and Quantum Information,''
  Rev.\ Mod.\ Phys.\  {\bf 88}, 15002 (2016),
%  [Rev.\ Mod.\ Phys.\  {\bf 88}, 15002 (2016)],
%  doi:10.1103/RevModPhys.88.015002
  [arXiv:1409.1231 [hep-th]].c
  %%CITATION = doi:10.1103/RevModPhys.88.015002;%%
  %48 citations counted in INSPIRE as of 15 Sep 2016


%\cite{Carroll:2004st}
\bibitem{Carroll:2004st} 
  S.~M.~Carroll,
  ``Spacetime and geometry: An introduction to general relativity,''
  San Francisco, USA: Addison-Wesley (2004) 513 p.c
  %203 citations counted in INSPIRE as of 21 Sep 2016

%\cite{Mukhanov:2007zz}
\bibitem{Mukhanov:2007zz} 
  V.~Mukhanov and S.~Winitzki,
  ``Introduction to quantum effects in gravity.''c
  %%CITATION = INSPIRE-775909;%%
  %56 citations counted in INSPIRE as of 29 Oct 2016


%\cite{Jacobson:2003vx}
\bibitem{Jacobson:2003vx} 
  T.~Jacobson,
  ``Introduction to quantum fields in curved space-time and the Hawking effect,''
%  doi:10.1007/0-387-24992-32
  gr-qc/0308048.c
  %%CITATION = doi:10.1007/0-387-24992-3_2;%%
  %93 citations counted in INSPIRE as of 29 Oct 2016

%\cite{Bigatti:1997jy}
\bibitem{Bigatti:1997jy} 
  D.~Bigatti and L.~Susskind,
  ``Review of matrix theory,''
  In *Cargese 1997, Strings, branes and dualities* 277-318
  [hep-th/9712072].c
  %%CITATION = HEP-TH/9712072;%%
  %187 citations counted in INSPIRE as of 01 Jun 2017

%\cite{Susskind:1997cw}
\bibitem{Susskind:1997cw} 
  L.~Susskind,
  ``Another conjecture about M(atrix) theory,''
  hep-th/9704080.c
  %%CITATION = HEP-TH/9704080;%%
  %419 citations counted in INSPIRE as of 01 Jun 2017

%\cite{Seiberg:1997ad}
\bibitem{Seiberg:1997ad} 
  N.~Seiberg,
  ``Why is the matrix model correct?,''
  Phys.\ Rev.\ Lett.\  {\bf 79}, 3577 (1997),
%  doi:10.1103/PhysRevLett.79.3577
  [hep-th/9710009].c
  %%CITATION = doi:10.1103/PhysRevLett.79.3577;%%
  %409 citations counted in INSPIRE as of 01 Jun 2017

%\cite{Taylor:2001vb}
\bibitem{Taylor:2001vb} 
  W.~Taylor,
  ``M(atrix) theory: Matrix quantum mechanics as a fundamental theory,''
  Rev.\ Mod.\ Phys.\  {\bf 73}, 419 (2001),
%  doi:10.1103/RevModPhys.73.419
  [hep-th/0101126].c
  %%CITATION = doi:10.1103/RevModPhys.73.419;%%
  %278 citations counted in INSPIRE as of 01 Jun 2017

%\cite{Sen:1997we}
\bibitem{Sen:1997we} 
  A.~Sen,
  ``D0-branes on T**n and matrix theory,''
  Adv.\ Theor.\ Math.\ Phys.\  {\bf 2}, 51 (1998)
  [hep-th/9709220].c
  %%CITATION = HEP-TH/9709220;%%
  %317 citations counted in INSPIRE as of 01 Jun 2017

%\cite{Berenstein:2002jq}
\bibitem{Berenstein:2002jq} 
  D.~E.~Berenstein, J.~M.~Maldacena and H.~S.~Nastase,
  ``Strings in flat space and pp waves from N=4 superYang-Mills,''
  JHEP {\bf 0204}, 013 (2002),
%  doi:10.1088/1126-6708/2002/04/013
  [hep-th/0202021].c
  %%CITATION = doi:10.1088/1126-6708/2002/04/013;%%
  %1633 citations counted in INSPIRE as of 02 Jun 2017

%\cite{Blau:2001ne}
\bibitem{Blau:2001ne} 
  M.~Blau, J.~M.~Figueroa-O'Farrill, C.~Hull and G.~Papadopoulos,
  ``A New maximally supersymmetric background of IIB superstring theory,''
  JHEP {\bf 0201}, 047 (2002),
%  doi:10.1088/1126-6708/2002/01/047
  [hep-th/0110242].c
  %%CITATION = doi:10.1088/1126-6708/2002/01/047;%%
  %540 citations counted in INSPIRE as of 02 Jun 2017

%\cite{KowalskiGlikman:1984wv}
\bibitem{KowalskiGlikman:1984wv} 
  J.~Kowalski-Glikman,
  ``Vacuum States in Supersymmetric Kaluza-Klein Theory,''
  Phys.\ Lett.\  {\bf 134B}, 194 (1984).c
%  doi:10.1016/0370-2693(84)90669-5
  %%CITATION = doi:10.1016/0370-2693(84)90669-5;%%
  %156 citations counted in INSPIRE as of 02 Jun 2017

%\cite{Blau:2002dy}
\bibitem{Blau:2002dy} 
  M.~Blau, J.~M.~Figueroa-O'Farrill, C.~Hull and G.~Papadopoulos,
  ``Penrose limits and maximal supersymmetry,''
  Class.\ Quant.\ Grav.\  {\bf 19}, L87 (2002),
%  doi:10.1088/0264-9381/19/10/101
  [hep-th/0201081].c
  %%CITATION = doi:10.1088/0264-9381/19/10/101;%%
  %475 citations counted in INSPIRE as of 02 Jun 2017

\bibitem{penrose} 
 R.~Penrose,
 ``Any spacetime has a plane wave as a limit'', 
 Differential geometry and relativity, Reidel, Dordrecht, 1976, pp. 271-275.c

%\cite{Hull:1984vh}
\bibitem{Hull:1984vh} 
  C.~M.~Hull,
  ``Exact $p p$ Wave Solutions of Eleven-dimensional Supergravity,''
  Phys.\ Lett.\  {\bf 139B}, 39 (1984).c
%  doi:10.1016/0370-2693(84)90030-3
  %%CITATION = doi:10.1016/0370-2693(84)90030-3;%%
  %84 citations counted in INSPIRE as of 02 Jun 2017

%\cite{Taylor:1999gq}
\bibitem{Taylor:1999gq} 
  W.~Taylor and M.~Van Raamsdonk,
  ``Multiple D0-branes in weakly curved backgrounds,''
  Nucl.\ Phys.\ B {\bf 558}, 63 (1999),
%  doi:10.1016/S0550-3213(99)00431-9
  [hep-th/9904095].c
  %%CITATION = doi:10.1016/S0550-3213(99)00431-9;%%
  %164 citations counted in INSPIRE as of 02 Jun 2017

%\cite{Dasgupta:2002hx}
\bibitem{Dasgupta:2002hx} 
  K.~Dasgupta, M.~M.~Sheikh-Jabbari and M.~Van Raamsdonk,
  ``Matrix perturbation theory for M theory on a PP wave,''
  JHEP {\bf 0205}, 056 (2002),
%  doi:10.1088/1126-6708/2002/05/056
  [hep-th/0205185].c
  %%CITATION = doi:10.1088/1126-6708/2002/05/056;%%
  %184 citations counted in INSPIRE as of 02 Jun 2017


%\cite{Taylor:1998tv}
\bibitem{Taylor:1998tv} 
  W.~Taylor and M.~Van Raamsdonk,
  ``Supergravity currents and linearized interactions for matrix theory configurations with fermionic backgrounds,''
  JHEP {\bf 9904}, 013 (1999),
%  doi:10.1088/1126-6708/1999/04/013
  [hep-th/9812239].c
  %%CITATION = doi:10.1088/1126-6708/1999/04/013;%%
  %103 citations counted in INSPIRE as of 02 Jun 2017


%\cite{Myers:1999ps}
\bibitem{Myers:1999ps} 
  R.~C.~Myers,
  ``Dielectric branes,''
  JHEP {\bf 9912}, 022 (1999),
%  doi:10.1088/1126-6708/1999/12/022
  [hep-th/9910053].c
  %%CITATION = doi:10.1088/1126-6708/1999/12/022;%%
  %1185 citations counted in INSPIRE as of 02 Jun 2017


%\cite{Catterall:2009xn}
\bibitem{Catterall:2009xn} 
  S.~Catterall and T.~Wiseman,
  ``Extracting black hole physics from the lattice,''
  JHEP {\bf 1004}, 077 (2010),
%  doi:10.1007/JHEP04(2010)077
  [arXiv:0909.4947 [hep-th]].c
  %%CITATION = doi:10.1007/JHEP04(2010)077;%%
  %66 citations counted in INSPIRE as of 02 Jun 2017

%\cite{Catterall:2010fx}
\bibitem{Catterall:2010fx} 
  S.~Catterall, A.~Joseph and T.~Wiseman,
  ``Thermal phases of D1-branes on a circle from lattice super Yang-Mills,''
  JHEP {\bf 1012}, 022 (2010),
%  doi:10.1007/JHEP12(2010)022
  [arXiv:1008.4964 [hep-th]].c
  %%CITATION = doi:10.1007/JHEP12(2010)022;%%
  %47 citations counted in INSPIRE as of 02 Jun 2017

%\cite{Hanada:2008gy}
\bibitem{Hanada:2008gy} 
  M.~Hanada, A.~Miwa, J.~Nishimura and S.~Takeuchi,
  ``Schwarzschild radius from Monte Carlo calculation of the Wilson loop in supersymmetric matrix quantum mechanics,''
  Phys.\ Rev.\ Lett.\  {\bf 102}, 181602 (2009),
%  doi:10.1103/PhysRevLett.102.181602
  [arXiv:0811.2081 [hep-th]].c
  %%CITATION = doi:10.1103/PhysRevLett.102.181602;%%
  %71 citations counted in INSPIRE as of 02 Jun 2017


%\cite{Kadoh:2015mka}
\bibitem{Kadoh:2015mka} 
  D.~Kadoh and S.~Kamata,
  ``Gauge/gravity duality and lattice simulations of one dimensional SYM with sixteen supercharges,''
  arXiv:1503.08499 [hep-lat].c
  %%CITATION = ARXIV:1503.08499;%%
  %18 citations counted in INSPIRE as of 02 Jun 2017

%\cite{Berkowitz:2016jlq}
\bibitem{Berkowitz:2016jlq} 
  E.~Berkowitz, E.~Rinaldi, M.~Hanada, G.~Ishiki, S.~Shimasaki and P.~Vranas,
  ``Precision lattice test of the gauge/gravity duality at large-$N$,''
  Phys.\ Rev.\ D {\bf 94}, no. 9, 094501 (2016),
%  doi:10.1103/PhysRevD.94.094501
  [arXiv:1606.04951 [hep-lat]].c
  %%CITATION = doi:10.1103/PhysRevD.94.094501;%%
  %8 citations counted in INSPIRE as of 02 Jun 2017

%\cite{Aharony:2008ug}
\bibitem{Aharony:2008ug} 
  O.~Aharony, O.~Bergman, D.~L.~Jafferis and J.~Maldacena,
  ``N=6 superconformal Chern-Simons-matter theories, M2-branes and their gravity duals,''
  JHEP {\bf 0810}, 091 (2008),
%  doi:10.1088/1126-6708/2008/10/091
  [arXiv:0806.1218 [hep-th]].c
  %%CITATION = doi:10.1088/1126-6708/2008/10/091;%%
  %1592 citations counted in INSPIRE as of 02 Jun 2017

%\cite{Berkooz:1996is}
\bibitem{Berkooz:1996is} 
  M.~Berkooz and M.~R.~Douglas,
  ``Five-branes in M(atrix) theory,''
  Phys.\ Lett.\ B {\bf 395}, 196 (1997),
%  doi:10.1016/S0370-2693(97)00014-2
  [hep-th/9610236].c
  %%CITATION = doi:10.1016/S0370-2693(97)00014-2;%%
  %137 citations counted in INSPIRE as of 02 Jun 2017

%\cite{Asano:2016xsf}
\bibitem{Asano:2016xsf} 
  Y.~Asano, V.~G.~Filev, S.~Kováčik and D.~O'Connor,
  ``The flavoured BFSS model at high temperature,''
  JHEP {\bf 1701}, 113 (2017),
%  doi:10.1007/JHEP01(2017)113
  [arXiv:1605.05597 [hep-th]].c
  %%CITATION = doi:10.1007/JHEP01(2017)113;%%
  %1 citations counted in INSPIRE as of 02 Jun 2017

%\cite{Azuma:2014cfa}
\bibitem{Azuma:2014cfa} 
  T.~Azuma, T.~Morita and S.~Takeuchi,
  ``Hagedorn Instability in Dimensionally Reduced Large-N Gauge Theories as Gregory-Laflamme and Rayleigh-Plateau Instabilities,''
  Phys.\ Rev.\ Lett.\  {\bf 113}, 091603 (2014),
%  doi:10.1103/PhysRevLett.113.091603
  [arXiv:1403.7764 [hep-th]].
  %%CITATION = doi:10.1103/PhysRevLett.113.091603;%%
  %7 citations counted in INSPIRE as of 02 Jun 2017

%\cite{Kim:2003rza}
\bibitem{Kim:2003rza} 
  N.~Kim, T.~Klose and J.~Plefka,
  ``Plane wave matrix theory from N=4 superYang-Mills on R x S**3,''
  Nucl.\ Phys.\ B {\bf 671}, 359 (2003),
%  doi:10.1016/j.nuclphysb.2003.08.019
  [hep-th/0306054].
  %%CITATION = doi:10.1016/j.nuclphysb.2003.08.019;%%
  %88 citations counted in INSPIRE as of 02 Jun 2017

%\cite{Kim:2002if}
\bibitem{Kim:2002if} 
  N.~Kim and J.~Plefka,
  ``On the spectrum of PP wave matrix theory,''
  Nucl.\ Phys.\ B {\bf 643}, 31 (2002),
%  doi:10.1016/S0550-3213(02)00738-1
  [hep-th/0207034].
  %%CITATION = doi:10.1016/S0550-3213(02)00738-1;%%
  %72 citations counted in INSPIRE as of 02 Jun 2017



%\cite{Kim:2011cr}
\bibitem{Kim:2011cr} 
  S.~W.~Kim, J.~Nishimura and A.~Tsuchiya,
  ``Expanding (3+1)-dimensional universe from a Lorentzian matrix model for superstring theory in (9+1)-dimensions,''
  Phys.\ Rev.\ Lett.\  {\bf 108}, 011601 (2012)
%  doi:10.1103/PhysRevLett.108.011601
  [arXiv:1108.1540 [hep-th]].
  %%CITATION = doi:10.1103/PhysRevLett.108.011601;%%
  %71 citations counted in INSPIRE as of 02 Jul 2017


%\cite{Kim:2011ts}
\bibitem{Kim:2011ts} 
  S.~W.~Kim, J.~Nishimura and A.~Tsuchiya,
  ``Expanding universe as a classical solution in the Lorentzian matrix model for nonperturbative superstring theory,''
  Phys.\ Rev.\ D {\bf 86}, 027901 (2012)
%  doi:10.1103/PhysRevD.86.027901
  [arXiv:1110.4803 [hep-th]].
  %%CITATION = doi:10.1103/PhysRevD.86.027901;%%
  %35 citations counted in INSPIRE as of 02 Jul 2017

%\cite{Nishimura:2012rs}
\bibitem{Nishimura:2012rs} 
  J.~Nishimura and A.~Tsuchiya,
  ``Local field theory from the expanding universe at late times in the IIB matrix model,''
  PTEP {\bf 2013}, 043B03 (2013)
%  doi:10.1093/ptep/ptt015
  [arXiv:1208.4910 [hep-th]].
  %%CITATION = doi:10.1093/ptep/ptt015;%%
  %17 citations counted in INSPIRE as of 02 Jul 2017


%\cite{Kim:2012mw}
\bibitem{Kim:2012mw} 
  S.~W.~Kim, J.~Nishimura and A.~Tsuchiya,
  ``Late time behaviors of the expanding universe in the IIB matrix model,''
  JHEP {\bf 1210}, 147 (2012)
%  doi:10.1007/JHEP10(2012)147
  [arXiv:1208.0711 [hep-th]].
  %%CITATION = doi:10.1007/JHEP10(2012)147;%%
  %38 citations counted in INSPIRE as of 02 Jul 2017


%\cite{Nishimura:2013moa}
\bibitem{Nishimura:2013moa} 
  J.~Nishimura and A.~Tsuchiya,
  ``Realizing chiral fermions in the type IIB matrix model at finite N,''
  JHEP {\bf 1312}, 002 (2013)
%  doi:10.1007/JHEP12(2013)002
  [arXiv:1305.5547 [hep-th]].
  %%CITATION = doi:10.1007/JHEP12(2013)002;%%
  %16 citations counted in INSPIRE as of 02 Jul 2017

%\cite{Ito:2013ywa}
\bibitem{Ito:2013ywa} 
  Y.~Ito, S.~W.~Kim, Y.~Koizuka, J.~Nishimura and A.~Tsuchiya,
  ``A renormalization group method for studying the early universe in the Lorentzian IIB matrix model,''
  PTEP {\bf 2014}, no. 8, 083B01 (2014)
%  doi:10.1093/ptep/ptu101
  [arXiv:1312.5415 [hep-th]].
  %%CITATION = doi:10.1093/ptep/ptu101;%%
  %13 citations counted in INSPIRE as of 02 Jul 2017

%\cite{Aoki:2014cya}
\bibitem{Aoki:2014cya} 
  H.~Aoki, J.~Nishimura and A.~Tsuchiya,
  ``Realizing three generations of the Standard Model fermions in the type IIB matrix model,''
  JHEP {\bf 1405}, 131 (2014)
%  doi:10.1007/JHEP05(2014)131
  [arXiv:1401.7848 [hep-th]].
  %%CITATION = doi:10.1007/JHEP05(2014)131;%%
  %14 citations counted in INSPIRE as of 02 Jul 2017

%\cite{Ito:2015mxa}
\bibitem{Ito:2015mxa} 
  Y.~Ito, J.~Nishimura and A.~Tsuchiya,
  ``Power-law expansion of the Universe from the bosonic Lorentzian type IIB matrix model,''
  JHEP {\bf 1511}, 070 (2015)
%  doi:10.1007/JHEP11(2015)070
  [arXiv:1506.04795 [hep-th]].
  %%CITATION = doi:10.1007/JHEP11(2015)070;%%
  %7 citations counted in INSPIRE as of 02 Jul 2017

%\cite{Ito:2015mem}
\bibitem{Ito:2015mem} 
  Y.~Ito, J.~Nishimura and A.~Tsuchiya,
  ``Large-scale computation of the exponentially expanding universe in a simplified Lorentzian type IIB matrix model,''
  PoS LATTICE {\bf 2015}, 243 (2016)
  [arXiv:1512.01923 [hep-lat]].
  %%CITATION = ARXIV:1512.01923;%%
  %2 citations counted in INSPIRE as of 08 Jul 2017

%\cite{Nishimura:2011xy}
\bibitem{Nishimura:2011xy} 
  J.~Nishimura, T.~Okubo and F.~Sugino,
  ``Systematic study of the SO(10) symmetry breaking vacua in the matrix model for type IIB superstrings,''
  JHEP {\bf 1110}, 135 (2011)
%  doi:10.1007/JHEP10(2011)135
  [arXiv:1108.1293 [hep-th]].
  %%CITATION = doi:10.1007/JHEP10(2011)135;%%
  %26 citations counted in INSPIRE as of 10 Jul 2017

%\cite{Aoki:1998vn}
\bibitem{Aoki:1998vn} 
  H.~Aoki, S.~Iso, H.~Kawai, Y.~Kitazawa and T.~Tada,
  ``Space-time structures from IIB matrix model,''
  Prog.\ Theor.\ Phys.\  {\bf 99}, 713 (1998)
%  doi:10.1143/PTP.99.713
  [hep-th/9802085].
  %%CITATION = doi:10.1143/PTP.99.713;%%
  %191 citations counted in INSPIRE as of 10 Jul 2017

%\cite{Nishimura:2000ds}
\bibitem{Nishimura:2000ds} 
  J.~Nishimura and G.~Vernizzi,
  ``Spontaneous breakdown of Lorentz invariance in IIB matrix model,''
  JHEP {\bf 0004}, 015 (2000)
%  doi:10.1088/1126-6708/2000/04/015
  [hep-th/0003223].
  %%CITATION = doi:10.1088/1126-6708/2000/04/015;%%
  %74 citations counted in INSPIRE as of 11 Jul 2017


%\cite{Nishimura:2000wf}
\bibitem{Nishimura:2000wf} 
  J.~Nishimura and G.~Vernizzi,
  ``Brane world from IIB matrices,''
  Phys.\ Rev.\ Lett.\  {\bf 85}, 4664 (2000)
%  doi:10.1103/PhysRevLett.85.4664
  [hep-th/0007022].
  %%CITATION = doi:10.1103/PhysRevLett.85.4664;%%
  %61 citations counted in INSPIRE as of 11 Jul 2017

%\cite{Austing:2001bd}
\bibitem{Austing:2001bd} 
  P.~Austing and J.~F.~Wheater,
  ``The Convergence of Yang-Mills integrals,''
  JHEP {\bf 0102}, 028 (2001)
%  doi:10.1088/1126-6708/2001/02/028
  [hep-th/0101071].
  %%CITATION = doi:10.1088/1126-6708/2001/02/028;%%
  %49 citations counted in INSPIRE as of 11 Jul 2017


%\cite{Austing:2001pk}
\bibitem{Austing:2001pk} 
  P.~Austing and J.~F.~Wheater,
  ``Convergent Yang-Mills matrix theories,''
  JHEP {\bf 0104}, 019 (2001)
%  doi:10.1088/1126-6708/2001/04/019
  [hep-th/0103159].
  %%CITATION = doi:10.1088/1126-6708/2001/04/019;%%
  %60 citations counted in INSPIRE as of 11 Jul 2017

%\cite{Ambjorn:2000bf}
\bibitem{Ambjorn:2000bf} 
  J.~Ambjorn, K.~N.~Anagnostopoulos, W.~Bietenholz, T.~Hotta and J.~Nishimura,
  ``Large N dynamics of dimensionally reduced 4-D SU(N) superYang-Mills theory,''
  JHEP {\bf 0007}, 013 (2000)
%  doi:10.1088/1126-6708/2000/07/013
  [hep-th/0003208]. See also \cite{Ambjorn:2000dx}.
  %%CITATION = doi:10.1088/1126-6708/2000/07/013;%%
  %111 citations counted in INSPIRE as of 11 Jul 2017


%\cite{Ambjorn:2000dx}
\bibitem{Ambjorn:2000dx}
  J.~Ambjorn, K.~N.~Anagnostopoulos, W.~Bietenholz, T.~Hotta and J.~Nishimura,
  ``Monte Carlo studies of the IIB matrix model at large N,''
  JHEP {\bf 0007}, 011 (2000)
  [arXiv:hep-th/0005147].
  %%CITATION = JHEPA,0007,011;%%


%\cite{Burda:2000mn}
\bibitem{Burda:2000mn}
  Z.~Burda, B.~Petersson and J.~Tabaczek,
  ``Geometry of reduced supersymmetric 4-D Yang-Mills integrals,''
  Nucl.\ Phys.\ B {\bf 602} (2001) 399
%  doi:10.1016/S0550-3213(01)00114-6
  [hep-lat/0012001].
  %%CITATION = doi:10.1016/S0550-3213(01)00114-6;%%
  %46 citations counted in INSPIRE as of 11 Jul 2017


%\cite{Ambjorn:2001xs}
\bibitem{Ambjorn:2001xs} 
  J.~Ambjorn, K.~N.~Anagnostopoulos, W.~Bietenholz, F.~Hofheinz and J.~Nishimura,
  ``On the spontaneous breakdown of Lorentz symmetry in matrix models of superstrings,''
  Phys.\ Rev.\ D {\bf 65}, 086001 (2002)
%  doi:10.1103/PhysRevD.65.086001
  [hep-th/0104260].
  %%CITATION = doi:10.1103/PhysRevD.65.086001;%%
  %49 citations counted in INSPIRE as of 11 Jul 2017


%\cite{Anagnostopoulos:2001yb}
\bibitem{Anagnostopoulos:2001yb} 
  K.~N.~Anagnostopoulos and J.~Nishimura,
  ``New approach to the complex action problem and its application to a nonperturbative study of superstring theory,''
  Phys.\ Rev.\ D {\bf 66}, 106008 (2002)
%  doi:10.1103/PhysRevD.66.106008
  [hep-th/0108041].
  %%CITATION = doi:10.1103/PhysRevD.66.106008;%%
  %101 citations counted in INSPIRE as of 11 Jul 2017

%\cite{Nishimura:2001sq}
\bibitem{Nishimura:2001sq} 
  J.~Nishimura,
  ``Exactly solvable matrix models for the dynamical generation of space-time in superstring theory,''
  Phys.\ Rev.\ D {\bf 65}, 105012 (2002)
%  doi:10.1103/PhysRevD.65.105012
  [hep-th/0108070].
  %%CITATION = doi:10.1103/PhysRevD.65.105012;%%
  %49 citations counted in INSPIRE as of 11 Jul 2017

%\cite{Nishimura:2004ts}
\bibitem{Nishimura:2004ts} 
  J.~Nishimura, T.~Okubo and F.~Sugino,
  ``Gaussian expansion analysis of a matrix model with the spontaneous breakdown of rotational symmetry,''
  Prog.\ Theor.\ Phys.\  {\bf 114}, 487 (2005)
%  doi:10.1143/PTP.114.487
  [hep-th/0412194].
  %%CITATION = doi:10.1143/PTP.114.487;%%
  %29 citations counted in INSPIRE as of 11 Jul 2017

%\cite{Anagnostopoulos:2011cn}
\bibitem{Anagnostopoulos:2011cn} 
  K.~N.~Anagnostopoulos, T.~Azuma and J.~Nishimura,
  ``A practical solution to the sign problem in a matrix model for dynamical compactification,''
  JHEP {\bf 1110}, 126 (2011)
%  doi:10.1007/JHEP10(2011)126
  [arXiv:1108.1534 [hep-lat]].
  %%CITATION = doi:10.1007/JHEP10(2011)126;%%
  %13 citations counted in INSPIRE as of 11 Jul 2017


%\cite{Stevenson:1981vj}
\bibitem{Stevenson:1981vj} 
  P.~M.~Stevenson,
  ``Optimized Perturbation Theory,''
  Phys.\ Rev.\ D {\bf 23}, 2916 (1981).
%  doi:10.1103/PhysRevD.23.2916
  %%CITATION = doi:10.1103/PhysRevD.23.2916;%%
  %1073 citations counted in INSPIRE as of 11 Jul 2017

%\cite{Aoyama:2010ry}
\bibitem{Aoyama:2010ry} 
  T.~Aoyama, J.~Nishimura and T.~Okubo,
  ``Spontaneous breaking of the rotational symmetry in dimensionally reduced super Yang-Mills models,''
  Prog.\ Theor.\ Phys.\  {\bf 125}, 537 (2011)
%  doi:10.1143/PTP.125.537
  [arXiv:1007.0883 [hep-th]].
  %%CITATION = doi:10.1143/PTP.125.537;%%
  %15 citations counted in INSPIRE as of 11 Jul 2017


%\cite{Patera:1976ud}
\bibitem{Patera:1976ud} 
  J.~Patera, R.~T.~Sharp, P.~Winternitz and H.~Zassenhaus,
  ``Invariants of Real Low Dimension Lie Algebras,''
  J.\ Math.\ Phys.\  {\bf 17}, 986 (1976).
%  doi:10.1063/1.522992
  %%CITATION = doi:10.1063/1.522992;%%
  %81 citations counted in INSPIRE as of 14 Jul 2017


%\cite{Steinacker:2011wb}
\bibitem{Steinacker:2011wb} 
  H.~Steinacker,
  ``Split noncommutativity and compactified brane solutions in matrix models,''
  Prog.\ Theor.\ Phys.\  {\bf 126}, 613 (2011)
%  doi:10.1143/PTP.126.613
  [arXiv:1106.6153 [hep-th]].
  %%CITATION = doi:10.1143/PTP.126.613;%%
  %28 citations counted in INSPIRE as of 14 Jul 2017


%\cite{Chatzistavrakidis:2011su}
\bibitem{Chatzistavrakidis:2011su} 
  A.~Chatzistavrakidis,
  ``On Lie-algebraic solutions of the type IIB matrix model,''
  Phys.\ Rev.\ D {\bf 84}, 106010 (2011)
%  doi:10.1103/PhysRevD.84.106010
  [arXiv:1108.1107 [hep-th]].
  %%CITATION = doi:10.1103/PhysRevD.84.106010;%%
  %13 citations counted in INSPIRE as of 14 Jul 2017


%\cite{VilKli}
\bibitem{VilKli} 
  N.~Ja.~Vilenkin and A.~U.~Klimyk,
  ``Representation of Lie groups and special functions,''
  volume 1, Kluwer Academic Publishers (1991).



%\cite{Stern:2014aqa}
\bibitem{Stern:2014aqa} 
  A.~Stern,
  ``Matrix Model Cosmology in Two Space-time Dimensions,''
  Phys.\ Rev.\ D {\bf 90}, no. 12, 124056 (2014)
%  doi:10.1103/PhysRevD.90.124056
  [arXiv:1409.7833 [hep-th]].
  %%CITATION = doi:10.1103/PhysRevD.90.124056;%%
  %7 citations counted in INSPIRE as of 15 Jul 2017


%\cite{Chaney:2015mfa}
\bibitem{Chaney:2015mfa} 
  A.~Chaney, L.~Lu and A.~Stern,
  ``Lorentzian Fuzzy Spheres,''
  Phys.\ Rev.\ D {\bf 92}, no. 6, 064021 (2015)
%  doi:10.1103/PhysRevD.92.064021
  [arXiv:1506.03505 [hep-th]].
  %%CITATION = doi:10.1103/PhysRevD.92.064021;%%
  %8 citations counted in INSPIRE as of 15 Jul 2017


%\cite{Chaney:2016npa}
\bibitem{Chaney:2016npa} 
  A.~Chaney and A.~Stern,
  ``Fuzzy $CP^2$ spacetimes,''
  Phys.\ Rev.\ D {\bf 95}, no. 4, 046001 (2017)
%  doi:10.1103/PhysRevD.95.046001
  [arXiv:1612.01964 [hep-th]].
  %%CITATION = doi:10.1103/PhysRevD.95.046001;%%

%\cite{Chaney:2015ktw}
\bibitem{Chaney:2015ktw} 
  A.~Chaney, L.~Lu and A.~Stern,
  ``Matrix Model Approach to Cosmology,''
  Phys.\ Rev.\ D {\bf 93}, no. 6, 064074 (2016)
%  doi:10.1103/PhysRevD.93.064074
  [arXiv:1511.06816 [hep-th]].
  %%CITATION = doi:10.1103/PhysRevD.93.064074;%%
  %4 citations counted in INSPIRE as of 15 Jul 2017



%\cite{Freedman:2004xg}
\bibitem{Freedman:2004xg} 
  D.~Z.~Freedman, G.~W.~Gibbons and M.~Schnabl,
  ``Matrix cosmology,''
  AIP Conf.\ Proc.\  {\bf 743}, 286 (2005)
%  doi:10.1063/1.1848334
  [hep-th/0411119].
  %%CITATION = doi:10.1063/1.1848334;%%
  %32 citations counted in INSPIRE as of 15 Jul 2017


%\cite{Das:2005vd}
\bibitem{Das:2005vd} 
  S.~R.~Das and J.~Michelson,
  ``pp wave big bangs: Matrix strings and shrinking fuzzy spheres,''
  Phys.\ Rev.\ D {\bf 72}, 086005 (2005)
%  doi:10.1103/PhysRevD.72.086005
  [hep-th/0508068].
  %%CITATION = doi:10.1103/PhysRevD.72.086005;%%
  %42 citations counted in INSPIRE as of 15 Jul 2017

%\cite{Craps:2005wd}
\bibitem{Craps:2005wd} 
  B.~Craps, S.~Sethi and E.~P.~Verlinde,
  ``A Matrix big bang,''
  JHEP {\bf 0510}, 005 (2005)
%  doi:10.1088/1126-6708/2005/10/005
  [hep-th/0506180].
  %%CITATION = doi:10.1088/1126-6708/2005/10/005;%%
  %134 citations counted in INSPIRE as of 15 Jul 2017




%\cite{Matsuo:2008yd}
\bibitem{Matsuo:2008yd} 
  T.~Matsuo, D.~Tomino, W.~Y.~Wen and S.~Zeze,
  ``Quantum gravity equation in large N Yang-Mills quantum mechanics,''
  JHEP {\bf 0811}, 088 (2008)
%  doi:10.1088/1126-6708/2008/11/088
  [arXiv:0807.1186 [hep-th]].
  %%CITATION = doi:10.1088/1126-6708/2008/11/088;%%
  %6 citations counted in INSPIRE as of 15 Jul 2017

%\cite{Ishino:2006nx}
\bibitem{Ishino:2006nx} 
  T.~Ishino and N.~Ohta,
  ``Matrix string description of cosmic singularities in a class of time-dependent solutions,''
  Phys.\ Lett.\ B {\bf 638}, 105 (2006)
%  doi:10.1016/j.physletb.2006.05.029
  [hep-th/0603215].
  %%CITATION = doi:10.1016/j.physletb.2006.05.029;%%
  %42 citations counted in INSPIRE as of 15 Jul 2017


%\cite{Martinec:2006ak}
\bibitem{Martinec:2006ak} 
  E.~J.~Martinec, D.~Robbins and S.~Sethi,
  ``Toward the end of time,''
  JHEP {\bf 0608}, 025 (2006)
%  doi:10.1088/1126-6708/2006/08/025
  [hep-th/0603104].
  %%CITATION = doi:10.1088/1126-6708/2006/08/025;%%
  %43 citations counted in INSPIRE as of 15 Jul 2017


%\cite{Chen:2005mga}
\bibitem{Chen:2005mga} 
  B.~Chen,
  ``The Time-dependent supersymmetric configurations in M-theory and matrix models,''
  Phys.\ Lett.\ B {\bf 632}, 393 (2006)
%  doi:10.1016/j.physletb.2005.10.021
  [hep-th/0508191].
  %%CITATION = doi:10.1016/j.physletb.2005.10.021;%%
  %45 citations counted in INSPIRE as of 15 Jul 2017

%\cite{Li:2005sz}
\bibitem{Li:2005sz} 
  M.~Li,
  ``A Class of cosmological matrix models,''
  Phys.\ Lett.\ B {\bf 626}, 202 (2005)
%  doi:10.1016/j.physletb.2005.08.099
  [hep-th/0506260].
  %%CITATION = doi:10.1016/j.physletb.2005.08.099;%%
  %51 citations counted in INSPIRE as of 15 Jul 2017




%\cite{Klammer:2009ku}
\bibitem{Klammer:2009ku} 
  D.~Klammer and H.~Steinacker,
  ``Cosmological solutions of emergent noncommutative gravity,''
  Phys.\ Rev.\ Lett.\  {\bf 102}, 221301 (2009)
%  doi:10.1103/PhysRevLett.102.221301
  [arXiv:0903.0986 [gr-qc]].
  %%CITATION = doi:10.1103/PhysRevLett.102.221301;%%
  %35 citations counted in INSPIRE as of 15 Jul 2017


%\cite{Lee:2010zf}
\bibitem{Lee:2010zf} 
  J.~Lee and H.~S.~Yang,
  ``Quantum Gravity from Noncommutative Spacetime,''
  J.\ Korean Phys.\ Soc.\  {\bf 65}, 1754 (2014)
%  doi:10.3938/jkps.65.1754
  [arXiv:1004.0745 [hep-th]].
  %%CITATION = doi:10.3938/jkps.65.1754;%%
  %26 citations counted in INSPIRE as of 15 Jul 2017


%\cite{She:2005mt}
\bibitem{She:2005mt} 
  J.~H.~She,
  ``A Matrix model for Misner universe,''
  JHEP {\bf 0601}, 002 (2006)
%  doi:10.1088/1126-6708/2006/01/002
  [hep-th/0509067].
  %%CITATION = doi:10.1088/1126-6708/2006/01/002;%%
  %53 citations counted in INSPIRE as of 15 Jul 2017

%\cite{Rivelles:2002ez}
\bibitem{Rivelles:2002ez} 
  V.~O.~Rivelles,
  ``Noncommutative field theories and gravity,''
  Phys.\ Lett.\ B {\bf 558}, 191 (2003)
%  doi:10.1016/S0370-2693(03)00271-5
  [hep-th/0212262].
  %%CITATION = doi:10.1016/S0370-2693(03)00271-5;%%
  %112 citations counted in INSPIRE as of 16 Jul 2017


%\cite{Steinacker:2010rh}
\bibitem{Steinacker:2010rh} 
  H.~Steinacker,
  ``Emergent Geometry and Gravity from Matrix Models: an Introduction,''
  Class.\ Quant.\ Grav.\  {\bf 27}, 133001 (2010)
%  doi:10.1088/0264-9381/27/13/133001
  [arXiv:1003.4134 [hep-th]].
  %%CITATION = doi:10.1088/0264-9381/27/13/133001;%%
  %110 citations counted in INSPIRE as of 16 Jul 2017

%\cite{Steinacker:2016vgf}
\bibitem{Steinacker:2016vgf} 
  H.~C.~Steinacker,
  ``Emergent gravity on covariant quantum spaces in the IKKT model,''
  JHEP {\bf 1612}, 156 (2016)
%  doi:10.1007/JHEP12(2016)156
  [arXiv:1606.00769 [hep-th]].
  %%CITATION = doi:10.1007/JHEP12(2016)156;%%
  %6 citations counted in INSPIRE as of 16 Jul 2017


%\cite{Steinacker:2007dq}
\bibitem{Steinacker:2007dq} 
  H.~Steinacker,
  ``Emergent Gravity from Noncommutative Gauge Theory,''
  JHEP {\bf 0712}, 049 (2007)
%  doi:10.1088/1126-6708/2007/12/049
  [arXiv:0708.2426 [hep-th]].
  %%CITATION = doi:10.1088/1126-6708/2007/12/049;%%
  %118 citations counted in INSPIRE as of 16 Jul 2017



%\cite{Yang:2008fb}
\bibitem{Yang:2008fb} 
  H.~S.~Yang,
  ``Emergent Spacetime and The Origin of Gravity,''
  JHEP {\bf 0905}, 012 (2009)
%  doi:10.1088/1126-6708/2009/05/012
  [arXiv:0809.4728 [hep-th]].
  %%CITATION = doi:10.1088/1126-6708/2009/05/012;%%
  %38 citations counted in INSPIRE as of 16 Jul 2017

%\cite{Yang:2006dk}
\bibitem{Yang:2006dk} 
  H.~S.~Yang,
  ``Emergent Gravity from Noncommutative Spacetime,''
  Int.\ J.\ Mod.\ Phys.\ A {\bf 24}, 4473 (2009)
%  doi:10.1142/S0217751X0904587X
  [hep-th/0611174].
  %%CITATION = doi:10.1142/S0217751X0904587X;%%
  %60 citations counted in INSPIRE as of 16 Jul 2017


%\cite{Yang:2009pm}
\bibitem{Yang:2009pm} 
  H.~S.~Yang and M.~Sivakumar,
  ``Emergent Gravity from Quantized Spacetime,''
  Phys.\ Rev.\ D {\bf 82}, 045004 (2010)
%  doi:10.1103/PhysRevD.82.045004
  [arXiv:0908.2809 [hep-th]].
  %%CITATION = doi:10.1103/PhysRevD.82.045004;%%
  %20 citations counted in INSPIRE as of 16 Jul 2017

%\cite{Kawai:2016wfh}
\bibitem{Kawai:2016wfh} 
  H.~Kawai, K.~Kawana and K.~Sakai,
  ``A note on graviton exchange in the emergent gravity scenario,''
  PTEP {\bf 2017}, no. 4, 043B06 (2017)
%  doi:10.1093/ptep/ptx036
  [arXiv:1610.09844 [hep-th]].
  %%CITATION = doi:10.1093/ptep/ptx036;%%


%\cite{Seiberg:1999vs}
\bibitem{Seiberg:1999vs} 
  N.~Seiberg and E.~Witten,
  ``String theory and noncommutative geometry,''
  JHEP {\bf 9909}, 032 (1999)
%  doi:10.1088/1126-6708/1999/09/032
  [hep-th/9908142].
  %%CITATION = doi:10.1088/1126-6708/1999/09/032;%%
  %3802 citations counted in INSPIRE as of 16 Jul 2017

%\cite{Douglas:2001ba}
\bibitem{Douglas:2001ba} 
  M.~R.~Douglas and N.~A.~Nekrasov,
  ``Noncommutative field theory,''
  Rev.\ Mod.\ Phys.\  {\bf 73}, 977 (2001)
%  doi:10.1103/RevModPhys.73.977
  [hep-th/0106048].
  %%CITATION = doi:10.1103/RevModPhys.73.977;%%
  %1665 citations counted in INSPIRE as of 16 Jul 2017


%\cite{Ydri:2016dmy}
\bibitem{Ydri:2016dmy} 
  B.~Ydri,
  ``Lectures on Matrix Field Theory,''
  Lect.\ Notes Phys.\  {\bf 929}, pp.1 (2017)
%  doi:10.1007/978-3-319-46003-1
  [arXiv:1603.00924 [hep-th]].
  %%CITATION = doi:10.1007/978-3-319-46003-1;%%
  %3 citations counted in INSPIRE as of 17 Jul 2017



%\cite{Ydri:2004vq}
\bibitem{Ydri:2004vq} 
  B.~Ydri,
  ``Noncommutative U(1) gauge theory as a non-linear sigma model,''
  Mod.\ Phys.\ Lett.\  {\bf 19}, 2205 (2004)
%  doi:10.1142/S0217732304015531
  [hep-th/0405208].
  %%CITATION = doi:10.1142/S0217732304015531;%%
  %7 citations counted in INSPIRE as of 17 Jul 2017

%\cite{Yang:2007rg}
\bibitem{Yang:2007rg} 
  H.~S.~Yang,
  %``Noncommutative Electromagnetism As A Large N Gauge Theory,''
  Eur.\ Phys.\ J.\ C {\bf 64}, 445 (2009)
%  doi:10.1140/epjc/s10052-009-1117-9
  [arXiv:0704.0929 [hep-th]].
  %%CITATION = doi:10.1140/epjc/s10052-009-1117-9;%%
  %23 citations counted in INSPIRE as of 17 Jul 2017

%\cite{Bietenholz:2004as}
\bibitem{Bietenholz:2004as} 
  W.~Bietenholz, F.~Hofheinz and J.~Nishimura,
  ``On the relation between non-commutative field theories at theta = infinity and large N matrix field theories,''
  JHEP {\bf 0405}, 047 (2004)
%  doi:10.1088/1126-6708/2004/05/047
  [hep-th/0404179].
  %%CITATION = doi:10.1088/1126-6708/2004/05/047;%%
  %20 citations counted in INSPIRE as of 17 Jul 2017


%\cite{EK}
\bibitem{EK} 
  J.~Ehlers, W.~Kundt,
  ``Exact solutions of the gravitational field equations,''
  in ``Gravitation: An introduction to current research'', Ed.~L.~Witten, 1962: New York, USA, pp. 49-101.

%\cite{Balachandran:2005ew}
\bibitem{Balachandran:2005ew}
  A.~P.~Balachandran, S.~Kurkcuoglu and S.~Vaidya,
  ``Lectures on fuzzy and fuzzy SUSY physics,''
  arXiv:hep-th/0511114.
  %%CITATION = HEP-TH/0511114;%%

%\cite{O'Connor:2003aj}
\bibitem{O'Connor:2003aj}
  D.~O'Connor,
  ``Field theory on low dimensional fuzzy spaces,''
  Mod.\ Phys.\ Lett.\  A {\bf 18}, 2423 (2003).
  %%CITATION = MPLAE,A18,2423;%%

%\cite{Ydri:2001pv}
\bibitem{Ydri:2001pv} 
  B.~Ydri,
  ``Fuzzy physics,''
  hep-th/0110006.
  %%CITATION = HEP-TH/0110006;%%
  %38 citations counted in INSPIRE as of 06 Feb 2016

%\cite{Kurkcuoglu:2004gf}
\bibitem{Kurkcuoglu:2004gf} 
  S.~Kurkcuoglu,
  ``Explorations in fuzzy physics and non-commutative geometry,''
  UMI-31-60408.
  %%CITATION = UMI-31-60408;%%



%\cite{Balachandran:2002ig}
\bibitem{Balachandran:2002ig} 
  A.~P.~Balachandran,
  ``Quantum space-times in the year 2002,''
  Pramana {\bf 59}, 359 (2002)
%  doi:10.1007/s12043-002-0128-y
  [hep-th/0203259].
  %%CITATION = doi:10.1007/s12043-002-0128-y;%%
  %33 citations counted in INSPIRE as of 06 Feb 2016





%\cite{Steinacker:2004mq}
\bibitem{Steinacker:2004mq} 
  H.~Steinacker,
  ``Field theoretic models on covariant quantum spaces,''
  hep-th/0408125.
  %%CITATION = HEP-TH/0408125;%%
  %1 citations counted in INSPIRE as of 06 Feb 2016


%\cite{Abe:2010an}
\bibitem{Abe:2010an} 
  Y.~Abe,
  ``Construction of Fuzzy Spaces and Their Applications to Matrix Models,''
  arXiv:1002.4937 [hep-th].
  %%CITATION = ARXIV:1002.4937;%%
  %8 citations counted in INSPIRE as of 05 Feb 2016


%\cite{Karabali:2004xq}
\bibitem{Karabali:2004xq} 
  D.~Karabali, V.~P.~Nair and S.~Randjbar-Daemi,
  ``Fuzzy spaces, the M(atrix) model and the quantum Hall effect,''
  In *Shifman, M. (ed.) et al.: From fields to strings, vol. 1* 831-875
  [hep-th/0407007].
  %%CITATION = HEP-TH/0407007;%%
  %32 citations counted in INSPIRE as of 06 févr. 2016



%\cite{Grosse:1995ar}
\bibitem{Grosse:1995ar} 
  H.~Grosse, C.~Klimcik and P.~Presnajder,
  ``Towards finite quantum field theory in noncommutative geometry,''
  Int.\ J.\ Theor.\ Phys.\  {\bf 35}, 231 (1996)
%  doi:10.1007/BF02083810
  [hep-th/9505175].
  %%CITATION = doi:10.1007/BF02083810;%%
  %173 citations counted in INSPIRE as of 05 Feb 2016

%\cite{Grosse:1996mz}
\bibitem{Grosse:1996mz} 
  H.~Grosse, C.~Klimcik and P.~Presnajder,
  ``On finite 4-D quantum field theory in noncommutative geometry,''
  Commun.\ Math.\ Phys.\  {\bf 180}, 429 (1996)
%  doi:10.1007/BF02099720
  [hep-th/9602115].
  %%CITATION = doi:10.1007/BF02099720;%%
  %133 citations counted in INSPIRE as of 09 Aug 2016



%\cite{Alekseev:1999bs}
\bibitem{Alekseev:1999bs} 
  A.~Y.~Alekseev, A.~Recknagel and V.~Schomerus,
  ``Noncommutative world volume geometries: Branes on SU(2) and fuzzy spheres,''
  JHEP {\bf 9909}, 023 (1999)
%  doi:10.1088/1126-6708/1999/09/023
  [hep-th/9908040].
  %%CITATION = doi:10.1088/1126-6708/1999/09/023;%%
  %207 citations counted in INSPIRE as of 05 Feb 2016


%\cite{Hikida:2001py}
\bibitem{Hikida:2001py} 
  Y.~Hikida, M.~Nozaki and Y.~Sugawara,
  ``Formation of spherical 2D brane from multiple D0 branes,''
  Nucl.\ Phys.\ B {\bf 617}, 117 (2001)
%  doi:10.1016/S0550-3213(01)00473-4
  [hep-th/0101211].
  %%CITATION = doi:10.1016/S0550-3213(01)00473-4;%%
  %34 citations counted in INSPIRE as of 05 Feb 2016

%\cite{Ahluwalia:1993dd}
\bibitem{Ahluwalia:1993dd} 
  D.~V.~Ahluwalia,
  ``Quantum measurements, gravitation, and locality,''
  Phys.\ Lett.\ B {\bf 339}, 301 (1994)
%  doi:10.1016/0370-2693(94)90622-X
  [gr-qc/9308007]. See quotation from \cite{Schwinger:1951xk}.
  %%CITATION = doi:10.1016/0370-2693(94)90622-X;%%
  %146 citations counted in INSPIRE as of 31 Aug 2016


%\cite{Schwinger:1951xk}
\bibitem{Schwinger:1951xk} 
  J.~S.~Schwinger,
  ``The Theory of quantized fields. 1.,''
  Phys.\ Rev.\  {\bf 82}, 914 (1951).
%  doi:10.1103/PhysRev.82.914
  %%CITATION = doi:10.1103/PhysRev.82.914;%%
  %447 citations counted in INSPIRE as of 31 Aug 2016



%\cite{Snyder:1946qz}
\bibitem{Snyder:1946qz} 
  H.~S.~Snyder,
  ``Quantized space-time,''
  Phys.\ Rev.\  {\bf 71}, 38 (1947).
%  doi:10.1103/PhysRev.71.38
  %%CITATION = doi:10.1103/PhysRev.71.38;%%
  %1302 citations counted in INSPIRE as of 09 Aug 2016

%\cite{Yang:1947ud}
\bibitem{Yang:1947ud} 
  C.~N.~Yang,
  ``On quantized space-time,''
  Phys.\ Rev.\  {\bf 72}, 874 (1947).
%  doi:10.1103/PhysRev.72.874
  %%CITATION = doi:10.1103/PhysRev.72.874;%%
  %270 citations counted in INSPIRE as of 09 Aug 2016






%\cite{Doplicher:1994tu}
\bibitem{Doplicher:1994tu} 
  S.~Doplicher, K.~Fredenhagen and J.~E.~Roberts,
  ``The Quantum structure of space-time at the Planck scale and quantum fields,''
  Commun.\ Math.\ Phys.\  {\bf 172}, 187 (1995)
%  doi:10.1007/BF02104515
  [hep-th/0303037].
  %%CITATION = doi:10.1007/BF02104515;%%
  %865 citations counted in INSPIRE as of 09 Aug 2016



%\cite{Frohlich:1993es}
\bibitem{Frohlich:1993es}
  J.~Frohlich and K.~Gawedzki,
  ``Conformal field theory and geometry of strings,''
  In *Vancouver 1993, Proceedings, Mathematical quantum theory, vol. 1* 57-97, and Preprint - Gawedzki, K. (rec.Nov.93) 44 p
  [hep-th/9310187].
  %%CITATION = HEP-TH/9310187;%%
  %63 citations counted in INSPIRE as of 10 Aug 2016




%\cite{Connes:1994yd}
\bibitem{Connes:1994yd}
  A.~Connes,
  ``Noncommutative geometry,''
   Academic Press,London, 1994.
%\href{http://www.slac.stanford.edu/spires/find/hep/www?irn=6658121}{SPIRES entry}


%\cite{Sniatycki:1995sn}
\bibitem{Sniatycki:1995sn} 
  J.~Sniatycki,
  ``Geometric quantization and quantum mechanics,''
  New York, USA: Springer (1980) 230 p. (Applied mathematical sciences, 30)

%\cite{Woodhouse:1992de}
\bibitem{Woodhouse:1992de} 
  N.~M.~J.~Woodhouse,
  ``Geometric quantization,''
  New York, USA: Clarendon (1992) 307 p. (Oxford mathematical monographs)
  %3 citations counted in INSPIRE as of 09 Feb 2016


%\cite{Nair:2005iw}
\bibitem{Nair:2005iw} 
  V.~P.~Nair,
  ``Quantum field theory: A modern perspective,''
  New York, USA: Springer (2005) 557 p
  %2 citations counted in INSPIRE as of 09 Feb 2016

%\cite{Hu:2001ue}
\bibitem{Hu:2001ue} 
  S.~Hu,
  ``Lecture notes on Chern-Simons-Witten theory,''
  River Edge, USA: World Scientific (2001) 200 p

%\cite{Murray:2006pi}
\bibitem{Murray:2006pi} 
  S.~Murray and C.~Saemann,
  ``Quantization of Flag Manifolds and their Supersymmetric Extensions,''
  Adv.\ Theor.\ Math.\ Phys.\  {\bf 12}, no. 3, 641 (2008)
%  doi:10.4310/ATMP.2008.v12.n3.a5
  [hep-th/0611328].
  %%CITATION = doi:10.4310/ATMP.2008.v12.n3.a5;%%
  %27 citations counted in INSPIRE as of 09 Feb 2016


%\cite{CastroVillarreal:2005uu}
\bibitem{CastroVillarreal:2005uu} 
  P.~Castro-Villarreal, R.~Delgadillo-Blando and B.~Ydri,
  ``Quantum effective potential for U(1) fields on S**2(L) x S**2(L),''
  JHEP {\bf 0509}, 066 (2005)
%  doi:10.1088/1126-6708/2005/09/066
  [hep-th/0506044].
  %%CITATION = doi:10.1088/1126-6708/2005/09/066;%%
  %16 citations counted in INSPIRE as of 05 Feb 2016



%\cite{Azuma:2005pm}
\bibitem{Azuma:2005pm} 
  T.~Azuma, S.~Bal, K.~Nagao and J.~Nishimura,
  ``Perturbative versus nonperturbative dynamics of the fuzzy S**2 x s**2,''
  JHEP {\bf 0509}, 047 (2005)
%  doi:10.1088/1126-6708/2005/09/047
  [hep-th/0506205].
  %%CITATION = doi:10.1088/1126-6708/2005/09/047;%%
  %24 citations counted in INSPIRE as of 05 Feb 2016


%\cite{Behr:2005wp}
\bibitem{Behr:2005wp} 
  W.~Behr, F.~Meyer and H.~Steinacker,
  ``Gauge theory on fuzzy S**2 x S**2 and regularization on noncommutative R**4,''
  JHEP {\bf 0507}, 040 (2005)
%  doi:10.1088/1126-6708/2005/07/040
  [hep-th/0503041].
  %%CITATION = doi:10.1088/1126-6708/2005/07/040;%%
  %33 citations counted in INSPIRE as of 05 Feb 2016



%\cite{Imai:2003ja}
\bibitem{Imai:2003ja} 
  T.~Imai and Y.~Takayama,
  ``Stability of fuzzy S**2 x S**2 geometry in IIB matrix model,''
  Nucl.\ Phys.\ B {\bf 686}, 248 (2004)
%  doi:10.1016/j.nuclphysb.2004.03.008
  [hep-th/0312241].
  %%CITATION = doi:10.1016/j.nuclphysb.2004.03.008;%%
  %31 citations counted in INSPIRE as of 05 Feb 2016



%\cite{Vaidya:2003ew}
\bibitem{Vaidya:2003ew} 
  S.~Vaidya and B.~Ydri,
  ``On the origin of the UV-IR mixing in noncommutative matrix geometry,''
  Nucl.\ Phys.\ B {\bf 671}, 401 (2003)
  [hep-th/0305201].
  %%CITATION = HEP-TH/0305201;%%
  %30 citations counted in INSPIRE as of 30 Sep 2015

%\cite{Kaneko:2005pw}
\bibitem{Kaneko:2005pw} 
  H.~Kaneko, Y.~Kitazawa and D.~Tomino,
  ``Stability of fuzzy S**2 x S**2 x S**2 in IIB type matrix models,''
  Nucl.\ Phys.\ B {\bf 725}, 93 (2005)
%  doi:10.1016/j.nuclphysb.2005.07.009
  [hep-th/0506033].
  %%CITATION = doi:10.1016/j.nuclphysb.2005.07.009;%%
  %20 citations counted in INSPIRE as of 05 Feb 2016



%\cite{Alexanian:2001qj}
\bibitem{Alexanian:2001qj} 
  G.~Alexanian, A.~P.~Balachandran, G.~Immirzi and B.~Ydri,
  ``Fuzzy CP**2,''
  J.\ Geom.\ Phys.\  {\bf 42}, 28 (2002)
%  doi:10.1016/S0393-0440(01)00070-5
  [hep-th/0103023].
  %%CITATION = doi:10.1016/S0393-0440(01)00070-5;%%
  %108 citations counted in INSPIRE as of 05 Feb 2016



%\cite{Azuma:2004qe}
\bibitem{Azuma:2004qe} 
  T.~Azuma, S.~Bal, K.~Nagao and J.~Nishimura,
  ``Dynamical aspects of the fuzzy CP**2 in the large N reduced model with a cubic term,''
  JHEP {\bf 0605}, 061 (2006)
%  doi:10.1088/1126-6708/2006/05/061
  [hep-th/0405277].
  %%CITATION = doi:10.1088/1126-6708/2006/05/061;%%
  %46 citations counted in INSPIRE as of 05 Feb 2016

%\cite{Grosse:2004wm}
\bibitem{Grosse:2004wm} 
  H.~Grosse and H.~Steinacker,
  ``Finite gauge theory on fuzzy CP**2,''
  Nucl.\ Phys.\ B {\bf 707}, 145 (2005)
%  doi:10.1016/j.nuclphysb.2004.11.058
  [hep-th/0407089].
  %%CITATION = doi:10.1016/j.nuclphysb.2004.11.058;%%
  %63 citations counted in INSPIRE as of 05 Feb 2016





%\cite{Dou:2007in}
\bibitem{Dou:2007in} 
  D.~Dou and B.~Ydri,
  ``Topology change from quantum instability of gauge theory on fuzzy CP**2,''
  Nucl.\ Phys.\ B {\bf 771}, 167 (2007)
%  doi:10.1016/j.nuclphysb.2007.02.010
  [hep-th/0701160].
  %%CITATION = doi:10.1016/j.nuclphysb.2007.02.010;%%
  %7 citations counted in INSPIRE as of 08 Feb 2016


%\cite{Grosse:1999ci}
\bibitem{Grosse:1999ci} 
  H.~Grosse and A.~Strohmaier,
  ``Towards a nonperturbative covariant regularization in 4-D quantum field theory,''
  Lett.\ Math.\ Phys.\  {\bf 48}, 163 (1999)
%  doi:10.1023/A:1007518622795
  [hep-th/9902138].
  %%CITATION = doi:10.1023/A:1007518622795;%%
  %79 citations counted in INSPIRE as of 05 Feb 2016


%\cite{Kitazawa:2002xj}
\bibitem{Kitazawa:2002xj} 
  Y.~Kitazawa,
  ``Matrix models in homogeneous spaces,''
  Nucl.\ Phys.\ B {\bf 642}, 210 (2002)
%  doi:10.1016/S0550-3213(02)00682-X
  [hep-th/0207115].
  %%CITATION = doi:10.1016/S0550-3213(02)00682-X;%%
  %57 citations counted in INSPIRE as of 05 Feb 2016



%\cite{Imai:2003jb}
\bibitem{Imai:2003jb} 
  T.~Imai, Y.~Kitazawa, Y.~Takayama and D.~Tomino,
  ``Effective actions of matrix models on homogeneous spaces,''
  Nucl.\ Phys.\ B {\bf 679}, 143 (2004)
%  doi:10.1016/j.nuclphysb.2003.11.038
  [hep-th/0307007].
  %%CITATION = doi:10.1016/j.nuclphysb.2003.11.038;%%
  %47 citations counted in INSPIRE as of 05 Feb 2016



%\cite{Dolan:2006tx}
\bibitem{Dolan:2006tx} 
  B.~P.~Dolan, I.~Huet, S.~Murray and D.~O'Connor,
  ``Noncommutative vector bundles over fuzzy CP**N and their covariant derivatives,''
  JHEP {\bf 0707}, 007 (2007)
%  doi:10.1088/1126-6708/2007/07/007
  [hep-th/0611209].
  %%CITATION = doi:10.1088/1126-6708/2007/07/007;%%
  %32 citations counted in INSPIRE as of 05 Feb 2016


%\cite{Balachandran:2001dd}
\bibitem{Balachandran:2001dd} 
  A.~P.~Balachandran, B.~P.~Dolan, J.~H.~Lee, X.~Martin and D.~O'Connor,
  ``Fuzzy complex projective spaces and their star products,''
  J.\ Geom.\ Phys.\  {\bf 43}, 184 (2002)
%  doi:10.1016/S0393-0440(02)00020-7
  [hep-th/0107099].
  %%CITATION = doi:10.1016/S0393-0440(02)00020-7;%%
  %116 citations counted in INSPIRE as of 16 Aug 2016






%\cite{Karabali:2002im}
\bibitem{Karabali:2002im} 
  D.~Karabali and V.~P.~Nair,
  ``Quantum Hall effect in higher dimensions,''
  Nucl.\ Phys.\ B {\bf 641}, 533 (2002)
%  doi:10.1016/S0550-3213(02)00634-X
  [hep-th/0203264].
  %%CITATION = doi:10.1016/S0550-3213(02)00634-X;%%
  %115 citations counted in INSPIRE as of 09 Feb 2016

%\cite{Karabali:2003bt}
\bibitem{Karabali:2003bt} 
  D.~Karabali and V.~P.~Nair,
  ``The effective action for edge states in higher dimensional quantum Hall systems,''
  Nucl.\ Phys.\ B {\bf 679}, 427 (2004)
%  doi:10.1016/j.nuclphysb.2003.11.020
  [hep-th/0307281].
  %%CITATION = doi:10.1016/j.nuclphysb.2003.11.020;%%
  %42 citations counted in INSPIRE as of 09 Feb 2016


%\cite{Karabali:2004km}
\bibitem{Karabali:2004km} 
  D.~Karabali and V.~P.~Nair,
  ``Edge states for quantum Hall droplets in higher dimensions and a generalized WZW model,''
  Nucl.\ Phys.\ B {\bf 697}, 513 (2004)
%  doi:10.1016/j.nuclphysb.2004.07.014
  [hep-th/0403111].
  %%CITATION = doi:10.1016/j.nuclphysb.2004.07.014;%%
  %34 citations counted in INSPIRE as of 09 Feb 2016

%\cite{Szabo:2006wx}
\bibitem{Szabo:2006wx} 
  R.~J.~Szabo,
  ``Symmetry, gravity and noncommutativity,''
  Class.\ Quant.\ Grav.\  {\bf 23}, R199 (2006)
%  doi:10.1088/0264-9381/23/22/R01
  [hep-th/0606233].
  %%CITATION = doi:10.1088/0264-9381/23/22/R01;%%
  %137 citations counted in INSPIRE as of 09 Aug 2016

%\cite{Szabo:2009tn}
\bibitem{Szabo:2009tn} 
  R.~J.~Szabo,
  ``Quantum Gravity, Field Theory and Signatures of Noncommutative Spacetime,''
  Gen.\ Rel.\ Grav.\  {\bf 42}, 1 (2010)
%  doi:10.1007/s10714-009-0897-4
  [arXiv:0906.2913 [hep-th]].
  %%CITATION = doi:10.1007/s10714-009-0897-4;%%
  %65 citations counted in INSPIRE as of 09 Aug 2016







%\cite{Dolan:2003kq}
\bibitem{Dolan:2003kq} 
  B.~P.~Dolan and D.~O'Connor,
  ``A Fuzzy three sphere and fuzzy tori,''
  JHEP {\bf 0310}, 060 (2003)
%  doi:10.1088/1126-6708/2003/10/060
  [hep-th/0306231].
  %%CITATION = doi:10.1088/1126-6708/2003/10/060;%%
  %31 citations counted in INSPIRE as of 09 Aug 2016



%\cite{Medina:2012cs}
\bibitem{Medina:2012cs} 
  J.~Medina, I.~Huet, D.~O'Connor and B.~P.~Dolan,
  ``Scalar and Spinor Field Actions on Fuzzy $S^4$: fuzzy $CP^3$ as a $S^2_F$ bundle over $S^4_F$,''
  JHEP {\bf 1208}, 070 (2012)
%  doi:10.1007/JHEP08(2012)070
  [arXiv:1208.0348 [hep-th]].
  %%CITATION = doi:10.1007/JHEP08(2012)070;%%
  %3 citations counted in INSPIRE as of 09 Aug 2016


%\cite{Medina:2002pc}
\bibitem{Medina:2002pc} 
  J.~Medina and D.~O'Connor,
  ``Scalar field theory on fuzzy S**4,''
  JHEP {\bf 0311}, 051 (2003)
%  doi:10.1088/1126-6708/2003/11/051
  [hep-th/0212170].
  %%CITATION = doi:10.1088/1126-6708/2003/11/051;%%
  %41 citations counted in INSPIRE as of 09 Aug 2016




%\cite{Steinacker:2016vgf}
\bibitem{Steinacker:2016vgf} 
  H.~C.~Steinacker,
  ``Emergent 4D gravity on covariant quantum spaces in the IKKT model,''
  arXiv:1606.00769 [hep-th]. See also \cite{Steinacker:2008ya}.
  %%CITATION = ARXIV:1606.00769;%%


%\cite{Steinacker:2015dra}
\bibitem{Steinacker:2015dra} 
  H.~C.~Steinacker,
  ``One-loop stabilization of the fuzzy four-sphere via softly broken SUSY,''
  JHEP {\bf 1512}, 115 (2015)
%  doi:10.1007/JHEP12(2015)115
  [arXiv:1510.05779 [hep-th]].
  %%CITATION = doi:10.1007/JHEP12(2015)115;%%
  %4 citations counted in INSPIRE as of 09 Aug 2016


%\cite{Steinacker:2008ya}
\bibitem{Steinacker:2008ya} 
  H.~Steinacker,
  ``Covariant Field Equations, Gauge Fields and Conservation Laws from Yang-Mills Matrix Models,''
  JHEP {\bf 0902}, 044 (2009)
%  doi:10.1088/1126-6708/2009/02/044
  [arXiv:0812.3761 [hep-th]].
  %%CITATION = doi:10.1088/1126-6708/2009/02/044;%%
  %25 citations counted in INSPIRE as of 24 Jul 2016





%\cite{Castelino:1997rv}
\bibitem{Castelino:1997rv} 
  J.~Castelino, S.~Lee and W.~Taylor,
  ``Longitudinal five-branes as four spheres in matrix theory,''
  Nucl.\ Phys.\ B {\bf 526}, 334 (1998)
%  doi:10.1016/S0550-3213(98)00291-0
  [hep-th/9712105].
  %%CITATION = doi:10.1016/S0550-3213(98)00291-0;%%
  %105 citations counted in INSPIRE as of 09 Aug 2016

%\cite{Kimura:2002nq}
\bibitem{Kimura:2002nq} 
  Y.~Kimura,
  ``Noncommutative gauge theory on fuzzy four sphere and matrix model,''
  Nucl.\ Phys.\ B {\bf 637}, 177 (2002)
%  doi:10.1016/S0550-3213(02)00469-8
  [hep-th/0204256].
  %%CITATION = doi:10.1016/S0550-3213(02)00469-8;%%
  %72 citations counted in INSPIRE as of 16 Aug 2016




%\cite{Ramgoolam:2001zx}
\bibitem{Ramgoolam:2001zx} 
  S.~Ramgoolam,
  ``On spherical harmonics for fuzzy spheres in diverse dimensions,''
  Nucl.\ Phys.\ B {\bf 610}, 461 (2001)
%  doi:10.1016/S0550-3213(01)00315-7
  [hep-th/0105006].
  %%CITATION = doi:10.1016/S0550-3213(01)00315-7;%%
  %123 citations counted in INSPIRE as of 09 Aug 2016

%\cite{Abe:2004sa}
\bibitem{Abe:2004sa} 
  Y.~Abe,
  ``Construction of fuzzy S**4,''
  Phys.\ Rev.\ D {\bf 70}, 126004 (2004)
%  doi:10.1103/PhysRevD.70.126004
  [hep-th/0406135].
  %%CITATION = doi:10.1103/PhysRevD.70.126004;%%
  %17 citations counted in INSPIRE as of 09 Aug 2016

%\cite{Valtancoli:2002sm}
\bibitem{Valtancoli:2002sm} 
  P.~Valtancoli,
  ``Projective modules over the fuzzy four sphere,''
  Mod.\ Phys.\ Lett.\ A {\bf 17}, 2189 (2002)
%  doi:10.1142/S021773230200868X
  [hep-th/0210166].
  %%CITATION = doi:10.1142/S021773230200868X;%%
  %9 citations counted in INSPIRE as of 14 Aug 2016






%\cite{CarowWatamura:2004ct}
\bibitem{CarowWatamura:2004ct} 
  U.~Carow-Watamura, H.~Steinacker and S.~Watamura,
  ``Monopole bundles over fuzzy complex projective spaces,''
  J.\ Geom.\ Phys.\  {\bf 54}, 373 (2005)
%  doi:10.1016/j.geomphys.2004.11.001
  [hep-th/0404130].
  %%CITATION = doi:10.1016/j.geomphys.2004.11.001;%%
  %39 citations counted in INSPIRE as of 16 Aug 2016


%\cite{Fulton}
\bibitem{Fulton} 
  W.~Fulton and J.~Harris,
  ``Representation Theory: A First Course,''
  Graduate Texts in Mathematics, Vol.129, Springer-Verlag, New York, 1991.

%\cite{badis}
\bibitem{badis} 
  B.~Ydri,
  ``Fuzzy ${\bf S}^4$,''
  unpublished notes $2002$.



%\cite{Ho:2001as}
\bibitem{Ho:2001as} 
  P.~M.~Ho and S.~Ramgoolam,
  ``Higher dimensional geometries from matrix brane constructions,''
  Nucl.\ Phys.\ B {\bf 627}, 266 (2002)
%  doi:10.1016/S0550-3213(02)00072-X
  [hep-th/0111278].
  %%CITATION = doi:10.1016/S0550-3213(02)00072-X;%%
  %56 citations counted in INSPIRE as of 18 Aug 2016

%\cite{Karczmarek:2015gda}
\bibitem{Karczmarek:2015gda} 
  J.~L.~Karczmarek and K.~H.~C.~Yeh,
  ``Noncommutative spaces and matrix embeddings on flat $\mathbb{R}^{2n+ 1}$,''
  JHEP {\bf 1511}, 146 (2015)
%  doi:10.1007/JHEP11(2015)146
  [arXiv:1506.07188 [hep-th]].
  %%CITATION = doi:10.1007/JHEP11(2015)146;%%
  %4 citations counted in INSPIRE as of 16 Aug 2016



%\cite{CastroVillarreal:2004vh}
\bibitem{CastroVillarreal:2004vh} 
  P.~Castro-Villarreal, R.~Delgadillo-Blando and B.~Ydri,
  ``A Gauge-invariant UV-IR mixing and the corresponding phase transition for U(1) fields on the fuzzy sphere,''
  Nucl.\ Phys.\ B {\bf 704}, 111 (2005)
%  doi:10.1016/j.nuclphysb.2004.10.032
  [hep-th/0405201].
  %%CITATION = doi:10.1016/j.nuclphysb.2004.10.032;%%
  %47 citations counted in INSPIRE as of 21 Aug 2016



%\cite{Blaschke:2011qu}
\bibitem{Blaschke:2011qu} 
  D.~N.~Blaschke and H.~Steinacker,
  ``On the 1-loop effective action for the IKKT model and non-commutative branes,''
  JHEP {\bf 1110}, 120 (2011)
%  doi:10.1007/JHEP10(2011)120
  [arXiv:1109.3097 [hep-th]].
  %%CITATION = doi:10.1007/JHEP10(2011)120;%%
  %15 citations counted in INSPIRE as of 21 Aug 2016


%\cite{Wald:1984rg}
\bibitem{Wald:1984rg} 
  R.~M.~Wald,
  ``General Relativity,''
  Chicago, Usa: Univ. Pr. ( 1984) 491p
%  doi:10.7208/chicago/9780226870373.001.0001
  %%CITATION = doi:10.7208/chicago/9780226870373.001.0001;%%
  %161 citations counted in INSPIRE as of 01 Sep 2016

%\cite{Aoyama:2002jd}
\bibitem{Aoyama:2002jd} 
  S.~Aoyama and T.~Masuda,
  ``The Fuzzy S4 by quantum deformation,''
  Nucl.\ Phys.\ B {\bf 656}, 325 (2003)
%  doi:10.1016/S0550-3213(03)00104-4
  [hep-th/0212214].
  %%CITATION = doi:10.1016/S0550-3213(03)00104-4;%%
  %5 citations counted in INSPIRE as of 05 Feb 2016




%\cite{Kimura:2003ab}
\bibitem{Kimura:2003ab} 
  Y.~Kimura,
  ``On Higher dimensional fuzzy spherical branes,''
  Nucl.\ Phys.\ B {\bf 664}, 512 (2003)
%  doi:10.1016/S0550-3213(03)00462-0
  [hep-th/0301055].
  %%CITATION = doi:10.1016/S0550-3213(03)00462-0;%%
  %33 citations counted in INSPIRE as of 05 Feb 2016



%\cite{Azuma:2004yg}
\bibitem{Azuma:2004yg} 
  T.~Azuma, S.~Bal, K.~Nagao and J.~Nishimura,
  ``Absence of a fuzzy S**4 phase in the dimensionally reduced 5-D Yang-Mills-Chern-Simons model,''
  JHEP {\bf 0407}, 066 (2004)
%  doi:10.1088/1126-6708/2004/07/066
  [hep-th/0405096].
  %%CITATION = doi:10.1088/1126-6708/2004/07/066;%%
  %18 citations counted in INSPIRE as of 05 Feb 2016







%\cite{Ramgoolam:2002wb}
\bibitem{Ramgoolam:2002wb} 
  S.~Ramgoolam,
  ``Higher dimensional geometries related to fuzzy odd dimensional spheres,''
  JHEP {\bf 0210}, 064 (2002)
%  doi:10.1088/1126-6708/2002/10/064
  [hep-th/0207111].
  %%CITATION = doi:10.1088/1126-6708/2002/10/064;%%
  %52 citations counted in INSPIRE as of 05 Feb 2016


%\cite{Kawahara:2005an}
\bibitem{Kawahara:2005an} 
  N.~Kawahara and J.~Nishimura,
  ``The Large N reduction in matrix quantum mechanics: A Bridge between BFSS and IKKT,''
  JHEP {\bf 0509}, 040 (2005)
%  doi:10.1088/1126-6708/2005/09/040
  [hep-th/0505178].
  %%CITATION = doi:10.1088/1126-6708/2005/09/040;%%
  %9 citations counted in INSPIRE as of 21 Jun 2016
















%\cite{Anagnostopoulos:2005cy}
\bibitem{Anagnostopoulos:2005cy} 
  K.~N.~Anagnostopoulos, T.~Azuma, K.~Nagao and J.~Nishimura,
  ``Impact of supersymmetry on the nonperturbative dynamics of fuzzy spheres,''
  JHEP {\bf 0509}, 046 (2005)
%  doi:10.1088/1126-6708/2005/09/046
  [hep-th/0506062].
  %%CITATION = doi:10.1088/1126-6708/2005/09/046;%%
  %22 citations counted in INSPIRE as of 05 Feb 2016







%\cite{Valtancoli:2002rx}
\bibitem{Valtancoli:2002rx} 
  P.~Valtancoli,
  ``Stability of the fuzzy sphere solution from matrix model,''
  Int.\ J.\ Mod.\ Phys.\ A {\bf 18}, 967 (2003)
%  doi:10.1142/S0217751X03014058
  [hep-th/0206075].
  %%CITATION = doi:10.1142/S0217751X03014058;%%
  %26 citations counted in INSPIRE as of 05 Feb 2016


%\cite{Azuma:2004ie}
\bibitem{Azuma:2004ie} 
  T.~Azuma, K.~Nagao and J.~Nishimura,
  ``Perturbative dynamics of fuzzy spheres at large N,''
  JHEP {\bf 0506}, 081 (2005)
%  doi:10.1088/1126-6708/2005/06/081
  [hep-th/0410263].
  %%CITATION = doi:10.1088/1126-6708/2005/06/081;%%
  %32 citations counted in INSPIRE as of 05 Feb 2016

%\cite{Azuma:2005bj}
\bibitem{Azuma:2005bj} 
  T.~Azuma, S.~Bal and J.~Nishimura,
  ``Dynamical generation of gauge groups in the massive Yang-Mills-Chern-Simons matrix model,''
  Phys.\ Rev.\ D {\bf 72}, 066005 (2005)
%  doi:10.1103/PhysRevD.72.066005
  [hep-th/0504217].
  %%CITATION = doi:10.1103/PhysRevD.72.066005;%%
  %29 citations counted in INSPIRE as of 05 Feb 2016


%\cite{Azuma:2004zq}
\bibitem{Azuma:2004zq} 
  T.~Azuma, S.~Bal, K.~Nagao and J.~Nishimura,
  ``Nonperturbative studies of fuzzy spheres in a matrix model with the Chern-Simons term,''
  JHEP {\bf 0405}, 005 (2004)
%  doi:10.1088/1126-6708/2004/05/005
  [hep-th/0401038].
  %%CITATION = doi:10.1088/1126-6708/2004/05/005;%%
  %81 citations counted in INSPIRE as of 05 Feb 2016


%\cite{Ishii:2008tm}
\bibitem{Ishii:2008tm} 
  T.~Ishii, G.~Ishiki, S.~Shimasaki and A.~Tsuchiya,
  ``Fiber Bundles and Matrix Models,''
  Phys.\ Rev.\ D {\bf 77}, 126015 (2008)
  [arXiv:0802.2782 [hep-th]].
  %%CITATION = ARXIV:0802.2782;%%
  %29 citations counted in INSPIRE as of 03 Nov 2013


%\cite{Ishiki:2009vr}
\bibitem{Ishiki:2009vr} 
  G.~Ishiki, S.~Shimasaki and A.~Tsuchiya,
  ``Large N reduction for Chern-Simons theory on S**3,''
  Phys.\ Rev.\ D {\bf 80}, 086004 (2009)
  [arXiv:0908.1711 [hep-th]].
  %%CITATION = ARXIV:0908.1711;%%
  %17 citations counted in INSPIRE as of 04 Nov 2013

%\cite{Ishiki:2008vf}
\bibitem{Ishiki:2008vf} 
  G.~Ishiki, K.~Ohta, S.~Shimasaki and A.~Tsuchiya,
  ``Two-Dimensional Gauge Theory and Matrix Model,''
  Phys.\ Lett.\ B {\bf 672}, 289 (2009)
  [arXiv:0811.3569 [hep-th]].
  %%CITATION = ARXIV:0811.3569;%%
  %9 citations counted in INSPIRE as of 04 Nov 2013


%\cite{Karabali:2006eg}
\bibitem{Karabali:2006eg} 
  D.~Karabali and V.~P.~Nair,
  ``Quantum Hall effect in higher dimensions, matrix models and fuzzy geometry,''
  J.\ Phys.\ A {\bf 39}, 12735 (2006)
%  doi:10.1088/0305-4470/39/41/S05
  [hep-th/0606161].
  %%CITATION = doi:10.1088/0305-4470/39/41/S05;%%
  %36 citations counted in INSPIRE as of 09 Aug 2016


%\cite{Ho:2000br}
\bibitem{Ho:2000br} 
  P.~M.~Ho and M.~Li,
  ``Fuzzy spheres in AdS / CFT correspondence and holography from noncommutativity,''
  Nucl.\ Phys.\ B {\bf 596}, 259 (2001)
%  doi:10.1016/S0550-3213(00)00594-0
  [hep-th/0004072].
  %%CITATION = doi:10.1016/S0550-3213(00)00594-0;%%
  %53 citations counted in INSPIRE as of 09 Aug 2016


%\cite{Hanada:2005vr}
\bibitem{Hanada:2005vr} 
  M.~Hanada, H.~Kawai and Y.~Kimura,
  ``Describing curved spaces by matrices,''
  Prog.\ Theor.\ Phys.\  {\bf 114}, 1295 (2006)
%  doi:10.1143/PTP.114.1295
  [hep-th/0508211].
  %%CITATION = doi:10.1143/PTP.114.1295;%%
  %53 citations counted in INSPIRE as of 09 Aug 2016

%\cite{DiFrancesco:1993cyw}
\bibitem{DiFrancesco:1993cyw} 
  P.~Di Francesco, P.~H.~Ginsparg and J.~Zinn-Justin,
  ``2-D Gravity and random matrices,''
  Phys.\ Rept.\  {\bf 254}, 1 (1995)
%  doi:10.1016/0370-1573(94)00084-G
  [hep-th/9306153].
  %%CITATION = doi:10.1016/0370-1573(94)00084-G;%%
  %644 citations counted in INSPIRE as of 19 Jul 2017

%\cite{DelgadilloBlando:2008vi}
\bibitem{DelgadilloBlando:2008vi} 
  R.~Delgadillo-Blando, D.~O'Connor and B.~Ydri,
  ``Matrix Models, Gauge Theory and Emergent Geometry,''
  JHEP {\bf 0905}, 049 (2009)
  doi:10.1088/1126-6708/2009/05/049
  [arXiv:0806.0558 [hep-th]].
  %%CITATION = doi:10.1088/1126-6708/2009/05/049;%%
  %25 citations counted in INSPIRE as of 19 Jul 2017


%\cite{DelgadilloBlando:2007vx}
\bibitem{DelgadilloBlando:2007vx} 
  R.~Delgadillo-Blando, D.~O'Connor and B.~Ydri,
  %``Geometry in Transition: A Model of Emergent Geometry,''
  Phys.\ Rev.\ Lett.\  {\bf 100}, 201601 (2008)
%  doi:10.1103/PhysRevLett.100.201601
  [arXiv:0712.3011 [hep-th]].
  %%CITATION = doi:10.1103/PhysRevLett.100.201601;%%
  %37 citations counted in INSPIRE as of 19 Jul 2017

%\cite{OConnor:2006iny}
\bibitem{OConnor:2006iny} 
  D.~O'Connor and B.~Ydri,
  ``Monte Carlo Simulation of a NC Gauge Theory on The Fuzzy Sphere,''
  JHEP {\bf 0611}, 016 (2006)
%  doi:10.1088/1126-6708/2006/11/016
  [hep-lat/0606013].
  %%CITATION = doi:10.1088/1126-6708/2006/11/016;%%
  %42 citations counted in INSPIRE as of 19 Jul 2017



%\cite{Ydri:2016daf}
\bibitem{Ydri:2016daf} 
  B.~Ydri,
  ``The multitrace matrix model: An alternative to Connes NCG and IKKT model in 2 dimensions,''
  Phys.\ Lett.\ B {\bf 763}, 161 (2016)
 % doi:10.1016/j.physletb.2016.10.043
  [arXiv:1608.02758 [hep-th]].
  %%CITATION = doi:10.1016/j.physletb.2016.10.043;%%
  %1 citations counted in INSPIRE as of 19 Jul 2017



%\cite{Ydri:2016osu}
\bibitem{Ydri:2016osu} 
  B.~Ydri, A.~Rouag and K.~Ramda,
  ``Emergent fuzzy geometry and fuzzy physics in four dimensions,''
  Nucl.\ Phys.\ B {\bf 916}, 567 (2017)
%  doi:10.1016/j.nuclphysb.2017.01.023
  [arXiv:1607.08296 [hep-th]].
  %%CITATION = doi:10.1016/j.nuclphysb.2017.01.023;%%
  %2 citations counted in INSPIRE as of 19 Jul 2017

%\cite{Ydri:2016kua}
\bibitem{Ydri:2016kua} 
  B.~Ydri, R.~Khaled and R.~Ahlam,
  ``Geometry in transition in four dimensions: A model of emergent geometry in the early universe,''
  Phys.\ Rev.\ D {\bf 94}, no. 8, 085020 (2016)
%  doi:10.1103/PhysRevD.94.085020
  [arXiv:1607.06761 [hep-th]].
  %%CITATION = doi:10.1103/PhysRevD.94.085020;%%
  %2 citations counted in INSPIRE as of 19 Jul 2017


%\cite{Ydri:2015vba}
\bibitem{Ydri:2015vba} 
  B.~Ydri, K.~Ramda and A.~Rouag,
  ``Phase diagrams of the multitrace quartic matrix models of noncommutative $\Phi^4$ theory,''
  Phys.\ Rev.\ D {\bf 93}, no. 6, 065056 (2016)
%  doi:10.1103/PhysRevD.93.065056
  [arXiv:1509.03726 [hep-th]].
  %%CITATION = doi:10.1103/PhysRevD.93.065056;%%
  %6 citations counted in INSPIRE as of 19 Jul 2017


%\cite{Ydri:2015zsa}
\bibitem{Ydri:2015zsa} 
  B.~Ydri, A.~Rouag and K.~Ramda,
  ``Emergent geometry from random multitrace matrix models,''
  Phys.\ Rev.\ D {\bf 93}, no. 6, 065055 (2016)
%  doi:10.1103/PhysRevD.93.065055
  [arXiv:1509.03572 [hep-th]].
  %%CITATION = doi:10.1103/PhysRevD.93.065055;%%
  %5 citations counted in INSPIRE as of 19 Jul 2017



%\cite{Ydri:2017riq}
\bibitem{Ydri:2017riq} 
  B.~Ydri, C.~Soudani and A.~Rouag,
  `Quantum Gravity as a Multitrace Matrix Model,''
  arXiv:1706.07724 [hep-th].
  %%CITATION = ARXIV:1706.07724;%%


%\cite{Kazakov:1985ds}
\bibitem{Kazakov:1985ds} 
  V.~A.~Kazakov,
  ``Bilocal Regularization of Models of Random Surfaces,''
  Phys.\ Lett.\  {\bf 150B}, 282 (1985).
%  doi:10.1016/0370-2693(85)91011-1
  %%CITATION = doi:10.1016/0370-2693(85)91011-1;%%
  %349 citations counted in INSPIRE as of 01 Aug 2017

%\cite{David:1984tx}
\bibitem{David:1984tx} 
  F.~David,
  ``Planar Diagrams, Two-Dimensional Lattice Gravity and Surface Models,''
  Nucl.\ Phys.\ B {\bf 257}, 45 (1985).
%  doi:10.1016/0550-3213(85)90335-9
  %%CITATION = doi:10.1016/0550-3213(85)90335-9;%%
  %565 citations counted in INSPIRE as of 01 Aug 2017


%\cite{Ambjorn:1985az}
\bibitem{Ambjorn:1985az} 
  J.~Ambjorn, B.~Durhuus and J.~Frohlich,
  ``Diseases of Triangulated Random Surface Models, and Possible Cures,''
  Nucl.\ Phys.\ B {\bf 257}, 433 (1985).
%  doi:10.1016/0550-3213(85)90356-6
  %%CITATION = doi:10.1016/0550-3213(85)90356-6;%%
  %613 citations counted in INSPIRE as of 01 Aug 2017

%\cite{Koplik:1977pf}
\bibitem{Koplik:1977pf} 
  J.~Koplik, A.~Neveu and S.~Nussinov,
  ``Some Aspects of the Planar Perturbation Series,''
  Nucl.\ Phys.\ B {\bf 123}, 109 (1977).
%  doi:10.1016/0550-3213(77)90344-3
  %%CITATION = doi:10.1016/0550-3213(77)90344-3;%%
  %109 citations counted in INSPIRE as of 01 Aug 2017


%\cite{Polyakov:1981rd}
\bibitem{Polyakov:1981rd} 
  A.~M.~Polyakov,
  ``Quantum Geometry of Bosonic Strings,''
  Phys.\ Lett.\  {\bf 103B}, 207 (1981).
%  doi:10.1016/0370-2693(81)90743-7
  %%CITATION = doi:10.1016/0370-2693(81)90743-7;%%
  %2550 citations counted in INSPIRE as of 01 Aug 2017

%\cite{DHoker:1988pdl}
\bibitem{DHoker:1988pdl} 
  E.~D'Hoker and D.~H.~Phong,
  ``The Geometry of String Perturbation Theory,''
  Rev.\ Mod.\ Phys.\  {\bf 60}, 917 (1988).
%  doi:10.1103/RevModPhys.60.917
  %%CITATION = doi:10.1103/RevModPhys.60.917;%%
  %430 citations counted in INSPIRE as of 01 Aug 2017


%\cite{Brezin:1990rb}
\bibitem{Brezin:1990rb} 
  E.~Brezin and V.~A.~Kazakov,
  ``Exactly Solvable Field Theories of Closed Strings,''
  Phys.\ Lett.\ B {\bf 236}, 144 (1990).
%  doi:10.1016/0370-2693(90)90818-Q
  %%CITATION = doi:10.1016/0370-2693(90)90818-Q;%%
  %1116 citations counted in INSPIRE as of 01 Aug 2017

%\cite{Douglas:1989ve}
\bibitem{Douglas:1989ve} 
  M.~R.~Douglas and S.~H.~Shenker,
  ``Strings in Less Than One-Dimension,''
  Nucl.\ Phys.\ B {\bf 335}, 635 (1990).
%  doi:10.1016/0550-3213(90)90522-F
  %%CITATION = doi:10.1016/0550-3213(90)90522-F;%%
  %1025 citations counted in INSPIRE as of 01 Aug 2017


%\cite{Gross:1989vs}
\bibitem{Gross:1989vs} 
  D.~J.~Gross and A.~A.~Migdal,
  ``Nonperturbative Two-Dimensional Quantum Gravity,''
  Phys.\ Rev.\ Lett.\  {\bf 64}, 127 (1990).
%  doi:10.1103/PhysRevLett.64.127
  %%CITATION = doi:10.1103/PhysRevLett.64.127;%%
  %951 citations counted in INSPIRE as of 01 Aug 2017

%\cite{Ginsparg:1993is}
\bibitem{Ginsparg:1993is} 
  P.~H.~Ginsparg and G.~W.~Moore,
  ``Lectures on 2-D gravity and 2-D string theory,''
  Yale Univ. New Haven - YCTP-P23-92 (92,rec.Apr.93) 197 p. Los Alamos Nat. Lab. - LA-UR-92-3479 (92,rec.Apr.93) 197 p. e: LANL hep-th/9304011
  [hep-th/9304011].
  %%CITATION = HEP-TH/9304011;%%
  %456 citations counted in INSPIRE as of 02 Aug 2017


%\cite{Ferretti:1995zn}
\bibitem{Ferretti:1995zn} 
  G.~Ferretti,
  ``On the large N limit of 3-d and 4-d Hermitian matrix models,''
  Nucl.\ Phys.\ B {\bf 450}, 713 (1995)
%  doi:10.1016/0550-3213(95)00382-3
  [hep-th/9504013].
  %%CITATION = doi:10.1016/0550-3213(95)00382-3;%%
  %16 citations counted in INSPIRE as of 02 Aug 2017


%\cite{Nishigaki:1996np}
\bibitem{Nishigaki:1996np} 
  S.~Nishigaki,
  ``Wilsonian approximated renormalization group for matrix and vector models in 2 < d < 4,''
  Phys.\ Lett.\ B {\bf 376}, 73 (1996)
%  doi:10.1016/0370-2693(96)00277-8
  [hep-th/9601043].
  %%CITATION = doi:10.1016/0370-2693(96)00277-8;%%
  %16 citations counted in INSPIRE as of 02 Aug 2017


%\cite{Ydri:2013zya}
\bibitem{Ydri:2013zya} 
  B.~Ydri and R.~Ahmim,
 ``Matrix model fixed point of noncommutative $ϕ^4$ theory,''
  Phys.\ Rev.\ D {\bf 88}, no. 10, 106001 (2013)
%  doi:10.1103/PhysRevD.88.106001
  [arXiv:1304.7303 [hep-th]].
  %%CITATION = doi:10.1103/PhysRevD.88.106001;%%
  %2 citations counted in INSPIRE as of 02 Aug 2017



\end{thebibliography}
\end{document}